\pdfoutput=1
\documentclass[11pt,twoside,a4paper,cmspaper,final,collab]{cms-tdr}
\def\svnVersion{a749408}\def\svnDate{2024/12/09}\def\cmsCernNoTag{CERN-EP-2024-057}\def\cmsCernDate{\today}\def\cmsMessage{Published in Physics Reports as \href{http://dx.doi.org/10.1016/j.physrep.2024.11.007}{\doi{10.1016/j.physrep.2024.11.007}.}}
\begin{document}\cmsNoteHeader{HIN-23-011}

\newlength\cmsTabSkip\setlength{\cmsTabSkip}{1ex}

\newcommand{\mumu}      {\ensuremath{\PGmp\PGmm}\xspace}
\newcommand{\gaga}      {\ensuremath{\PGg\PGg}\xspace}
\newcommand{\mleplep}   {\ensuremath{m_{\Plp\Plm}}\xspace}
\newcommand{\upsOne}    {\ensuremath{\PGUP{1S}}\xspace}
\newcommand{\chicOne}   {\ensuremath{\chi_{{\PQc} 1}}\xspace}
\newcommand{\chicTwo}   {\ensuremath{\chi_{{\PQc} 2}}\xspace}
\newcommand{\chibnP}    {\ensuremath{\chi_{\mathrm{\PQb J}}{(\mathrm{nP})}}\xspace}
\newcommand{\tautau}    {\ensuremath{\PGtp\PGtm}\xspace}
\newcommand{\threeprong}{\ensuremath{\PGt_{\text{3prong}}}\xspace}
\newcommand{\muonic}    {\ensuremath{\PGt_{\PGm}}\xspace}

\newcommand{\gammagammatautau}{\ensuremath{\gaga\to\tautau}\xspace}
\newcommand{\gammagammamumu}{\ensuremath{\gaga\to\mumu}\xspace}
\newcommand{\tautripi}  {\ensuremath{\PGt\to\PGppm\PGpmp\PGppm\PGnGt}\xspace}
\newcommand{\taumunu}   {\ensuremath{\PGt\to\PGm\PAGnGm\PGnGt}\xspace}
\newcommand{\xsec}      {\ensuremath{\sigma(\gammagammatautau)}\xspace}

\newcommand{\rootsNN}   {\ensuremath{\sqrt{\smash[b]{s_{_{\mathrm{NN}}}}}}\xspace}
\newcommand{\roots}     {\ensuremath{\sqrt{\smash[b]{s}}}\xspace}
\newcommand{\sqrts}     {\ensuremath{\sqrt{\smash[b]{s}}}\xspace}
\newcommand{\pPb}       {\ensuremath{\Pp\mathrm{Pb}}\xspace}
\newcommand{\Pbp}       {\ensuremath{\mathrm{Pb}\Pp}\xspace}
\newcommand{\pp}        {\ensuremath{\Pp\Pp}\xspace}
\newcommand{\ppbar}     {\ensuremath{\Pp\PAp}\xspace}
\newcommand{\NN}        {\ensuremath{\PN\PN}\xspace}
\newcommand{\AonA}      {\ensuremath{\mathrm{AA}}\xspace}
\newcommand{\pA}        {\ensuremath{\Pp\mathrm{A}}\xspace}
\newcommand{\dAu}       {\ensuremath{\text{dAu}}\xspace}
\newcommand{\XeXe}      {\ensuremath{\mathrm{Xe}\mathrm{Xe}}\xspace}
\newcommand{\PbPb}      {\ensuremath{\mathrm{Pb}\mathrm{Pb}}\xspace}
\newcommand{\PhotonP}   {\ensuremath{{\PGg\Pp}}\xspace}
\newcommand{\PhotonPb}  {\ensuremath{{\PGg}\text{Pb}}\xspace}
\newcommand{\PhotonA}   {\ensuremath{\PGg\mathrm{A}}\xspace}
\newcommand{\LbL}       {LbL\xspace}

\newcommand{\SC}        {\ensuremath{SC(n,m)}}
\newcommand{\SCnm}[2]   {\ensuremath{SC(#1,#2)}}
\newcommand{\ket}       {\ensuremath{KE_{\mathrm{T}}}\xspace}
\newcommand{\ra}        {\rangle}
\newcommand{\lan}       {\langle}
\newcommand{\mean}[1]   {\lan #1 \ra}
\newcommand{\ETfour}    {\ensuremath{\ET^{4<\abs{\eta}<5.2}}\xspace}
\newcommand{\KETavg}    {\ensuremath{\left<{KE_{\mathrm{T}}}\right>}\xspace}
\newcommand{\QTwo}      {\ensuremath{Q^2}\xspace}

\newcommand{\dndeta}    {\ensuremath{\rd{}N/\rd{}\eta}\xspace}
\newcommand{\dnchdeta}  {\ensuremath{\rd{}N_{\text{ch}}/\rd{}\eta}\xspace}
\newcommand{\deta}      {\ensuremath{\Delta\eta}\xspace}
\newcommand{\dphi}      {\ensuremath{\Delta\phi}\xspace}
\newcommand{\Deltar}    {\ensuremath{\Delta r}\xspace}
\newcommand{\Rhodr}     {\ensuremath{P(\Deltar)}\xspace}
\newcommand{\dphiOneTwo}{\ensuremath{\dphi_{\mathrm{1,2}}}\xspace}
\newcommand{\phii}      {\ensuremath{\phi_{i}}\xspace}
\newcommand{\ylab}      {\ensuremath{y_{\text{lab}}}\xspace}
\newcommand{\ycm}       {\ensuremath{y_{\text{cm}}}\xspace}
\newcommand{\AJ}        {\ensuremath{A_{\mathrm{J}}}\xspace}

\newcommand{\ptave}     {\ensuremath{p_\mathrm{T}^{\text{ave}}}}
\newcommand{\pti}       {\ensuremath{p_{\mathrm{T}}^{i}}\xspace}
\newcommand{\ptOne}     {\ensuremath{p_{\mathrm{T,1}}}\xspace}
\newcommand{\ptTwo}     {\ensuremath{p_{\mathrm{T,2}}}\xspace}
\newcommand{\mpt}       {\ensuremath{p_{\mathrm{T}}^{\shortparallel}\hspace{-1.02em}/\kern 0.5em}\xspace}
\newcommand{\avempt}    {\ensuremath{\langle p_{\mathrm{T}}^{\shortparallel}\hspace{-1.02em}/\kern 0.5em\rangle}\xspace}
\newcommand{\pparatrk}  {\ensuremath{p_{\shortparallel}^{\text{track}}}\xspace}
\newcommand{\pttrk}     {\ensuremath{\pt^{\text{trk}}}\xspace}
\newcommand{\ptj}       {\ensuremath{\pt^\text{jet}}\xspace}
\newcommand{\pj}        {\ensuremath{p^\text{jet}}\xspace}
\newcommand{\ptgvec}    {\ensuremath{\vec{p}_{\mathrm{T}}^{\PGg}}\xspace}
\newcommand{\pttrkvec}  {\ensuremath{\vec{p}_{\mathrm{T}}^{\text{trk}}}\xspace}
\newcommand{\pTassoc}   {\ensuremath{{p\mathstrut}^{\text{assoc}}_{\mathrm{T}}}\xspace}
\newcommand{\pTtrig}    {\ensuremath{{p\mathstrut}^{\text{\mathstrut{trig}}}_{\mathrm{T}}}\xspace}
\newcommand{\ptg}       {\ensuremath{\pt,\cPgg}\xspace}
\newcommand{\etg}       {\ensuremath{\ET^{\PGg}}\xspace}

\newcommand{\vTwo}      {\ensuremath{v_2}\xspace}
\newcommand{\vThree}    {\ensuremath{v_3}\xspace}
\newcommand{\vFour}     {\ensuremath{v_4}\xspace}
\newcommand{\vSix}      {\ensuremath{v_6}\xspace}
\newcommand{\vN}        {\ensuremath{v_{n}}\xspace}
\newcommand{\VnDelta}   {\ensuremath{V_{n\Delta}}\xspace}
\newcommand{\VoneDelta} {\ensuremath{{V_{1\Delta}}}\xspace}
\newcommand{\VtwoDelta} {\ensuremath{{V_{2\Delta}}}\xspace}
\newcommand{\VthreeDelta}{\ensuremath{{V_{3\Delta}}}\xspace}
\newcommand{\vtwo}[1]   {\ensuremath{v_2\{#1\}}}
\newcommand{\vtworatio}[2]{\ensuremath{v_2\{#1\}/v_2\{#2\}}}

\newcommand{\npart}     {\ensuremath{N_\text{part}}\xspace}
\newcommand{\npartave}  {\ensuremath{\langle N_\text{part}\rangle}\xspace}
\newcommand{\ncoll}     {\ensuremath{N_\text{coll}}\xspace}
\newcommand{\noff}      {\ensuremath{N_\text{trk}^\text{offline}}\xspace}
\newcommand{\ntrcorr}   {\ensuremath{N_\text{trk}^\text{corrected}}\xspace}
\newcommand{\Ntracks}   {\ensuremath{N_\text{tracks}}\xspace}
\newcommand{\Ntrackavg} {\ensuremath{\langle N_{\text{trk}}^{\text{offline}} \rangle}\xspace}
\newcommand{\Ntrackcorravg} {\ensuremath{\langle N_{\text{trk}}^{\text{corrected}} \rangle}\xspace}
\newcommand{\Ntrig}     {\ensuremath{N_\text{trig}}\xspace}
\newcommand{\Npair}     {\ensuremath{N^\text{pair}}\xspace}
\newcommand{\nq}        {\ensuremath{n_\cPq}\xspace}

\newcommand{\xjave}     {\ensuremath{\langle x_{\mathrm{J}} \rangle}\xspace}
\newcommand{\xj}        {\ensuremath{x_{\mathrm{J}}}\xspace}
\newcommand{\XJ}        {\ensuremath{x_{\mathrm{j}}}\xspace}
\newcommand{\xjg}       {\ensuremath{\xj\cPgg}\xspace}
\newcommand{\atau}      {\ensuremath{a_{\PGt}}\xspace}
\newcommand{\aphi}      {\ensuremath{\mathrm{A}_{\phi}}\xspace}

\newcommand{\NAA}       {\ensuremath{N_{\mathrm{AA}}}\xspace}
\newcommand{\TAA}       {\ensuremath{T_{\mathrm{AA}}}\xspace}
\newcommand{\RAA}       {\ensuremath{R_{\mathrm{AA}}}\xspace}
\newcommand{\RAAStar}   {\ensuremath{R_{\mathrm{AA}}^{*}}\xspace}
\newcommand{\RpA}       {\ensuremath{R_{\Pp\mathrm{A}}}\xspace}
\newcommand{\RpPb}      {\ensuremath{R_{\Pp\Pb}}\xspace}

\newcommand{\xijet}     {\ensuremath{\xi^{\text{jet}}}\xspace}
\newcommand{\xigamma}   {\ensuremath{\xi^{\PGg}_{\mathrm{T}}}\xspace}
\newcommand{\LcDratio}  {\ensuremath{\PcgLp/ \PDz}\xspace}
\newcommand{\Lcdecay}   {\ensuremath{\PcgLp\to\Pp\PKm\Pgpp}\xspace}
\newcommand{\xthree}    {\HepParticleResonance{X}{3872}{}{}\Xspace}
\newcommand{\BsBpratio} {\ensuremath{\PBzs / \PBp }\xspace}
\newcommand{\JEWEL}     {{\textsc{jewel}}\xspace}
\newcommand{\HYBRID}    {{\textsc{hybrid}}\xspace}
\newcommand{\MARTINI}   {{\textsc{martini}}\xspace}
\newcommand{\LBT}       {{\textsc{lbt}}\xspace}
\newcommand{\CCNU}      {{\textsc{ccnu}}\xspace}
\newcommand{\EPOS}      {{\textsc{epos}}\xspace}
\newcommand{\PYTHIAHYDJET}{{\textsc{pythia}+\textsc{hydjet}}\xspace}
\newcommand{\Starlight} {\textsc{STARlight}\xspace}

\newcommand{\rTwo}      {\ensuremath{r_2}\xspace}
\newcommand{\rThree}    {\ensuremath{r_3}\xspace}
\newcommand{\rN}        {\ensuremath{r_{n}}\xspace}
\newcommand{\FEtaTwo}   {\ensuremath{{F}^{{\eta}}_{\mathrm{2}}}\xspace}
\newcommand{\FEtaN}     {\ensuremath{{F}^{{\eta}}_{\PN}}\xspace}
\newcommand{\chiFour}   {\ensuremath{\chi_{422}}\xspace}
\newcommand{\chiFive}   {\ensuremath{\chi_{523}}\xspace}
\newcommand{\chiSTTT}   {\ensuremath{\chi_{6222}}\xspace}
\newcommand{\chiSTT}    {\ensuremath{\chi_{633}}\xspace}
\newcommand{\chiSeven}  {\ensuremath{\chi_{7223}}\xspace}
\newcommand{\Rinvt}     {\ensuremath{R_\text{inv}}\xspace}
\newcommand{\qinv}      {\ensuremath{q_\text{inv}}\xspace}
\newcommand{\RS}        {\ensuremath{R_\text{S}}\xspace}
\newcommand{\RL}        {\ensuremath{R_\text{L}}\xspace}
\newcommand{\RO}        {\ensuremath{R_\text{O}}\xspace}
\newcommand{\muFive}    {\ensuremath{\mu_5}\xspace}
\newcommand{\PsiRP}     {\ensuremath{\Psi_\mathrm{RP}}\xspace}
\newcommand{\PsiPP}     {\ensuremath{\Psi_\mathrm{PP}}\xspace}
\newcommand{\PsiTwo}    {\ensuremath{\Psi_2}\xspace}
\newcommand{\PsiThree}  {\ensuremath{\Psi_3}\xspace}
\newcommand{\rTwoNorm}  {\ensuremath{{r}^{\text{norm}}_2}\xspace}
\newcommand{\rThrNorm}  {\ensuremath{{r}^{\text{norm}}_3}\xspace}
\newcommand{\ATrueCh}   {\ensuremath{{A}^{\text{true}}_{\text{ch}}}\xspace}
\newcommand{\vTwoSub}   {\ensuremath{v_2^{\text{sub}}}\xspace}

\newcommand{\FigureFrom}[1]{\text{(Figure adapted from Ref.~\cite{#1}.)}}
\newcommand{\FigureCompiled}[1]{\text{(Figure adapted from Refs.~\cite{#1}.)}}
\newcommand{\FigureCompiledSingular}[1]{\text{(Figure adapted from Ref.~\cite{#1}.)}}
\newcommand{\FiguresFrom}[1]{\text{(Figures adapted from Ref.~\cite{#1}.)}}
\newcommand{\FiguresCompiled}[1]{\text{(Figures adapted from Refs.~\cite{#1}.)}}
\newcommand{\FiguresAdaptedFrom}[1]{\text{(Figures adapted from Ref.~\cite{#1}.)}}
\newcommand{\FigureAdaptedFrom}[1]{\text{(Figure adapted from Ref.~\cite{#1}.)}}

\cmsNoteHeader{HIN-23-011}
\title{Overview of high-density QCD studies with the CMS experiment at the LHC}

\date{\today}

\abstract{We review key measurements performed by CMS in the context of its heavy ion physics program, using event samples collected in 2010--2018 with several collision systems and energies. These studies provide detailed macroscopic and microscopic probes of the quark-gluon plasma (QGP) created at the LHC energies, a medium characterized by the highest temperature and smallest baryon-chemical potential ever reached in the laboratory. Numerous observables related to high-density quantum chromodynamics (QCD) were studied, leading to some of the most impactful and qualitatively novel results in the 40-year history of the field. Using a dedicated high-multiplicity trigger in the first pp run, CMS discovered that small collision systems can exhibit signs of collectivity, a generic phenomenon with significant implications and presently understood to affect essentially all soft physics processes. This observation opened new paths to understand how fluidity and plasma properties emerge in QCD matter as a function of system size. Measurements of jet quenching  have reached a completely new level of detail by directly assessing, for the first time, the medium modification of parton showers, as opposed to simply observing leading hadrons or di-hadrons. The first fully reconstructed beauty hadron and heavy-flavor jet nuclear modifications were also measured. The large size of the event samples, the precision of the measurements, and the extension of the probed kinematical phase space, allowed many other hard probes of the QGP medium to be explored in detail, leading to multiple groundbreaking findings. In particular, the seminal measurements of bottomonium suppression patterns answer fundamental questions that have been actively pursued, both theoretically and experimentally, by the community since the mid-1980s. We conclude by outlining the opportunities offered by the continuation of this physics program at the LHC.}

\hypersetup{%
pdfauthor={CMS Collaboration},%
pdftitle={Overview of high-density QCD studies with the CMS experiment at the LHC},%
pdfsubject={CMS},
pdfkeywords={CMS, heavy ion physics, QGP, overview}}

\maketitle
\tableofcontents
\clearpage

\section{Introduction}
\label{sec:Introduction}

The Compact Muon Solenoid~(CMS) detector~\cite{CMS:2008xjf}, optimized for studies of high transverse-momentum~(\pt) particle production in high-luminosity proton-proton~(\pp) collisions at the CERN Large Hadron Collider~(LHC), has proven to be a versatile tool with significant potential in other areas of research. From the very beginning~\cite{CMS:1992tji}, it was realized that its muon acceptance and calorimetric $4\pi$ coverage, along with appropriate adaptations of its data acquisition system and online event filtering, could lead to significant contributions in the field of ultrarelativistic heavy ion~(HI) collisions. In particular, the CMS detector is capable of performing detailed investigations~\cite{Baur:2000wn,CMS:2007eug} of the thermodynamic and transport  properties of the quark-gluon plasma~(QGP), a deconfined state of quarks and gluons~\cite{Pasechnik:2016wkt,Busza:2018rrf} formed in such collisions. These studies are based on numerous experimental probes measured in an extended \pt range from about 0.3\GeV up to a few \TeV. 
Along with the ALICE~\cite{ALICE:2022wpn}, ATLAS~\cite{ATLAS:2008xda} and LHCb~\cite{LHCb:2008vvz} experiments, CMS results cover a very broad range of experimental studies of ultrarelativistic HI collisions at the LHC. 

The study of proton-nucleus (\pA{}) interactions, which are intermediate in terms of the system size between \pp and HI collisions, was initially considered a means to provide a baseline of ``cold nuclear matter'', a non-QGP system considered to be formed in such collisions, to better understand the observations made in HI collisions~\cite{Salgado:2011wc}. Studies of \pA and high multiplicity \pp collisions have since become a fascinating area of research in their own right.  It has become increasingly evident that the likely partonic systems produced in \pA collisions and in very high multiplicity events formed in \pp collisions offer valuable information and play a crucial role in unraveling the small-size limit of the QGP. Additionally, the capabilities of the CMS detector have allowed research into numerous non-QGP-related physics phenomena, such as photon-photon~(\gaga) and photon-nucleus ($\PGg\PN$) interactions. 

This article provides a comprehensive overview of the multiyear efforts undertaken by the CMS Collaboration in the field of HI physics, summarizing the key physics results and discoveries. It delves into the partonic structure of the nuclei and properties of the QGP, the unexpected QGP-like effects observed in small collision systems, and other important findings. The article concludes with possible future directions for the CMS experiment to address open questions in the field in a unique or complementary way. The article is organized as follows.

\begin{itemize}
    \item This Introduction presents the physics motivations behind studying HI collisions at LHC energies, and in particular how data from the CMS detector can be used to address fundamental questions related to the high-density matter that is created in these strong interactions, governed by quantum chromodynamics~(QCD).
    \item Section~\ref{sec:ExperimentalMethods} outlines the experimental challenges associated with the overall data-taking strategy and the subsequent reconstruction of HI events. It also describes the various detectors, filters (triggers), and offline software techniques used to select the events of interest and to provide data samples for physics  analysis. 
    \item Section~\ref{sec:InitialState} focuses on studies of the initial state of the collisions, in particular on constraints of nuclear parton distribution functions (nPDFs), and on searches for gluon saturation phenomena at small values of parton fractional momenta (Bjorken $x$), by means of particles produced in processes with a large squared momentum transfer \QTwo (``hard probes'') and diffractive photoproduction of vector mesons. 
    \item Section~\ref{sec:softQGP} describes the bulk observables that provide information on the geometry, entropy, energy density, and other collective properties of the medium produced in HI collisions for various collision systems and LHC energies. Searches for chiral symmetry restoration are also discussed. 
    \item Section~\ref{sec:hardQGP} presents the interesting physics in the high-\QTwo realm, in particular how hadrons, jets, and particles containing heavy quarks can act as ``tomographic'' probes of the medium formed in the HI reactions.
    \item Section~\ref{sec:SmallSystems} highlights the paradigm-shifting findings made by CMS in studies of small collision systems, such as the emergence of collectivity for unidentified, light- and heavy-flavor hadrons. 
    \item Section~\ref{sec:QEDBSM} details the results related to photon-induced processes in electromagnetic~(EM) nucleus-nucleus (\AonA) interactions.
    \item Finally, Section~\ref{sec:Summary} provides a comprehensive summary of the results obtained by CMS in the first two LHC runs and discusses an outlook for future discoveries using the CMS detector.
    \item Appendix~\ref{app:Glossary} contains a glossary defining all of the acronyms used in this review.
\end{itemize}

\subsection{Evolution of the field: From first studies to CMS}
\label{sec:IntroductionOverview}

Soon after the demonstration of asymptotic freedom as a fundamental property of QCD gauge theory, it was realized that strongly interacting matter at finite temperature and/or net-baryon density can exist in different thermodynamic phases~\cite{Collins:1974ky,Cabibbo:1975ig}. Early predictions based on finite-temperature simulations of QCD on the lattice~\cite{Engels:1980ty} indicated a transition to a new phase of QCD matter, the QGP~\cite{Shuryak:1977ut}, in which the quarks and gluons are no longer confined in hadrons and chiral symmetry, a global symmetry of QCD, is restored. More recently, lattice QCD calculations considering realistic light-quark masses have shown that chiral symmetry is restored in a crossover transition at a vanishingly small net-baryon density and at a precisely determined ``pseudocritical'' temperature (as opposed to a critical temperature in fixed order transitions) of 
$T_{\mathrm{c}}=156$--158\MeV~\cite{HotQCD:2018pds,Borsanyi:2020fev}. These same calculations estimated that the critical energy density at $T_{\mathrm{c}}$ is $\epsilon_{\mathrm{c}}=0.3$--0.4\GeV/fm$^3$. 

It had already been conjectured in Ref.~\cite{Shuryak:1977ut} that this regime of the QCD phase transition could be accessible by investigating high-energy HI collisions, where extremely hot and dense QCD matter could be produced by concentrating a substantial amount of energy in the center-of-mass overlap region of the two nuclei. For center-of-mass energies larger than 5\GeV (achieved at the AGS accelerator at BNL), the initial energy density in a large overlap area of two heavy nuclei is expected to significantly exceed 
$\epsilon_{\mathrm{c}} \approx 0.4$\GeV/fm$^3$~\cite{Gross:2022hyw}. Studying collisions of nuclei at high energies looked attractive for at least two reasons: i) the large amount of energy liberated
in the collision of highly
Lorentz-contracted nuclei is distributed over a very small volume, as defined by their transverse size, and ii) these interactions are expected to provide very high densities of soft partons. From \pA and \AonA interactions at high energies, one may therefore gain insight into many aspects of the strong interaction in the QCD framework. Despite the fact that QCD theory remains unchallenged since the discovery of asymptotic freedom~\cite{Gross:2022hyw}, a complete and deep understanding of the nature of its phase transitions is still lacking. 

Almost four decades of experimental and theoretical research have followed a scientific strategy aimed at the discovery and characterization of the QGP. Significant progress in the experimental field was made in the early period using fixed-target experiments at the BNL AGS~\cite{Schmidt:1992ge} and the CERN SPS~\cite{Heinz:2000bk}. The BNL RHIC, a dedicated HI collider that provides interactions up to a center-of-mass energy per nucleon pair~(\NN) of $\rootsNN = 200\GeV$, started operating in the year 2000. With their much higher achievable center-of-mass energies, HI colliders are better suited to provide the appropriate conditions for the study of highly excited quark-gluon matter. An integral part of the CERN LHC experimental program has been devoted to HI collisions, such as lead-lead~(\PbPb) and proton-lead (\pPb) collisions up to \rootsNN of 5.36 and 8.16\TeV, respectively. Since the inauguration of RHIC and the LHC (year 2009), these facilities and their experiments have continuously upgraded their 
capabilities, collecting a wealth of data
across a wide range of collision energies and various colliding systems~\cite{Muller:2006ee,Muller:2012zq}.

Although the experimental conditions in HI collisions are tailored to create a hot and dense system, this medium cannot be formed under controlled thermodynamical conditions (such as the formation of water vapor from liquid water), but rather follows a dynamical trajectory across the temperature and net-baryon density axes of the QCD phase diagram~\cite{Heinz:2004qz}. Contrary to what was initially believed, the medium produced in HI collisions has been shown to not be a nearly free gas of quarks and gluons~\cite{Braun-Munzinger:2007edi,Shuryak:2007qs,Busza:2018rrf,Harris:2023tti}. Instead, a ``strongly coupled'' quark-gluon plasma was discovered~\cite{BRAHMS:2004adc, PHENIX:2004vcz, PHOBOS:2004zne, STAR:2005gfr}, exhibiting properties akin to conventional (electromagnetic) plasmas~\cite{MIHALCEA20231}. These plasmas often demonstrate liquid or solid-like behavior, characterized by inter-particle potential energies exceeding the particles' kinetic energy. Within a time of the order of 1\unit{fm} (in natural units), the conditions that prevail in a HI collision lead to the formation of a strongly coupled liquid with a nonuniform energy density, which then evolves according to the laws of viscous relativistic hydrodynamics~\cite{Gale:2013da}.  Hydrodynamics converts initial spatial anisotropies into momentum space via pressure gradients~\cite{Heinz:2013th}. These anisotropies persist because of the small specific viscosity of the QGP~\cite{Bernhard:2019bmu}. The observation of a strong anisotropy known as ``elliptic flow'', in which final-state hadrons exhibit preferential emission with respect to a specific azimuthal angle, not only contributed to shedding light on the existence of the QGP but also characterized it as ``the most perfect fluid known''~\cite{Herrmann:1999wu}. A wealth of experimental data, such as unidentified particle and heavy-flavor quark \pt distributions, as well as two-particle and multiparticle correlations, have elucidated the ``bulk'' properties of this medium across a variety of energy- and volume-dependent colliding systems~\cite{Song:2017wtw,Dusling:2015gta,Nagle:2018nvi}. The strongly coupled nature of the QGP is also investigated on a wide range of length scales by taking advantage of the rapid ``energy loss'' by highly energetic partons traversing it (more accurately, a redistribution of energy to the surrounding medium)~\cite{Accardi:2004gp,Apolinario:2022vzg}.

Over the past few years, experiments at RHIC and the LHC, including CMS, performed increasingly precise ``soft'' (low-\QTwo processes) and hard (high-\QTwo processes) measurements of the bulk properties and jet modifications in HI collisions. Concurrently, advances in both the macroscopic (long-wavelength dynamical description) and microscopic (short-wavelength dynamical description) theories of interactions within the QGP are beginning to transform the abundant data into insights regarding the structure and properties of this state of matter. Notable success has been achieved in data-to-theory comparisons, particularly within well-defined frameworks in subfields such as the physics of soft probes~\cite{Gale:2013da,Heinz:2013th,Bernhard:2019bmu,Nijs:2021clz}. These achievements guide ongoing and future studies of the QGP nature, on a wide range of length scales~\cite{Andres:2016iys}. Such an exploration is facilitated by the extensive nuclear data sets currently available or to be collected in the near future.

\subsection{HI operation at the LHC}
\label{sec:IntroductionLHC}

In most years since the start of its operations in 2009, the LHC has been reconfigured for a month-long HI run~\cite{Jowett:2019whe}. The first two \PbPb collision runs were performed during LHC Run~1 (years 2010--2013) at $\sqrtsNN=2.76\TeV$ in 2010~\cite{Arduini:2011zzz} and 2011~\cite{Jowett:1492972}. 

Following these initial runs, the 
LHC HI physics community requested that the next run during this one-month period provide \pPb collisions.
Asymmetric collisions were not included in the LHC design, and are nontrivial since the ``2-in-1'' magnet design requires the two beams to have identical rigidity. This leads to unequal beam energies in the laboratory frame, since the ratio of the atomic over mass numbers for the lead nucleus, $Z/A$, is only about 40\% of that for protons. The original physics case~\cite{Salgado:2011wc} was based on a target luminosity of $1.15\ten{29}\percms$ 
at a beam energy of $7 Z\TeV$ (``design'' parameters). However, to better match \sqrtsNN for future \PbPb runs, a value of $4 Z\TeV$ was chosen, with the exact number determined by accelerator requirements.
Following a feasibility test and pilot physics fill in October 2011 and September 2012, respectively, 
the first \pPb run took place in January 2013. This one month data-taking period provided the four major LHC experiments with approximately 36\nbinv of \pPb integrated luminosity 
at $\sqrtsNN=5.02\TeV$~\cite{Jowett:2013uka}. This first \pPb run represented a gain of about a factor of 25 in collision energy compared to previous asymmetric collisions studied at RHIC, one of the largest energy leaps in the history of particle accelerators. Together with a ``reference'' \pp run at $\sqrts = 2.76\TeV$ (\ie, at the same \sqrtsNN as the previously collected \PbPb data), these runs were the last physics operations before the first LHC shutdown in years 2013 and 2014.

The LHC Run 2 (2015--2018) operational period with HI collisions started with a reference \pp run with beams of 2.51\TeV\ to obtain the same center-of-mass energy as in the \pPb run of 2013 ($\sqrtsNN=5.02\TeV$). For the same reason, the ensuing \PbPb operation in November--December 2015 was carried out at $\sqrtsNN=5.02\TeV$, that is, an energy of $6.37 Z\TeV$~\cite{Jowett:2016ctg} (slightly less than the maximum possible value at that time of $6.5 Z\TeV$). A second \pPb run at $\sqrtsNN=5.02\TeV$ occupied part of the period devoted to HI physics in November--December 2016.

Based on a range of crucial physics questions that emerged from the earlier \pPb data, as well as the 
opportunity to measure various heavy elementary particles, the remainder of the 2016 \pPb run was at a higher \sqrtsNN. Despite the complex strategy for repeated recommissioning and operation of the LHC, a plan to satisfy most requirements was implemented~\cite{Jowett:2017dqj} by successfully exploiting the different beam lifetimes at two \sqrtsNN values of 5.02 and 8.16\TeV. In the latter case, the peak luminosity surpassed the design value by a factor of almost~8.

\begin{figure*}[ht]
\centering
 \includegraphics[width=0.55\textwidth]{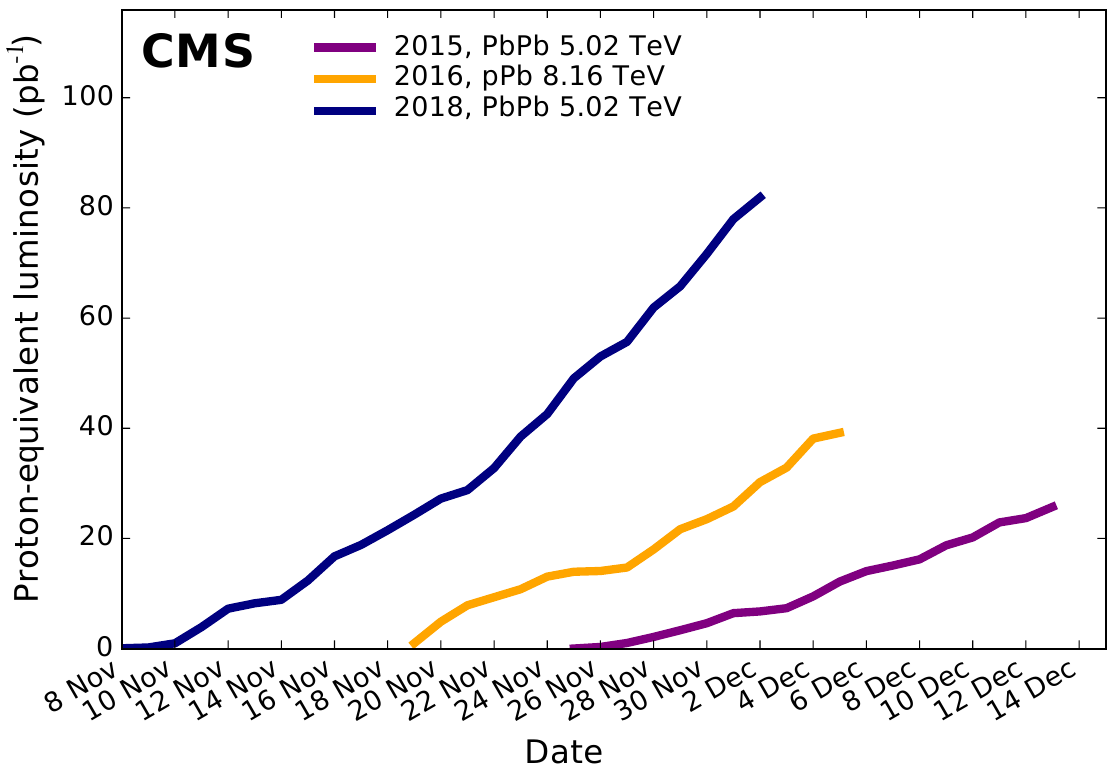}
 \caption{Integrated luminosity delivered to the CMS experiment with \PbPb and \pPb collisions at $\rootsNN=5.02$~\cite{CMS-PAS-LUM-18-001} and 8.16\TeV~\cite{CMS-PAS-LUM-17-002}, respectively, as a function of time during the LHC Run~2 period. The years of data collection shown correspond to 2015 (purple), 2016 (orange), and 2018 (navy blue). This plot shows the proton-equivalent luminosity, \ie, the values for the \PbPb data have been scaled by $A^2=208^2$ and the values for the \pPb data by $A=208$.\label{fig:CMS_lum}}
\end{figure*}

As shown in Fig.~\ref{fig:CMS_lum}, the subsequent 2018 \PbPb run~\cite{Jowett:2019jni} provided more than three times the integrated luminosity than was collected in 2015, bringing the LHC one step closer to its high-luminosity era (HL-LHC) with heavy ions. A series of improvements, both in the LHC and in its injector chain, including an increase in the average colliding bunch intensity and a decrease in the nominal bunch
spacing, resulted in reaching about six times higher instantaneous luminosity than
the design value of $1.0 \ten{27}\percms$, and delivering to the CMS experiment an integrated luminosity of 1.89\nbinv of \PbPb data (note the multiplicative factor mentioned in the caption of Fig.~\ref{fig:CMS_lum}).

\subsection{CMS detector design and implementation}
\label{sec:IntroductionExperiment}

The CMS detector~\cite{CMS:2008xjf} is one of the two general-purpose detectors at the LHC and is located at interaction point~5 (IP5). 
It has an overall length of 22\unit{m}, a diameter of 15\unit{m}, and weighs 14\,000\, tons.
The detector uses a right-handed coordinate system, with the origin at the nominal interaction point, the $x$ axis pointing to the
center of the LHC ring, the $y$ axis pointing up (perpendicular to the LHC plane) and
the $z$ axis along the counterclockwise beam direction. The azimuthal angle $\phi$ is measured in the $x$--$y$ plane, 
with $\phi=0$ along the positive $x$ axis, and $\phi=\pi/2$ along the positive $y$ axis.
The radial coordinate in this plane is denoted by $r$, while the polar angle $\theta$ is defined
in the $r$--$z$ plane with respect to the $z$ axis. The pseudorapidity is given by $\eta = -\ln\left(\tan\left(\theta/2\right)\right)$. For particles whose momentum and \pt are much higher than their invariant mass, $\eta\approx y$, where $y=\frac{1}{2}\ln{\frac{E+p_z}{E-p_z}}$ is the rapidity and $E$ and $p_z$ the energy and the particle momentum parallel to the $z$ axis, respectively. The shape of any distribution as a function of $y$ is invariant under Lorentz boosts in the beam direction.
The component of the momentum transverse to the $z$ axis is denoted by \pt, whereas the missing transverse momentum \ptmiss is the vectorial sum of the undetectable transverse momenta of the particles. The transverse energy is defined as $\ET=E\sin\theta$.

The key elements of the detector are as follows.

\begin{itemize}
\item A large solenoidal magnetic field magnetic field of 3.8\unit{T} to measure the momentum of charged particles and to separate the calorimeter energy deposits of charged and neutral particles.
\item A fine-grained tracker providing efficient reconstruction of
charged particle trajectories.
\item A highly segmented electromagnetic calorimeter~(ECAL) allowing energy deposits of charged hadrons, neutral hadrons, and photons to be clearly separated
from each other.
The ECAL combines
efficient photon identification with high resolution in both energy and position.
\item A hermetic hadron calorimeter~(HCAL) with modest energy resolution and coarse segmentation but sufficient to separate
charged and neutral hadron energy deposits. 
\item A muon tracking system delivering efficient and high-purity muon identification, regardless of the surrounding particle density.
\item Two forward subdetectors, the Centauro And STrange Object Research~(CASTOR) and Zero Degree Calorimeter~(ZDC), 
enhance the hermeticity of the CMS detector during HI data-taking periods.
\end{itemize}

\begin{figure}[ht]
\centering
\includegraphics[width=0.75\columnwidth]{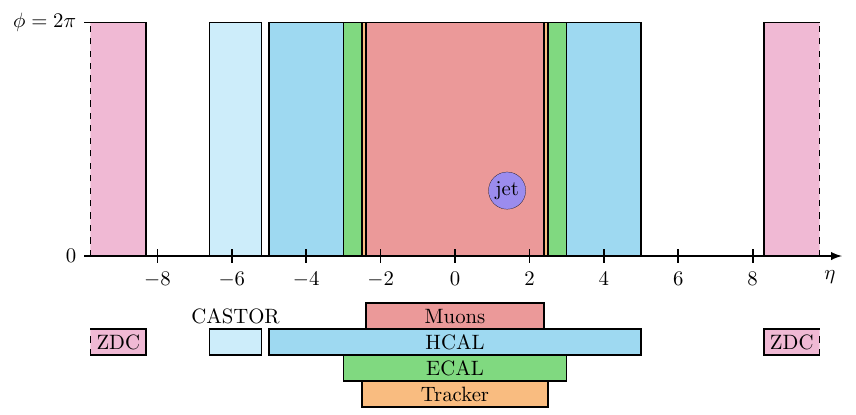}
\caption[A sketch of the \texttt{CMS} detector]{\label{fig:CMSSlice} A simplified sketch of the acceptance in $\eta$ and $\phi$ for the tracking, calorimetry (ECAL, HCAL, CASTOR, and ZDC) and muon identification (``Muons'') components of the CMS detector. In the lower section, the central elements (that is, excluding ZDC and CASTOR) are arranged based on their proximity to the beam, with the tracker being the closest element of the central detectors, and the muon detectors positioned farthest away. The size of a jet cone with $R = 0.5$ (to be discussed in Section~\ref{sec:ExperimentalMethods_JetMET}) is also depicted for illustration. \FigureFrom{CMS:2007eug}}
\end{figure}

Figure~\ref{fig:CMSSlice} displays a simplified sketch of the acceptance of the various components of the CMS detector in $\eta$--$\phi$ coordinates. The characteristics of the CMS subdetectors are described in more detail in Section~\ref{sec:ExperimentalMethods}.

The CMS general-purpose detector is designed to explore the standard model~(SM) and to search for physics beyond the SM~(BSM) at the \TeV scale. In addition, it is equally capable of studying the properties of
strongly interacting matter produced in nuclear collisions at the highest energy densities ever reached in the laboratory. The detector subsystems were designed with a resolution and granularity adapted to cope with the high number of simultaneous collisions per bunch crossing that occur during high-luminosity \pp running. As a result, the detector is also able to deal with the very large charged particle multiplicities that can be created in a single \PbPb collision. Therefore, CMS provides a range of remarkable capabilities, some of which are unique, to the HI effort at the LHC~\cite{CMS:2007eug}.
\begin{itemize}
    \item Acceptance: Broad coverage near midrapidity ($\abs{\eta} < 2.5$, full $\phi$ coverage) for layered detection of charged and neutral hadrons as well as muons, electrons, and photons, over a wide range of \pt (from about 100 MeV to hundreds of GeV).
    \item Resolution: Exceptional dimuon mass resolution, leading to a clean separation of the various heavy quarkonium states and an improved signal-over-background ratio, also coupled with excellent charged particle momentum resolution over a wide \pt and $\eta$ range. At $\abs{y} \approx 0$, the relative dimuon mass resolution is 0.6\%, or 20\MeV for the \JPsi and 70\MeV for the \upsOne. Integrated over the rapidity ranges used in the analyses reported in this paper, it becomes around 1.3\%, in \pp and \PbPb collisions (even in the most central collisions)~\cite{CMS-PAS-MUO-21-001}.
    \item Calorimetry: Full electromagnetic and hadronic calorimetry for complete jet triggering and reconstruction over a very large solid angle, leading to large event samples for measurements of individual jets (discussed in Section~\ref{sec:ExperimentalMethods_JetMET}) and $\text{jet} + X$ channels along with the recoiling \pt in them, where $X$ could be another jet, an electroweak~(EW) boson, etc.
    \item Forward coverage: Excellent forward physics and global event capabilities thanks to the forward hadronic~(HF) calorimeters ($3 < \abs{\eta} < 5$), CASTOR ($-6.6 < \eta < -5.2$), and the ZDCs ($\abs{\eta}>8.3$).
    \item Optimized online and offline operation: The data acquisition~(DAQ) system is capable of delivering almost every \PbPb event to a two-tier trigger system, allowing maximum flexibility to select ``bulk'' and rare probes. 
\end{itemize}

\subsection{Initial physics goals of the CMS heavy ion program}
\label{sec:IntroductionQCD}

This section presents a concise historical overview, both experimental and phenomenological, of the observables that were considered to be essential measurements for CMS prior to the inauguration of the LHC. Capitalizing on previous discoveries, the LHC experimental program with \PbPb and
\pPb collisions has significantly advanced the state of the art in both the soft- and hard-physics sectors. Sections~\ref{sec:InitialState}--\ref{sec:QEDBSM} describe in detail the physics analyses motivated by the priorities introduced here, as well as significant extensions that far surpassed these initial goals. 

At the LHC, the \pPb and \PbPb center-of-mass energies exceed those at RHIC by factors of roughly 25, thereby accessing a completely uncharted regime. Initial expectations were primarily driven by the fact that this regime could be characterized by the following properties.
\begin{itemize}
    \item An initial state dominated by high-density parton distributions. The relevant range of the parton momentum fraction $x$ reaches as low as ${\sim}10^{-4}$ (at midrapidity) for values of the squared momentum transfer \QTwo as large as $\QTwo \approx 10^6\GeV{}^2$. This \mbox{small-$x$} range, where the gluon density becomes so high that perturbation theory breaks down even for a small coupling constant, is expected to be dominated by nonlinear gluon dynamics~\cite{Ducloue:2019ezk} governed by a characteristic saturation scale that is a factor of 2--3 times larger than that probed at RHIC~\cite{dEnterria:2007jwb}. Important aspects of particle production and the early time evolution of the system were expected to be governed by classical chromodynamics, as described, \eg, in the color-glass condensate~(CGC) framework~\cite{Iancu:2003xm}, an approximation of the quantum theory (``effective field theory'') of the dense initial-state partonic wavefunctions. 
    \item Since the initial energy density, temperature, volume, and lifetime of the QGP were expected to be much larger than those at RHIC, parton dynamics were expected to drive the evolution of the medium~\cite{Proceedings:2007ctk,Armesto:2009ug}. Partonic degrees of freedom should thus dominate the QGP expansion and the collective features of the resulting hadronic final state.
    \item The higher yield of hard probes (\eg, prompt EW bosons, jets produced by hard-scattered partons, high-\pt hadrons, heavy-flavor hadrons) should provide direct information on the nPDFs of the colliding ions. Since their production cross sections can be calculated with high accuracy using perturbative QCD (pQCD), these probes provide a ``calibrated'' reference for final-state interactions in the medium. Any observed attenuation would give precise ``tomographic'' information about the QGP and its eventual disintegration into hadrons.
    \item Because of the very large electric charge of the Pb ions, the induced EM processes provide unique possibilities for studying high-energy \gaga and \PhotonA interactions in unexplored regions of phase space, thus complementing the QCD physics aspects listed above. 
\end{itemize}

The initial emphasis was placed on measurements that could assist in clarifying some of the previously unresolved issues.
For example, before the start up of RHIC, predictions for the charged-particle multiplicity per unit of rapidity at midrapidity $\rd{}N/\rd{}y\big|_{y=0}$ (largely based on extrapolations of SPS measurements) varied widely. This observable is related
to the entropy density produced in the collision, which impacts the global properties of the medium. The initial predictions for RHIC were mostly overestimates. The expected values for the maximum design energy of the LHC (5.5\TeV) were later refined to $\rd{}N/\rd{}\eta\big|_{\eta=0}\approx1.5\ten{3}$~\cite{Proceedings:2007ctk,Armesto:2009ug} for the most central \PbPb collisions. These lower particle multiplicities at the LHC would be more easily manageable by the CMS detector. 

Measurements of the properties of momentum anisotropies 
in \PbPb collisions at the LHC were deemed of primary importance to confirm or reject the interpretation of the fluid-like state
found at RHIC. At LHC energies, the contribution from the QGP phase
to the collective momentum anisotropy was expected to be more dominant than at RHIC. Consequently, interpretations of the properties of the QGP might be less dependent on the details of the later hadronic phase. 

Among the most exciting results of the RHIC physics program was the observation of a large suppression
of the yields of high-\pt hadrons in head-on AuAu collisions, compared to the expectations from an incoherent superposition of \pp collisions~\cite{dEnterria:2005gfc}. The capability to fully reconstruct jets at the LHC was expected to result in a better understanding of the mechanisms leading to high-\pt hadron suppression. The measurement of jets recoiling opposite prompt EW bosons as well as performing high-precision studies in
the heavy-flavor sector (using both jets and identified particles) were also of utmost importance. These observations could both clarify some apparently conflicting results at RHIC (\eg, on the energy loss flavor dependence) and also provide accurate information on the transport properties of the QCD matter. The LHC measurements were also crucial for resolving surprising findings, such as the similar amount of \PJGy yield modification observed at SPS and RHIC energies~\cite{Djordjevic:2003qk,Lourenco:2006sr}, and the rapidity dependence of that modification~\cite{PHENIX:2011img}. In addition, the more abundant production of particles containing bottom quarks, including the \PGU states, at LHC energies, coupled with the excellent reconstruction capabilities of the CMS detector, would enable a unique opportunity to extend the study of the behavior of heavy quarks in the QGP. 

Forward coverage was considered crucial for measuring low-$x$ PDFs, particularly the gluon densities, in protons and nuclei. Initial studies in \pp collisions involving perturbative probes, such as Drell--Yan and jet production within and beyond the HF acceptance, laid the groundwork for extending measurements to other hard probes, including inclusive high-\pt hadron or photon production in nuclear collisions, where gluon saturation effects are expected to be more pronounced. Additionally, hadron production at forward rapidity in nuclear collisions at LHC energies presented interesting connections to cosmic-ray physics, providing data necessary for calibrating the models used to study ultra-high energy cosmic ray interactions in the upper atmosphere. To complement the physics program in the baryon-free midrapidity region, the unique design of the CASTOR detector was implemented to further investigate exotic (``Centauro'') cosmic-ray events.

Finally, the high ``photon fluxes'' produced by the large electric charge of the relativistic Pb nuclei~\cite{Grabiak:1987uf,Hencken:1995me} also opened up possibilities for \gaga as well as \PhotonA studies, reaching
energies that had not been explored at colliders prior to the LHC~\cite{Bertulani:1987tz,Krauss:1997vr,Baur:1998ay}. Interesting physics within the SM, including both QCD and quantum electrodynamics~(QED) studies, would thus be possible~\cite{Baur:2001tj,Bertulani:2005ru,Baltz:2007kq}. Additionally,
states with high invariant mass could be explored, where the detection of new particles could potentially fall within the phase space region probed by CMS. In general, events with far-grazing collisions, so-called ``ultraperipheral collisions''~(UPC), characterized by relatively small outgoing particle multiplicities and a small background, offered a very wide range of possibilities~\cite{Baur:2001jj,Natale:1994nb,Baltz:2002pp,Klein:2003vd,Abraham:1990wm,dEnterria:2009cwl,Klein:2000dk,Ahern:2000jn,Natale:1995np,Rau:1989fv,Almeida:1990idx,Greiner:1992fz}. 

\subsection{Major achievements of the CMS heavy ion physics program}

The heavy ion physics program of CMS, based on analyses of data collected from 2010 to 2018 across various collision systems and energies, has provided groundbreaking insights into high-density QCD, in general, and the nature of the QGP state, in particular. These discoveries, briefly summarized in this section, have fundamentally advanced the field, setting new benchmarks for the study of QCD matter.

\begin{itemize}
    \item \textbf{Collective behavior in small collision systems}
    \item[] A groundbreaking discovery by CMS was the observation of collective behavior in small collision systems, such as \pp and \pPb collisions. This phenomenon, known as ``collectivity", had previously only been associated with larger systems, such as \PbPb collisions. This breakthrough, supported by measurements made in many different classes of observables, suggests that fluidity and plasma properties may emerge in QCD matter in systems with very different sizes, opening new paths for theoretical and experimental studies aimed at understanding the origins of these collective effects.

    \item \textbf{Properties of the quark-gluon plasma}
    \item[] Several analyses of CMS data provided both a macroscopic and a microscopic characterization of the highest temperature and smallest baryon-chemical potential QCD medium ever produced in a laboratory setting. Thanks to the unprecedented precision of these measurements, the results have significantly advanced the understanding of the QGP's thermodynamical and transport properties, including its fluid-like behavior and the dynamics of its formation and evolution.

    \item \textbf{Jet quenching and medium modification of parton showers}
    \item[] The direct observation of jet quenching established new standards in the field, significantly extending previous studies beyond the leading hadrons and di-hadrons to assess the medium modification of entire parton showers.
    The results reveal the intricate mechanisms by which the QGP alters the energy and structure of high-energy jets passing through it, offering deep insights into the interactions between hard probes and the QCD medium.

    \item \textbf{Heavy-flavor hadron and jet nuclear modifications}
    \item[] CMS conducted pioneering studies of fully reconstructed beauty hadrons and of heavy-flavor jets (\ie, jets containing charm or bottom quarks).
    These measurements, performed for the first time in the harsh environment produced in nucleus-nucleus collisions,
    provide critical information on the interactions of heavy quarks with the QGP and on how they lose energy or are scattered by the medium, shedding light on the role of the quark mass in quark energy loss mechanisms.

    \item \textbf{Suppression patterns of five S-wave quarkonium states}
    \item[] For the first time, CMS measured the centrality dependence of the nuclear suppression of all five S-wave quarkonia, including the elusive \PGUP{3S}, showing that the suppression patterns follow a sequential hierarchy reflecting the binding energies of the quarkonium states: the more strongly bound is the considered meson, the hotter must be the medium before we see its suppression. Interestingly, the loosely-bound \Pgy meson is significantly suppressed even in the most peripheral \PbPb collisions probed by the CMS data. Using the distance between the dimuon vertex and the primary collision vertex, we could also measure the suppression pattern of nonprompt charmonia, an indirect measurement of the effects of the QCD medium on \PB mesons.
\end{itemize}

\clearpage

\section{Experimental challenges}
\label{sec:ExperimentalMethods}

Experimental measurements in ultrarelativistic HI collisions have reached a precision era. As shown in Sections~\ref{sec:InitialState}--\ref{sec:QEDBSM}, 
in order to effectively discriminate between the various theoretical models attempting to describe the phenomena at play in these collisions, there are two 
experimental challenges. First, it is important to collect large samples of very specific types of events in order to reduce the statistical uncertainties. 
Second, large event samples and improved analysis techniques are needed to minimize the systematic uncertainties in these measurements. Accomplishing both 
of these goals is required to perform quantitative data-theory comparisons. Of particular importance in HI physics is to carry out multiple differential 
analyses in bins of centrality (defined in Section~\ref{sec:ExperimentalMethods_Centrality}).

\begin{figure}[ht]
    \centering
    \includegraphics[width=0.8\textwidth]{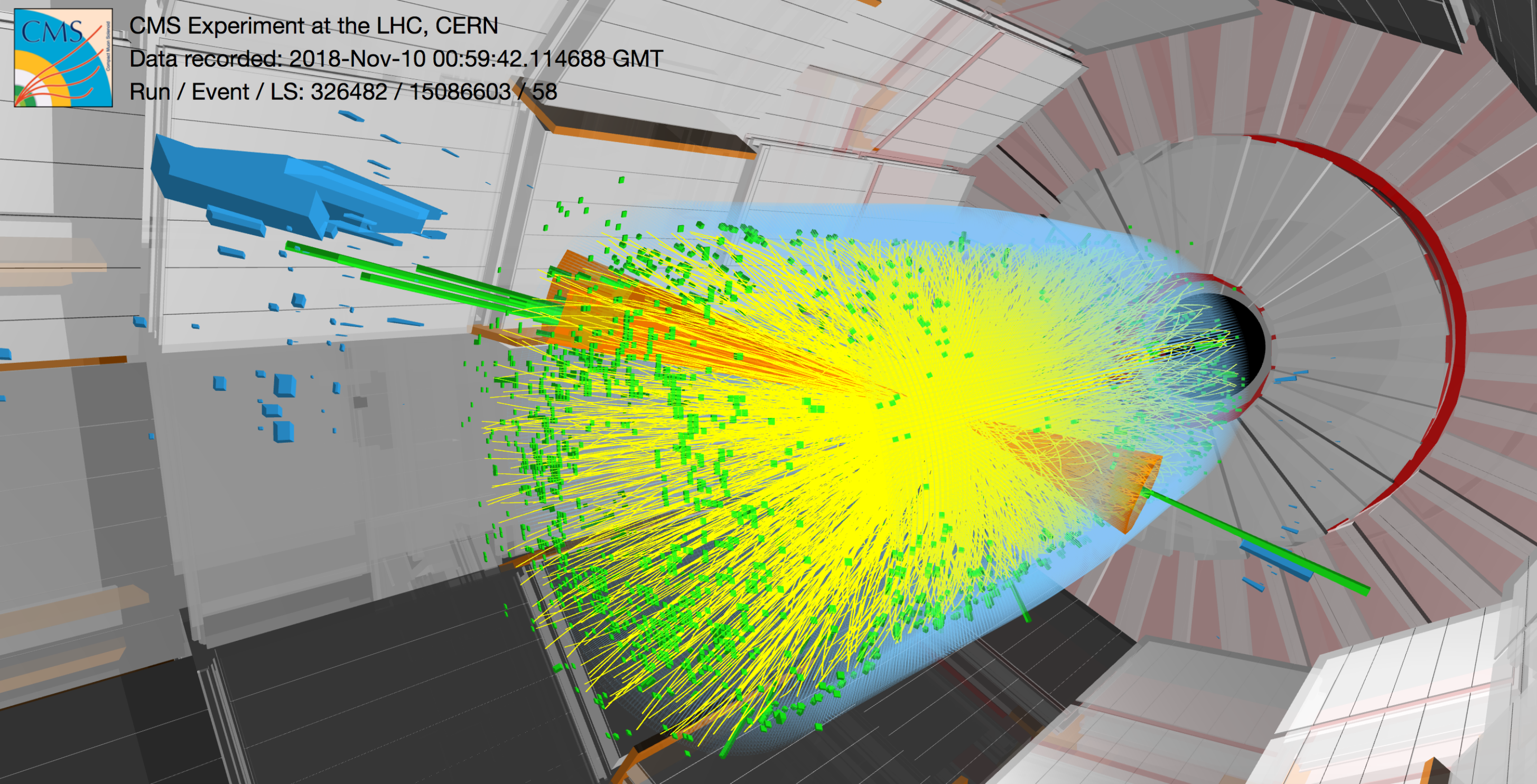}
    \caption{An almost head-on collision event selected from the 2018 \PbPb data set. The yellow lines show the huge number of charged-particle tracks 
    and the two cones show nearly back-to-back candidate jets originating from bottom quarks. \FigureFrom{display}}
    \label{fig:event_display}
\end{figure}

This section describes the hardware and techniques used by the CMS Collaboration to select and classify events, detect the properties of the produced 
particles, and extract various physics observables. These tasks are particularly difficult in the extremely high particle multiplicity environment of 
HI collisions at LHC energies, as illustrated by the event display shown in Fig.~\ref{fig:event_display}.

\subsection{The CMS detector}
\label{subsec:CMSapparatus}

The central feature of the CMS apparatus is a superconducting solenoid of 6\unit{m} internal diameter, providing a magnetic field of 3.8\unit{T}. 
Within the solenoid volume, as shown in Fig.~\ref{fig:CMSSlice}, are a silicon pixel and strip tracker, a lead tungstate crystal ECAL, and a brass 
and scintillator HCAL, each composed of a barrel and two endcap sections. Hadron forward~(HF) calorimeters, made of steel and quartz-fibers, extend 
the pseudorapidity coverage provided by the barrel and endcap detectors. Muons are measured in gas-ionization detectors embedded in the steel 
flux-return yoke outside the solenoid. The procedure followed for aligning the detector is described in Ref.~\cite{CMS:2021ime}. 
A more detailed description of the CMS detector can be found in Ref.~\cite{CMS:2008xjf}.

The silicon tracker used until 2016 measured charged particles within the range $\abs{\eta} < 2.5$. For nonisolated particles of $1 < \pt < 10\GeV$ 
and $\abs{\eta} < 1.4$, the track resolutions were typically 1.5\% in \pt and 25--90 (45--150)\mum in the transverse (longitudinal) impact 
parameter~\cite{CMS:2014pgm}. At the start of 2017, a new pixel detector was installed~\cite{CMSTrackerGroup:2020edz}; the upgraded tracker measured 
particles up to $\abs{\eta} < 3.0$ with typical resolutions of 1.5\% in \pt and 20--75\mum in the transverse impact parameter~\cite{DP-2020-049} 
for nonisolated particles of $1 < \pt < 10\GeV$. More details about the tracking algorithms are discussed in Section~\ref{sec:ExperimentalMethods_Tracking}.

In the region $\abs{\eta} < 1.74$, the HCAL cells have widths of 0.087 in pseudorapidity and 0.087 in azimuth~($\phi$). In the $\eta$--$\phi$ plane, 
and for $\abs{\eta} < 1.48$, the HCAL cells map on to $5{\times}5$ arrays of ECAL crystals to form calorimeter towers projecting radially outwards 
from close to the nominal interaction point. For $1.74 < \abs{\eta} < 3.0$, the coverage of the towers increases progressively to a maximum of 0.174 
in $\Delta \eta$ and $\Delta \phi$. Within each tower, the energy deposits in ECAL and HCAL cells are summed to define the calorimeter tower energies, 
which are used as inputs to the jet-finding algorithms determining the energies and directions of hadronic jets.

The two halves of the HF calorimeter are located 11.2\unit{m} from the interaction region, one on each end, and together they provide coverage 
in the range $3.0<\abs{\eta}<5.2$. They also serve as luminosity monitors. Two subdetectors, CASTOR and ZDC, enhanced the hermeticity of the CMS 
detector during the HI data-taking periods by extending the rapidity coverage. The single CASTOR detector, a Cherenkov sampling calorimeter with 
no segmentation, was located 14.37\unit{m} from IP5 and about 1\unit{cm} from the beam line, covering the region $-6.6< \eta<-5.2$. 
The two ZDCs, made of quartz fibers and plates embedded in tungsten absorbers, are installed at $0^{\circ}$ relative to the beam direction 
at the interaction point and between the two beam lines 140\unit{m} away from IP5. Their location behind the first bending magnets of the LHC 
allows for detecting neutrons from nuclear dissociation events in the range $\abs{\eta}>8.3$. The first section of each ZDC is segmented 
into 5 vertical slices, giving some information about particle direction.

Muons are measured in the pseudorapidity range $\abs{\eta} < 2.4$, with four detection planes made using three technologies: drift tubes~(DT), 
cathode strip chambers~(CSC), and resistive-plate chambers~(RPC). The reconstruction of muon tracks is described in Section~\ref{sec:ExperimentalMethods_Muon}. 
During Run~1, matching muons to tracks measured in the silicon tracker results in a relative transverse momentum resolution, for muons 
with $20 <\pt < 100\GeV$, of 1.3--2.0\% in the barrel and better than 6\% in the endcaps. The \pt resolution in the barrel is better 
than 10\% for muons with \pt up to 1\TeV~\cite{CMS:2012nsv}. For Run~2, the corresponding numbers are \pt resolutions of 1\% in the 
barrel and 3\% in the endcaps for muons with \pt up to 100\GeV and better than 7\% in the barrel for muons with \pt up to 1\TeV~\cite{CMS:2018rym}.

The global event reconstruction (also called particle-flow~(PF) event reconstruction~\cite{CMS:2017yfk}) aims to reconstruct and identify each 
individual particle in an event, with an optimized combination of all subdetector information. In this process, the identification of the particle 
type (photon, electron, muon, charged hadron, neutral hadron) plays an important role in the determination of the particle direction and energy. 
Photons (\eg, coming from \PGpz decays or from electron bremsstrahlung) are identified as ECAL energy clusters not linked to the extrapolation 
of any charged-particle trajectory to the ECAL. Electrons (\eg, coming from photon conversions in the tracker material or from \PQb~hadron 
semileptonic decays) are identified as a primary charged-particle track, and potentially several ECAL energy clusters corresponding to this 
track extrapolation to the ECAL and to possible bremsstrahlung photons emitted along the way through the tracker material. Muons (\eg, from 
quarkonium and EW boson decays) are identified as tracks in the central tracker consistent with either a track or several hits in the muon 
system, and associated with calorimeter deposits compatible with the muon hypothesis. Charged hadrons are identified as charged-particle 
tracks neither identified as electrons, nor as muons. Finally, neutral hadrons are identified as HCAL energy clusters not linked to any 
charged-hadron trajectory, or as a combined ECAL and HCAL energy excess with respect to the expected charged hadron energy deposit. 
Details on how the PF objects are used as input in jet reconstruction can be found in Section~\ref{sec:ExperimentalMethods_JetMET}.

The energy of photons is obtained from the ECAL measurement. The energy of electrons is determined from a combination of the track momentum at 
the main interaction vertex, the corresponding ECAL cluster energy, and the energy sum of all bremsstrahlung photons attached to the track. 
The energy of charged hadrons is determined from a combination of the track momentum and the corresponding ECAL and HCAL energies, corrected 
for the response function of the calorimeters to hadronic showers. Finally, the energy of neutral hadrons is obtained from the corresponding corrected ECAL and HCAL energies.
Details about the methods to reconstruct and identify electrons and photons can be found in Section~\ref{sec:ExperimentalMethods_EGamma}.

During Run~1, the barrel section of the ECAL achieved an energy resolution of about 1\% for unconverted or late-converting photons in 
the tens of \GeVns energy range. The energy resolution of the remaining barrel photons is about 1.3\% up to $\abs{\eta} = 1$, 
increasing to about 2.5\% at \mbox{$\abs{\eta} = 1.4$}. In the endcaps, the energy resolution is about 2.5\% for unconverted or late-converting 
photons, and between 3 and 4\% for photons converting in the tracker material before reaching the ECAL~\cite{CMS:2015myp}. 

For Run~2, the electron momentum is estimated by combining the energy measurement in the ECAL with the momentum measurement in the 
tracker. The momentum resolution for electrons with $\pt \approx 45\GeV$ from $\PZ \to \Pe \Pe$ decays ranges from 1.6 to 5\%. 
It is generally better in the barrel region than in the endcaps, and also depends on the bremsstrahlung energy emitted by the 
electron as it traverses the material in front of the ECAL~\cite{CMS:2020uim, CMS-DP-2020-021}. 

Jets are reconstructed offline from the energy deposits in the calorimeter towers or PF objects, clustered using the anti-\kt 
algorithm~\cite{Cacciari:2008gp, Cacciari:2011ma,Giammanco:2014bza,Abdullin:2011zz} with a distance parameter of 0.4 (discussed in Section~\ref{sec:ExperimentalMethods_JetMET}). 
When combining information from the entire detector, the jet energy resolution in \pp collisions amounts typically to 15--20\% 
at 30\GeV, 10\% at 100\GeV, and 5\% at 1\TeV~\cite{CMS:2016lmd}.

\subsection{Luminosity measurement techniques and luminosity-dependent corrections}
\label{sec:ExperimentalMethods_Luminosity}

For many analyses of the HI program performed at the LHC, a precise
knowledge of the integrated luminosity is essential to achieve the high-precision cross sections needed to both test and, 
in many cases, improve relevant theory calculations. Although a key parameter for any particle
collider, the task of calibrating the absolute scale of the luminosity has proven to be 
particularly challenging at hadron colliders. Knowing the instantaneous luminosity during data acquisition is also 
essential to monitor the beam condition and accelerator performance. 

To determine the absolute luminosity, a technique based on varying the separation of the two LHC beams is used, 
the so-called ``van der Meer scans''~\cite{CMS:2021xjt}. 
These scans allow for a determination of the luminosity per colliding bunch pair directly 
from the machine parameters. With the exception of special, reference proton and HI runs, 
these scans are performed either under carefully controlled conditions, with tailored beam parameters, or during normal physics operation. 
In both cases, the target precision is set to $\mathcal{O}(1\text{--}2\%)$~\cite{CMS:2021xjt}.

The CMS experiment used a system consisting of up to five detectors to monitor and measure the luminosity delivered by the LHC 
during LHC Runs~1 and~2. Real-time monitoring of the luminosity was achieved with three of them: the fast beam conditions monitor, 
HF, and pixel luminosity telescope detectors, each with its own high-rate data acquisition system.
Two additional systems, the silicon pixel detector and the drift tube chambers, feature very low occupancy and good stability over time. 
The absolute luminosity is determined by integrating the subdetector rate of these two systems as a function of beam separation, 
which corresponds to an approximate van der Meer scan precision of $\mathcal{O}(1\%)$. The dominant uncertainty in the absolute 
luminosity is typically related to how well the beam bunch density profiles can be factorized. For the measurements highlighted 
in this report, a small total uncertainty of 
$\mathcal{O}(1.5\text{--}2.0\%)$~\cite{CMS-PAS-LUM-16-001,CMS-PAS-LUM-17-002,CMS-PAS-LUM-18-001,CMS-PAS-LUM-19-001} was achieved, 
including the time stability of the van der Meer calibrated subdetector response.

\subsection{Online event selection}

\label{sec:ExperimentalMethods_trigger}

During Runs~1 and~2, the CMS detector was used to collect a large quantity of data for the different colliding systems provided 
by the LHC as part of its HI program. A beam of Pb nuclei in one LHC ring collided with either a second counterrotating Pb beam 
or with a proton beam. A small event sample of colliding xenon~(Xe) nuclei was also collected. The resulting integrated luminosities 
(\lumi) are summarized in Table~\ref{tab:tabHIN}. A comparison of the integrated luminosity delivered by the LHC 
($5^\text{th}$ column) to that collected by the CMS detector (last column) shows that the detector acquired more than 95\% of the 
available data (with the exception of the very first run in 2010, for which the fraction was 90\%). 

\begin{table}[ht]
\topcaption{Summary of HI data-taking periods during Runs~1 and~2.}
\centering
\renewcommand{\arraystretch}{1.1}
\begin{tabular}{cccccc}
  \hline
  LHC Run   & Year & Colliding system   & \sqrtsNN ({\TeVns}) & Delivered \lumi ($\text{nb}^{-1}$) & Recorded \lumi ($\text{nb}^{-1}$) \\[\cmsTabSkip]
  \hline
            & 2010      & \PbPb         & 2.76              & $9.69\ten{-3}$& $8.70\ten{-3}$ \\
  Run~1     & 2011      & \PbPb         & 2.76              & 0.184         & 0.174 \\
            & 2013      & \pPb          & 5.02              & 36.14         & 35.5 \\
  \hline
            & 2015      & \PbPb         & 5.02              & 0.59          & 0.56 \\ 
            & 2016      & \pPb          & 5.02              & 0.530         & 0.509 \\
  Run~2     & 2016      & \pPb          & 8.16              & 188.3         & 180.2 \\
            & 2017      & \XeXe         & 5.44              & $3.50\ten{-3}$& $3.42\ten{-3}$ \\
            & 2018      & \PbPb         & 5.02              & 1.89          & 1.79 \\
\hline
\label{tab:tabHIN}
\end{tabular}
\end{table}
\begin{table}[ht]
\topcaption{Summary of reference \pp data-taking periods during Runs~1 and~2. To compare with the nucleon-nucleon-equivalent 
luminosities from Table~\ref{tab:tabHIN}, it is important to note that the listed integrated luminosities should be divided 
by factors of either $\mathrm{A}^2$ (for the \PbPb case) or A (for the \pPb case), where $\mathrm{A}=208$ is the Pb mass number.}
\centering
\begin{tabular}{cccc}
  \hline
  LHC Run & Year & \sqrts ({\TeVns}) & Recorded \lumi ($\text{pb}^{-1}$) \\
  \hline
  \multirow{ 2}{3em}{Run~1} & 2011 & 2.76 & 0.200 \\
                            & 2013 & 2.76 & 5.40 \\
  \hline
  \multirow{ 2}{3em}{Run~2} & 2015 & 5.02 & 27.4 \\ 
                            & 2017 & 5.02 & 304 \\
\hline
\label{tab:tabHINppref}
\end{tabular}
\end{table}

In addition to the HI runs shown 
in Table~\ref{tab:tabHIN}, special \pp runs were required by the HI program in order to provide the so-called ``\pp reference data'' 
for comparison to the HI results. They are listed in Table~\ref{tab:tabHINppref} and were taken with detector conditions similar to those used in the 
HI runs, and with the same colliding energy per nucleon pair (which is lower than that used in the high-energy \pp program). 
In addition, \pp data were collected under conditions that yielded a very small number (${\ll}1$) of concurrent interactions in 
the same bunch crossing (pileup,~PU). These low-PU data were required for studies searching for the possible existence of a hot 
and dense medium in events which produced a very large number of charged particles, the so-called ``high-multiplicity events.''
Table~\ref{tab:tabHINpplowPU} summarizes the integrated luminosities collected for the low-PU \pp collisions.

\begin{table}[ht]
\topcaption{Summary of low-PU \pp data-taking periods during Runs~1 and~2.}
\centering
\begin{tabular}{cccc}
  \hline
  LHC Run & Year & \sqrts ({\TeVns}) & Recorded \lumi ($\text{pb}^{-1}$) \\
  \hline
  Run~1  & 2010 & 7 & 6.2 \\
  \hline
  \multirow{ 4}{*}{Run~2}   & 2015 & 5.02   & 1.0   \\ 
                            & 2015 & 13     & 2.0   \\
                            & 2017 & 13     & 1.3   \\
                            & 2018 & 13     & 10.2  \\
\hline
\label{tab:tabHINpplowPU}
\end{tabular}
\end{table}

To deal with the large amount of data delivered by the LHC, online event selection is performed by the CMS trigger system~\cite{Dasu:2000ge} 
that was developed to quickly and precisely select the events of interest. The trigger operates in two stages, the first of which (level-1 or L1) 
is a hardware-based trigger that examines every collision using a set of event selections implemented directly in the firmware. 
The L1 trigger uses energy deposited in the towers of the calorimetry system and signals from the muon detectors to construct various physics 
objects, and then uses that information to make the initial event selection.
In order to suppress noncollision-related noise, cosmic rays, prefiring triggers~\cite{CMS:2020cmk}, and beam backgrounds, the L1 trigger is 
required to accept events in coincidence with the presence of both colliding ion bunches in the interaction region. 
The next stage is the high-level trigger~(HLT), a software-based trigger running on a computer farm composed of $\approx$30\unit{k} computer 
central processing units~(CPUs). The HLT uses information from the L1 trigger and performs additional event filtering using sophisticated computer 
algorithms executed similarly to those used for the offline physics analyses. This processing in the HLT takes $\approx$250\unit{ms} and 
$\approx$350\unit{ms} on average for \pp and \PbPb events, respectively. In \pp collisions, the L1 trigger reduces the event rate from 40\unit{MHz} 
to 100\unit{kHz}, while the HLT further reduces the rate from 100\unit{kHz} to around 2\unit{kHz} on average in \pp collisions. In \PbPb and 
\pPb collisions on average the rates are reduced to 8\unit{kHz} and 20\unit{kHz}, respectively. During Run~2, event rates of around 30\unit{kHz} 
were delivered for \PbPb collisions, while around 100\unit{kHz} were delivered for both \pp reference and \pPb collisions. The trigger system 
output rates are optimized to fill as much bandwidth as the DAQ can support (around 6\unit{GB/s} maximum during Run~2)~\cite{Cittolin:578006}.
Between Runs~1 and 2, the L1 trigger system was updated significantly and restructured~\cite{Tapper:1556311}.

\begin{figure}[t]
\centering
\includegraphics[width=0.49\linewidth]{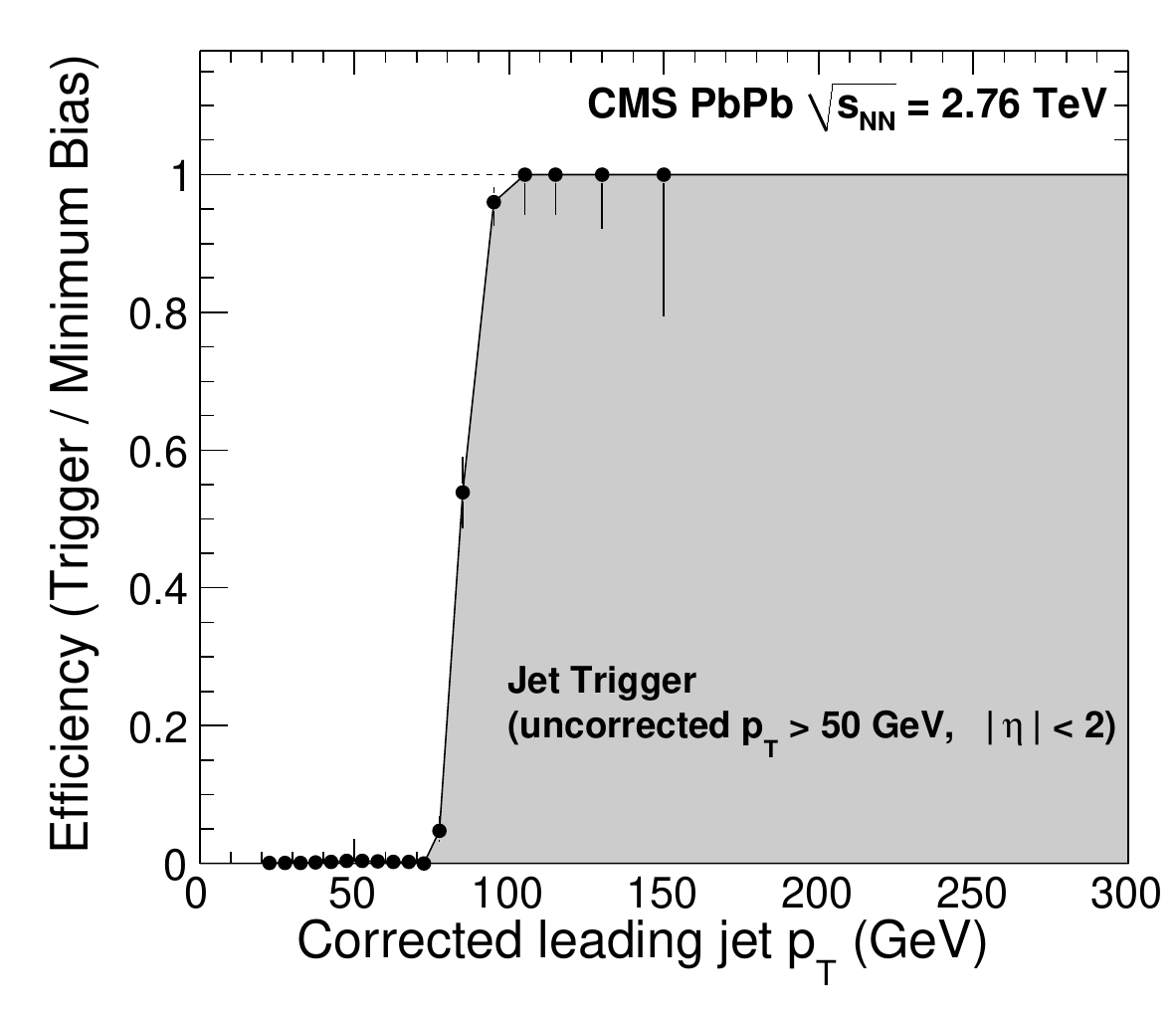}
\includegraphics[width=0.45\linewidth]{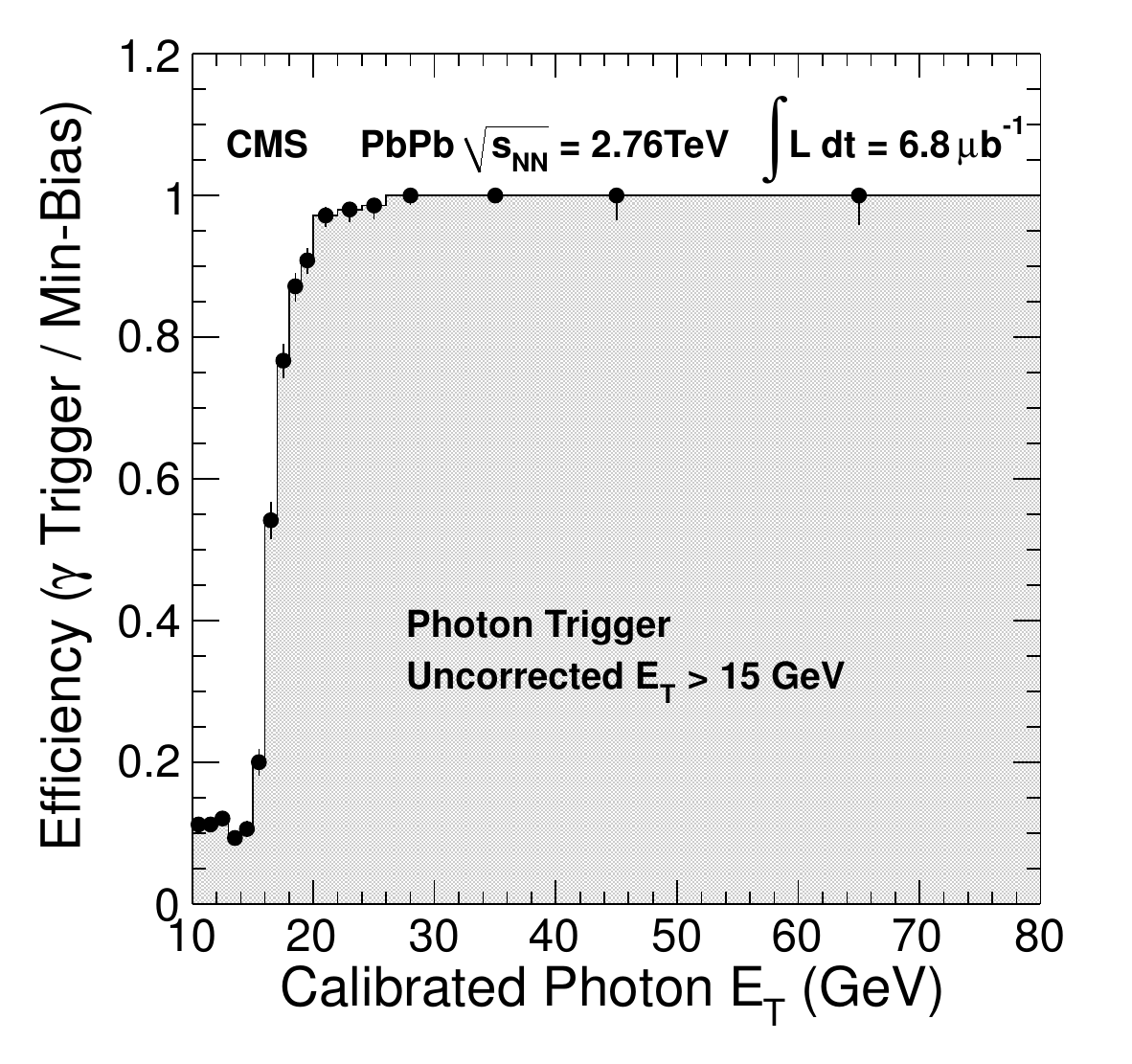}
\caption{Left: Efficiency for the 50\GeV single-jet trigger as a function of the corrected leading jet transverse momentum in \PbPb collisions at 2.76\TeV. 
Right: Efficiency for the 15\GeV photon trigger as a function of the corrected photon transverse energy in \PbPb 
collisions at 2.76\TeV. \FigureCompiled{CMS:2011iwn, CMS:2012oiv}}
\label{fig:JetPhotoneff}
\end{figure}

In order to optimize the performance of the trigger system in HI collisions, 
several modifications and additions are made to the L1 and HLT setups used for \pp collisions~\cite{CMSTrigger:2005yhe,Virdee:1043242,Choudhury:2021soo}. 
In HI collisions, minimum bias~(MB) triggers require at least one channel in both sides of HF to have 
deposited energies that exceed a certain threshold. By triggering on the total energy deposited in the HF calorimeters, 
events can be selected with different centrality. 
The rate of the MB triggers in 2018 \PbPb collisions were around 6\unit{kHz}, where one out of three MB events were recorded. In addition, 
ultraperipheral events, where one of the colliding nuclei remains intact, 
can be triggered by requiring activity on only one side of the HF.
The jet background subtraction at L1 is optimized for triggering on jets in
the higher multiplicity HI environment. To suppress the influence of underlying event~(UE) fluctuations, the average UE contribution to the 
jet energy is estimated by summing the energy over full calorimeter $\phi$-rings sharing the jet position in $\eta$. The sum is rescaled by 
the number of towers in the jet compared to the number of towers in the calorimeter $\phi$-ring and then subtracted from the jet energy~\cite{Hasan:2022uuo}. 
For the electron/photon triggers, the isolation and selections were relaxed and the shape requirements were bypassed at L1 to maintain 
a high efficiency in the high-occupancy environment of HI collisions.
Figure~\ref{fig:JetPhotoneff} shows the efficiency of the 50\GeV single-jet and 15\GeV photon triggers during Run~1 \PbPb data taking.
In the HLT configuration for HI collisions, the track selection requirements are loosened to be the same as the offline ones in order 
to reach higher efficiency for the triggers using track reconstruction (described in Section~\ref{sec:ExperimentalMethods_Tracking}). 
Changes are also made for the muon triggers, where only outside-in tracking (Section~\ref{sec:ExperimentalMethods_Muon}) is 
implemented, and for the heavy-flavor triggers, in which the selection criteria are tightened to reduce the processing time.
The typical \pt thresholds of the jet, photon, electron, and muon triggers in \PbPb collisions were 100, 40, 20, and 12\GeV, respectively.

\begin{figure} [!hbpt]
\centering
\includegraphics[width=0.45\linewidth]{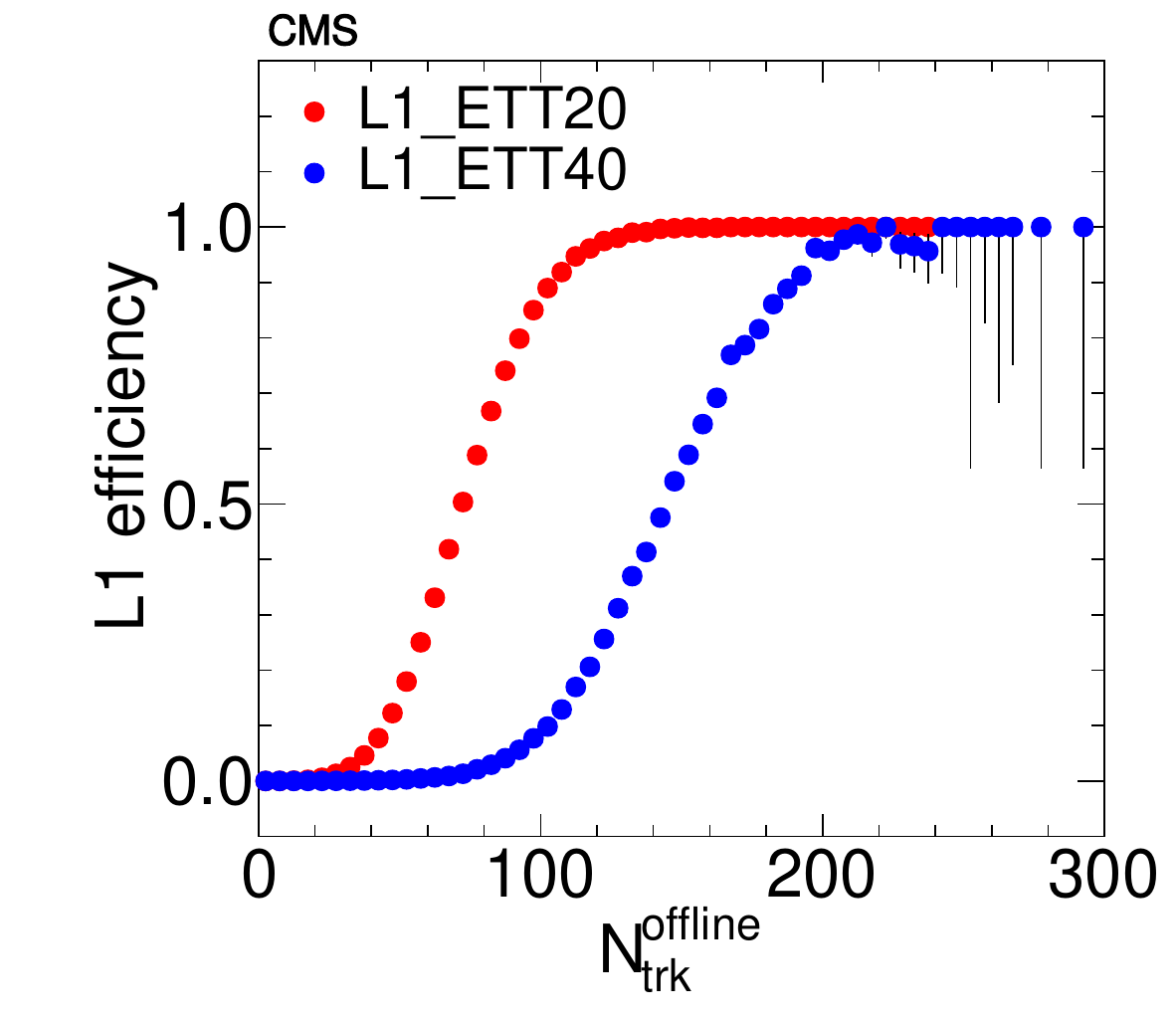}
\includegraphics[width=0.45\linewidth]{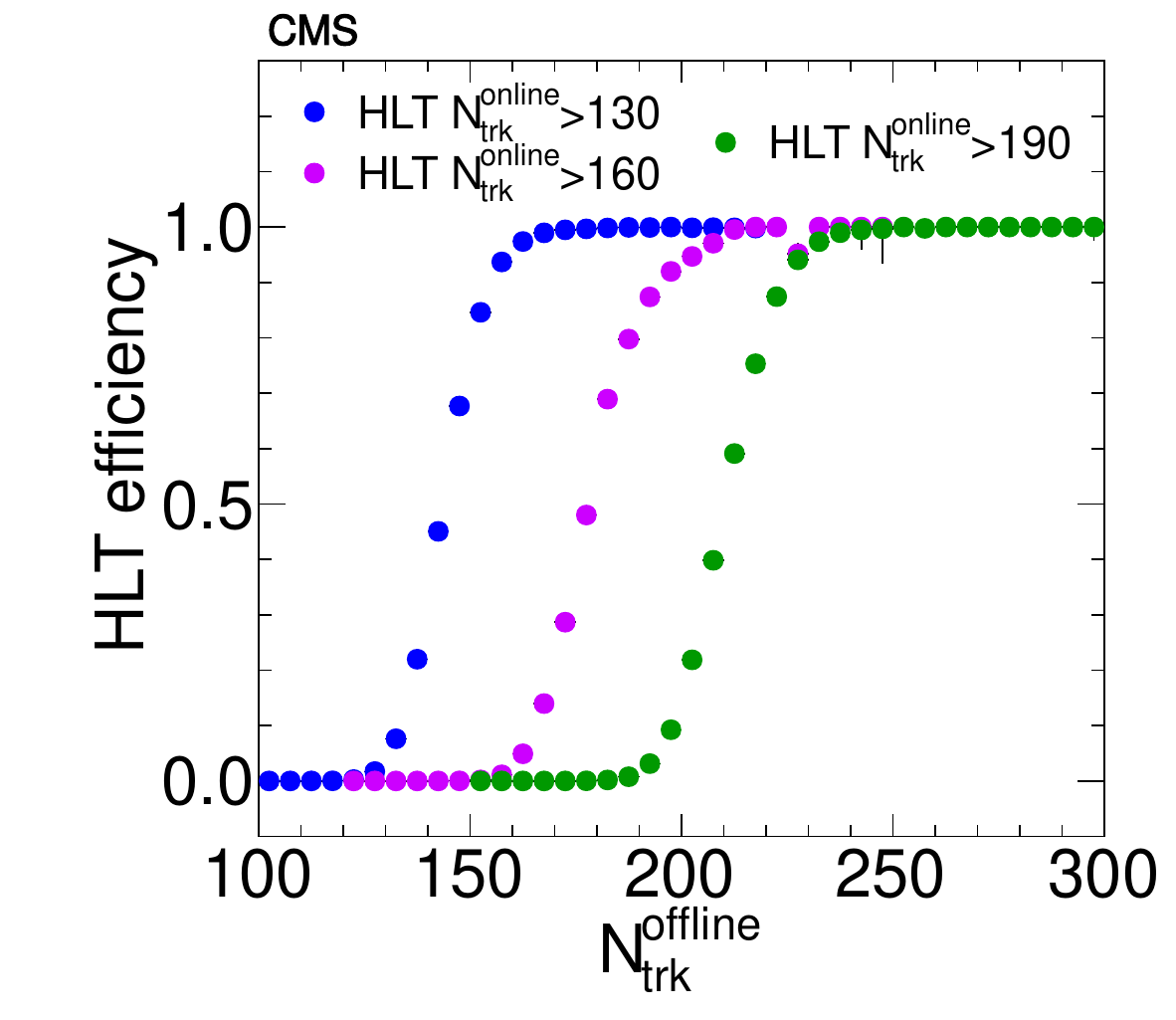}
\caption{The L1 and HLT trigger efficiencies for the high-multiplicity triggers as functions of \noff 
for 5.02\TeV \pPb collision data taking in the year of 2013. \FigureFrom{Chatrchyan:2013nka}}
\label{fig:HMeff}
\end{figure}

With the goal of studying the properties of high-multiplicity \pPb and \pp collisions (discussed in 
Section~\ref{sec:SmallSystems_Collectivity}), dedicated triggers were designed and implemented to capture the rare events with a 
large number of produced particles. An L1 trigger which filters on the scalar sum of total transverse momentum over the calorimeters 
(ETT, including both ECAL and HCAL) is used to select events. Those events are passed to the HLT where track reconstruction is performed using 
the pixel tracker and the number of found pixel tracks is used to select high-multiplicity events.
Figure~\ref{fig:HMeff} shows the L1 and HLT trigger efficiencies as functions of the number of tracks 
reconstructed offline (\noff) for 5.02\TeV \pPb collision data taken in 2013. 

\subsection{Minimum bias event selection}
\label{sec:ExperimentalMethods_MinBias}

Hadronic interactions of HIs can occur over a broad range of overlaps of the two nuclei, from head-on collisions (most central) 
to just barely grazing (most peripheral). Investigating the full range of possibilities requires a MB event sample. The selection 
procedure for this sample includes both an online trigger (discussed in Section~\ref{sec:ExperimentalMethods_trigger}) as well as 
offline quality criteria. The optimal selection maximizes the overall efficiency for the total inelastic hadronic cross section, 
while mitigating contamination from non-hadronic collision sources, including beam-gas collisions and electromagnetic interactions 
in peripheral and ultraperipheral collisions~(UPC). The first three filters listed below, namely the primary vertex~(PV), cluster 
compatibility, and HF coincidence filters were used as standard event selection for \PbPb Run~2 analyses.

\subsubsection{Primary vertex filter}
\label{sec:ExperimentalMethods_VertexFilter}

The definition of MB interactions is limited to only those events in which at least one PV containing at least two reconstructed 
tracks is found. Track reconstruction and vertex finding are detailed in Section~\ref{sec:ExperimentalMethods_Tracking}, whereas 
the centrality description is given in Section~\ref{sec:ExperimentalMethods_Centrality}. For most of the events, only tracks satisfying 
the high-purity quality criterion~\cite{CMS:2014pgm}, with the additional restrictions $\pt>0.7\GeV$ and $d_0 < 2\mm$ (where $d_0$ is 
the transverse impact parameter of a track with respect to the beam), are included in the PV filter. In more peripheral events, all 
reconstructed tracks are used because the high-purity selection is too restrictive. For the most central collisions, the minimum \pt 
requirement was increased to $\pt>1.0\GeV$. These restrictions keep the maximum number of fitted tracks less than about 40--60, thereby 
ensuring a time-efficient reconstruction. The requirement of an accepted vertex removes a large fraction of the background events, 
especially beam-gas interactions, which can have large HF energy deposits but very few pixel hits.

\subsubsection{Cluster compatibility filter}
\label{sec:ExperimentalMethods_ClusterFilter}

A particle traversing a pixel module at some angle leaves a cluster with a width proportional to its angle of incidence. That angle, and 
hence the expected width of the cluster, can be determined by the particle pseudorapidity and the position of the collision vertex 
along the beam direction ($z$). This information can be used to determine the number of clusters in an event that have a width compatible 
with particles originating from the vertex position. Alternatively, this compatibility can be investigated without a predetermined vertex 
by scanning the $z$ axis to determine how many clusters are compatible with a vertex at each value of $z$. This technique can be used 
to locate the most likely $z$ position of the collision without the need for any reconstructed tracks or, instead, to determine if the 
collision likely occurred outside the interaction region. The cluster size is proportional to $\abs{\sinh{\eta}}$, where the $\eta$ 
of the cluster is computed with respect to the reconstructed vertex. 
A selection on this variable is performed as a function of~$\eta$~\cite{Chatrchyan:2011pb}.

\subsubsection{Other filters}
\label{sec:ExperimentalMethods_OtherFilters}

The HF coincidence filter requires at least two towers in both of the HF calorimeters (one on each side of the interaction point) 
with a deposited energy above 4\GeV. This requirement removes approximately 99\% of the UPC events 
(discussed in Section~\ref{sec:ExperimentalMethods_EM_Contamination}).

Finally, the beam scintillation counter~(BSC)~\cite{CMS-DN-2010-018} is a set of large-area scintillators mounted in front of HF to 
provide beam halo information. A dedicated filter excluded events where any of the BSC halo L1 trigger bits were set. 
This filter was only used for the Run~1 data samples.

\subsubsection{MB event selection efficiency}
\label{sec:ExperimentalMethods_MinBiasEffic}

Along with the unwanted background, the event selection criteria described above also remove some valid hadronic collision events. 
The selection efficiency is defined as the fraction of valid events that pass the MB criteria applied to the data, and is found using 
a method based on Monte Carlo~(MC) simulations. The distribution in data, either the total energy deposited in the HF calorimeters 
(which is not saturated in the more peripheral events) or the number of pixel hits which pass the cluster size requirement, is compared 
to the respective distribution found using simulated events. The two distributions are normalized in the higher-multiplicity region and 
any differences at lower multiplicity are indicative of selection inefficiencies. The uncertainty in this estimate is determined 
by varying the MC simulation parameters, particularly those affecting the average multiplicity, and also by varying the normalization 
range. The overall trigger and event selection efficiency for MB HI collisions is estimated to be $97\pm3\%$ in \PbPb collisions at 
2.76\TeV (Run~1)~\cite{CMS:2011iwn} using events from the \textsc{AMPT} 1.26t5 simulation~\cite{Lin:2004en}, and $97\pm1.5\%$ in 
\PbPb collisions at 5.02\TeV (Run~2) using \HYDJET 1.9 simulations~\cite{Lokhtin:2005px}.

\subsubsection{Electromagnetic contamination}
\label{sec:ExperimentalMethods_EM_Contamination}

When two nuclei pass each other with transverse impact parameters larger than the sum of their nuclear radii, their hadronic interaction 
cross section is vanishingly small, but the nuclei can still interact through their large EM fields. The EM interactions occurring in the 
UPC events induce a level of contamination that is studied by using generated events from \Starlight~2.2 + 
\textsc{dpmjet}~3.0~\cite{Klein:2016yzr, Roesler:1996qw} or \Starlight+\PYTHIA{}8.2~\cite{Klein:2016yzr, Sjostrand:2014zea},
passing them through a simulation of the CMS detector response based on \GEANTfour~\cite{Agostinelli:2002hh}, and then applying the same 
reconstruction procedure as for the data.
In order to estimate this contamination in the MB sample, the number of events surviving different event selections are determined. 
Event rates obtained by applying the PV plus cluster compatibility filters are used as a baseline. The HF coincidence filter is then 
applied in data and simulations, but with varying requirements on the number of towers and their energies. This variation is needed 
because the simulations do not include UPC events. Any differences seen between the data and MC distributions with the varying HF 
filters imply the presence of remaining UPC contamination in the MB sample. This allows for the determination of the best HF coincidence 
filter to remove the overwhelming majority of the UPC events and also to estimate the remaining EM contamination. A \Starlight sample 
simulating photoproduction events (including only single photon events) was used for the efficiency determination. For Run~2 \PbPb collisions, 
only 0.8\% of the EM events survive the selection criteria. The total expected photoproduction cross section in \PbPb collisions at 
5.02\TeV is 34\unit{b}, implying an EM contamination cross section of 0.27\unit{b}. Using the MB event selection efficiency of 98\%, 
combined with the hadronic cross section of 7.7\unit{b}, gives 7.5\unit{b} of hadronic events. The ratio of the 0.27\unit{b} of EM 
contamination and the total hadronic plus EM cross section of 7.77\unit{b} implies that the EM contamination in the selected MB sample is about 3.5\%. 

\subsection{Centrality determination in nucleus-nucleus collisions}
\label{sec:ExperimentalMethods_Centrality}

The degree of overlap of two colliding nuclei (called ``centrality''), or equivalently their transverse impact parameter ($b$), is a 
critical component in the study of HI collisions. 
Many properties of the interaction vary significantly with centrality. These include the shape of the overlap region (varying from 
lenticular for peripheral collisions to roughly spherical for central ones) and its size (or equivalently the number of participating 
nucleons, \npart). Other quantities of interest include the average distance each nucleon traverses while passing through the other nucleus, 
from which one can calculate the average number of interactions per nucleon and the total number of binary nucleon-nucleon~(\NN) 
collisions, \ncoll. The nuclear overlap function, \TAA, is then defined as \ncoll divided by the inelastic \NN cross section 
$\sigma_\NN^{\text{inel}}$. The value of \TAA can be interpreted as equivalent to the \NN integrated luminosity per HI collision. 
These quantities are essential for comparing data from different collision systems or different experiments, and also for comparing data to 
theoretical calculations. The quantitative measure of centrality is defined as the percentage of the total inelastic hadronic nucleus-nucleus 
cross section, with 0\% corresponding to full overlap and 100\% corresponding to the nuclei just barely missing each other. 
Values of $b$, \npart, and \ncoll cannot be experimentally measured, and so they need to be deduced from the data using Glauber MC 
models~\cite{Miller:2007ri,Loizides:2017ack}. To obtain these values, experiments rely on other observables, such as the multiplicity 
of outgoing particles (or the energy in the forward region), which are roughly proportional to \npart.

\subsubsection{Glauber Monte Carlo model}

The Glauber MC model estimates geometric quantities such as \npart, \ncoll, and the impact parameter. This method is used by experiments at both 
RHIC and the LHC~\cite{Loizides:2017ack} and is related to, but distinct from, the original so-called ``Glauber model'', which first used a 
variation of the optical model of scattering theory to analytically derive the properties of collisions of protons with nuclei (as described 
in Ref.~\cite{Miller:2007ri} and references therein). The Glauber MC model first assumes that, at ultrarelativistic energies, individual nucleons 
in one nucleus carry enough momentum to not be deflected as they traverse the other colliding nucleus. The collisions are assumed to happen over 
such a short time scale that there is no movement within the nuclei, so the constituent nucleons move in independent linear trajectories parallel 
to the beam direction. Finally, nucleons from opposite nuclei are assumed to interact if their relative transverse distance is less than the 
``ball diameter'', \ie, $D = \sqrt{\smash[b]{\sigma_{\NN}^{\text{inel}}/\pi}}$, where $\sigma_{\NN}^{\text{inel}}$ is the inelastic 
hadronic \NN cross section at the \NN center of mass energy. With these hypotheses, a MC model can be used to find values for the interaction 
cross section of the two nuclei, as well as \npart and \ncoll, in terms of the basic \NN interaction. Individual events in the MC involve 
randomly distributing the nucleons within each nucleus and then following their trajectories.
 
The probability density used to place the nucleons is based on Woods--Saxon distributions (\ie, Fermi--Dirac distributions applied to describe the nuclear density),
\begin{equation}
\rho(r)=\rho_{0}\frac{1+w(r/R)^{2}}{1+\exp{(\frac{r-R}{a}})},
\end{equation}
where $\rho_{0}$ is the nucleon density in the center of the nucleus, $R$ is the nuclear radius, $a$ is the skin depth, and $w$ represents the 
deviation of the nucleus from a spherical shape. For $^{208}$Pb nuclei, the parameter $\rho_{0}$ is an overall normalization and $w$ is equal to zero.
The impact parameter of the collisions is distributed by using $\rd N/\rd b \approx b$, considering a $b_{\max}$ around 20\unit{fm}. 
A new parameter is introduced to require a minimum inter-nucleon distance between the centers of the nucleons. Specifically, nucleons 
are distributed on a uniform 3-dimensional lattice with a minimum nodal separation $d_{\text{node}}=0.4\unit{fm}$. In this way, the 
position of each nucleon is determined stochastically, event-by-event, and the geometric quantities are estimated by averaging over multiple events.

\subsubsection{Centrality determination}

The primary detector for centrality determination is the HF, which covers the forward rapidities $3<\abs{\eta}<5.2$, as described in 
Section~\ref{subsec:CMSapparatus}. The transverse energies on both sides of this detector are summed to give a variable that is 
monotonically increasing with charged-particle multiplicity. 

This sum of transverse energy \et in the HF calorimeters was the default centrality variable during Runs~1 and 2~\cite{CMS:2011iwn}.
A set of \et boundaries was determined, which divided the full HF distribution into 200 centrality bins, corresponding to centrality 
classes of width 0.5\% of the total inelastic hadronic cross section. The bin boundaries are calculated from a sample of MB events with 
the trigger and event selections applied. To consider the event selection efficiency and possible EM and UPC contamination, a \HYDJET MC 
simulation~\cite{Lokhtin:2005px} was used for events with HF \et less than a threshold, above which the HF \et distributions for the data 
and \HYDJET MC have identical shapes. The MC distribution is scaled to match that of the data in the high \et (central collision) region 
by minimizing the $\chi^{2}$ goodness-of-fit. This scaled MC distribution can then be used to determine centrality bins in the low-\et 
(peripheral collision) region, where inefficiencies and/or EM contamination can distort the distribution in data.
The centrality class for a selection of events is given as a range in percentage of the total inelastic hadronic cross section.

\begin{figure}[ht]
   \centering
   \includegraphics[width=0.60\linewidth]{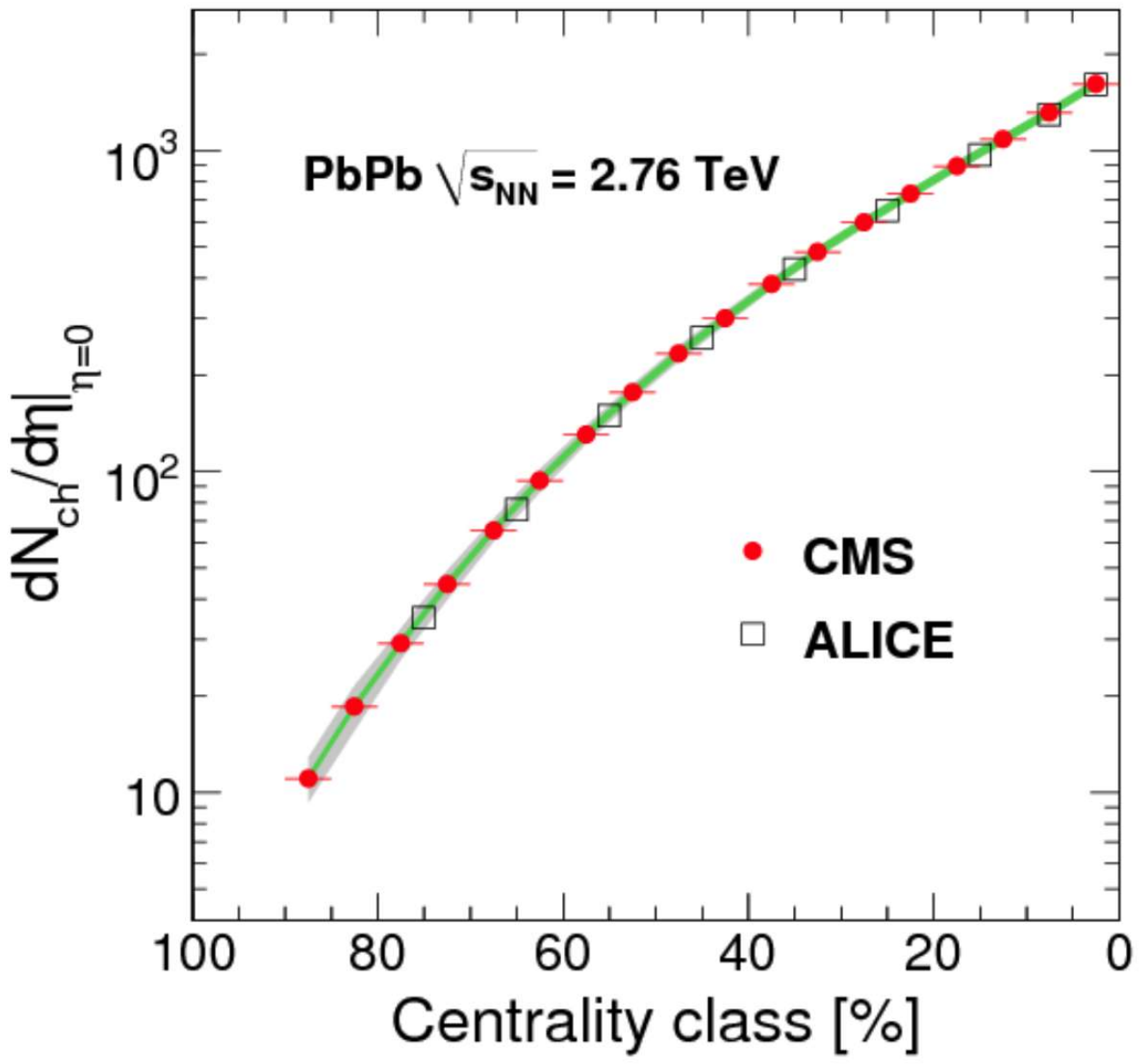}
   \caption{Charged hadron multiplicity density at mid-rapidity ($\rd N_{\text{ch}} / \rd \eta |_{\eta=0}$) as a function of 
   centrality class in \PbPb collisions at $\sqrtsNN=2.76\TeV$ from the CMS (solid circles) and ALICE~\cite{Aamodt:2010cz} (open squares) 
   experiments. The inner green band shows the measurement uncertainties affecting the scale of the measured distribution, while the 
   outer grey band represents the full systematic uncertainty, \ie, affecting both the scale and the slope. \FigureFrom{Chatrchyan:2011pb}}
   \label{fig:Multiplicity_centrality}
\end{figure}

The dependence of the charged-particle multiplicity density at midrapidity on the centrality class is shown in Fig.~\ref{fig:Multiplicity_centrality}.
As expected, the multiplicity increases monotonically with increasing overlap of the two nuclei (\ie, decreasing centrality percentage). 
Note the logarithmic scale for the $y$~axis, indicating that the multiplicity increases more rapidly as the collisions get more central.

The number of pixel hits is also considered in the centrality determination. Having the same monotonic dependence as the energy measured in 
the HF, this quantity provides a good cross-check. In addition, the ZDC detector is used to test the quality of both variables. 
The total energy in the ZDC is correlated to the number of spectator neutrons released in the interaction, thus providing a variable 
that is correlated (anti-correlated) with the multiplicity of events in peripheral (more central) events~\cite{Miller:2007ri}. 

\subsubsection{Uncertainties in the centrality determination}

The two major contributions to the centrality uncertainties are from the Glauber model parameters and the estimate of the event selection efficiency.
The uncertainties in the Glauber parameters are extracted by using the procedure described in Ref.~\cite{Loizides:2017ack}, 
where Glauber MC samples were produced with varying values of the parameters. The systematic uncertainties in the four input parameters to 
the Glauber MC calculation are shown in Tables~\ref{tbl:PbPbGlauberVariation} and~\ref{tbl:PbPbGlauberCentwise} for \PbPb collisions at 
5.02\TeV, whereas similar conclusions hold for the other \rootsNN values.

\begin{table}[h]
\centering
\topcaption{Input parameters and their uncertainties for the \PbPb 5.02\TeV Glauber MC~\cite{PHOBOS:2007vdf}.}
\label{tbl:PbPbGlauberVariation}
\begin{tabular}{lc}
\hline
Glauber parameter & Value and uncertainty \\
\hline
$R$ (nuclear radius)    & $\phantom{0}6.68 \pm 0.02\unit{fm}$ \\
$a$ (skin depth)        & $0.447 \pm 0.01\unit{fm}$ \\
$d_{\text{node}}$       & $\phantom{00}0.4 \pm 0.4\unit{fm}\phantom{0}$ \\
$\sigma_{\NN}^{\text{inel}}\vspace{0.03cm}$ & $\phantom{0}67.6 \pm 0.6\unit{mb}\phantom{0}$ \\
\hline
\end{tabular}
\end{table}

\begin{table}[htb]
\centering
 \topcaption{Geometric quantities and their systematic uncertainties averaged over centrality ranges in \PbPb collisions at 5.02\TeV.}
\label{tbl:PbPbGlauberCentwise}
\begin{tabular}{cccc}
\hline
\multicolumn{1}{c}{Centrality interval} &
  \multicolumn{1}{c}{$\mean\ncoll$} &
  \multicolumn{1}{c}{$\mean\npart$} &
  \multicolumn{1}{c}{$\mean\TAA$ (\unit{mb$^{-1}$})} \\ \hline
0--5\%   & $1737\pm32$ (1.8\%)      & $382.3\pm1.5$ (0.4\%)     & $25.70\pm0.47$ (1.8\%)    \\
5--10\%  & $1379\pm27$ (2.0\%)      & $331.5\pm1.2$ (0.4\%)     & $20.40\pm0.40$ (2.0\%)      \\
0--10\%  & $1558\pm28$ (1.8\%)      & $356.9\pm0.9$ (0.3\%)     & $23.05\pm0.42$ (1.8\%)    \\
10--20\% & $973\pm21$ (2.1\%)       & $262.3\pm1.3$ (0.5\%)     & $14.39\pm0.30$ (2.1\%)    \\
20--30\% & $595\pm15$ (2.5\%)       & $188.2\pm1.4$ (0.8\%)     & $8.80\pm0.22$ (2.5\%)   \\
30--40\% & $346\pm11$ (3.1\%)       & $131.0\pm1.4$ (1.1\%)     & $5.12\pm0.16$ (3.1\%)   \\
40--50\% & $187.7\pm7.2$ (3.8\%)    & $87.2\pm1.3$ (1.5\%)      & $2.78\pm0.11$ (3.8\%)   \\
50--80\% & $50.4\pm2.5$ (5.0\%)     & $33.8\pm0.8$ (2.4\%)      & $0.745\pm0.037$ (5.0\%) \\[\cmsTabSkip]
0--100\% & $382\pm27$ (2.0\%)       & $113.7\pm0.8$ (0.7\%)     & $5.65\pm0.12$ (2.2\%)   \\ \hline
\end{tabular}   
\end{table}

\subsection{Event classification methods in \texorpdfstring{\pp}{pp} and \texorpdfstring{\pPb}{pPb} collisions}
\label{sec:ExperimentalMethods_Classification}

\begin{table*}[ht]
\centering
\topcaption{\label{tab:flowntrkbinning} Fraction of MB triggered events after event selections
in each multiplicity bin, and the average multiplicity of reconstructed tracks per bin
with $\abs{\eta}<2.4$ and $\pt >0.4\GeV$, before (\noff) and after (\ntrcorr) acceptance and efficiency
corrections, for \pPb and \PbPb collisions at 5.02 and 2.76\TeV, respectively \cite{Chatrchyan:2013nka}.}
\begin{tabular}{ c  c  c  c  c  c  c }
\hline
\multicolumn{1}{l}{} & \multicolumn{3}{c}{\pPb data at 5.02\TeV} & \multicolumn{3}{c}{\PbPb data at 2.76\TeV} \\[\cmsTabSkip]
\hline
\noff bin & Fraction &           $\left< \noff \right>$ & $\left<\ntrcorr\right>$ & $\langle \text{Centrality} \rangle$ & $\left< \noff \right>$ & $\left<\ntrcorr\right>$ \\
&  & & & $\pm$ RMS (\%) & & \\
\hline
0--$\infty$&   1.00       & 40    & $50\pm2$   & & &\\[\cmsTabSkip]
0--20    &   0.31       &  10   & $12\pm1$   & $92\pm4$   &  10   & $13\pm1$    \\
20--30   &   0.14       &  25   & $30\pm1$   & $86\pm4$   &  24   & $30\pm1$    \\
30--40   &   0.12       &  35   & $42\pm2$   & $83\pm4$   &  34   & $43\pm2$    \\
40--50   &   0.10       &  45   & $54\pm2$   & $80\pm4$   &  44   & $55\pm2$    \\
50--60   &   0.09       &  54   & $66\pm3$   & $78\pm3$   &  54   & $68\pm3$    \\
60--80   &   0.12       &  69   & $84\pm4$   & $75\pm3$   &  69   & $87\pm4$    \\
80--100  &   0.07       &  89   & $108\pm5$  & $72\pm3$   &  89   & $112\pm5$   \\
100--120 &   0.03       &  109  & $132\pm6$  & $70\pm3$   &  109  & $137\pm6$   \\
120--150 &   0.02       &  132  & $159\pm7$  & $67\pm3$   &  134  & $168\pm7$   \\
150--185 &   $4\ten{-3}$ &  162 & $195\pm9$  & $64\pm3$   &  167  & $210\pm9$   \\
185--220 &   $5\ten{-4}$ &  196 & $236\pm10$ & $62\pm2$   &  202  & $253\pm11$  \\
220--260 &   $6\ten{-5}$ &  232 & $280\pm12$ & $59\pm2$   &  239  & $299\pm13$  \\
260--300 &   $3\ten{-6}$ &  271 & $328\pm14$ & $57\pm2$   &  279  & $350\pm15$  \\
300--350 &   $1\ten{-7}$ &  311 & $374\pm16$ & $55\pm2$   &  324  & $405\pm18$  \\
\hline
\end{tabular}
\end{table*}

For \pp and \pPb collision data, an event selection similar to that described for \PbPb events was adopted. The integrated values for the combined 
trigger and event selection efficiency for \pPb collisions at 5.02\TeV are 90.7\% based on \textsc{epos lhc}~\cite{Pierog:2013ria} 
simulations, and 95.0\% when using \HIJING v2.1~\cite{Wang:1991hta}, both with a systematic uncertainty of 3\%. 

In \pp collisions, the events are characterized by bins in \noff, which is the multiplicity of 
high-purity quality tracks within $\abs{\eta}<2.4$ and $\pt > 0.4\GeV$. 
To mitigate effects from backgrounds, a reconstructed track was considered as a primary-track candidate if the transverse impact 
parameter significance (the value divided by its uncertainty) and the significance of the separation between the track and the best 
reconstructed PV along the beam direction both have an absolute value less than 3. In order to remove tracks with poor momentum 
estimates, the relative uncertainty of the momentum measurement was required to be less than 10\%. 

For \pPb collisions, the HF calorimeters could, in principle, be used for a centrality measurement in the same way as was done 
for \PbPb collisions. However, this is not particularly useful since simulations show that the correlation between the HF energy 
and the number of participating nucleons is extremely broad~\cite{CMS:2014qvs}.
So, data for \pPb collisions are, for the most part, also binned in \noff. The average multiplicity 
values, \Ntrackavg and \Ntrackcorravg, where the latter 
are corrected for efficiency and acceptance, are listed in Table~\ref{tab:flowntrkbinning} for each \noff interval. 
A number of analyses have attempted to compare \pPb and \PbPb results for similar system ``size'' by using identical bins 
in \noff. Table~\ref{tab:flowntrkbinning} also shows the corresponding average \PbPb collision 
centrality (as determined by the total energy deposited in the HF calorimeters), as well as the same average 
multiplicities in bins of \noff.

\subsection{Tracking and vertex reconstruction}
\label{sec:ExperimentalMethods_Tracking}

The CMS Collaboration uses two approaches for offline reconstruction of charged particles in HI collisions. One employs an 
iterative approach based on the combinatorial Kalman filter~(CKF), resulting in a set of tracks called ``general tracks'' \cite{CMS:2014pgm}, 
whereas the other method uses a single iteration based on the pixel detector only, known as ``pixel detector tracks'' \cite{CMS-DP-2023-011}. 
The first method, similar to the one applied in \pp collisions, considers both strips and pixel detectors, and covers the charged-particle 
transverse momentum region above a few hundreds of \MeV. The pixel detector tracking is designed to reach the lowest possible 
transverse momentum. It accomplishes this by having a very reduced rate of misreconstructed tracks as compared to the general tracks, 
even in events with very high charged-particle multiplicities. 

For \PbPb data taken during Runs~1 and 2, the total efficiency $\epsilon_{\text{eff}}$ of the general tracks (track reconstruction, 
and track selection), folded with the detector acceptance, varied as a function of the collision centrality and the transverse 
momentum of the particles in the range 10--75\% (lower values in central collisions and the low-\pt region down to 0.5\GeV). 
Similar trends are observed for pixel tracking with total efficiency in the range 15--70\% for $0.3<\pt<1.0\GeV$. 
For misreconstructed tracks, the general tracks have $\epsilon_{\text{misID}}$ values (ratio between the number of reconstructed 
tracks that do not share more than 75\% of their hits with a generator-level track and the total number of reconstructed tracks) in 
the range 1--20\% (higher in central collisions and the low-\pt region down to 0.5\GeV). Again, the results for pixel detector 
tracking are similar, with $\epsilon_{\text{misID}}$ values in the range 2--10\% for $0.3<\pt<1.0\GeV$. The performance for \pp, 
\pPb, and \XeXe data samples is similar, especially when comparing events with similar track multiplicity, as described in Refs.~\cite{CMS:2014pgm, CMS:2018yyx}. 

In Run~1, the general track collection in \PbPb collisions was built using a variation of the iterative tracking procedure used 
for \pp data, including modified code and a different number of steps in the algorithm, as described in Ref.~\cite{CMS:2012zex}. 
As described in Ref.~\cite{CMS:2014pgm}, the analyses in Run~2 used the same code for track reconstruction for all colliding systems, 
as well as almost the same number of iterations. An exception to this similarity is the algorithm dedicated to reconstructing 
tracks whose origin was significantly displaced from the PV, \eg, particles created in heavy-flavor hadron decay. However, the 
\PbPb collision environment is considerably different from that found in ``typical'' \pp collisions with an average PU of 20, 
being much denser in number of tracks (details are given in Sections~\ref{sec:ExperimentalMethods_Classification} and~\ref{sec:QGP_Thermodynamics}). 
Therefore, the tracking parameters used in the algorithms shared with \pp analysis needed to be tuned for \PbPb collisions 
in order to have a good performance in terms of CPU time, storage, tracking efficiency, and the rate of misreconstructed tracks. 
The sharing of algorithms between the various colliding systems is very important because now all the developments implemented 
for \pp collisions are generally
incorporated for \PbPb data taking and vice-versa. This commonality will become increasingly important since a central \PbPb 
collision has a track density similar to that which will be created by PU in \pp collisions in future high-luminosity LHC running. 

The offline PV reconstruction used in CMS analysis is based on a two-step procedure: vertex finding using a deterministic 
annealing algorithm to produce clusters of tracks coming from the same interaction vertex, followed by vertex fitting using 
the adaptive vertex fit to compute the best estimate of the vertex position and the corresponding parameters of its associated 
tracks~\cite{CMS:2014pgm}. The main challenge in the vertexing procedure is to avoid vertex merging and splitting (combining 
two independent vertices into one, or generating two separate vertices out of a single collision). Compared to the method 
used for \pp collisions, an additional track \pt requirement ($\pt > 1\GeV$) was applied for the 20\% most central \PbPb collisions, 
with no additional selections added for other centrality classes. This requirement was used to reduce vertex merging and splitting 
in central \PbPb collisions, while maintaining high vertex reconstruction efficiency in peripheral collisions.

\subsection{Muon reconstruction}
\label{sec:ExperimentalMethods_Muon}

Final states containing muons are important components of many HI physics analyses, including studies of quarkonia and electroweak bosons. 
The excellent performance of the CMS muon detection system enables the reconstruction and identification of muons with high efficiency 
and accuracy over a wide momentum range~\cite{CMS:2012nsv, CMS:2018rym, CMS:2021yvr}. Details regarding the performance of CMS muon 
reconstruction across collision systems during Run~2 can be found in Ref.~\cite{CMS-PAS-MUO-21-001}. This section summarizes the main 
findings related to offline muon reconstruction, stressing the challenges arising in the high-multiplicity HI collision environment.

\begin{figure}[ht]
    \centering
    \includegraphics{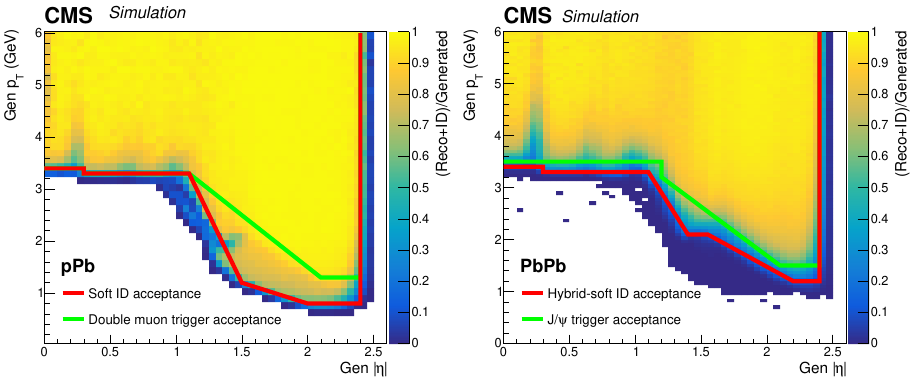}
    \caption{Muon reconstruction and identification efficiencies as functions of the simulated muon pseudorapidity and \pt in \pPb 
    (left) and \PbPb (right) collisions. The lines delimit the acceptance regions used for physics analyses: the red curves for 
    measurements not relying on a dedicated muon trigger while the green ones are for analyses using the muon trigger information, 
    \ie, for most of the quarkonia results presented in this paper. \FigureFrom{CMS-PAS-MUO-21-001}}
    \label{fig:muonPerf_pPbPbPb}
\end{figure}

Similar tracking algorithms~\cite{CMS:2018rym} are employed for all collision systems in Runs~1 and 2 as mentioned in Section~\ref{sec:ExperimentalMethods_Tracking}.
Based on the output of track reconstruction in the silicon tracker (``tracker track'') and in the muon system 
(``standalone muon track'') independently, muons are reconstructed following two complementary approaches.
In the ``tracker muon'' reconstruction, muons correspond to extrapolated tracker tracks matching at least one segment (local 
tracks built within each CSC or DT chamber) in the muon system; while the ``global muon'' reconstruction combines standalone 
muon tracks with hits from tracker tracks via an outside-in fit procedure~\cite{CMS:2012nsv, CMS:2018rym}.
Muon candidates are then fed into the PF algorithm~\cite{CMS:2017yfk} for a complete description of all individual particles 
per event, as discussed in Section~\ref{subsec:CMSapparatus}. 
The nominal muon detection acceptance of the CMS apparatus is defined by the minimum momentum needed to traverse the material 
in front of the first muon detector layer. As shown in Fig.~\ref{fig:muonPerf_pPbPbPb}, the \pt threshold is about 3.2\GeV 
for $\abs{\eta}\lesssim 1$ and decreases to about 0.5\GeV for $2.0 \lesssim \abs{\eta} < 2.4$. 

Muon reconstruction is found to be highly efficient from \pp to \pPb to \PbPb collisions, with the exception of the highest 
multiplicity \PbPb events, particularly those with more than about 1000 tracks~\cite{CMS-PAS-MUO-21-001}.
The tracker muon approach is inherently more efficient than the global one at low momentum since the latter starts from a 
standalone track reconstructed with segments in two or more muon detector layers. However, the inside-out approach has 
disadvantages that are amplified in HI events.
Because all of the tracker tracks are propagated to the muon system as potential muon candidates, the reconstruction 
efficiency---including the matching with muon detector segments---degrades with the occupancy in the silicon tracker layers.
Moreover, most of low-\pt muons only reach the innermost station and thus only match one segment. This increases the 
probability of misidentification from charged hadrons either produced promptly in the collision or by hadron showers developed within the HCAL.
These effects are dramatically enhanced with the large number of pions produced in nucleus-nucleus collisions.
For Run~1 and 2 analyses of \PbPb collision data, this misidentification rate was partially mitigated by selecting muon objects 
reconstructed as global muon tracks, at the cost of a reduced fiducial region for the measurements, as shown in 
the right panel of Fig.~\ref{fig:muonPerf_pPbPbPb}.
This selection is complemented by a set of identification criteria called \emph{hybrid-soft~muon~ID}, a version of 
the \emph{soft~muon~ID} optimized for HI events~\cite{CMS-PAS-MUO-21-001}.
Those two sets differ mainly in the addition of the global muon requirement and the removal of the selection of high-purity 
quality tracks~\cite{CMS:2014pgm} for hybrid-soft ID.
The physics analyses use the soft muon identification to select low-\pt (${<}20\GeV$) muons in \pp, \pPb, and 
ultraperipheral \PbPb collisions, and the hybrid-soft version for hadronic \PbPb events.

\subsection{\texorpdfstring{High-\ET}{High-ET} electron and photon reconstruction}
\label{sec:ExperimentalMethods_EGamma}

Electrons are found by combining information from the ECAL with charged-particle tracks. Photons are found using only ECAL information, 
but tracking information is also used to reject electrons or other sources of misidentified photon candidates.

\subsubsection{Electron reconstruction}

The electron reconstruction uses information from the pixel and strip tracker as well as the ECAL. Electrons traversing the silicon 
tracker can emit bremsstrahlung photons, which can also deposit energy in the ECAL. This causes a significant spread of the signals 
in the azimuthal direction. An algorithm for creating superclusters, which are clusters of signals from all particles passing through 
the ECAL, is used to estimate the proper energy of electrons and photons in the HI environment~\cite{CMS:2012oiv}. 

For Run~1 data, a Gaussian-sum filter algorithm, which combines ECAL superclusters with information from the pixel and strip tracker 
considering the bremsstrahlung emissions, is used to reconstruct electrons~\cite{Adam_2005}. Standard algorithms and 
identification criteria~\cite{CMS:2013lxn} were used for \pp and \pPb data, resulting in a reconstruction efficiency larger than 
95\%. For \PbPb collisions, the electron reconstruction efficiency is smaller, approximately 85\% for electrons from 
\Z~boson decays with $\pt>20\GeV$ and $\abs{\eta}<1.44$, because the tracking algorithm optimized for high-multiplicity 
events has a lower track reconstruction efficiency than that used for \pp collisions~\cite{CMS:2014dyj}. The requirements 
used to reduce background (Ref.~\cite{CMS:2013lxn} contains the variable definitions) include selections on: the energy-momentum 
combination between the supercluster and the track, the $\eta$ and $\phi$ spatial matching between the track and the 
supercluster, the supercluster shower shape width, the hadronic leakage (the ratio of energy deposited in the HCAL and 
ECAL, $H/E$), and a transverse distance of the closest approach to the measured PV. These selections eliminate most 
of the background while reducing the single-electron efficiency by only about 10 (5)\% in \PbPb (\pp) collisions. 

For Run~2 \PbPb data, electrons are identified as ECAL superclusters~\cite{CMS:2015xaf} matched in position and energy to 
tracks reconstructed in the tracker, using the PF algorithm as discussed in Section~\ref{subsec:CMSapparatus}. The electrons 
must have $\pt>20\GeV$ and their supercluster must be within the acceptance of the trigger, $\abs{\eta}<2.1$. The reconstruction 
efficiency is ${>}95\%$, whereas a multivariate discriminant, optimized using the \textsc{tmva} package~\cite{Hocker:2007ht}, 
selects electrons with a working point corresponding to 90\% identification efficiency and 80\% rejection of misreconstructed electrons~\cite{CMS:2015xaf}. 

\subsubsection{Photon reconstruction}

In analyses using Run~1 and Run~2 data, photons are reconstructed offline in \PbPb collisions using an island energy clustering 
algorithm~\cite{CMS:2012oiv} that is optimized for high-multiplicity HI events. The island algorithm builds ECAL superclusters in two steps:
\begin{itemize}
\item Defines clusters by adding energy of adjacent reconstructed hits in the ECAL using some building stopping criteria 
 (\eg, if the corrected energy of reconstructed hits is below some threshold or if the hits were already included in other clusters); 
\item Combines the clusters from previous step into superclusters. The criterion for merging the clusters requires a minimum 
value of its transverse energy of 1\GeV and the clusters should be located in a spatial strip of $\deta=0.07$ and $\dphi=0.8$.
\end{itemize}
The photon momentum is calculated with respect to the location of the reconstructed primary interaction vertex. If multiple 
vertices are reconstructed, the vertex with the largest scalar sum of the transverse momenta of the associated tracks is 
selected. For \pp data, the island algorithm is used for Run~1, while the global event description~(GED) algorithm~\cite{CMS:2015myp} 
is used for Run~2. The GED uses a similar idea as the island method to build the superclusters~\cite{CMS:2015xaf}. 
It uses additional variables with respect to the discrimination between converted and unconverted photons. In addition, 
there are considerable differences in the procedure for applying corrections to the energy of the clusters.

\begin{figure}[ht]
    \centering
    \includegraphics[width=0.95\linewidth]{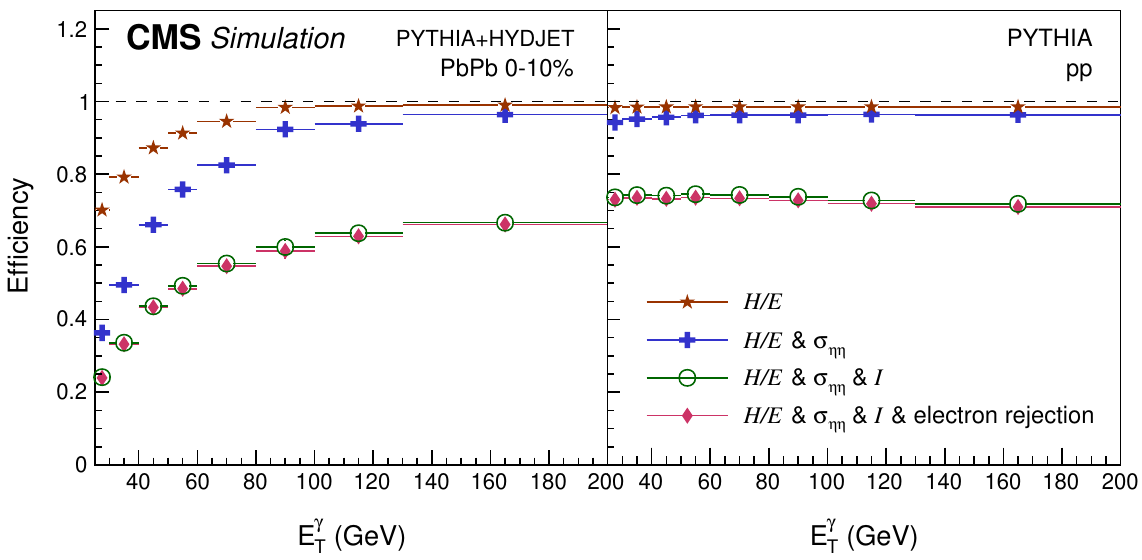}
    \caption{Isolated photon detection efficiency in $\abs{\eta}<1.44$ as a function of \etg obtained 
    from MC simulations. Left: \PbPb collisions in the 0--10\% centrality range. Right: \pp collisions. Both the \PbPb 
    and \pp collisions are at 5.02\TeV. The different colors represent efficiencies reached for successive application 
    of the listed selection criteria: ratio of HCAL over ECAL energies $H/E<0.1$, EM shower shape variable $\sigma_{\eta\eta}<0.01$, 
    isolation variable $I<1\GeV$, and electron rejection criterion. \FigureFrom{CMS:2020oen}}
    \label{fig:performance_photons}
\end{figure}

Additional criteria to reject electrons that are misidentified as photons and misidentified photons caused by highly 
ionizing particles interacting directly with the silicon avalanche photodiodes in the ECAL barrel readout are applied 
following the procedure described in Ref.~\cite{CMS:2012oiv}.
Several additional criteria are applied: corrections for UE contamination and the effects from the material in front 
of the ECAL, selections to eliminate high-\pt hadron contamination, and an isolation ($I$) requirement~\cite{CMS:2020oen}. 
The latter one is defined as the sum of transverse energies in the ECAL and HCAL (excluding the photon component) and 
the transverse momenta of all reconstructed tracks with $\pt>2\GeV$ inside the cone $\Delta R=\sqrt{\smash[b]{(\deta)^2+(\dphi)^2}}=0.4$. 
The efficiency to detect isolated photons as a function of their transverse energy (\etg), extracted from 
MC simulations, is shown in Fig.~\ref{fig:performance_photons}.

\subsection{Jet reconstruction}
\label{sec:ExperimentalMethods_JetMET}

Jet reconstruction for CMS data takes PF objects as the input set of constituents for the iterative recombination 
algorithms encoded in the \FASTJET software package~\cite{Cacciari:2011ma,Cacciari:2006sm}. The algorithm can combine 
either the PF objects themselves or, instead, a set of objects modified by subtraction of their UE contributions. 
The subtraction method used for HI collision data, which differs from the approach taken for high-PU \pp collisions, 
is detailed in Section~\ref{subsubsec:ExperimentalMethods_UE_Jet}.

The iterative recombination family of jet-finding algorithms takes as a starting point a set of particles or particle ``proxies'' 
(such as calorimeter towers or PF objects). The algorithm proceeds through all combinations of two entries in the list of input 
objects and determines whether or not to merge a given pair by finding the minimum values of $d_{ij}$ and $d_{iB}$, defined as
\begin{equation}
\label{eqn:iterRecombAlgoMetric1}
d_{ij} = \text{min}((p_{\mathrm{T},i})^{2p}, (p_{\mathrm{T},j})^{2p})\frac{\Delta R_{ij}^{2}}{R^2}\quad 
d_{iB} = (p_{\mathrm{T},i})^{2p}
\end{equation}
where $p_{\mathrm{T},i}$ is of particle $i$, $\Delta R_{ij} = \sqrt{\smash[b]{(y_i - y_j)^2 + (\phi_i - \phi_j)^2}}$ 
is the 2-dimensional distance between the two objects in rapidity and azimuthal angle, $R$ is the so-called ``distance'' parameter, and 
the parameter $p$ typically takes values of $p=1$, $p=0$, or $p=-1$. If $\text{min}(d_{ij}) < \text{min}(d_{iB})$, the $i$-th 
and $j$-th objects are combined in a 4-vector sum, the resulting combination replaces the two particles in the list, and the list 
of objects is scanned again. Otherwise, if $\text{min}(d_{iB}) < \text{min}(d_{ij})$, object $i$ is removed from the list as 
a final-state jet. Iterations continue until the list is exhausted, and the set of objects removed using the $d_{iB}$ criterion contains the resulting jets.

The typical choices for the value of parameter $p$, $p=1$, $p=0$, and $p=-1$, correspond to the \kt, Cambridge--Aachen, and 
anti-\kt algorithms, respectively. The \kt algorithm preferentially clusters soft particles nearby in $\eta$ and $\phi$, the 
Cambridge--Aachen one clusters the closest particles irrespective of their momentum, and the anti-\kt choice preferentially 
clusters the hardest particles with all nearby particles. As a result, anti-\kt jets have a regular, cone-like shape. 
For experimental reasons relating to background subtraction and energy calibration, anti-\kt is the preferred choice for jet 
reconstruction in HI data. In specific studies of jet substructure (discussed in Section~\ref{ssec:Substructure_JetSubstructure}), 
jets originally clustered with the anti-\kt may have their resulting constituent set reclustered using another algorithm but, 
outside of this exception, the anti-\kt algorithm is used.

\subsection{Treating the underlying event in physics object reconstruction}
\label{sec:ExperimentalMethods_UE}

Compared to \pp collisions, one of the primary additional challenges faced by HI analyses is the large UE produced by the many 
binary \NN collisions that occur when nuclei collide. As one example, in order to extract the properties of the fragmenting 
hard-scattered parton generating a jet, corrections must be made for the significant additional energy the UE can add to the 
reconstructed jet. A similar problem occurs in \pp collisions because of the high rates of PU, requiring correction for the 
additional event activity produced. However, in \PbPb collisions, the additional activity/underlying event to be subtracted is
\begin{enumerate}
\item{typically a much larger fraction of the signals of interest},
\item{anisotropic in azimuthal angle due to collective flow (as discussed in Section~\ref{sec:HydroQGP})},
\item{originates from a volume comparable to the diameter of a Pb nucleus ($\approx$10\unit{fm}), as opposed to being spread 
out across many vertices along the beam direction, which is the case for PU in \pp}.
\end{enumerate}
As a result of these three differences, techniques developed to correct for the UE contributions in \pp data, such as vertexing 
techniques for the removal of PU contributions to jet energy, are frequently ineffective in the HI environment. 
The following subsections detail how physics object reconstruction is modified to account for these differences. 

\subsubsection{Correcting for the underlying event in jet reconstruction}
\label{subsubsec:ExperimentalMethods_UE_Jet}

As discussed in Section~\ref{sec:ExperimentalMethods_JetMET}, jets are typically reconstructed using PF objects as discussed 
in Section~\ref{subsec:CMSapparatus}. In \pp collisions, the additional activity energy primarily comes from PU collisions which 
are separable by longitudinal vertex position. The PF objects not originating from the vertex of the hard-scattering that produced 
the jets can be identified and removed prior to clustering. This technique is called charged-hadron subtraction~\cite{CMS-JME-18-001}. 
In HI events, both the hard scattering and the additional energy from the UE share a single vertex and there is no 
possibility of determining which PF objects come from the hard scattering.
As a result, a two-part approach to correcting for UE energy contributions is typically employed for HI collision data. 
A determination of the UE contribution as a function of $\eta$, $\phi$, and centrality is followed by an algorithm by which the UE is subtracted. 

The first half of this two-part approach follows the iterative pedestal PU subtraction procedure~\cite{Kodolova_JetReco} 
modified to account for the azimuthal modulation introduced by hydrodynamic flow (introduced in Section~\ref{sec:IntroductionOverview} 
and discussed in detail in Section~\ref{sec:HydroQGP}), and is similar to the ALICE event-by-event fitting method~\cite{ALICE:2015efi}. 
This procedure first estimates the energy from the UE by taking the average energy in $\eta$ strips defined by the HCAL tower geometry. 
As this estimator is known to be biased by the presence of jets, jet finding and correction is performed using the biased estimator, 
and regions identified as containing jets are excluded for a second iteration of determining the UE contribution. 
Because the HI UE is asymmetric in $\phi$ due to the presence of hydrodynamic flow, a $\phi$-dependence is added to this $\eta$-dependent 
UE estimation via event-by-event fits of track multiplicities in $\phi$ following the ALICE example. 
Note that this $\phi$-dependent correction was added to jet reconstruction late in the Run~2 period, and 
therefore many analyses detailed in this document do not include this step.

The second half of the UE correction procedure follows the jet-by-jet constituent subtraction method~(CS)~\cite{CSsub1,CSsub2}. 
In this approach, ``ghost'' particles, or four-vectors of infinitesimally small energies, are randomly distributed in $\eta$-$\phi$ 
space and the number that are included in a particular jet is used to calculate its area. 
These ghosts can be clustered into the jets without modifying their kinematics. Once this clustering is complete, the 
ghosts are assigned an UE energy (found using the estimator in step one) according to their $\eta$-$\phi$ position, 
and that energy is removed from the jets they are part of. If the total energy of the ghosts exceeds the total jet energy, 
the jet is removed rather than being assigned a negative energy, as it is taken to be a misidentified jet coming from the UE.

\begin{figure}[ht!]
    \centering
    \includegraphics[width=0.95\linewidth]{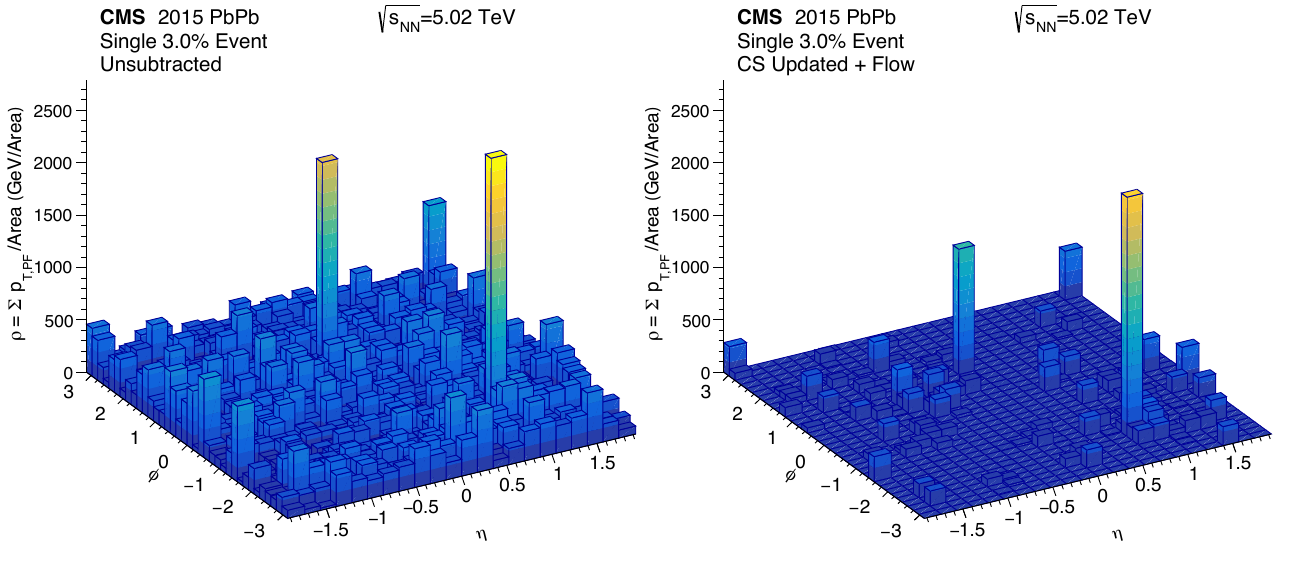}
    \caption{Left: Distribution of PF pseudotowers in $\eta$-$\phi$ in a single central (top 3\%) event in \PbPb 
    collisions before subtraction, with the $z$ axis showing the corresponding tower energy per unit tower area. 
    Right: The same event after full subtraction with flow modulation is applied. \FigureFrom{CMS-DP-2018-024}.}
    \label{fig:jetFinding}
\end{figure}

\begin{figure}[ht!]
    \centering
    \includegraphics[width=0.45\linewidth]{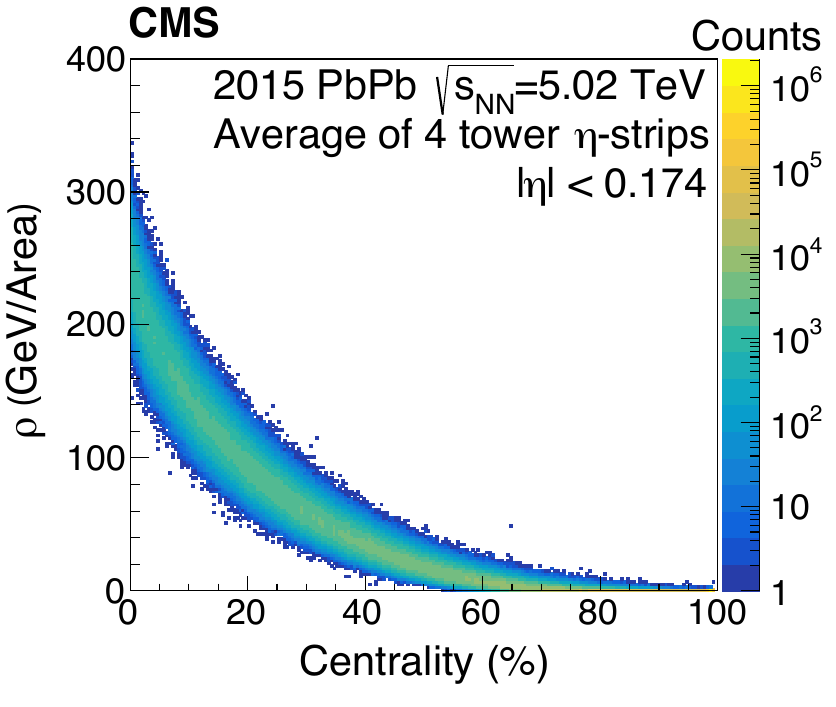}
    \caption{Distribution of $\rho$, the UE energy per unit area, as a function of centrality, 
    found using the central-$\eta$ strip of PF pseudotowers. \FigureFrom{CMS-DP-2018-024}.}
    \label{fig:jetRho}
\end{figure}

\begin{figure}[ht!]
    \centering
    \includegraphics[width=0.6\linewidth]{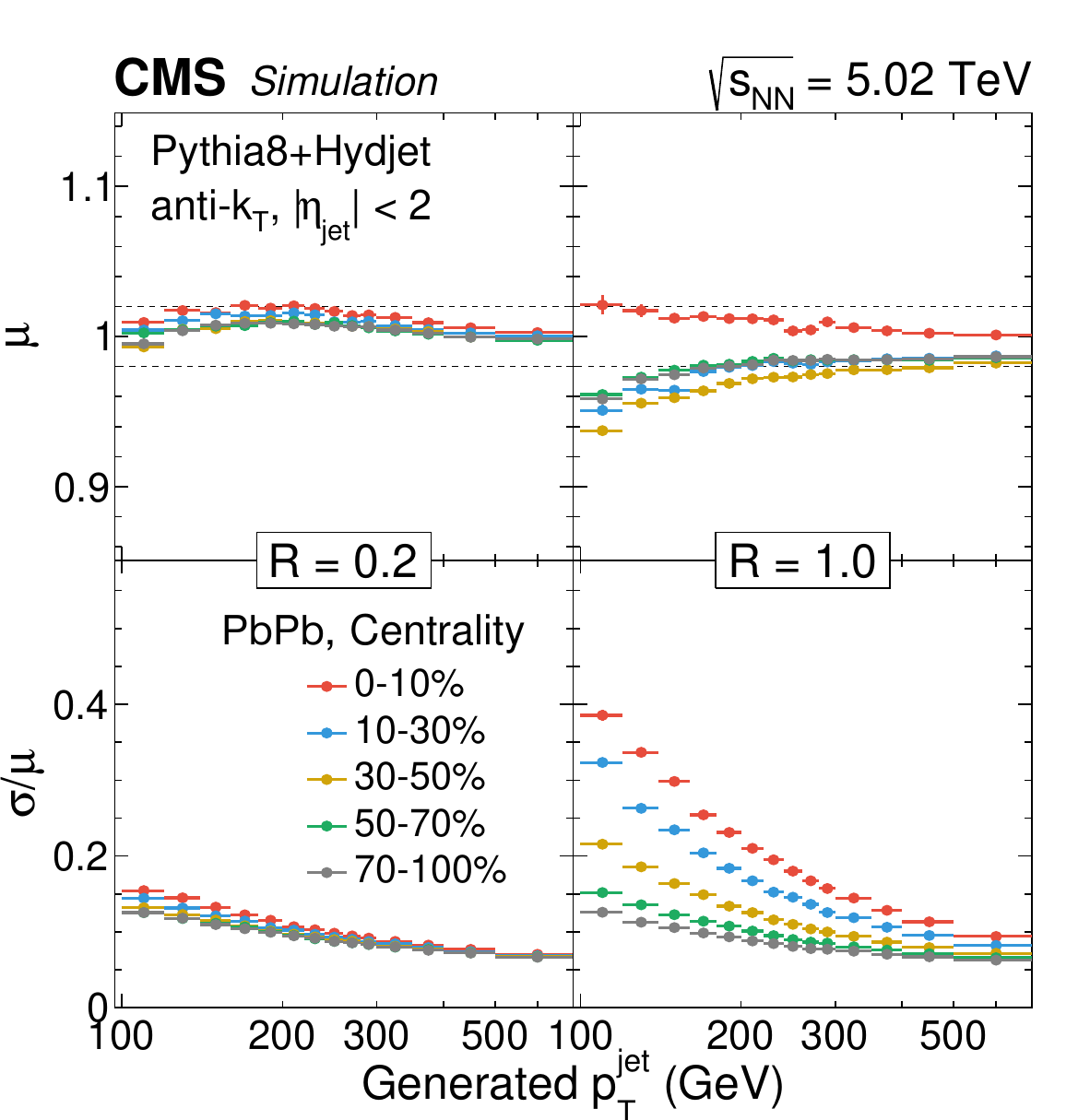}
    \caption{Performance of jet reconstruction in the HI environment for jet distance parameters of $R=0.2$ (left) and $R=1.0$ (right). 
    The jet energy scale is shown in the upper panels, while the jet energy resolution is plotted in the lower panels. \FigureFrom{CMS:2021vui}}
    \label{fig:jesAndJERForR0p2and1p0}
\end{figure}

The resulting performance of this hybrid iterative pedestal/CS subtraction is documented in Ref.~\cite{CMS-DP-2018-024}. 
Figure~\ref{fig:jetFinding} (left and right panels) shows the full subtraction as applied to a single central \PbPb event. 
As PF objects do not have well-defined areas, they have first been combined into pseudotowers in this illustration, 
with their \ET sums restricted in $\eta$-$\phi$ as defined by the HCAL tower geometry. Figure~\ref{fig:jetRho} shows the 
distribution of UE energy per unit area ($\rho$) as a function of the centrality class. Here, $\rho$ is estimated by averaging 
the energy over an area spanning a central-$\eta$ strip corresponding to four HCAL tower widths and covering the full detector in $\phi$. 
The increasing value of $\rho$ with centrality is illustrative of the difficulty of accounting for the UE contributions in the 
most central \PbPb events. Figure~\ref{fig:jesAndJERForR0p2and1p0} is adapted from Ref.~\cite{CMS:2021vui} and shows the resulting 
jet energy scale and resolution after the application of this UE subtraction procedure for both small and large jet distance parameters, 
$R=0.2$ and $R=1.0$. The performance degrades as the distance parameter $R$ increases, as the greater transverse area of the jet cone 
increases the contribution of the UE. To mitigate this degrading performance, jets with large $R$ are only studied at higher \pt values, 
thereby reducing the fractional contribution of the UE.

\subsubsection{Correcting for the UE in photon isolation}
\label{ssec:ExperimentalMethods_UE_Pho}

High-energy isolated photons are produced mostly in hard quark-gluon scatterings (in contrast to nonisolated ones which arise from hadron 
decays and parton fragmentation)~\cite{dEnterria:2012kvo}, and are identified by the absence of other particles produced within a cone surrounding the photon candidate.
The presence of a large UE in \PbPb collisions poses a challenge because of the presence of many other soft processes.
To determine the UE contribution around a photon candidate, the local surroundings within a cone in pseudorapidity and azimuthal angle 
around the centroid of the photon is examined to identify any hadronic activity that surpasses a specific veto threshold 
(typically 5\GeV for isolated photon analyses). 

When measuring $I$ (Section~\ref{sec:ExperimentalMethods_EGamma}) in \PbPb data, the UE contribution is removed by subtracting 
the average value of the energy in a rectangular area with a length of $2\Delta R$ in the $\eta$-direction around a photon candidate 
and a width of $2\pi$ in the $\phi$-direction. However, no UE correction is applied in \pp data.

\subsection{Heavy-flavor hadron reconstruction and identification}
\label{sec:Heavy_flavor_hadron_reco}

In HI collisions, heavy-flavor hadrons can be identified in several decay modes involving charged hadrons 
(pions, kaons, and protons) and/or leptons. 
During Run~1, these heavy-flavor hadrons were identified by performing a reconstruction step to identify a secondary vertex~(SV) 
that was the origin of the decay particles (without identifying them as pions, kaons, or protons). This was followed by simple 
selections on individual topologically motivated variables. These included the impact parameter of the reconstructed momentum 
vector of the decaying hadron with respect to the PV, as well as the angle between that momentum vector and a line connecting 
the PV and SV~\cite{Sirunyan:2017plt}. 

For Run~2, machine learning approaches started to be incorporated into the identification procedure. The primary method uses 
boosted decision trees (trained using simulations) from the \textsc{TMVA} package. Unfortunately, systematic uncertainties in 
final results are dominated by uncertainties from these ML procedures~\cite{CMS:2020bnz, CMS:2021qqk, CMS:2022vfn}, primarily 
because the MC simulations used in the training do not describe well the kinematics of heavy-flavor hadrons. 

\begin{figure}[ht]
    \centering
    \includegraphics[width=0.5\textwidth]{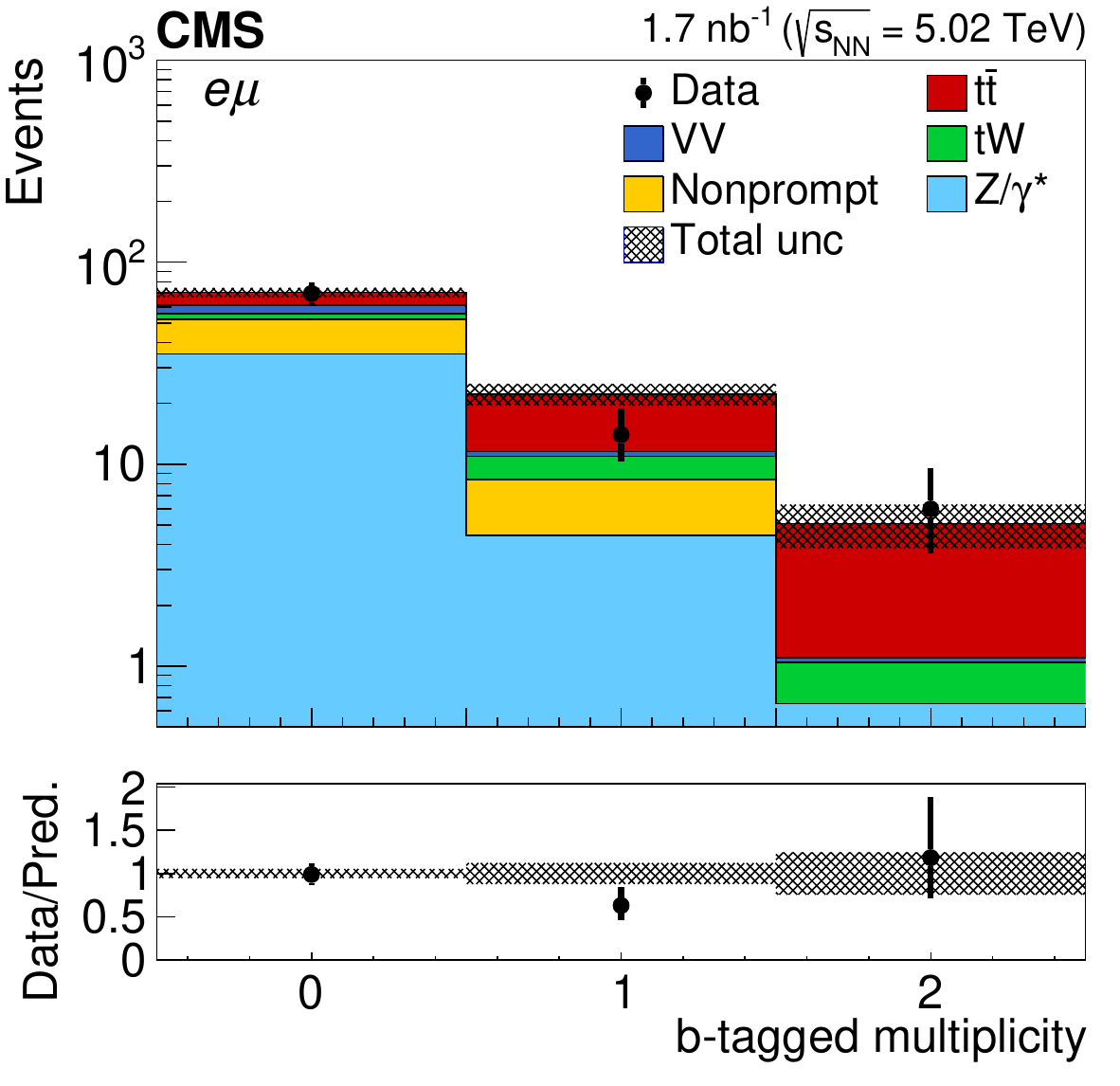}
    \caption{Multiplicity distribution of the \PQb-tagged jets in a top quark pair enriched final state using \PbPb collisions. 
    The distribution of the main background is taken from the data. Backgrounds and \ttbar signal are shown with the filled 
    histograms and data are shown with the markers. The vertical bars on the markers represent the statistical uncertainties 
    in data. The hatched regions show the uncertainties in the sum of \ttbar signal and backgrounds. The lower panel displays 
    the ratio of the data to the predictions with bands representing the uncertainties in the predictions. \FigureFrom{CMS:2020aem}}
    \label{fig:ttbar_bjets}
\end{figure}

Similarly to heavy-flavor hadrons, the identification of \PQb jets is based on kinematic variables related to the relatively long 
lifetime and large mass of \PQb hadrons. Indeed, heavy-flavor jet identification techniques exploit the properties of the hadrons 
in the jet to discriminate between jets originating from bottom or charm quarks and those originating from light-flavor quarks or gluons. 
Several improvements have been made in heavy-flavor jet identification algorithms for Run~2 data, including multivariate 
analysis developments. For jets with \pt in the range found in simulated top quark pair events, an efficiency of 70\% for the correct 
identification of a \PQb jet, along with a probability of 1\% of misidentifying a light-flavor jet, was achieved. The improvement 
in relative efficiency is about 15\% (at the same misidentification probability) compared to previous algorithms~\cite{CMS:2017wtu}. 
Figure~\ref{fig:ttbar_bjets} shows the number of jets ``tagged'' as originating from \PQb quarks (referred to as ``\PQb-tagged jets'') 
in events progressively enriched with top quarks, \ie, going from no \PQb-tagged jets up to a \PQb-tagged jet multiplicity of 
2~\cite{CMS:2020aem}. The application of sophisticated \PQb-tagging algorithms is therefore found to enhance the signal 
(depicted in red in Fig.~\ref{fig:ttbar_bjets}) over background ratio in \PbPb collisions, as is the case for standard \pp analyses. 

\clearpage

\section{The initial state of the collisions}
\label{sec:InitialState}

This section discusses a number of measurements by the CMS Collaboration that constrain our understanding of the initial state 
in \pp, \pPb, and \PbPb collisions. The initial state comprises the spacetime (or energy-momentum) distribution of parton and nucleon 
constituents just before a collision. The large number of nucleons present in a HI, such as lead, gives rise to interesting emergent 
properties of the initial state in high energy collisions. For example, a proposed saturation of the parton density may have measurable 
consequences at energies achievable at current accelerator facilities using HI probes. While the initial state of an isolated nucleon, 
or a nucleon that exist within a nucleus, is of interest in its own right, this state has a profound influence on the evolution of a 
nuclear collision, including the formation and properties of the QGP created in \PbPb collisions at the LHC.

Many CMS Run~1 and 2 measurements have helped define the initial-state properties relevant for experiments at the LHC. Of particular 
interest are the small- and high-$x$ partonic distributions in protons and nuclei. These measurements have also provided a test 
of the Glauber model that is used to simulate the initial geometry of heavy ion collisions. Here, we highlight the diverse experimental 
program, including a wide range of measurements designed to access the properties of the initial state, extending from heavy EW boson 
and high-\pt jet production in hadronic collisions to heavy-flavor photoproduction in UPCs.

\subsection{Constraining nuclear parton distribution functions with hard probes}
\label{sec:InitialState_EW}

Parton distribution functions~(PDFs) are key quantities used in the description of the initial state of a hadronic collisions. 
They describe the fraction $x$ of the total momentum of an isolated nucleon that each parton carries. When the nucleon resides within a 
nucleus, these distributions are known as nPDFs. Calculation of nPDFs from first principles is challenging because of their intrinsic 
nonperturbative nature, so experimental input is required to establish reference points at different values of $x$ and of the momentum 
transfer scale, \QTwo. Global fits of these reference data and the Dokshitzer--Gribov--Lipatov--Altarelli--Parisi~(DGLAP) evolution 
equations~\cite{Altarelli:1977zs,Dokshitzer:1977sg,Gribov:1972ri} can then be employed to infer the values of nPDFs at a given value 
of $x$ and \QTwo. The accuracy of these fits and extrapolations is largely dictated by the precision and $(x,\QTwo)$ coverage of the 
input experimental data, meaning that high-quality precision measurements in a large kinematic range are extremely valuable. 
Studies of EW boson production are examples of such measurements. 

Massive EW bosons, \ie, \PW and \PZ bosons, have lifetimes on the order of 0.1\unit{fm} and can decay to final states that include 
highly energetic leptons. These leptons do not have a QCD color charge and, consequently, do not interact strongly with other particles. 
Thus, massive EW bosons and their decay products should be relatively unmodified by the existence of any hot nuclear medium. 
Consequently, they encode information about the earliest stages of a HI collision and can be used to constrain the corresponding nPDFs and, 
by extension, the collision's initial conditions.

The \pPb collision system is an ideal environment to carry out measurements of nPDFs using EW bosons. One advantage of using 
proton-lead collisions is that one can use the better known PDF of the proton to cleanly probe the ``target'' nuclear PDF. 
In \PbPb collisions, a mixture of two unknown nuclear PDFs is used, which makes the nPDF constraint less strong. 
The asymmetric collision system provides access to two different regions of initial parton $x$ for a given value 
of $\abs{\eta}$, greatly expanding the $x$ coverage of several measurements as compared to a symmetric collision system. 
We adopt the convention that positive $\eta$ values indicate the proton-going, or ``forward'', side of the detector, \ie, 
the side that preferentially selects low-$x$ partons in the target Pb nucleus. Similarly, negative $\eta$ values denote 
the lead-going, or ``backward'', side which preferentially measures high-$x$ partons in the target nucleus. 
During \pPb collisions at the LHC, there is an asymmetry in the per-nucleon energy in each beam that causes an offset of 0.465 
units of rapidity between the laboratory and nucleon-nucleon center-of-mass reference frames. Results are presented as 
functions of the pseudorapidity $\eta_{\mathrm{CM}}$ and rapidity $y_{\mathrm{CM}}$ calculated in the center-of-mass frame. 
An additional benefit of \pPb collisions is that they typically have less event activity than \AonA collisions at a similar 
center-of-mass energy, allowing more precise lepton identification and reconstruction.

\begin{figure}[ht]
    \centering
    \includegraphics[width=0.45\linewidth]{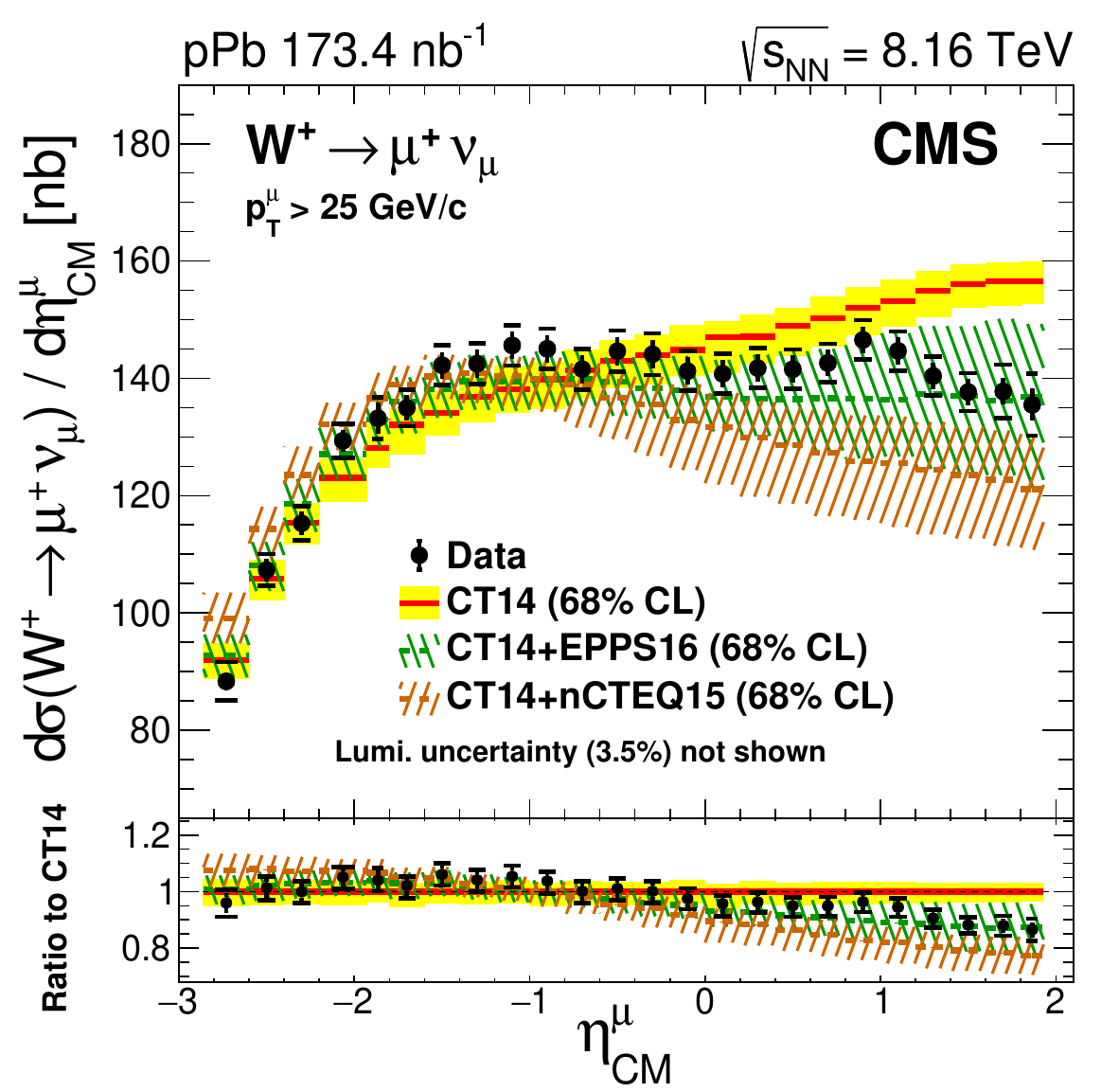}
    \includegraphics[width=0.48\linewidth]{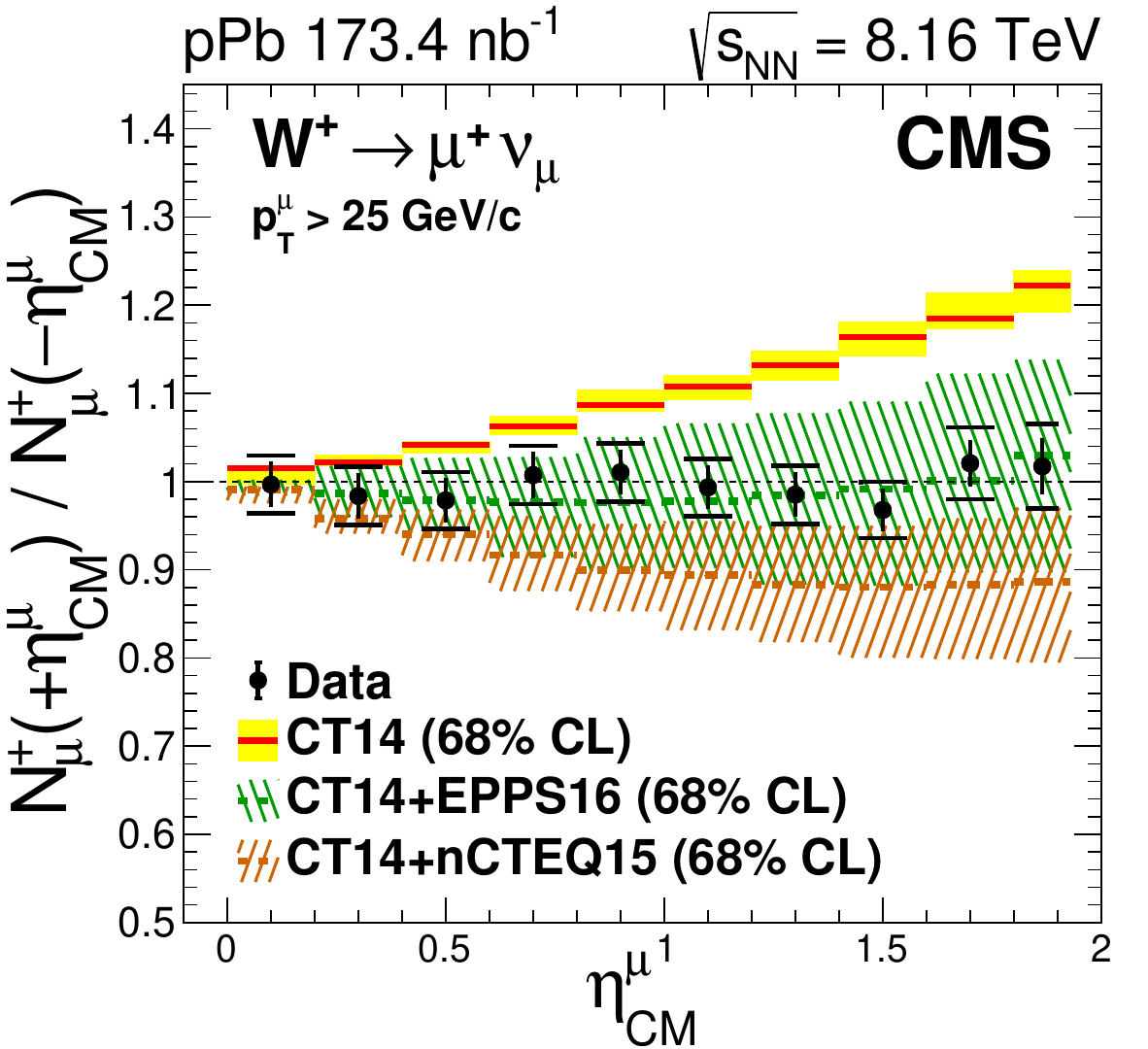}
    \caption{The differential cross sections (left) and forward-backward ratio for decay muon yields (right) for the 
    process $\PWp\to\PGmp\PGnGm$ versus muon pseudorapidity in the center-of-mass frame ($\eta_{\mathrm{CM}}$). Black horizontal 
    lines above and below the data points represent the quadrature sum of statistical and systematic uncertainties, 
    whereas the vertical bars show the statistical uncertainties only. The NLO calculations with CT14 PDF, and CT14+nCTEQ15 
    and CT14+EPPS16 nPDFs are displayed, including their 68\% confidence interval uncertainty bands. The ratios of data, 
    CT14+nCTEQ15 and CT14+EPPS16 with respect to CT14 are shown in the lower left panel. A global integrated luminosity 
    uncertainty of 3.5\% in the cross section is not shown. (Figure compiled from Ref.~\cite{CMS:2019leu}.)}
    \label{fig:8TeVWpPb}
\end{figure}

At leading order~(LO) in perturbation theory, \PW bosons are produced primarily through the annihilation of quarks and 
antiquarks, \eg, $\PQu\PAQd\to\PWp$ and $\PQd\PAQu\to\PWm$. Thus, measurements of the \PW boson production can give 
access to the light quark and antiquark nPDFs. Additionally, the charge asymmetry of \PW boson production enables 
disentangling of the \PQu and \PQd quark nPDF contributions individually. The left panel of Fig.~\ref{fig:8TeVWpPb} 
shows a measurement of the differential cross section of \PW boson production in $\sqrtsNN = 8.16\TeV$ \pPb collisions as 
a function of the decay lepton $\eta_{\mathrm{CM}}$~\cite{CMS:2019leu}. The result is for the $\PWp\to\PGmp\PGnGm$ process. 
Also shown, as shaded bands, are next-to-leading order~(NLO) pQCD MC predictions calculated with the MCFM v8.0~\cite{Boughezal:2016wmq} 
program interfaced with the CT14~\cite{Dulat:2015mca} free proton PDF, as well as the combined CT14+EPPS16~\cite{Eskola:2016oht} 
and CT14+nCTEQ15~\cite{Kovarik:2015cma} nPDFs. In general, the data agree better with the predictions using nPDFs than with 
those using the free proton PDFs, with a slight enhancement in the backward region and a suppression in the forward region. 
These trends correspond to the ``antishadowing'' and ``shadowing''~\cite{Armesto:2006ph} regions of the nPDF, respectively. 
The results are consistent with earlier CMS analyses of \PW boson production in \pPb collisions at 
the lower collision energy of $\sqrtsNN = 5.02\TeV$~\cite{CMS:2015ehw}.

A more precise test of the nPDF predictions was performed by taking forward-backward ratios of the yields of muons resulting 
from \PW boson decays to enable cancellation of systematic uncertainties. This quantity is shown in the right panel of 
Fig.~\ref{fig:8TeVWpPb} for \PWp bosons. The data clearly favor models containing nuclear effects. 
Similar conclusions were reached with \PWm bosons. The comparison of this measurement to various models is the first clear 
demonstration of the nuclear modification of quark PDFs using EW bosons in nuclear collisions~\cite{CMS:2019leu}. Because 
of the unprecedented precision of these measurements (as can be seen by comparing the measurement uncertainties to the model 
uncertainties in the right panel of Fig.~\ref{fig:8TeVWpPb}), state-of-the-art nPDF models such as EPPS21~\cite{Eskola:2021nhw}, 
nCTEQ15WZ~\cite{Kusina:2020lyz}, 
nNNPDF2.0~\cite{AbdulKhalek:2020yuc},
nNNPDF3.0~\cite{AbdulKhalek:2022fyi}, and TUJU21~\cite{Helenius:2021tof} have all incorporated 
these results into their global fits to extract the parton densities in heavy nuclei.

\begin{figure}[ht!]
    \centering
    \includegraphics[width=0.55\linewidth]{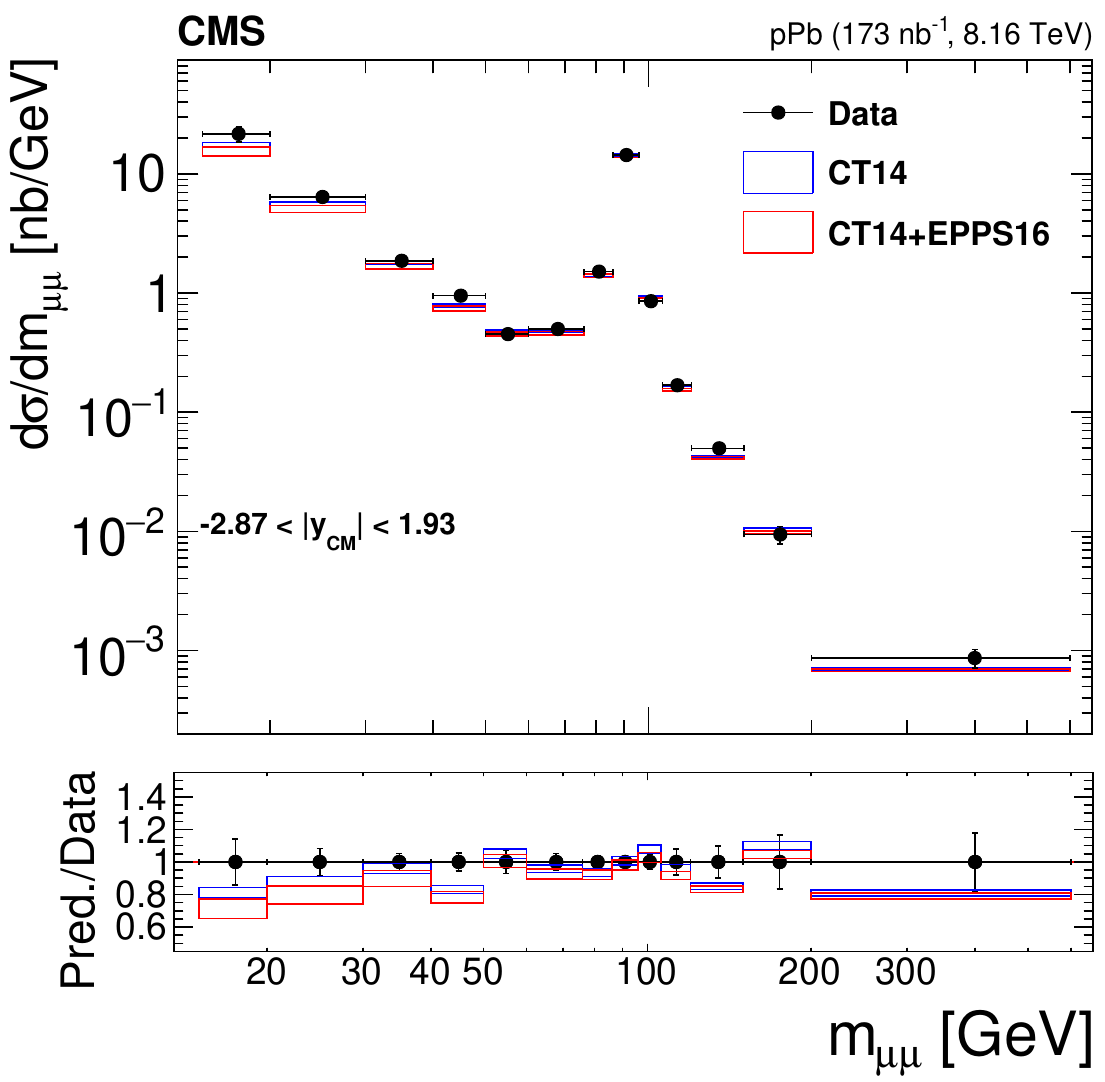}\\
    \includegraphics[width=0.49\linewidth]{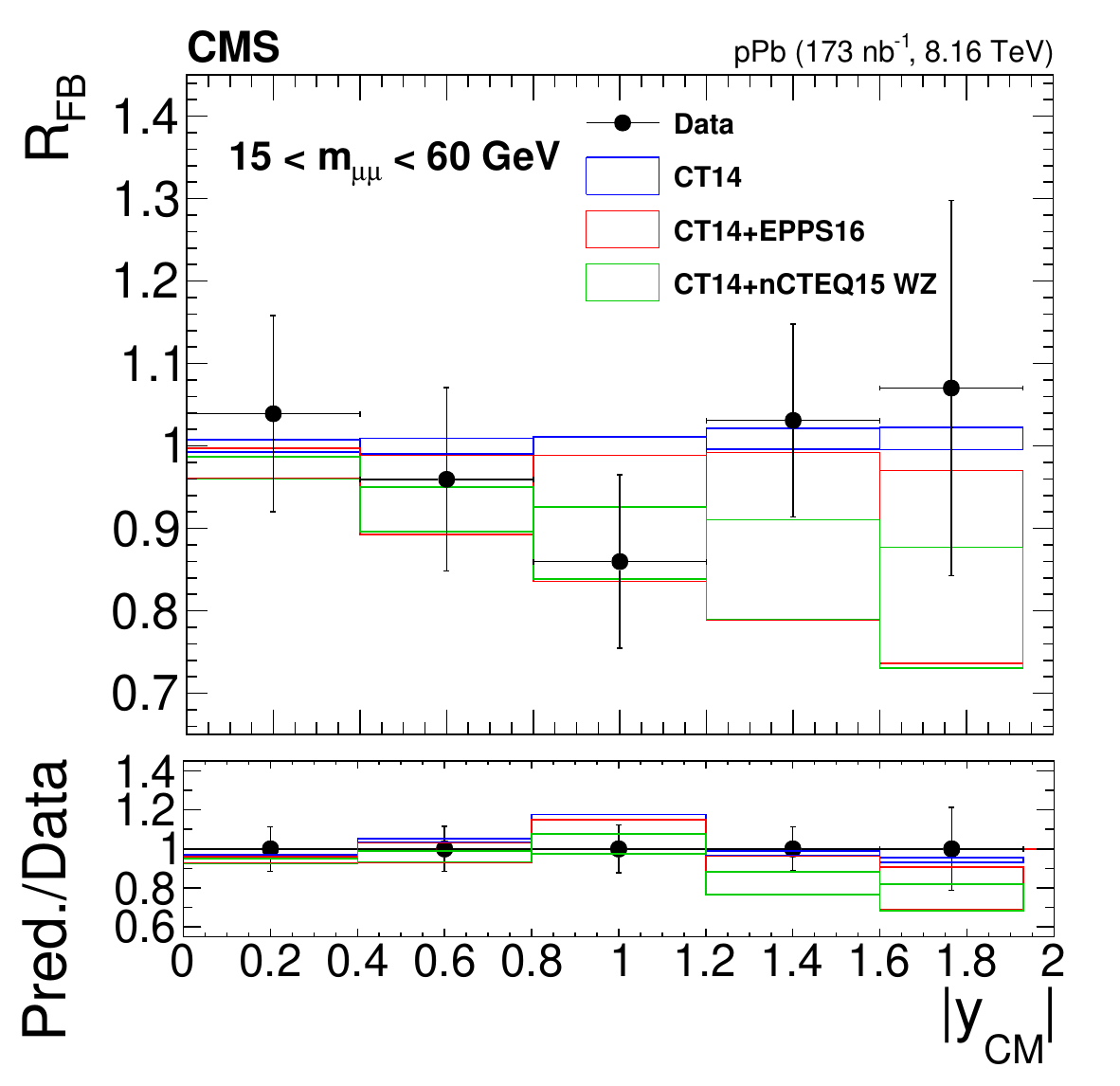}
    \includegraphics[width=0.49\linewidth]{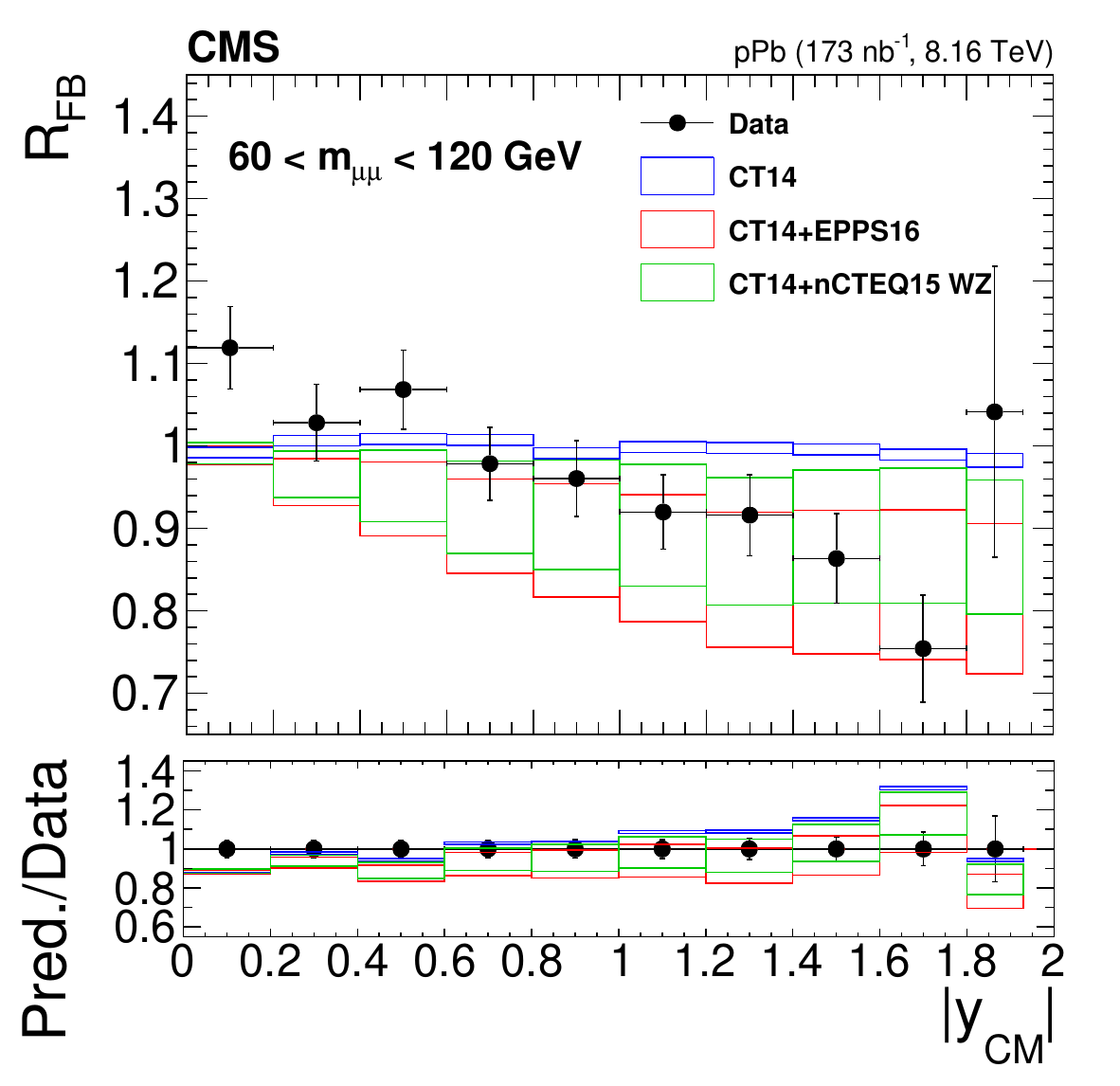}
    \caption{Differential cross section for the Drell--Yan process measured in the muon channel as a function of the dimuon invariant 
    mass (upper) and the forward-backward ratios for $15 < m_{\mumu} < 60\GeV$ (lower left) and $60 < m_{\mumu} < 120\GeV$ 
    (lower right). Error bars represent the total measurement uncertainty. Theory predictions from the \POWHEG NLO generator 
    using the CT14 PDF (blue), or the CT14+EPPS16 (red) or CT14+nCTEQ15WZ (green) nPDF sets are shown. The standard deviation of 
    the nPDF uncertainties are shown by the boxes. Ratios of theory predictions over data are shown in the lower panels. 
    \FigureCompiledSingular{CMS:2021ynu}}
    \label{fig:8TeVDYpPb}
\end{figure}

The production of oppositely charged lepton pairs via $\PQq\PAQq$ annihilation in the $s$-channel through the exchange of a \PZ boson or 
virtual photon $\PGg^*$ is known as the neutral Drell--Yan~(DY) process. Like \PW boson production, this process is sensitive to 
quark nPDFs. The upper panel of Fig.~\ref{fig:8TeVDYpPb} shows a measurement of the differential cross section of DY dimuons as a 
function of their invariant mass for \pPb collisions at $\sqrtsNN = 8.16\TeV$~\cite{CMS:2021ynu}. This measurement probes a large 
range of invariant mass from 15 to 600\GeV. A clear peak in the data can be seen, corresponding to the \PZ boson mass. 
For DY dimuons in this mass range, a measurement of the cross section as a function of the dimuon rapidity shows features similar to 
those observed for \PW bosons, \ie, enhancement compared to the CT14 PDF in the backward antishadowing region and suppression in the 
shadowing region. The forward-backward ratio of this cross section as a function of dimuon rapidity $\abs{y_{\mathrm{CM}}}$ is shown 
in the lower right panel of Fig.~\ref{fig:8TeVDYpPb} for a mass selection around the \PZ boson peak ($60 < m_{\mumu} < 120\GeV$). 
The error bars represent the total measurement uncertainties. Theoretical predictions from the \POWHEG NLO~\cite{Alioli:2010xd,Alioli:2008gx} 
generator using the CT14~\cite{Dulat:2015mca} free proton PDF, as well as the CT14+EPPS16~\cite{Eskola:2016oht} and 
CT14+nCTEQWZ~\cite{Kusina:2020lyz} nPDFs, are shown by blue, red, and green boxes, respectively. The data strongly deviate 
from the CT14 prediction for large values of $\abs{y_{\mathrm{CM}}}$ but are consistent with the nPDF models. 
Similar trends were observed in earlier CMS measurements of \PZ boson production in $5.02\TeV$ \pPb collisions~\cite{CMS:2015zlj}. 
The precision of the measurement is better than the nPDF model uncertainties, including the nCTEQWZ model, which was already 
updated to include the previously discussed CMS \PW boson data. Thus, the DY data are expected to further improve the 
understanding of quark and antiquark nPDFs.

At lower invariant masses, the dynamics of the DY process are increasingly dictated by virtual photon exchange and, therefore, 
probe a region of lower $x$ and energy scale \QTwo compared to the production of \PZ or \PW bosons. For the first time in collisions 
of nuclei, similar cross section and forward-backward ratio measurements were performed in a lower mass region of 
$15 < m_{\mumu} < 60\GeV$, as shown in the lower left panel of Fig.~\ref{fig:8TeVDYpPb}. Although the precision of the measurement 
does not currently allow for strong constraints of the various models, these measurements represent an important step 
towards expanding the kinematic region in which the DY process can be used to understand nPDF effects. For example, 
Ref.~\cite{Helenius:2021tof} found that the inclusion of next-to-NLO~(NNLO) corrections can significantly increase 
the ability of nPDF models to describe these low-mass data. 

\begin{figure}[ht]
    \centering
    \includegraphics[width=0.75\textwidth]{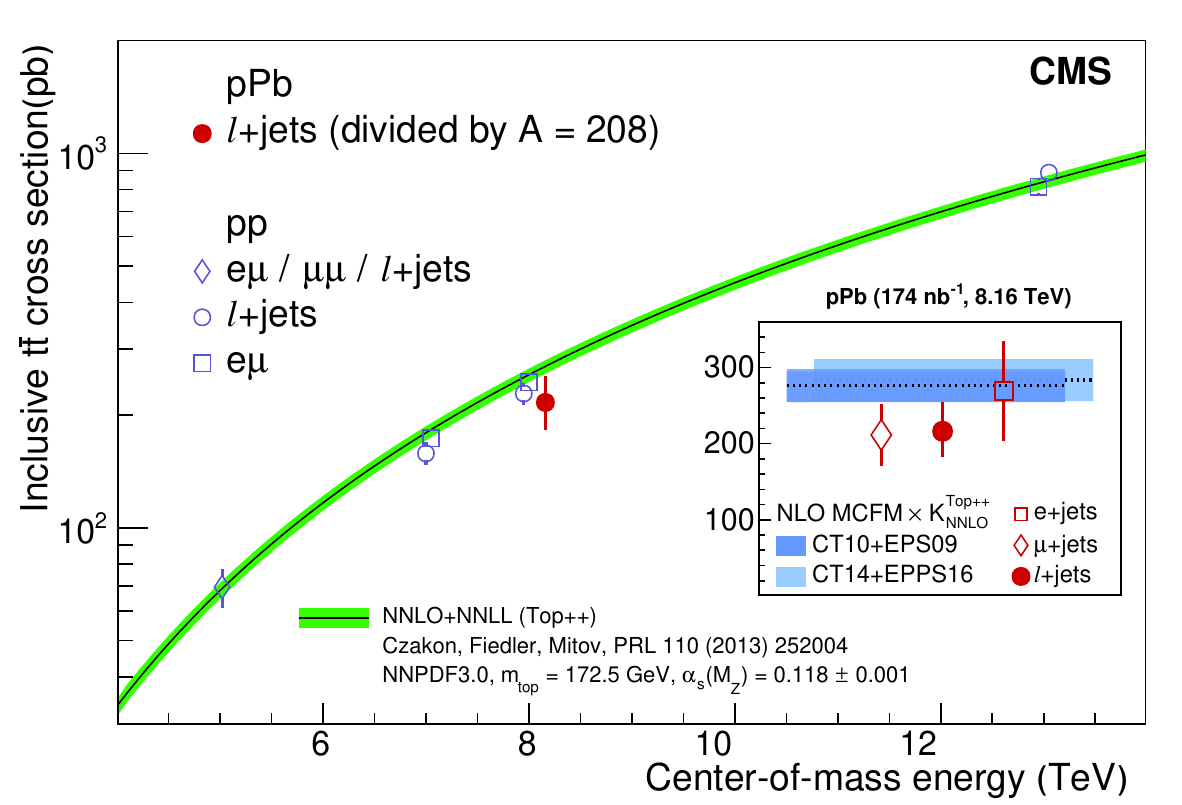}
    \caption{Top quark pair production cross section in \pp and \pPb collisions as a function of the center-of-mass energy per 
    nucleon pair; the CMS results at different center-of-mass energies in the dilepton and semileptonic channels. The measurements are 
    compared to the NNLO+NNLL QCD theory predictions~\cite{Czakon:2013goa,Campbell:2010ff,Albacete:2017qng}. (Figure adapted from Ref.~\cite{CMS:2017hnw}.)}
    \label{fig:top_pPb}
\end{figure}

The production of top quark pairs in nuclear collisions probes the gluon nPDFs at high-$x$, and is therefore complementary 
to EW boson measurements primarily probing quark PDFs~\cite{dEnterria:2015mgr}. The first observation and evidence of top 
quark production in \pA and \AonA collisions, respectively, was performed by the CMS collaboration~\cite{CMS:2017hnw,CMS:2020aem}. 
As displayed in Fig.~\ref{fig:top_pPb}, the measured cross section is consistent with the expectations from scaled \pp data, 
as well as pQCD calculations at NNLO,
with soft gluon resummation at next-to-next-to-leading logarithmic~(NNLL) accuracy~\cite{Czakon:2013goa,Campbell:2010ff,Albacete:2017qng}. 
The difference in the inclusive cross section computed with the PDF for free protons and for bound nucleons is small. 
A net overall enhancing (antishadowing) effect increases the total top quark pair cross section by approximately 5\% in \pPb 
relative to \pp collisions. Such a difference is too small to be observed in the data with the current experimental uncertainties. 
However, this first measurement paves the way for future detailed investigations of top quark production in nuclear interactions. 
In particular, top quark pair production provides a new tool for studying the strongly interacting matter created in 
\AonA collisions~\cite{CMS:2020aem} (as discussed in Section~\ref{sec:top_PbPb}).

Another well-known probe of nPDFs is the production of high transverse momentum jets. Both quark and gluon nPDFs can be studied over 
a wide range of $x$ values based on jet production. For the \pt range of 50--200\GeV probed in \pPb collisions, it is expected that 
jets at central rapidity mostly constrain gluon nPDFs at mid- to high-$x$. This helps cover swathes of phase space that are more challenging 
to constrain with \PW or \PZ boson production in \pPb collisions. Unlike EW bosons, jets can be produced via processes involving only QCD interactions. 
This is advantageous because it leads to jet events being produced quite copiously at the LHC. However, jet observables come with 
experimental challenges as well; the CMS detector resolution in jet \pt and pointing angle tends to be larger than for other objects. 
This means that a careful assessment of the bin-to-bin migration effects, and their corrections, are required.

\begin{figure}[t!]
    \centering
    \includegraphics[width=0.9\textwidth]{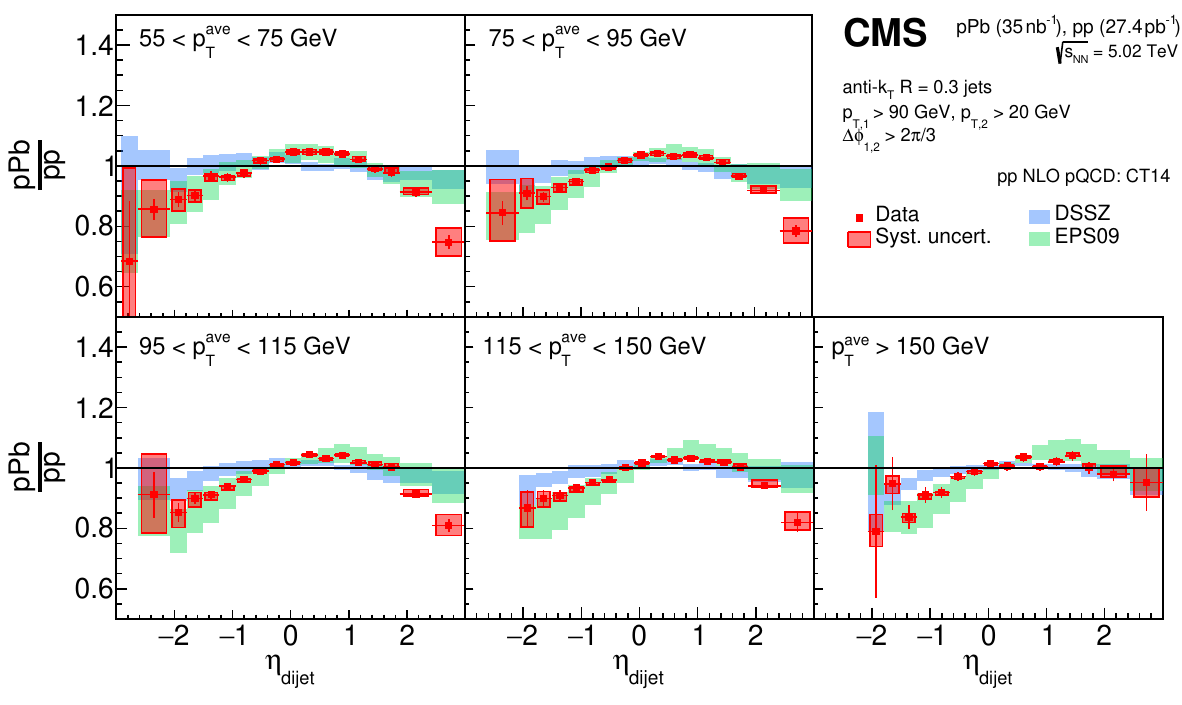}
    \caption{The ratio of the dijet $\eta$ spectra for \pPb and \pp data in a selection of \ptave ranges. 
    Theoretical predictions are from the NLO pQCD calculations of DSSZ~\cite{deFlorian:2011fp} 
    and EPS09~\cite{Eskola:2009uj} are shown. Red boxes and bars indicate the systematic and statistical uncertainties in data, respectively. 
    Green and blue boxes represent nPDF uncertainties. \FigureFrom{CMS:2018jpl}}
    \label{fig:dijetEtapPbDists}
\end{figure}

\begin{figure}[ht]
    \centering
    \includegraphics[width=0.5\textwidth]{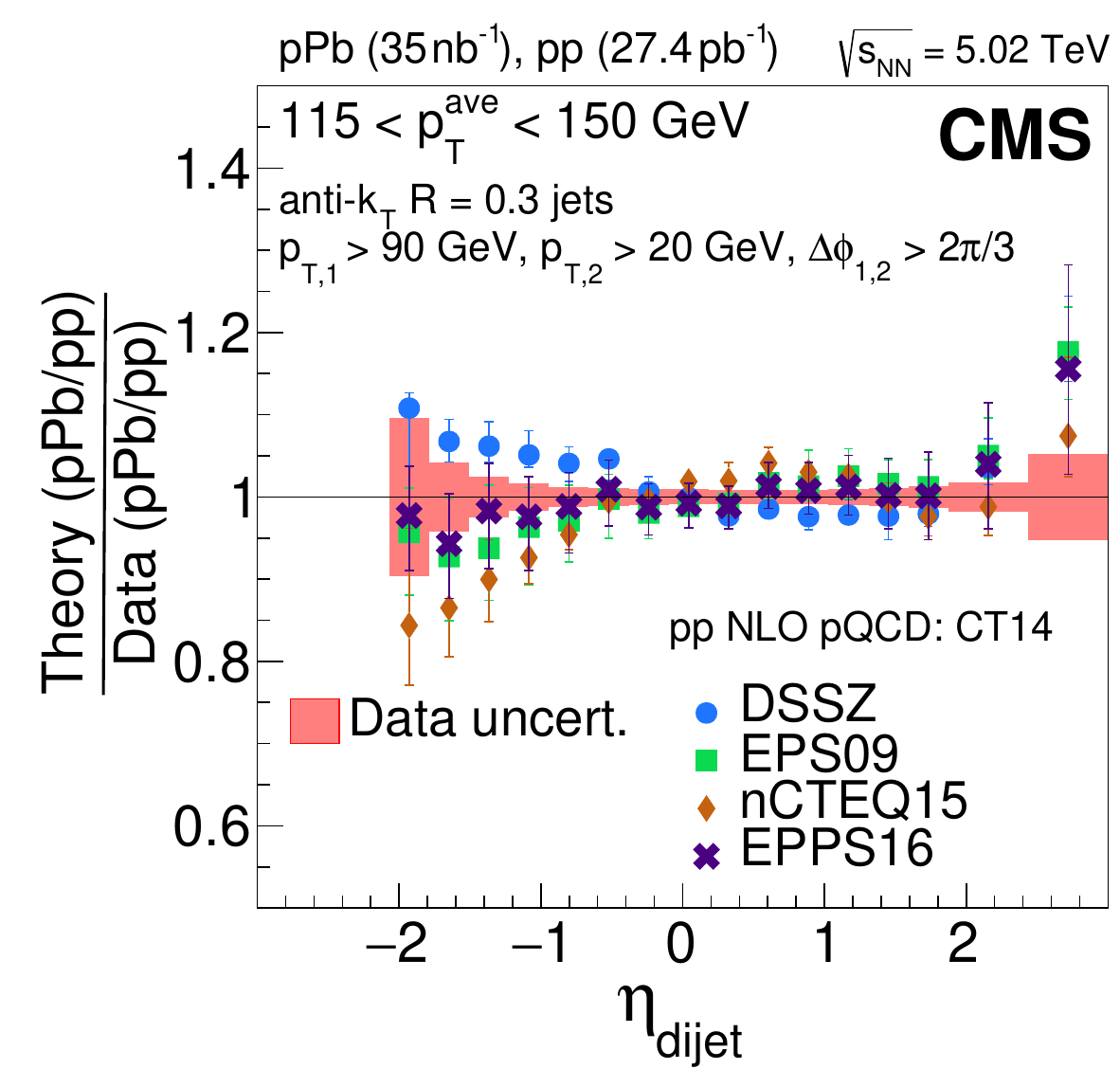}
    \caption{The ratio of theoretical predictions to CMS data for the ratio of the \pPb to \pp dijet $\eta$ spectra for 
    $115<\ptave<150\GeV$. Theoretical predictions are from the NLO pQCD calculations of DSSZ~\cite{deFlorian:2011fp}, 
    EPS09~\cite{Eskola:2009uj}, nCTEQ15~\cite{Kovarik:2015cma}, and EPPS16~\cite{Eskola:2016oht} nPDFs, using CT14~\cite{Dulat:2015mca} 
    as the baseline PDFs. Red boxes indicate the total uncertainties in data and the error bars on the points represent 
    nPDF uncertainties. \FigureFrom{CMS:2018jpl}}
    \label{fig:dijetEtapPb}
\end{figure}

\begin{figure}[ht]
\centering
\includegraphics[width=0.7\textwidth]{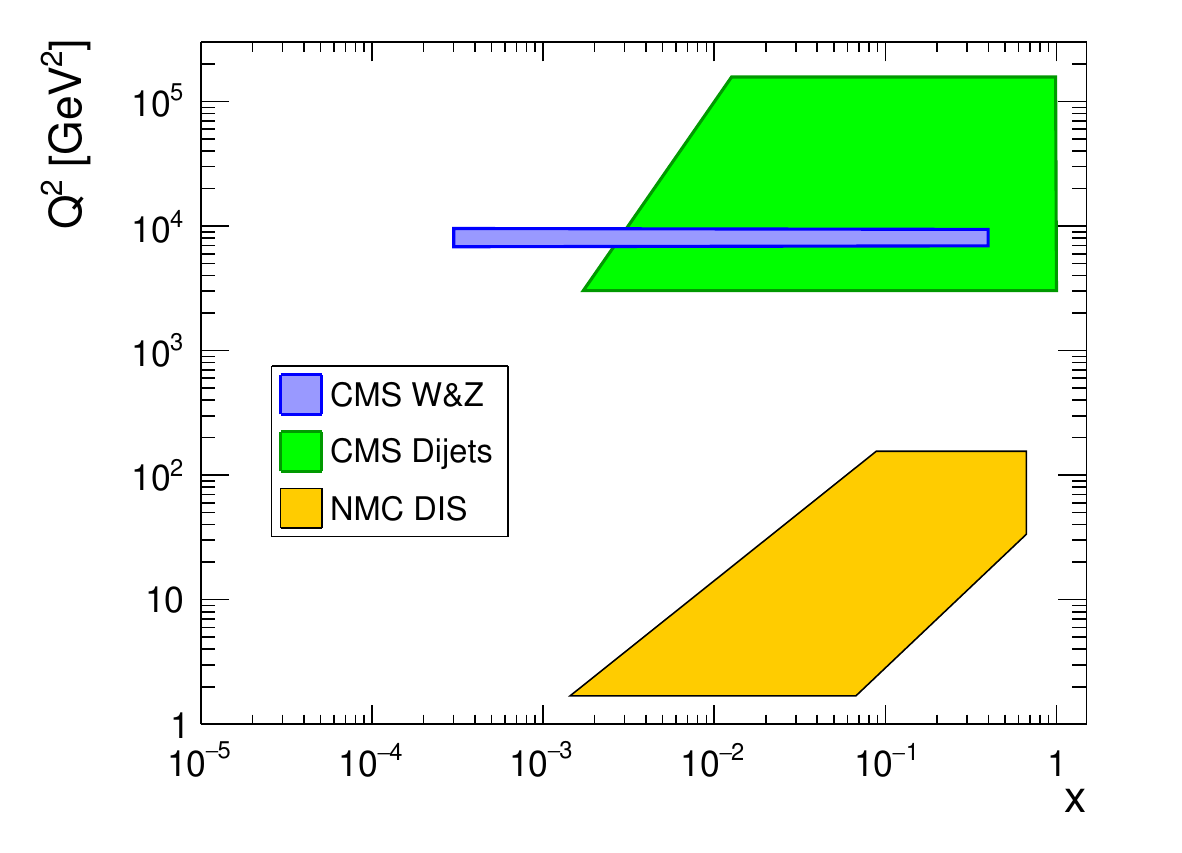}
\caption{Schematic representation of the phase space regions, in the $x$ and \QTwo plane, covered by the CMS measurements of dijets (green area) and electroweak bosons (blue area).
They cover a much higher \QTwo region than previous measurements from fixed-target experiments also included in the EPPS21 analysis~\cite{Eskola:2021nhw} (orange area).}
\label{fig:Q2XPlane}
\end{figure}

Measurements of the inclusive dijet pseudorapidity spectra at 5.02\TeV have been performed by the CMS Collaboration~\cite{CMS:2014qvs,CMS:2018jpl}, 
with the most recent measurement shown in Fig.~\ref{fig:dijetEtapPbDists}. To make potential nPDF effects more visible, 
the result is presented as a ratio between the \pPb and \pp data. The measurements, which are differential in dijet 
pseudorapidity, $\eta_{\text{dijet}} = (\eta_\text{jet 1}+\eta_\text{jet 2})/2$, and in average dijet transverse 
momentum, \ptave, help constrain nPDFs for a wide range of $x$ and \QTwo.
The measurements show tension with the nPDF sets that were available when they were first presented, as can be seen in the ratios 
of theoretical predictions to the data from Ref.~\cite{CMS:2018jpl} in Fig.~\ref{fig:dijetEtapPb}. In particular, discrepancies 
were observed for values of $\eta_{\text{dijet}}>1.5$ and $\eta_{\text{dijet}}<-0.5$, which correspond to low $x$ and 
intermediate $x$ suppression of the nPDF relative to the proton PDF (known as the shadowing and the EMC~\cite{EuropeanMuon:1983wih} 
regions of nPDFs, respectively). This measurement was also the first-ever evidence that \textit{gluons} in the nuclei featured 
antishadowing (an enhancement of the nPDF at $x\approx0.1$) compared to the proton densities~\cite{Eskola:2018sxu,Eskola:2021nhw}. 
Previous measurements of the antishadowing and EMC effects had only probed nuclear quark densities. Before these measurements, 
nPDFs did not have input from dijet data at LHC energies. In recent years, the PDF collaborations have incorporated these 
data sets~\cite{Eskola:2021nhw,AbdulKhalek:2022fyi}, which has significantly improved the gluon PDFs across a wide $x$ range, demonstrating 
the unique constraining power of these measurements. 
The $x$ and \QTwo two-dimensional regions constrained by the CMS measurements of dijets and electroweak bosons are presented in Fig.~\ref{fig:Q2XPlane}.

\subsection{Tests of the Glauber model and \texorpdfstring{$\ncoll$}{NColl} scaling using electroweak bosons}
\label{ssec:testsOfGlauberModelEWKBosons}

The nuclear modification factor \RAA is a common observable that is used when studying the QGP produced in \AonA collisions. It is defined as
\begin{equation}
\label{eqn:RAA}
  \RAA = \frac{1}{\langle\TAA\rangle} \frac{\ddinline{\NAA}{\pt}}{\ddinline{\sigma_{\pp}}{\pt}} = \frac{1}{\langle\ncoll\rangle} \frac{\ddinline{\NAA}{\pt}}{\ddinline{N_{\pp}}{\pt}},
\end{equation}
where \NAA is the corresponding yield of the particle species of interest in \AonA collisions, and $N_{\pp}$ ($\sigma_{\pp}$) 
is the corresponding yield (cross section) in \pp collisions. The average values of the nuclear overlap function $\langle\TAA\rangle$ and $\langle\ncoll\rangle$ are typically 
calculated for a given centrality range with a Glauber MC model~\cite{Alver:2008aq,Loizides:2017ack,dEnterria:2020dwq}, which uses parameters 
such as the nuclear radius, deformation, and skin depth as input (as discussed in Section~\ref{sec:ExperimentalMethods_Centrality}). 
A typical interpretation of the nuclear modification factor is that an \RAA value of unity indicates the absence of nuclear effects, 
\ie, that the collision at a given centrality can be treated as a superposition of \ncoll independent nucleon-nucleon collisions. 
This expectation is known as \ncoll scaling.

As previously discussed, the EW bosons (photons, \PW, and \PZ bosons) do not interact strongly with a QGP, and the \PW and \PZ 
bosons decay to leptons in the earliest stages of the collision. Additionally, nPDF effects on EW boson production in \PbPb collisions 
are fairly well understood and expected to be relatively small (of the order of 5\% or less at midrapidity) compared to the nuclear 
modifications observed for color-charged particles~\cite{Helenius:2021tof}. These factors make EW bosons ideal probes to test the 
\ncoll scaling hypothesis and, by extension, the Glauber model used to calculate \ncoll and \TAA values. 

\begin{figure}[ht]
    \centering
    \includegraphics[width=0.7\textwidth]{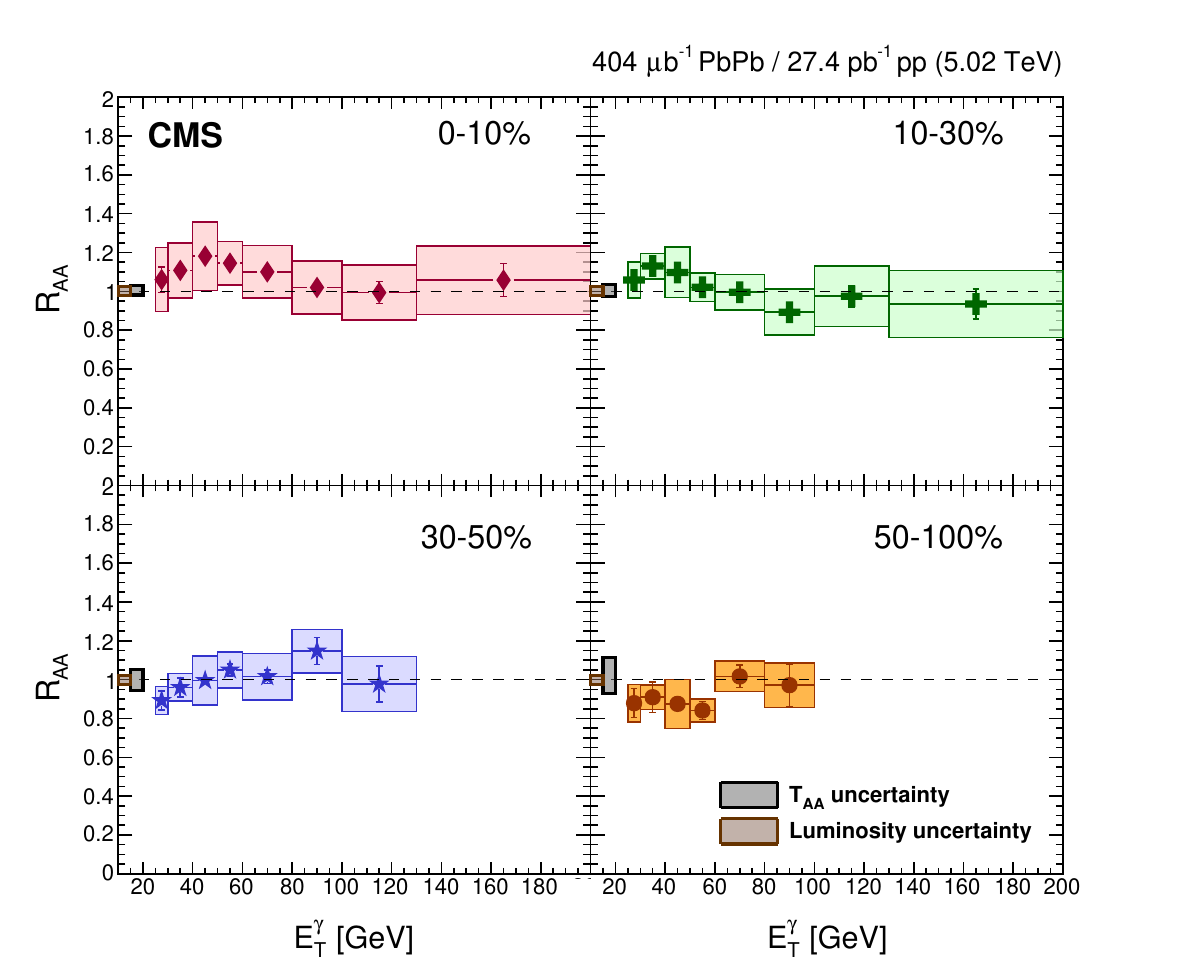}
    \caption{The photon \RAA versus photon \etg in four centrality ranges for 5.02\TeV \PbPb collisions. The error bars indicate 
    the statistical uncertainties and the systematic uncertainty, excluding \TAA uncertainties, are shown by the colored boxes. 
    The \TAA uncertainties are common to all points in a given centrality range, and are indicated by a gray box on the left side of 
    each panel. Similarly, a 2.3\% \pp collision integrated luminosity uncertainty is shown with a brown box. \FigureFrom{CMS:2020oen}}
    \label{fig:5TeVphotonRAA}
\end{figure}

Figure~\ref{fig:5TeVphotonRAA} shows a measurement of the midrapidity ($\abs{\eta}<1.44$) isolated photon \RAA as a function of the 
photon transverse energy \etg in four \PbPb centrality ranges using 5.02\TeV \PbPb and \pp data~\cite{CMS:2020oen}. 
For the measured range $25 < \etg < 200\GeV$, \RAA is consistent with unity, supporting the \ncoll scaling hypothesis. 
An earlier analysis of 2.76\TeV \PbPb data reached a similar conclusion~\cite{CMS:2012oiv}. The 5.02\TeV data are found to be 
consistent with NLO calculations from the \textsc{jetphox} version~1.3.1\_4~\cite{Aurenche:2006vj} MC generator, indicating a 
solid understanding of isolated photon production in \AonA collisions.

\begin{figure}[t]
    \centering
    \includegraphics[width=0.75\textwidth]{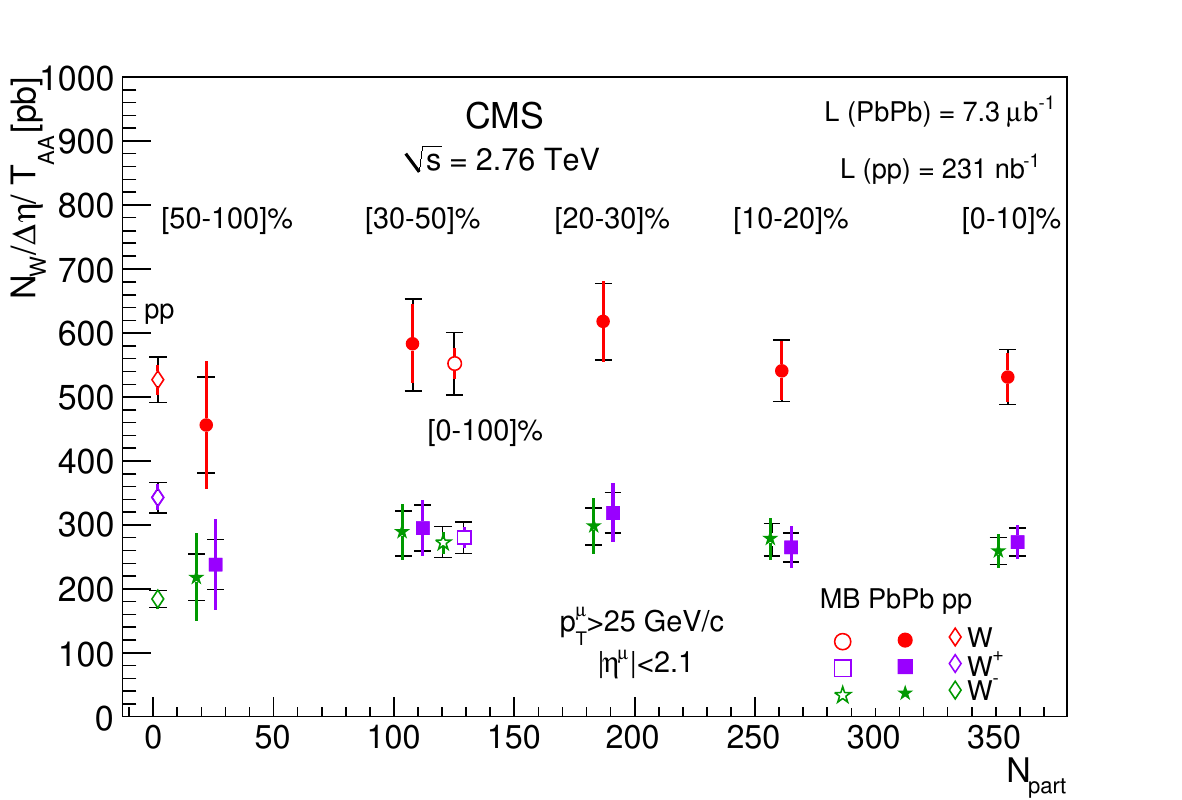}
    \caption{Normalized yields (per \NN integrated luminosity and per unit rapidity) of $\PW\to\PGm\PGn$ production in 2.76\TeV \PbPb collisions, 
    shown for inclusive \PW~(red), \PWp~(violet), and \PWm~(green). The open symbols at $\npart=120$ represent values for MB collisions. 
    At $\npart = 2$ the corresponding cross sections, divided by the muon pseudorapidity acceptance \deta, for \pp collisions at the
    same energy are displayed. For clarity the \PWp and \PWm points are slightly shifted horizontally. Error bars represent statistical 
    uncertainties and horizontal lines show systematic uncertainties. \FigureFrom{CMS:2012fgk} }
    \label{fig:276TeVW}
\end{figure}

Measurements of massive EW bosons complement measurements of isolated photons because they can access similar information about 
the initial state without being sensitive to photon reconstruction effects, \PGpz and \PGh decay contamination, and fragmentation 
photon backgrounds. Figure~\ref{fig:276TeVW} shows a measurement of \PW bosons in $\sqrtsNN = 2.76\TeV$ \PbPb collisions~\cite{CMS:2012fgk}. 
The quantity displayed is the yield of \PW bosons divided by the \TAA value calculated for the centrality range of interest (\ie, per 
\NN integrated luminosity, $1/\TAA$, at a given \PbPb impact parameter), which is then scaled by the muon $\eta$ acceptance 
($\deta = 4.2$). The measurements are plotted as a function of \npart for five centrality selections, shown by the solid markers, 
with an inclusive selection plotted with open markers around $\npart=120$. For comparison, the same quantity for \pp collisions at the 
same collision energy is shown by the open markers at $\npart=2$. For all centrality selections in \PbPb collisions, the results for 
\PWp and \PWm bosons are consistent with each other. This is not the case for \pp collisions, where \PWp bosons are produced at 
nearly twice the rate of \PWm bosons. This difference is attributed to the combination of two effects. Because of isospin, the presence 
of neutrons within the lead nucleus affects the production of $\PQu\PAQd\to\PWp$ and $\PQd\PAQu\to\PWm$. In addition, conservation of 
angular momentum results in preferential emission of $\PWp\to\Plp\PGnl$ and $\PWm\to\Plm\PAGnl$ toward midrapidity and more forward 
rapidities, respectively. When summing over both \PW charge states, the normalized yields in \PbPb collisions are consistent with 
those in \pp collisions for all centrality selections. The measured centrality-inclusive \RAA value for \PW bosons is determined to 
be $1.04\pm0.07\stat\pm0.12\syst$, a value which strongly supports the assumption of binary scaling of the production of hard probes. 

\begin{figure}[ht!]
    \centering
    \includegraphics[width=0.5\textwidth]{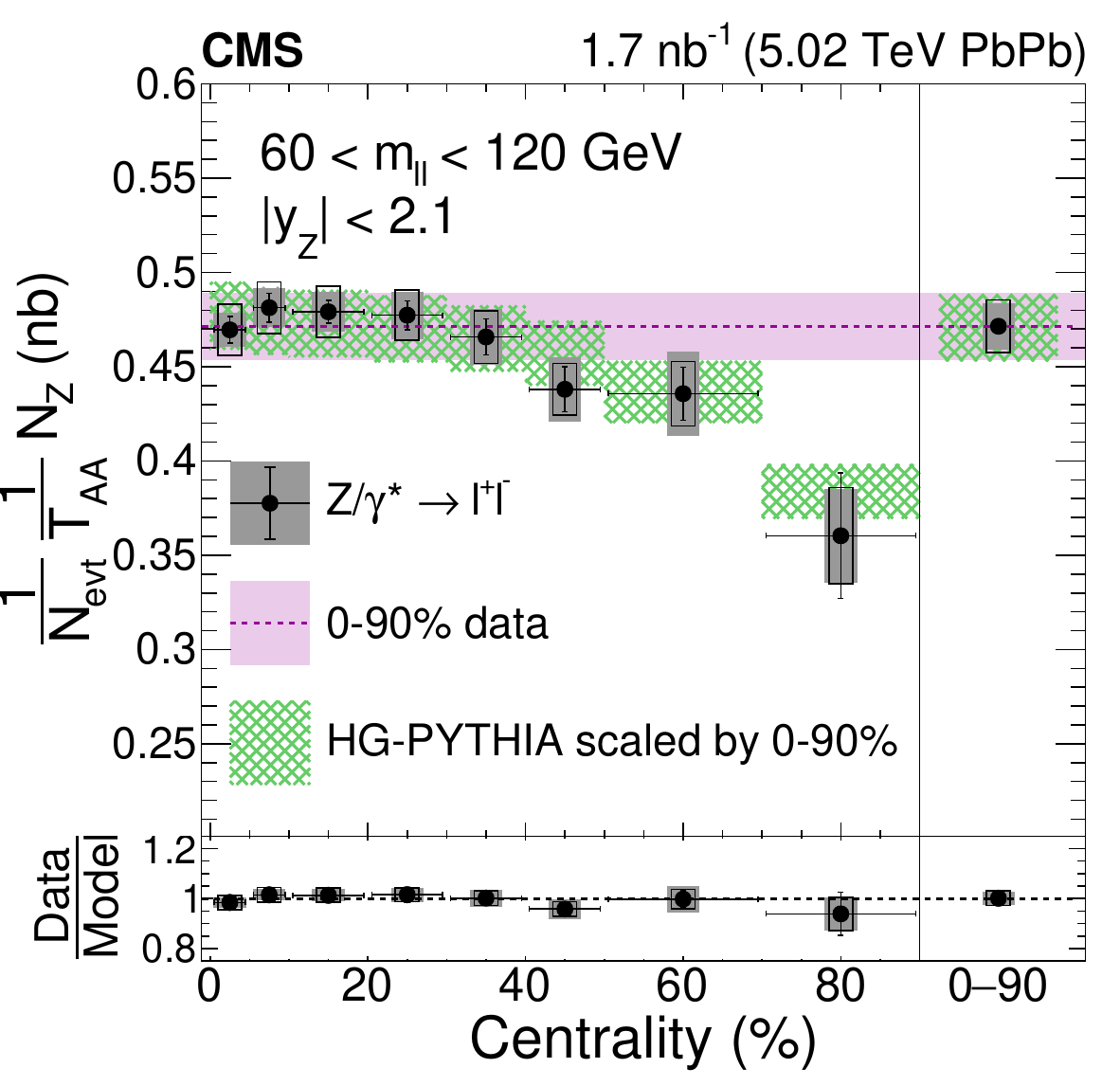}
    \caption{The \TAA-normalized yields of \PZ bosons versus centrality for 5.02\TeV \PbPb collisions. The error bars, open boxes, and solid gray boxes represent 
    the statistical, systematic, and \TAA uncertainties, respectively. The value of the 0--90\% data (pink) and the scaled \textsc{hg-pythia} model (green) 
    are displayed. The width of the bands represents the contribution from the total 0--90\% data point uncertainty. \FigureFrom{CMS:2021kvd}}
    \label{fig:5TeVZ}
\end{figure}

Early analyses of \PZ boson production at $\sqrtsNN = 2.76\TeV$~\cite{CMS:2011zfr,CMS:2014dyj} produced similar conclusions to those from the \PW boson 
and photon measurements. However, these measurements had large uncertainties, which prevented detailed examination of peripheral (50--100\%) collisions. 
The larger integrated luminosities achieved for 5.02\TeV \PbPb collisions during the LHC Run~2 enabled much more precise studies of these peripheral events. 
Figure~\ref{fig:5TeVZ} shows the per-event yields normalized per \NN integrated luminosity for \PZ bosons decaying to two muons or electrons in 
5.02\TeV \PbPb collisions~\cite{CMS:2021kvd}. The data points in the 0--40\% centrality range are consistent with the inclusive centrality selection, 
supporting the \ncoll scaling hypothesis in this centrality region. However, the increased precision of the measurement, as compared to previous measurements, 
reveals a falling trend in the 40--90\% centrality range. In particular, the data for the 40--90\% and 70--90\% centrality ranges deviate from inclusive 
0--90\% data point by 1.6 and 2.2 standard deviations, respectively. The green boxes show a prediction from the \textsc{hg-pythia} model~\cite{Loizides:2017sqq}, 
which agrees with the measurement. This model incorporates the \ncoll scaling hypothesis, but accounts for additional event selection effects and correlations 
between the centrality calibration and the hard process being measured when predicting \RAA for a hard, colorless probe, such as the \PZ boson. The 
agreement of these data with \textsc{hg-pythia} implies that, even if binary scaling of hard probes production is correct, the absence of final-state effects 
does not guarantee an \RAA of unity for very peripheral collisions, which in turn could affect the interpretation of similar measurements of color-charged particles. 
Therefore, this deviation from unity cannot be interpreted as violation of binary scaling, but instead points to additional selection effects 
in peripheral collisions which must be accounted for in addition to \ncoll scaling~\cite{dEnterria:2020dwq}. 

\begin{figure}[ht]
    \centering
    \includegraphics[width=0.45\textwidth]{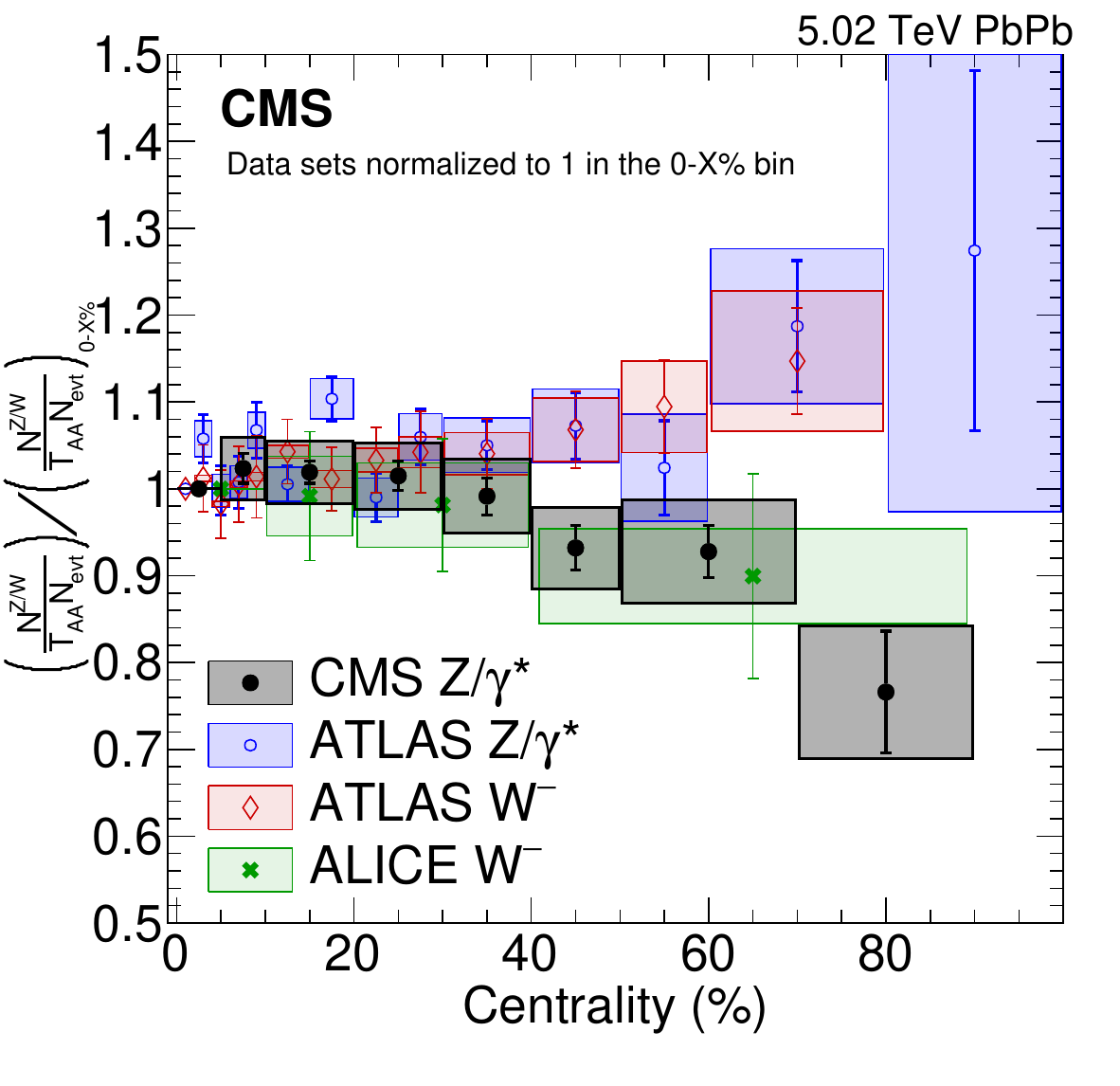}
    \includegraphics[width=0.45\textwidth]{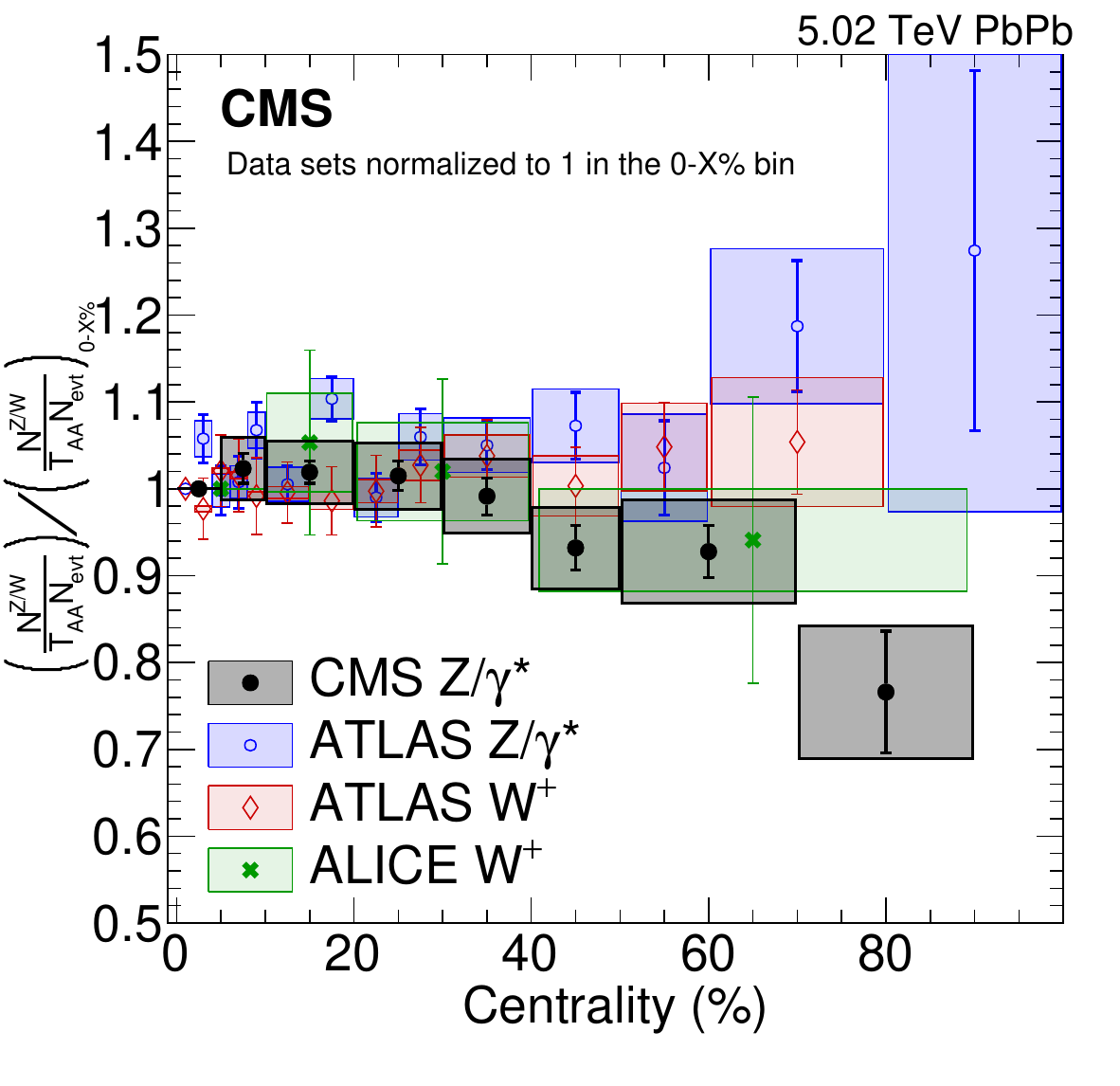}
    \caption{A comparison of results from the ALICE~\cite{ALICE:2022cxs}, ATLAS~\cite{ATLAS:2019maq}, and CMS~\cite{CMS:2021kvd} Collaborations for \PZ 
    and \PW production in \PbPb collisions. The data have been normalized so that the most central data point equals unity to enable comparison of 
    the shape of the distribution. The left (right) panel shows data for \PWm (\PWp) and \PZ bosons. For the ATLAS \PW data, the error bars 
    represent the combined statistical and systematic uncertainty, while the boxes show \TAA-related uncertainties. For all other data sets, the error 
    bars display statistical uncertainties and the boxes show combined systematic and \TAA uncertainties.}
    \label{fig:5TeVZSummary}
\end{figure}

Figure~\ref{fig:5TeVZSummary} shows a comparison of the CMS 5.02\TeV \PbPb collision \PZ boson data with EW boson measurements from the ATLAS~\cite{ATLAS:2019maq} 
and ALICE~\cite{ALICE:2022cxs} Collaborations at the same collision energy. To remove overall scale and isospin effects and to allow for comparison of 
the centrality dependence of the measurements, each data set has been normalized such that the most central point equals unity. A scale uncertainty common 
to all points in a data set, resulting from the normalization by the most central data point, is not shown on the figure. These uncertainties are 3.7\% 
(CMS $\PZ/\PGg^*$), 3.2\% (ATLAS $\PZ/\PGg^*$), 4.1\% (ATLAS \PWp), 3.9\% (ATLAS \PWm), 9.6\% (ALICE \PWp), and 7.5\% (ALICE \PWm). For the 
data shown here, \TAA values calculated with \textsc{TGlauberMC} v3.2 (as opposed to the earlier v2.4) are used by all experiments to ensure a fair comparison. 
A difference in the trends of the CMS \PZ~\cite{CMS:2021kvd}, and ATLAS \PZ and \PW data is apparent. The rising trend in the peripheral ATLAS data 
has been interpreted as a shadowing of the total \NN cross section~\cite{Eskola:2020lee}, which is a key input parameter in MC Glauber simulations. 
This interpretation is not consistent with the CMS \PZ and ALICE \PW boson results. 
 
In summary, from these measurements of EW bosons it is clear that the \ncoll-scaling hypothesis, which is a key component of interpreting observables such 
as \RAA, is well supported in the \mbox{0--40}\% centrality range. In the 40--90\% centrality range the situation appears to be more complex. Although the 
\ncoll scaling hypothesis cannot be definitely refuted in this centrality region, a combination of \ncoll scaling, as well as selection and centrality effects 
such as those included in HG-PYTHIA may be needed to adequately describe the data. 

\subsection{\texorpdfstring{Small-$x$}{Small-x} nuclear structure}
 
As a first approximation, the small-$x$ evolution of the nuclear wave function is dominated by gluon splitting $\Pg \to \Pg\Pg$. The gluon splitting contribution 
is incorporated in the DGLAP evolution equations of pQCD~\cite{Dokshitzer:1977sg,Altarelli:1977zs,Gribov:1972ri}, which resum at all-orders all diagrams that 
lead to logarithmic enhancements with the four-momentum transfer of the collision. While the DGLAP evolution equations capture some of the dominant contributions 
that compensate for the small value of the strong coupling constant \alpS via logarithms of $1/x$, a dedicated resummation is needed to properly account 
for these logarithmically-enhanced terms at small-$x$. This can be done using the Balitsky--Fadin--Kuraev--Lipatov 
(BFKL)~\cite{Kuraev:1977fs,Balitsky:1978ic,Lipatov:1985uk,Fadin:1998py} evolution equations of pQCD, which resum terms of the form $\alpS \ln(1/x)$ 
to all orders in the perturbative expansion. One of the key predictions from the BFKL equation is that the gluon density grows at small-$x$, following a 
power law, with the exponent given by the running of \alpS. Since both BFKL and DGLAP equations incorporate parton splitting contributions, it is predicted 
that the gluon densities should only increase at smaller $x$. Thus, gluon splitting alone leads to unitarity violation for cross sections. However, 
at high enough gluon occupancy numbers, it is expected that gluon recombination $\Pg\Pg \to \Pg$ also plays a role. The evolution equations that 
incorporate both splitting and recombination in the nuclear wave function are the Balitsky--Kovchegov~(BK) evolution 
equations~\cite{Balitsky:1995ub,Balitsky:1998kc,Balitsky:1998ya,Kovchegov:1999ua}. The characteristic energy 
scale at which both the splitting and recombination mechanisms are in balance is known as the parton saturation scale.
Since the initial state of the collision is a crucial ingredient for predictions in HI collisions, it is imperative to measure the splitting and recombination 
behavior at small-$x$ in controlled environments. Establishing the existence of parton saturation effects is a long-standing problem in hadronic physics, 
since it relates to the quantum mechanical behavior of gluons at high density. Parton saturation effects are expected to be universal, and are expected 
to manifest in both the structure of protons and nuclei at small-$x$. However, the advantage of studying this effect with HIs is that their parton density 
is much larger than that in protons, and therefore the critical energy scale below which gluon saturation manifests itself is larger, and thus more accessible experimentally.

A natural way of constraining the small-$x$ gluon nPDFs is by extending the measurements discussed in previous sections to the forward rapidity region. 
In particular, forward jets with low \pt offer insights into the parton densities and their evolution at small $x$ because at lowest order in \alpS, 
the $\eta$- and \pt-dependences of jets are correlated to the momentum fraction $x$ carried by the incoming parton, which can be estimated with 
$x \approx (\pt/\!\sqrt{s}\,) \exp(\pm \eta)$. The nominal acceptance for jet reconstruction in CMS extends over the range $\abs{\eta} < 5.2$, limited by 
the acceptance of the HF calorimeters. However, the acceptance for forward particle production has been extended to $-6.6 < \eta < -5.2$ using the CASTOR 
calorimeter during special runs. This detector allows for the detection and reconstruction of jets with a minimum \pt of approximately 3\GeV. Therefore, 
the study of jets using CASTOR provides an opportunity to explore the low-$x$ regime and examine perturbative nonlinear parton evolution effects. In 
\pPb collisions where the incoming Pb ion is in the direction pointing to CASTOR, the jets detected in the acceptance range of that detector allow for 
measurements highly sensitive to the small-$x$ region of the Pb nucleus down to $x\sim10^{-6}$. The most challenging aspect of this measurement is the calibration of the forward 
jets detector in CASTOR. An energy-dependent correction factor is used to account for the noncompensating behavior of the detector. These energy-dependent 
calibration functions are obtained from simulation by matching particle-level jets (with a particle-level jet isolation requirement) to the detector-level 
jets. More details of the jet calibration and reconstruction are presented in Ref.~\cite{ThesisKlundert}.

\begin{figure}[ht]
    \centering
    \includegraphics[width=0.49\textwidth]{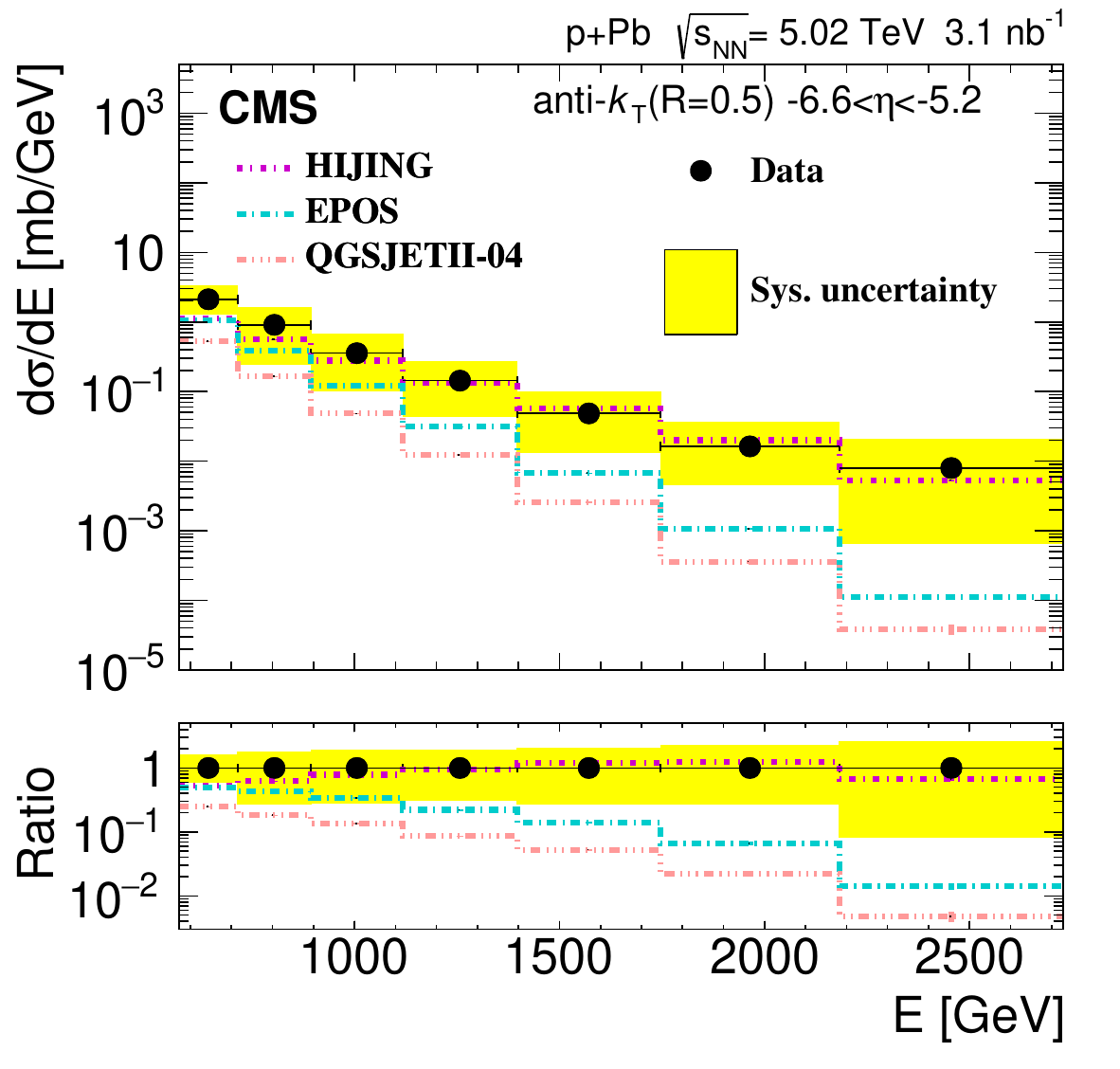}
    \includegraphics[width=0.49\textwidth]{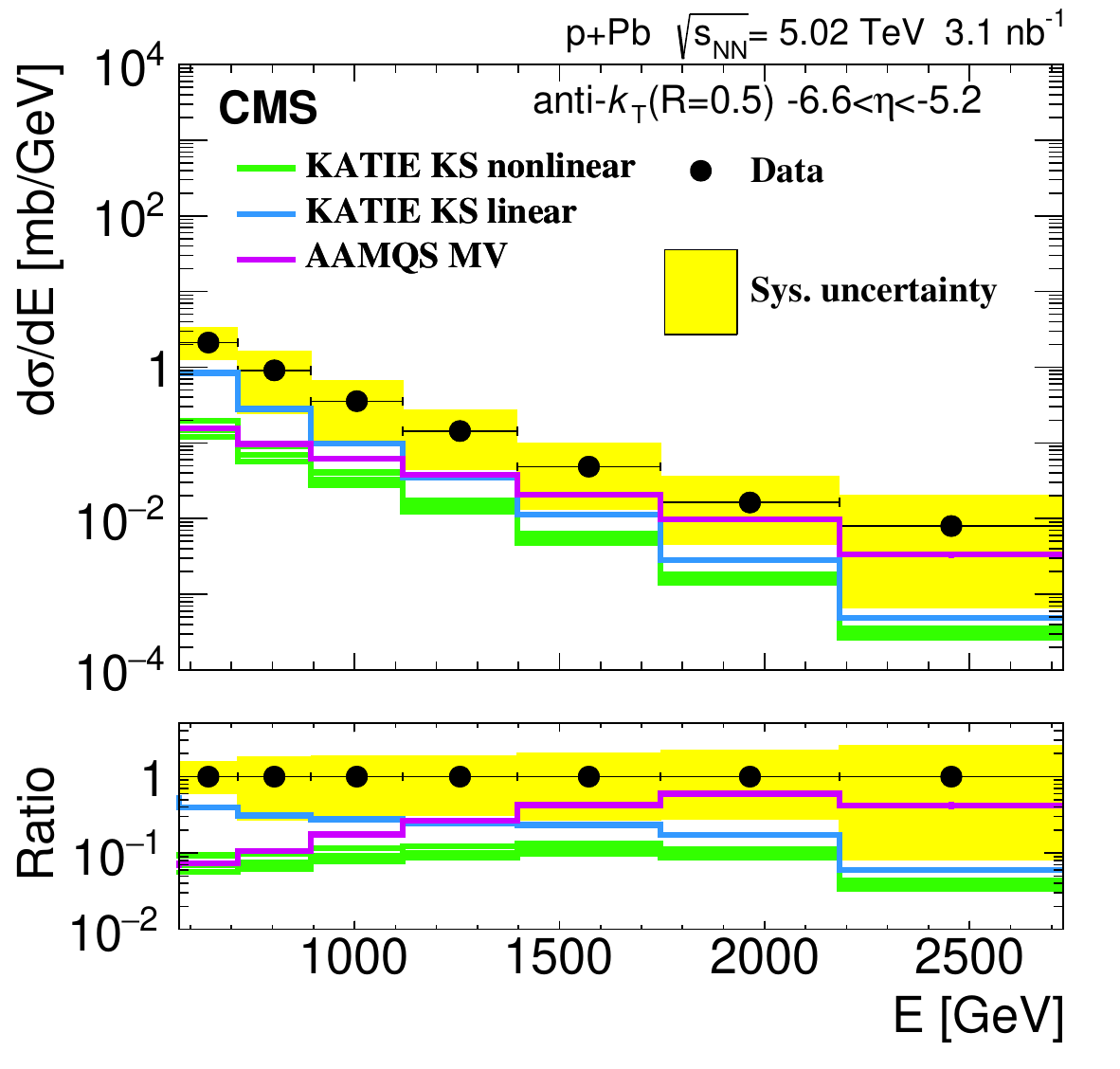}
    \caption{Forward jet differential cross section, where forward jet is in the proton-going direction, as a function of jet energy in \pPb 
    collisions at 5.02\TeV. The kinematics of the collision allows us to probe the small-$x$ wave function in the Pb nucleus with a high-$x$ parton 
    from the proton. This measurement is  compared with different Monte Carlo event generators, \textsc{epos-lhc}~\cite{Pierog:2013ria}, \HIJING~\cite{Wang:1991hta}, and 
    \textsc{qgsjetii-04}~\cite{Ostapchenko:2004ss} (left) and  predictions of the KATIE~\cite{vanHameren:2016kkz} and AAMQS~\cite{Albacete:2016tjq} saturation models (right). 
    \FiguresFrom{CMS:2018yhi} }
    \label{fig:JetEnergypPb}
\end{figure}

\begin{figure}[ht]
    \centering
    \includegraphics[width=0.49\textwidth]{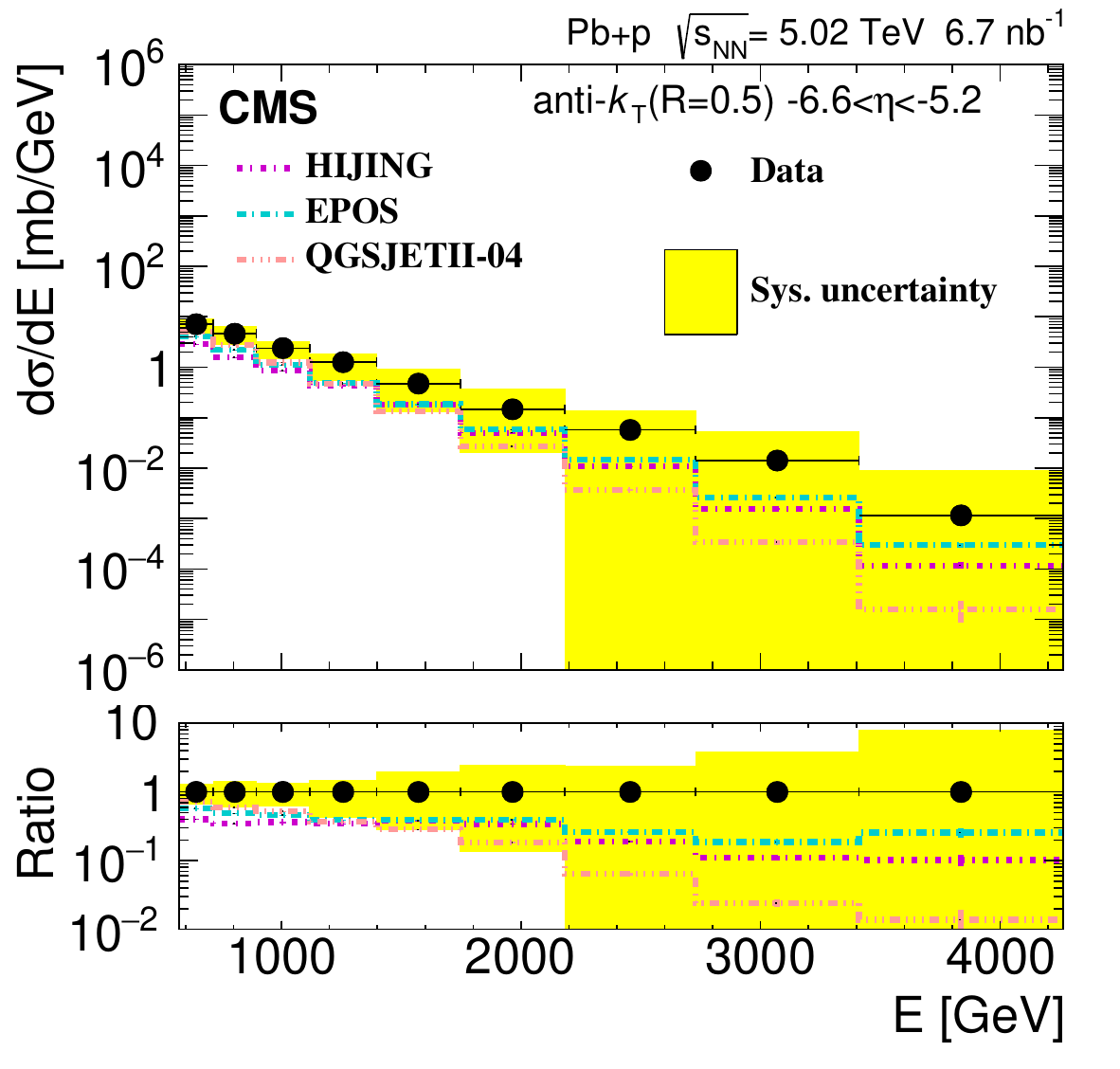}
    \includegraphics[width=0.49\textwidth]{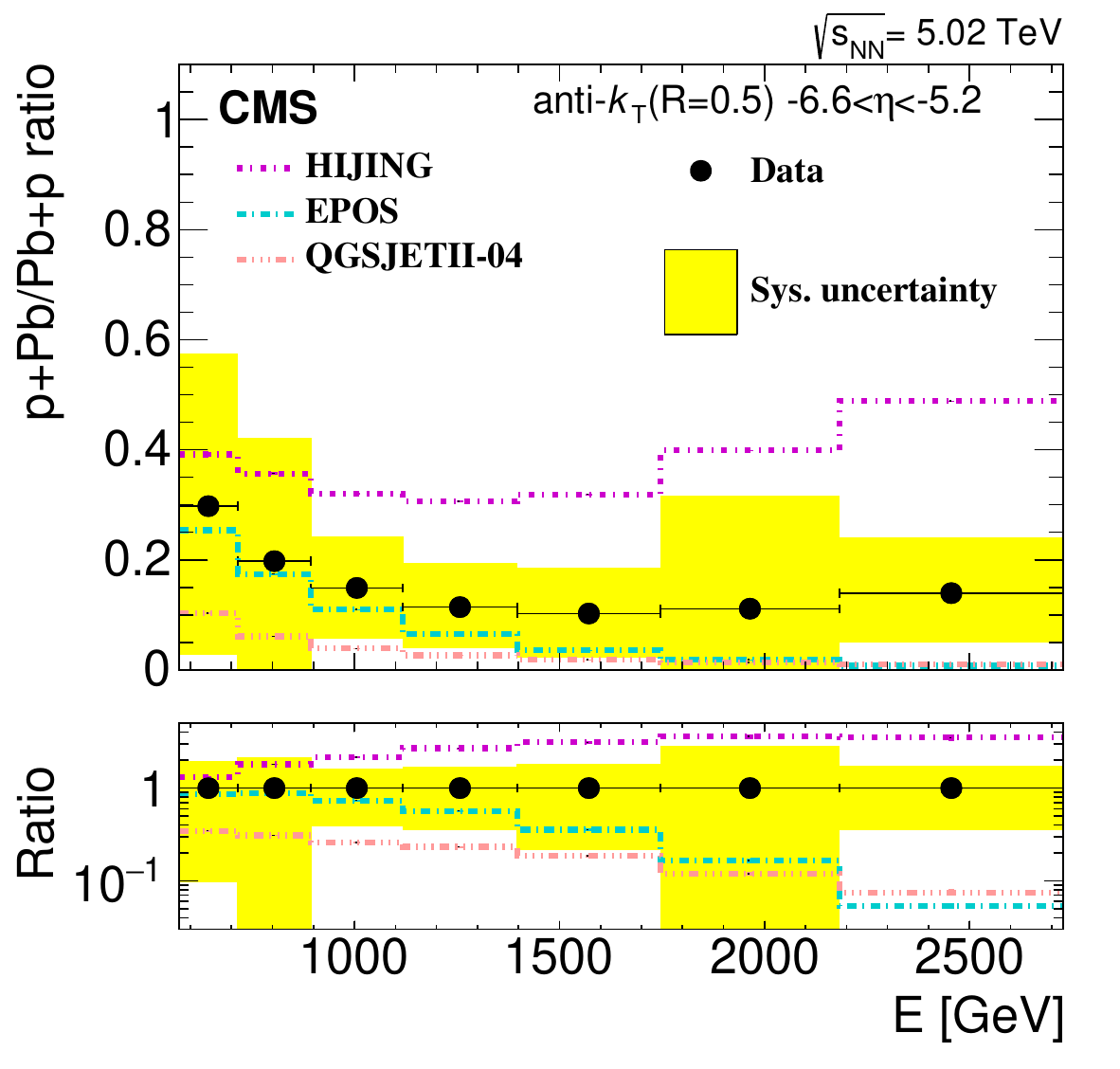}
    \caption{(Left) Forward jet differential  cross section, where the forward jet is in the Pb-going direction, as a function of the jet 
    energy in \pPb collisions at 5.02\TeV. The kinematic properties of the collision probe the small-$x$ wave function of the proton with a high-$x$ 
    parton from the Pb nucleus. The data are compared with different Monte Carlo event generators: 
\textsc{epos-lhc}~\cite{Pierog:2013ria}, \HIJING~\cite{Wang:1991hta}, and \textsc{qgsjetii-04}~\cite{Ostapchenko:2004ss}. (Right) The ratio of the inclusive 
jet cross sections; the numerator (denominator) of the ratio corresponds to the case where the jet is measured in the proton-going (Pb-going) direction.
    \FigureFrom{CMS:2018yhi}}
    \label{fig:JetEnergyPbp}
\end{figure}

\begin{figure}[ht]
    \centering
    \includegraphics[width=0.55\textwidth]{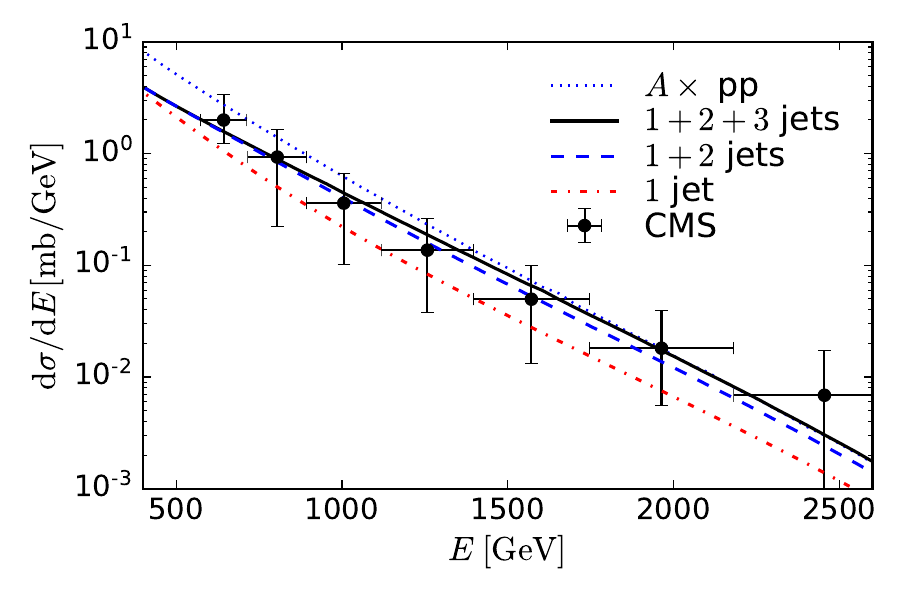}
    \caption{Forward jet differential cross section, where the forward jet ($-6.6<\eta<-5.2$) is in the \Pp-going direction, as a function of the jet energy 
    in \pPb collisions at 5.02\TeV. The kinematics of the collision allows us to probe the small-$x$ wave function of the Pb nucleus with a high-$x$ 
    parton from the proton. The data points are from Ref.~\cite{CMS:2018yhi}, with the error bars denoting the total uncertainty in the measurement.
    \FigureFrom{Mantysaari:2019nnt}}
    \label{fig:JetEnergyPbp_Heikki}
\end{figure}

The CMS experiment has measured differential cross sections for inclusive forward jet production in \pPb collisions at a center-of-mass energy of 
$\sqrtsNN = 5.02\TeV$ using the CASTOR detector~\cite{CMS:2018yhi}. This measurement was performed as a function of jet energy in hadronic, nondiffractive \pPb 
collisions, as presented in Figs.~\ref{fig:JetEnergypPb} and~\ref{fig:JetEnergyPbp}. The experimental uncertainty of this measurement is dominated by 
the jet energy scale calibration for jets in CASTOR. Also, since CASTOR lacks segmentation in $\eta$, other effects add to the uncertainty, such as the merging of particles from beam-beam remnants or two jets merged into a single jet at the detector level. 
The comparisons with numerous MC event generators presented in Fig.~\ref{fig:JetEnergypPb}, 
excluding (left panel) or including (right panel) parton saturation effects, 
show that none of the models studied can describe all the features observed in the experimental data. 
The MC predictions including saturation effects are consistent with the data 
within the uncertainties in the absolute cross sections 
(this is for small-$x$ evolution for the proton). 
Data obtained with the reversed beam (``\Pbp'') configuration are compared 
to \textsc{epos lhc}, \textsc{hijing}, and \textsc{qgsjetii-04} in Fig.~\ref{fig:JetEnergyPbp}~(left). 
This is the region with significant contributions from nuclear remnants. 
The \textsc{epos lhc} and \textsc{hijing} models describe the shape of the distribution reasonably well, 
but are too low in normalization. 
The \textsc{qgsjetii-04} model yields a spectrum that is too soft. 
Since the measurement is dominated by experimental uncertainties, 
the ratio of the \pPb and \Pbp systems is also reported in the right panel of Fig.~\ref{fig:JetEnergyPbp}. 
This ratio allows for large cancellations of correlated experimental uncertainties, with the trade-off that one is comparing systems 
with different rapidity boosts with respect to the laboratory frame. With this in mind, one can readily see that the predictions from MC-generated events 
cannot simultaneously describe the cross section ratio and the absolute cross sections. After the publication of this measurement, the theory interpretation has 
been further refined. In Ref.~\cite{Mantysaari:2019nnt}, an updated set of predictions were presented. Such a set of predictions includes the contribution of 
virtual $1\to 2$ splittings in the forward region, as well as an improved modeling of forward jets from multiple parton-parton interactions. The jets 
from those mechanisms merge with the forward jets from the hard scattering in the forward region due to the coarse calorimeter granularity of CASTOR, 
so accounting for them has an important numerical effect in the prediction. This is all in addition to small-$x$ nonlinear evolution of interest. As 
seen in Fig.~\ref{fig:JetEnergyPbp_Heikki}, where the theory predictions from Ref.~\cite{Mantysaari:2019nnt} are compared to the data, the prediction with 
these additional contributions lead to a better agreement with the \pPb data.

The advantage of measuring jets in the forward region is that it allows for the use of similar experimental techniques as previously employed in dijet 
studies at central pseudorapidity. From the theory point of view, the use of collinear PDFs (\ie, PDFs with impact parameter and momentum degrees of 
freedom integrated) is also well justified. The challenge in the very forward region comes from the contribution from other higher-order corrections and 
nonperturbative corrections that are not related to the initial state in order to arrive at a clean theoretical conclusion, as discussed in the previous paragraph. 
Thus, additional probes of small-$x$ nuclear structure are needed. A complementary study involves measuring exclusive final states, as discussed in the next section.

\subsection{Photoproduction of vector mesons}
\label{sec:InitialState_VM}

In ultraperipheral collisions~(UPCs) of HI collisions, where the impact parameter of the two colliding nuclei is greater than the sum of the two nuclear 
radii, hadronic interactions are highly suppressed compared to central collisions and the strong electromagnetic fields surrounding the nuclei give rise 
to \gaga and \PhotonA interactions. Such electromagnetic fields are highly Lorentz-contracted and can be treated as linearly polarized quasi-real 
photons with a flux that depends on the square of the electric charge of the emitting nucleus~\cite{vonWeizsacker:1934nji,Williams:1934ad}. These quasi-real 
photons can fluctuate into a quark-antiquark pair, essentially a color dipole, that interacts with the target nucleus or proton via two-gluon color-singlet 
exchange. The quark-antiquark pair eventually hadronizes into a vector meson~(VM). These interactions are usually classified depending on whether the 
projectile photon interacts with the target ion as a whole (coherently) or if it interacts with a single nucleon inside the ion (incoherently). Coherent photoproduction 
of heavy VM is of particular interest given that at lowest order in pQCD, the cross section is directly proportional to the square of the gluon PDF of 
the target at small $x$~\cite{Brodsky:1994kf,Martin:2007sb}. The mass of VMs sets an energy scale large enough to be studied in the framework of pQCD. 
In addition, at LHC energies, coherent photoproduction of VMs opens a special window to the poorly known low-$x$ region, allowing studies of shadowing effects 
towards the high energy limit of QCD. 

To probe the internal structure of the proton at small $x$, one can study the exclusive photoproduction of VMs, using the lead ion electromagnetic field 
as a source of quasi-real photons. The CMS experiment has studied the exclusive photoproduction of $\PGUP{n\mathrm{S}}$ and $\rho^0$ mesons in \pPb collisions 
at $\sqrtsNN= 5.02\TeV$~\cite{CMS:2018bbk, CMS:2019awk}. The advantage of using different VMs is that they give different effective sizes of the color 
dipoles probing the structure of the proton (or nucleus), which have different sensitivities to potential nonlinear evolution effects at small-$x$ 
and low \QTwo. If nonlinear evolution effects are present, they are expected to also depend on the probe. Thus, they should manifest differently for 
a variety of VMs as well. The general analysis strategy relies on the identification of the two oppositely-charged particles from the VM decay in an 
otherwise empty detector. Additional exclusivity criteria are also applied using the HF calorimeters, requiring energy deposits to be below noise thresholds 
in order to suppress contributions from nondiffractive hadronic interactions. The signal is separated from the main background (for example, the QED 
continuum) by fitting the dimuon invariant mass distribution of two charged particles. These raw signal yields contain contributions from different processes, 
such as coherent (where the emitted photon interacts with the whole nuclear ``target'') and incoherent (where the \PGg interacts with individual nucleons) 
photoproduction of VM, and also and also VM resulting from resonance decay feed-down. The coherent (incoherent) photoproduction have characteristically low 
(high) dimuon transverse momentum distributions, so their individual yields can be separated by means of multidimensional template fits.

The advantage of using asymmetric \pPb collisions for these measurements is that one can (to a large extent) unambiguously unfold the cross section in the 
laboratory frame to the cross section in the photon-proton center-of-mass frame. The center-of-mass energy of the photon-proton system, 
$$W_{\PhotonP}=\sqrt{\rootsNN M_{\text{VM}}\exp{(\pm y)}} ,$$
with $y$ the rapidity of the vector meson,
is strongly correlated with the parton momentum fraction $x$: smaller (larger) $W_{\PhotonP}$ corresponds to high $x$ (small $x$). This mapping 
allows for more direct comparisons between measurements at the DESY HERA in electron-proton collisions with those at the LHC in \pPb collisions. 

\begin{figure}[ht]
    \centering
    \includegraphics[width=0.7\textwidth]{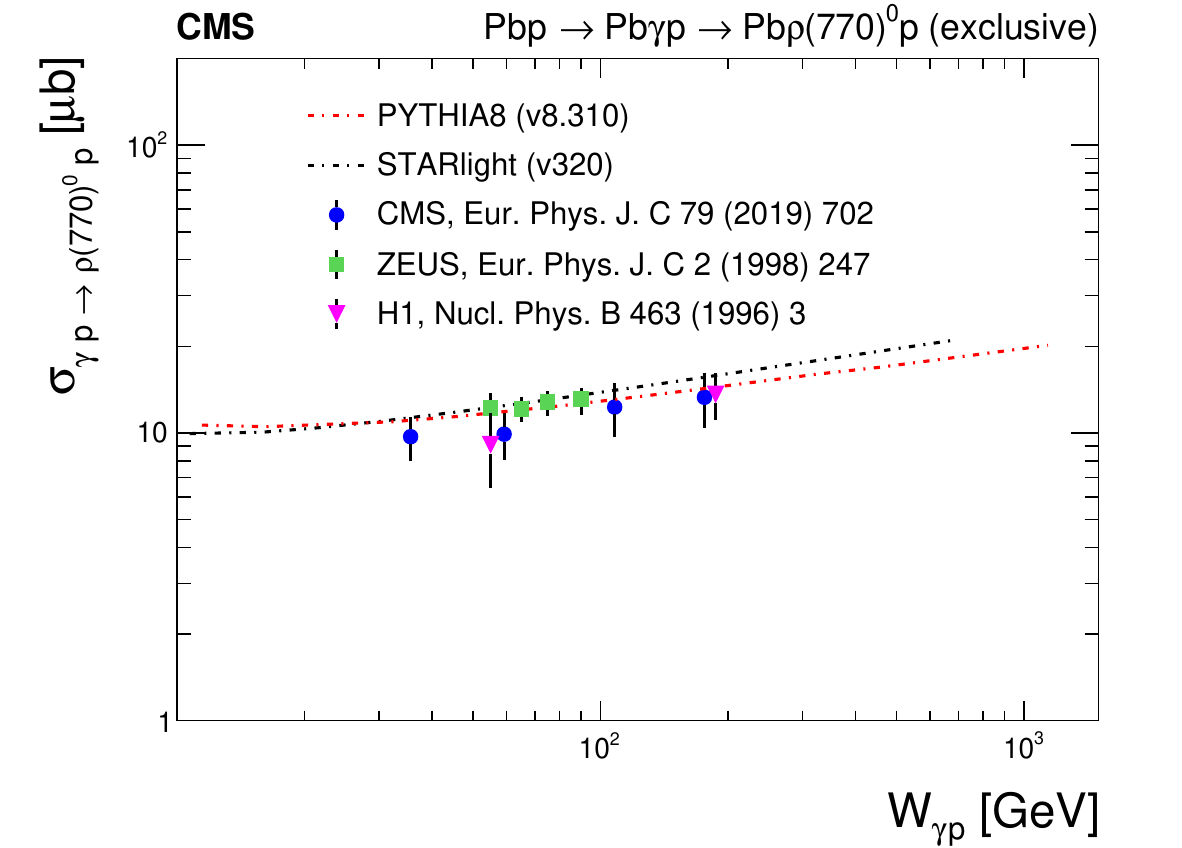}
    \caption{The cross section in the \PhotonP center-of-mass frame $\sigma(\PGg^{*}\Pp \to \Pgr^0 \Pp)$ for 
    exclusive $\Pgr^0$ VM photoproduction as a function of $W_{\PhotonP}$. CMS measurements during Run~2 extend up to $W_{\PhotonP} = 1\TeV$. 
    The CMS data points are from Ref.~\cite{CMS:2019awk}. The H1 and ZEUS data in electron-proton collisions are shown in the same panel. 
    The data points are compared to predictions from \PYTHIA{}8~\cite{Sjostrand:2014zea} and \Starlight~\cite{Klein:2016yzr}. \FigureFrom{CMS:2019awk}}
    \label{fig:exclusive_Rho}
\end{figure}

\begin{figure}[ht]
    \centering
    \includegraphics[width=0.7\textwidth]{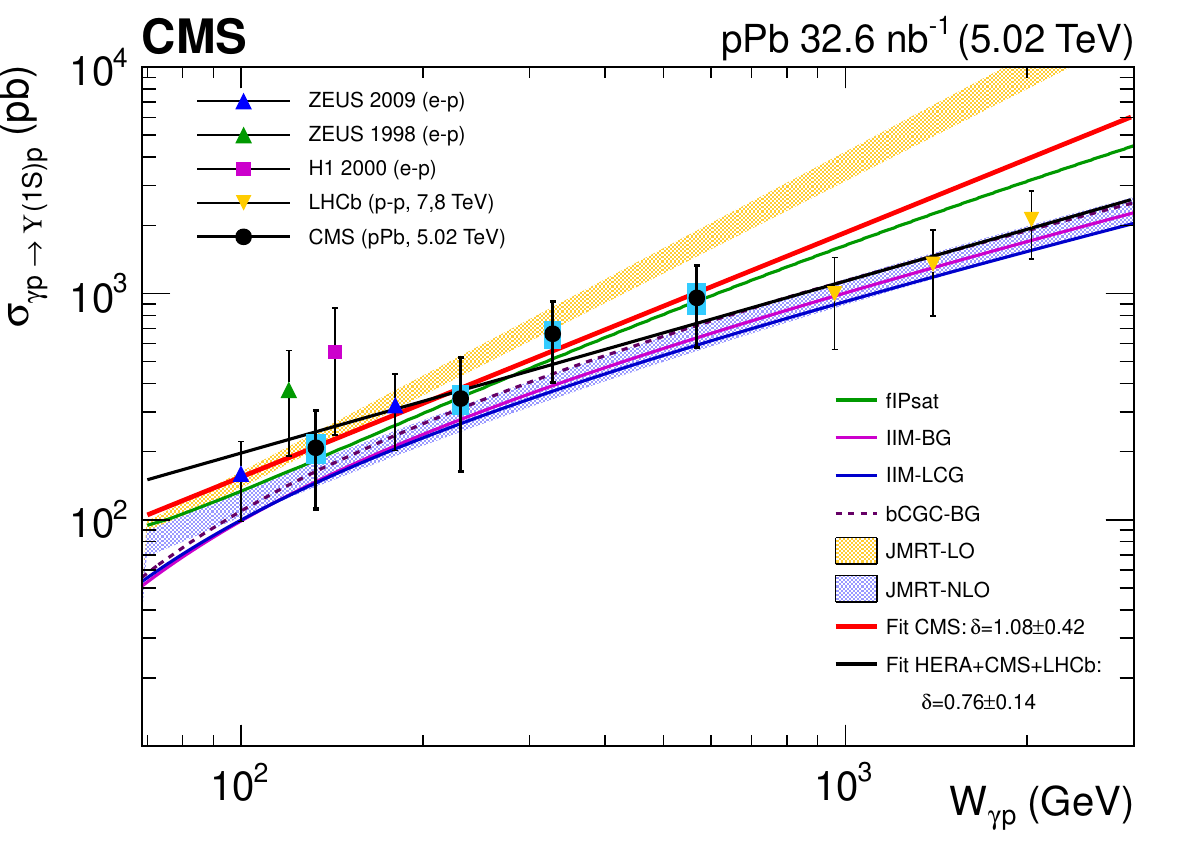}
    \caption{Photoproduction cross section in the photon-proton center-of-mass frame $\sigma(\PGg^* \Pp \to \upsOne \Pp)$ 
    for exclusive \upsOne VM photoproduction as a function of $W_{\PhotonP}$. The data are compared with different calculations 
    with different implementations of nonlinear evolution in the parton distributions.
    \FigureFrom{CMS:2018bbk}}
    \label{fig:exclusive_Upsilon}
\end{figure}

Figures~\ref{fig:exclusive_Rho} and~\ref{fig:exclusive_Upsilon} compare the photoproduction cross sections at HERA and the LHC. 
Notably, this is one of the few instances where one can compare cross sections from completely different colliding configurations in the same plots. 
It is observed that the CMS measurements complement the kinematical reach of the ones by HERA and from the LHCb experiment. The measurements have been 
compared with calculations based on BFKL, DGLAP, and BK evolution equations. It seems that, within the experimental sensitivity, no clear distinction between 
the nonlinear and linear evolution can be established with these measurements alone. These data in principle can be used as input for global collinear PDF 
fits~\cite{Flett:2019pux}. They are expected to constrain the small-$x$ gluon distribution in the $\QTwo \approx m_{\text{VM}}^2$ region, complementary to 
the small-$x$ reach at HERA. Information about the distribution of gluons in impact parameter space can be obtained via the \pt distribution of the 
VM, as the two variables are Fourier conjugates. Thus, this process can be used not only to learn about small-$x$ evolution, but also how such evolution is 
linked with the spatial distribution of partons within the proton or the nucleus.

\begin{figure}[ht]
    \centering
    \includegraphics[width=0.5\textwidth]{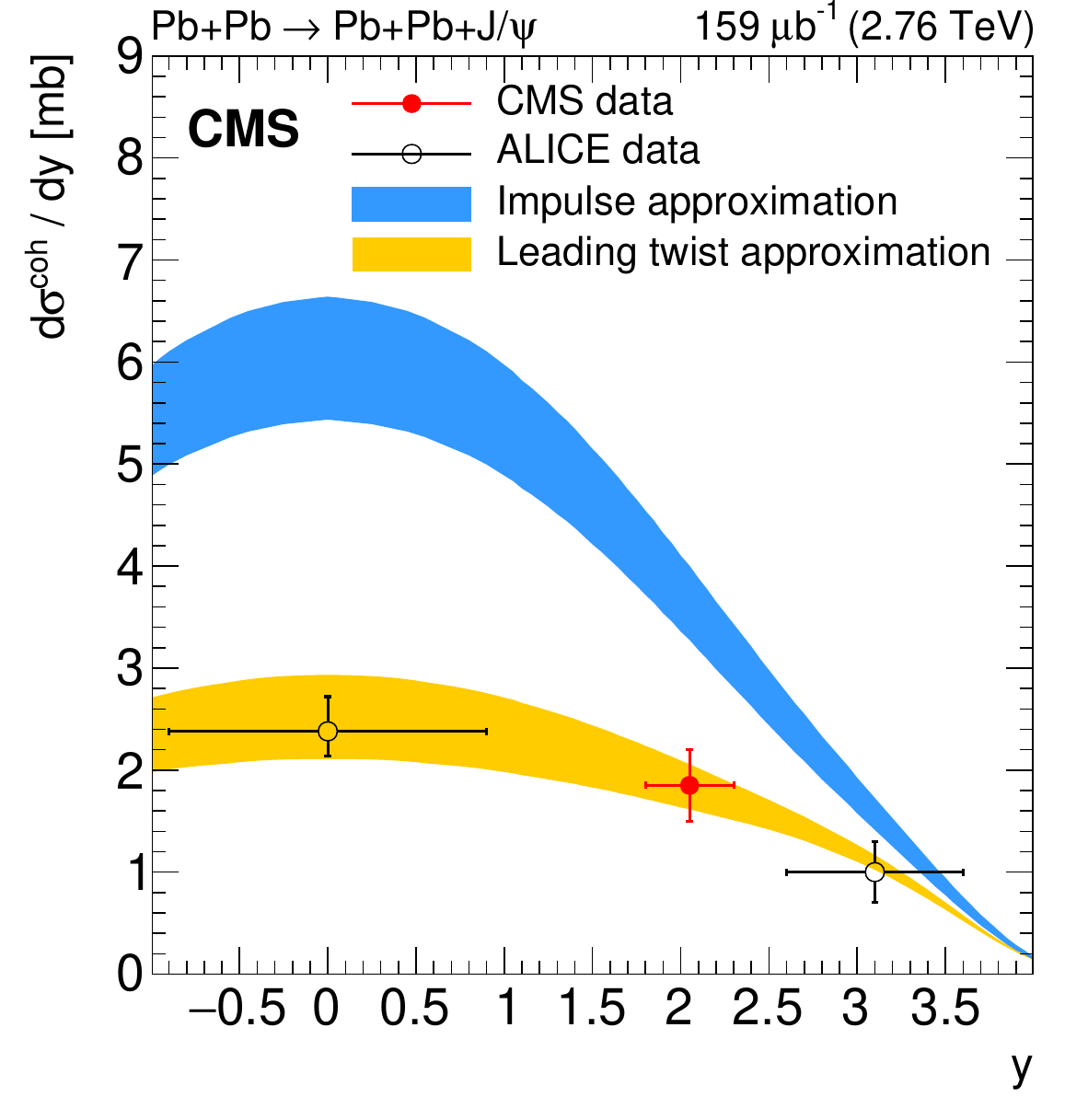}
    \caption{Differential \JPsi meson photoproduction cross section as a function of rapidity in \PbPb collisions at $\sqrtsNN = 2.76\TeV$ measured 
    by ALICE~\cite{ALICE:2012yye,ALICE:2013wjo} and CMS~\cite{CMS:2016itn}. Data are compared with the leading twist~\cite{Guzey:2013jaa} and the 
    impulse approximation~\cite{Guzey:2013jaa,Guzey:2013xba} predictions. The leading twist approximation is a perturbative QCD calculation that takes 
    into account nuclear shadowing effects from multinucleon interference. \FigureFrom{CMS:2016itn} }
    \label{fig:Jpsi276}
\end{figure}

A more promising way to establish if there is a manifestation of nonlinear evolution is by using nuclei as targets instead of protons. As mentioned 
earlier in this section, the advantage of the use of HIs is that the manifestation of nonlinear evolution effects can occur at lower collision energies 
than what is expected in proton-proton collisions. Figure~\ref{fig:Jpsi276} shows the CMS results for exclusive \JPsi production in \PbPb collisions at 
$\sqrtsNN= 2.76\TeV$~\cite{CMS:2016itn}. The calculation labeled ``impulse approximation'' simply scales the prediction for \PhotonP collisions 
by the number of nucleons, without any nuclear modification effects. The experimental cross sections are significantly smaller than this simple 
prediction, demonstrating the presence of strong nuclear modification effects that suppress the cross section relative to pure scaling expectations. 
The CMS acceptance in rapidity for low-\pt~\JPsi particle production is complementary to that of the ALICE experiment~\cite{ALICE:2012yye,ALICE:2013wjo}. 
This measurement demonstrates that there is a suppression relative to calculations that include pure scaling due to the larger number of nucleons 
in \PbPb collisions. However, in order to establish if this shadowing is a result of nonlinear evolution at small-$x$, one has to do the mapping 
from the laboratory-frame cross sections to the \PhotonA center-of-mass frame. Unfortunately, unlike the asymmetric \pPb collisions case, 
in a symmetric \PbPb collision either ion can be the emitter or the target nucleus, hence it is (at face value) not possible to identify the 
contributions from low- and high-energy photons. One can make educated guesses in certain kinematical regions (for example, at central 
rapidities one can extract an average of the high energy and low energy photon contributions). However, if we want to do a comparison as 
is done for \PhotonP collisions, some additional input is needed.

\begin{figure}[ht]
\centering
\includegraphics[width=1\textwidth]{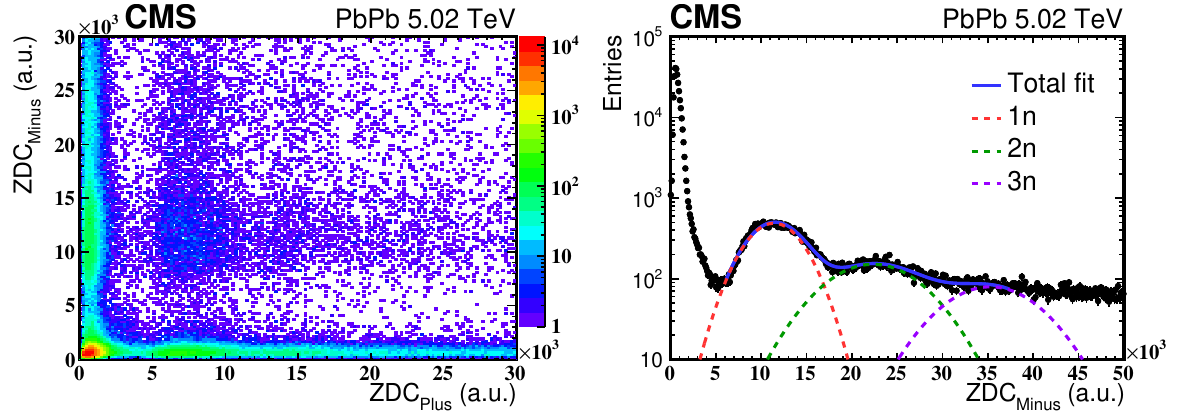}
\caption{The left panel shows the correlation between energy distributions of the Minus and Plus ZDC detectors (one entry per event), while the 
right panel shows a multi-Gaussian function fit to the Minus ZDC energy distribution. The different ``peaks'' in the ZDC energy distribution can 
be assigned to different forward neutron multiplicities, the first peak is detector noise, which corresponds to no detected neutrons, the second 
peak can be associated with one neutron, and so on. \FiguresFrom{CMS:2020skx}}
\label{fig:zdc}
\end{figure}

Indeed, it was proposed in Refs.~\cite{Guzey:2013jaa,Guzey:2016piu} that one can set additional constraints in a way that makes it possible 
to obtain cross sections in the \PhotonA frame. Such additional constraints are obtained by detecting forward neutrons emitted by 
virtue of the deexcitation of giant dipole resonances of the Pb nuclei~\cite{Berman:1975tt}. These giant dipole resonances take place as 
a consequence of additional soft-photon interactions between colliding ions, which are absorbed by the HI. When the excited HI relaxes, 
it is accompanied by the emission of forward collinear neutrons. This phenomenon is well-known from low-energy nuclear physics, and it 
turns out that it can be exploited in order to tag certain geometrical configurations of the colliding lead ions at the LHC. Indeed, the 
more ``central'' (smaller impact parameter) the UPC is, the more likely it is to have additional soft-photon emissions. Thus, by directly counting 
the number of neutrons in the forward region, one is effectively filtering UPCs in a way analogous to the centrality classification in conventional 
head-on \PbPb collisions. To detect these forward neutrons, dedicated calorimeters are installed in \PbPb collisions, known as the Zero Degree 
Calorimeters~(ZDCs) with an acceptance of $\abs{\eta} > 8.3$, with a ZDC located on either side of the interaction point. The neutron multiplicity 
is determined by the energies deposited in the ZDCs~\cite{CMS:2020skx}. For single neutrons, the relative energy resolution of each ZDC is 22--26\%, 
while the detection efficiency is close to 100\% in simulated events~\cite{Suranyi:2021ssd}. Based on neutron peaks observed in the total ZDC 
energy distribution, as shown in Fig.~\ref{fig:zdc}, coherent \JPsi meson photoproduction events are classified as having no neutrons (0\Pn) or 
with at least one neutron (X\Pn, $\PX \geq 1$) in each ZDC. 

\begin{figure}[ht]
    \centering
    \includegraphics[width=0.49\textwidth]{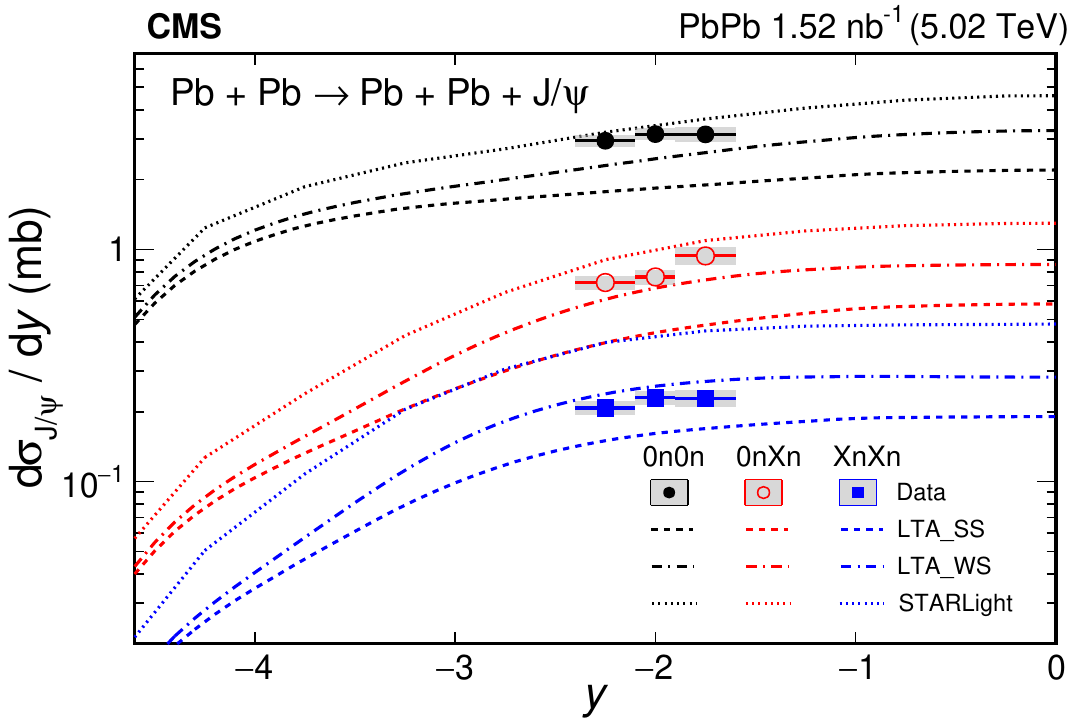}
    \includegraphics[width=0.49\textwidth]{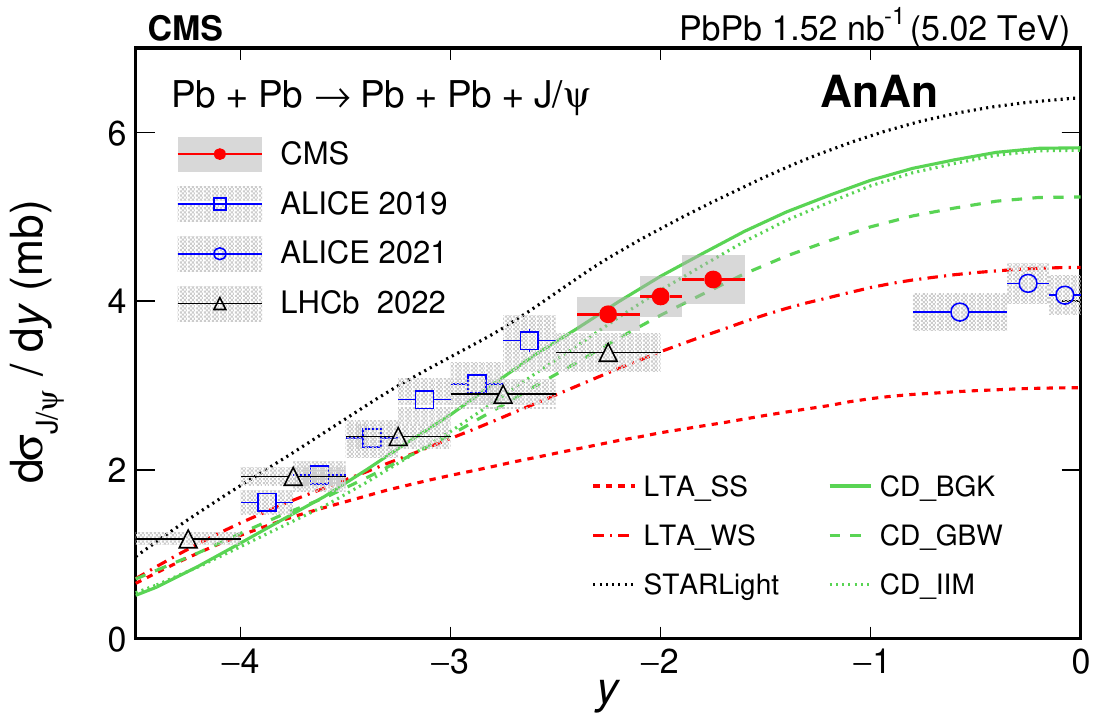}
    \caption{The differential coherent \JPsi meson photoproduction cross section as a function of rapidity, in different neutron multiplicity 
    classes (left): $0\Pn0\Pn$, $0\Pn\PX\Pn$ and $\PX\Pn\PX\Pn$ ($\PX \geq 1$); 
    (right): $\mathrm{A}\Pn\mathrm{A}\Pn$ (inclusive in the number of neutrons detected in the ZDC). The small vertical 
    bars and shaded boxes represent the statistical and systematic uncertainties, respectively. The horizontal bars represent the bin widths. Theoretical 
    predictions from LTA weak/strong shadowing~\cite{Guzey:2016piu}, color dipole models (CD\_BGK, CD\_BGW, and CD\_IIM)~\cite{Luszczak:2019vdc}, 
    and \Starlight~\cite{Klein:2016yzr} are shown. \FiguresFrom{CMS:2023snh}}
    \label{fig:HIN22022_rapidity}
\end{figure}

The measured coherent \JPsi meson photoproduction differential cross sections with and without neutron selection over the rapidity range $1.6<\abs{y}<2.4$ 
are reported in the left and right panels of Fig.~\ref{fig:HIN22022_rapidity}, respectively~\cite{CMS:2023snh}. Theoretical results based on the 
leading twist approximation~(LTA)~\cite{Guzey:2016piu} and color dipole models~\cite{Luszczak:2019vdc} are also shown for comparison. The leading 
twist approximation \cite{Guzey:2016piu} is a pQCD calculation
that accounts for nuclear shadowing effects using multinucleon interference. In each neutron multiplicity class, the LTA predictions tend to be 
lower than the CMS results. For the case of no neutron selection (AnAn), the data follow the trend of the forward-rapidity measurements of ALICE~\cite{Acharya:2019vlb}, 
but over a new $y$ region. None of the models describe the results with or without neutron selection over the full rapidity range, which may indicate 
that there are key ingredients missing from the theoretical understanding of high-energy \PhotonA scattering processes.

\begin{figure}[ht]
    \centering
    \includegraphics[width=0.75\textwidth]{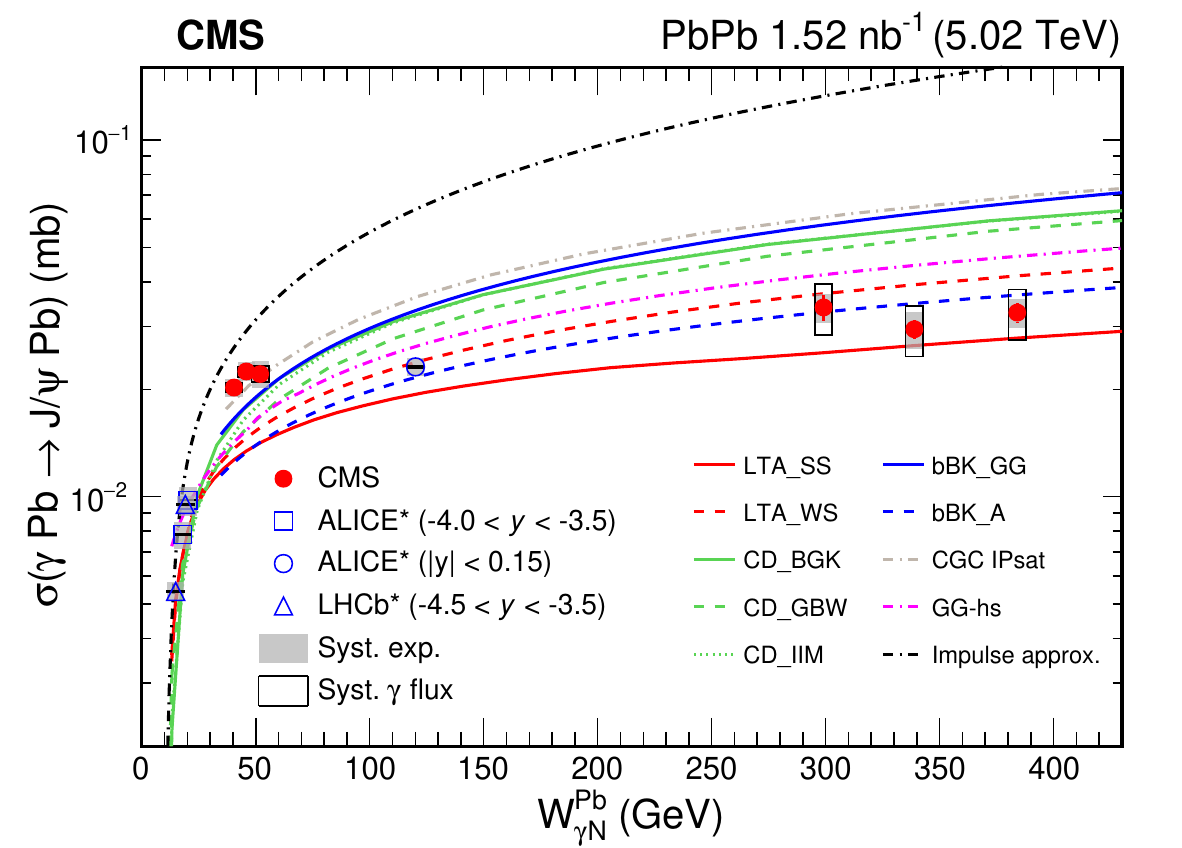}
    \caption{Total coherent \JPsi meson photoproduction cross section as a function of $W^{\mathrm{Pb}}_{\PGg\PN}$ in \PbPb UPCs at $\sqrtsNN = 5.02\TeV$. 
    The vertical bars and the shaded and open boxes represent the statistical, experimental, and theoretical (photon flux) uncertainties, respectively. 
    The predictions from various theoretical calculations~\cite{Guzey:2013jaa,Mantysaari:2017dwh,Cepila:2017nef,Guzey:2016piu,Bendova:2020hbb,Luszczak:2019vdc} 
    are shown by the curves. \FigureFrom{CMS:2023snh}}
    \label{fig:HIN22022_disentangled}
\end{figure}

The measured coherent \JPsi meson photoproduction cross section as a function of $\PGg \PN$ energy ($W^{\mathrm{Pb}}_{\PGg\PN}$) 
up to ${\approx}400\GeV$ is shown in Fig.~\ref{fig:HIN22022_disentangled}, after decomposing the two-way ambiguity with the differential cross 
sections split in different neutron multiplicity classes. The results show that the coherent \JPsi meson photoproduction cross section as a function 
of $W^{\mathrm{Pb}}_{\PGg\PN}$ increases, as it does for the \PhotonP case, but the slope of the cross section quickly changes 
at energies around 40--100\GeV, at which point the growth occurs at a different rate. This follows the qualitative expectation that at smaller $x$, 
the fast growth slows down due to nonlinear evolution effects. None of the theoretical predictions are able to reproduce the trends at high or 
small $x$. While the predictions qualitatively predict a change of shape as is observed in data, none of the theoretical predictions agree with 
the data in the full range explored in the measurement. To establish if this is due to genuine small $x$ nonlinear evolution effects, it is important 
to continue exploring these measurements for other vector mesons. Indeed, nonlinear evolution effects are expected to be universal, and thus they 
should not depend strongly on the VM used to probe the nuclear structure. Since different vector mesons have different masses, which can be related 
to the effective size of their color dipoles, they will have different sensitivities to the properties of high partonic densities at small $x$. 
Also, significant advances have been made on the theoretical side. Notably, in recent years, the fixed-order next-to-leading order corrections 
to the hard scattering have been provided for exclusive quarkonium production in \PbPb collisions~\cite{Eskola:2022vpi, Mantysaari:2022kdm, Eskola:2022vaf}. 
It is found that the quark-antiquark channel has a nonnegligible contribution, and large cancellations between the real and virtual contributions 
for the two-gluon channel are found for \JPsi meson production. For the data to be used in global nPDF fits, an understanding of these corrections, 
as well as additional measurements using other VMs, will be important.

\subsection{Summary of results for the initial state}
\label{sec:ChapterSummary}

During LHC Runs 1 and 2, the CMS Collaboration used \pPb data to make significant strides in constraining nPDFs, particularly through the study of EW gauge bosons, dijets, and top quark pairs. These measurements have provided crucial input for nPDF models and have resulted in substantial improvements in accurately reproducing experimental results, especially at medium and high Bjorken-$x$ values. Furthermore, studies of EW bosons in \PbPb collisions have confirmed that colorless hard probes, such as photons and \PZ bosons, are not significantly modified by the QGP in central collisions, offering the potential opportunity to use these probes for improved event selection and centrality calibration, especially in peripheral collisions.

For the low-$x$ regions, measurements at LHC energies have primarily focused on the evolution of gluon nPDFs. Forward inclusive jet cross sections in \pPb collisions, along with exclusive VM production in both \pPb and \PbPb collisions, have been instrumental in constraining models that incorporate gluon recombination alongside gluon splitting within their small-$x$ perturbative QCD evolution. Notably, very forward jet measurements by the CASTOR detector probe the phase space in a region with  exceptionally low $x$ ($\approx 10^{-5}$) and \QTwo ($\approx 10 \GeV^2$).  These measurements have shown that existing predictions, which also apply to cosmic ray physics, severely underestimate jet cross sections by factors of 10--100, underscoring the potential of these data to refine theoretical models.

Exclusive VM production has demonstrated that \pPb collisions can effectively function as \PhotonP colliders, providing valuable constraints on gluon PDFs for protons in the small-$x$ region at low $\QTwo$. Similarly, \PbPb collisions offer the opportunity to constrain gluon nPDFs in comparable kinematic regimes. However, analyzing exclusive VM production in symmetric \PbPb collisions presents a unique challenge because of the ambiguity in identifying which nucleus emits the quasireal photon and which the pomeron. The use of the ZDCs has been pivotal in resolving this ambiguity by distinguishing between low- and high-energy photon contributions, enabling the determination of the exclusive VM cross section in the \PhotonA center-of-mass frame for the first time.

The energy dependence of exclusive VM production in the \PhotonA frame reveals a marked suppression compared to the scaling behavior expected from \PhotonP cross sections, consistent with parton saturation effects. However, the overall trend is not fully captured by current theoretical models, suggesting that
further research is necessary to definitively attribute this nuclear suppression to the expected short-distance mechanism of gluon recombination.

\clearpage

\section{Bulk properties and novel phenomena}
\label{sec:softQGP}

Understanding the bulk thermodynamic and hydrodynamic properties of the QGP formed in ultrarelativistic heavy ion collisions is crucial for gaining insights into the fundamental degrees of freedom of this medium and its transport dynamics.
This section reviews studies of the bulk properties of the QGP by the CMS Collaboration. The results are based on measurements that use the large pseudorapidity coverage of the CMS detector. Charged-hadron densities and the correlations among particles widely separated in pseudorapidity are presented and discussed in context of the initial-state geometry. Measurements employing femtoscopy techniques of the size and shape of particle emitting sources at the last stage of the system evolution are also performed for different collision systems and LHC energies. Searches for novel phenomena related to chiral anomalous transport effects are also reviewed.

\subsection{Initial-state entropy and energy densities}
\label{sec:QGP_Thermodynamics}

The multiplicity and energy distributions of the primary charged particles (discussed in Section~\ref{subsec:CMSapparatus}) that emerge from HI collisions are basic observables that inform on the initial entropy and energy density and the medium evolution. 
At lower energies, these rapidity distributions are generally consistent with Landau hydrodynamics~\cite{Landau:1953gs}. The LHC experiments can test if a hydrodynamic description continues to be valid at \TeV energies. 
In an early measurement, CMS established the $\eta$ dependence of charged-particle production in \PbPb collisions at $\rootsNN= 2.76\TeV$~\cite{Chatrchyan:2011pb}. For the 5\% most central collisions, a charged-particle density per unit of pseudorapidity (\dnchdeta) of $1612\pm55$ was found. This value is consistent with a similar measurement by the ALICE Collaboration~\cite{Aamodt:2010cz}, and is twice the value found at RHIC~\cite{Adler:2004zn}.

\begin{figure}[ht]
    \centering
    \includegraphics[width=0.42\linewidth]{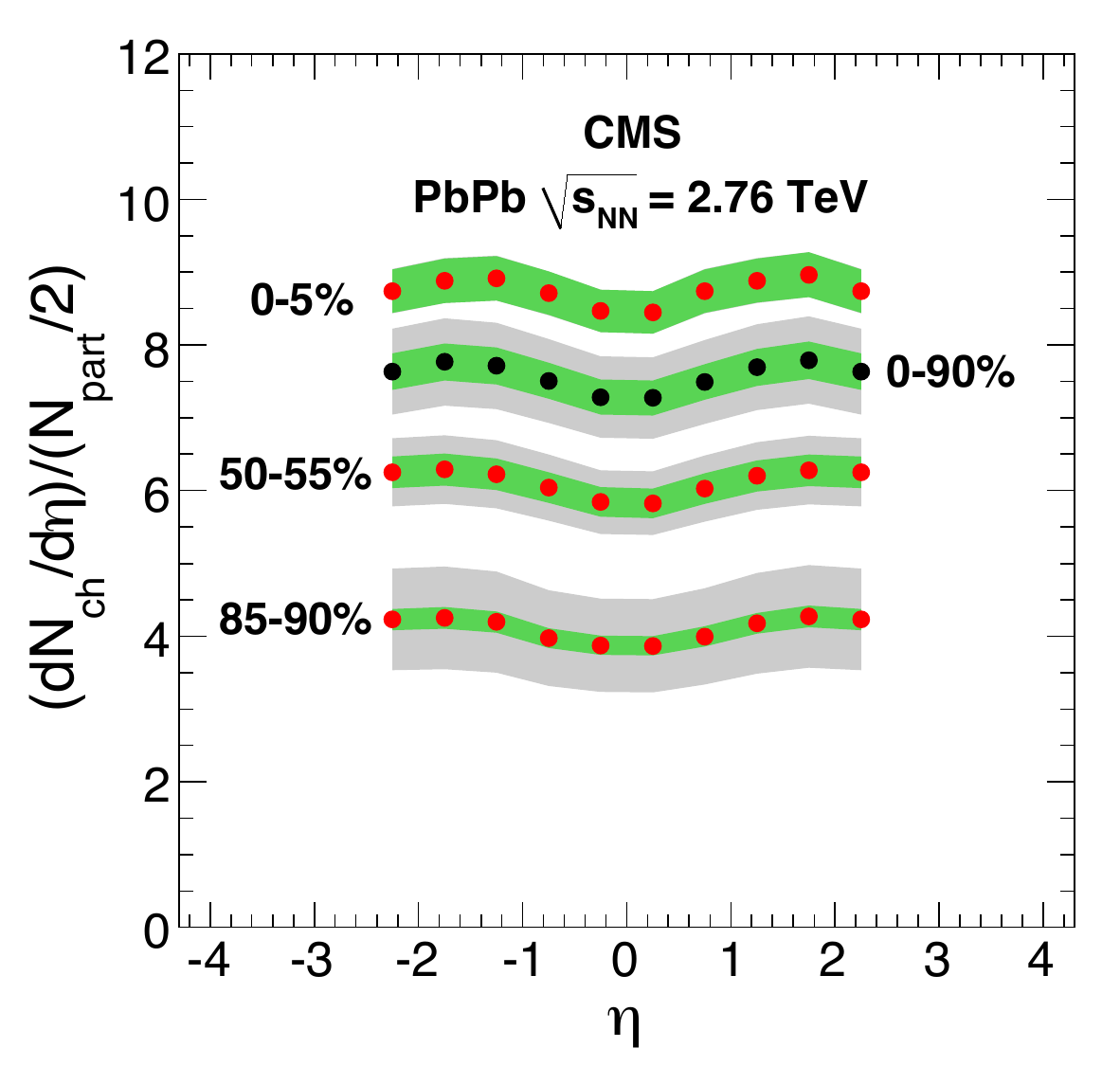}
    \includegraphics[width=0.42\linewidth]{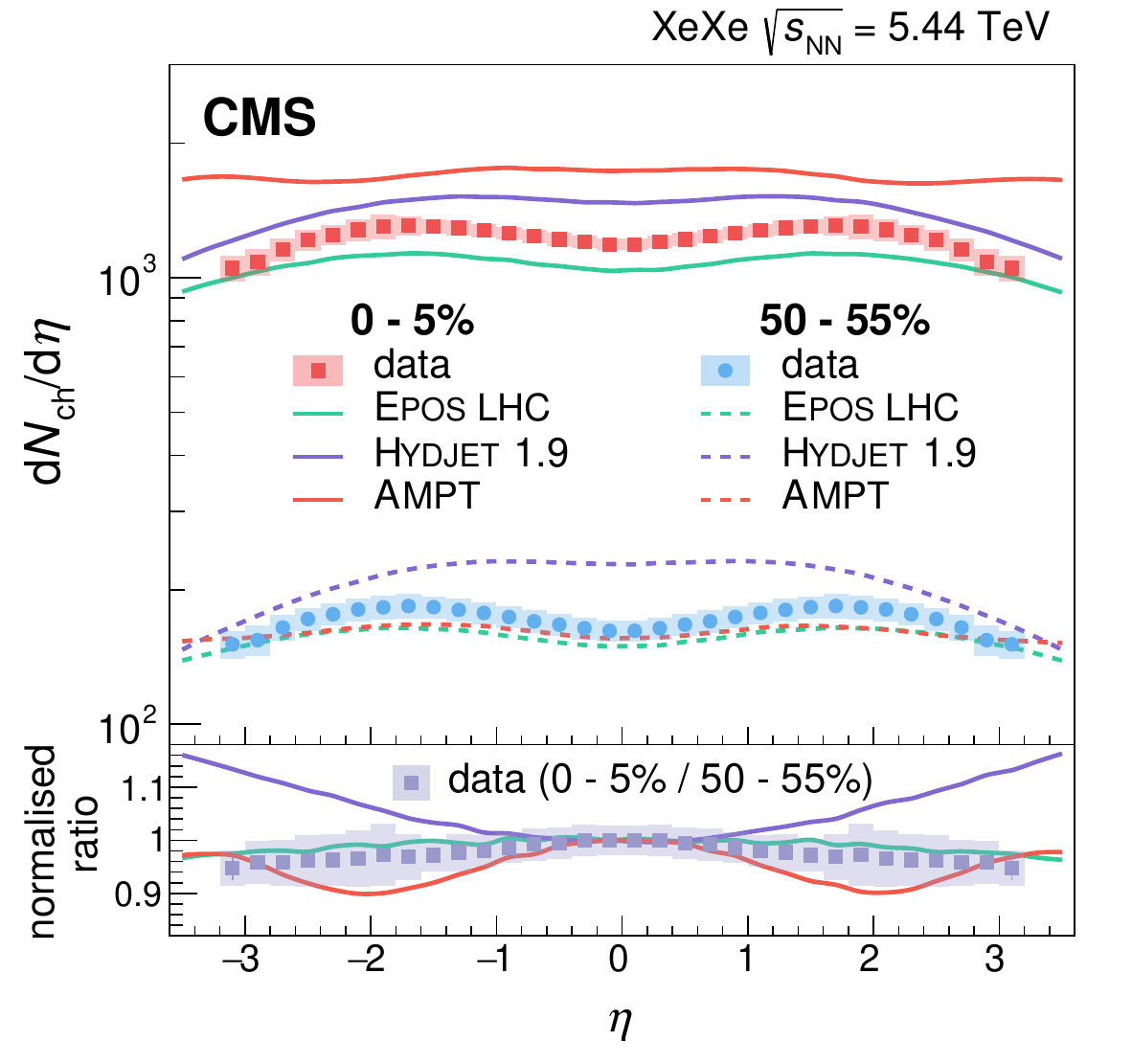}
    \includegraphics[width=0.45\linewidth]{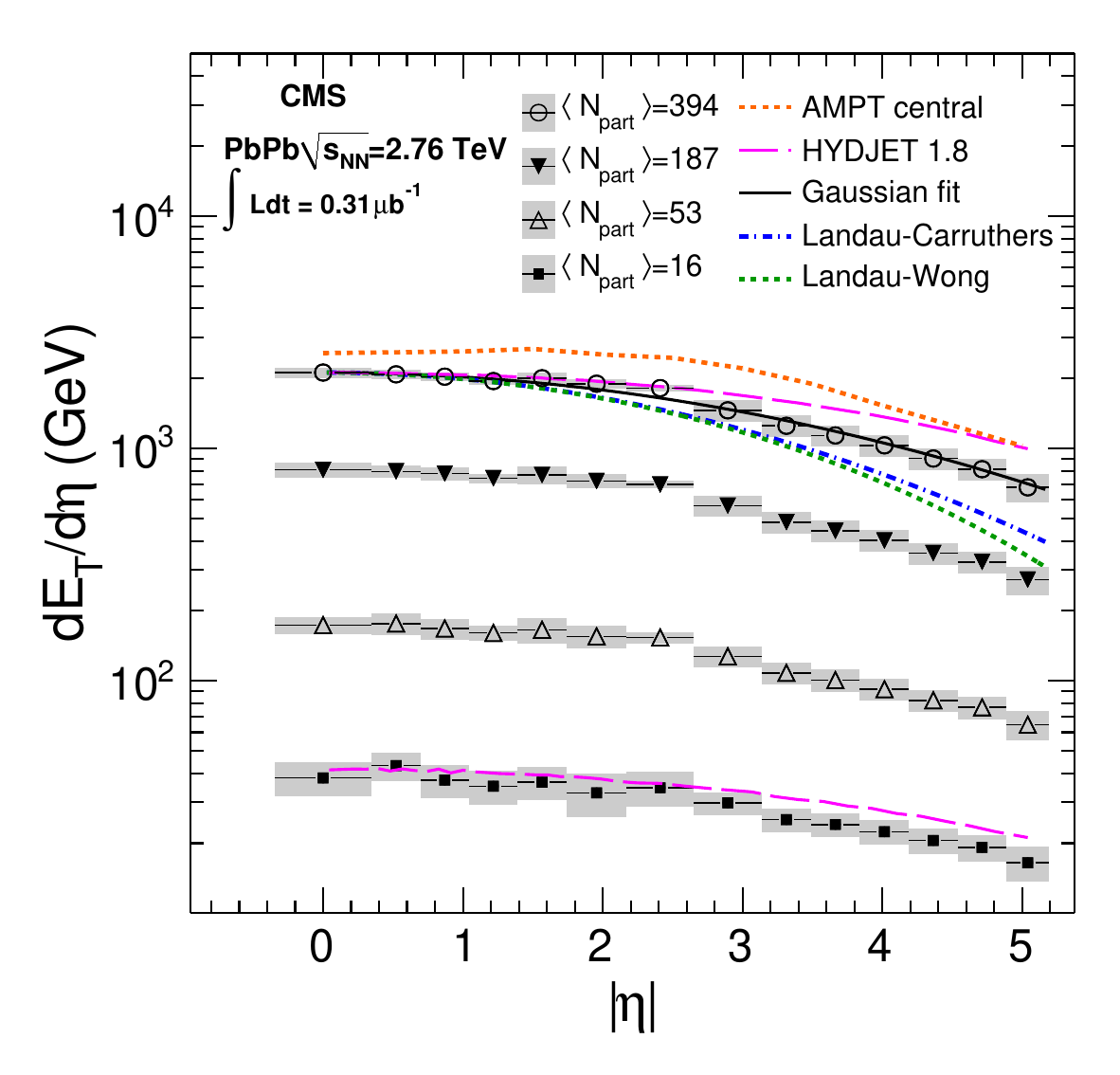}
    \caption{Distributions $(\dnchdeta)/(\npart/2)$ in 2.76\TeV \PbPb (top left) and $\dnchdeta$ in 5.44\TeV XeXe (top right) collisions, and $\rd\ET/\rd\eta$ in 2.76\TeV \PbPb collisions (bottom) as functions of $\eta$ in various centrality bins. The inner green band in the left panel shows the measurement uncertainties affecting the scale of the measured distribution, while the outer gray band shows the full systematic uncertainty, \ie affecting both the scale and the slope. \FiguresCompiled{Chatrchyan:2011pb,CMS:2019gzk,CMS:2012krf}}
    \label{fig:dNdetaAA}
\end{figure}

The top panels of Fig.~\ref{fig:dNdetaAA} show, for different centrality ranges,
the $\eta$ dependence of the charged-particle densities $\dnchdeta$ 
in \PbPb (left) and XeXe (right) collisions, the former case scaled by $\npart/2$, 
where \npart is the number of nucleons that participate in the collisions.
The $\eta$ dependence of the results is weak, varying by less than 10\% over the range $\abs{\eta} < 2.4$. The slight dip at $\eta = 0$ is a trivial (Jacobian) kinematic effect resulting from the use of $\eta$ rather than rapidity $y$ for the independent variable. This dip is absent in the $\rd\ET/\rd\eta$ distribution (bottom panel of Fig.~\ref{fig:dNdetaAA}), where \ET is the measured transverse energy. This latter distribution can be described by a Gaussian function of width $\sigma_\eta=3.4\pm0.1$ for central collisions, which is larger than predicted by Landau hydrodynamics. Indeed, none of the standard LHC event generators, including \textsc{ampt}, \HYDJET, and \EPOS, have been able to fully describe either the measured charged-particle multiplicity or the transverse energy distributions~\cite{Chatrchyan:2011pb, CMS:2019gzk, CMS:2012krf}. That means that the new LHC results provide important constraints on models and generators that
characterize multiparticle production in HI collisions at high energies.

\begin{figure}[ht]
  \centering
    \includegraphics[width=0.4\textwidth]{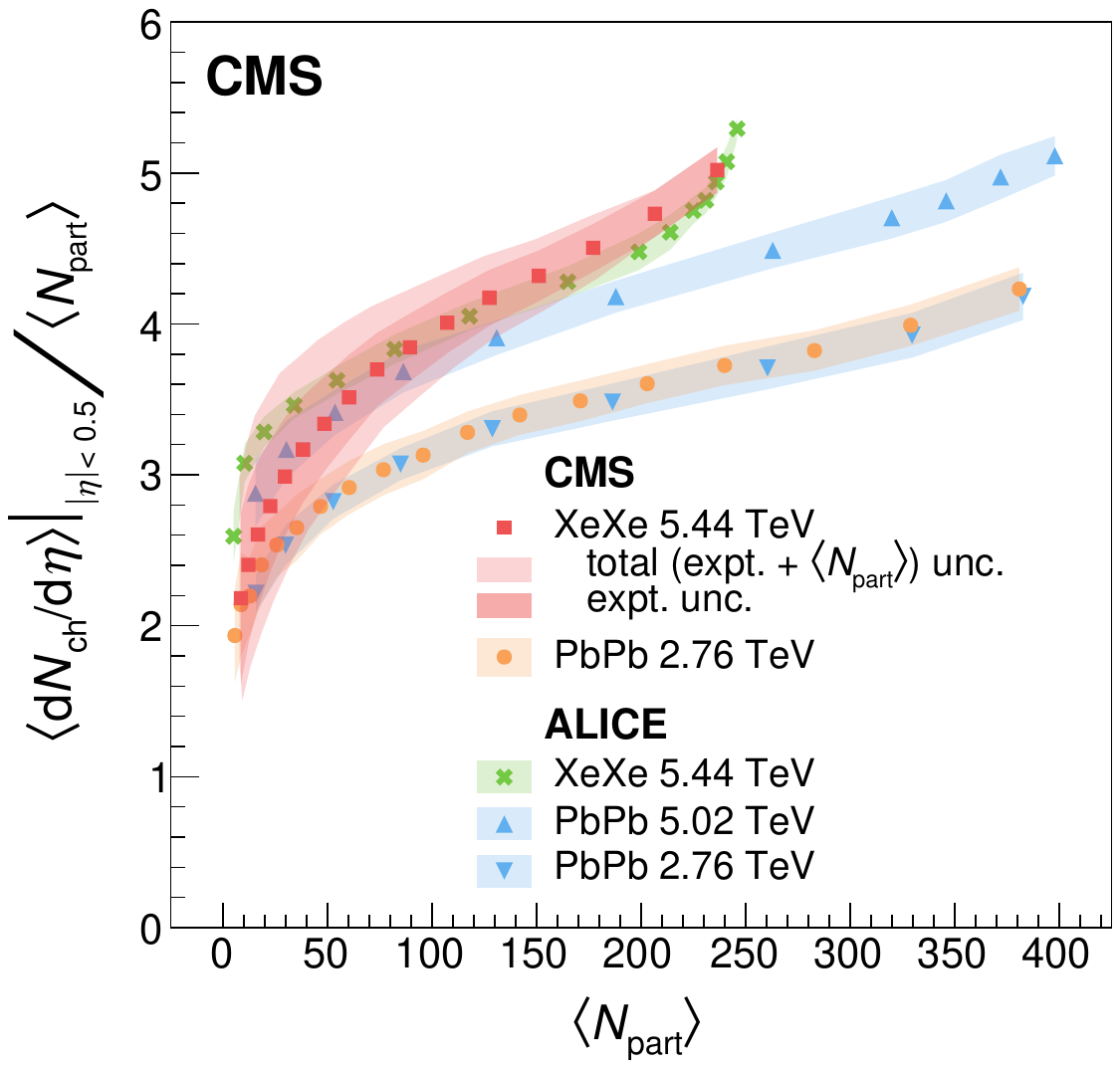}
    \includegraphics[width=0.4\textwidth]{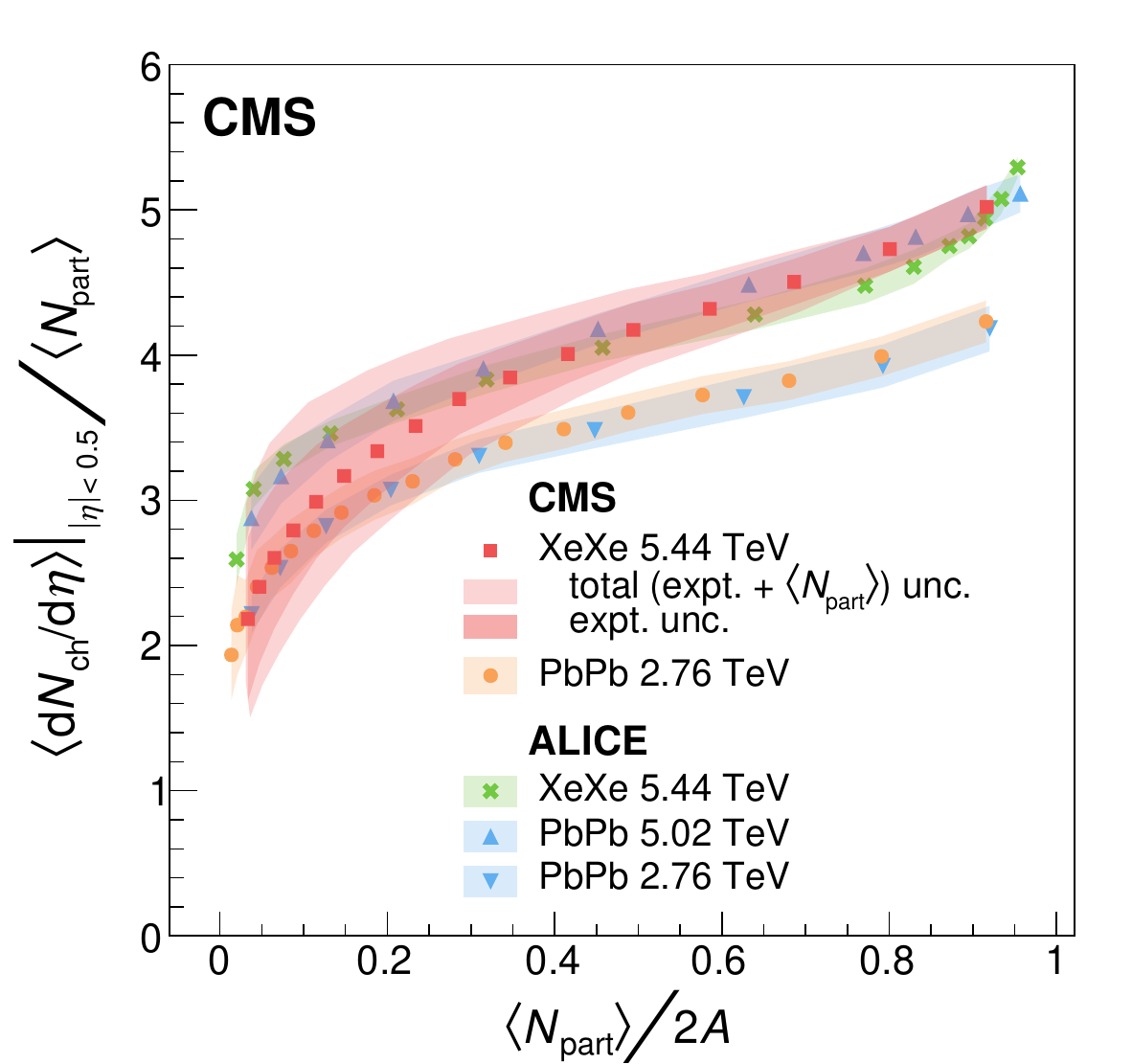}
    \caption{Average \dnchdeta\ at midrapidity normalised by \npartave, shown as
    a function of \npartave (left) and $\npartave/2A$ (right), where
    $A$ is the mass number of the nuclei. \FiguresFrom{CMS:2019gzk}}
    \label{fig:dNdetaNpart}
\end{figure}

In order to further study the system size dependence of particle density distributions, we have measured \dnchdeta\ values in the smaller \XeXe system~\cite{CMS:2019gzk}. 
The per-participant multiplicities for \XeXe and \PbPb collisions with similar
\npartave, and consequently corresponding to different centrality classes, are inconsistent in the two collision systems, as shown in Fig.~\ref{fig:dNdetaNpart} (left). 
This is most apparent for $\npartave \approx 236$. 
However, as shown in
Fig.~\ref{fig:dNdetaNpart} (right), where $\npartave/2A$ is used as a proxy for
centrality, the per-participant charged-hadron multiplicities for
different colliding nuclei are equal within uncertainties when the geometry
(centrality) and energy of the compared systems are the same.
This mirrors a lower-energy result obtained at RHIC that the particle production is dependent on the collision geometry in addition to the system size and collision energy~\cite{Alver:2007aa}. 

\begin{figure}[ht]
    \centering
    \includegraphics[width=0.4\linewidth]{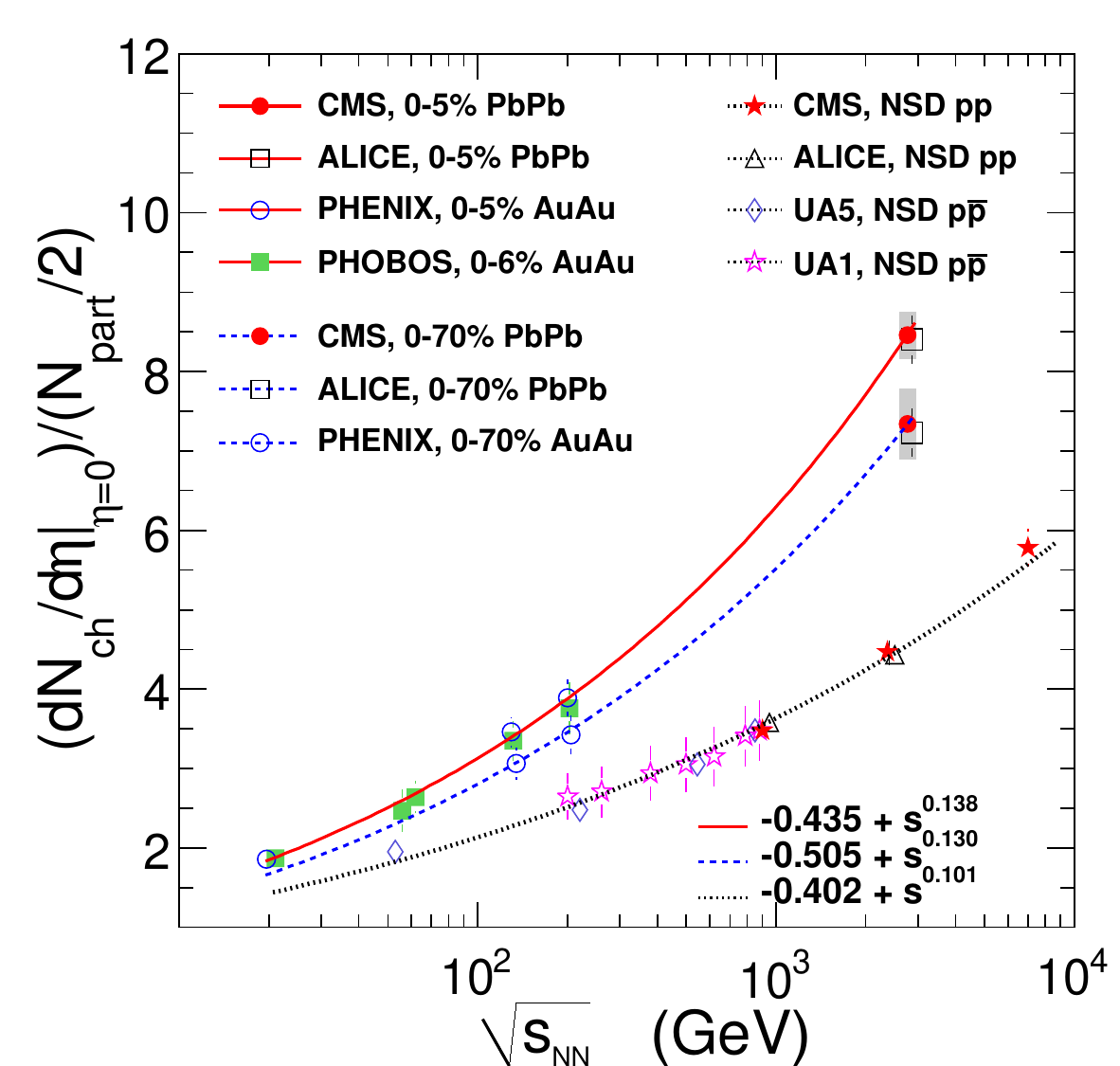}
    \includegraphics[width=0.4\linewidth]{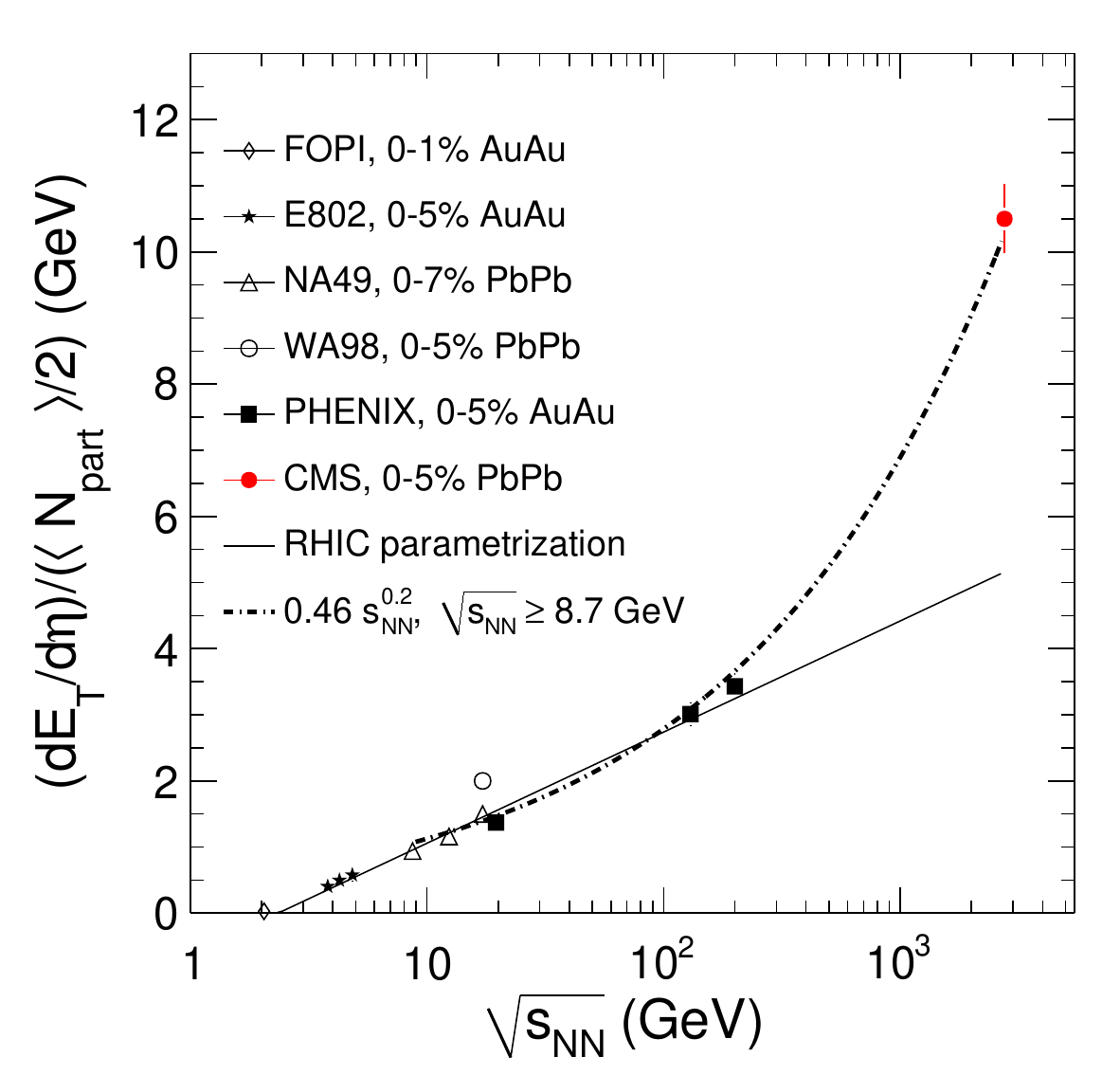}
    \caption{Normalized charged-particle pseudorapidity (left, figure adapted from Ref.~\cite{Chatrchyan:2011pb}) and transverse energy density (right, figure adapted from Ref.~\cite{CMS:2012krf}) at $\eta= 0$ as functions of center-of-mass energy, from various experiments. The fits to power-law functions are shown by lines.}
    \label{fig:dNdetaFit}
\end{figure}

Figure~\ref{fig:dNdetaFit} shows the dependence on the center-of-mass energy per nucleon pair \rootsNN of the charged-particle multiplicities (left panel) at midrapidity ($\eta=0$). The \AonA results from CMS, ALICE~\cite{Aamodt:2010cz}, PHENIX~\cite{Adler:2004zn}, and PHOBOS~\cite{Alver:2010ck}, and the non-single-diffractive~(NSD) \pp results (excluding events with significantly high particle density on one rapidity side only) from CMS~\cite{Khachatryan:2010us, Khachatryan:2010xs}, ALICE~\cite{Aamodt:2010ft}, UA5~\cite{Alner:1986xu}, and UA1~\cite{Albajar:1989an} experiments, are shown. 
The dependence is modeled by a power-law function $a + {s^n_{_{\mathrm{NN}}}}$, with observed value $n = 0.13$ for \PbPb and $n = 0.10$ for NSD \pp events. This shows that the multiplicity increases more rapidly with the center-of-mass energy than the logarithmic dependence used to describe data up to $\sqrtsNN = 200\GeV$~\cite{Adler:2004zn}.
A similar study has been performed for the transverse energy distribution (right panel), where $n \approx 0.2$, showing that the transverse energy density increases faster with collision energy than the charged-particle multiplicity. 
Furthermore, for the 5\% most central collisions, CMS has measured the transverse energy per charged-particle at $\eta=0$ of $1.25 \pm 0.08\GeV$ at $\rootsNN= 2.76\TeV$.
The corresponding value at $\rootsNN = 200\GeV$ was found to be $0.88 \pm 0.07\GeV$~\cite{Adler:2004zn}, indicating a significant increase of transverse energy per particle at the higher beam energy. This increase reflects a higher initial energy density at the LHC, compared to RHIC, as transverse energy is closely related to the  energy deposited in the medium.

\subsection{The paradigm of a nearly perfect liquid}
\label{sec:HydroQGP}

In a noncentral HI collision, the overlap region 
has a lenticular initial shape, and the interacting nucleons in this region are
known as ``participants''. The ``participant plane'' is defined by the beam
direction and the short axis of the participating nucleon distribution.
Because of fluctuations that arise from the finite number of nucleons,
the impact parameter vector typically does not coincide with
the short axis of this lenticular region. Strong partonic
rescatterings of the system may lead to local thermal
equilibrium and the build-up of anisotropic pressure gradients, driving a
collective expansion that is anisotropic with a faster expansion along the short axis of the lenticular overlap region.
As a result, the eccentricity of the initial collision geometry
translates in an anisotropic azimuthal distribution of the final-state particles.
This final-state anisotropy is typically characterized by the Fourier harmonic coefficients
(\vN) in the azimuthal angle ($\phi$) distribution of the hadron yield,
\begin{equation}
\label{eq:fourier}
\rd{}N/\rd\phi \propto 1+2 \sum_{n}^{} v_{n}\cos[n(\phi-\Psi_\PN)],   
\end{equation}
\noindent where $\Psi_\PN$ is the event-by-event azimuthal angle of the event plane, defined as the direction of
maximum final-state particle density. 
The second-order Fourier component~(\vTwo)
is known as the ``elliptic flow'' harmonic, and its event plane angle $\Psi_{2}$
corresponds, approximately, to the short axis direction of the lenticular region.
Because of event-by-event fluctuations, higher-order eccentricities
can also arise, leading to higher-order Fourier
harmonics (\vN, $n \ge 3$) in the final state with respect to their corresponding event plane angles,
$\Psi_\PN$~\cite{Heinz:2013th}. In hydrodynamic models, the \vN coefficients are related to the response of the QGP
medium to the initial geometry and its fluctuations~\cite{Heinz:2013th}. As such, these Fourier
components can provide insight into the fundamental transport properties of the medium. 

A wide $\eta$ coverage gives the CMS Collaboration an opportunity to correlate particles with large $\eta$ difference and thus significantly suppress short-range correlations. Taking this advantage extensive
studies have been performed of the particle anisotropy developed through collective flow using
several techniques based on particle correlations over a wide 
phase space~\cite{CMS:2011cqy,CMS:2012xss,CMS:2012zex,Chatrchyan:2013kba,CMS:2013bza,CMS:2015xmx} to extract the \vN coefficients. These techniques include using correlations of two-particle pairs over long ranges in $\eta$~\cite{CMS:2011cqy,CMS:2012xss,CMS:2013bza,CMS:2015xmx} and the scalar-product or, in earlier studies, event plane method that correlates individual particles in one region of phase with an event plane angle $\Psi_\PN$ established in another ~\cite{CMS:2012zex,Chatrchyan:2013kba}. Correlations among multiple particles (four or more), known as the "cumulants," have also been studied. These multiparticle correlations are particularly sensitive to event-by-event fluctuations of the \vN coefficients~\cite{CMS:2012zex,Chatrchyan:2013kba}. Measurements of event-by-event \vTwo probability distributions provide
a direct means to constrain the elliptic-flow fluctuations~\cite{CMS:2017glf}.
Collectively, the particle correlation studies have played a vital role in constraining 
the initial state and transport properties of the QGP medium,
leading to the paradigm of a
nearly perfect QCD liquid formed in ultrarelativistic 
nuclear collisions. 

\subsubsection{Transport properties and ripples in the QGP}
\label{sec:TransportProperties}

Two-particle correlations provide a powerful quantitative tool
to study the collective aniso\-tro\-py of
final-state particles from HI collisions.
In this section we review results based on two-particle correlations to demonstrate how transport properties of the QGP can be constrained by experimental data. Each pair of particles can have its constituents chosen from the same
or different \pt ranges, denoted as \pTtrig (or ``trigger'') and \pTassoc (or ``associated''), within the CMS tracker
acceptance of $\abs{\eta} < 2.5$.
The two-dimensional (2D) two-particle correlations as functions of the relative pseudorapidity (\deta) and azimuthal angle (\dphi) between the two particles of a pair is given by
\begin{equation}
\label{2pcorr_incl}
\frac{1}{\Ntrig}\frac{\rd^{2}\Npair}{\rd\deta\, \rd\dphi}
= B(0,0) \frac{S(\deta,\dphi)}{B(\deta,\dphi)},
\end{equation}
\noindent where \Ntrig is the number of trigger particles in an event and \Npair is the total number of hadron pairs for the event. The signal distribution, $S(\deta,\dphi)$, is constructed by 
taking particle pairs from the same event, while
the background distribution, $B(\deta,\dphi)$,
is obtained by pairs of particles taken from
different events with similar topology. The ratio $B(0,0)/B(\deta,\dphi)$ represents
the correction for pair-acceptance effects.

\begin{figure}[ht]
    \centering
    \includegraphics[width=\linewidth]{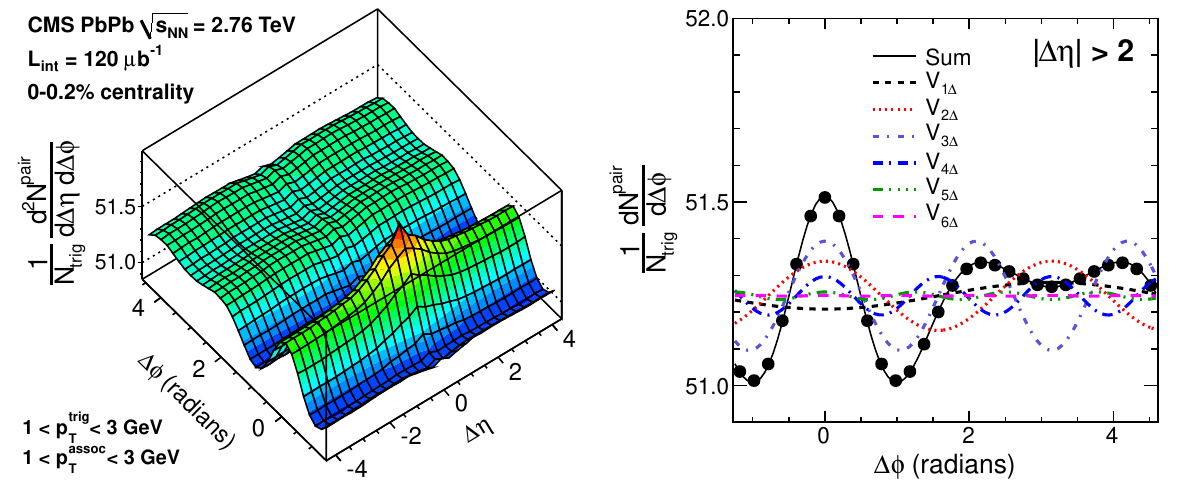}
    \caption{The 2D (left) and 1D \dphi\ (right) two-particle correlation functions for
    $1<\pt<3\GeV$ in 0--0.2\% central \PbPb collisions at
    $\rootsNN = 2.76\TeV$.~\FiguresFrom{CMS:2013bza}}
    \label{fig:UCCCF}
\end{figure}

Figure~\ref{fig:UCCCF} (left) shows the two-particle correlation function for both particles with $1<\pt<3\GeV$ in 
0--0.2\% central \PbPb events at $\rootsNN = 2.76\TeV$.
On the near side ($\dphi \sim 0$) of the correlation function, a long-range structure
extending over the entire \deta\ region is evident. 
This feature of long-range rapidity correlations
has been observed across multiple centrality ranges~\cite{CMS:2011cqy,CMS:2012xss},
corresponding to different initial 
size and geometry of the system. The one-dimensional (1D) \dphi correlation function is shown in Fig.~\ref{fig:UCCCF} (right) with a requirement of $\abs{\deta}>2$ to exclude
noncollective effects from other sources of correlations, such as jet fragmentation.
By fitting the 1D \dphi correlation function
by a Fourier series (as indicated by dashed curves), 
\begin{equation}
\label{fourier_2pcorr}
\frac{1}{\Ntrig}\frac{\rd \Npair}{\rd\dphi} \propto 1+\sum\limits_{n=1}^{\infty} 2V_{n\Delta}(\pTtrig,\pTassoc) \cos (n\dphi),
\end{equation}
\noindent 
where $V_{n\Delta}(\pTtrig,\pTassoc)$ are the two-particle Fourier coefficients. Assuming that $V_{n\Delta}(\pTtrig,\pTassoc)$ can be factorized into a product of single-particle, global \vN coefficients (as defined in Eq.~(\ref{eq:fourier})), as both particles share a common event plane $\Psi_\PN$~\cite{Aamodt:2011by},
\begin{equation}
\label{eq:factorization}
V_{n\Delta}=v_{n}(\pTtrig) \, v_{n}(\pTassoc).
\end{equation}
Hence, the anisotropy flow coefficients as functions of \pt can be extracted. 

\begin{figure}[ht]
    \centering
    \includegraphics[width=0.5\linewidth]{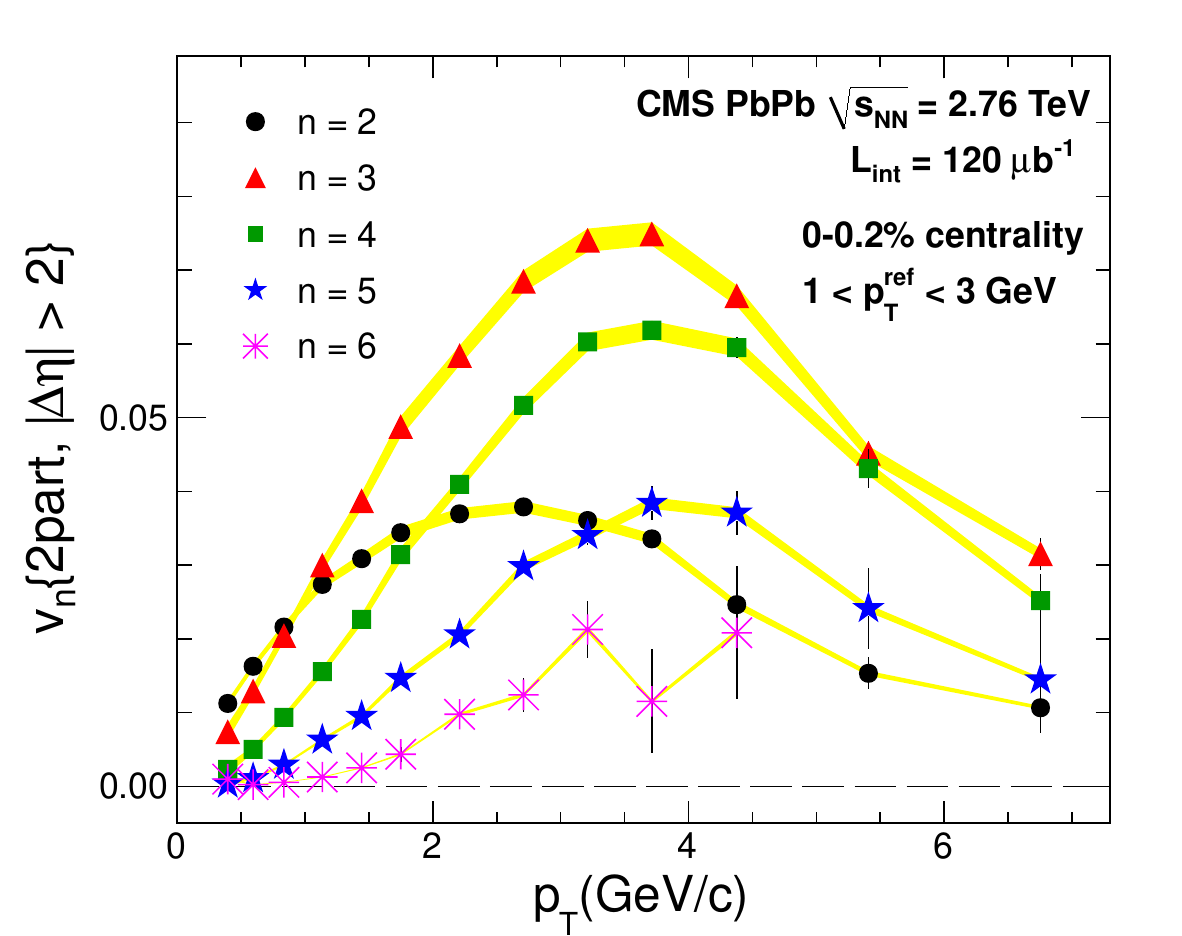}
    \includegraphics[width=0.42\linewidth]{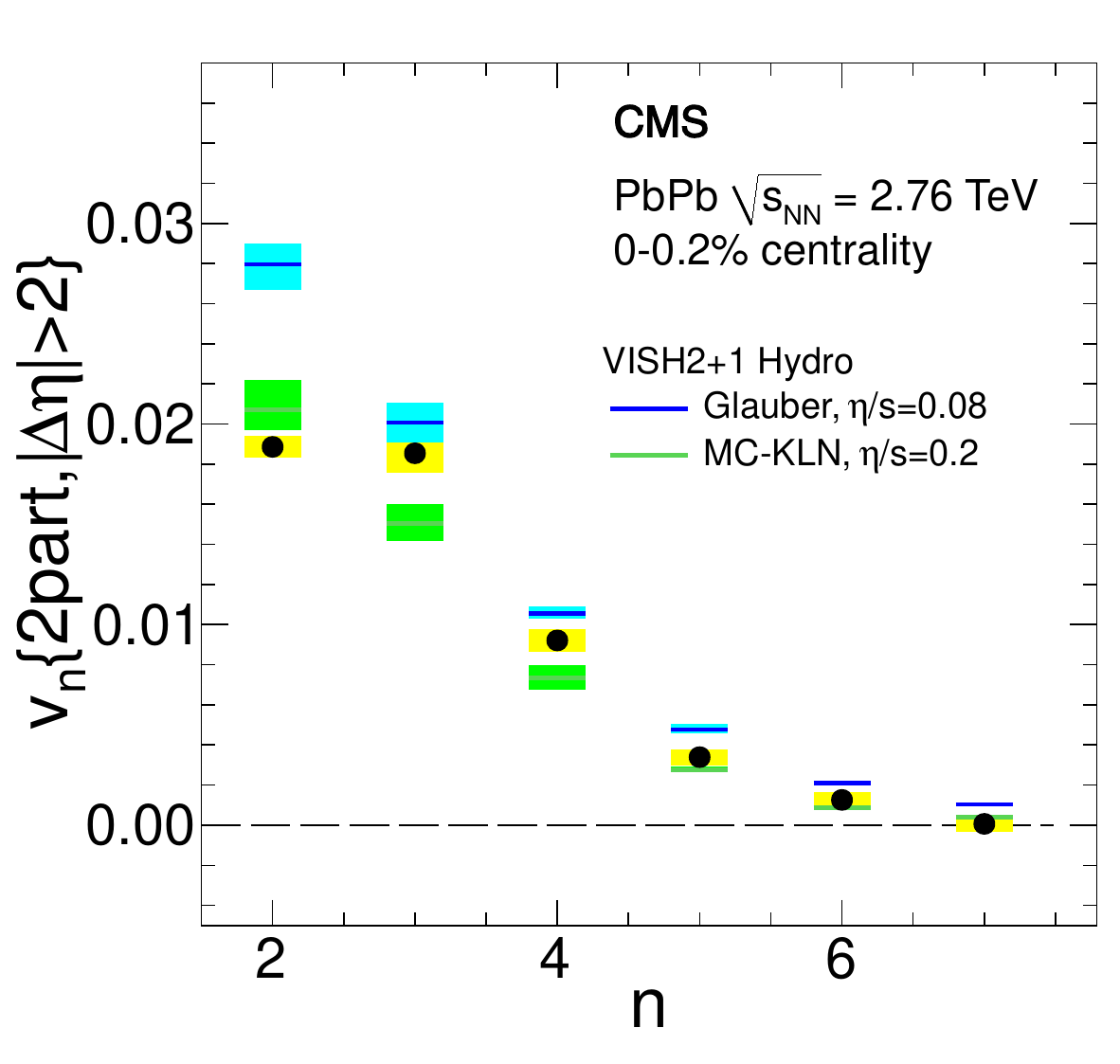}
    \caption{Left: the \vTwo to \vSix values as functions of \pt in
    0--0.2\% central \PbPb collisions at $\rootsNN = 2.76\TeV$. Right: Comparison of \pt-integrated (0.3--3.0\GeV) \vN data with VISH2+1D hydrodynamic calculations for Glauber initial condition with $\eta/s=0.08$ (blue) and MC-KLN initial condition with $\eta/s=0.2$ (green), in 0--0.2\% central \PbPb collisions at $\rootsNN = 2.76\TeV$. Error bars denote the statistical uncertainties, while the shaded color bands correspond to the systematic uncertainties.~\FiguresFrom{CMS:2013bza}}
    \label{fig:UCCvn}
\end{figure}

The single-particle azimuthal anisotropy coefficients, from \vTwo to \vSix, as functions of \pt
extracted in 0--0.2\% central \PbPb collisions at $\rootsNN\ = 2.76\TeV$, are shown in Fig.~\ref{fig:UCCvn} (left). 
Different orders of \vN
harmonics show very different dependences on \pt. At low \pt ($\pt < 1\GeV$), the \vTwo
harmonic coefficient that corresponds to an elliptical anisotropy has the greatest magnitude. However, this coefficient
becomes smaller than the \vThree coefficient at $\pt\approx 1\GeV$, and even smaller than the $v_5$ coefficient for $\pt>3\GeV$. This intriguing \pt dependence can be used to quantitatively constrain hydrodynamics models of HI collisions with fluctuating initial conditions.

The \pt-averaged \vN values from 0.3 to 3.0\GeV
are presented in Fig.~\ref{fig:UCCvn} (right) as functions of $n$ up to \mbox{$n=7$}, and compared with hydrodynamic model calculations~\cite{Shen:2015qta}. As the collisions
are extremely central, the initial eccentricities
for all orders are mostly driven
by event-by-event participant fluctuations and are of similar values~\cite{Shen:2015qta}. Therefore, the diminishing \vN values
towards higher orders reflects damping effects of 
viscous dissipation (typically characterized by the shear viscosity to entropy density ratio, $\eta/s$, which is a dimensionless quantity~\cite{Heinz:2013th})
that tends to suppress higher-order
deformations more strongly. As shown in Fig.~\ref{fig:UCCvn} (right), the CMS \vN data for all orders except for $n=2$
lie between the two hydrodynamic calculations
with only differences in initial-state models (MC-Glauber and MC-KLN) and $\eta/s$ values (0.08 and 0.2). The `tension' for $n=2$ has been largely resolved in more recent calculations with improved modeling of the initial state~\cite{Nijs:2021clz}. Therefore, these studies have imposed a stringent constraint on the allowed $\eta/s$ value for the QGP. The observed value, 
in the range 0.08--0.2, suggests that the QGP behaves
like a nearly perfect liquid (close to the theoretical lower bound by quantum fluctuations of $\eta/s= 1/4\pi$~\cite{Son:2007vk}) with little frictional momentum dissipation.

\subsubsection{Direct constraints on initial-state fluctuations}

As a consequence of the initial geometry fluctuations, flow harmonic magnitudes vary significantly event-by-event. This is also the case for the elliptic flow \vTwo coefficient that, for noncentral events, has its origin in the shape of the overlapping area of the colliding nuclei. The second-order eccentricity $\epsilon_n$ of the medium responsible for the azimuthal particle density asymmetry is affected by fluctuations of participant positions in the colliding nuclei which, in turn, results in fluctuations in the observed \vTwo values. Different methods for measuring azimuthal anisotropy, which essentially have different ways of averaging anisotropy over many events, give different \vTwo values. Comparison of flow coefficients measured by different methods is a direct probe of the initial-state conditions.

\begin{figure}[ht]
    \centering
    \includegraphics[width=0.9\linewidth]{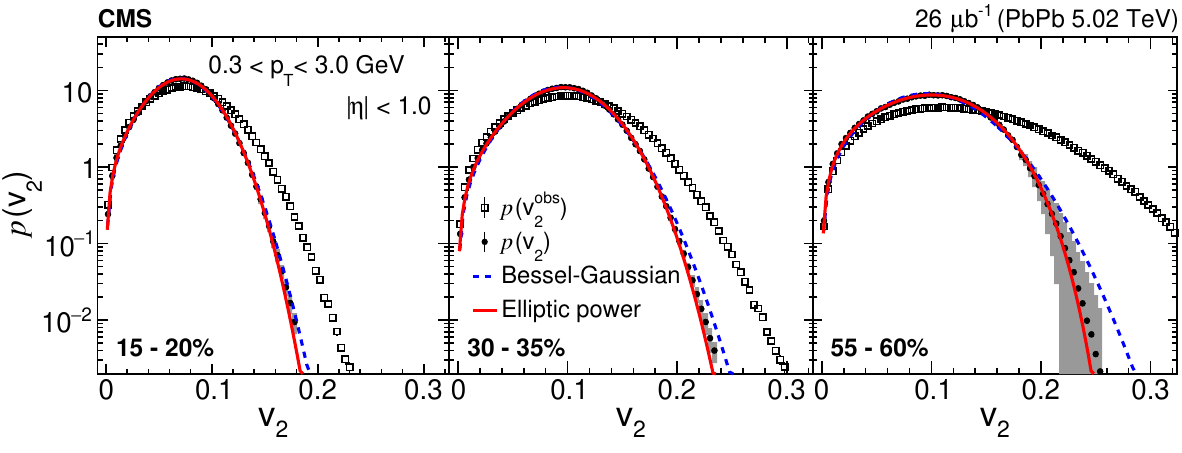}
    \caption{Representative final unfolded $p(\vTwo)$ distributions (closed black circles)
             in three centrality bins (15--20\%, 30--35\%, and 55--60\%). Respective observed $p(\vTwo^\text{obs})$ distributions
             (open black squares) are shown to illustrate the statistical resolution present in each centrality bin
             prior to unfolding. Distributions are fitted with Bessel-Gaussian
             and elliptic power functions to infer information
             on the underlying $p(\varepsilon_2)$ distributions.~\FigureFrom{CMS:2017glf}}
    \label{fig:pv2}
\end{figure}

The CMS Collaboration has directly studied the probability distribution functions of the magnitudes of the \vN values, $p(\vN)$, through an unfolding technique~\cite{CMS:2017glf}. The particles within an event are used to construct the ``observed'' ${p}(\vTwo^{\text{obs}})$ distributions, while residual contributions from multiplicity-related fluctuations and nonflow effects are estimated and subtracted by considering $p(\vN)$ difference between two symmetric subevents based on pseudorapidity. This difference should not contain ``real'' flow, given that $\vTwo(\eta)$ is symmetric about
$\eta=0$, on average, for the symmetric \PbPb system.

Figure~\ref{fig:pv2} shows the $p(\vTwo)$ distribution in \PbPb collisions for three centrality classes. In addition, ${p}(\vTwo^{\text{obs}})$ distributions are plotted for each centrality
to illustrate the statistical resolution effects present prior to unfolding. 
The elliptic power and Bessel-Gaussian parametrizations are used to fit the data (a discussion of the parameterizations can be found in Ref.~\cite{Yan:2014nsa}). The elliptic power $\chi^2/\mathrm{dof}$ values vary between 0.8 and 1.5 from central to peripheral collisions, while the Bessel--Gaussian $\chi^2/\mathrm{dof}$ values vary between 3 and 9. Both models assume linear response between eccentricity and flow, with $p(\epsilon_2)=k_2 p(\vTwo)$, but only the elliptic power function allows for a nonzero skewness (asymmetry of the distribution), hence being able to better fit the data. 
The $k_2$ parameter of the elliptic power function is expected to have only a weak dependence on the initial conditions and the measured value is consistent with the hydrodynamic calculation with Glauber initial conditions and an $\eta/s$ value of 0.19~\cite{Yan:2014nsa}. 

In addition to \PbPb results, a short run with Xe nuclei in 2017 gave the LHC experiments a chance to probe the scaling of the hydrodynamics and initial-state effects with system size. 
Fluctuations of the initial state are proportional to $A^{-1/2}$, where $A$ is the atomic mass, and, therefore, one can expect a larger fluctuation component for \XeXe collisions than for \PbPb collisions~\cite{Bhalerao:2011bp}.
However, the viscosity, which tends to decrease the azimuthal anisotropy, is thought to be proportional to $A^{-1/3}$~\cite{Romatschke:2007mq} and is, therefore, also expected to be larger for \XeXe collisions. 
Also, the quadrupole deformation of the Xe nuclei can cause two colliding systems in the same centrality class to have different geometries.

\begin{figure}[ht]
    \centering
    \includegraphics[width=0.9\linewidth]{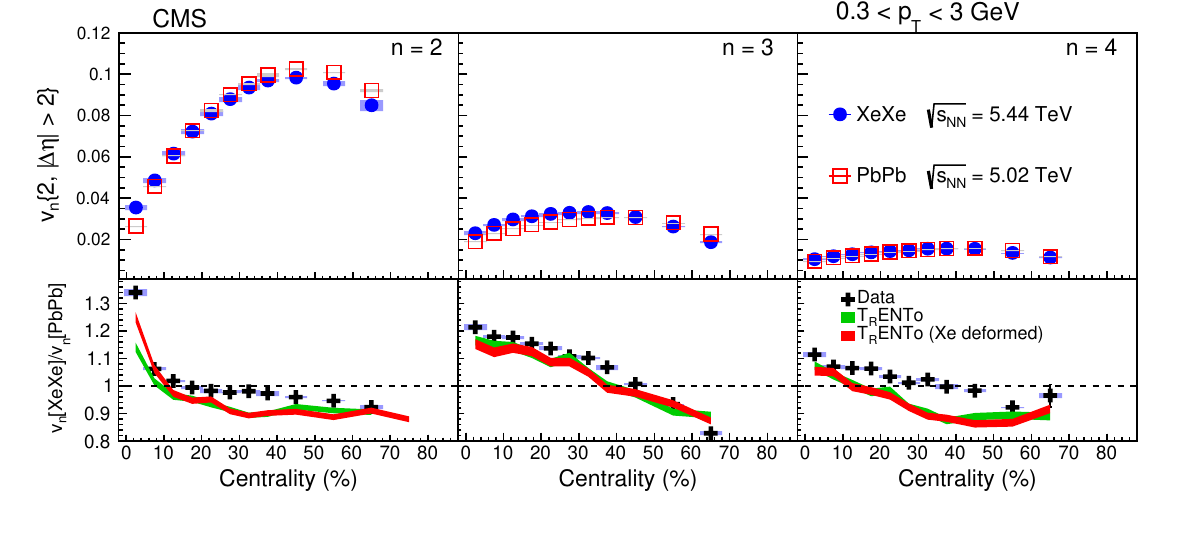}
    \caption{Centrality dependence of the \vTwo, \vThree, and \vFour harmonic coefficients from two-particle correlations method for $0.3 < \pt < 3.0\GeV$ for \XeXe collisions at $\sqrtsNN=5.44\TeV$ and \PbPb collisions at 5.02\TeV. The lower panels show the ratio of the results for the two systems. Theoretical predictions from Ref.~\cite{Giacalone:2017uqx} are compared to the data. The model calculation is done for the \pt range $0.2 < \pt < 5.0\GeV$.~\FigureFrom{CMS:2019cyz}}
    \label{fig:vnXeXe}
\end{figure}

Figure~\ref{fig:vnXeXe} compares the spectrum-weighted \vTwo, \vThree, and \vFour values
 with $ 0.3 < \pt < 3.0\GeV$ for the \XeXe and \PbPb systems.
For all three harmonics, the \XeXe values are higher in central collisions, while the \PbPb results become larger for more peripheral events.
The ordering of the measured harmonics between the two systems is consistent
with participant fluctuations having a dominant role in central collisions, and viscosity effects
becoming more important for mid-central and peripheral collisions.
The largest difference between the two systems is found for the \vTwo coefficients corresponding to the most central events,
where the \XeXe results are larger by a factor of about 1.3.
The hydrodynamic model calculations with Trento initial conditions~\cite{Moreland:2014oya} for both spherical and deformed Xe shape, performed for the \pt range $0.2 < \pt < 5.0\GeV$, are shown in the lower panel.
The xenon nuclear deformation is found to only have a significant effect on the model \vTwo values for the most central collisions, where the calculation with deformed nuclei is closer to the data. For all measured harmonics, the model values lie below the experimental results, although qualitatively the behavior is similar.

\subsubsection{A new window to the full (3+1)-dimensional space-time and dynamical evolution}
\label{sec:Hydro3D}

It was thought originally that the factorization relation in Eq.~(\ref{eq:factorization}) holds for correlations arising from collective hydrodynamic flow, where emitted particles share a common event plane $\Psi_\PN$. However, in a first analysis of its kind, the CMS Collaboration has observed and studied the factorization breaking in anisotropic flow measurements using two-particle correlations~\cite{CMS:2012xss}. Because of initial-state local fluctuations, the final-state event plane depends on the 
particle kinematics, instead of being a global property of the phase space~\cite{Gardim:2012im,Heinz:2013bua}.
Comprehensive studies were conducted in Ref.~\cite{CMS:2015xmx}, where the \pt-dependent factorization ratio,
\begin{equation}
\label{eq:rn}
r_{n} = \frac{V_{n\Delta}(\pt^{a},\pt^{b})}{\sqrt{V_{n\Delta}(\pt^{a},\pt^{a})V_{n\Delta}(\pt^{b},\pt^{b})}} ,
\end{equation}
served as a quantitative measure of the factorization breaking. 
This ratio probes the relative
fluctuations of flow vectors with particles from two different \pt ranges~\cite{Gardim:2012im,Heinz:2013bua}, denoted $a$ and $b$. Hence, if the factorization holds exactly, this ratio is expected to be unity. However, with the presence of \pt-dependent flow fluctuations, this ratio typically becomes smaller than~1. 

\begin{figure}[ht]
    \centering
    \includegraphics[width=0.8\linewidth]{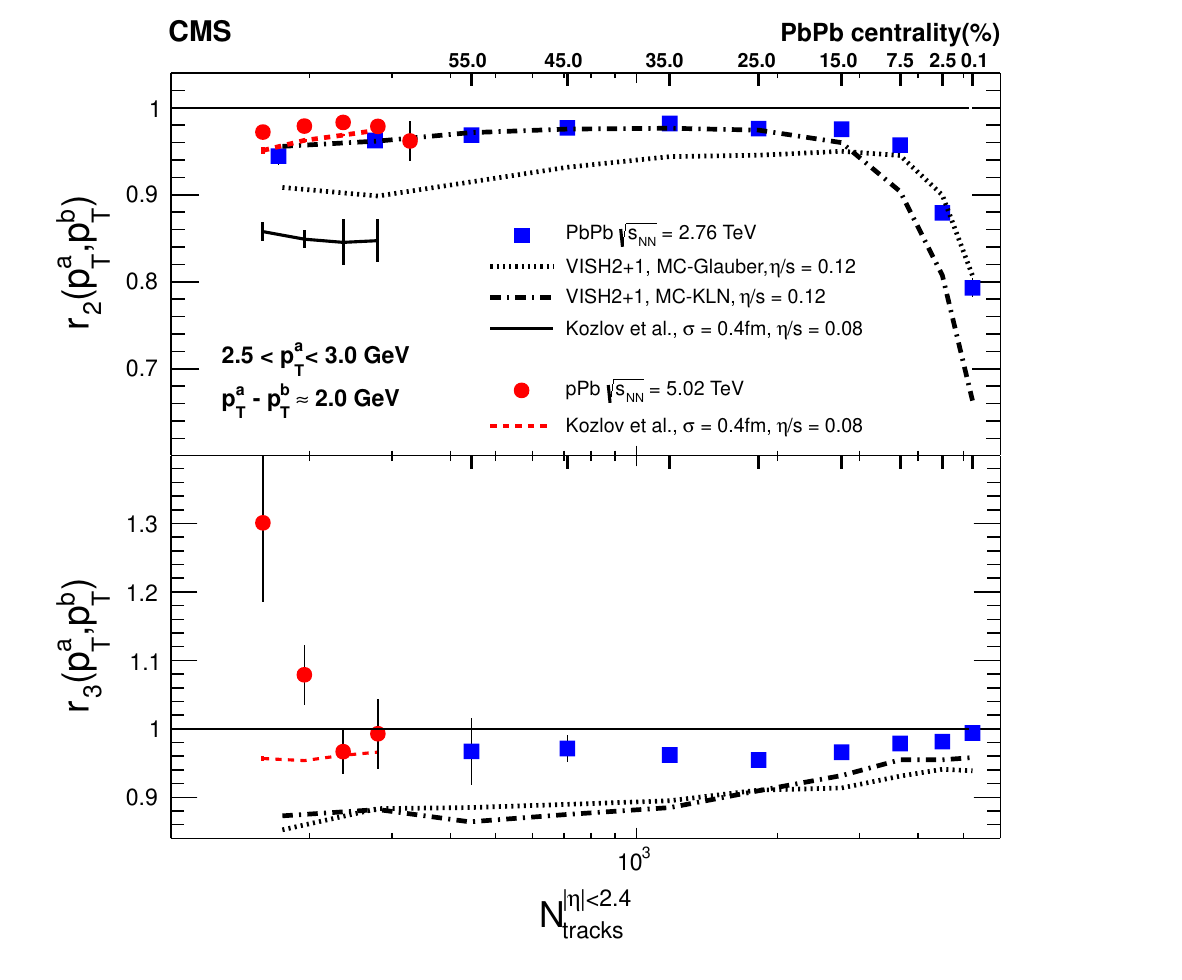}
    \caption{The \pt-dependent factorization ratios, \rTwo and \rThree, as functions of event
multiplicity in \pPb and \PbPb collisions. The lines represent different hydrodynamics calculations.~\FigureFrom{CMS:2015xmx}}
    \label{fig:factbreakptMult}
\end{figure}

\begin{figure}[ht]
    \centering
    \includegraphics[width=\linewidth]{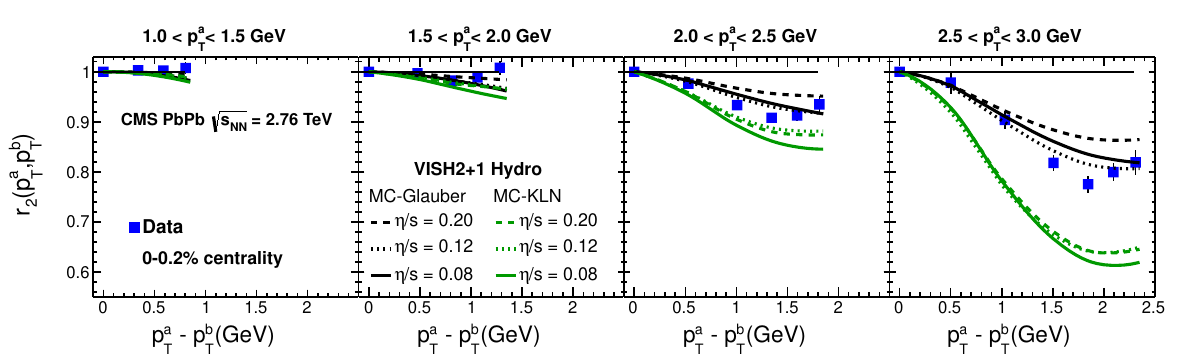}
    \caption{The \pt-dependent factorization ratios, $\rTwo(\pt)$, in very central (0--0.2\% centrality) \PbPb collisions. The lines represent hydrodynamics calculations for different initial conditions and different values of $\eta/s$. \FigureFrom{CMS:2015xmx}}
    \label{fig:factbreakptPt}
\end{figure}

Figure~\ref{fig:factbreakptMult} shows $r_n$ results for $2.5 < \pt^a < 3.0\GeV$ and $0.3 < \pt^b < 0.5 \GeV$ in \pPb and \PbPb collisions as functions of event multiplicity (the corresponding centrality scale for \PbPb events is shown by the upper $x$ axis)~\cite{CMS:2015xmx}. Factorization breaking is clearly observed for both the second and third harmonic. The \rTwo value deviates from unity by 2--5\% for midcentral and peripheral events, but suddenly increases to $\approx$20\% for 0--0.2\% centrality events. The \rThree value, in comparison, stays at a 2--3\% level for the entire centrality range. For a similar multiplicity range, the \rTwo value in \pPb collisions is slightly higher than for \PbPb collisions, but with the two values within statistical uncertainties. An \rThree value larger than~1, as found for low-multiplicity \pPb events, corresponds to the presence of nonflow effects. Alternative hydrodynamic model calculations 
in \PbPb collisions, using either MC-Glauber or MC-KLN~\cite{Dumitru:2011wq} initial conditions, are also shown. Neither set of initial conditions leads to quantitative agreement with the 
data over the entire centrality range, although the qualitative trend is reproduced.

For very central events, where the factorization breaking effect is the strongest, the calculations using different values of $\eta/s$ are compared to the data in Fig.~\ref{fig:factbreakptPt}~\cite{CMS:2015xmx}. 
For each initial-state model, the \rTwo values are found to be largely insensitive to different values of $\eta/s$. This observation, as well as the centrality dependence of the \rTwo values, is consistent with the flow fluctuations in \pt 
being driven primarily by local fluctuations in the initial energy density distribution. Thus, $r_n$ measurements can provide unique constraints on the initial-state modeling.

An equivalent approach for studying the \pt-dependence 
of the event plane and factorization breaking is the principal component analysis,
where measured two-particle Fourier coefficients
as functions of $\pt^a$ and $\pt^b$ can be
expressed in terms of an orthogonal basis of leading and subleading flow modes, as detailed in Ref.~\cite{CMS:2017mzx}.

Most of the earlier studies on collective flow have focused on the transverse expansion in the midrapidity region. Leveraging the wide coverage of the CMS detector,
the CMS Collaboration has now explored
the longitudinal dynamics of the QGP to establish, for the first time,
a full three-dimensional picture of the system
evolution. By studying the decorrelation of flow 
harmonic vectors measured at different rapidities, 
the CMS Collaboration aims to address two key 
questions related to the (3+1)D dynamics of a QGP: 
(1)~How is the initial entropy deposited in 3-D space, and how does it fluctuate event-by-event?
(2)~What is the role of the longitudinal pressure gradient?

\begin{figure}[ht]
    \centering
    \includegraphics[width=\linewidth]{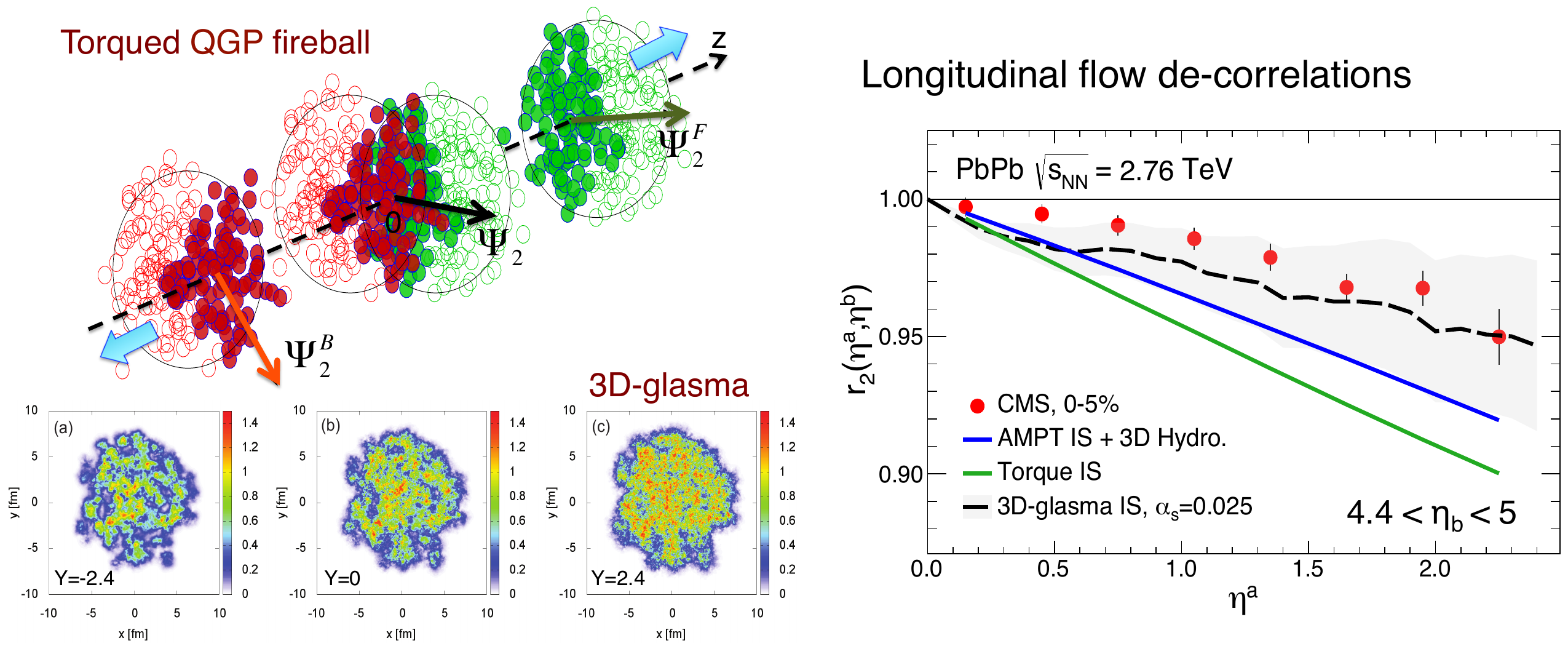}
    \caption{Left: illustration of flow event plane
  decorrelations as functions of rapidity in the wounded nucleon picture 
  (or ``torqued QGP fireball'')~\cite{Bozek:2010vz} 
  and 3D color glass condensate model~\cite{Schenke:2016ksl}.
  Right: measurement of elliptic flow decorrelations 
  as functions of pseudorapidity in 0--5\% central \PbPb collisions at 2.76\TeV 
  from CMS~\cite{CMS:2015xmx}, with comparison to 
  theoretical calculations~\cite{Bozek:2010vz,Schenke:2016ksl}.}
    \label{fig:decorrelations}
\end{figure}

A rapidity-dependent event plane twist decorrelation has been
predicted, as illustrated in Fig.~\ref{fig:decorrelations} (left). 
Based on a ``wounded" nucleon model~\cite{Bozek:2010vz}, particles in the forward rapidity regions are predominantly produced
from one of the projectile nuclei. As a result, the flow orientation angle (or event plane) 
at forward and backward rapidities can be slightly twisted event-by-event, 
creating a torqued QGP along the rapidity direction. Additionally, in the color glass condensate model~\cite{Schenke:2016ksl}, fluctuating granularity of the gluon field at different rapidities
can also lead to rapidity-correlated flow fluctuations.

Rapidity-dependent flow decorrelations have been observed by the CMS Collaboration using a novel observable based on two flow vectors, $\vec{V}_{n}(\eta^{a})=v_{n}(\eta^{a})\exp{(-in\Psi_{n}(\eta^{a}))}$ and $\vec{V}_{n}(\eta^{b})=v_{n}(\eta^{b})\exp{(-in\Psi_{n}(\eta^{b}))}$, measured in different rapidity regions, 
\begin{equation}
r_{n} \equiv \frac{\left<\vec{V}_{n}(-\eta^{a})\vec{V}^{*}_{n}(\eta^{b})\right>}{\left<\vec{V}_{n}(\eta^{a})\vec{V}^{*}_{n}(\eta^{b})\right>}. 
\end{equation}
This \rN ratio is similar to that used to measure the \pt dependent decorrelation, but now
designed to approximate the 
decorrelation between two event plane angles separated by a large gap of 
$2\eta_{a}$, $\left<\cos n\left[\Psi_{n}(\eta^{a})-\Psi_{n}(-\eta^{a})\right]\right>$,
as shown in Fig.~\ref{fig:decorrelations} (right) for elliptic flow
in 0--5\% central \PbPb collisions, while avoiding the contamination of short-range nonflow correlation.  
The data are compared to 
several initial-state models, including the torqued QGP model, the \textsc{ampt} initial state 
followed by a (3+1)D hydrodynamics, and the 3D CGC glasma model. All of the initial state models are able to qualitatively reproduce the data. It is worth noting that almost all
of the rapidity decorrelation effect is determined by the initial state. The
addition of (3+1)D hydrodynamic evolution is found to have little impact 
on the \rN ratio. This underlines the importance of incorporating
a rapidity-dependent modeling of initial-state fluctuations in hydrodynamic calculations.

\begin{figure}[ht]
    \centering
    \includegraphics[width=0.6\linewidth]{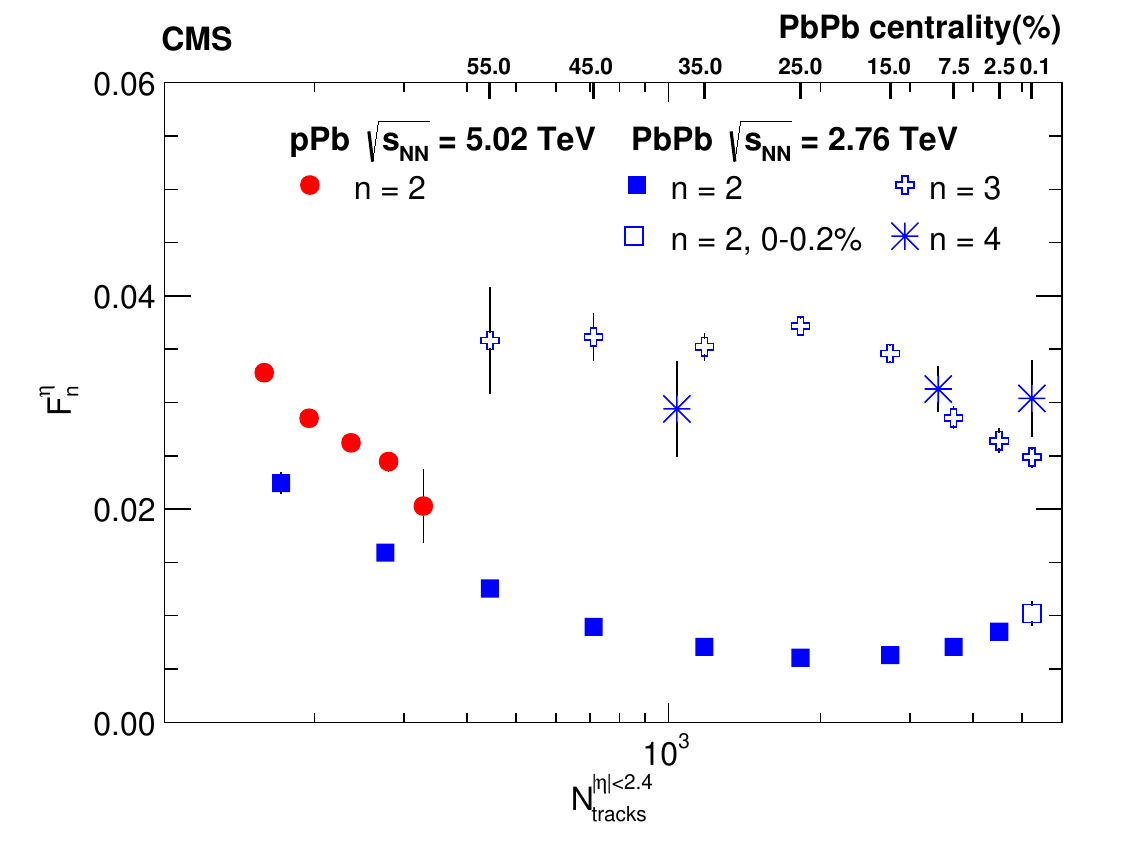}
\caption{The \FEtaN parameter as a function of event multiplicity in \PbPb collisions at \mbox{$\rootsNN=2.76\TeV$} for \mbox{$n=2$--$4$} and \pPb collisions at \mbox{$\rootsNN = 5.02\TeV$} for \mbox{$n=2$}. \FigureFrom{CMS:2015xmx}}
    \label{fig:decorrmult}
\end{figure}

The slope in the $\eta$ dependence of the 
\rN ratio, parameterized as \FEtaN, is plotted as a function of event multiplicity in Fig.~\ref{fig:decorrmult}
for \pPb collisions with $n=2$ and for \PbPb collisions with $n=2$--4. For \PbPb collisions, the \FEtaTwo
value reaches its minimum value near $\sim$20\% centrality (\ie, midcentral events), and increases significantly
for more peripheral or central \PbPb events and also for \pPb events, where flow fluctuations become more dominant~\cite{Chatrchyan:2013kba}. At a similar multiplicity,
the magnitude of the \FEtaTwo parameter is significantly larger for \pPb collisions than for \PbPb collisions.
In \PbPb collisions, a much stronger $\eta$-dependent factorization breakdown is
seen for higher-order harmonics than for the second-order harmonic, as shown by the
$F^{\eta}_{3}$ and $F^{\eta}_{4}$ parameters.
There is little centrality dependence for $n=3$, except for the most central 0--20\% \PbPb collisions.

\subsubsection{Nonlinear evolution and novel hydrodynamic observables}
\label{sec:nonlinearhydro}

Developing precise constraints for the transport properties of the QGP is one of the principal goals of the HI physics programs. While the 
\vTwo and \vThree flow coefficients reflect the transport properties, their values also depend on the initial-state geometry and its fluctuations. Additional observables are needed to disentangle the various contributions to these coefficients.

Higher-order flow coefficients \vN with $n\geq4$ can arise from
initial-state anisotropies in the same-order harmonic (linear response) or can be induced by
lower-order harmonics (nonlinear response)~\cite{Yan:2015jma, Qian:2016fpi, Teaney:2012ke}.
Based on the notation of Eq.~(\ref{eq:fourier}), complex anisotropic flow coefficients can be defined for different harmonics $n$, with $V_n = v_n \exp{(i n \Psi_n)}$. The $V_n$ coefficients should not be confused with the previously defined two-particle Fourier coefficients $V_{n\Delta}$.
Expressed in terms of their linear- and nonlinear-response components~\cite{Yan:2015jma, Qian:2016fpi},
\begin{equation}
\label{eq:QGP:decompositionVn}
\begin{split}
  V_4 & = V_{4 L} + \chi_{422} V_2^2,\\
  V_5 & = V_{5 L} + \chi_{523} V_2 V_3,\\
  V_6 & = V_{6 L} + \chi_{624} V_2 V_{4L} + \chi_{633} V_3^2 + \chi_{6222}  V_2^3,\\
  V_7 & = V_{7 L} + \chi_{725} V_2 V_{5L} + \chi_{734} V_3 V_{4L} + \chi_{7223}  V_2^2 V_3,
\end{split}
\end{equation}
where $V_{nL}$ denotes the part of $V_n$ that is not induced
by lower-order harmonics~\cite{Teaney:2012ke, 1703.04077, Giacalone:2018wpp}, and
the $\chi$\space are the nonlinear response coefficients.
Each nonlinear-response coefficient has its associated mixed harmonic,
which is $V_n$ measured
with respect to the lower-order symmetry plane angle.
The $V_{nL}$ component can be obtained by subtracting the nonlinear term from $V_n$.

As one can see from Eq.~(\ref{eq:QGP:decompositionVn}), the nonlinear-response coefficients are dimensionless values that are ratios of 
different flow coefficients. To illustrate, taking $V_5$ as an example, 
if we multiply both sides of the equation $V_5 = V_{5 L} + \chi_{523} V_2 V_3$ 
by the complex conjugate terms $V^{*}_2V^{*}_3$, and assume that the two terms on the right-hand side of this equation are uncorrelated~\cite{Yan:2015jma}, $\chi_{523}$ can be expressed as $V_5V^{*}_2V^{*}_3/(V^2_2V^2_3)$.
Therefore, the nonlinear-response coefficients are not strongly sensitive to the initial-state anisotropies~\cite{Yan:2015jma, Qian:2016fpi, Giacalone:2018wpp, Zhao:2017yhj}. As
a result, their experimental values can serve as
unique and robust probes of hydrodynamic behavior
of the QGP~\cite{Giacalone:2018wpp}.

\begin{figure}[ht]
    \centering
    \includegraphics[width=0.9\linewidth]{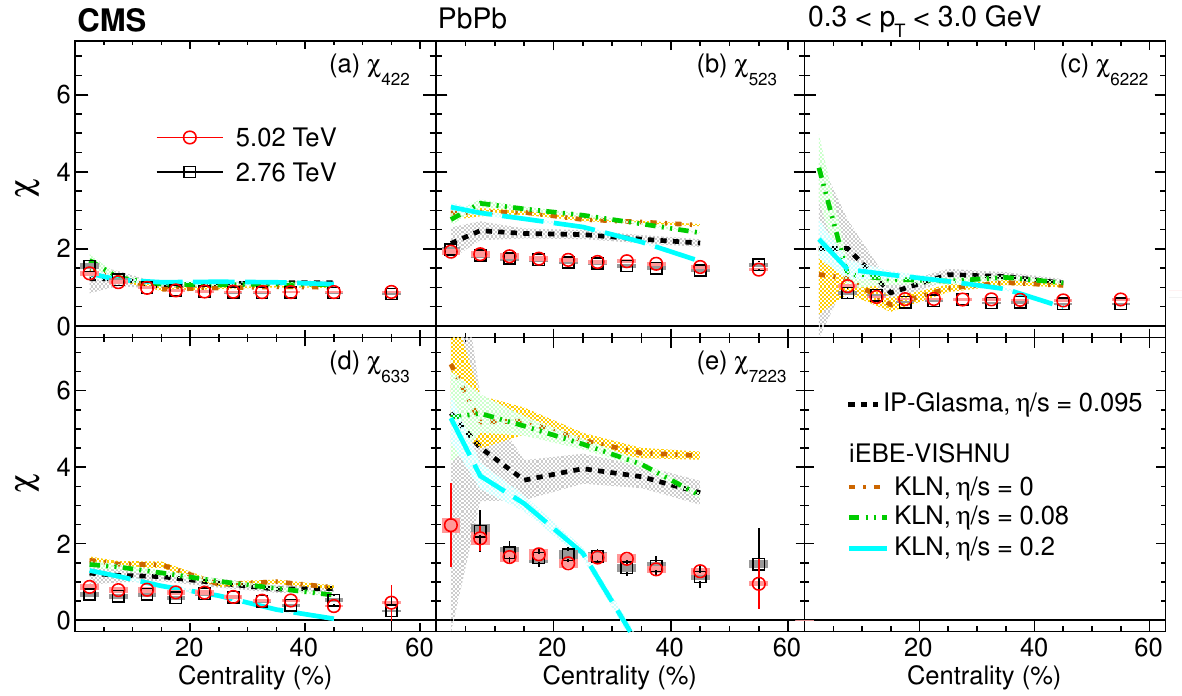}
    \caption{Nonlinear-response coefficients, \chiFour,
\chiFive, \chiSTTT, \chiSTT, and \chiSeven 
at 2.76 and 5.02\TeV, as functions of centrality.
The results are compared with predictions from a hydrodynamics $+$ hadronic cascade
hybrid approach with the IP-Glasma initial conditions
using \mbox{$\eta/s = 0.095$}~\cite{McDonald:2016vlt} at 5.02\TeV
and from iEBE-VISHNU hydrodynamics with the KLN
initial conditions using $\eta/s = 0$, 0.08, and 0.2~\cite{Qian:2016fpi} at 2.76\TeV.~\FigureFrom{CMS:2019nct}}
    \label{fig:QGP:CMS-HIN-17-005_Figure_006}
\end{figure}

The nonlinear-response coefficients, \chiFour,
\chiFive, \chiSTTT, \chiSTT, and \chiSeven at 2.76 and 5.02\TeV are presented as functions of centrality in Fig.~\ref{fig:QGP:CMS-HIN-17-005_Figure_006}. The results are also compared with the predictions
from a hybrid model of hydrodynamics and a hadronic cascade that uses IP-Glasma initial
conditions with $\eta/s = 0.095$~\cite{McDonald:2016vlt} at 5.02\TeV
and from iEBE-VISHNU hydrodynamics with the KLN initial conditions
using $\eta/s = 0$, 0.08, and 0.2~\cite{Qian:2016fpi} at 2.76\TeV.
All calculations describe the \chiFour centrality dependence well, but none of them
give a good description of the \chiFive and \chiSeven centrality dependences.
The model calculations of \chiSeven vary for the different initial conditions and
$\eta/s$ values, which suggests that the measurement of
\chiSeven could provide strong constraints on models.

\subsection{System space-time evolution via femtoscopy}
\label{sec:femtoscopt}

Femtoscopy is a powerful tool to infer the shape and size of the particle emitting region formed in high-energy collisions by measuring two-particle correlation functions in terms of the momentum difference of particle pairs~\cite{Lisa:2005dd}. The method reflects the quantum statistics governing identical particles, \ie, Bose--Einstein correlations~(BEC) for bosons (the situation for almost all cases discussed in this section), or Fermi--Dirac correlations for fermions. Nevertheless, it is also sensitive to final-state interactions, \eg, the Coulomb interaction for charged particles or the strong force between emitted hadrons. This technique, originally proposed for estimating stellar dimensions \cite{HanburyBrown:1954amm, Brown:1956zza, HanburyBrown:1956bqd}, was accidentally discovered in high-energy collisions in 1960~\cite{Goldhaber:1960sf} and has since been applied to a multitude of different high-energy analyses, both for small colliding systems, such as $\Pep\Pem$, \pp, and for \AonA collisions, with the measurements covering a wide energy spectrum. It was early thought that BEC data could provide a signature of QGP formation~\cite{Bjorken:1982qr} and this signature was searched for at the AGS, SPS, and RHIC~\cite{Lisa:2005dd}. In 2010, at the beginning of the LHC era, CMS made the first BEC correlation measurement for \pp collisions at $\sqrts =0.9$ and 2.36\TeV in terms of the 1D invariant relative momentum of particle pairs, $q_\text{inv}=\sqrt{-(p_1-p_2)^\mu(p_1-p_2)_\mu}$, with $p_{1,2}$ being the individual four-momenta of the particles in the pair, 
to establish the invariant radius \Rinvt~\cite{CMS:2010qkf}. 

In high-energy collisions, the femtoscopic correlation function can be defined by a single ratio~(SR) of signal over reference pair distributions, $C(q) = [\rd N_{\mathrm{sig}}(q)/\rd q]/[\rd N_{\mathrm{ref}}(q)/\rd q]=SR$, with $[\rd N_{\mathrm{sig}}(q)/\rd q]$ constructed by pairing same-sign~(SS) particles from the same event, and $[\rd N_{\mathrm{ref}}(q)/\rd q]$ built as a reference sample (ideally containing all pair correlations that are present in the signal sample, except for those arising from femtoscopic effects, such as quantum statistics and final-state interactions). The most common form of defining this reference distribution is by pairing particles from different events. In principle, the SR would yield a correlation function containing femtoscopic effects only. In case of correlations involving charged pairs, the Coulomb final-state interaction has to be taken into account. For \pp collisions, the approximation represented by the Gamow factor~\cite{CMS:2010qkf,CMS:2011nlc,CMS:2017mdg} can be employed to the final-state charged hadrons. In addition, other effects may still distort the signal, such as minijets or resonances, generically called background contributions~\cite{CMS:2017mdg}, requiring additional techniques for removing such spurious correlations. In CMS, several techniques have been adopted for this purpose, with details given in Refs.~\cite{CMS:2010qkf, CMS:2011nlc, CMS:2017mdg, CMS:2019fur, CMS:2023xyd}.

\begin{figure}[t]
\centering
  \includegraphics[width=0.45\textwidth]{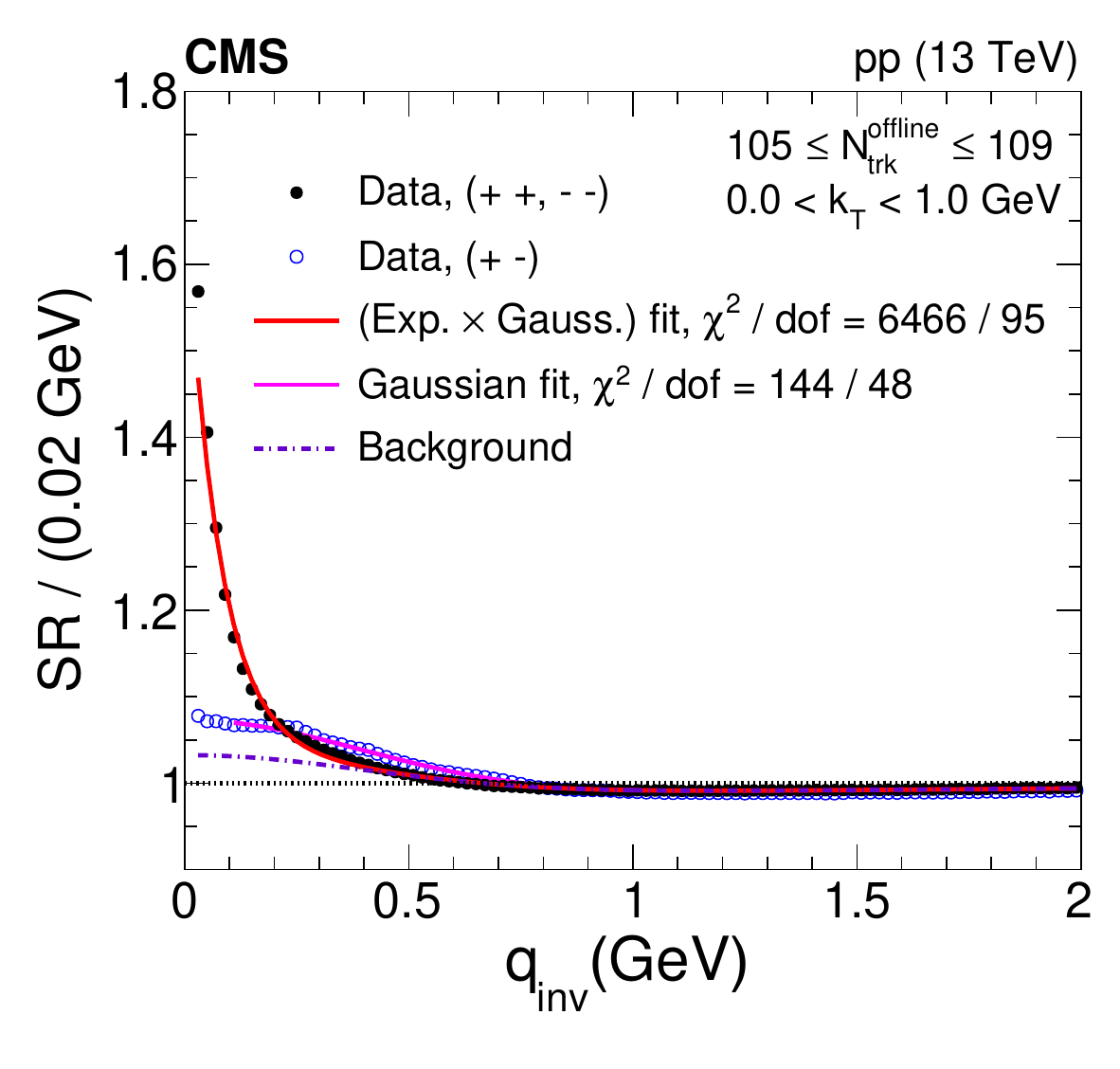} 
  \includegraphics[width=0.45\textwidth]{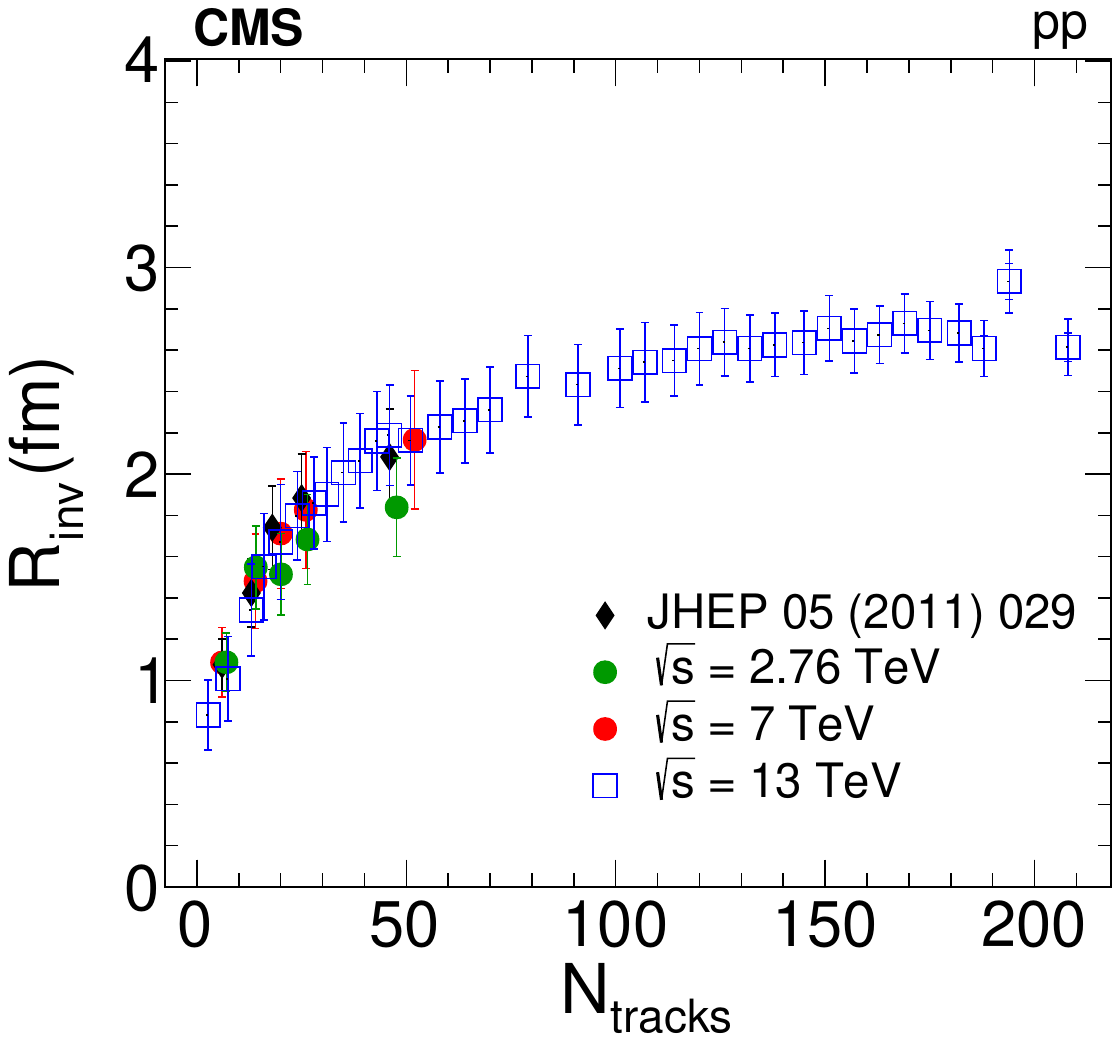} 
  \includegraphics[width=0.45\textwidth]{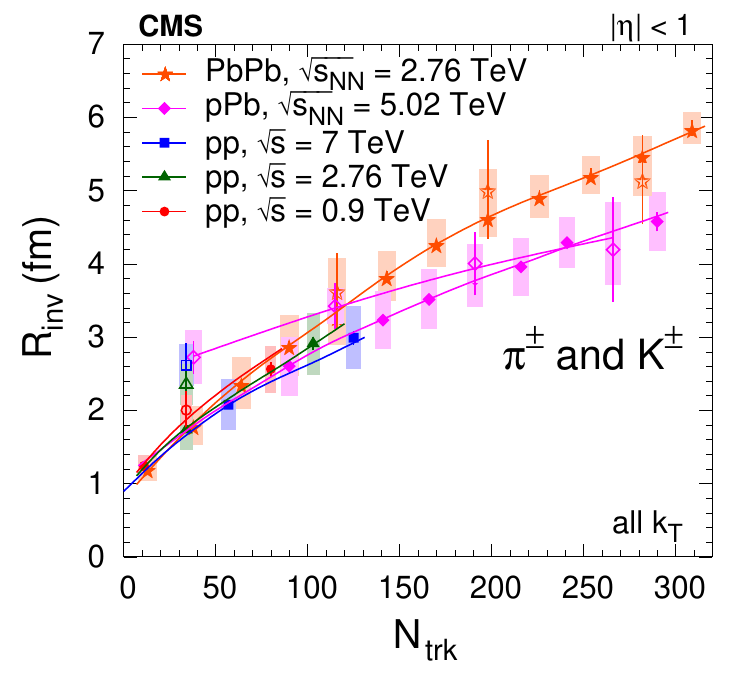} 
  \caption{Top left: Illustration of a typical BEC as functions of \qinv, for \pp collisions at 13\TeV, for opposite-sign pairs (no BEC), used to estimate the background contribution, and for same-sign pairs, together with the fits to both cases.
  Top right: Results for femtoscopic correlations of unidentified charged hadrons from \pp collisions at various LHC energies and in different multiplicity ranges. 
  Bottom: The plot shows results for identified pions (filled markers) and kaons (open markers) for different colliding systems and at several LHC energies. The error bars correspond to the statistical uncertainties, the colored boxes to the systematic uncertainties. The lines are cubic spline interpolations, added to guide the eye. \FigureCompiled{CMS:2011nlc,CMS:2017mdg,CMS:2019fur}}
 \label{fig:RinvVsNtrk}
\end{figure}

For extracting the information about the effective source sizes revealed by the femtoscopic technique, a function is fitted to the pair correlation function, which can be parametrized by a generic L\'evy stable distribution~\cite{Csorgo:2003uv}, as employed in Refs.~\cite{CMS:2010qkf, CMS:2011nlc, CMS:2017mdg, CMS:2019fur}, 
 \begin{equation}      
   C_{\text{BE}} (q_\text{inv}) = C [1 + \lambda \re^{ - (q_\text{inv} R_{\text{inv}})^\alpha } ]  \; ( 1 + \epsilon \;q_\text{inv}). 
    \label{eq:1d-levy}
    \end{equation}
In Eq.~(\ref{eq:1d-levy}), $C_{\text{BE}} (\qinv)$ refers to the two-particle Bose--Einstein correlation, $C$ is a constant; \Rinvt and $\lambda$ are the radius (also called the length of homogeneity) and intercept (correlation intensity) parameters, respectively. The exponent $\alpha$ is the L\'evy index of stability satisfying $0 < \alpha \le 2$. If treated as a free parameter in the fit, $\alpha$ usually returns a number between the value characterizing an exponential function ($\alpha=1$) and that for a Gaussian distribution ($\alpha = 2$). More details can be found in Ref.~\cite{Csorgo:2003uv}. The additional term, linear in \qinv and proportional to a fitting constant $\epsilon$, is introduced to account for possible long-range nonfemtoscopic correlations. An example of a typical correlation function versus \qinv is shown in Fig.~\ref{fig:RinvVsNtrk} (top left) for high-multiplicity \pp events at 13\TeV. This illustrates the Gaussian-type fit to the opposite-sign~(OS) (background-type contribution) correlation function and the exponential+background fit to the SS correlation function that contains the BECs. 

The results for the invariant source radius \Rinvt from a follow up measurement conducted in 2010 in \pp collisions 
at 7\TeV~\cite{CMS:2011nlc} are shown in Fig.~\ref{fig:RinvVsNtrk} (top right). The abscissa is the particle multiplicity of the events, \Ntracks, after correction for the detector acceptance and efficiency. 
Figure~\ref{fig:RinvVsNtrk} (top right) also shows the results from another analysis performed years later, employing the same analysis framework in terms of the 1D correlation function versus 
\qinv, 
 using data from \pp collisions at 2.76\TeV and a significantly larger sample of new data at 7\TeV~\cite{CMS:2017mdg}. The data show a steady rise as the number of produced tracks increases, with no clear dependence on the collision energy. 

Simultaneously, another analysis was conducted, in \pp collisions at different LHC energies, as well as in \pPb and peripheral \PbPb collisions at $\sqrtsNN = 5.02$ and 2.76\TeV, respectively, using a special tracker condition that allowed for identifying pions and kaons with high purity~\cite{CMS:2017mdg}. 
The resulting femtoscopic correlations of identified pions and kaons in different colliding systems and energies are shown in Fig.~\ref{fig:RinvVsNtrk} (bottom). A continuous rise with multiplicity can also be seen in this case, in a larger range of multiplicity in \pPb and peripheral \PbPb collisions. 

More recently, an additional 1D analysis was performed for \pp collisions at $\sqrts =13\TeV$ that covered a very wide range of particle multiplicities~\cite{CMS:2019fur}. Tracks with $\pt > 0.4\GeV$ were selected for events with multiplicities ranging from only a few tracks and up to $\langle \Ntracks \rangle \sim 250$ charged particles. This is a range of event activity similar to that for \pPb and peripheral \PbPb collisions. Recording such a large range in \Ntracks in \pp collisions was made possible with the help of very efficient high-multiplicity triggers available at CMS~\cite{CMS:2017kcs}. The main motivation for this study was to investigate if a continuous increase with \Ntracks would be observed for the femtoscopic radius \Rinvt, as expected from hydrodynamical models, or if the rise would saturate at some point, as suggested by the CGC theory~\cite{Bzdak:2013zma, McLerran:2013oju}. Three different techniques were employed to guarantee the independence of the results on the adopted analysis method, all three returning compatible values for \Rinvt~\cite{CMS:2019fur}. The results, illustrated using values from one of the methods, are shown in Fig.~\ref{fig:RinvVsNtrk} (top right) by the blue square markers. The values of \Rinvt increase with multiplicity and seem to saturate at higher values of \Ntracks, as suggested by the CGC model~\cite{Bzdak:2013zma, McLerran:2013oju}, although a continuous rise, as suggested by hydrodynamics, cannot be dismissed.

\begin{figure}[ht]
\centering
\includegraphics[width=0.42\textwidth]{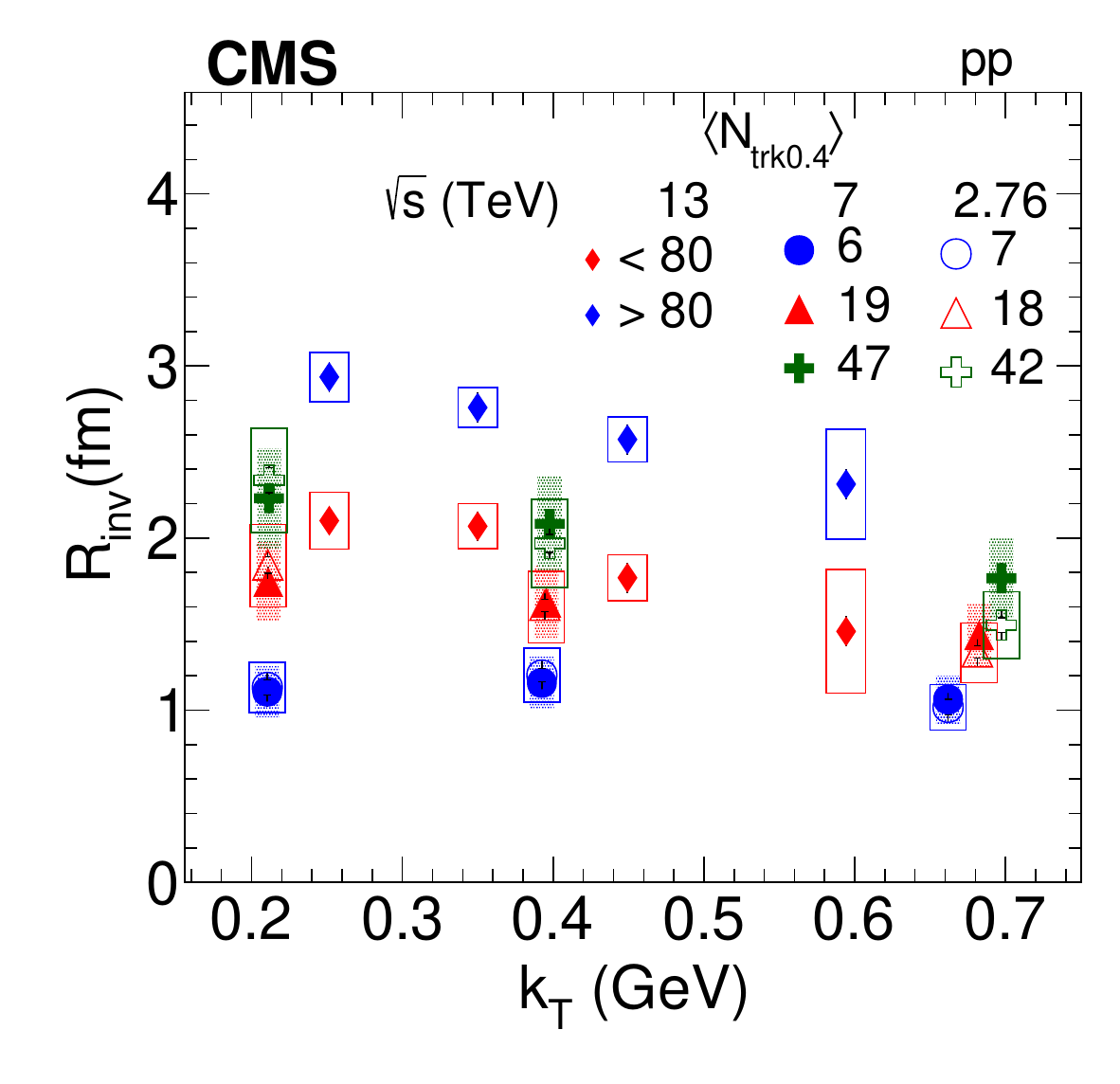}
\includegraphics[width=0.44\textwidth]{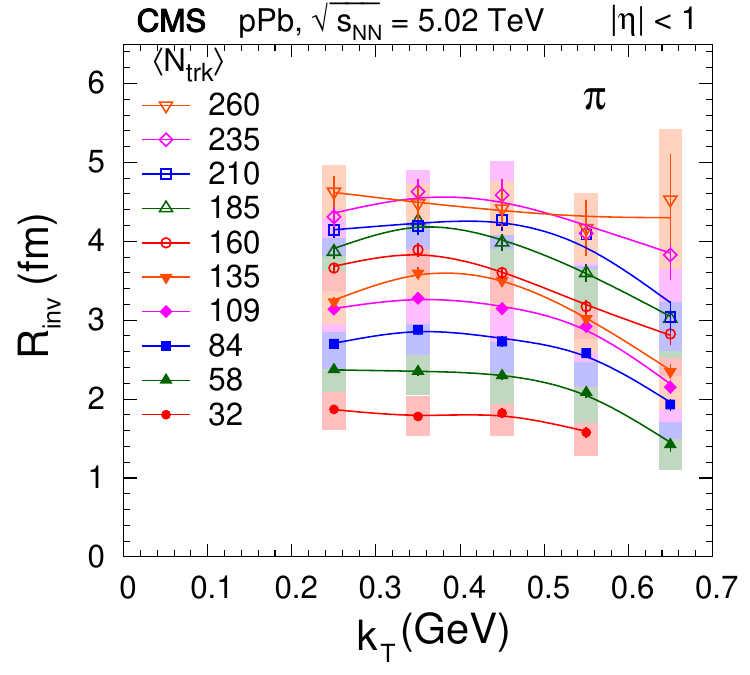} 
\caption{Left: Results for \Rinvt are shown as a function of \kt for \pp collisions at different energies and multiplicity ranges. \FigureCompiled{CMS:2017mdg,CMS:2019fur} Right: Similarly, \Rinvt values versus \kt are shown for \pPb collisions at 5.02\TeV. The error bars correspond to the statistical uncertainties, the colored boxes to the systematic uncertainties. The lines are cubic spline interpolations, added to guide the eye.
\FigureFrom{CMS:2017mdg}}
 \label{fig:RinvVskT}
\end{figure}

The 1D investigation on the behavior of the invariant radius parameter was also conducted for \Rinvt as a function of the average transverse momentum of the pair, $\vec{k}_\mathrm{T} = (\vec{p}_\mathrm{T,1} + \vec{p}_\mathrm{T,2})/2$. This study is important to explore the dynamics involved in the system evolution: a static system is not expected to show a $\kt (= \abs{\vec{k}_\mathrm{T}}$) dependence, whereas such a dependence would be expected for an expanding system subjected to
flow. The results are shown in Fig.~\ref{fig:RinvVskT} for some of the systems and energies mentioned above. 

Figure \ref{fig:RinvVskT} (left) shows results of \Rinvt as a function of \kt for \pp collisions at 2.76, 7, and 13\TeV. In this plot the data points are shown at the average values of \kt, taken in each bin of variable width. The latter is shown for two ranges of multiplicity: the lower values corresponding to events with multiplicity smaller than 80 tracks and the higher values to multiplicities greater than 80 tracks. The results for identified pions from \pPb collisions $\sqrtsNN = 5.02\TeV$ are shown in Fig.~\ref{fig:RinvVskT} (right). In this plot the data points are shown at the bin center. In all cases we see that the values of \Rinvt decrease with \kt, a behavior normally seen in data and more clearly illustrated by the results from \pp at 13\TeV. This behavior indicates that the system expands before decoupling. 

\begin{figure}[t]
\centering
 \includegraphics[width=0.32\textwidth]{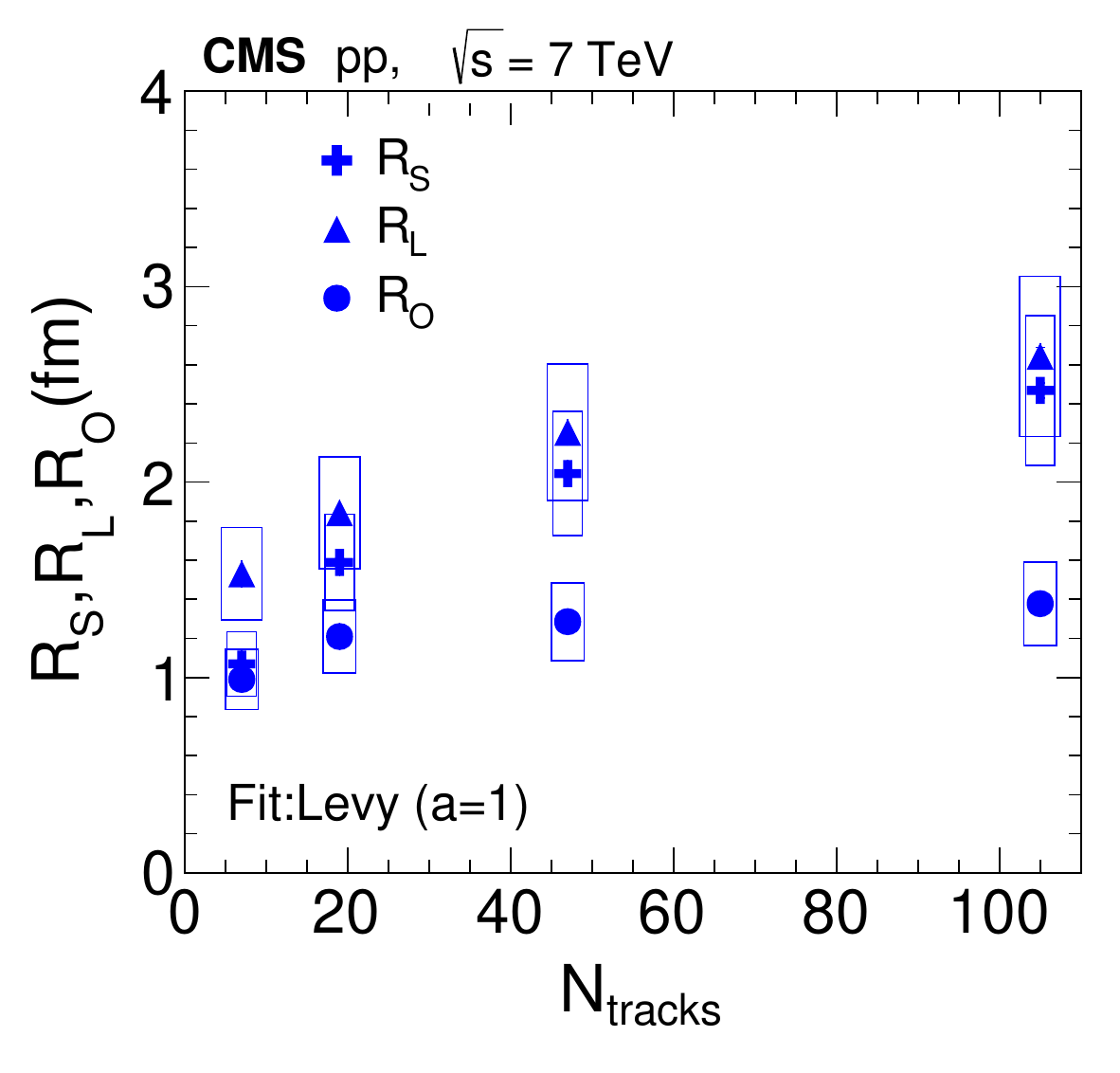}
 \includegraphics[width=0.33\textwidth]{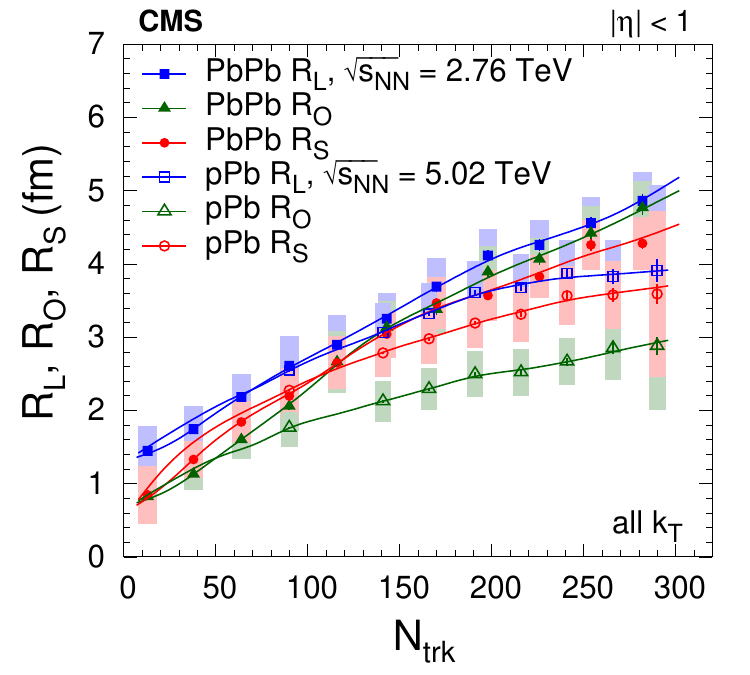}
 \includegraphics[width=0.32\textwidth]{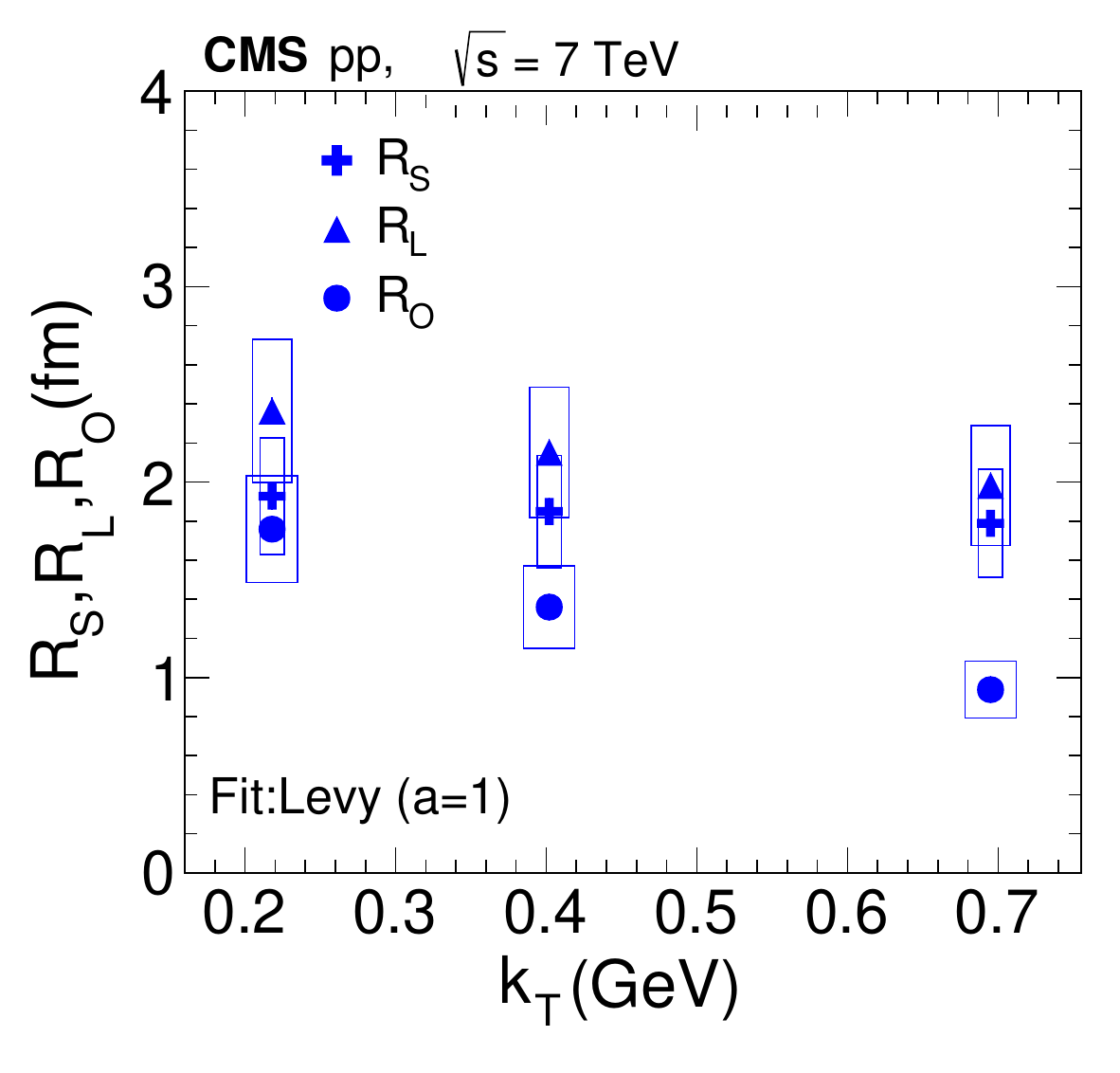} 
 \caption{Left: The femtoscopic Bertsch--Pratt radius components in different directions (\RS, \RL, \RO) are shown as functions of multiplicity for charged hadrons from \pp collisions at 7\TeV. 
Middle: The three variables are shown for pions from the \pPb and \PbPb systems at 2.76 and 5.02\TeV, respectively. The lines are cubic spline interpolations, added to guide the eye.
A similar tendency of increasing radius parameters with multiplicity is seen in each of the three directions, for all cases. Right: The variation of these components with \kt is shown for charged hadrons from \pp collisions at 7\TeV. 
(Figures adapted from Ref.~\cite{CMS:2017mdg}).}
 \label{fig:RsRlRo}
\end{figure}

In addition, an extension of the previous analyses was developed for investigating the source in three different directions (3D case), in terms of the Bertsch--Pratt variables (\RS, \RL, \RO)~\cite{Lisa:2005dd}, where \RL is the component along the beam (longitudinal) direction, \RO is transverse to \RL and parallel to the direction of \kt and reflects different emission times, and \RS is transverse to the directions of both \RL and \RO. The results for these variables as functions of multiplicity are shown in Fig.~\ref{fig:RsRlRo} for both unidentified charged hadrons from \pp collisions (left) and for charged pions from \pPb collisions (middle). These radial components, in general, show similar behavior as functions of \Ntracks as in the 1D case. The right panel of Fig.~\ref{fig:RsRlRo} also illustrates the behavior of (\RS, \RL, \RO) versus \kt (the data points are shown at the average values of \kt, taken in each bin of variable width). All three components tend to decrease with \kt, as expected for expanding sources. 

Furthermore, the results for (\RS, \RL, \RO) were found to be similar for \pp and \pPb collisions, with $\RL > \RS > \RO$ in both cases~\cite{CMS:2017mdg}. However, in \PbPb collisions, a different relation is observed, showing similar values for the three variables, \ie, $\RL \approx \RS \approx \RO$~\cite{CMS:2017mdg}. In other words, the shape of the system formed in \pp and \pPb collisions is elongated in the longitudinal direction, whereas the system formed in \PbPb collisions is more spherical in shape. 

The findings from femtoscopic correlation measurements performed in \pp collisions at $\sqrts=13\TeV$ demonstrate the complex and complementary behavior of the systems formed in such collisions: under certain conditions they behave similarly to systems formed in high-energy $\Pep\Pem$ collisions. This is reflected in an anticorrelation (values of $C_\mathrm{BE}$ below unity) seen in the 1D double ratios away from the BEC peak, which is more pronounced in the lower multiplicity ranges~\cite{CMS:2011nlc, CMS:2017mdg, CMS:2019fur}. This anticorrelation is usually quantified in terms of a model ($\tau$ model~\cite{Csorgo:1990up,L3:2011kzb}), in which particle production has a broad distribution in proper time and the phase space distribution of the emitted particles is dominated by strong correlations of the space-time coordinate and momentum components. The depth of the anticorrelation has been quantified~\cite{CMS:2011nlc, CMS:2017mdg, CMS:2019fur} and shown to decrease with increasing $\langle \Ntracks \rangle$ and \kt (except for large values of multiplicities, in the latter case). 
Although this observation in minimum bias \pp collisions suggests that such a structure could be associated with small systems, 
another investigation~\cite{PhysRevC.71.044902} reveals a linear relation between the fitted $\Rinvt^{-2}$ values 
and the transverse mass of the pair, $\mT = \sqrt{m^2 + \kt^2}$, 
with a slope proportional to the square of a Hubble constant divided by the system freeze-out temperature 
(assumed to be $T_\text{fo}\sim 500$\MeV)~\cite{CMS:2019fur}. 
This result matches expectations from hydrodynamics: 
the formed system undergoes a strong expansion characterized by a Hubble-type flow, as found in high-energy nuclear collisions.
The analogy with cosmology is made because hydrodynamical calculations show a behavior compatible with the Hubble law, 
$v = H \, r$, where $v$ is the fluid velocity at the fluid position $r$ and $H=\dot{R}/R$ is the Hubble constant, $R$ being a scale parameter. 
Towards the end of the fluid expansion, when the acceleration coming from pressure gradients is negligible, 
$\dot{R}$ tends to a constant value and $R \sim \dot{R} \, t$, so that $H \sim 1/t$, with $t$ representing time~\cite{PhysRevC.71.044902}. 
From the slope of $\Rinvt^{-2}$ versus $\mT$, the Hubble constant of the collision has been determined in two multiplicity ranges~\cite{PhysRevC.71.044902,CMS:2019fur},
$H_\text{MB} = 0.298 \pm 0.004\stat\unit{fm}^{-1}$ for minimum bias events ($\Ntracks \lesssim 80$) and 
$H_\text{HM} = 0.17 \pm 0.04\stat\unit{fm}^{-1}$ for high-multiplicity events ($\Ntracks \gtrsim 80$).
These values are compatible with those obtained for peripheral and central AuAu collisions at 
RHIC~\cite{Adler:2004rq,Adams:2004yc,Abelev:2009tp,Adare:2015bcj}, respectively, 
implying that the expansion is faster in peripheral collisions than in more central ones. 
These values correspond to a directionally-averaged Hubble constant. 
A detailed discussion can be found in Ref.~\cite{CMS:2019fur}. 

More recently, two-particle BEC functions have been measured in \PbPb collisions at $\sqrtsNN = 5.02\TeV$ in different centrality and transverse momentum classes~\cite{CMS:2023xyd}, and compared to theoretical models based on parametric L\'evy sources, incorporating the Coulomb effect~\cite{Csanad:2019lkp}. The value of the $\alpha$ parameter, describing the source shape, was found to be approximately 50\% larger than that found in 200\GeV AuAu collisions at RHIC~\cite{PHENIX:2017ino}. This difference in the $\alpha$ values found at RHIC and LHC may stem from a larger mean free path at lower collision energies, reflecting a larger deviation from normal diffusion (whose processes lead to a Gaussian distribution) in systems formed at lower energies, resulting in a heavy tail; the further the deviation, the heavier the tail, and the smaller is the $\alpha$ value. This is consistent with the observed centrality dependence of $\alpha$: it is closer to 2 in case of the most central collisions, and decreases to values close to 1.6 for peripheral collisions. Moreover, it was found that the $R$ parameter (describing the homogeneity length, similarly to \Rinvt, and representing the final state) scales as the cube root of the average number of participating nucleons in the collision, a proxy for the initial size. Furthermore, a linear dependence of $1/R^2$ on the pair transverse mass \mT was observed. This is consistent with a hydrodynamic scaling, predicted for Gaussian sources. From this linear dependence, the Hubble constant of the collisions was determined to increase from 0.11 to $0.18\unit{fm}^{-1}$ from central to peripheral collisions, comparable to those values found at RHIC~\cite{PHENIX:2017ino} or in high-multiplicity \pp collisions at $\sqrts=13\TeV$~\cite{CMS:2019fur}, as mentioned above. Taken together, these results can be interpreted as the hadron emitting source having a shape consistent with a L\'evy distribution in 5.02\TeV \PbPb collisions. 

Besides revealing valuable insights on the space-time dimensions of particle-emitting sources created in high-energy collisions, as discussed above, femtoscopy has also been used in high-energy experiments to extract parameters related to Coulomb and strong force final-state interactions~\cite{Lisa:2005dd}. In particular, the strong force interaction scattering parameters, such as the scattering length and effective range, can be extracted with this technique. For instance, femtoscopy of strange \PGL baryons can add significant information about baryon-baryon interactions and, depending on the values found for the scattering length and effective range, could indicate the potential formation of exotic bound states, such as the H-dibaryon~\cite{PhysRevLett.38.195}. In addition, studies of $\PKzS\PKzS$ and $\PGL\PGL$ correlations offer information about the interactions of strange hadrons, thus providing valuable guidance to model the composition of neutron stars~\cite{Kaplan:1987sc, Schaffner-Bielich:2000nft, Morita:2014kza}. Different fitting functions and analysis procedures are adopted in this case (as discussed in Ref.~\cite{CMS:2023jjt}). The CMS Collaboration conducted $\PKzS\PKzS$, $\PGL\PKzS$, and $\PGL\PGL$ femtoscopy studies using \PbPb collision data at $\sqrtsNN = 5.02\TeV$~\cite{Morita:2014kza}. The $\PKzS\PKzS$ correlation is measured in six centrality bins, covering the 0--60\% range~\cite{CMS:2023jjt}. The source size extracted from the $\PKzS\PKzS$ correlation shows the expected decreasing trend from central to peripheral collisions. The negative scattering length extracted from $\PGL\PKzS$ correlations indicates that the strong interaction between \PGL and \PKzS is repulsive. On the other hand, the positive scattering length extracted from $\PGL\PGL$ correlations indicates that the interaction between \PGL particles is attractive and disfavors the existence of a possible bound H-dibaryon state.

\subsection{Searches for chiral magnetic effects and early-stage short-lived electromagnetic fields}
\label{sec:QGP_ChiralityEMFields}

An object is \textit{chiral}, either left- or right-handed, if it is not invariant under the parity ($P$) transformation.
In a chiral system, the imbalance of right- and left-handed objects can
be characterized by a chiral chemical potential (\muFive). It has been predicted that in a system of charged chiral fermions with a finite \muFive value, an electric current density ($\overrightarrow{J_\mathrm{e}}$) can be induced when an external magnetic field ($\overrightarrow{B}$) is applied, with the current density along the direction of the magnetic field,
\begin{equation}
\overrightarrow{J_e} \propto \mu_5\overrightarrow{B}.
\label{eq:CME}
\end{equation}
This phenomenon is known as the chiral magnetic effect~(CME)~\cite{Kharzeev:2004ey,Kharzeev:2007jp}. 

In relativistic nuclear collisions, the chiral symmetry is expected to be restored in a QGP, rendering nearly massless or chiral quarks.
If the topological solutions of the SU(3) gauge group of QCD are chiral, they can transfer chirality to quarks via a chiral anomaly,
forming local chiral domains with finite \muFive values in the initial stage (more details in Ref.~\cite{Li:2020dwr} and references therein). Within each domain, there is an imbalance of right- and left-handed chiral quarks. Meanwhile, extremely strong magnetic fields ($B \sim 10^{15} \unit{T}$) can be formed in noncentral HI events, mostly by energetic spectator protons. The presence of the parity-even $\overrightarrow{B}$ 
field and parity-odd \muFive is 
predicted to lead to an electric current along the direction of $\overrightarrow{B}$, namely the CME.
Observing a CME signal in nuclear collisions 
would have profound impacts on many aspects of 
fundamental physics,
including the topological phases of QCD, chiral symmetry
restoration, and QGP evolution with strong electromagnetic fields.
The CME and related phenomena, such as the chiral magnetic wave, emerge when applying a fluid dynamics description to a combined QED+QCD system influenced by the chiral anomaly. While the theoretical basis for these effects is well established, the potential magnitude of a CME signal is highly model-dependent, as it is significantly affected by the initial conditions, which are not well known.

In this section, we review the progress in
searching for the CME in high-energy nuclear collisions
made by the CMS Collaboration.

\subsubsection{Searches for chiral magnetic effects}
\label{sec:CME}

The charge separation induced by the CME can be manifested as the first parity-odd ($P$-odd) sine term ($a_{1}$)
in a Fourier decomposition of the charged-particle azimuthal distribution~\cite{Voloshin:2004vk},
\begin{equation}
\label{azimuthal}
\frac{\rd N}{\rd\phi} \propto 1 + 2\sum_{n} \bigl\{v_{n}\cos[n(\phi-\Psi_\mathrm{RP})] + a_{n}\sin[n(\phi-\Psi_\mathrm{RP})]\bigr\},
\end{equation}
where $\phi - \PsiRP$ represents the particle azimuthal angle
with respect to the reaction plane angle \PsiRP (determined by the impact parameter and beam axis), and \vN and $a_{n}$ denote the $P$-even and $P$-odd Fourier coefficients, respectively. Experimentally, the \PsiRP is approximated by the second-order event
plane, $\Psi_{2}$, of the elliptic flow. As any $P$-odd term will vanish after averaging over events, the most commonly investigated observable is an azimuthal three-particle correlator, $\gamma_{112}$, which measures $\left\langle a_1^{2} \right\rangle$~\cite{Voloshin:2004vk},
\begin{equation}\label{2pcorrelatorEP}
\gamma_{112} \equiv \left\langle \cos(\phi_{\alpha} + \phi_{\beta} - 2\Psi_{2}) \right\rangle = \left\langle \cos(\phi_{\alpha}-\Psi_{2})\cos(\phi_{\beta}-\Psi_{2}) \right\rangle
- \left\langle \sin(\phi_{\alpha}-\Psi_{2})\sin(\phi_{\beta}-\Psi_{2}) \right\rangle.
\end{equation}
Here, $\alpha$ and $\beta$ denote particles with the same or opposite electric
charge sign and the angle brackets reflect an averaging over particles and events.
The first term on the right-hand side of Eq.~(\ref{2pcorrelatorEP})
becomes $\left\langle v_{1,\alpha}v_{1,\beta} \right\rangle$, which is generally small
and independent of the charge~\cite{STAR:2009tro}, while the second term is sensitive to the
charge separation and can be expressed as $\left\langle a_{1,\alpha}a_{1,\beta} \right\rangle$. By taking a difference between OS (where $\alpha$ and $\beta$ have OS electric charge) and SS (where $\alpha$ and $\beta$ have SS electric charge) $\gamma$ correlators,
\begin{equation}
\Delta \gamma \equiv \gamma^{\mathrm{OS}} - \gamma^{\mathrm{SS}}, 
\end{equation}
all charge-independent effects are canceled.

Despite having a relatively simple observable, the existence of the CME in nuclear collisions remained inconclusive after more than a decade of experimental searches. While observations were consistent with the existence of a CME, these could also be interpreted as resulting from background contributions, such as local charge conservation from resonance decays embedded in an elliptic-flow background.
Because of the nonperturbative nature of background processes, theory is not able to provide a quantitative estimate of their importance. By applying an approach based on control samples in data to control the signal strength, while keeping the backgrounds constant,
CMS has made two key contributions that have convincingly demonstrated that the CME signal at LHC energies, even if it exists, is too small to observe. 

\begin{figure}[ht]
    \centering
    \includegraphics[width=\linewidth]{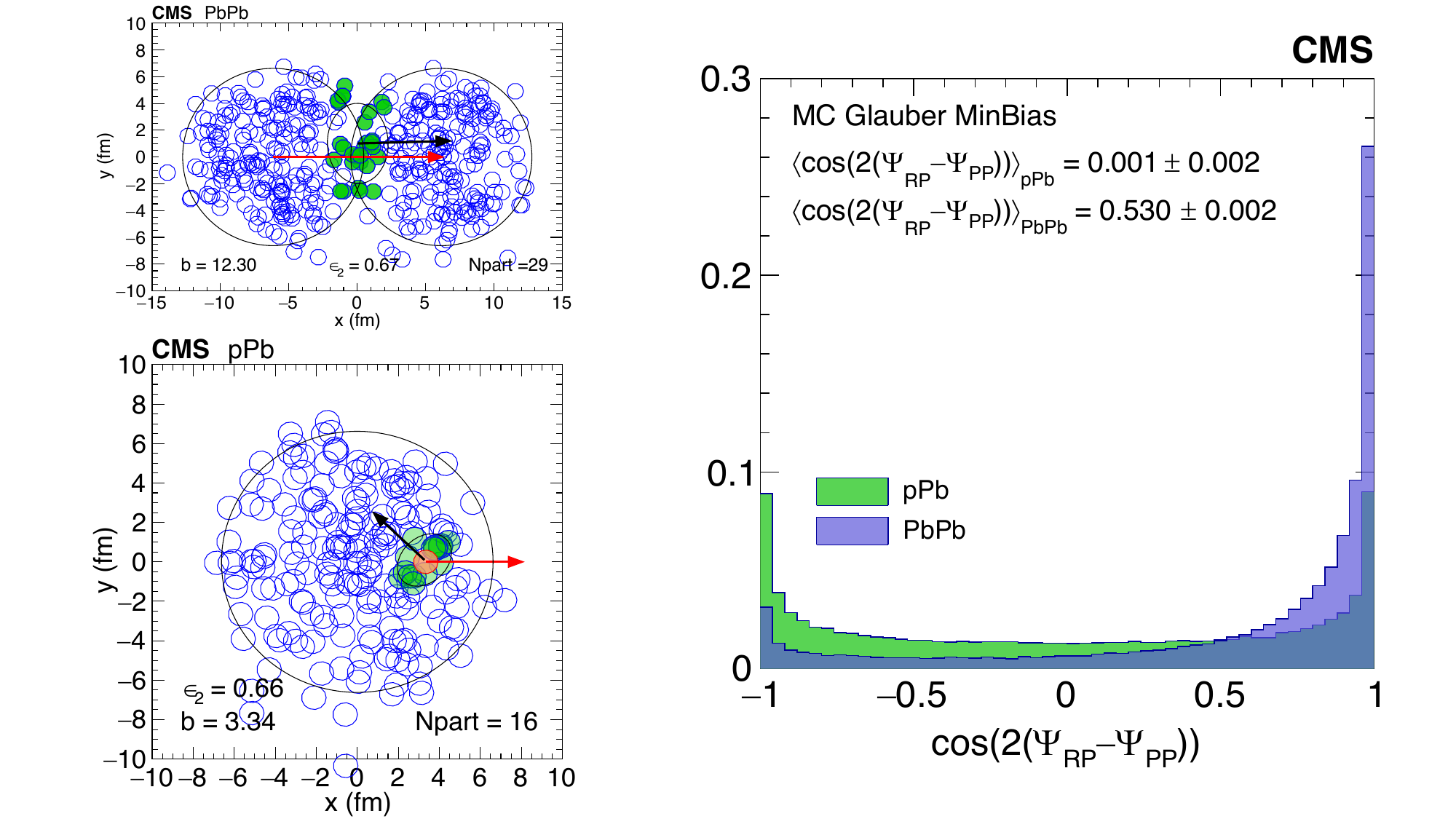}
    \caption{Left: Event geometry of one peripheral \PbPb and one central \pPb event using MC Glauber simulation at at $\rootsNN\ = 5.02\TeV$. The red and black arrows point in the direction of the reaction and participant plane angle, respectively. Right: The cosine of the relative angle between the reaction plane and the participant plane.~\FiguresFrom{CMS:2016wfo}}
    \label{fig:CMECartoon}
\end{figure}

High-multiplicity \pp and \pPb collisions have been shown to generate large final-state azimuthal anisotropies, comparable to those in \AonA collisions~\cite{CMS:2010ifv,CMS:2015fgy,CMS:2016fnw,CMS:2012qk,Khachatryan:2014jra,Chatrchyan:2013nka,CMS:2015yux,Dusling:2015gta}.
However, the CME contribution to any charge-dependent signal is expected to be negligible
in a high-multi\-pli\-city \pPb collision. 
As illustrated in Fig.~\ref{fig:CMECartoon} (left) based on MC Glauber calculations~\cite{Alver:2008aq},
while the angle between the magnetic field direction, which is given approximately by the direction of the reaction plane (red arrow in the figure), and the event plane of elliptic anisotropy (black arrow in figure) is strongly correlated in \PbPb collisions, it is expected to be mostly random in \pPb collisions. Figure~\ref{fig:CMECartoon} (right) shows the correlation between the reaction plane angle~(\PsiRP) and participant plane angle~(\PsiPP, approximating the event plane) in terms of the distribution of $\cos(2(\PsiRP-\PsiPP))$
for \pPb and \PbPb collisions. The event-averaged value of 
$\cos(2(\PsiRP-\PsiPP))$ is consistent with zero for \pPb collisions, while a significant correlation is observed for \PbPb collisions.
With a random field orientation, the CME contribution to any
charge-dependent signal is expected to be small in \pPb collisions.

\begin{figure}[ht]
    \centering
    \includegraphics[width=0.56\linewidth]{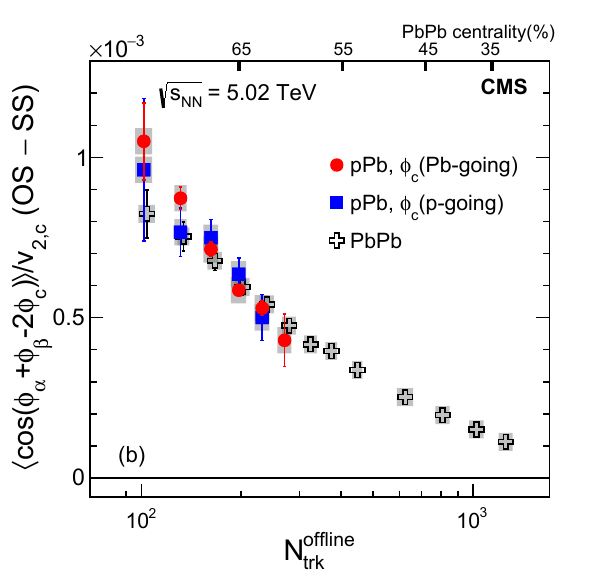}
    \includegraphics[width=0.42\linewidth]{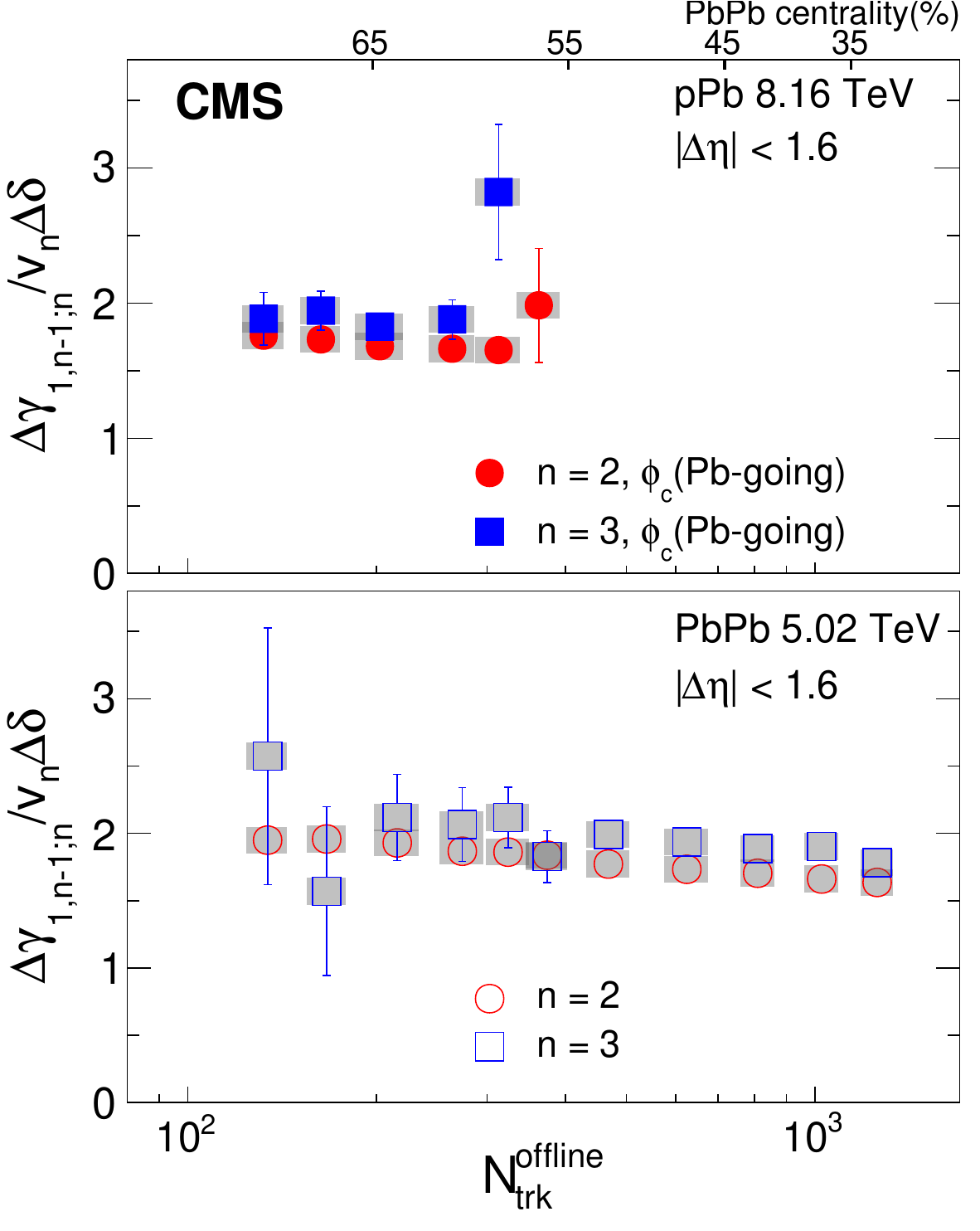}    
    \caption{Left: the difference of the opposite sign~(OS) and same sign~(SS) three-particle correlators as a function of \noff in \pPb and \PbPb collisions at $\rootsNN = 5.02\TeV$. \FigureFrom{CMS:2016wfo} Right: ratio of $\Delta\gamma_{112}$ and $\Delta\gamma_{123}$ to the product of \vN and $\delta$ in \pPb collisions for the Pb-going direction at $\rootsNN = 8.16\TeV$ and \PbPb collisions at 5.02\TeV. \FigureFrom{Sirunyan:2017quh}}
    \label{fig:CME_a}
\end{figure}

The high-multiplicity \pPb data sample collected by CMS gives access to multiplicities comparable 
to those in peripheral
\PbPb\ collisions, allowing for a direct comparison of the two systems with very different CME contributions.
Figure~\ref{fig:CME_a} (left) shows the difference of the charge-dependent three-particle correlator for the OS and SS cases, as a function of multiplicity for \pPb and \PbPb collisions at $\rootsNN = 5.02\TeV$. Within uncertainties, the \pPb and \PbPb data show nearly identical values.
The striking similarity in the observed charge-dependent
azimuthal correlations strongly suggests a common physical origin. In \PbPb
collisions, it was suggested that the charge dependence of the three-particle
correlator is an indication of the charge separation
effect due to the CME signal~\cite{STAR:2009tro, ALICE:2012nhw}.
However, as argued earlier, a strong charge separation signal from the CME
is not expected in a very high-multiplicity \pPb collision.
Therefore, the similarity seen between high-multiplicity \pPb and peripheral \PbPb collisions
presents a significant challenge to the attribution of the observed charge-dependent correlations to the CME.

Furthermore, the charge separation effect from the CME is only expected along the direction
of the induced magnetic field normal to the reaction plane, approximated by the second-order event plane, \PsiTwo.
As the symmetry plane of the third-order Fourier term (``triangular flow"~\cite{Alver:2010gr}), \PsiThree, is expected to have a weak
correlation with \PsiTwo~\cite{ATLAS:2014ndd}, the charge separation effect with respect to
\PsiThree is also expected to be negligible. By constructing a charge-dependent
correlator with respect to the third-order event plane,
\begin{equation}
\label{eq:gamma123}
\gamma_{123} \equiv \left\langle \cos(\phi_{\alpha}+2\phi_{\beta}-3\Psi_{3}) \right\rangle,
\end{equation}
charge-dependent background effects unrelated to the CME can be explored.
In particular, in the context of the local charge conservation mechanism, the
$\gamma_{123}$ correlator is also expected to have a background contribution,
similar to that for the $\gamma_{112}$ correlator, but proportional to \vThree, instead of \vTwo.
After scaled by \vTwo and \vThree, respectively, the 
$\gamma_{112}$ and $\gamma_{123}$ correlators 
are expected to be similar, largely independent of harmonic event plane orders, as shown in Fig.~\ref{fig:CME_a} (right). 
This similarity, seen in high-multiplicity \pPb and peripheral \PbPb collisions for both
$\Delta\gamma_{112}$ and $\Delta\gamma_{123}$, again challenges the attribution of the
observed charge-dependent correlations to the CME.

To set a quantitative limit on the existence of the CME signal, CMS has applied the event shape engineering~(ESE) technique~\cite{Schukraft:2012ah}.
This technique involves establishing a direct link between the $\gamma$ correlators and \vN coefficients. By applying ESE in a specific range of centrality or multiplicity, where the magnetic field remains relatively constant, events are further categorized based on the magnitude of the \vN coefficient measured in the forward rapidity region. In each event category, measurements of $\gamma$ correlators and \vN values are compared to assess the linear relationship, and observed dependence is extrapolated to the $\vN=0$ region. A non-zero value of the $\gamma$ correlators at that point would reflect the strength of the CME. 

\begin{figure}[ht]
    \centering
    \includegraphics[width=0.7\linewidth]{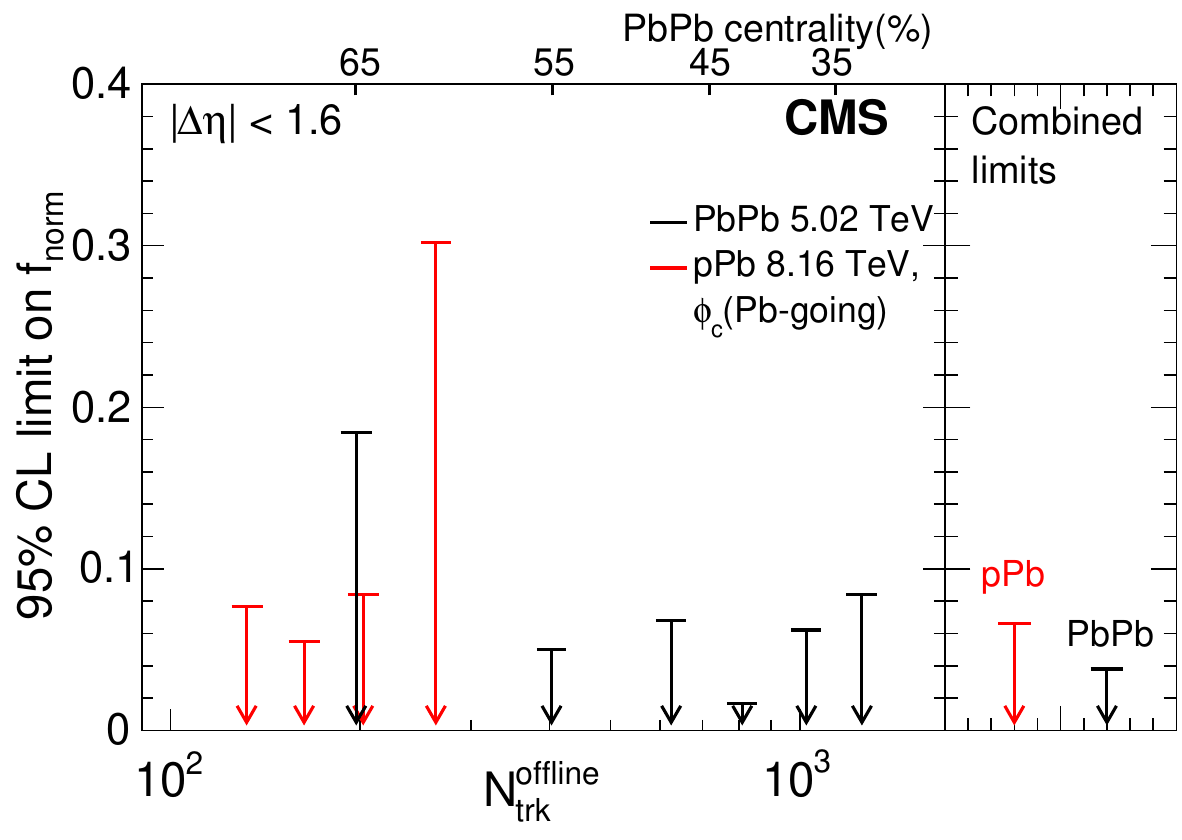}    
    \caption{Upper limits of the fraction of \vTwo-independent $\gamma_{112}$
  correlator component
  as a function of \noff in \pPb collisions at \mbox{$\rootsNN = 8.16\TeV$} and
  \PbPb collisions at 5.02\TeV. \FigureFrom{Sirunyan:2017quh}}
    \label{fig:CME}
\end{figure}

Based on the assumption of a nonnegative CME signal, the upper limit of the \vTwo-independent fraction $f_{\text{norm}}$ in the $\Delta\gamma_{112}$ correlator is obtained with the measured statistical and systematic uncertainties. In Fig.~\ref{fig:CME}, the upper limit of $f_{\text{norm}}$ is presented at 95\% confidence level (\CL)
as a function of event multiplicity. The combined limits from all presented multiplicities and centralities are also shown in \pPb and \PbPb collisions. An upper limit on the \vTwo-independent fraction of the three-particle correlator, or possibly the CME signal contribution, is
estimated to be 13\% in \pPb and 7\% in \PbPb collisions, at 95\% \CL.

The data presented here provide new stringent constraints on
the nature of the background contribution to the charge-dependent azimuthal
correlations, and establish a new baseline for the search for the CME in HI collisions.

\subsubsection{Searches for chiral magnetic waves}
\label{sec:CMW}

The chiral magnetic wave~(CMW) is a phenomenon similar to the CME.
The chiral separation effect
(CSE) is a process where the separation of the chiral charges
along the magnetic field will be induced by a finite density of the initial net electric charges~\cite{Burnier:2011bf}.
The coupling of electric and chiral charge densities and currents leads to a long-wavelength collective excitation, known as the chiral magnetic wave~\cite{Kharzeev:2010gd}. It is worth noting that a lack of experimental evidence for the CME~\cite{CMS:2016wfo,Sirunyan:2017quh}
does not necessarily imply the absence of the CMW effect, as the CME requires an initial chirality
imbalance from topological QCD charges,
whereas a CMW only requires an initial net electric charge density~\cite{Kharzeev:2010gd,Burnier:2011bf}.
Therefore, the CME and CMW deserve independent experimental investigations.

The propagation of the CMW leads to an electric quadrupole moment,
where additional positive (negative) charges are accumulated away
from (close to) the reaction plane~\cite{Burnier:2011bf}.
Following a hydrodynamic evolution of the medium formed in \AonA collisions,
this electric quadrupole moment is expected to result in a charge-dependent variation
of the second-order anisotropy coefficient~(\vTwo) in the Fourier expansion of the final-state particle azimuthal distribution. More specifically,
the \vTwo coefficient will exhibit a linear dependence on the observed event
charge asymmetry~\cite{Burnier:2011bf}, $A_\text{ch}\equiv (N_{+}-N_{-})/(N_{+}+N_{-})$,
where $N_{+}$ and $N_{-}$ denote the number of positively and negatively charged
hadrons in each event,
\begin{equation}
\label{f.ellipticflow}
v_{2,\pm} = v_{2,\pm}^\text{base} \mp rA_\text{ch}.
\end{equation}
\noindent Here $v_{2,\pm}^\text{base}$ represents the value in the absence of a charge
quadrupole moment from the CMW for positively ($+$)
and negatively ($-$) charged particles,
and $r$ denotes the slope parameter. In the presence of a CMW, the difference of \vTwo values between positively and negatively charged particles would be proportional to $A_\text{ch}$.
Similar charge-dependent effects from the CMW are not expected for
the third-order anisotropy coefficient~(\vThree)~\cite{Kharzeev:2015znc}.

\begin{figure}[ht]
    \centering
    \includegraphics[width=0.49\linewidth]{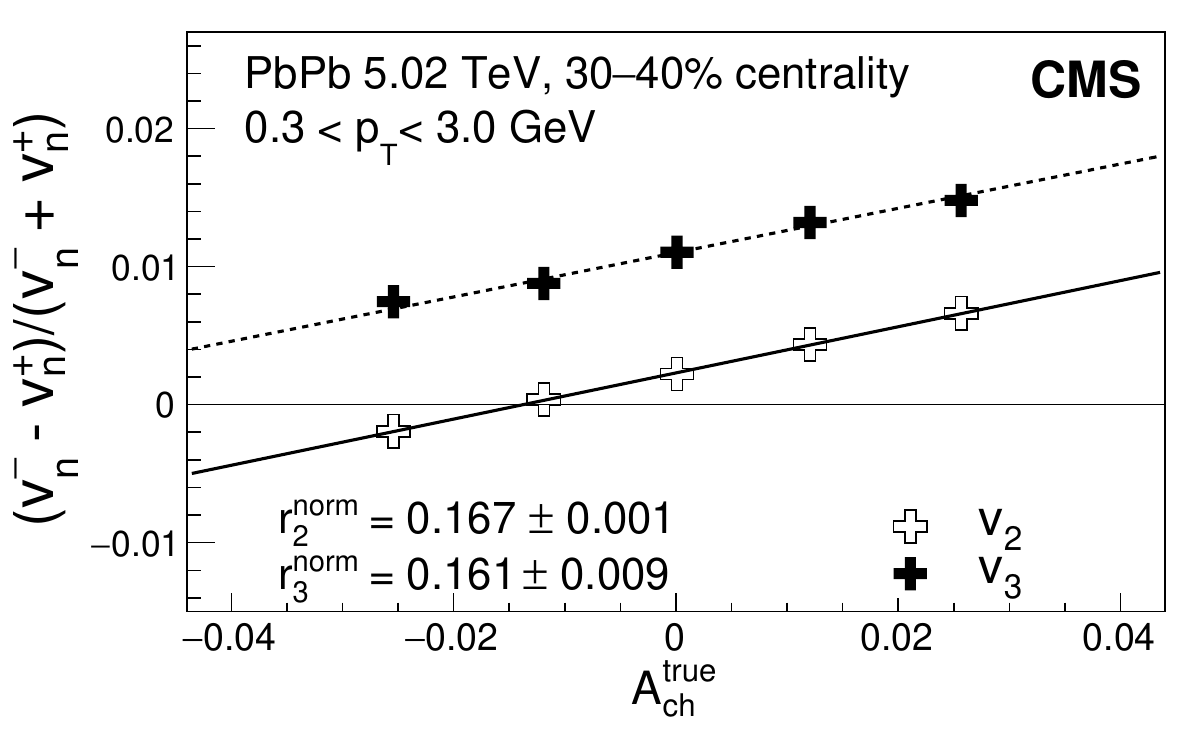}
    \includegraphics[width=0.49\linewidth]{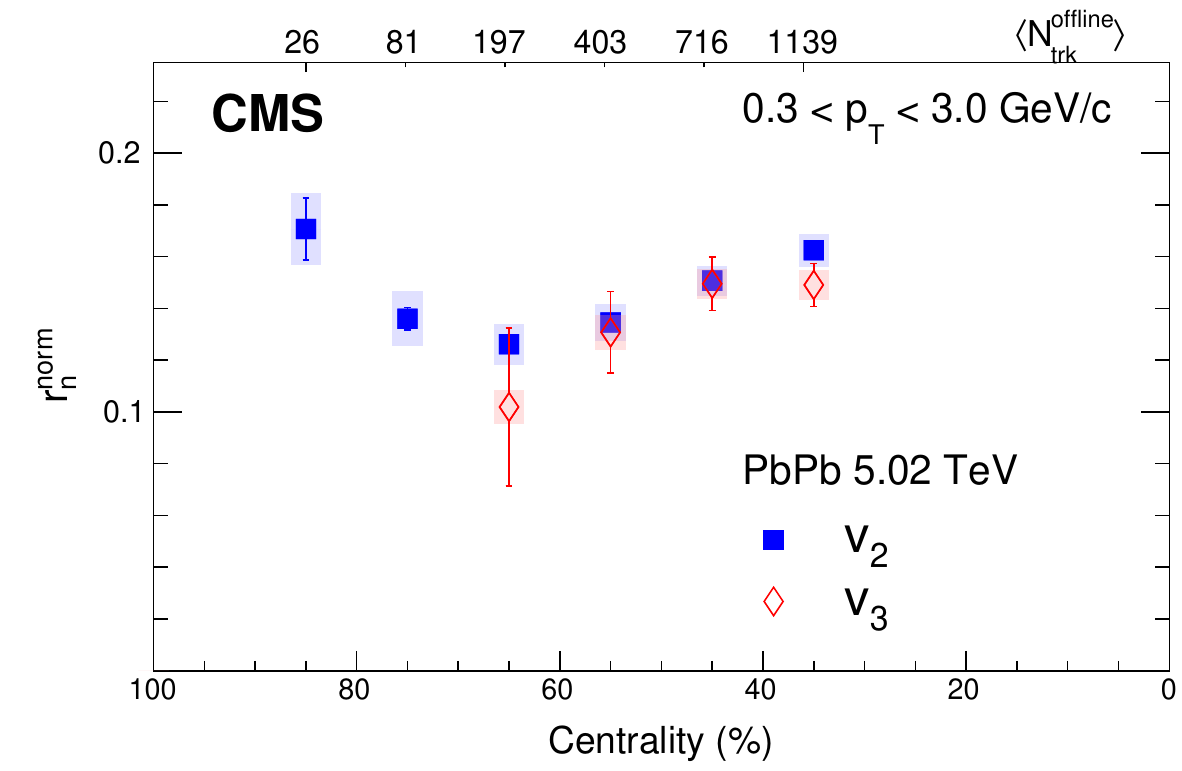}    
    \caption{Left: The normalized difference in
\vN, $(v^{-}_{n} - v^{+}_{n})/(v^{-}_{n} + v^{+}_{n})$,
for $n=2$ and 3, as a function of true event charge asymmetry
for the 30--40\% centrality class in \PbPb collisions at $\sqrtsNN = 5.02\TeV$. Right: The linear slope parameters, \rTwoNorm and \rThrNorm,
as functions of the centrality class in \PbPb collisions.~\FiguresFrom{CMS:2017pah}}
   \label{fig:CMW}
\end{figure}

The normalized \vThree difference,
$(v^{-}_3-v^{+}_3)/(v^{-}_3+v^{+}_3)$, is derived as
a function of true event charge asymmetry~(\ATrueCh), obtained by correcting the observed value for the detector acceptance and tracking efficiency, in \PbPb collisions and compared with
that for \vTwo in Fig.~\ref{fig:CMW} (left). The normalized slope
parameter of \vThree, \rThrNorm, agrees well with \rTwoNorm within statistical uncertainties.
Once normalized, no difference is observed for the \ATrueCh dependence between the charge-dependent \vTwo and \vThree values. 

The \rTwoNorm and \rThrNorm values of \PbPb collisions at $\sqrtsNN = 5.02\TeV$ are shown in Fig.~\ref{fig:CMW} (right), as functions of centrality in the range 30--90\%. As found for \rTwoNorm, a moderate centrality dependence of \rThrNorm
is observed. Over the centrality range studied in this analysis, the \rTwoNorm and
\rThrNorm slope parameters are consistent with each other within uncertainties.
The CMW effect is expected with respect to the reaction plane, which is approximated by the second-order event plane in \AonA collisions, but highly suppressed with respect to the third-order event plane~\cite{Kharzeev:2015znc}.
The observation of the harmonic-order independence, reflected in the similar \rTwoNorm
and \rThrNorm values, indicates an underlying physics mechanism unrelated to the CMW effect and, instead,
can be qualitatively explained by late local-charge conservation~\cite{Bzdak:2013yla}.

\subsubsection{Searches for the electromagnetic conductivity in QGP}
\label{sec:EMFields}

Very strong and short-lived EM fields might be created in the early stages of relativistic HI collisions. The configuration of these electromagnetic fields is not trivial to predict because they receive contributions from several sources that involve the spectators and participants in the collision. In some theoretical predictions, the net magnetic or Coulomb fields are expected to generate significant rapidity-odd (rapidity-even) contributions to \vN coefficients, with $n$ odd (even) \cite{Gursoy:2018yai}. The lifetime of the EM fields is expected to depend on the electric conductivity of the medium~\cite{Dubla:2020bdz}. Therefore, measuring such effects in the \vN coefficients as functions of rapidity would not only point to the existence of strong EM fields created in the collisions, but also constrain the properties of the QGP, such as its electric conductivity.

Heavy-flavor quarks are expected to be produced primarily in the initial stages of a collision (order of ${\sim}0.1\unit{fm}$) and to pass through the medium with a lower probability of annihilation as compared to light-flavor quarks \cite{Braun-Munzinger:2007fth, Liu:2012ax}. The EM fields are, at least in some theoretical approaches, expected to have a maximum magnitude on a time scale below 0.2\unit{fm}. As a consequence, the impact of EM fields on \vN values as a function of rapidity is predicted to be much stronger for \PDz mesons (containing charm quarks) than for the abundantly produced charged hadrons~\cite{Das:2016cwd}. 

\begin{figure}[ht]
    \centering
    \includegraphics[width=0.6\linewidth]{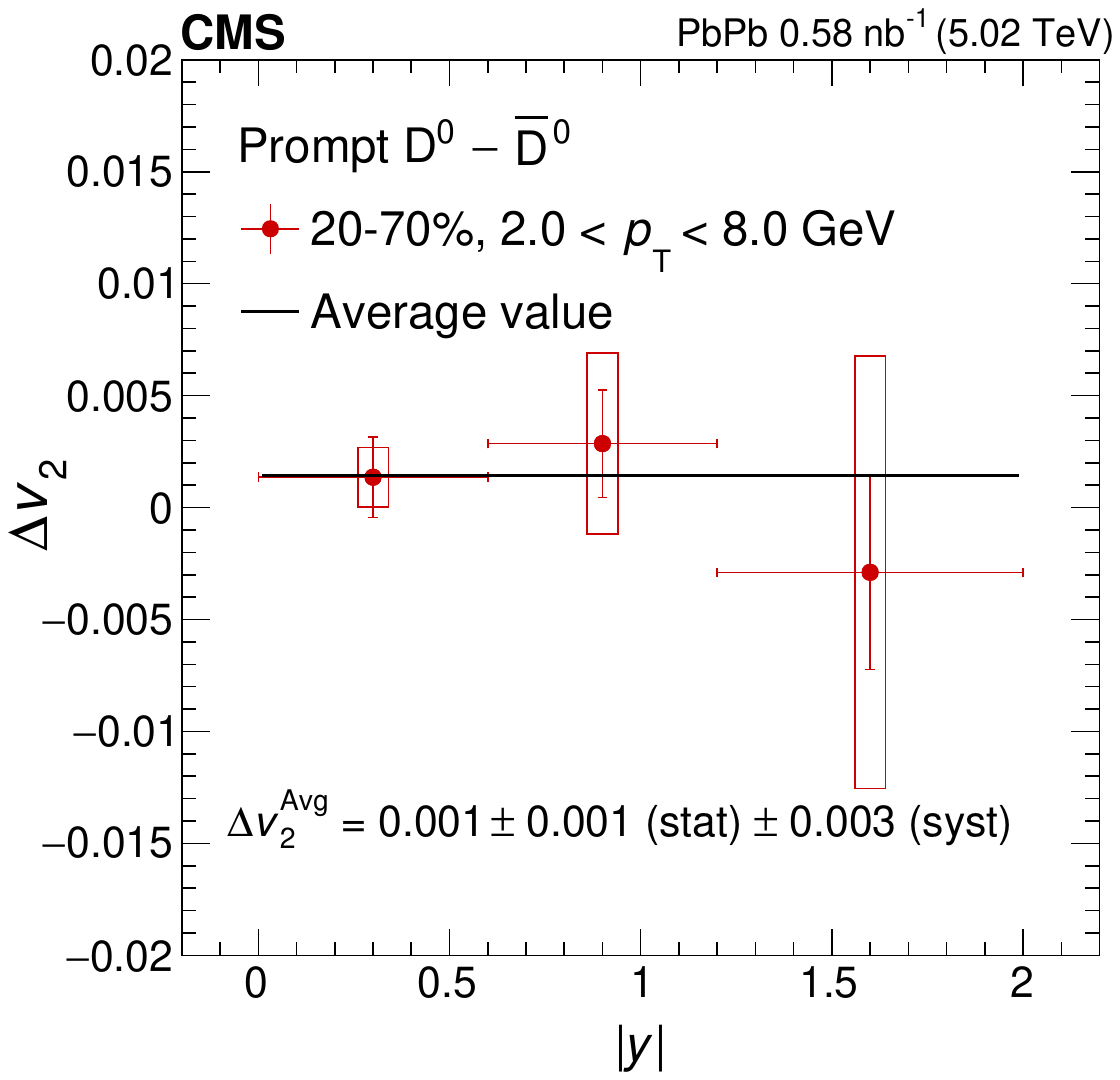}
    \caption{Difference of \vTwo between \PDz and \PADz mesons as a function of rapidity. The average value ($\Delta \vTwo^{\text{Avg}}$) is extracted by fitting the data considering the statistical uncertainties only. The systematic uncertainty of the $\Delta \vTwo^{\text{Avg}}$ is estimated by shifting the each point up and down by its systematic uncertainty.~\FigureFrom{CMS:2020bnz}}
    \label{fig:QGP_EMFields}
\end{figure}

In light of these predictions, the CMS Collaboration measured the \vTwo difference ($\Delta \vTwo$) between \PDz and \PADz mesons as a function of rapidity to search for the effect of a possible strong Coulomb field created by the collision participants~\cite{CMS:2020bnz}. The results, with an average value of $\Delta \vTwo^{\text{avg}} = 0.001\pm0.001\stat\pm0.003\syst$, are shown in Fig.~\ref{fig:QGP_EMFields}. The expected magnitude for charged hadrons is $\Delta \vTwo \sim - 0.001$~\cite{Gursoy:2018yai}, \ie, with the same magnitude, but with a different sign. Given the present uncertainties, the measurement sensitivity is not sufficient to clarify if charm hadron collective flow is affected or not by the strong Coulomb field created in ultrarelativistic heavy ion collisions. Significant improvements in both statistical and systematic uncertainties for the $\Delta \vTwo$ measurement are expected with future large data samples made possible with the upgraded CMS detector.

\subsection{Summary of results for bulk properties and novel phenomena}

The CMS Collaboration has leveraged the extensive phase space coverage of its detector to explore QGP properties and to probe fundamental aspects of the strong force. The full coverage in $\phi$ and the large $\eta$ range of the CMS detector have enabled precise measurements of particle densities and correlations, offering deep insights into the behavior of the QGP.

In the most central \PbPb collisions at the LHC, the charged-particle density and average transverse energy per particle are significantly higher than those observed at RHIC. This suggests that the QGP formed at the LHC is denser, hotter, and longer-lived, while reaffirming the notion of a ``nearly perfect liquid’’ with minimal viscosity. Detailed studies of azimuthal correlations, particularly the elliptic and higher-order Fourier coefficients, have constrained the shear viscosity to entropy density ratio ($\eta/s$) to within 0.08--0.2, close to the theoretical lower bound.

Data from CMS have also been used to challenge the assumption that two-particle correlations can be factorized with respect to a common event plane. The observed \pt- and $\eta$-dependent factorization breaking provides new insights into initial-state fluctuations and the longitudinal dynamics of the QGP, enabling a three-dimensional view of the medium evolution. Nonlinear response coefficients, derived from high-order Fourier analyses, offer robust probes of the QGP's hydrodynamic behavior, independent of initial-state anisotropies.

Femtoscopy techniques have revealed that the size and shape of particle-emitting sources vary with collision system and energy. In \PbPb collisions, the sources are more spherical, while in \pp and \pPb collisions, they are elongated along the longitudinal direction. The Levy-type shape observed in \PbPb collisions at 5.02\TeV suggests a shorter mean free path and a closer approach to normal diffusion at LHC energies, compared to the more Gaussian-like distributions at RHIC energies. These findings are supported by the similarity in the measured Hubble constant between low-multiplicity \pp events and peripheral AuAu collisions at RHIC. Investigations using femtoscopy to study the interactions between strange hadrons have provided valuable data on these  interactions, further enriching our understanding of QGP dynamics.

In the search for CME and CMW signals, CMS has pioneered the development of a series of new observables and their application to small systems to set unique constraints on background contributions to the CME and CMW measurements. The results show unambiguously that the CME and CMW signals in relativistic nuclear collisions are too small to be observed at LHC energies with the current data set. The most stringent upper limit to date has been set on these signals.

\clearpage

\section{Hard probes in heavy ion collisions and sensitivity to quark-gluon plasma}
\label{sec:hardQGP}

In ultrarelativistic HI collisions, rare hard scatterings of the parton constituents of the nucleons can produce a suite of energetic final states, known collectively as ``hard probes''. Heavy-flavor quarks, jets, photons, weak bosons, and even top quarks are all hard probes measured by CMS during Runs 1 and 2 in \PbPb and \pPb collisions, as well as in \pp collision data at the same energy as the other two systems to be used as a reference. Produced predominantly during the initial collision prior to the formation of the QGP, the production mechanisms and vacuum propagation of these particles are strongly constrained by studies in the experimentally cleaner \pp collision environment, with theoretical control via pQCD calculations. As a result, hard probes can be used to tag the initial momentum scale of a hard scattering (in the case of photons and similarly colorless probes), to determine the strength and nature of the medium interactions (in the case of QCD color-charge carriers such as quarks and gluons), and thoroughly map the QCD medium interactions across a suite of topologies and kinematic extremes (via jet substructure, the more-massive top quarks, and the highest-\pt jets). The following sections will provide detailed information about these and other phenomena, as studied with CMS data.

\subsection{Observations of parton quenching}
\label{sec:InclusiveJetObservables}

The current section details the first observations of partonic energy loss, manifesting experimentally as ``quenching'', using inclusive jet production dominated by hard-scattered light quarks and gluons. Observations of enhanced dijet asymmetry, transverse momentum imbalance, and the suppression of both jet and high-\pt hadron spectra with respect to \pp data are discussed.

\subsubsection{Dijet asymmetry and relative energy loss}
\label{ssec:InclusiveJetObservables_Dijets}

The suppression of high-\pt hadrons, indicating modifications to hard-scattered partons induced by the QGP, was initially observed at RHIC by both the PHENIX and STAR experiments~\cite{PHENIX:2001hpc,STAR:2002ggv}. With the start of beams at the LHC and data taking by general-purpose experiments with nearly $4\pi$ calorimetric coverage, the study of partonic energy loss with fully reconstructed jets became possible. The first observations by ATLAS~\cite{ATLAS:2010isq} and CMS~\cite{CMS:2011iwn} were of a substantial, centrality-dependent enhancement of dijet asymmetry~(\AJ), defined as
\begin{linenomath}
\begin{equation}
\label{eqn:dijetAJ}
\AJ = \frac{\ptOne - \ptTwo}{\ptOne + \ptTwo},
\end{equation}
\end{linenomath}
where \ptOne corresponds to the highest \pt jet (``leading jet'') in the event and \ptTwo corresponds to the second-highest \pt jet (``subleading jet'') in the event. 

\begin{figure}[ht]
    \centering
    \includegraphics[width=0.95\linewidth]{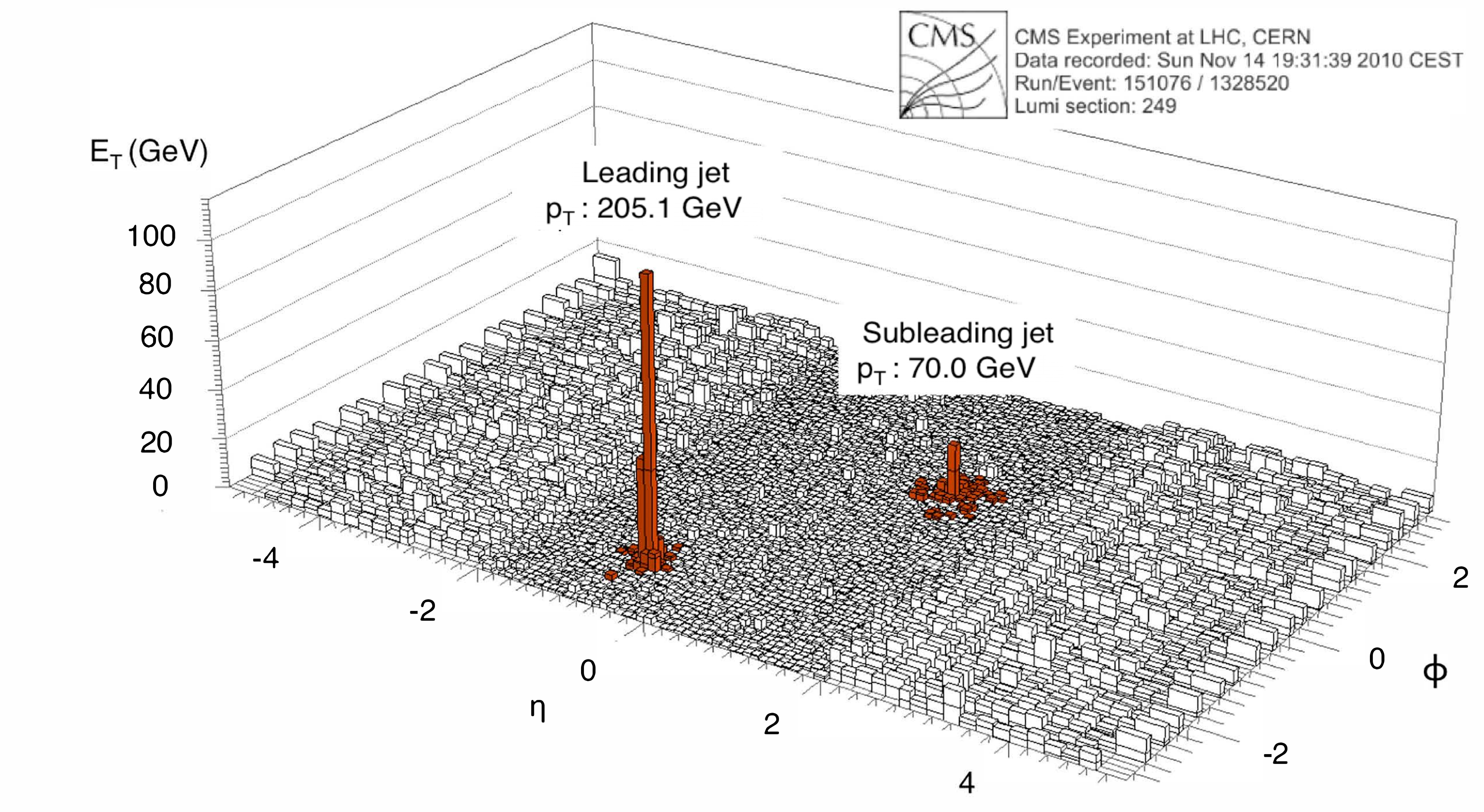}
    \caption{An ``unrolled'' calorimeter display of energy deposition in an event containing an unbalanced dijet pair in a $\rootsNN=2.76\TeV$ \PbPb collision, as recorded by the CMS detector in 2010. The tower-by-tower transverse energy sum combining the measurement in electromagnetic and hadronic calorimeters is plotted as a function of $\eta$ and $\phi$. The fully corrected transverse momenta of the unbalanced dijet pair are labeled and their position in $\eta$-$\phi$ indicated with the red-highlighted constituent towers. \FigureAdaptedFrom{CMS:2011iwn}}
    \label{fig:dijetEvtDisplay}
\end{figure}

In \pp collisions, \AJ is typically used for jet energy calibration and observed dijet pairs with significant transverse momentum asymmetry, after accounting for effects such as finite jet energy resolution and initial-state/final-state radiation, typically indicate the presence of a third jet to conserve the transverse momentum of the system~\cite{cmsPPJetCalibration}. However, as shown in calorimeter event displays, such as the example in Fig.~\ref{fig:dijetEvtDisplay}, frequently there is no such compensating third jet present in \PbPb collisions. Instead, an energetic leading jet is observed back-to-back in azimuthal angle with a substantially less energetic subleading jet, and no third jet is visible in the event display by inspection.

\begin{figure}[ht]
    \centering
    \includegraphics[width=0.95\linewidth]{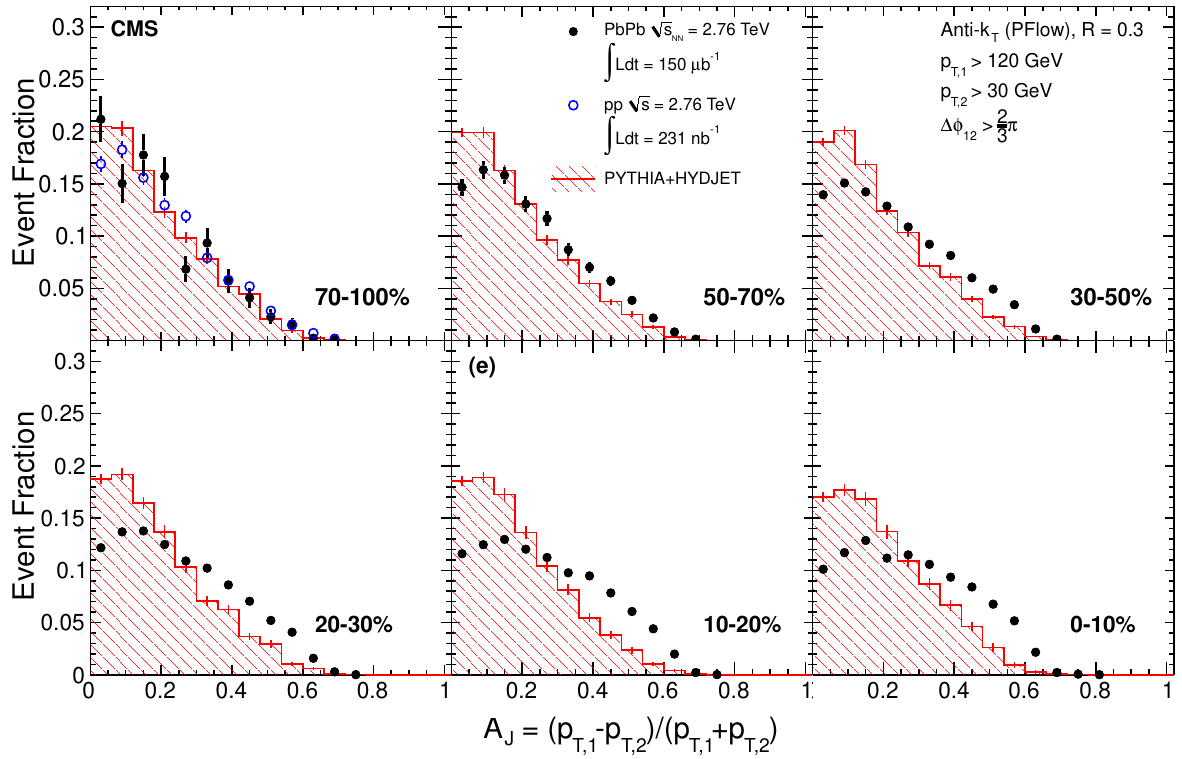}
    \caption{The \AJ distributions for jet pairs with a leading jet of $\ptOne > 120\GeV$ and subleading jet of $\ptTwo > 30\GeV$, presented for different event centrality classes. The dijet pair is required to fulfill a back-to-back requirement in azimuthal angle of $\dphiOneTwo > 2\pi/3$. Black filled points represent the \PbPb data, while the red hatched histogram shows the \PYTHIAHYDJET simulation results. The open blue circles in the upper left panel are the results from $\roots = 2.76\TeV$ \pp collisions, acting as an unquenched reference in conjunction with the simulations. Vertical bars represent statistical uncertainties only. \FigureAdaptedFrom{CMS:2012ulu}}
    \label{fig:276TeVAJ}
\end{figure}

Subsequent studies of dijet pairs produced in \PbPb collisions quantified the magnitude and centrality dependence (Section~\ref{sec:ExperimentalMethods_Centrality}) of the observed enhancement in \AJ. Figure~\ref{fig:276TeVAJ} shows \AJ as a function of centrality class for \PbPb data and \PYTHIAHYDJET simulation, with \AJ in \pp collisions shown in the first panel only~\cite{CMS:2012ulu}. All systems correspond to $\rootsNN = 2.76\TeV$, and both the \pp data and the \PYTHIAHYDJET simulations act as a reference for what the \AJ distribution would be in the absence of parton-medium interactions. In addition, the \PYTHIAHYDJET simulation includes the impact of the broadening of the measured energy distribution as a result of the degrading jet energy resolution caused by the underlying event background increasing as a function of event centrality. In the peripheral centrality selection of 70--100\%, the \PbPb data is qualitatively consistent with both the \pp data and the simulations. However, moving from peripheral to more central selections increases the observed dijet \AJ value in data beyond the expected changes from energy resolution effects modeled in \PYTHIAHYDJET, with the greatest observed discrepancy occurring in the 0--10\% centrality class. Similar trends were also observed for inclusive dijets produced at the higher collision energy of $\rootsNN=5.02\TeV$~\cite{CMS:2018dqf}.

The observed enhancement in dijet asymmetry is interpreted as a signature of differential jet-energy loss, whereby the leading jet has lost less energy to parton-medium interactions than the subleading jet. There are multiple possible causes of the jet energy loss being differential between leading and subleading jets; a few examples include a difference in path-length through the QGP, a color-charge factor governing interactions differently for quark- and gluon-initiated jets, or the highly stochastic nature of parton-medium interactions resulting in significant biases when selecting final-state leading and subleading jets (the \pt-dependent studies of \AJ in Ref.~\cite{CMS:2012ulu} suggest this last option). Independently of the underlying mechanism, one can additionally characterize the energy loss with a missing transverse momentum observable (\mpt), defined as
\begin{linenomath}
\begin{equation}
\label{eqn:mpt}
\mpt = -\sum\limits_{i} \pti \cos(\phii - \phi_{\text{leadingJ}}),
\end{equation}
\end{linenomath}
where the index $i$ is the $i^{\mathrm{th}}$ reconstructed track in the event, \pti its transverse momentum, \phii its azimuthal position, and $\phi_{\text{leadingJ}}$ is the azimuthal position of the leading jet. Note that by this definition, particles in the direction of the leading jet will have a negative contribution. 

\begin{figure}[ht]
    \centering
    \includegraphics[width=0.9\linewidth]{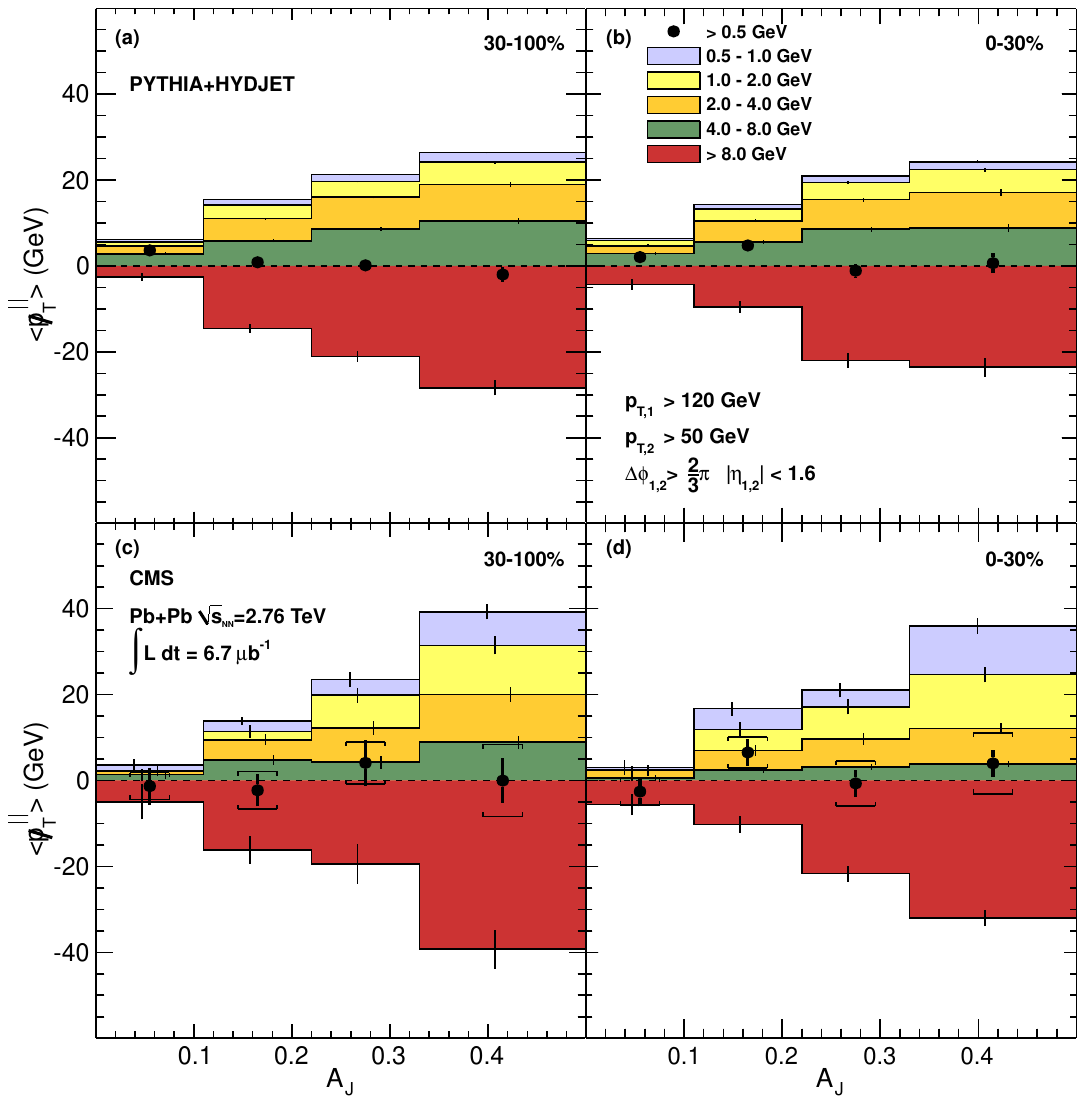}
    \caption{The \avempt values as a function of \AJ for tracks with $\pt>0.5\GeV$. Dijets are selected with $\ptOne > 120\GeV$, $\ptTwo > 50\GeV$, and $\dphiOneTwo > 2\pi/3$. The left panels are for peripheral, 30--100\% centrality events, and the right panels are for central, 0--30\% events. The upper row shows the results in \PYTHIAHYDJET simulation (lacking quenching) while the lower row shows the result in \PbPb data. Both data and simulation are for $\rootsNN = 2.76\TeV$. Solid circles show the total average \mpt while individual color-filled histograms	show contributions from	particles of \pt ranging from 0.5--1.0\GeV to larger than 8.0\GeV. Vertical bars represent statistical uncertainties while the horizontal bars surrounding the solid black circles represent systematic uncertainties. \FigureAdaptedFrom{CMS:2011iwn} }
   \label{fig:dijetMissingPt2010}
\end{figure}

The average of the \mpt observable over events passing back-to-back dijet selections, \avempt, is related to the relative distribution of energy in the leading and subleading jet hemispheres. This average is shown in Fig.~\ref{fig:dijetMissingPt2010} as a function of \AJ for two different centrality classes (30--100\% and 0--30\%). The color-filled histograms show the \avempt contributions for track \pt ranges of 0.5--1.0, 1.0--2.0, 2.0--4.0, 4.0--8.0, and greater than 8.0\GeV. From the figure, in both \PYTHIAHYDJET simulations and \PbPb data, the leading jet dominates the greater than 8.0\GeV \pt bin (red histogram, negative $\textit{y}$ axis contributions). However, relative to the simulations, which lack a quenching mechanism, the balancing spectra of particles corresponding to those in the subleading jets direction (positive $\textit{y}$ axis contributions) are softer in \PbPb data. A substantial excess of particles is observed in the 0.5--1.0 and 1.0--2.0\GeV \pt ranges (light-blue and yellow histograms). Furthermore, this excess increases with the \AJ category, here used as a proxy for the strength of the relative energy lost to the medium between the jets of the dijet pair. Finally, there is an observable depletion in the relative contribution of the semi-hard category of particles of 4.0--8.0\GeV in central \PbPb events compared to both peripheral \PbPb events and central simulations. Later studies performed by CMS using a data set 25 times larger reached similar conclusions and were able to establish that these trends hold for a variety of jet distance parameters from $R = 0.2$ to 0.5~\cite{CMS:2015hkr}, as long as particles up to an $\eta$-$\phi$ distance of 2.0 in $\Delta R$ from the jet axis are included. In summary, studies of dijet asymmetry indicate differential jet energy loss, with a depletion of hard particles in subleading jets and a corresponding enhancement of soft constituent particles.

\subsubsection{Suppression of jet spectra in \texorpdfstring{\PbPb}{PbPb} collisions}
\label{ssec:InclusiveJetObservables_JetRAA}

While the measurements described in Section~\ref{ssec:InclusiveJetObservables_Dijets} firmly establish that parton-medium interactions modify both the energy and radiation patterns of final-state jets, the observables \AJ and \avempt are constructed so that all such statements can only be made \textit{relatively}, \ie, the quenching effects are only seen comparing a subleading jet to a leading jet. Both jets however are potentially quenched, as both are produced prior to medium formation and traverse the medium over some path length. Therefore, to observe energy loss in absolute terms for inclusive jet production, CMS and other HI experiments use the nuclear modification factor \RAA, as defined by Eq.~(\ref{eqn:RAA}) in Section~\ref{ssec:testsOfGlauberModelEWKBosons}. In the absence of medium effects, hard probes, such as jets and high-\pt tracks, are expected to scale with the number of binary nucleon-nucleon collisions and therefore give an \RAA of one. An \RAA larger than unity, such as that observed from anti-shadowing effects in \pPb (as discussed in Section~\ref{sec:SmallSystems_JetQuenching}), indicates enhancement, while an \RAA smaller than unity indicates suppression.

\begin{figure}[ht]
    \centering
    \includegraphics[width=0.95\linewidth]{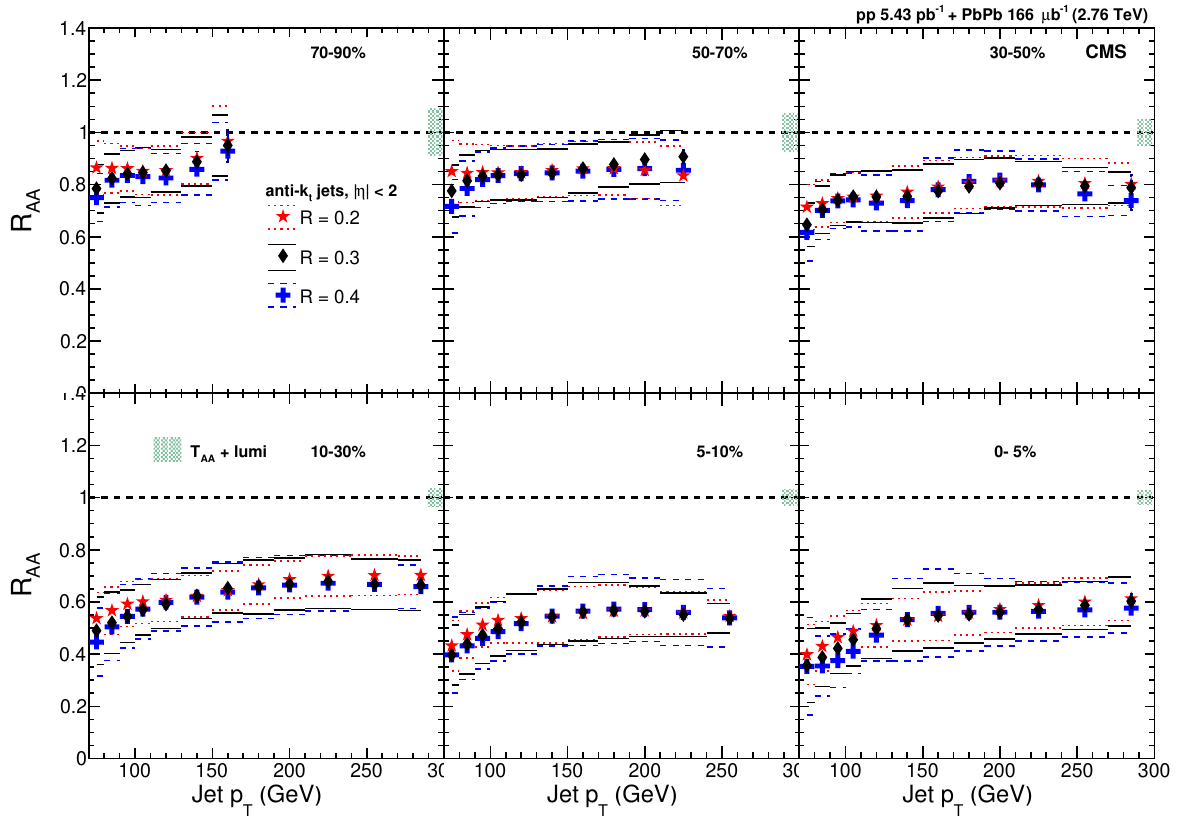}
    \caption{Inclusive jet \RAA plotted	as a function of the jet \pt for $\abs{\eta}<2.0$. Each	panel corresponds to a different centrality class (upper left) 70--90\%, (upper middle) 50--70\%, (upper right) 30--50\%, (lower left) 10--30\%, (lower middle) 5--10\%, and (lower right) 0--5\%. Results for three jet distance parameters, $R = 0.2$, 0.3, and 0.4, are overlaid as red stars, black diamonds, and blue crosses, respectively. Vertical bars (typically smaller than the markers) represent the statistical uncertainty, while horizontal bars around each point are the nonglobal systematic uncertainties. Finally, the combined global systematic uncertainty coming from \TAA and the integrated luminosity measurement is plotted as a shaded green bar on the horizontal black-dashed unity line. \FigureAdaptedFrom{CMS:2016uxf}}
    \label{fig:276TeVJetRAA}
\end{figure}

The first measurement of inclusive jet \RAA by CMS~\cite{CMS:2016uxf} observed a significant suppression for jets produced in \PbPb collisions at $\rootsNN = 2.76\TeV$ for three jet distance parameters ($R=0.2$, 0.3, and 0.4), spanning a jet \pt range of 70--300\GeV and in six centrality classes covering 70--90\% through 0--5\%, as shown in Fig.~\ref{fig:276TeVJetRAA}. A strong centrality dependence is observed, with largest suppression occurring in the 0--5\% centrality class with an \RAA of 0.35 at a jet \pt value of 70\GeV. However, as functions of jet $R$, the central values for \RAA are consistent within the reported uncertainties across all centrality classes. The lack of $R$ dependence is curious given the expectation that some fraction of the observed lost jet energy is recoverable when looking beyond the jet cone, as observed in the initial and subsequent \avempt studies discussed in Section~\ref{ssec:InclusiveJetObservables_Dijets}~\cite{CMS:2011iwn,CMS:2015hkr}. 

The paired observations of \avempt recovery at large $\Delta R$ angular distances from the jet axis~\cite{CMS:2015hkr} and the jet \RAA insensitivity to the distance parameter $R$ for $R \leq 0.4$~\cite{CMS:2016uxf} motivated a study of jets with even larger jet distance parameters. In addition, partonic quenching models, such as \JEWEL~\cite{Zapp:2013vla}, show a jet \RAA dependence at large $R$ that is subject to assumptions on how the missing jet energy is deposited into the larger medium, making such studies a model-dependent test of the response of the medium~\cite{KunnawalkamElayavalli:2017hxo}.

\begin{figure}[ht]
    \centering
    \includegraphics[width=0.95\linewidth]{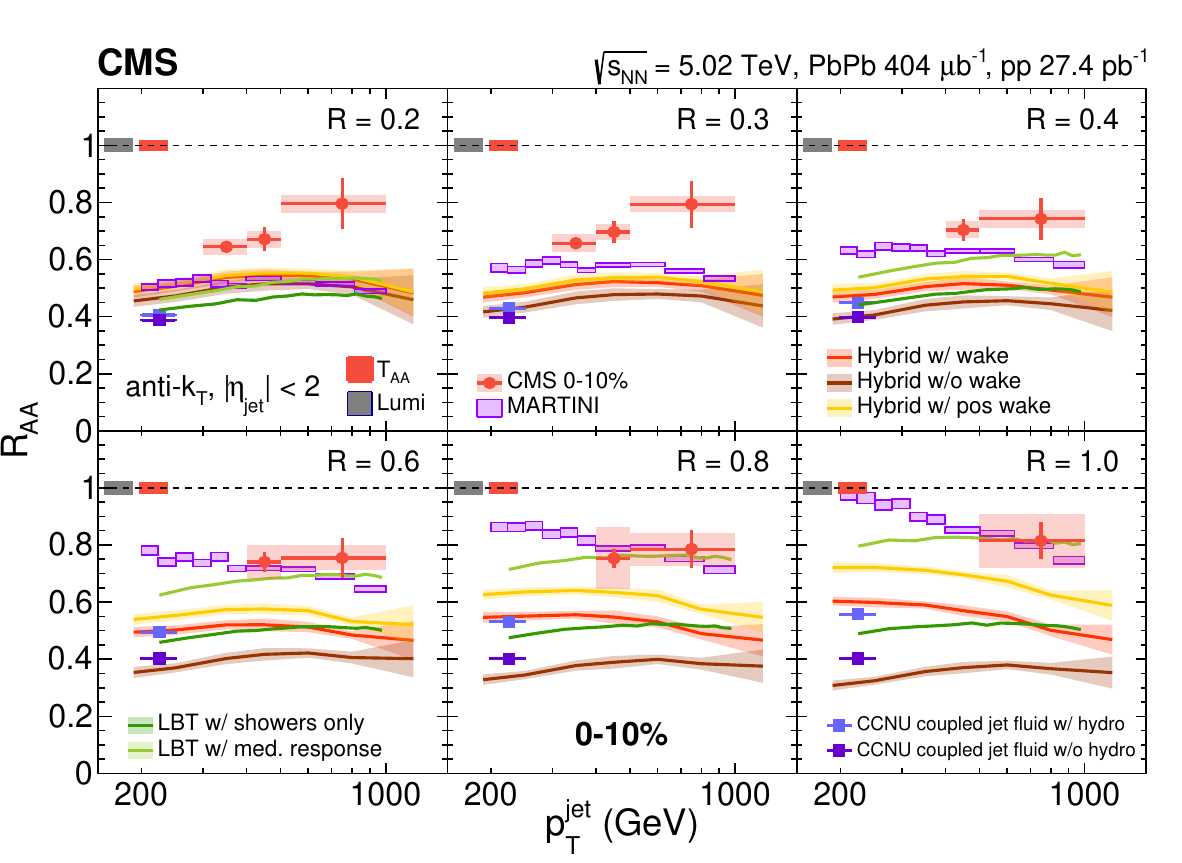}
    \caption{Jet \RAA in the 0--10\% centrality class as a function of jet \pt for jets with $\abs{\eta}<2.0$. Each panel corresponds to a different distance parameter $R$, as indicated. Filled red circle markers represent the data, with vertical red lines representing statistical uncertainties and horizontal red lines representing bin widths. The shaded red boxes around the	points represent systematic uncertainties. Integrated luminosity (for \pp collisions) and $\langle \TAA \rangle$ (for \PbPb collisions) global uncertainties are shown as shaded boxes around the dashed horizontal line for $\RAA = 1$. Predictions for the \HYBRID~\cite{Casalderrey-Solana:2014bpa,Pablos:2019ngg}, \MARTINI~\cite{Schenke:2009gb}, \LBT~\cite{He:2018xjv}, and \CCNU~\cite{Tachibana:2017syd,Chang:2016gjp,Chang:2019sae} models are plotted for comparison. \FigureFrom{CMS:2021vui}}
    \label{fig:5TeVJetRAA}
\end{figure}

The CMS Collaboration has reconstructed jets up to an $R$ parameter of 1.0, for the first time in HI collisions, using the 5.02\TeV \PbPb (corresponding to an integrated luminosity of 404\mubinv) and \pp reference (27.4\pbinv) data sets collected in 2015. The \RAA values are determined as a function of jet \pt. To mitigate the growing impact and associated uncertainties arising from the underlying event background on jet energy as the jet $R$ parameter is increased, stringent thresholds are implemented on the minimum \pt of the jets considered, with respect to $R$. Specifically, for $R=1.0$, only jets with $\pt > 500\GeV$ are taken into consideration. The results for six $R$ values in the centrality class 0--10\% are plotted in Fig.~\ref{fig:5TeVJetRAA}. 
All six \RAA values for the highest jet \pt bin are close to 0.8, with no significant dependence on $R$, within systematic uncertainties, even after accounting for the correlated uncertainties. There is no indication that \RAA approaches unity, even for $R=1$, the largest $R$ analyzed using Run~2 data, as would be expected if the quenched energy transported to large angles is fully recovered.

Figure~\ref{fig:5TeVJetRAA} also compares a number of jet quenching models to the experimental results. While the models encode different assumptions about the physics of parton-medium and jet quenching interactions, one interesting detail is how assumptions regarding the medium response affect the results. In the context of jet quenching, medium response is the collective motions and modes induced by the jet energy deposited in parton-medium interactions, resulting in nontrivial structures such as a wake of energetic and depleted regions. Both the \HYBRID and \LBT models allow for alternative assumptions of the medium response. As an example, the \HYBRID model provides three different curves: a red curve corresponding to the full model of the medium wake, a brown curve corresponding to zero medium wake, and a golden yellow curve that only includes positive contributions and ignores energy depletion resulting from the medium wake~\cite{Pablos:2019ngg}. While each of these curves roughly match within reported systematic uncertainties for small jet $R$ (0.2--0.4), at large $R$ there are significant differences. The CMS data indicates a preference for the \HYBRID implementation featuring a positive wake, while it is worth highlighting that all curves exhibit a tendency to underestimate the measured values. Likewise, of the two curves provided by the \LBT model, the prediction with showers only is disfavored and the prediction incorporating a medium response is favored~\cite{He:2018xjv}. The prediction from \MARTINI does not include alternative assumptions for the medium response effect and, therefore, no comment on the medium impact can be made~\cite{Schenke:2009gb}. Finally, the \CCNU prediction is restricted to a jet \pt below what was measured for large $R$~\cite{Tachibana:2017syd,Chang:2016gjp,Chang:2019sae}. These observations motivate future jet measurements, at low \pt and using a large $R$ parameter, similar to the one reported in Ref.~\cite{ALICE:2023waz}, to further constrain the modeling of the medium response.

\subsubsection{Suppression of high-\texorpdfstring{\pt}{pt} hadron production}
\label{ssec:InclusiveJetObservabeles_TrackRAA}

High-\pt hadrons are produced via fragmentation and hadronization mechanisms initiated by the same partonic hard scatterings that result in final-state jets. Thus, high-\pt hadron spectra can also be used to probe the strength of parton energy loss in the QGP medium. As compared to jet spectra, theoretical predictions of hadron spectra (and the resulting \RAA values) are more sensitive to details of the fragmentation and hadronization model used. However, high-\pt charged hadrons are produced in larger numbers than jets and can have their \pt measured with excellent resolution. These experimental considerations have made the \RAA of charged hadrons a touchstone measurement of parton energy loss effects for over two decades, and was key to the discovery of jet quenching at RHIC~\cite{PHENIX:2001hpc,STAR:2002ggv}.

\begin{figure}[ht]
    \centering
    \includegraphics[width=0.95\linewidth]{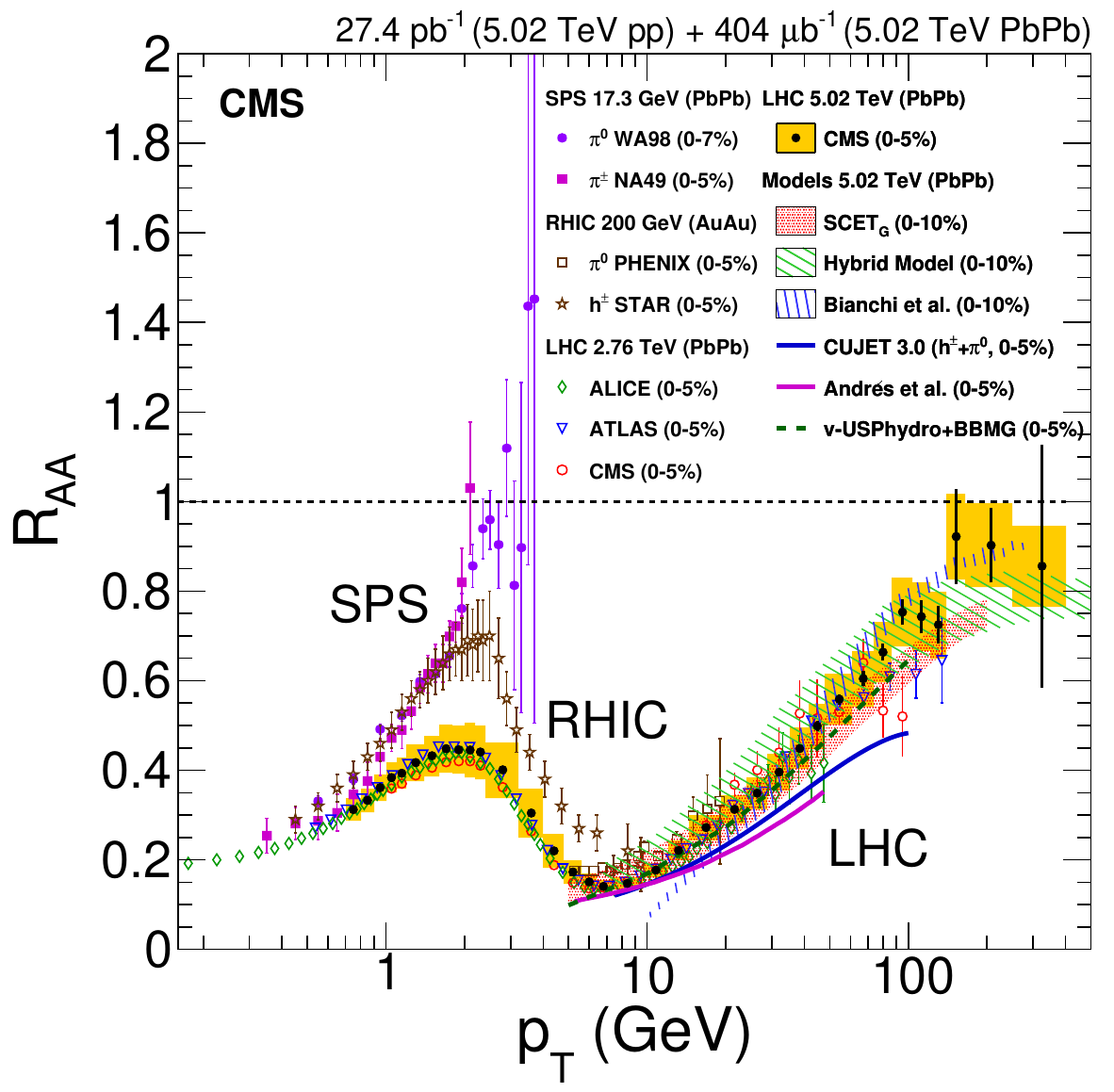}
    \caption{Measurements of \RAA in central heavy ion collisions at four different center-of-mass energies, for neutral pions (SPS, RHIC), charged hadrons ($h^{\pm}$) (SPS, RHIC), and charged particles (LHC). Data are taken from Refs.~\cite{WA98:2001lnw,dEnterria:2004cly,PHENIX:2012jha,STAR:2003fka,NA49:2007gga,ALICE:2012aqc,ATLAS:2015qmb,CMS:2012aa,CMS:2016xef}. Predictions of six models for $\sqrtsNN=5.02\TeV$ \PbPb collisions are shown~\cite{Chien:2015vja,Casalderrey-Solana:2014bpa,Bianchi:2017wpt,Xu:2015bbz,Andres:2016iys,Noronha-Hostler:2016eow}. The error bars represent the statistical uncertainties and the yellow boxes around the 5.02\TeV CMS data show systematic uncertainties. The \TAA uncertainties, which are small, are not shown. 
 \FigureFrom{CMS:2016xef}}
    \label{fig:5TeVhadronRAA}
\end{figure}

Figure~\ref{fig:5TeVhadronRAA} shows a compilation comparing CMS measurements of the charged-particle \RAA at $\sqrtsNN = 2.76\TeV$~\cite{CMS:2012aa} (red points) and 5.02\TeV~\cite{CMS:2016xef} (black points) to other experimental results~\cite{WA98:2001lnw,dEnterria:2004cly,PHENIX:2012jha,STAR:2003fka,NA49:2007gga,ALICE:2012aqc,ATLAS:2015qmb}. The CMS 5.02\TeV data spans nearly three orders of magnitude of \pt, revealing an oscillating structure having a local minimum around 7\GeV. For \pt values under 3\GeV, where effects such as parton energy loss, initial-state effects, and radial flow can all have significant contributions, an approximate ordering with collision energy is observed. The 17.3\GeV SPS data has higher values than the 200\GeV RHIC data, which are in turn higher than the \TeV-scale LHC data. For higher \pt values, parton energy loss is expected to be the dominant effect, resulting in a strong suppression that is remarkably similar at RHIC and the LHC around $\pt=7\GeV$, despite the order of magnitude difference in collision energy. This similarity can be explained by the shape of the underlying hadron spectra. Flatter spectra, which are observed at higher collision energies, demand a greater absolute energy loss to achieve similar \RAA suppression values. This implies that energy loss effects are stronger at higher collision energies. 

As \pt increases, \RAA becomes less suppressed, and at \pt values above 150\GeV, \RAA is consistent with unity. This contrasts with the trend observed for the inclusive jet \RAA shown in Fig.~\ref{fig:276TeVJetRAA}, which is nearly independent of jet \pt for jet $\pt>150\GeV$. 
One potential explanation for the difference between the jet and the charged-particle \RAA trends is related to selection effects coming from the requirement of a high-\pt hadron being present in the event.
Color coherence effects tend to reduce radiative energy loss as parton energy increases~\cite{Gyulassy:1993hr}, because of the Landau--Pomeranchuck--Migdal~(LPM) mechanism (in which consecutive gluon emissions in the parton shower exhibit quantum interference) and because the effective number of radiating color sources is suppressed in the parton shower~\cite{Casalderrey-Solana:2012evi}.
The requirement for a high-\pt hadron in an event could preferentially select configurations where the LPM effect is stronger than average for a jet at any given \pt. 
Six theoretical predictions of the charged-hadron \RAA are also overlaid with the data in Fig.~\ref{fig:5TeVhadronRAA}~\cite{Chien:2015vja,Casalderrey-Solana:2014bpa,Bianchi:2017wpt,Xu:2015bbz,Andres:2016iys,Noronha-Hostler:2016eow}. Most of the models are able to accurately predict the strong suppression observed at the local minimum of \RAA, but the models predict a large range of values at higher \pt. Thus, these data can be used to constrain energy loss models for events containing hard jet fragmentation patterns that result in a very energetic hadron and are complementary to measurements of the inclusive jet \RAA, which are not strongly biased towards any particular fragmentation pattern.

\subsection{Strength of energy loss}
\label{sec:StrengthOfELoss}

This section details the Run 1 and 2 measurements by CMS that quantify the strength of interactions between color-charge carriers and the QGP through studies of the path-length dependence of parton energy loss, measurements of the absolute (as opposed to relative) energy loss of jets, and the QCD color-charge dependence of parton energy loss effects. Also discussed is the first measurement of top quarks in HI collisions, which may lead to new methods for probing the energy loss of partons at different stages of the QGP evolution.

\subsubsection{Path-length dependence of energy loss}
\label{ssec:StrengthOfELoss_PathLength}

The amount of energy lost by a parton in the QGP is thought to depend on its path length through the medium. Although the average energy lost is expected to increase as the average path length increases, a quantitative understanding of this dependence can provide insight into the relative strengths of collisional and radiative energy loss. Collisional energy loss is expected to scale linearly with the path length $L$ in a static medium. Because radiated gluons can also lose energy, radiative energy loss processes are expected to scale with an approximately $L^2$ dependence in a static medium~\cite{Baier:1996kr}. Although additional effects such as the expansion of the medium, the LPM effect, and color coherence effects can reduce the power of the anticipated path-length dependence, in general radiative energy loss is expected to scale faster with $L$ compared to collisional energy loss. Multiple experimental techniques are available to explore this topic.

\begin{figure}
    \centering
    \includegraphics[width=0.4716\linewidth]{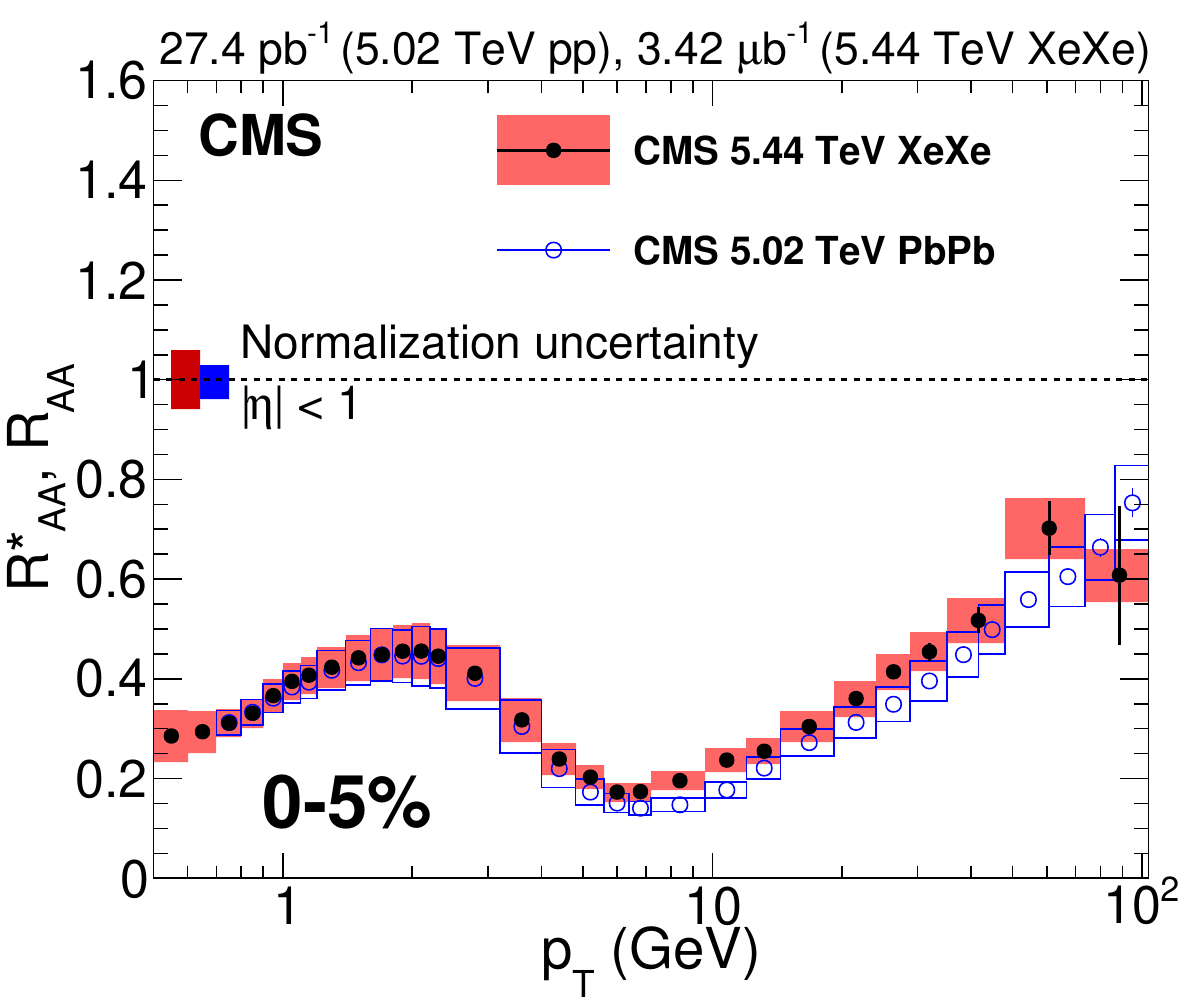}
    \raisebox{-0.25ex}{\includegraphics[width=0.45\linewidth]{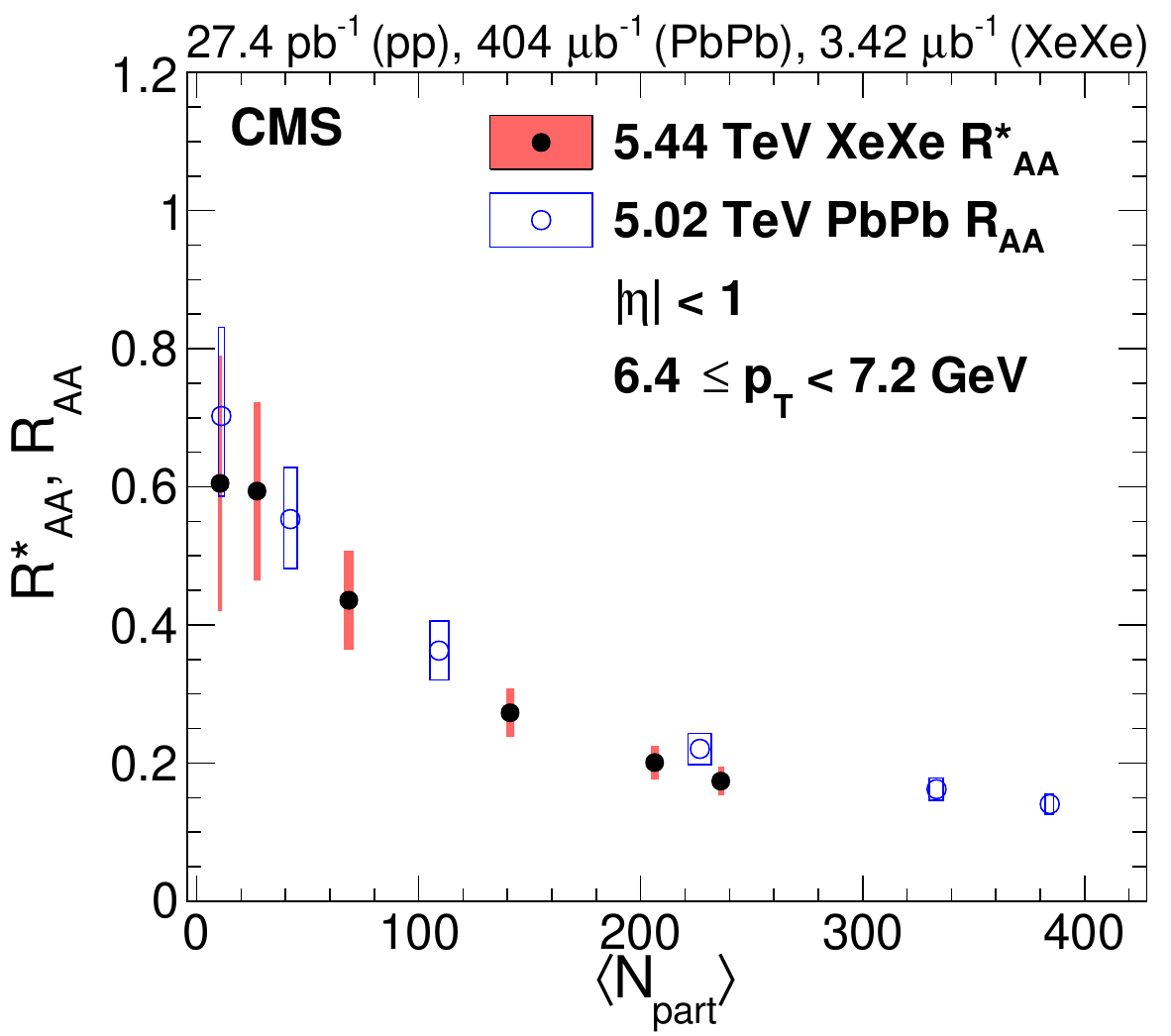}}
    \caption{The charged-particle \RAAStar for \XeXe collisions at $\sqrtsNN=
 5.44\TeV$~\cite{CMS:2018yyx} and \RAA for \PbPb collisions at 5.02\TeV~\cite{CMS:2016xef}. The asterisk in \RAAStar indicates that the 5.44\TeV \pp reference has been calculated by extrapolating a measured 5.02\TeV \pp spectrum. The solid pink and open blue boxes represent the systematic uncertainties of the \XeXe and \PbPb data, respectively. The left panel shows the result as a function of particle \pt for a 0--5\% centrality selection. In the right panel, the results for the $6.4<\pt<7.2\GeV$ range are plotted as functions of average \npart. \FiguresCompiled{CMS:2016xef,CMS:2018yyx}}
    \label{fig:XeXeRAA}
\end{figure}

The first technique involves studying the strength of the energy loss by hard probes traveling through different volumes of QGP. One way of experimentally varying the volume of the QGP produced in HI collisions is to collide different ion species. The left panel of Fig.~\ref{fig:XeXeRAA} shows a comparison of the charged-hadron \RAA for central \PbPb collisions at $\sqrtsNN = 5.02\TeV$~\cite{CMS:2016xef}, collected in 2015, and \RAAStar for \XeXe collisions at $\sqrtsNN = 5.44\TeV$~\cite{CMS:2018yyx}, collected in 2017. The asterisk in \RAAStar indicates that a MC-based extrapolation procedure was used to adjust a measured 5.02\TeV \pp reference spectrum to the appropriate reference energy of 5.44\TeV for the \XeXe measurement. As discussed in Section~\ref{ssec:InclusiveJetObservabeles_TrackRAA}, the relatively small difference in collision energy of these two systems is not expected to strongly affect the magnitude of the \RAA suppression. However, the radius of the Xe nucleus is ${\approx}~5.4\unit{fm}$, while that of the Pb nucleus is ${\approx}6.6\unit{fm}$~\cite{Loizides:2014vua}. Thus, a smaller volume of QGP is produced in collisions of \XeXe compared to collisions of \PbPb at the same centrality. The values of the \XeXe charged-hadron \RAAStar in the range $\pt>5\GeV$, where energy loss effects dominate, are clearly less suppressed than those of \PbPb collisions, which is consistent with partons experiencing less energy loss in the smaller collision system. These data have been used to estimate that the path-length dependence of energy loss scales as $L^{1.3\pm0.5}$~\cite{Arleo:2019qzc}. 

The right panel of Fig.~\ref{fig:XeXeRAA} displays the charged-hadron \RAA and \RAAStar for these two collision systems near the local minimum of \RAA at $6.4<\pt<7.2\GeV$ as a function of the average \npart. The data from the two collision systems seem to follow a common decreasing trend. A value of $\npartave\approx230$ corresponds to the most central (0--5\%) \XeXe collisions and semicentral (10--30\%) \PbPb collisions. Still, for this \npart value, the two systems have very similar \RAA and \RAAStar values. This implies, that at a given center-of-mass energy, systems containing similar volumes of QGP produce similar values of energy loss for hard probes, regardless of the initial colliding ion species or impact parameter. Interestingly, both the \PbPb and \XeXe data show a significant suppression in the most peripheral events examined (around $\npartave\approx10$). The values of \RAA and \RAAStar are consistent with the suppression observed for color-neutral \PZ bosons produced in 5.02\TeV \PbPb collisions, as discussed in Section~\ref{ssec:testsOfGlauberModelEWKBosons}. Therefore, the \RAA suppression observed in this centrality region cannot be interpreted as a signature of parton energy loss. Other potential event selection or centrality calibration effects may fully explain this observation~\cite{Loizides:2017sqq}. Such effects are expected to be negligible for central and semicentral collisions and therefore do not strongly affect any conclusions regarding energy loss in larger volumes of QGP.

Another technique for assessing the path-length dependence of parton energy loss makes use of inherent spatial anisotropies within the initial state of HI collisions. The initial partons resulting from a hard scattering are not expected to have any preferred azimuthal direction. Because of the 
initial-state geometry of HI collisions, the overlapping transverse area between the two ions, where the QGP is expected to form, typically has a nonzero eccentricity that is related to the impact parameter, and therefore the collision centrality. This is important because the relatively large transverse eccentricities present in semicentral HI collisions can give rise to substantially different path lengths for partons traveling parallel and transverse to the direction of the collision impact parameter. Furthermore, comparable effects may arise even in events lacking significant initial-state eccentricity due to the influence of fluctuations of the initial-state geometry. Given that partons encountering a greater path length within the medium are expected to experience comparatively higher energy loss on average, this phenomenon can induce an azimuthal anisotropy of final-state high-\pt particles. This anisotropy can be expressed in terms of Fourier coefficients \vN (as shown by Eq.~(\ref{eq:fourier}) in Section~\ref{sec:HydroQGP}). We note that this picture of path length dependent energy loss does not account for possible large jet-by-jet fluctuations in energy loss, as discussed in Section~\ref{ssec:InclusiveJetObservables_Dijets}.

\begin{figure}[t]
    \centering
    \includegraphics[width=0.85\linewidth]{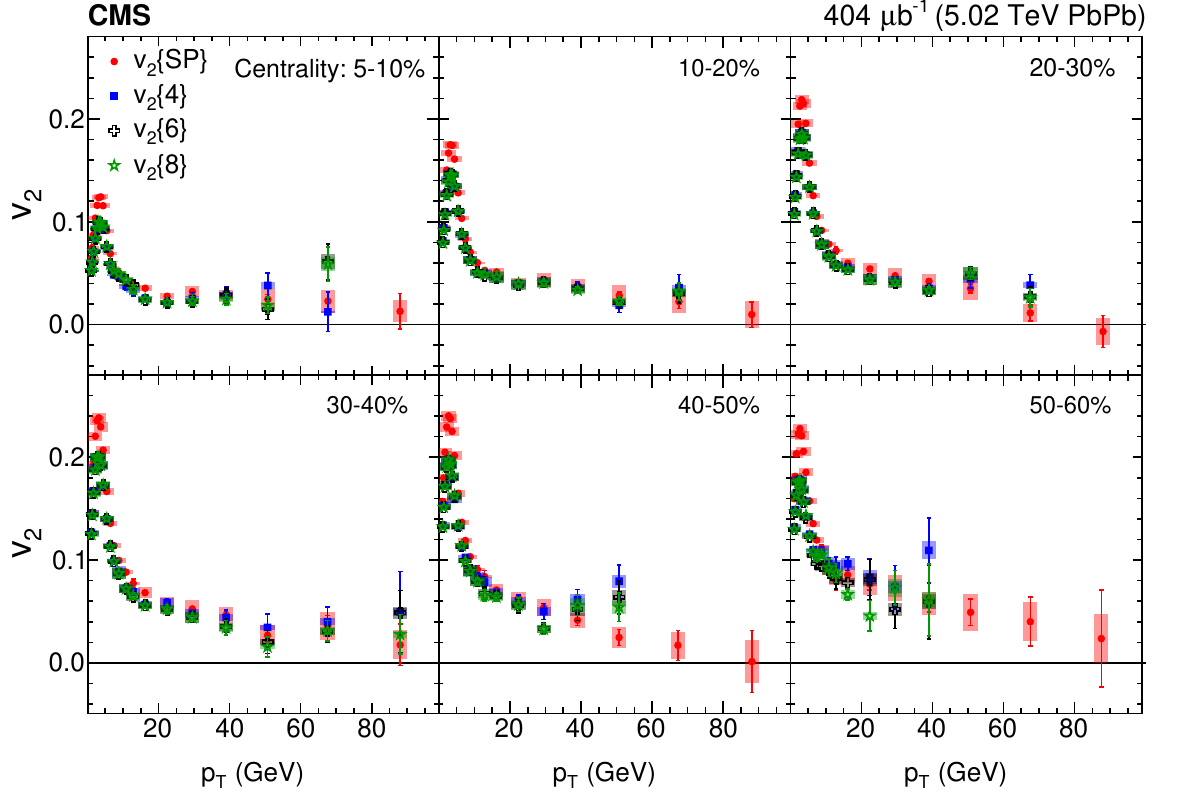}
    \caption{Comparison between charged-hadron \vTwo results from various methods as a function of \pt in six centrality selections from 0--5\% to 50--60\%. The vertical bars represent the statistical uncertainties, while the shaded boxes represent systematic uncertainties. \FigureAdaptedFrom{CMS:2017xgk}}
    \label{fig:highPtV2Hadrons}
\end{figure}

Figure~\ref{fig:highPtV2Hadrons} shows a measurement of the \vTwo coefficient as a function of \pt for charged hadrons in 5.02\TeV \PbPb collisions~\cite{CMS:2017xgk}. For low \pt values (\ie, $\pt < 3\GeV$), hydrodynamic flow is believed to dominate multiparticle correlations, with the \vTwo value reflecting the eccentricity of the particle-emitting region. Four different methods for determining \vTwo values are shown, including the scalar-product method \vtwo{\mathrm{SP}} and four, six, and eight particle correlators \vtwo{4}, \vtwo{6}, \vtwo{8}, respectively (Section~\ref{sec:QGP_Thermodynamics} presents the methods). 
The method originally used for bulk particle production has been adapted and modified for the high-\pt regime. Specifically, the \vtwo{8}(80\GeV) is not derived from the correlation of eight particles with \pt around 80\GeV, but rather from correlating one 80\GeV particle with softer particles. 
Significant positive \vTwo values are observed up to $\pt=80\GeV$ for most centrality selections investigated. Furthermore, the \vTwo values in the range $10<\pt<50\GeV$ tend to be larger for centrality selections containing a larger average initial-state eccentricity, as indicated by the magnitude of the \vTwo values at low \pt. These results do not seem to strongly depend on the method used to extract the \vTwo values. In particular, the results using multiparticle correlation methods, which tend to suppress non-flow contributions to the \vTwo values, strongly imply a connection between \vTwo and the initial-state geometry and its event-by-event fluctuations. A similar conclusion was reached when analyzing 2.76\TeV \PbPb collision data~\cite{CMS:2012tqw}. In the analysis of 5.02\TeV collisions, positive values of \vThree were also observed up to \pt values of around 20\GeV, but were found to be consistent with zero for higher \pt values.

\begin{figure}[t]
    \centering
    \includegraphics[width=0.95\linewidth]{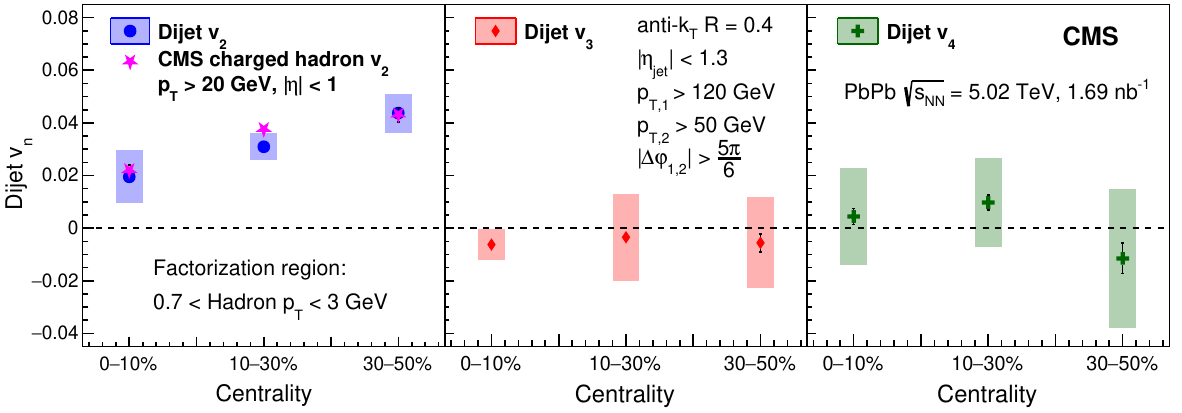}
    \caption{The dijet \vTwo (left), \vThree (middle), and \vFour (right) measured as functions of collision centrality in 5.02\TeV \PbPb collisions. The dijet \vTwo results are compared to CMS high-\pt hadron \vTwo results. The shaded boxes represent systematic uncertainties, while the vertical bars show statistical uncertainties. \FigureFrom{CMS:2022nsv}}
    \label{fig:dijet_v2}
\end{figure}

A similar measurement of \vTwo using fully reconstructed dijets in 5.02\TeV \PbPb collisions is shown in Fig.~\ref{fig:dijet_v2}~\cite{CMS:2022nsv}. The results are calculated by correlating each jet in a dijet pair with hadrons having large $\eta$ separation from the jet. A significant positive \vTwo is observed in all three centrality selections examined, with the magnitude of \vTwo increasing as the average initial-state eccentricity increases. This indicates more jets are observed in the azimuthal direction parallel to the event plane, as compared to the perpendicular direction. The results for the \vTwo of dijets are compatible with those for individual high-\pt charged hadrons. These results strongly imply a path-length dependence of parton energy loss and can be used to constrain parton energy loss models. Measurements of dijet \vThree and \vFour were also performed, but the values were found to be consistent with zero, implying that jets are not strongly affected by event-by-event initial-state geometry fluctuations. 

\subsubsection{Measurements of absolute jet energy loss}
\label{ssec:StrengthOfELoss_AbsELoss}

\begin{figure}[ht]
    \centering
    \includegraphics[width=0.70\linewidth]{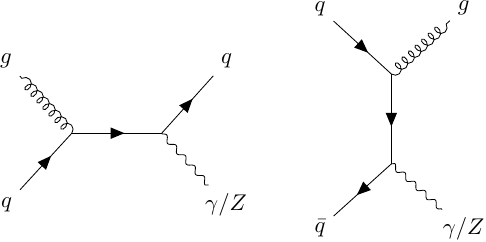}
    \caption{Feynman diagrams depicting two leading-order processes producing a photon or a \cPZ boson with a jet balancing the transverse momentum in the final state. The first diagram shows the outgoing jet to be initiated by a quark, while the other shows the outgoing jet to be initiated by a gluon. These rare hard scatterings have been used to study jet quenching in a number of CMS analyses~\cite{CMS:2017ehl}.}
   \label{fig:feynLO}
\end{figure}

As noted in Section~\ref{ssec:InclusiveJetObservables_JetRAA}, many of the initial observations of jet quenching relied on observables, such as the dijet \AJ, where both jets were quenched. To determine the total amount of energy a color-charge carrying parton loses to medium interactions in the QGP, as opposed to its relative energy loss, it is necessary to determine the initial parton energy. This can be done by studying rarer hard-scattering processes with a boson + jet in the final state, such as the LO processes depicted in Fig.~\ref{fig:feynLO}. The boson, typically a photon or \cPZ boson for the purposes of these studies, is colorless and therefore is not modified by strong interactions in the medium. Thus, the boson can be used to determine the initial energy of the companion parton that eventually results in a jet. By comparing the jet energies tagged in this manner with those similarly tagged in \pp reference collisions, the absolute impact of medium interactions on partons is observable.

\begin{figure}[ht]
    \centering
    \includegraphics[width=0.95\linewidth]{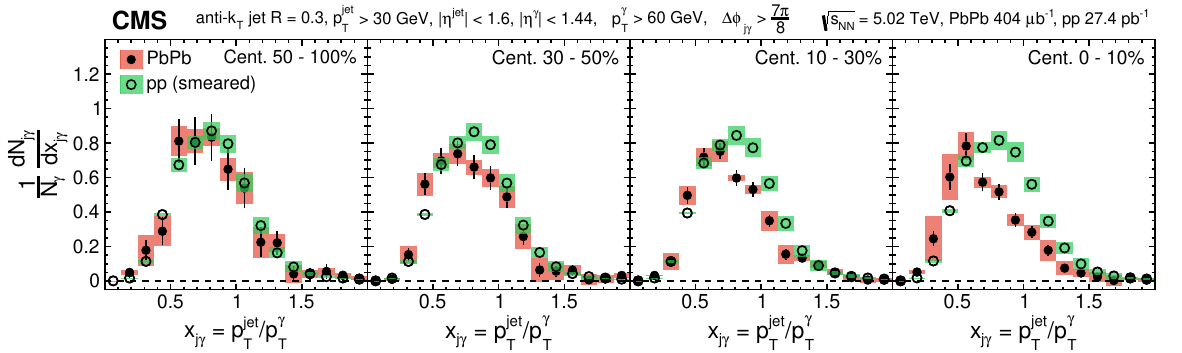}
    \caption{The \pt balancing observable \xjg for $\PGg+\text{jet}$ pairs is plotted as a function of centrality class panel-by-panel, with the leftmost panel corresponding to the 50--100\% peripheral selection, progressing to the 0--10\% central selection in the rightmost panel. The distribution is normalized by the number of photons in a \pp reference (open markers) and \PbPb (full markers) data, per centrality class. Vertical lines display the statistical uncertainties while the shaded bars around the points (red for \PbPb, green for \pp) show the systematic uncertainties. The statistical uncertainties of the \pp data are smaller than the markers for many data points. \FigureAdaptedFrom{CMS:2017ehl}}
    \label{fig:gammaJet}
\end{figure}

Such studies were carried out with the limited integrated luminosity of Run 1 with photon-tagged jets~\cite{CMS:2012oiv} and with the cleaner but statistically limited \cPZ-tagged jets with Run 2 data from 2015~\cite{CMS:2017eqd}. The currently most precise measurement of absolute jet energy loss by CMS uses photon-tagged jets taken at \rootsNN = 5.02\TeV in 2015 with integrated luminosities of 404\mubinv for \PbPb collisions and 27.4\pbinv for the \pp reference collisions~\cite{CMS:2017ehl}. Photons are required to be isolated, as discussed in Section~\ref{ssec:ExperimentalMethods_UE_Pho}, reducing contributions from photons produced in fragmenting jets or from resonance decays. In each centrality class (50--100\%, 30--50\%, 10--30\%, 0--10\%), photons are correlated with all jets opposite in azimuthal angle and for each photon-jet pair the balancing observable \xjg, defined as \xjg = \ptj / \ptg, is calculated. The distributions are then normalized by the number of photons found in each centrality class (and in the \pp reference). The result of this measurement is shown in Fig.~\ref{fig:gammaJet}. 

On inspection, it is clear that quenching effects do not modify the \xjg distribution in the peripheral 50--100\% centrality class compared to the \pp reference spectrum beyond the currently reported uncertainties. For this comparison the \pp reference spectrum is modified to have a similar resolution as the \PbPb distribution, \ie, ``smeared''. However, moving to a semiperipheral selection of 30--50\% centrality already shows a modest depletion in balanced photon-jet configurations, corresponding to \xjg larger than 0.75. Going to the semicentral 10--30\% centrality selection, the observed depletion of balanced configurations increases substantially, and in the central 0--10\% selection even extends down to $\xjg \approx 0.6$. While there are hints of a corresponding enhancement for $\xjg < 0.6$, this enhancement is at the edge of the reach allowed by the statistical and systematic uncertainties. The integral areas of the \PbPb distributions decrease as collisions become more central, resulting from a relatively larger fraction of jets quenched below the \ptj threshold of 30\GeV.

\subsubsection{Color-charge dependence of energy loss}

For the hard probes sector, specifically jets, constraints based on control samples in data are provided by CMS for theoretical calculations of jet transport coefficients, energy loss parametric dependencies, and in-medium shower evolution. The interactions in the QGP medium are expected to vary with the color charge of the energetic parton traversing it; gluons will interact more strongly than quarks given their larger Casimir color factor~\cite{Apolinario:2020nyw,Brewer:2020och}. Thus, significant efforts have been made to develop observables that preserve sensitivity in the final state to the identity of the parton initiating the observed process. 

\begin{figure}[b]
    \centering
    \includegraphics[width=0.45\linewidth]{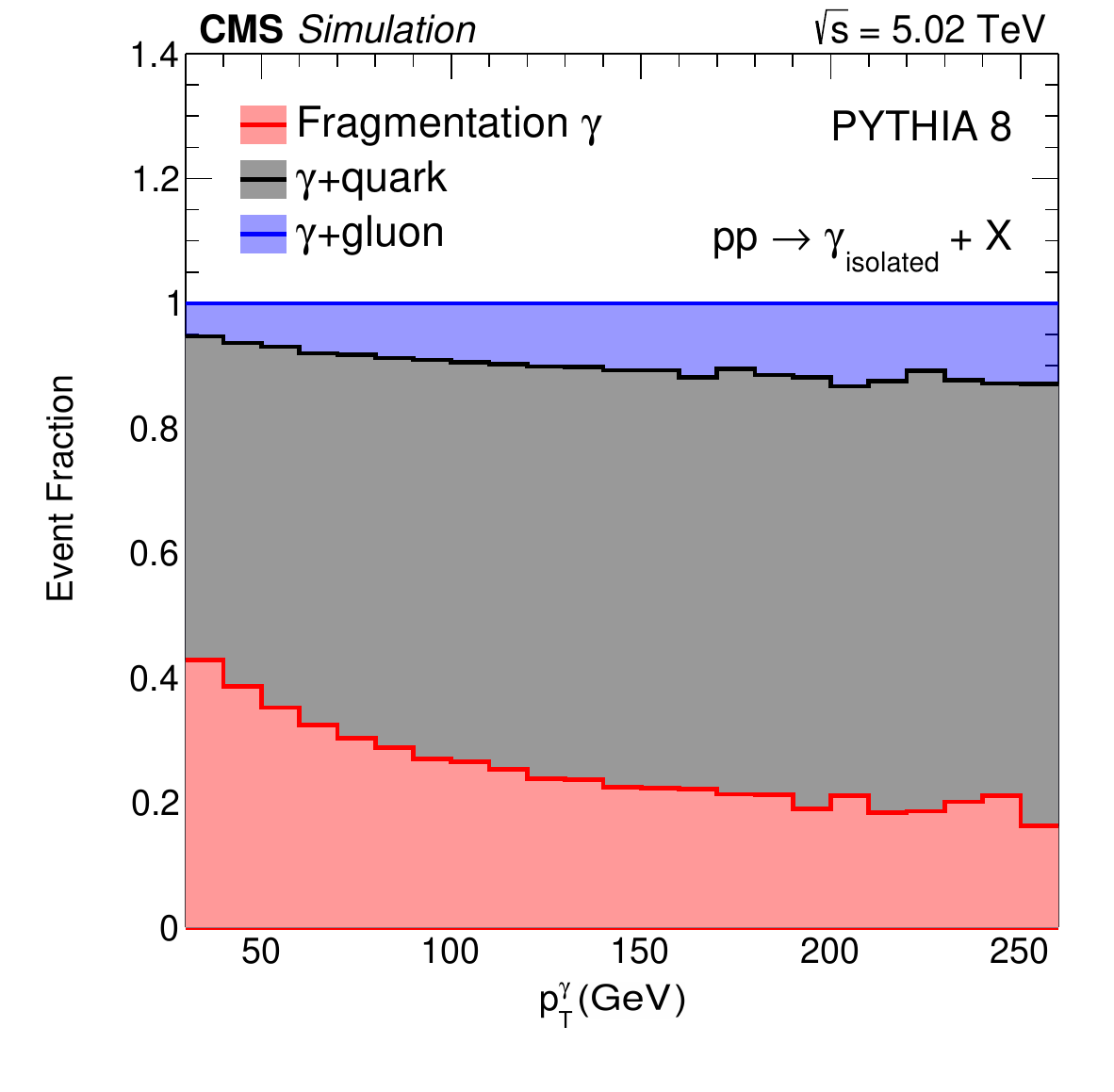}
    \caption{Relative contributions of fragmentation (red), photon+quark jet (grey), and photon+gluon jet (blue) processes to the production of isolated photons in \PYTHIA{}8 events. The requirement of an isolated photon in the event increases the fraction of quark-initiated jets relative to an inclusive jet sample. \FigureAdaptedFrom{CMS:2017ehl}}
    \label{fig:gammaJetqg}
\end{figure}
 
Aside from small contributions from heavy-flavor quarks, the overwhelming majority of jets from HI collision data represent a mix of light-quark and gluon contributions, with no identification of partonic origin possible on a per-jet basis. Even statistical discrimination methods for light-quark and gluon jet tagging remain challenging. Several standard tagging tools are used as light-quark versus gluon jet discriminators. These tools include jets produced in coincidence with EW bosons as discussed in Section~\ref{ssec:StrengthOfELoss_AbsELoss}, and the jet charge that serves as a proxy for the electric charge of the parent parton. 

It is important to note that within the framework of common searches for hot nuclear effects, where a specific measurement from \AonA collisions is compared to a \pp reference, none of the aforementioned discriminators allows the \textit{isolation} of color-charge effects in jet quenching studies. Instead, they provide various combinations of potential selection biases. For example, the photon-tagged jet sample used in previously discussed studies of absolute energy loss has a significant fraction of initial quark jets, as illustrated using a MC generator in Fig.~\ref{fig:gammaJetqg}~\cite{CMS:2017ehl}, which is higher than the initial quark jet fraction of an inclusive jet sample. Additionally, in the QGP medium, the requirement of a nonstrongly interacting boson as a ``trigger'' potentially alters the survival (or surface) bias of the studied jets. Thus, future comparisons of boson-tagged jet samples against inclusive jet samples may yield insights about the interplay between these selection effects and the different energy loss behavior of the various light-parton flavors.

For the jet charge (or other jet constituent-based observables), complications may arise from the medium response that is inevitably reconstructed as part of the jet shower; additionally, the jet charge is not infrared- or collinear-safe. Despite its potential limitations, the jet charge is an experimentally available observable that enables the evaluation of the quark-gluon composition of the jet sample, albeit relying on MC modeling.
Introduced in the 1970s as one of the earliest jet substructure observables, the measurement of the jet charge has served as a method for measuring the electric charge of a quark~\cite{Field:1977fa}. The technique was first used in deep inelastic scattering experiments at Fermilab~\cite{Fermilab-Serpukhov-Moscow-Michigan:1979zgc,Berge:1980dx}, CERN~\cite{Aachen-Bonn-CERN-Munich-Oxford:1981lfk,allen1982,albanese1984,Amsterdam-Bologna-Padua-Pisa-Saclay-Turin:1981hcw}, and Cornell University~\cite{Erickson:1979wa}. The jet charge is defined as the transverse momentum-weighted sum of the charges of particles within the jet cone, and experiments have used a variation of weight exponents and momentum thresholds to maximize its discriminating power. While initially used to study the substructure of nucleons, it has been suggested that the jet charge may also yield insights into the properties of the QGP~\cite{Li:2019dre}.

\begin{figure}[ht]
    \centering
    \includegraphics[width=\linewidth]{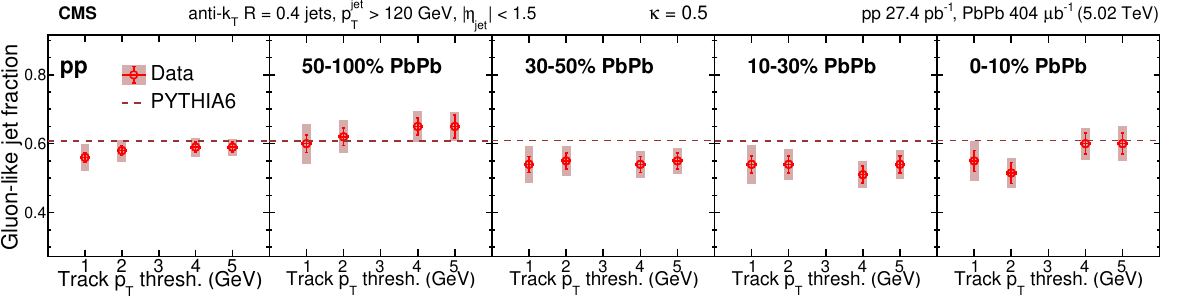}
    \caption{Results for the gluon-like jet fractions in \pp and \PbPb data shown for different track \pt threshold values and event centrality selections in \PbPb collisions. The systematic and statistical uncertainties are represented by the shaded regions and vertical bars, respectively. The predictions for the gluon jet fractions from \PYTHIA 6 are shown in dashed red lines. \FigureAdaptedFrom{CMS:2020plq}}
    \label{fig:jetCharge}
\end{figure}

CMS has conducted a measurement of jet charge distributions in HI collisions using 5.02\TeV \PbPb data~\cite{CMS:2020plq}. Surprisingly, the investigation revealed no significant differences from what is observed in \pp collisions. The widths, average values, and fractions of gluon-like jets, obtained through MC-based template fits, remained consistent with the reference \pp collision sample from peripheral to central \PbPb collisions. This stability is illustrated in Fig.~\ref{fig:jetCharge}, which displays the gluon-like jet fraction as a function of the track \pt threshold used in the measurement. Assuming that medium-induced effects remain charge-neutral on average and that gluon-initiated jets suffer more energy loss than quark-initiated jets on average, it could be expected that fewer gluon-like jets, as measured using the jet charge, might be observed in a sample at a given jet \pt when compared to a \pp reference. This result shows no evidence for such an effect and therefore calls for careful examination of possible selection biases in this (and other) measurements featuring quark or gluon tagging. Conversely, if no such biases are present, this observable may provide a way of enhancing the fraction of quark- or gluon-initiated jets in a sample on a statistical basis without being sensitive to medium effects. 

\begin{figure}[ht]
    \centering
    \includegraphics[width=\linewidth]{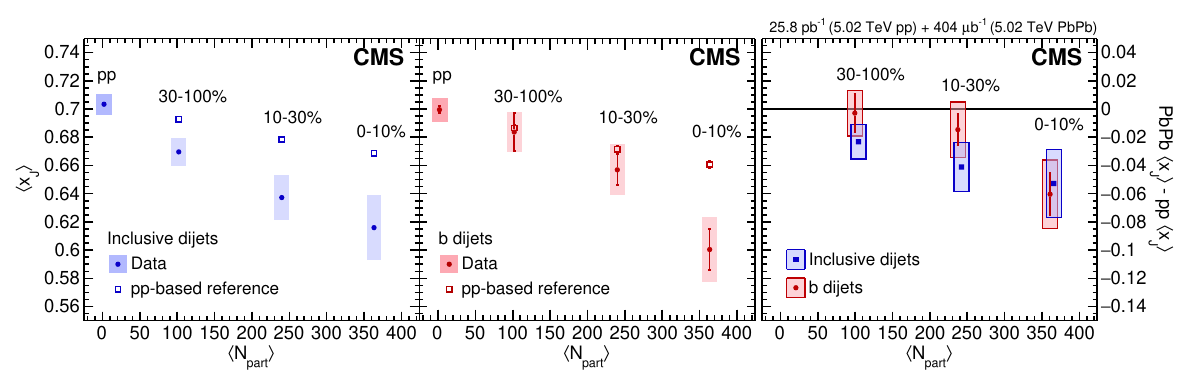}
    \caption{Dijet imbalance for inclusive (left) dijets and \cPqb dijets (center) in \pp collisions and for different centrality selections of 5.02\TeV \PbPb collisions. The right panel shows the difference in the \xjave values between \PbPb and the smeared \pp reference. Systematic uncertainties are shown as shaded boxes and statistical uncertainties are displayed as vertical lines. \FigureFrom{CMS:2018dqf}}
    \label{fig:bdijet}
\end{figure}

Heavy-flavor jets are rarer than light-flavor or gluon jets, but can provide a more unambiguous tag of the flavor of the parton that initiates a jet. In particular, a selection of dijets in which both jets are ``\cPqb tagged'', \ie, have a displaced secondary vertex or tracks displaced from the primary vertex that most likely results from the decay of a \PQb hadron, can heavily suppress gluon-initiated jet contributions. A measurement of the average \pt imbalance, \xjave, of \cPqb dijets~\cite{CMS:2018dqf} is shown in Fig.~\ref{fig:bdijet}, where \xj is defined as
\begin{linenomath}
\begin{equation}
    \xj = \ptTwo / \ptOne.
\end{equation} 
\end{linenomath}
The imbalance is found to increase as a function of collision centrality for both an inclusive dijet (left panel) and \cPqb dijet (middle panel) selection. The comparison of the two selections (right panel) reveals no significant differences between them, given the current measurement uncertainties. Although inclusive \cPqb jet measurements are expected to exhibit larger contributions from processes such as gluon splitting ($\Pg\to\PQb\PAQb$), when compared to \cPqb dijets, previous CMS measurements of the inclusive \cPqb jet \RAA in 2.76\TeV \PbPb collisions are consistent with the \RAA of inclusive jets~\cite{CMS:2013qak}. Both of these measurements, as well as the ATLAS \PGg and \cPqb jet measurements~\cite{ATLAS:2023iad,ATLAS:2022agz}, indicate that any potential differences between samples of light-quark or gluon initiated jets and \cPqb quark jets in the range $80<\pt<250\GeV$ are small when compared to the sensitivity of current measurements. Extending these measurements to lower jet \pt values, where flavor-dependent differences are expected to be larger but also where high rates of fake jets make the measurements difficult, may allow a clear distinction to be drawn between the dynamics of heavy-quark and light-quark or gluon initiated jets. 

\subsubsection{Prospects for measuring energy loss across various stages of QGP evolution} 
\label{sec:top_PbPb} 
 
All parton energy loss observables described so far are only sensitive to the properties of the QGP integrated over its lifetime of ${\sim}10^{-23}\unit{s}$, as the hard scattered partons are produced during the initial stages of the collision. In contrast, the top quark, the heaviest elementary particle known (and accessible in nucleus-nucleus collisions at the LHC), can provide complementary insights into the time structure of the QGP. Indeed, and as noted in Ref.~\cite{Apolinario:2017sob}, hadronically decaying \PW bosons, produced in top quark decays, provide a well calibrated ``time delay'' between the moment of the collision (when the top quarks are produced) and the moment when the \PW boson decay daughters start interacting with the QGP medium. The magnitude of the time delay can be determined by implementing a selection based on the reconstructed \pt of the top quark.
Using such a procedure, and the event samples expected to be collected by the end of the HL-LHC running period, we should be able to probe lifetime scales at the 1\,fm level.

\begin{figure}[ht]
    \centering
    \includegraphics[width=0.44\linewidth]{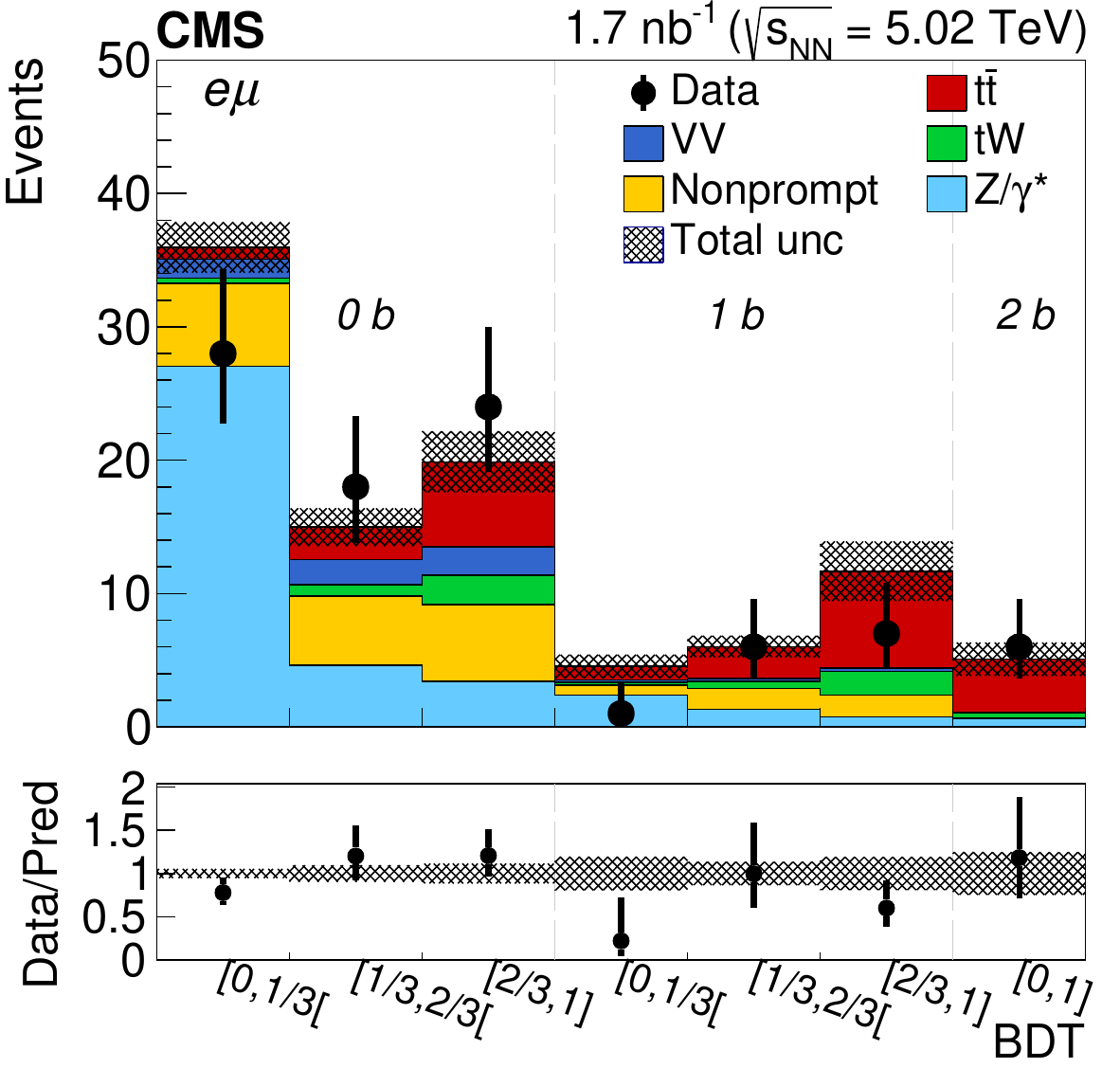}
    \includegraphics[width=0.553\linewidth]{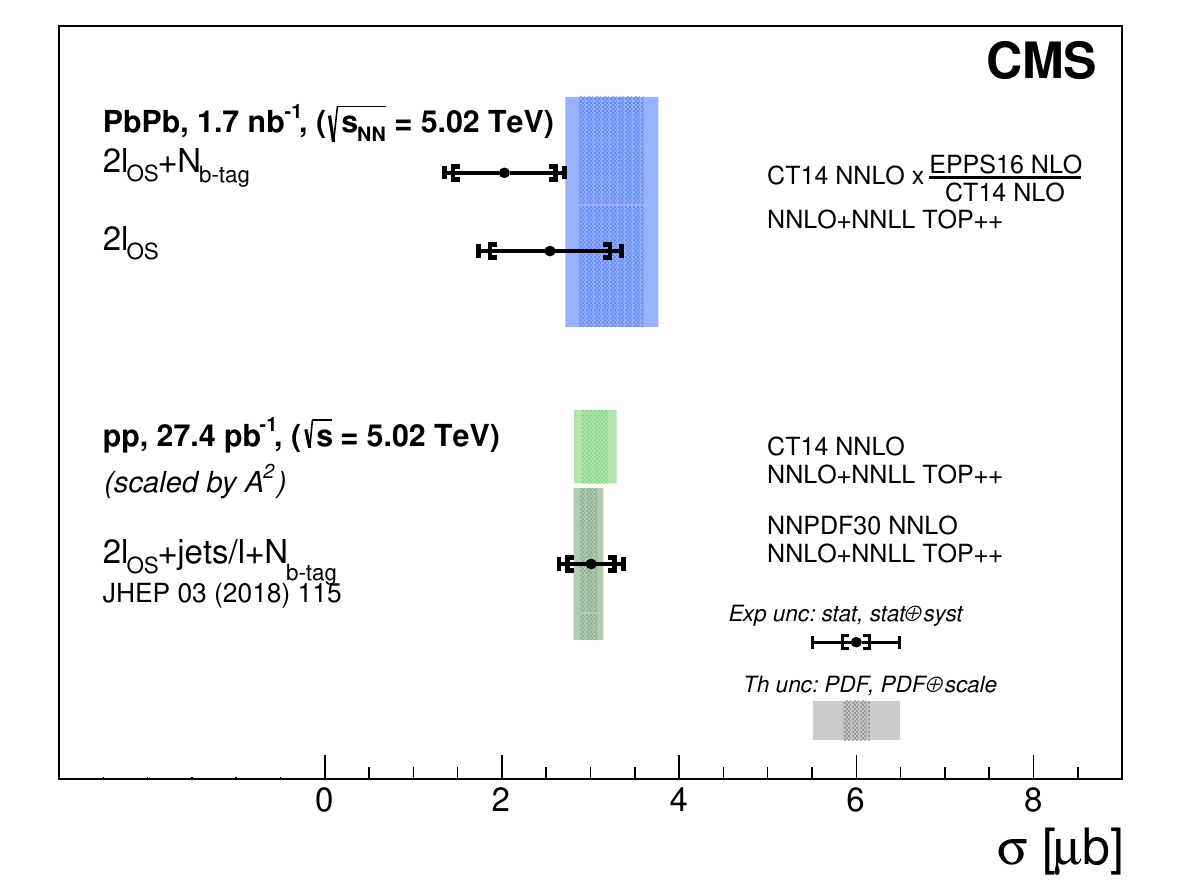}
    \caption{Left: Observed and postfit predicted BDT discriminator distributions in the $\Pepm\PGmmp$ final state separately in the 0\PQb-, 1\PQb-, and 2\PQb-tagged jet multiplicity categories.
    The data are shown with markers, and the signal and background processes with filled histograms.
    The vertical bars on the markers represent the statistical uncertainties in data. The hatched regions show the uncertainties in the sum of \ttbar\ signal and backgrounds.
    The lower panel displays the ratio between the data and the predictions, including the \ttbar\ signal, with bands representing the uncertainties in the postfit predictions.
    Right: Inclusive \ttbar cross sections measured with two methods in the combined $\Pepm\PGmmp$, \mumu, and $\Pep\Pem$ final states in \PbPb collisions at $\sqrtsNN=5.02\TeV$, and \pp results at $\roots=5.02\TeV$ (scaled by $A^2$). The measurements are compared with theoretical predictions at NNLO+NNLL accuracy in QCD. The inner (outer) experimental uncertainty bars include statistical (statistical and systematic, added in quadrature) uncertainties. The inner (outer) theoretical uncertainty bands correspond to nPDF or PDF (PDF and scale, added in quadrature) uncertainties. \FiguresAdaptedFrom{CMS:2020aem}}
    \label{fig:openhf_top}
\end{figure}

As an initial step toward conducting these types of investigations, CMS has presented the first evidence of top quark production in \AonA collision systems. This evidence is derived from 5.02\TeV \PbPb collision data~\cite{CMS:2020aem}. In particular, two methods are used to measure the cross section for top quark pair production via the decay into charged leptons (electrons or muons, leading to $\Pepm\Pgm^\mp$, \mumu, and $\Pep\Pem$ final states) and \cPqb quarks (separated into 0\PQb-, 1\PQb-, and 2\PQb-tagged jet multiplicity categories). One method relies on the leptonic information alone, while the other one exploits, in addition, the presence of \cPqb quarks. For both the dilepton-only and dilepton plus \cPqb-tagged jets methods, a boosted decision tree classifier is trained on the simulated \ttbar signal versus the overall $\cPZ/\PGg^\ast$ background (Fig.~\ref{fig:openhf_top}, left). This classifier is based exclusively on leptonic quantities to minimize effects from the imprecise knowledge of the jet properties in the HI environment. For both methods, the measured cross sections are compatible with, though somewhat lower than, the expectations from scaled \pp data and QCD predictions (Fig.~\ref{fig:openhf_top}, right). This measurement serves as a proof of concept for using the top quark as a novel probe, potentially enabling the investigation of energy loss and QGP dynamics at various stages during the temporal evolution of the system.

\subsection{Medium modifications to jet substructure and fragmentation}
\label{sec:Substructure}

Jets, as final-state multiscale composite objects initiated by hard-scattered partons, are characterized both by macroscopic quantities (such as their energy and direction) and by microscopic quantities that describe their internal structure. Therefore, the jet quenching phenomenon should be understood not only as a single overall medium-induced energy loss, but also as more detailed modifications to internal jet characteristics. Simultaneous measurements of jet properties at the macroscopic and microscopic scales can disentangle model scenarios with fundamentally different approaches to parton-medium interactions. These measurements should include characteristics such as the jet mass, angularity, and net charge, together with the longitudinal and transverse constituent distributions in the shower. Such an all-encompassing approach is necessary, realizing that using a jet to probe the medium is not equivalent to using a single parton that has a set (hard) perturbative scale. Rather, a jet is instead a continuous-scale dynamical process involving momentum exchanges with variable couplings. The following sections describe CMS studies of jet fragmentation functions (Section~\ref{ssec:Substructure_JetFF}), shapes (Section~\ref{ssec:Substructure_JetShapes}), and substructure (Section~\ref{ssec:Substructure_JetSubstructure}).

\subsubsection{Longitudinal structure of jets: fragmentation functions}
\label{ssec:Substructure_JetFF}

Implicit in the iterative recombination algorithms employed for jet identification is the notion of a jet constituent, which refers to the individual physics objects (such as PF or generator-level particles) clustered together within the resultant jet in the final state. One method to study the internal jet structure is to simply count the number of constituents within a jet according to their relative contribution to the overall jet energy,
\begin{linenomath}
\begin{equation}
\label{eq:jetFF}
    z = \frac{\pparatrk}{\pj}, \qquad \xi = \ln{\frac{1}{z}},
\end{equation}
\end{linenomath}
where \pparatrk is the track momentum projected onto the axis of the jet into which it is clustered and \pj is the momentum of the jet. The distribution of this final state constituent-by-constituent energy fraction for a given initiating parton energy, evaluated over an ensemble, is the fragmentation function. All tracks with $\Delta R$ less than the jet distance parameter $R = 0.3$ are used, where \dphi the difference in $\phi$ between track and jet and \deta is the corresponding difference in $\eta$. 

\begin{figure}[ht]
    \centering
    \includegraphics[width=0.95\linewidth]{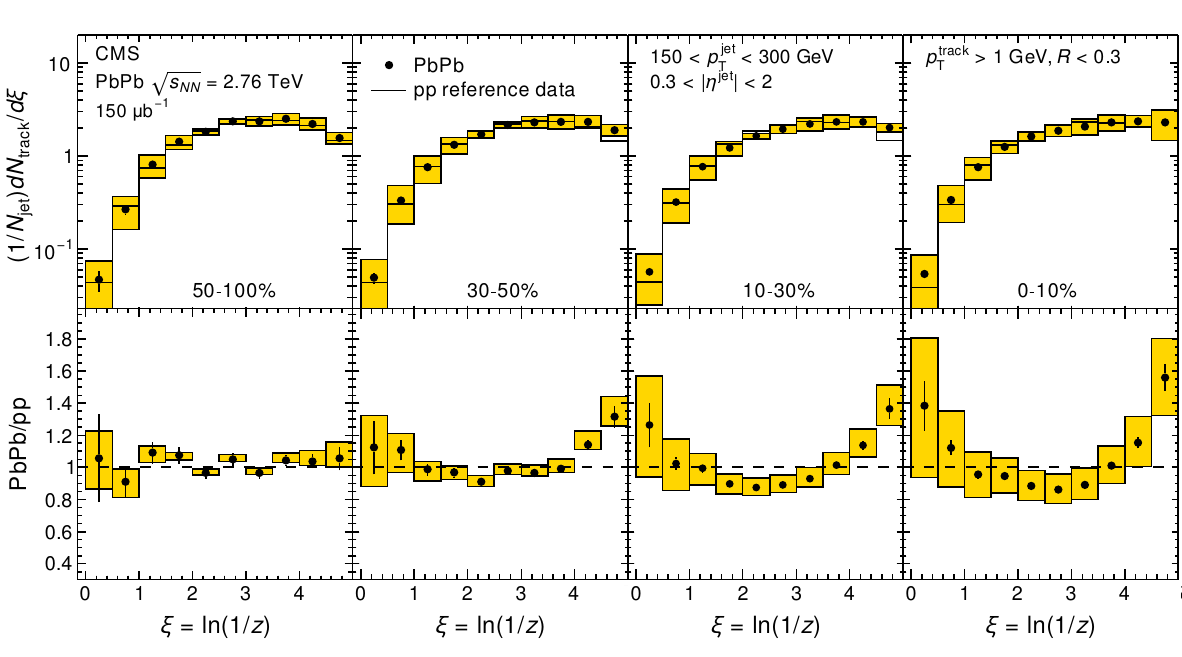}
    \caption{Upper: Fragmentation functions as a function of $\xi$ in bins of \PbPb centrality (left-to-right: 50--100\%, 30--50\%, 10--30\%, and 0--10\%) with the result from \pp reference data overlaid. Lower: Ratios of the \PbPb fragmentation functions over those for the \pp reference. Jets are selected in the \pt range 150 to 300\GeV and tracks with $\pt>1\GeV$. Vertical bars and shaded boxes represent the statistical and systematic uncertainties, respectively. \FigureAdaptedFrom{CMS:2014jjt}}
    \label{fig:PbPbJetFF}
\end{figure}

Figure~\ref{fig:PbPbJetFF} shows a high-precision measurement of inclusive jet fragmentation functions as the number of tracks in bins of $\xi$ normalized by bin width and per jet in 2.76\TeV \PbPb collisions, for a jet \pt selection of 150--300\GeV and track $\pt>1\GeV$~\cite{CMS:2014jjt}. The upper panels show fragmentation functions for \PbPb collisions, plotted in four centrality classes, along with their corresponding \pp reference data. The lower panels show the ratio between the two, which can be used to study possible medium-induced modifications of the fragmentation functions. In the leftmost panel, which corresponds to the most peripheral centrality selection of 50--100\%, the ratio displayed in the lower panel remains consistent with unity within one standard deviation after accounting for both statistical and systematic uncertainties. This suggests that there is no significant change to the fragmentation functions resulting from interactions between jets and the plasma for for these peripheral collisions. However, starting with the 30--50\% centrality class there is an observable enhancement in the high $\xi$ (or low track \pt) region. This is consistent with \mpt measurements, detailed in Section~\ref{ssec:InclusiveJetObservables_Dijets}, which show enhancement of soft particle production in the subleading (more quenched) jet direction. For the 10--30\% selection, in addition to the enhancement of low-\pt tracks, there is also an observable trend of depletion in the intermediate track \pt range ($\xi$ 1.5--3.0). The lowest $\xi$ point suggests an upward trend, but remains consistent with unity. For the most central events (0--10\% shown in the rightmost panels), the trends closely resemble those observed in the 10--30\% centrality range. The increased systematic uncertainty observed in this centrality range primarily stems from the significant correction needed for UE contributions.

The deviations observed in the fragmentation functions measured in \PbPb collisions relative to the reference \pp data can be attributed to several potential causes. First, it is expected from QCD that quarks and gluons have different color charge factors regulating the strength of parton-medium interactions. Consequently, it is likely that gluons will experience more energy loss while traversing the QGP than quarks. As quarks and gluons are known to have different fragmentation patterns in vacuum, any change in inclusive jet $\Pq/\Pg$ fraction as a result of quenching would cause changes to the fragmentation functions. In addition to this, parton-medium interactions can either modify the parton showers or induce a medium response that remains confined to the vicinity of the jet itself, consequently imitating an altered fragmentation pattern. Distinguishing between these three possibilities in a rigorous, systematic manner requires extensive study beyond the inclusive jet system.

This initial study measured inclusive jets in both \PbPb and \pp reference data at $\rootsNN = 2.76\TeV$ taken during Run 1. Similarly sized samples of \PGg-tagged jets in \PbPb and \pp collisions at $\rootsNN = 5.02\TeV$ were recorded during Run 2. As discussed in Section~\ref{ssec:StrengthOfELoss_AbsELoss}, one use of these samples is to explore the absolute energy loss of jets by using the photon energy as a proxy for the energy of the initial hard scattering. These events can also be used to study fragmentation functions for cases in which both the initial- and final-state energies of the jet are tagged. For the inclusive jet fragmentation functions shown in Fig.~\ref{fig:PbPbJetFF}, the \pt selection ($150 < \ptj < 300\GeV$) is applied to the jets in the final state, \ie, after quenching has changed the jet population substantially as compared to the \pp reference data. In contrast, \PGg-tagged jet samples avoid this bias by selecting on the \pt of the colorless photon. Using these tagged events, fragmentation functions can be extracted using both the traditional 
observable $\xi$ (renamed to \xijet in the results shown below) as well as a new observable \xigamma, defined as
\begin{linenomath}
\begin{equation}
\label{eq:gammaFF}
\xigamma = \ln{\frac{-\abs{\ptgvec}^2}{\pttrkvec \cdot \ptgvec}},
\end{equation}
\end{linenomath}
where \ptgvec and \pttrkvec are the transverse momentum vectors of the photons and tracks, respectively. The quantity \xigamma is similar to \xijet, except that the track momentum is now projected onto a direction opposite to that of the photon (presumably the unmodified jet direction). 

\begin{figure}[ht]
    \centering
    \includegraphics[width=0.65\linewidth]{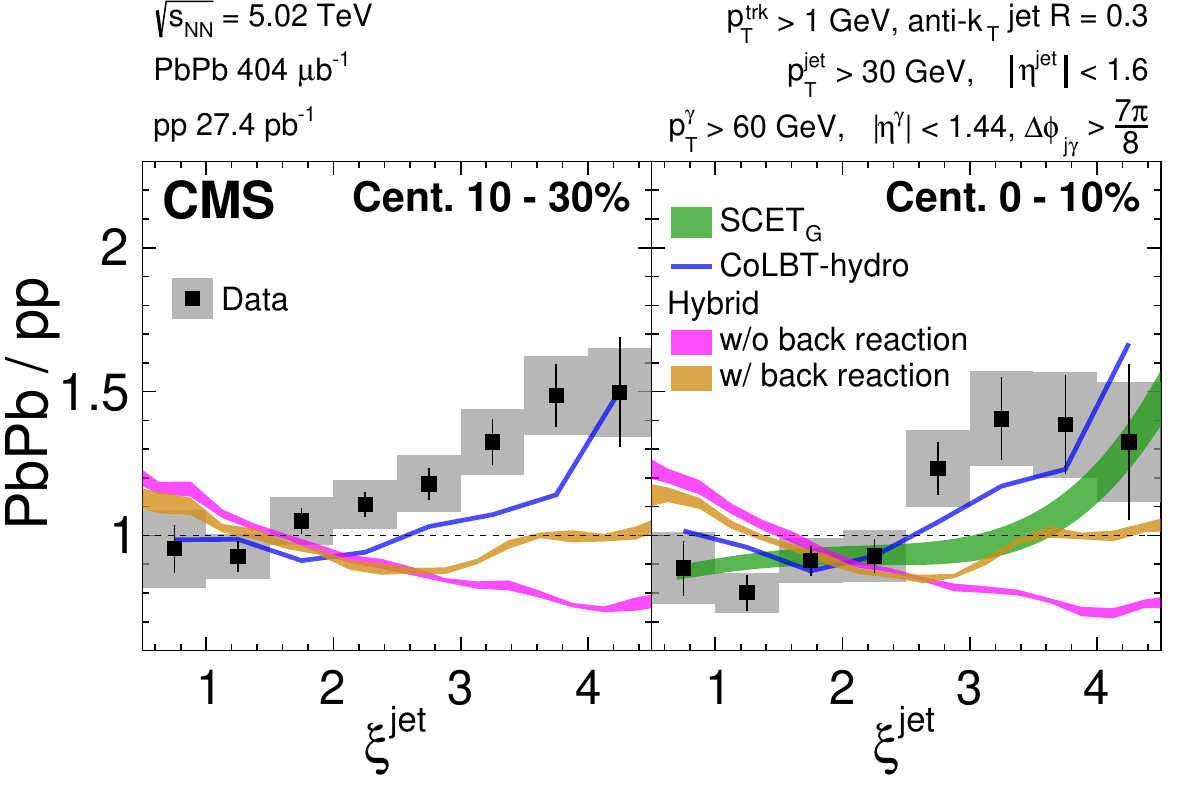}
    \includegraphics[width=0.65\linewidth]{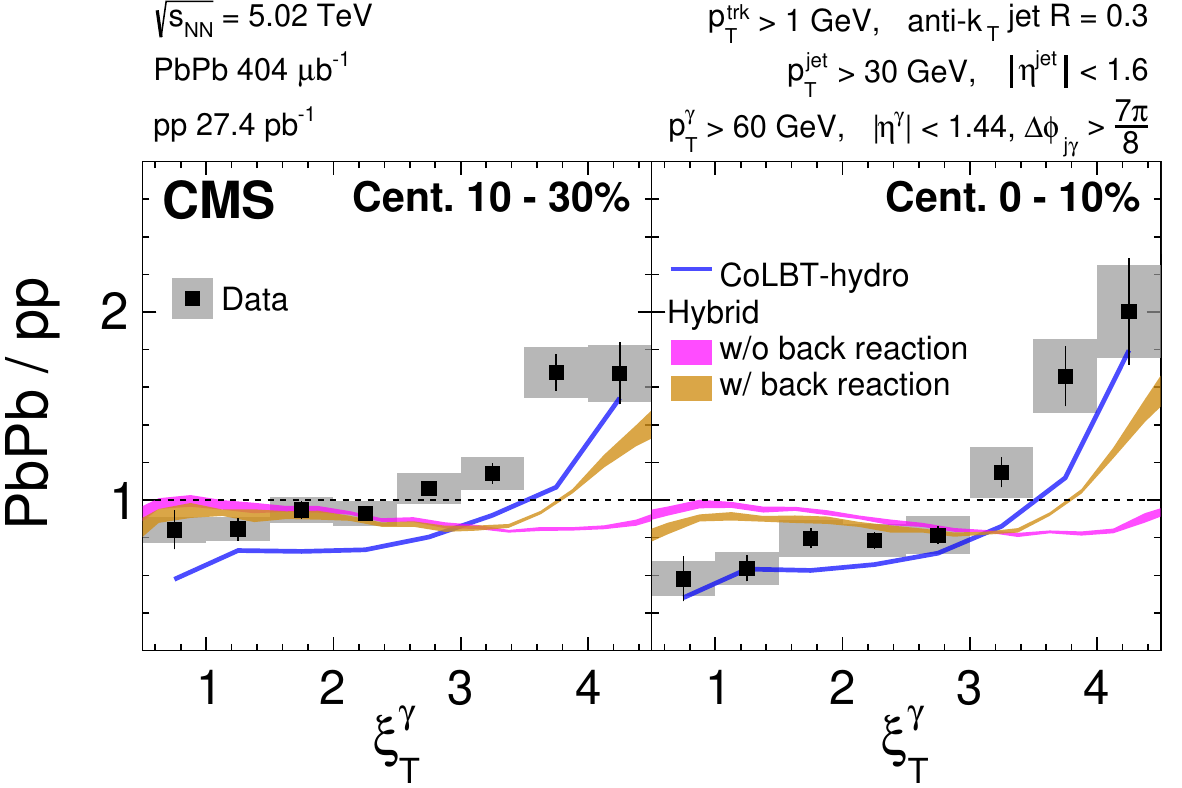}
    \caption{Comparison of \PGg-tagged fragmentation functions in centrality bins 10--30\% (left) and 0--10\% (right) as a function of the observables \xijet (upper), defined in Eq.~(\ref{eq:jetFF}), and \xigamma (lower), defined in Eq.~(\ref{eq:gammaFF}). For comparison, curves from the theoretical models SCETG~\cite{Chien:2015vja}, CoLBT-hydro~\cite{Wang:1996yh,Wang:1996pe,Wang:2013cia}, and \HYBRID~\cite{Casalderrey-Solana:2015vaa} are overlaid. The widths of the bands represent variations of the coupling strength in the SCETG case and of the dimensionless parameter $\kappa$ in the \HYBRID case. Vertical bars and shaded boxes represent the statistical and systematic uncertainties, respectively. \FiguresAdaptedFrom{CMS:2018mqn}} 
    \label{fig:PbPbPhotonJetFF}
\end{figure}

Figure~\ref{fig:PbPbPhotonJetFF} shows fragmentation functions in terms of both \xijet (upper panels) and \xigamma (lower panels) in \PGg-tagged jet events~\cite{CMS:2018mqn}. Results are presented for both semicentral (10--30\%, left panels) and central (0--10\%, right panels) events. The conventional fragmentation functions using \xijet display only a minor enhancement of soft particles alongside a corresponding reduction of hard particles (at large and small \xijet, respectively). In contrast, those using \xigamma 
demonstrate that once the influence of jet energy loss is eliminated, these effects become notably more pronounced, exhibiting a clear centrality dependence. Theory curves are plotted for the SCETG~\cite{Chien:2015vja}, CoLBT-hydro~\cite{Wang:1996yh,Wang:1996pe,Wang:2013cia}, and \HYBRID~\cite{Casalderrey-Solana:2015vaa} models. Of particular interest is the \HYBRID model, which only describes the data qualitatively well when incorporating the medium response (labeled \HYBRID with back reaction). This is an example of a model that can describe the energy loss of partons in the QGP without accounting for the response of the medium if we only consider what happens at the scale of the full jet but fail to do so at a smaller scale (at the jet substructure level), where the response of the medium needs to be taken into consideration. Nevertheless, this evidence for the existence of medium response remains indirect and model-dependent.

\subsubsection{Transverse structure of jets: jet shapes}
\label{ssec:Substructure_JetShapes}

In this section, a review of the measurements for the transverse shower profiles, known as ``jet shapes,'' is presented, complementing the CMS results on the modification of longitudinal jet substructure discussed in Section~\ref{ssec:Substructure_JetFF}. The same jet definitions and angular track associations are used as in the previous section. 

Measurements of the jet shape, $\rho(\Deltar)$, are obtained by identifying the jet constituents and examining the distribution of charged-particle tracks in rings around the jet axis, with each particle weighted by its corresponding track transverse momentum value \pttrk. Typically, shower particles are separated from those of the underlying event on a statistical basis. The transverse momentum profile \Rhodr of the jet is then defined as
\begin{linenomath}
\begin{equation}
\Rhodr = \frac{1}{\delta r} \frac{1}{N_{\text{jets}}} \Sigma_{\text{jets}} \Sigma_{\text{tracks} \in ( \Delta r_{\mathrm{a}}, \Delta r_{\mathrm{b}} )} \pttrk, \, \Deltar<1,
\end{equation}
\end{linenomath}
where $\Delta r_{\mathrm{a}}$ and $\Delta r_{\mathrm{b}}$ define the annular edges of \Deltar, and $\delta r = \Delta r_{\mathrm{b}} - \Delta r_{\mathrm{a}}$. 
The jet profile, which is normalized to unity within $\Deltar = 1$, is related to the \Rhodr distribution, with the jet shape definition
\begin{linenomath}
\begin{equation}
\rho(\Deltar) = \frac{\mathrm{P} (\Deltar)}{\Sigma_{\text{jets}} \Sigma_{\text{tracks}} \pttrk}.
\end{equation} 
\end{linenomath}
The $\rho(\Deltar)$ and \Rhodr distributions are sensitive to subsequent parton emissions by the initial hard-scattered parton. These distributions have been used historically in high-energy physics to provide robust tests of parton showering calculations in QCD. Together with calculations of hadronization and underlying event contributions, the $\rho(\Deltar)$ and \Rhodr distributions are also used for tuning MC event generators that account for parton showering effects. Jet shapes have been measured in elementary collisions ($\Pe\Pp$, \ppbar, \pp) at HERA~\cite{ZEUS:1998fts,H1:1995fzz}, the Tevatron~\cite{D0:1995nxp,CDF:1992cus}, and the LHC~\cite{ATLAS:2011bvd}.

In a HI collision environment, the jet shape measurements are particularly challenging given the high multiplicities encountered. Significant correlations with the underlying event lead to difficulties in differentiating the shower constituents from particles produced through other processes. The CMS Collaboration has addressed these issues in the first measurement of jet shapes using data from the LHC Run 1 \PbPb collisions at $\sqrtsNN =2.76\TeV$~\cite{CMS:2013lhm}. With this measurement, modifications of jet shower profiles while passing through the QGP medium were determined by comparing the measurements at different centralities of \PbPb data and \pp reference data. A clear modification of the in-cone ($\Deltar < 0.3$) jet constituent distributions was observed in the \PbPb data. As compared to the \pp reference, a greater fraction of the jet's transverse momentum is measured at a large \Deltar. This modification was shown to become stronger from peripheral to central collisions. 
However, these in-cone modifications were insufficient to explain the previously reported dijet momentum imbalance~\cite{CMS:2012ulu}, which does not account for a significant amount of the \mpt. Instead, the observed trends in the \PbPb to \pp jet shapes ratios suggested that the modifications are not limited to the small cone size used in the measurements. This cone size was chosen to have better control over the fluctuating background. Measurements of energy redistribution between the entire hemispheres of dijet-containing events~\cite{CMS:2011iwn,CMS:2015hkr} have shown that the energy flow is globally modified in HI events as a result of jet quenching and that the energy ``splash'' is felt at very large angles from the axis of a dijet.

Extending measurements of jet profiles to large angles was crucial to properly determine the dominant jet energy-loss mechanisms in various kinematic domains. The extended profiles are also needed to clearly establish the response of the QGP medium to the evolving jet. CMS has since performed a series of such studies~\cite{CMS:2016qnj, CMS:2016cvr, CMS:2018zze, CMS:2021nhn}, extending the angular range of jet constituent measurements with respect to a jet axis through a jet-track correlation technique. In this technique, tracks are classified by \pt and proximity to the jet axis in \deta and \dphi. They are also corrected for acceptance affects, tracking inefficiencies, uncorrelated backgrounds, and jet reconstruction biases. 

\begin{figure}[ht]
    \centering
    \includegraphics[width=0.95\linewidth]{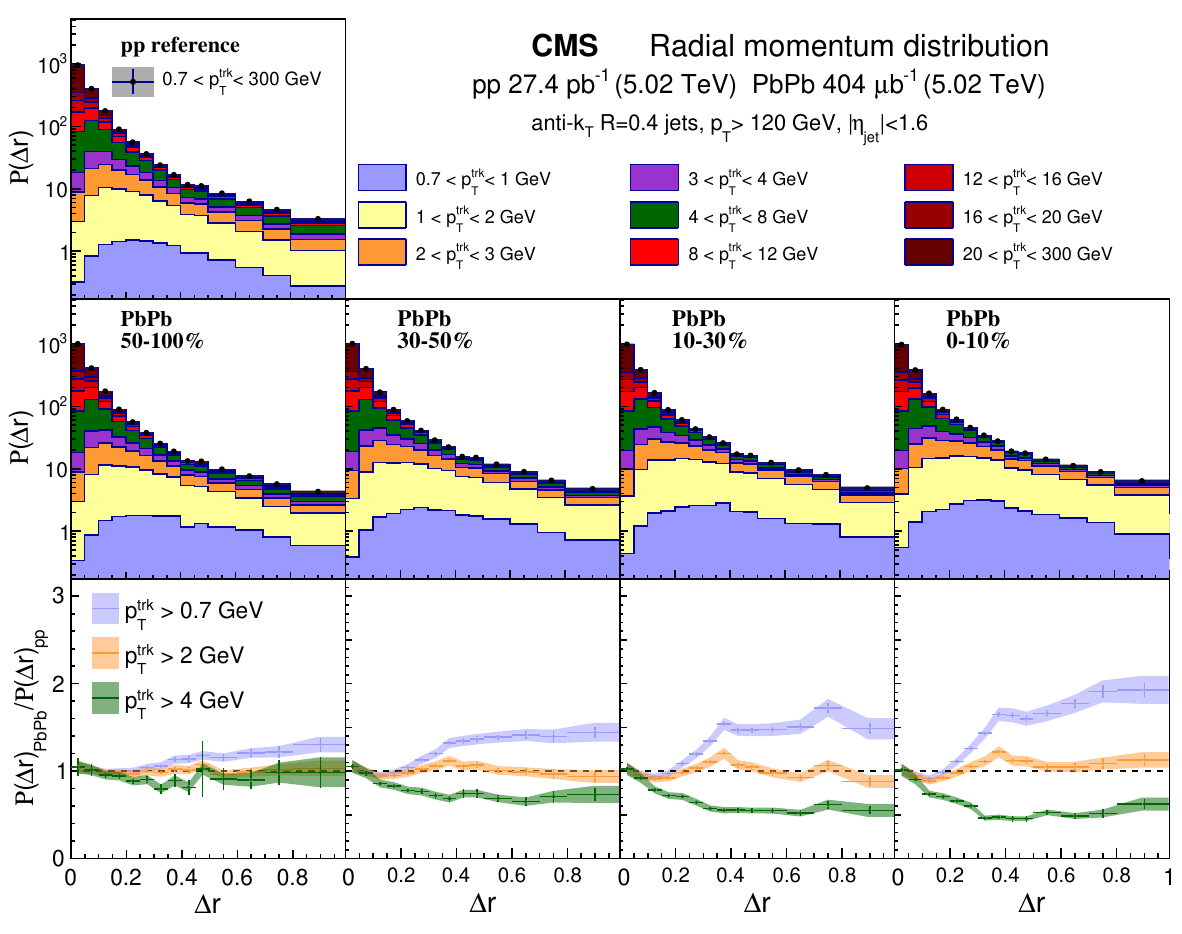}
    \caption{The angular jet momentum distribution \Rhodr of jets in \pp (upper) and \PbPb (middle) collisions. The \PbPb results are shown for different centrality regions. The lower row shows
the ratio between \PbPb and \pp data for the indicated intervals of \pttrk. The shaded bands show the
total systematic uncertainties. \FigureFrom{CMS:2018zze}}
    \label{fig:PbPbJetTrackCorrelations}
\end{figure}

The findings from these studies are illustrated in Fig.~\ref{fig:PbPbJetTrackCorrelations}, which shows measurements of transverse momentum profiles for inclusive jets from 5.02\TeV \pp and \PbPb collisions. The upper panel presents reference measurements using \pp data, detailing relative transverse momentum contributions of jet constituents at various distances from the jet axis. The second panel shows the same results for several centrality ranges of the \PbPb collision data. To facilitate the comparison, the lower panel shows ratios between \PbPb and \pp results for the indicated \pttrk intervals. The CMS jet shape measurements consistently demonstrate a redistribution of the jet energy inside and outside typical cone sizes. A significant excess of soft particles in \PbPb events relative to \pp events at intermediate to large angles from the jet axis is seen most prominently in central collisions, compensated for by a relative depletion at all track angles at high \pttrk. 

This two-sided modification has been argued to result from a combination of jet quenching in the medium and the medium response (or back reaction) to the propagating jet. The details of these interpretations remain nontrivial. For example, the narrowing of the hard "core" of the jet could be an artifact of a selection bias. At the same reconstructed momentum, a jet from a \PbPb collision is more likely to have originated from a higher initial energy parton than in a \pp event, where fragmentation occurs in vacuum. Alternatively, the hardening could arise from a selection bias caused by having a higher fraction of quark-initiated jets in the HI sample because of the expected color-charge effects on the energy loss~\cite{Chien:2015hda, Brewer:2017fqy}. Regarding the broadening or enhancement of the soft components within the jets, although models exhibit variations in the details of the jet-medium interactions, it has become evident that only models integrating medium feedback can replicate the significant excesses at very large \Deltar in the momentum profiles, particularly the substantial low-\pttrk excesses~\cite{KunnawalkamElayavalli:2017hxo,Casalderrey-Solana:2016jvj}. 

\begin{figure}[ht]
    \centering
    \includegraphics[width=0.5\linewidth]{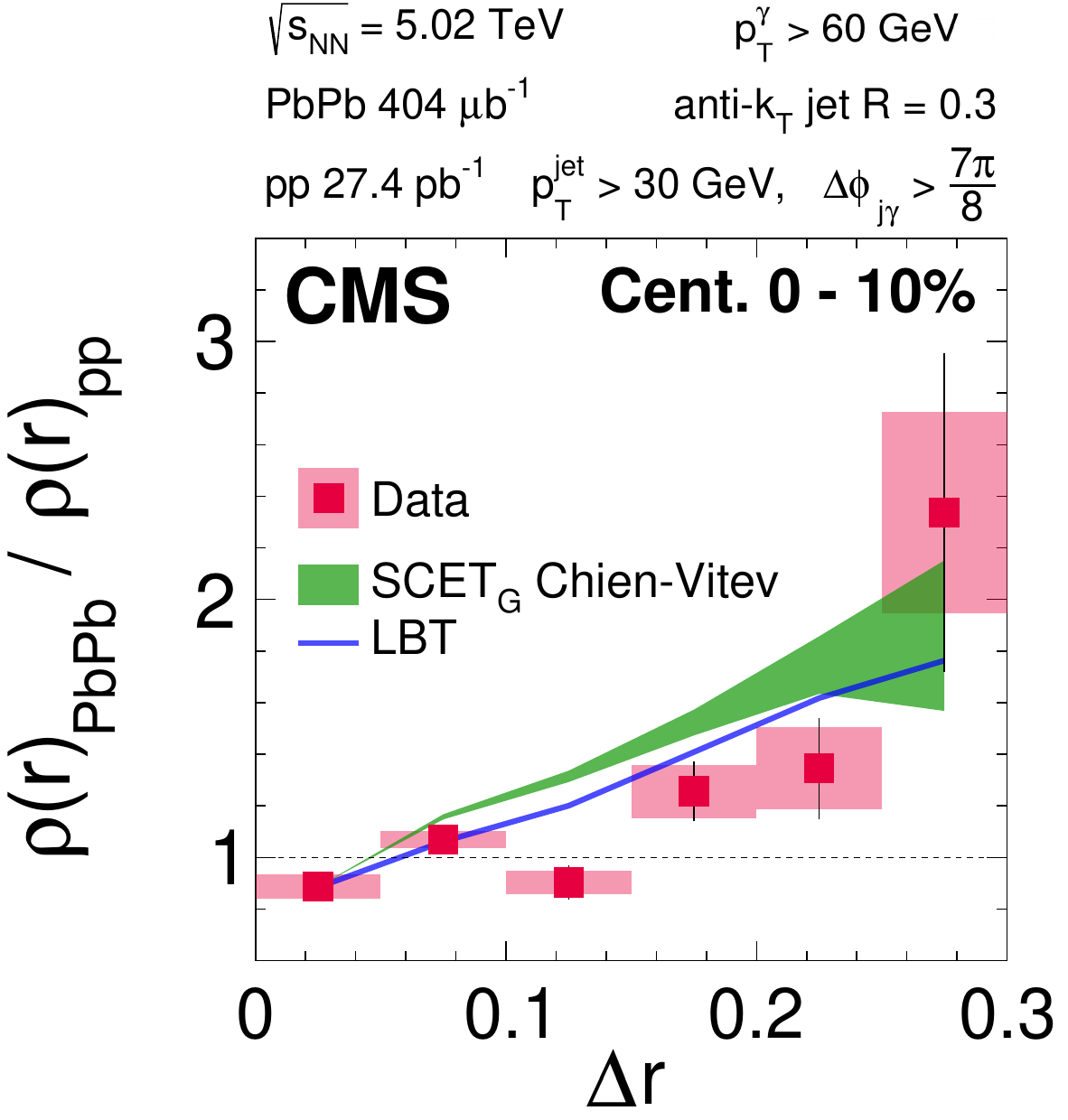}
    \caption{Ratio of the differential jet shape for jets associated with an isolated photon for 5.02\TeV 0--10\% \PbPb collisions and \pp reference data. The measurement is performed using jets having $\ptj > 30 \GeV$ and tracks with $\pttrk > 1 \GeV$. \FigureAdaptedFrom{CMS:2018jco}}
    \label{fig:PhotonJetShape}
\end{figure}

It is important to understand if the observed modifications are induced by the presence of the quark-gluon plasma or by a selection bias. 
CMS has explored a variety of experimental means to constrain possible selection biases in jet quenching studies. As discussed in Section~\ref{ssec:StrengthOfELoss_AbsELoss}, some biases can be alleviated by using jets produced in conjunction with an EW probe, such as a photon or \cPZ boson. Photon-jet events have substantially lower cross sections than inclusive jet production, and the reconstruction of an \textit{isolated} photon (as discussed in Section~\ref{ssec:ExperimentalMethods_UE_Pho}) required for such studies is experimentally challenging. 
The CMS Collaboration has successfully measured the fragmentation functions (as discussed in Section~\ref{ssec:Substructure_JetFF}) and the jet shapes for photon-jet events at 5.02\TeV in \pp and \PbPb collisions. Figure~\ref{fig:PhotonJetShape} illustrates the modifications of the differential jet shape for jets associated with an isolated photon found in the 10\% most central \PbPb events~\cite{CMS:2018jco}. The momentum carried by charged particles with $\pttrk > 1 \GeV$ for jets with $\ptj > 30 \GeV$, tagged by an isolated photon of $\ptg > 60 \GeV$, is redistributed towards larger angular distances from the jet axis. Minimal, if any, suppression is seen at the low/intermediate angles, where it stood more prominently for inclusive jet results. Similar measurements of events producing high-\pt hadrons in association with a \cPZ boson have been used to study this redistribution over an ever larger angular range~\cite{CMS:2021otx}. In central events, the redistribution of momentum carried by particles with $\pttrk > 1 \GeV$ appears to extend somewhat uniformly across nearly the full range of azimuthal angles with respect to the \cPZ boson. These boson-tagged measurements on one side establish unambiguously the broadening of the shape of the corresponding jets. Yet, the narrowing of the jet core is not strongly evident in the measurement. This could be explained by the intrinsically higher quark content of a boson-tagged jet sample as compared to the inclusive jet sample. Alternatively, the differences in the steepness of the transverse profile, which complicate the direct comparison of the jet shape ratio between different samples, could potentially explain the deviations in boson-tagged and inclusive jet samples. 

As mentioned, selection bias is one of the main difficulties for interpreting the existing experimental results. ``Survival bias'', another form of selection bias, is also a significant issue for comparing the \PbPb and \pp results. This bias is an unavoidable consequence of jet quenching itself: selected with any given momentum threshold, any sample of HI collision jets will always be biased towards those least quenched. To ``dig deeper'' into the medium, a set of jet shape measurements has been performed for dijets using the dijet momentum asymmetry as a quenching control parameter~\cite{CMS:2021nhn}. This study found that both sides of a dijet (the leading and subleading) appear to be modified through interactions with the QGP medium. Even for the most asymmetric scenario considered, the leading jets shape was modified compared to a reference from \pp collisions. Overall, the jet shape distributions for leading jets are the widest for events with balanced jet momenta. A possible interpretation, consistent with these observations, is that the leading jets traversing the largest average in-medium path lengths come from momentum-balanced events. In contrast, for subleading jets, the widest distributions were observed for the most unbalanced dijets. The relatively wider subleading jet shape distributions can be explained by this relationship between path length and survival bias. Alternative explanations are subject to active developments, with several recent works indicating that the dijet momentum asymmetry, dijet jet shape modifications, and even high-\pt $v_2$, as discussed in Section~\ref{ssec:StrengthOfELoss_PathLength}, could also result, at least in part, from energy loss fluctuations~\cite{Milhano:2015mng, Brewer:2018mpk}.  

\subsubsection{Parton-level substructure}
\label{ssec:Substructure_JetSubstructure}

A key feature of jet substructure is that the underlying physics can be factorized into a convolution of a pQCD-like probabilistic parton shower and non-pQCD effects including hadronization. This results in jets with a variety of topologies. In particular, jet substructure might be different in vacuum compared to jets in the QGP, where one expects a range of parton energy-loss effects. Variations in energy loss resulting from the possible impact of the in-medium path length and from fluctuations can result in jet modifications. The use of advanced algorithms that enable us to reconstruct the partonic structure of a jet can help in untangling these various processes, which is crucial for understanding the dependence of the jet topology on energy loss. The clustering tree of a reconstructed jet can be analyzed in more detail to get information about the parton splitting history. Recently, this sort of analysis has grown into an active area of research in both the experimental and theoretical regimes. 

Such studies emphasize the importance of quantifying the energy loss of jets as a function of their momentum and angular scales. Observables related to jet substructure typically start with the constituents of the jet (tracks or calorimeter towers in an experiment, particles in a theoretical study) originally found via the anti-\kt algorithm, followed by a reclustering with an alternative algorithm. The most common reclustering uses the Cambridge--Aachen algorithm, which enforces angular ordering in the pairing. Thus, in vacuum, one can directly associate specifically identified hard splittings found at later stages of the reclustering with wide-angle and early-time emissions in the parton shower. The process by which a particular hard splitting is selected is termed ``grooming'', which iteratively proceeds backwards through the clustering tree (\ie, declusters the jet) and requires some criterion to select a specific splitting. From a phenomenological standpoint, observables found using these groomed jets have a reduced contribution from soft, wide-angle radiation, particles from the underlying event, and multiparton interactions, all of which are theoretically more challenging to describe. Thus, we can compare the data more accurately with theoretical calculations. 

\begin{figure}[b]
    \centering
    \includegraphics[width=0.95\linewidth]{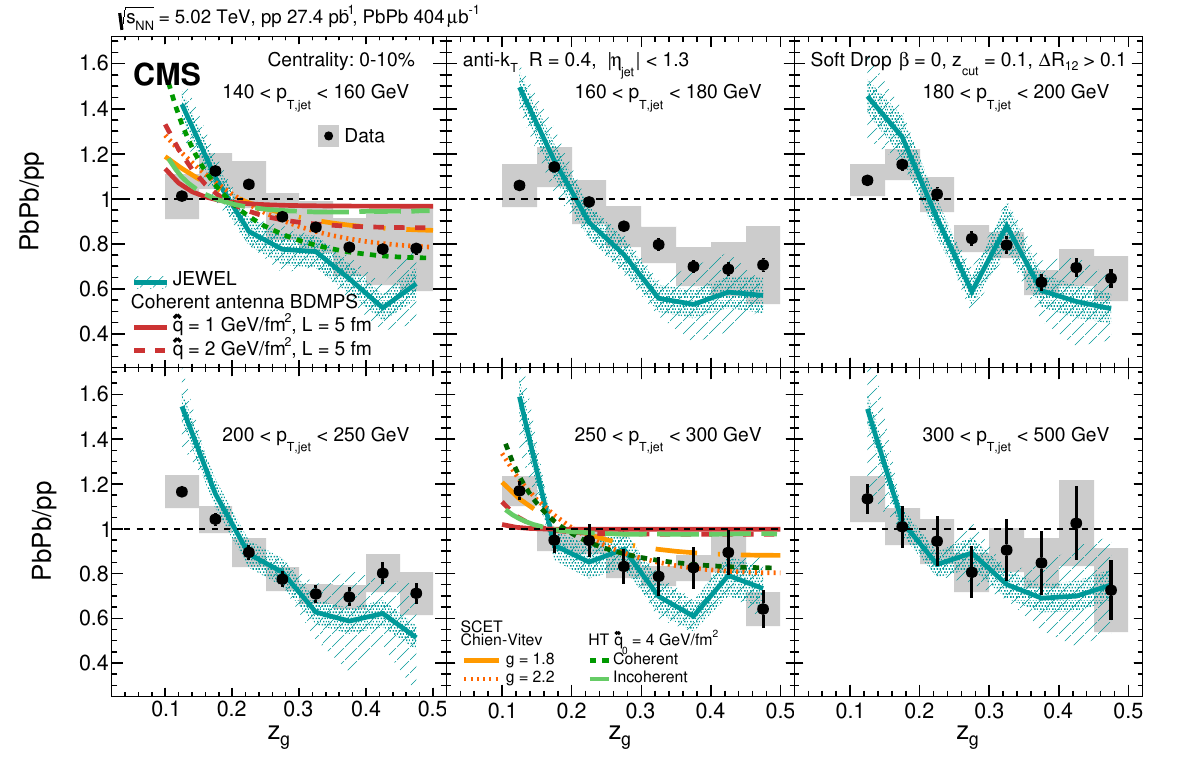}
    \caption{Distribution of the ratio of the groomed jet splitting fraction in central \PbPb data compared to a \pp reference. Each panel corresponds to a different jet \pt range and the different colored lines and bands are predictions from MC models. Statistical and systematic uncertainties in the data are shown by vertical bars and shaded boxes, respectively. \FigureFrom{CMS:2017qlm}}
    \label{fig:JetSplittingFunction}
\end{figure}

The first algorithm used to measure the groomed jet substructure in HI collisions was the soft-drop~\cite{Larkoski:2014wba} groomed momentum sharing, which uses the observable $z_{\mathrm{g}}$, defined as the momentum fraction of the first soft branch in the declustering satisfying the condition,
\begin{linenomath}
\begin{equation}
    z_{\mathrm{g}} = \frac{\mathrm{min}(p_{\mathrm{T},1},p_{\mathrm{T},2})}{p_{\mathrm{T},1} + p_{\mathrm{T},2}} > z_{\text{cut}} \Big(\frac{R_{12}}{R}\Big)^{\beta},
\end{equation}
\end{linenomath}
where the indices~1 and~2 identify the two branches in the clustering tree, and $z_{\text{cut}}$ and $\beta$ are the soft-drop parameters. The quantity $R_{12}$ is the angle between the two branches and $R$ is the jet distance parameter. The groomed jet distance parameter $R_{\mathrm{g}}$ is defined similarly as the angle between two such branches. For this very first HI study, $\beta$ was set to zero and $z_{\text{cut}}$ was set to 0.1, meaning that only the momentum fraction was required to select the first hard branch. Using data gathered in 2015 by the CMS experiment, this analysis involved \PbPb and \pp data samples, both obtained at $\sqrtsNN=5.02\TeV$~\cite{CMS:2017qlm}. The measured ratio of the $z_{\mathrm{g}}$ distribution in the most central \PbPb collisions compared to a \pp reference is shown in Fig.~\ref{fig:JetSplittingFunction}, where each panel shows a different jet \pt selection. Compared to expectations based on \pp data, the \PbPb data show a marked suppression for jets with a larger $z_{\mathrm{g}}\approx0.5$ and an enhancement at smaller values, $z_{\mathrm{g}}\approx0.1$. The ratios are also compared to different calculations and MC models, which are found to qualitatively describe the trend of the suppression. The enhancement of jets with more asymmetric splittings is consistent with expectations from jet energy loss, where jets with a single hard cluster are found more often than jets where the energy is more equally shared among multiple subclusters. 

\begin{figure}[ht]
    \centering
    \includegraphics[width=0.45\linewidth]{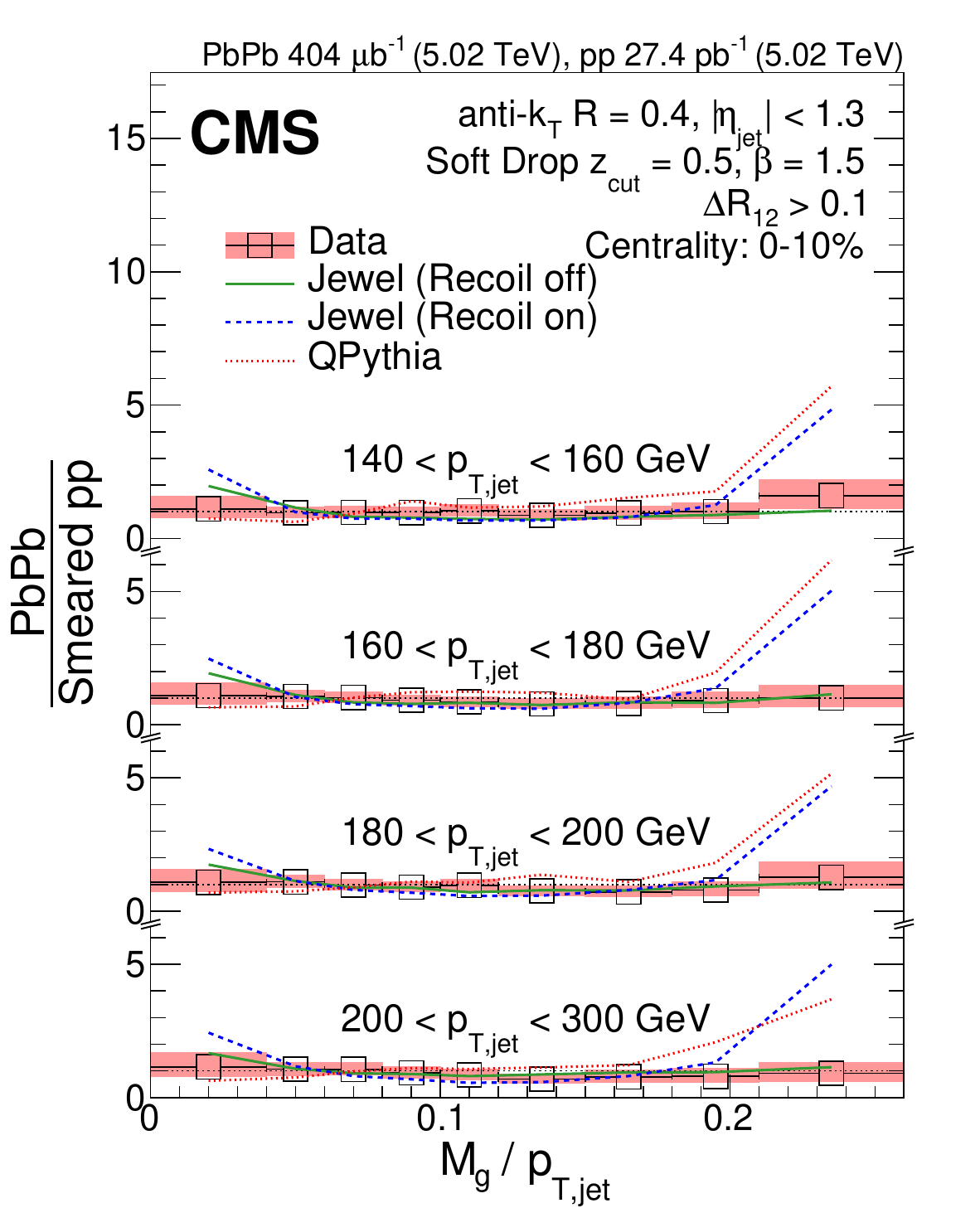}
    \includegraphics[width=0.45\linewidth]{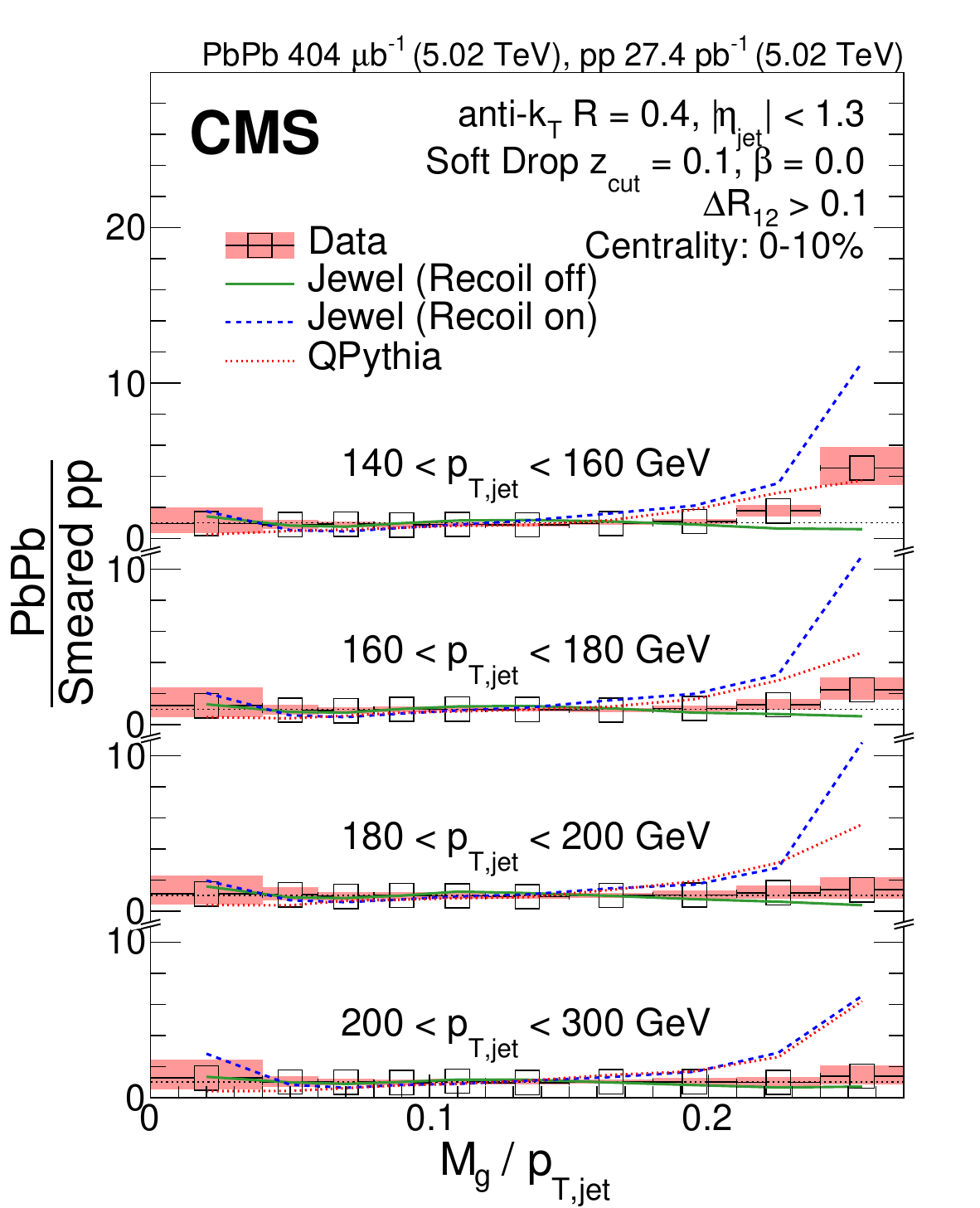}
    \caption{Distribution of the ratio of the groomed jet mass, $M_{\mathrm{g}}$, in central \PbPb data compared to the \pp reference for two different grooming criteria in four ranges of jet \pt. The left panel shows more stringent grooming criteria, while the right panel shows the same measurement for the default grooming requirements. The different lines represent MC predictions; they show deviations from the data at larger masses. Statistical and systematic uncertainties in the data are shown by vertical bars and shaded boxes, respectively. \FigureCompiledSingular{CMS:2018fof}}
    \label{fig:GroomedJetMass}
\end{figure}

The next study of jet substructure performed using CMS data varied the grooming procedure by employing a stronger angular-dependent grooming ($z_{\text{cut}} = 0.5$, $\beta = 1.5$) compared to the softer grooming previously used ($z_{\text{cut}} = 0.1$, $\beta = 0$)~\cite{CMS:2018fof}. Figure~\ref{fig:GroomedJetMass} shows the differential measurement of the groomed jet mass, $M_{\mathrm{g}}$, defined as the invariant mass of the system consisting of the two subjets (normalized by the jet \pt) for \PbPb data compared to a \pp reference, for jets in different ranges of \pt. The $M_{\mathrm{g}}$ variable is sensitive to both the parton splitting function and the opening angle between the two outgoing partons. The results are compared to the predictions of the \JEWEL and \textsc{qpythia} event generators. In the case of stronger grooming, no significant modification of the groomed jet mass is observed; however, there is a slight indication of enhancement for jets with larger masses in the context of weaker grooming. The indication of enhancement appears for configurations where the opening angle between the two subjets is large and one subjet has significantly more momentum than the other. The MC predictions qualitatively follow the trends in the data but significantly overestimate the enhancement effect. Note that, in contrast to what was observed for $z_{\mathrm{g}}$, any possibility of small modifications is only present for lower-\pt jets, and essentially disappears for jets of higher \pt. This is significant since the mass is an observable which convolutes both the momentum and angular scales, with sensitivity to the virtuality, where competing effects from both scales affect the distribution. 

\subsection{Studying wavelength behavior by varying parton flavor}
\label{sec:FlavorAndMass}

As discussed in previous sections, high-\pt jets are used to investigate the quenching of energetic partons traversing the medium, and varying flavors of the probe partons enables the study of the processes that dominate in different kinematic regions.
Because their mass is larger than the QCD perturbative scale, $m_{\mathrm{HQ}}\gg\Lambda_{\mathrm{QCD}}$, as well as the typical temperature reached in HI experiments, $m_{\mathrm{HQ}}\gg T_{\mathrm{QGP}}$, heavy quarks (charm and bottom) are mainly produced by hard scatterings. This feature provides sensitivity to transport properties of the QGP across a broad \pt range, and the comprehensive coverage of phase space offers distinctive insights into various structures within the QCD medium. In this section, the role of parton flavors, their dominant processes in different \pt regions, and the corresponding implications on the nature of the QGP are discussed. 

\subsubsection{Mass dependence of energy loss}

From intermediate to high \pt, within the framework of perturbative QCD, heavy quarks experience considerable energy loss through gluon radiation, similar to the situation observed for high-\pt light quarks; however, the magnitude of this effect is expected to vary depending on the quark mass. First, gluons have larger color charge than quarks and are therefore expected to experience stronger energy loss. In addition, the emission of gluons is predicted to be suppressed inside an angle proportional to the ratio of the quark mass to its energy, denoted as $m/E$~\cite{Dokshitzer:1991fd}. The color-charge effect and the ``dead-cone'' effect lead to a hierarchy of parton energy loss, where $\Delta E_{\Pg} > \Delta E_{\ell} > \Delta E_{\PQc} > \Delta E_{\PQb}$. This flavor dependence of energy loss is studied by comparing \RAA of hadrons containing light and heavy quarks. 

\begin{figure}[ht]
    \centering
    \includegraphics[width=0.55\linewidth]{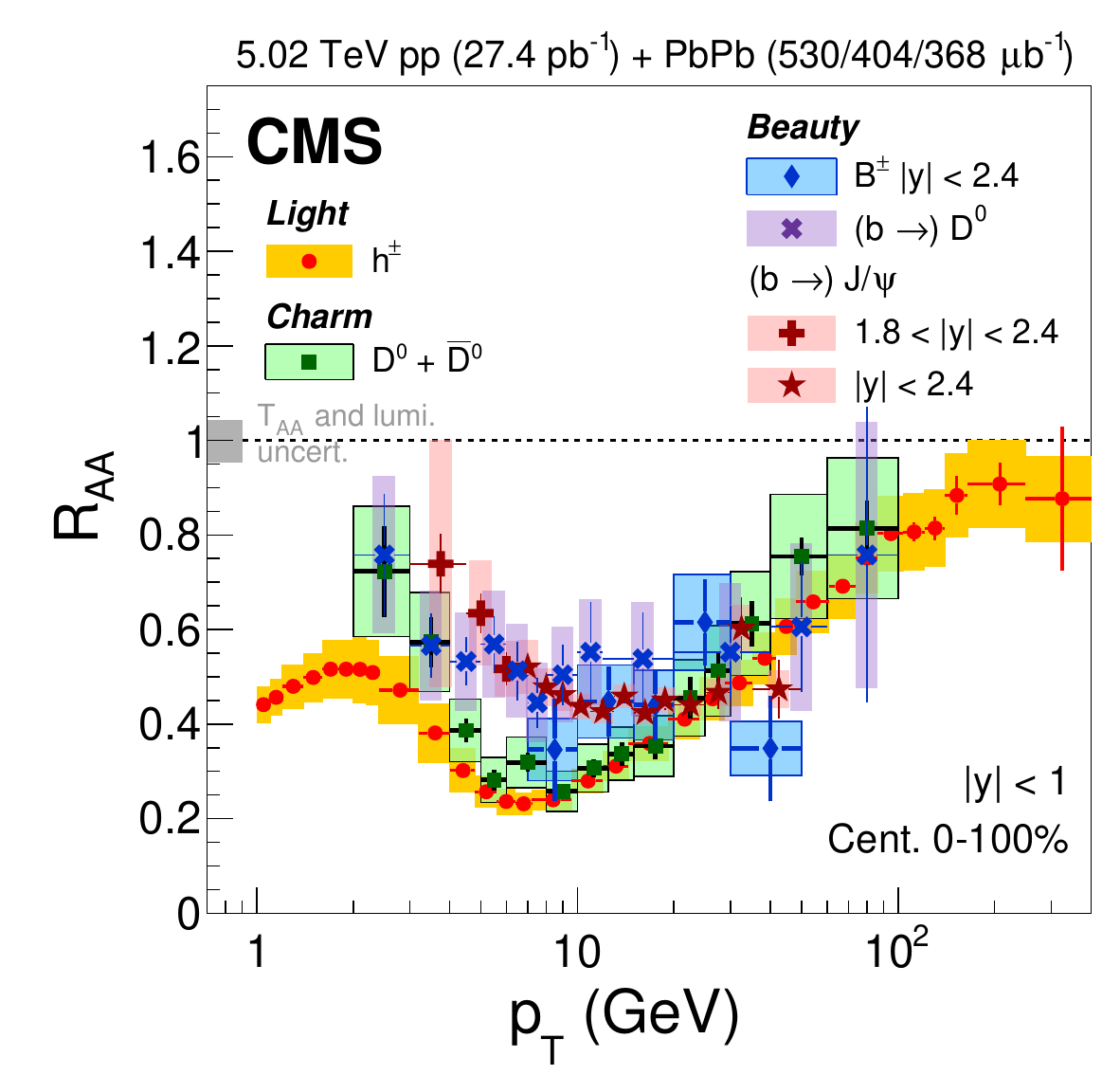}
    \caption{Nuclear modification factors of inclusive charged particles, prompt \PDz and \PBp, and nonprompt \PDz and \JPsi mesons, as a function of their \pt in \PbPb collisions. \FigureCompiled{CMS:2016xef,CMS:2017qjw,CMS:2017uoy,CMS:2018bwt,CMS:2017uuv}}
    \label{fig:openhf_raa}
\end{figure}

\begin{figure}[t]
    \centering
    \raisebox{-1.9ex}{\includegraphics[width=0.525\linewidth]{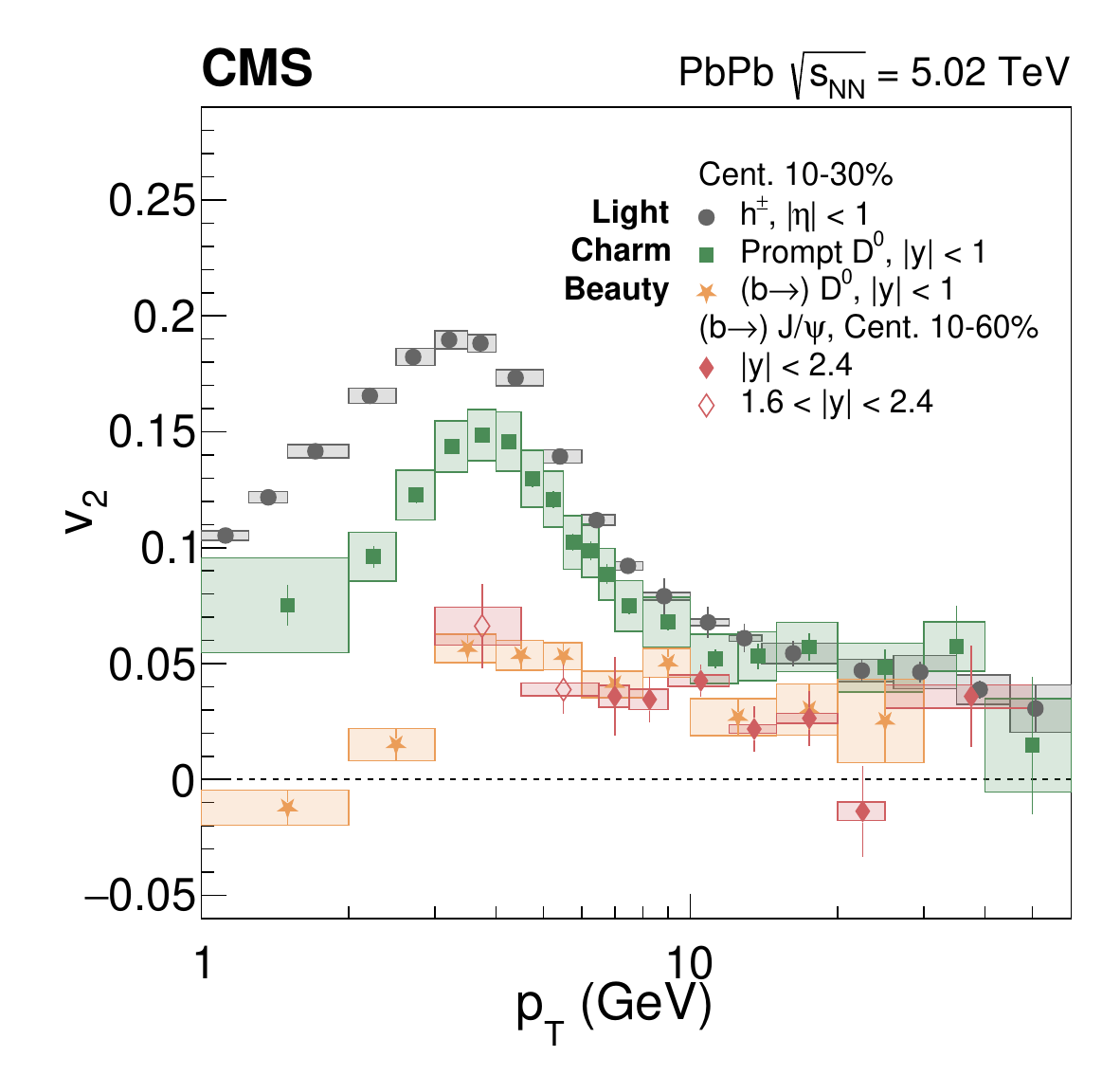}}
    \raisebox{1.4ex}{\includegraphics[width=0.455\linewidth]{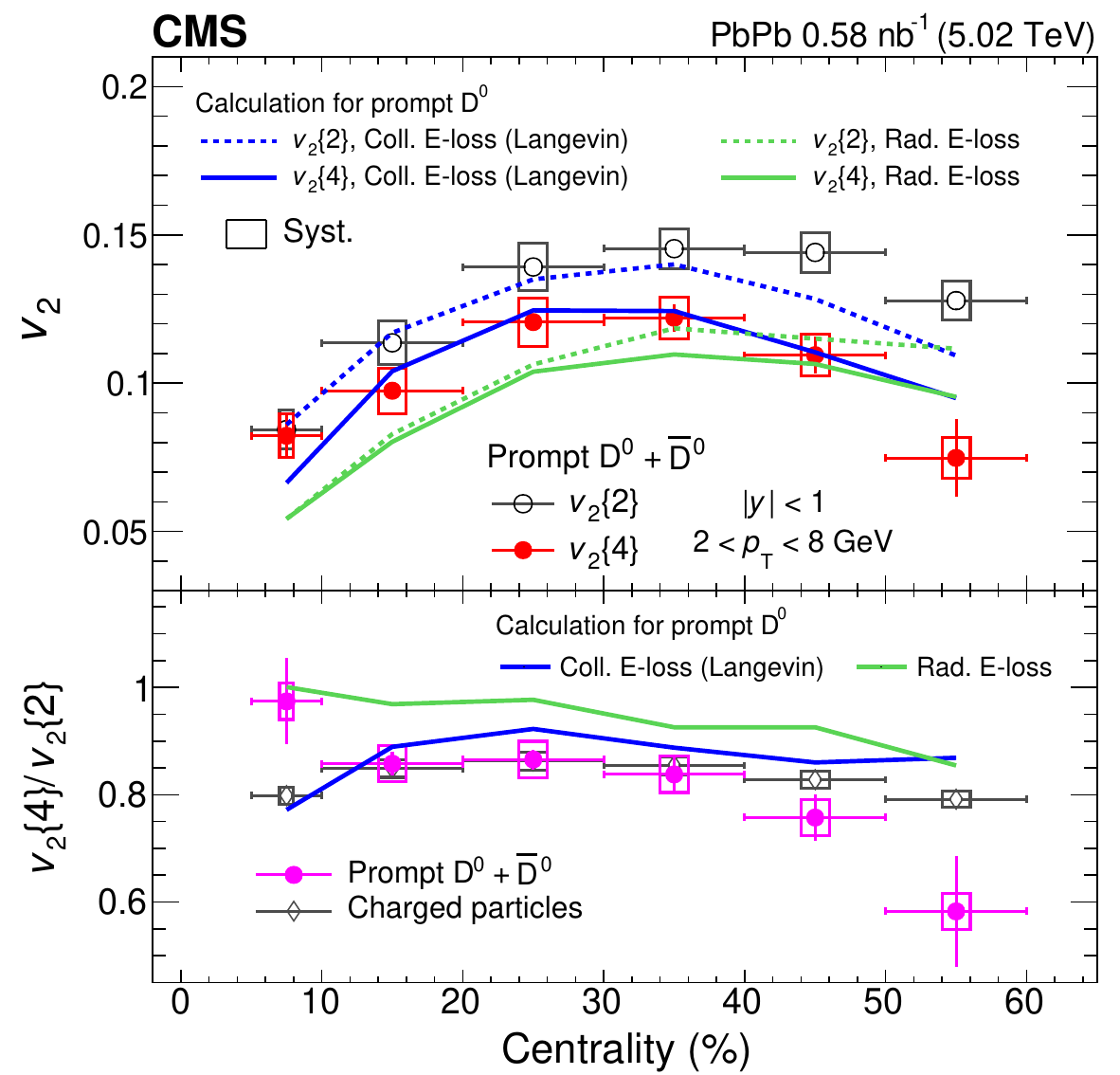}}
	\caption{Left: Azimuthal anisotropy coefficient \vTwo of inclusive charged particles, prompt \PDz, and nonprompt \PDz and \JPsi mesons as a function of \pt in \PbPb collisions. Right: Prompt \PDz meson \vtwo{2}, \vtwo{4} and their ratio as functions of centrality. 
 \FigureCompiled{CMS:2017xgk,CMS:2021qqk,CMS:2020bnz,CMS:2022vfn,CMS:2023mtk}}
    \label{fig:openhf_v2}
\end{figure}

As shown in Fig.~\ref{fig:openhf_raa}, the nuclear modification factor~(\RAA) of charged particles, \PDz mesons, and \PBp mesons, as well as \PDz and \JPsi mesons originating from \PQB hadron decays (nonprompt), have been measured by CMS. For $\pt > 20\GeV$, the \RAA values for all particle species are similar and noticeably smaller than unity. This result suggests that both light and heavy quarks are suppressed within the QGP, and the impact of quark mass becomes less pronounced when \pt greatly exceeds the parton mass. This behavior aligns with the expected outcome in the context of radiative energy loss. 
As shown in the left panel of Fig.~\ref{fig:openhf_v2}, the \vTwo values of hadrons with different flavors at high \pt are also similar. One possible interpretation is that the diminishing effect of the parton mass on energy loss at high \pt affects not only the overall magnitude of inclusive suppression but also the dependence on path length. However, in the intermediate-\pt range, the \RAA of nonprompt \JPsi mesons is notably larger than that of \PDz mesons and charged particles, implying a less pronounced suppression of bottom quarks compared to charm and light quarks.

\subsubsection{Elastic energy loss and diffusion}

In addition to gluon radiation, partons can also lose energy through elastic collisions with medium partons. Radiative and collisional energy loss effects dominate at high and low \pt, respectively, while intermediate-\pt values serve as a transition region. Therefore, the measurement of the \PDz meson \RAA values over a wide \pt range, as shown in Fig.~\ref{fig:openhf_raa}, is ideal for studying the underlying mechanisms of parton energy loss. Theoretical models incorporating collisional energy loss can qualitatively describe the shape of the \PDz \RAA distribution at low \pt, while the models that do not consider radiative energy loss fail to replicate the experimental results at high \pt. Therefore, the relative significance of both contributions is determined by varying the \pt. 
In addition to examining the average suppression, the fluctuations of energy loss can also be studied using \vTwo measurements. For light-flavor particles, event-by-event geometry fluctuations have been shown to result in a difference in \vTwo values based on two-particle correlations and those determined using multiparticle correlations~\cite{Voloshin:2007pc,Bilandzic:2010jr}. This difference is shown for charged particles and prompt \PDz mesons in the right panel of Fig.~\ref{fig:openhf_v2}. For heavy quarks, an additional contribution coming from energy-loss fluctuations has been suggested~\cite{Betz:2016ayq}.

With a large number of elastic collisions in the medium, low-\pt heavy quarks can undergo thermalization. On the one hand, thermalization of heavy quarks is delayed by effects related to the heavy quark mass $m_{\mathrm{HQ}}$. This results in a relaxation time that is comparable to the lifetime of the QGP medium produced in HI collisions. Hence, the extent of thermalization of heavy quarks when the QGP medium ceases to exist becomes an indicator of their coupling strength. On the other hand, since the mass of heavy quarks is larger than the typical temperature of the expanding medium, the momentum exchange between heavy quarks and medium partons remains limited. As a consequence, their behavior is similar to ``Brownian motion'' and can be described by the Fokker--Planck equation~\cite{Svetitsky:1987gq}. The transport properties of the QGP are encoded in the coefficients that vary with temperature and momentum. In particular, the spatial diffusion coefficient \PDs characterizes the long-wavelength behavior of heavy quarks and can be directly translated into the fundamental properties of the medium, such as shear viscosity (a review can be found in Ref.~\cite{He:2022ywp}). This theoretical insight implies a strong coupling of heavy quarks with the expanding medium, allowing them to behave collectively. Moreover, at low \pt, \vTwo exhibits a distinct flavor hierarchy. This observation indicates that heavy quarks do not completely undergo thermalization and retain sensitivity to the microscopic transport properties. The spatial diffusion coefficient \PDs has been constrained using measurements of the \PDz meson \RAA and \vTwo carried out by multiple experiments and using various transport models~\cite{Dong:2019byy,Apolinario:2022vzg}. For instance, the obtained \PDs values from a Langevin model-based Bayesian analysis in Ref.~\cite{Xu:2017obm} rule out the pQCD calculations that characterize the weak coupling limit, but align with the predictions from lattice QCD and AdS/CFT calculations, the latter representing the strong-coupling limit for a scale-invariant gauge theory using the conjectured equivalence between a weakly coupled gravitational and conformal field theory~\cite{He:2022ywp}. This consistency confirms the sensitivity of charm quarks to the non-perturbative structure of the QGP.

\begin{figure}[ht!]
    \centering
    \includegraphics[width=0.49\linewidth]{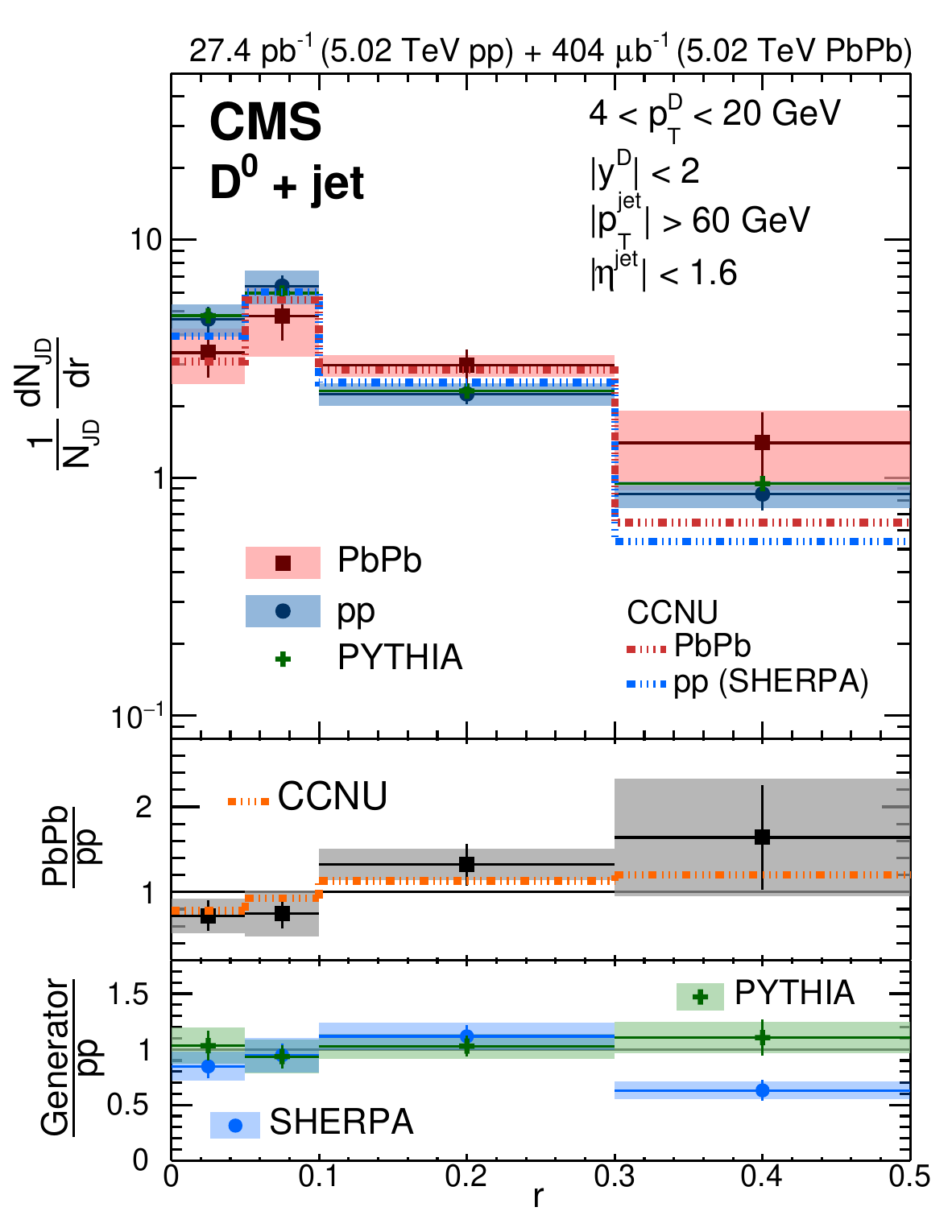} 
    \caption{Distributions of \PDz mesons in jets, as a function of the distance from the jet axis. The ratios of the \PDz meson radial distributions in \PbPb and \pp data are shown in the middle panel, whereas in the lower panel the ratios of the \PDz meson radial distributions of \pp over the two MC event generators are presented.
 \FigureFrom{CMS:2019jis}}
    \label{fig:DinJet}
\vglue4mm
    \centering
    \includegraphics[width=0.49\linewidth]{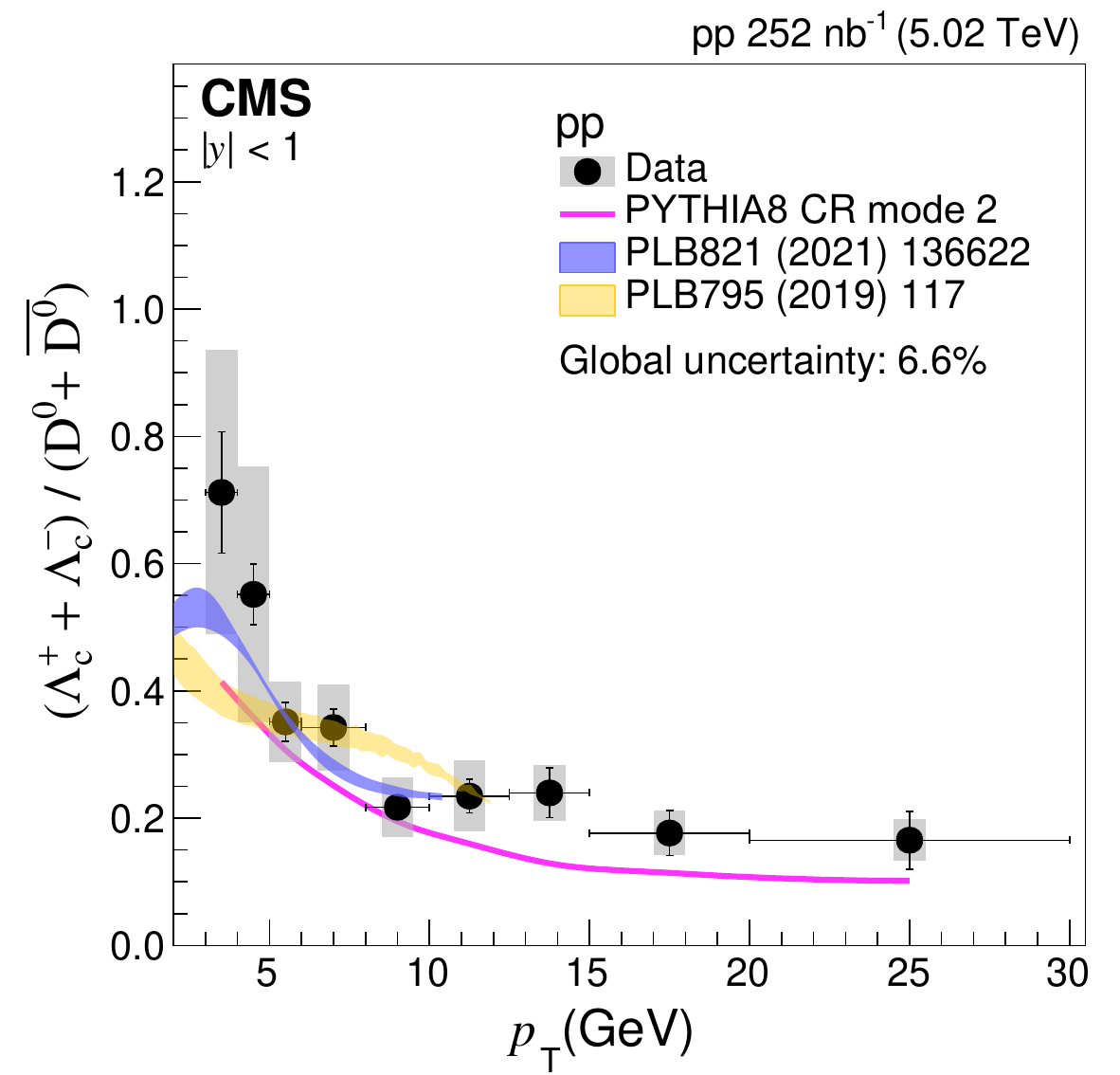}
    \includegraphics[width=0.49\linewidth]{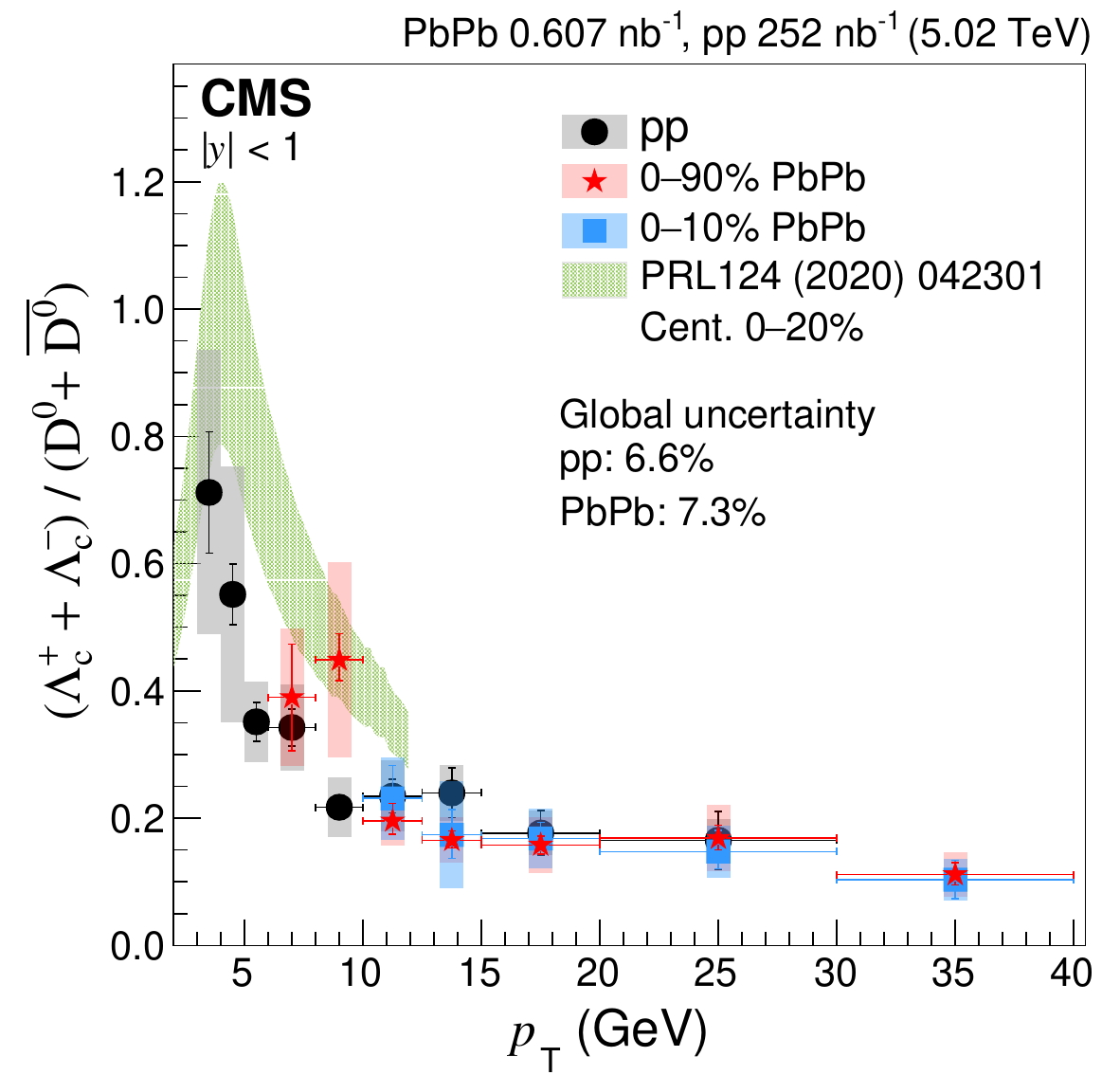}
    \caption{The ratio of the production cross sections of prompt \PcgLp to prompt \PDz versus \pt from \pp collisions. The data are compared to various models (left) and to similar measurements in \PbPb collisions (right). \FigureAdaptedFrom{CMS:2023frs}}
    \label{fig:openhf_lc}
\end{figure}

The radial distribution of \PDz mesons relative to jet axes was also studied to examine the alteration of heavy quarks. As shown in Fig.~\ref{fig:DinJet}, there is a subtle hint that \PDz mesons in the intermediate \pt range of 4--20\GeV that are associated with jets may be slightly farther away from the jet axes in \PbPb collisions than in \pp collisions~\cite{CMS:2019jis}. Such measurements can provide further constraints on the role of charm quark diffusion in the QGP medium.

\subsection{Studies of in-medium hadronization}

The correct interpretation of experimental data requires a thorough understanding of both the in-medium interactions and the subsequent hadronization processes. Parton fragmentation~\cite{Andersson:1983ia} is expected to be the form of hadronization in \pp collisions. In HI collisions, hadrons can be produced not only through parton fragmentation but also through other mechanisms, such as coalescence of partons inside or at the boundary of the QGP medium~\cite{Greco:2003xt,Greco_follow}. The production of hadrons through coalescence is predicted to be most prominent for the regions of low and intermediate \pt, where the density is highest for the precursor partons and decreases with increasing hadron \pt. At high \pt, the fragmentation process is anticipated to dominate hadron production. The coalescence effect is anticipated to be more visible for heavy-flavor hadrons containing strange quarks owing to the strangeness enhancement in the QGP medium~\cite{Rafelski:1982pu}. To study the heavy quark hadronization, comprehensive measurements of both charm and bottom hadrons were conducted at CMS using data taken during Runs~1 
and~2~\cite{CMS:2018eso,CMS:2023frs,CMS:2021mzx,CMS:2022sxl,CMS:2019uws,CMS:2021znk}. Some of the findings are highlighted in this section.

Measurements of \PcgLp baryons via the \Lcdecay channel~\cite{CMS:2023frs} are presented in Fig.~\ref{fig:openhf_lc}. The left panel displays the \LcDratio ratio as a function of \PcgLp \pt in \pp collisions at 5.02\TeV.
The right panel shows the same \pp data points, compared with \PbPb results at the same collision energy. For $\pt > 10\GeV$, the \LcDratio ratios for \pp and \PbPb collisions are consistent with each other, suggesting that the coalescence process does not significantly affect \PcgLp baryon production in this \pt region. The \LcDratio ratio in all collision systems are observed to be much higher than that in $\Pep\Pem$ collisions for $\pt < 30 \GeV$, which is around 0.1 with mild \pt dependence. To further understand the implications of the measurements for \PcgLp production, the results are compared with predictions of various models. The color reconnection mechanism~\cite{CRmodes} implemented in {\PYTHIA{}8} enhances baryon-to-meson ratios in \pp collisions by considering the string fragmentation to be color connected in a way that the total string length becomes as short as possible. This prediction (shown by the CR2 prediction displayed as a purple line in the left panel of Fig.~\ref{fig:openhf_lc}) is consistent with the \pp measurement for $\pt < 10\GeV$, but is systematically lower than the data at higher \pt. A model involving both coalescence and fragmentation in \pp collisions~\cite{Greco_pub_2}, shown by the blue band, describes the \pp measurements well after it is tuned using previous CMS measurements~\cite{CMS:2019uws}. It is worth mentioning that the Catania model~\cite{Greco_pub_2} assumes the formation of a QGP medium in \pp collisions. Another model, displayed as a yellow and green band in the left and right panels, respectively, adopts a statistical hadronization approach and takes into account decays of excited charm baryon states~\cite{Rapp_ppresult, Rapp_PbPb}. It provides a reasonable description of the measurements in \pp and \PbPb collisions for $\pt < 12\GeV$. 

\begin{figure}[ht]
    \centering
    \includegraphics[width=0.49\linewidth]{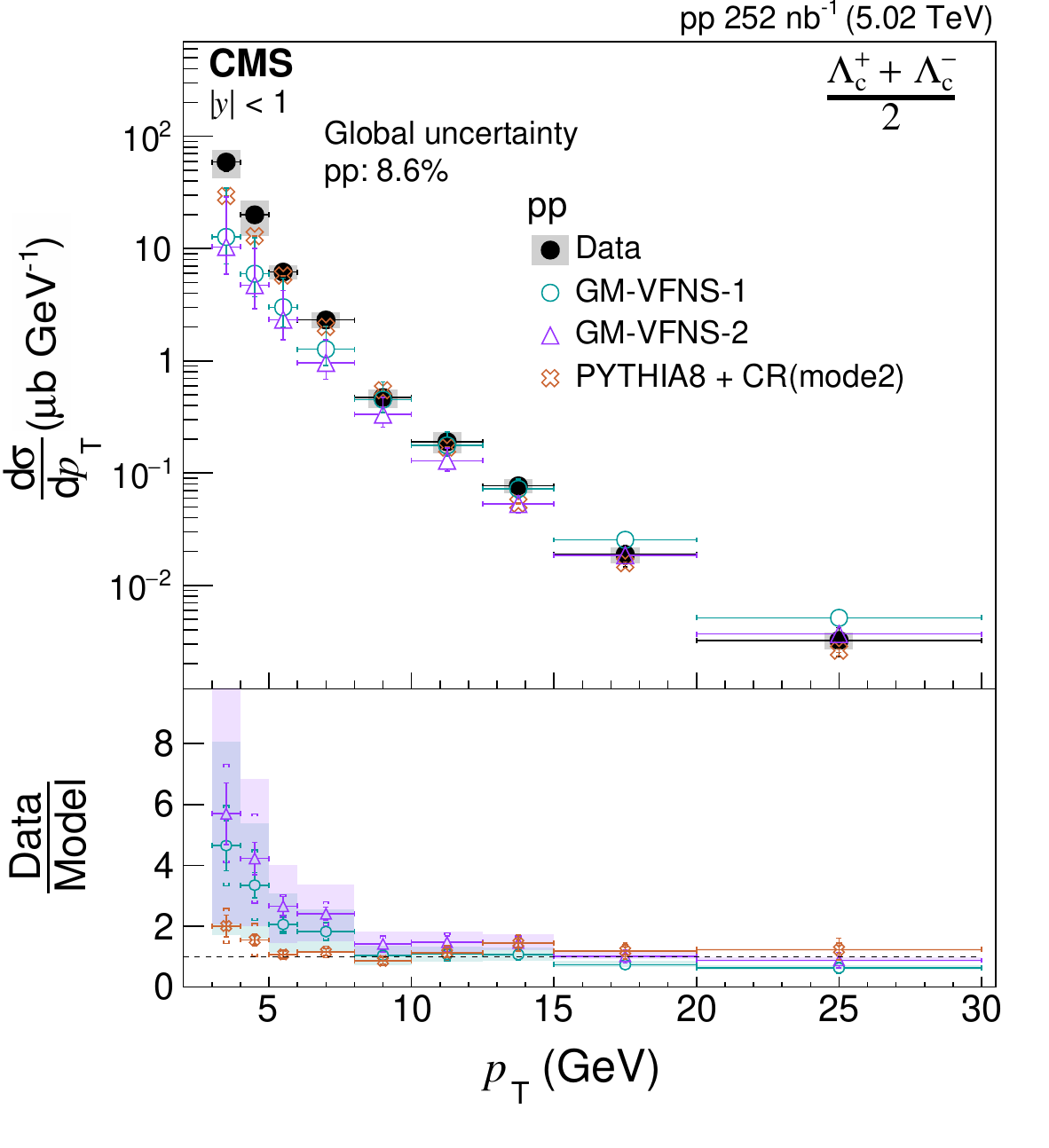}
    \caption{The \pt-differential cross sections for prompt \PcgLp baryon production in \pp collisions, together with model calculations. \FigureAdaptedFrom{CMS:2023frs}}
    \label{fig:openhf_lc_Spec}
\end{figure}

Figure~\ref{fig:openhf_lc_Spec} shows the \pt-differential cross sections for prompt \PcgLp production in \pp collisions\\~\cite{CMS:2023frs}. In addition to the previously discussed model of \PYTHIA{}8 with color reconnection, the results are compared to calculations using the general-mass variable-flavor-number scheme~(GM-VFNS)~\cite{gmvfns1, gmvfns2}. The GM-VFNS calculations use fragmentation functions derived from fits to measurements from the OPAL and Belle experiments. While these calculations accurately describe the \PDz cross section~\cite{CMS:2017qjw}, they significantly underestimate the \LcDratio ratios measured in \pp collisions.
The \PcgLp baryon yield in \pp collisions is also considerably higher than the GM-VFNS calculation, indicating a breakdown of the universality of charm quark fragmentation functions.

\begin{figure}[ht]
    \centering
    \includegraphics[width=0.4835\linewidth]{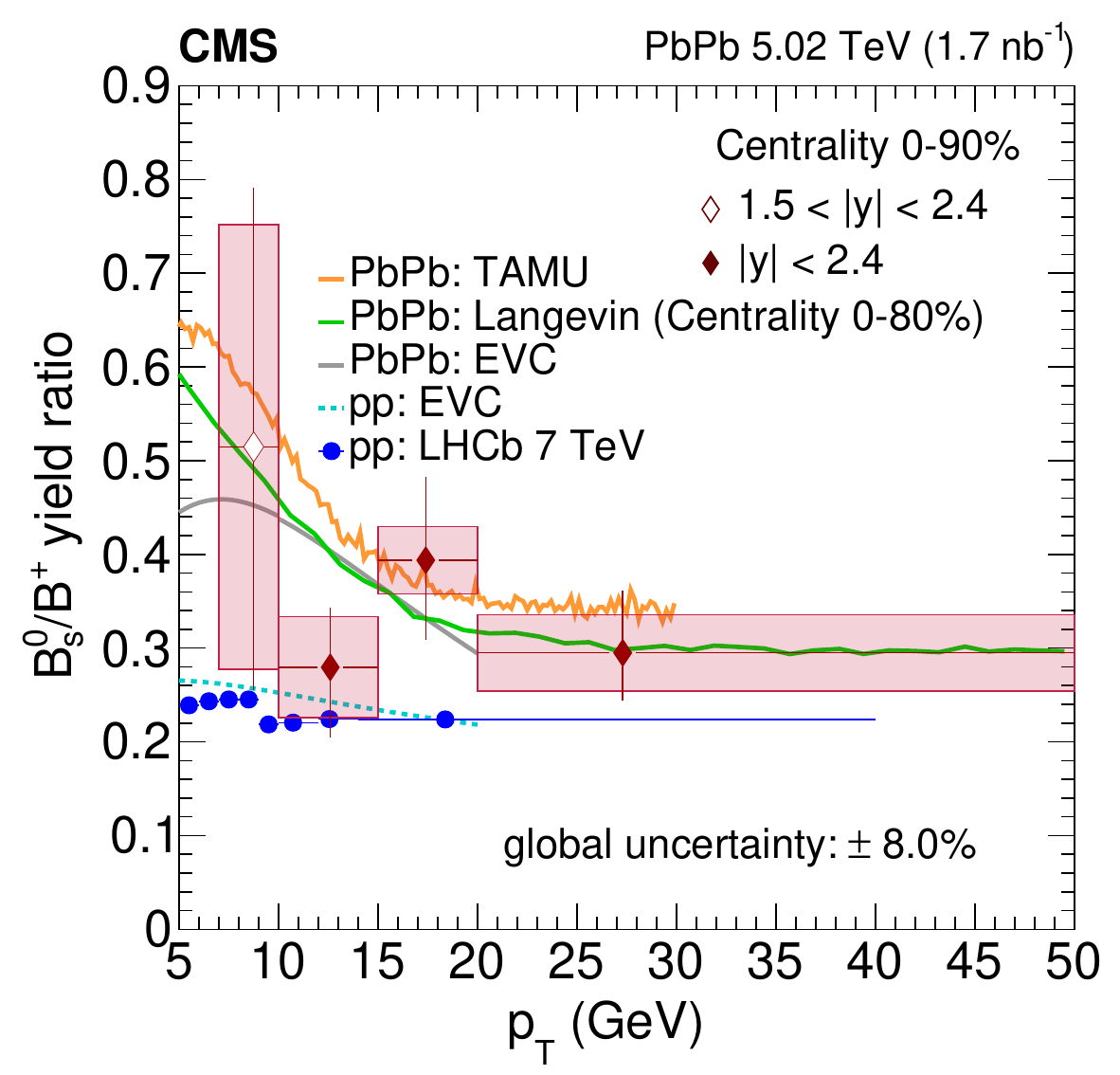}
    \raisebox{0.1ex}{\includegraphics[width=0.4835\linewidth]{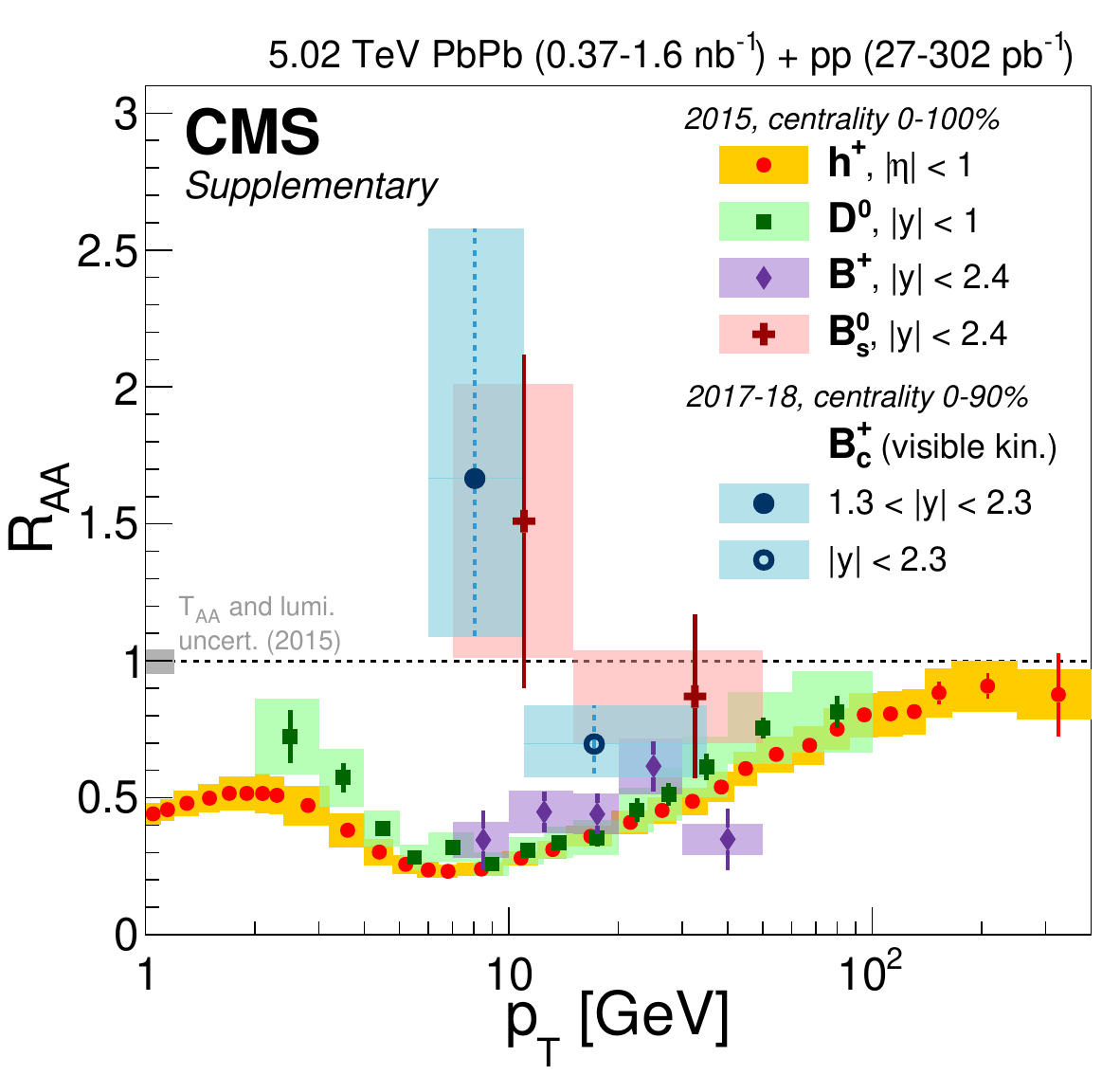}}
    \caption{Left: The ratio of \PBzs and \PBp production yields as a function of \pt in \pp and \PbPb collisions, together with model calculations. Right: The nuclear modification factor of \PBpc and other hadrons in \PbPb collisions. \FigureCompiled{CMS:2021mzx,CMS:2022sxl}}
    \label{fig:hadronization_bottom}
\end{figure}

\begin{figure}[ht]
    \centering
    \includegraphics[width=0.57\linewidth]{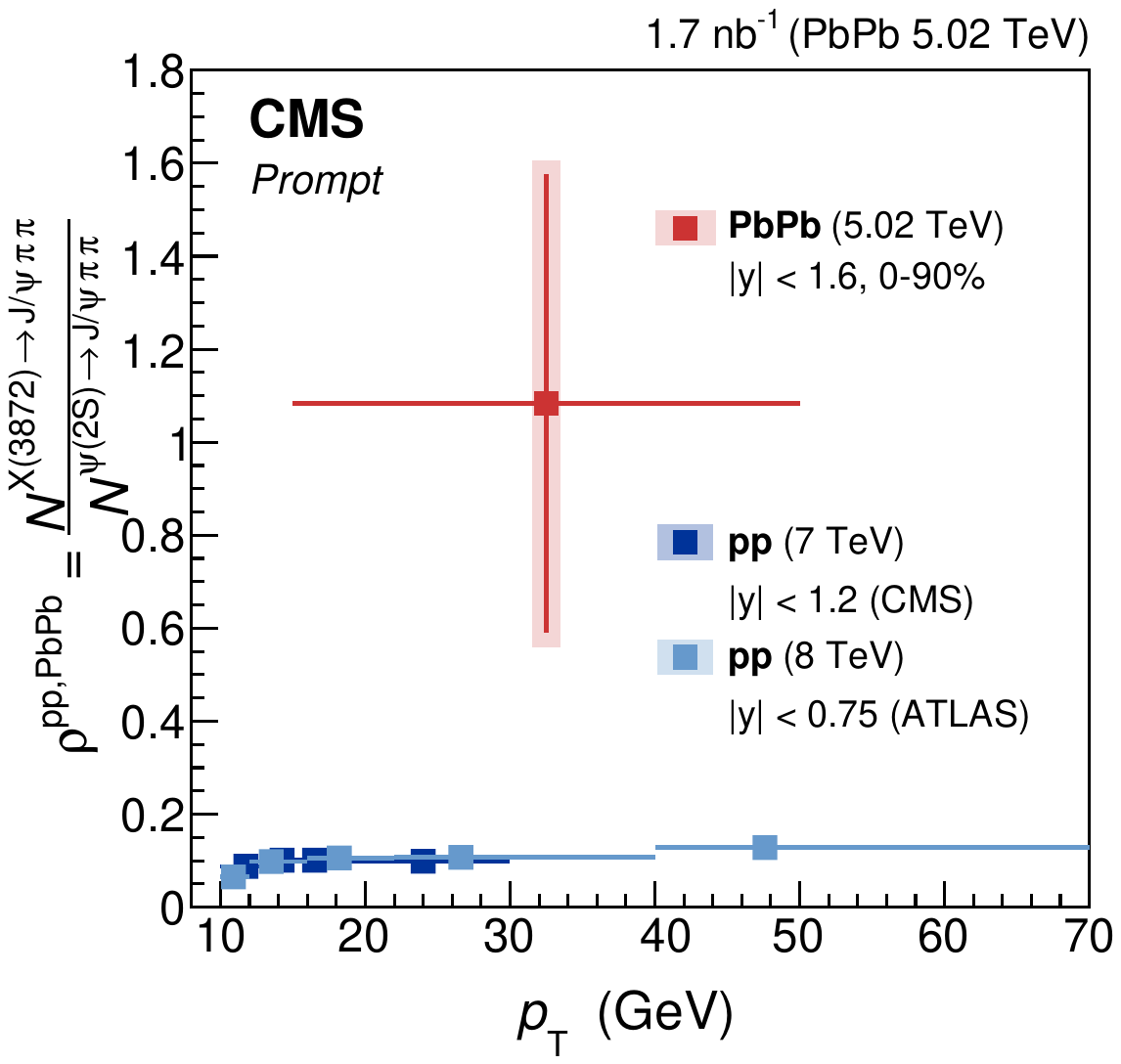}
    \caption{The yield ratio $\rho^{\PbPb}$ of prompt \xthree over \Pgy production in \pp and \PbPb collisions as a function of \pt. \FigureAdaptedFrom{CMS:2021znk}}
    \label{fig:x3872}
\end{figure}

CMS has performed extensive studies of flavor-identified \PQB hadrons, including the first observation of \PBzs and \PBp production, and evidence of \PBpc meson production, in \PbPb collisions. Measurements of \PBzs, \PBp, and \PBpc mesons~\cite{CMS:2022sxl} in \PbPb collisions are presented in Fig.~\ref{fig:hadronization_bottom}. The left panel displays the \BsBpratio yield ratio as a function of \pt in \PbPb collisions, together with the LHCb measurements in \pp collisions at 7\TeV~\cite{LHCb:2019lsv}, scaled by the branching fraction ratio serving as a baseline for \PbPb results. The \PbPb measurement is consistently above the \pp result; however, due to substantial uncertainties, no significant dependence on \pt and centrality can be established. 
It should be noted that the \pp ratio at forward rapidity, as measured by LHCb, may be different from that at mid-rapidity. 
The results are compared with several models, including a transport model based on a Langevin equation that incorporates collisional energy loss and heavy quark diffusion in the medium~\cite{He:2014cla}, an advanced Langevin hydrodynamic model~\cite{cao20}, and a quark combination model using the equal-velocity combination approximation~\cite{EVC}. All models predict a significant enhancement at low \pt attributed to the coalescence contribution to \PBzs production and are in agreement with the data. The right panel of Fig.~\ref{fig:hadronization_bottom} displays the nuclear modification factor of \PBpc meson in \PbPb collisions in comparison with other hadrons. The \PBpc meson, composed of a charm quark and a bottom antiquark, bridges the gap between the ground states of charmonia and bottomonia in size and binding energy. The modification of the \PBpc meson production in \PbPb collisions can offer additional insights into the interaction of heavy quarks with the medium. Due to the small production cross section of the \PBpc meson, the coalescence contribution could potentially be more significant. Although the suppression levels of \PBpc and \PBzs production align with those of other hadrons in the high-\pt region, both of them show a reduced level of suppression at low \pt. This consistency suggests an enhancement in their production through coalescence. To draw more conclusive insights into the \PQB meson hadronization process, future measurements with increased precision are needed.

To study the coalescence effect with a larger number of valence quarks, CMS has presented the first evidence for \xthree production in HI collisions, using the \JPsi\PGpp\PGpm decay channel~\cite{CMS:2021znk}. Although it is generally agreed that the \xthree state---the first discovered exotic hadron---is composed of four valence quarks, its internal structure remains under discussion and several options have been proposed for its composition, including charm-anticharm quark pairs, charm meson molecules, tetraquarks, and their mixtures. In HI collisions, the production yield of the \xthree state, which is affected by the rate of coalescence and dissociation, should depend on its internal structure~\cite{ExHIC:2011say, Zhang:2020dwn, Wu:2020zbx}. Figure~\ref{fig:x3872} depicts the ratio between the prompt \xthree and the \Pgy production yields in \PbPb collisions, alongside the \pp results~\cite{Chatrchyan:2013cld, Aaboud:2016vzw, Aaij:2013zoa}. The measurement suggests that, in \PbPb collisions, either the \xthree state is not suppressed with respect to its \pp production yield or it is suppressed at a level similar to that of the \Pgy.

\subsection{Quarkonium production and suppression in \texorpdfstring{\PbPb}{PbPb} collisions}
\label{sec:Quarkonia}

The suppression of quarkonium states in high-energy HI collisions was first proposed as a signature of QGP formation in a famous paper published by T.~Matsui and H.~Satz in 1986~\cite{Matsui:1986dk}.
The basic idea is rather simple: heavy quarkonium states (\ie, bound states of the charmonium and bottomonium families) should be produced less frequently as we move from small-system collisions (such as \pp, \pPb, or peripheral \PbPb collisions) to increasingly central HI collisions
because the color-charge distribution of the created QGP screens away the potential that binds together the two heavy quarks.
This conjecture leads to a definite and characteristic prediction:
for a given medium temperature, the level of suppression should be different for the various quarkonium states
and follow a sequential hierarchy, reflecting the different values of binding energy
(\ie, the difference between the mass of the particle and twice the mass of the lightest corresponding open-flavor meson)~\cite{DPS,KKS}.
In other words, the more strongly bound the considered quarkonium state is,
the hotter must be the medium before we start seeing signs of its suppression.
This means, in particular, 
that the more loosely bound states 
should be suppressed already in relatively peripheral HI collisions, 
while the states with the largest binding energies 
should only show signs of being suppressed in the most head-on nucleus-nucleus collisions.

While the prediction that deconfinement produces a sequential suppression of the quarkonium states is intuitively straightforward and potentially translates into qualitatively well-defined production patterns, 
in practice there are many challenges that need to be overcome before it can be reliably and cleanly compared with experimental measurements.

\subsubsection{Quarkonium suppression: context and challenges}

The first hurdle, already faced when interpreting the first charmonium suppression measurements from the SPS experiments NA38, NA50, and NA60~\cite{NA38,NA50-PbPb,NA60},
is that the quarkonium production yields also decrease in the absence of a QGP medium. For example, the production of \JPsi mesons has been firmly established in \pA collisions~\cite{NA50-pA400, NA50-pA450, E866} to increase at a rate that is less than linear with the mass number of the target nucleus. This behavior can seemingly be attributed to a range of cold nuclear matter~(CNM) effects, which include nuclear modifications of parton densities, multiple collisions of the final-state resonance with nucleons of the target nuclei resulting in the disintegration of the meson, and effects of energy loss and transverse momentum broadening.
These CNM effects vary with the rapidity and \pt of the produced meson, as well as with the center-of-mass energy of the collision.
Additionally 
these effects act distinctively on different quarkonium states, leading to a more pronounced absorption of the more loosely bound states. For example, the \Pgy meson production has been observed to experience a greater reduction than \JPsi production, even in the absence of a QGP medium~\cite{NA50-pA400, NA50-pA450, E866, PHENIX:2016vmz, LVW, PHENIX:2022nrm, LHCb:2022sxs}.

Another important challenge is that we do not yet know sufficiently well, 
even for the most copiously produced ground states, \JPsi and \PGUP{1S} mesons, 
and for the baseline reference \pp collisions, 
the fraction of the observed yields corresponding to \emph{directly} produced mesons,
as opposed to those created in feed-down decays of heavier quarkonia.
This aspect is of critical importance in any endeavor to interpret the quarkonium suppression measurements. 
For the sake of illustration, let us consider a scenario in which only 50\% of the observed \PGUP{1S} yield in \pp collisions can be attributed to direct production. This reasonable assumption is supported by LHCb measurements at forward rapidity~\cite{LHCb:2014ngh} and trends extrapolated from midrapidity LHC cross sections~\cite{FLAS,Boyd:2023ybk}.
Additionally, let us further restrict this scenario to the limiting case where the heavier (S- and P-wave~\cite{ParticleDataGroup:2022pth}) states, accounting for the other half of the \PGUP{1S} yield through feed-down decays, are no longer produced in central \PbPb collisions because they dissociate in the QGP (given their weaker binding). Then, we would only observe 50\% of the \PGUP{1S} yield and might be tempted to wrongly infer that there is a very strong suppression of the direct \PGUP{1S} production, contradicting expectations based on its relatively large binding energy.
Therefore, a reliable interpretation of the experimental results must carefully account for how the QCD medium 
impacts not only the directly produced particle under examination 
but also all the relevant feed-down sources, each with their specific characteristics.

One further complication affecting the theory-to-data comparisons has to do with the polarizations assumed for the different quarkonium states. 
In fact, there is a correlation between the assumed polarizations and the measured feed-down fractions. 
We can realistically assume, for example, that around 25\% of the observed (promptly produced) 
\JPsi mesons come from decays of the \chicOne and \chicTwo states~\cite{FLSW2006, Lansberg:2019adr}.
However, this assumption arises from measurements of the (\chicOne + \chicTwo)\,/\,\JPsi cross section ratio, and these results depend on the polarizations assumed for these three states,
which affect the calculation of the detection acceptances.
Reasonable variations in the assumed polarization scenarios, 
including nonnegligible changes as a function of \pt 
(\eg, from longitudinal to transverse polarizations when \pt increases)
can easily lead to \PGcc feed-down fractions significantly different from 25\%.

It should be clear by now that it is challenging to achieve compelling experimental evidence
that confirms (or rules out) the existence of a sequential suppression mechanism.
A crucial element in the path to reliably probe that prediction is to collect high-precision data,
both in \pp and in HI collisions,
of as many quarkonium states as possible,
including not only the S-wave vector states (\JPsi, \Pgy, \PGUP{1S}, \PGUP{2S}, and \PGUP{3S})
but also the \chicOne and \chicTwo states, 
and the corresponding \chibnP states in the bottomonium family.
To set the baseline reference, we need to measure, in \pp collisions, 
the double-differential cross sections, in rapidity and \pt,
as well as the corresponding polarizations, for as many states as possible.
This will allow us to define a detailed matrix with all of the feed-down fractions and polarizations.
We also need, naturally, detailed measurements of quarkonium production yields in \PbPb collisions, 
as functions of collision centrality, also paying attention to the explored kinematics phase space.
Measurements made with \pPb collisions should help in understanding the relevance of the CNM effects mentioned above.

Lastly, to probe the existence of the sequential quarkonium suppression signal, it is important to restrict the theory-to-data comparisons to events that are not substantially affected by background processes.
In particular, low-\pt charmonia should be excluded from our investigations because they might be 
dominated by mesons composed of quarks produced in uncorrelated nucleon-nucleon collisions,
a phenomenon known as coalescense, recombination, or regeneration~\cite{Thews:2000rj,Braun-Munzinger:2000csl}.

In the remainder of this section, we will present a selection of the CMS measurements that we consider as important milestones in the challenging path just mentioned.
More specifically, we summarize the \RAA patterns
that CMS measured for the \JPsi, \Pgy, \PGUP{1S}, \PGUP{2S}, and \PGUP{3S} quarkonia, for \PbPb collisions at $\sqrtsNN = 5.02\TeV$, as a function of collision centrality.
These suppression patterns, together with the feed-down and polarization inputs mentioned above, are the most crucial inputs for global analyses~\cite{FL} testing whether the binding energy hierarchy at the core of the sequential suppression conjecture provides a reliable explanation of the experimental data. In addition, pioneering, high-precision \vTwo measurements of various quarkonium states and results on the fragmentation of \JPsi mesons in jets are also discussed, providing crucial information about quarkonium formation. 

\subsubsection{Measurements of quarkonium suppression in \texorpdfstring{\PbPb}{PbPb} collisions}

The most recent CMS publication on charmonium production and suppression in \PbPb collisions~\cite{CMS:2017uuv}
provides a rather complete and diversified set of \RAA measurements 
based on \pp and \PbPb data collected in 2015.
Measurements are reported for the promptly produced \JPsi mesons 
(directly produced or coming from feed-down decays of heavier charmonia)
as well as for the nonprompt mesons, originating from decays of \cPqb hadrons.
The two components can be resolved thanks to the separation of the $\JPsi \to \mumu$ decay vertex from the primary collision vertex.
We can see, in particular, how the \RAA changes with the collision centrality 
(using the \npart variable),
as well as with the meson's \pt and rapidity,
for both prompt and nonprompt \JPsi mesons.
Among the more differential results, 
the CMS measurements include dependences with \pt and \npart for two rapidity ranges,
as well as with \pt for several \npart ranges and vice-versa, 
both for central and forward rapidity.

\begin{figure}[ht]
\centering
\includegraphics[width=\linewidth]{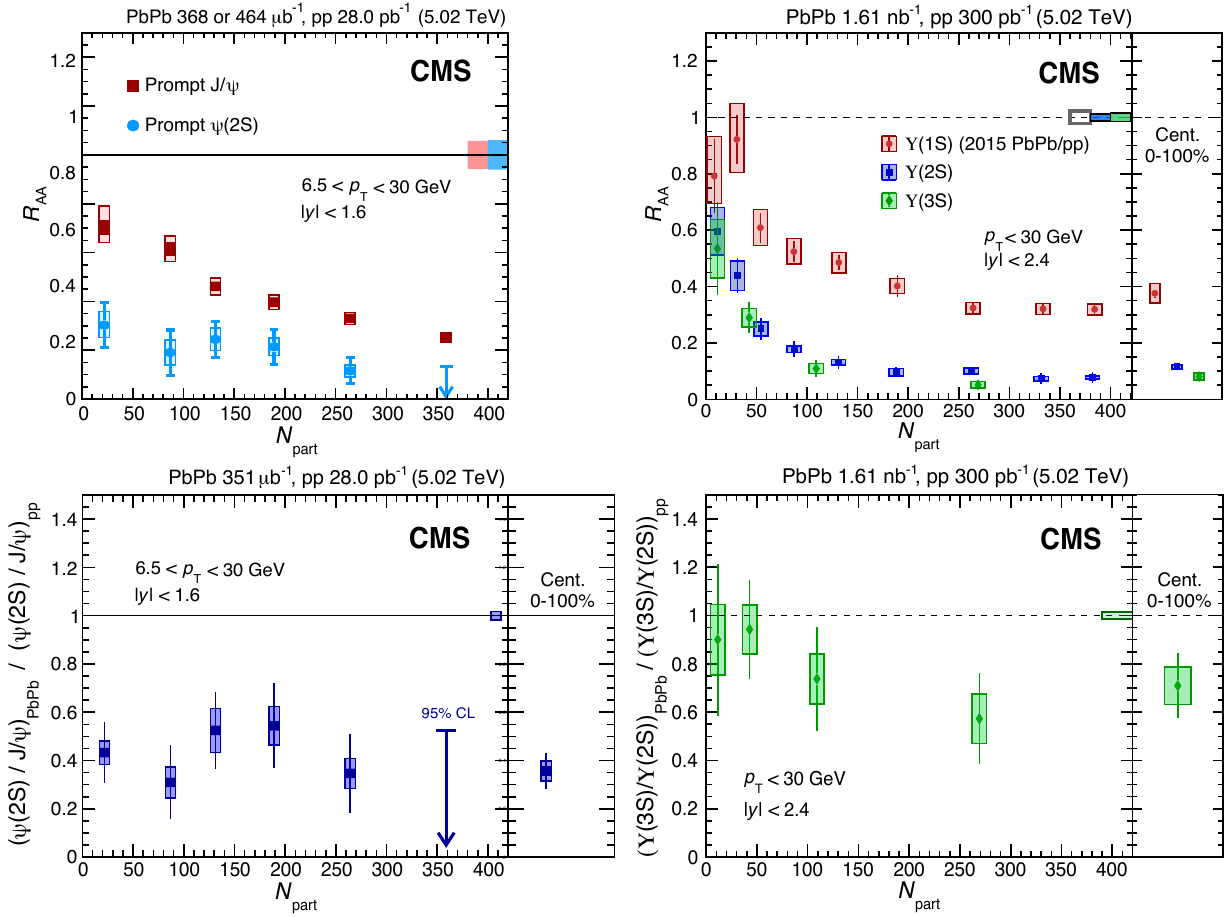}
\caption{Upper: Nuclear modification factors, as a function of the mean number of participants, 
for the promptly-produced \JPsi and \Pgy mesons (left), 
as well as for the \PGUP{1S}, \PGUP{2S}, and \PGUP{3S} (right),
as measured from \pp and \PbPb data at 5.02\TeV.
Lower: Corresponding \Pgy/\JPsi (left) and \PGUP{3S}/\PGUP{2S} (right) 
double-ratios. \FigureCompiled{CMS:2017uuv,HIN-21-007,HIN-16-004}}
\label{fig:quarkonia-RAA}
\end{figure}

Particularly interesting for the studies of sequential suppression, 
the \Pgy and \JPsi \RAA patterns are reported as a function of \npart
for $\pt > 6.5\GeV$, in the central rapidity range $\abs{y} < 1.6$,
as shown in the upper left panel of Fig.~\ref{fig:quarkonia-RAA}.
The \Pgy \RAA values are derived by using the \Pgy/\JPsi double ratio
previously reported in Ref.~\cite{HIN-16-004} and shown in the lower left panel.
No significant dependence on rapidity is observed, in the $\abs{y} < 2.4$ range.
Also the dependence on \pt is rather mild, at least in the $6.5 < \pt < 20\GeV$ range.
Instead, we see a rather strong centrality dependence, 
the production yields being increasingly suppressed as the collisions become more central,
for both the prompt \JPsi and \Pgy cases, as well as for the nonprompt \JPsi mesons.
Most crucially for the investigation of sequential suppression, it is evident that the yield of \Pgy mesons is notably more suppressed than that of the \JPsi mesons, with this stronger suppression becoming apparent even in the most peripheral \PbPb collisions covered by the collected data.

The upper right panel of Fig.~\ref{fig:quarkonia-RAA} shows the centrality-dependent \RAA patterns for the \PGUP{1S}, \PGUP{2S}, and \PGUP{3S} states. Contrary to the charmonium results, the \PGUP{nS} \pt reach starts at $\pt = 0\GeV$. This arises from the higher \PGUP{nS} mass, which enables the muons to reach the CMS muon detectors even if
their parent particle is generated at rest. 
It is to be noted that the likelihood of more than one pair of beauty quarks being produced in each \PbPb collision remains quite low, even at LHC collision energies, so that our results should be essentially unaffected by the coalescence process, even in the low-\pt region, which is more likely to influence the charmonium case. 

\begin{figure}[ht]
\centering
\includegraphics[width=0.5\linewidth]{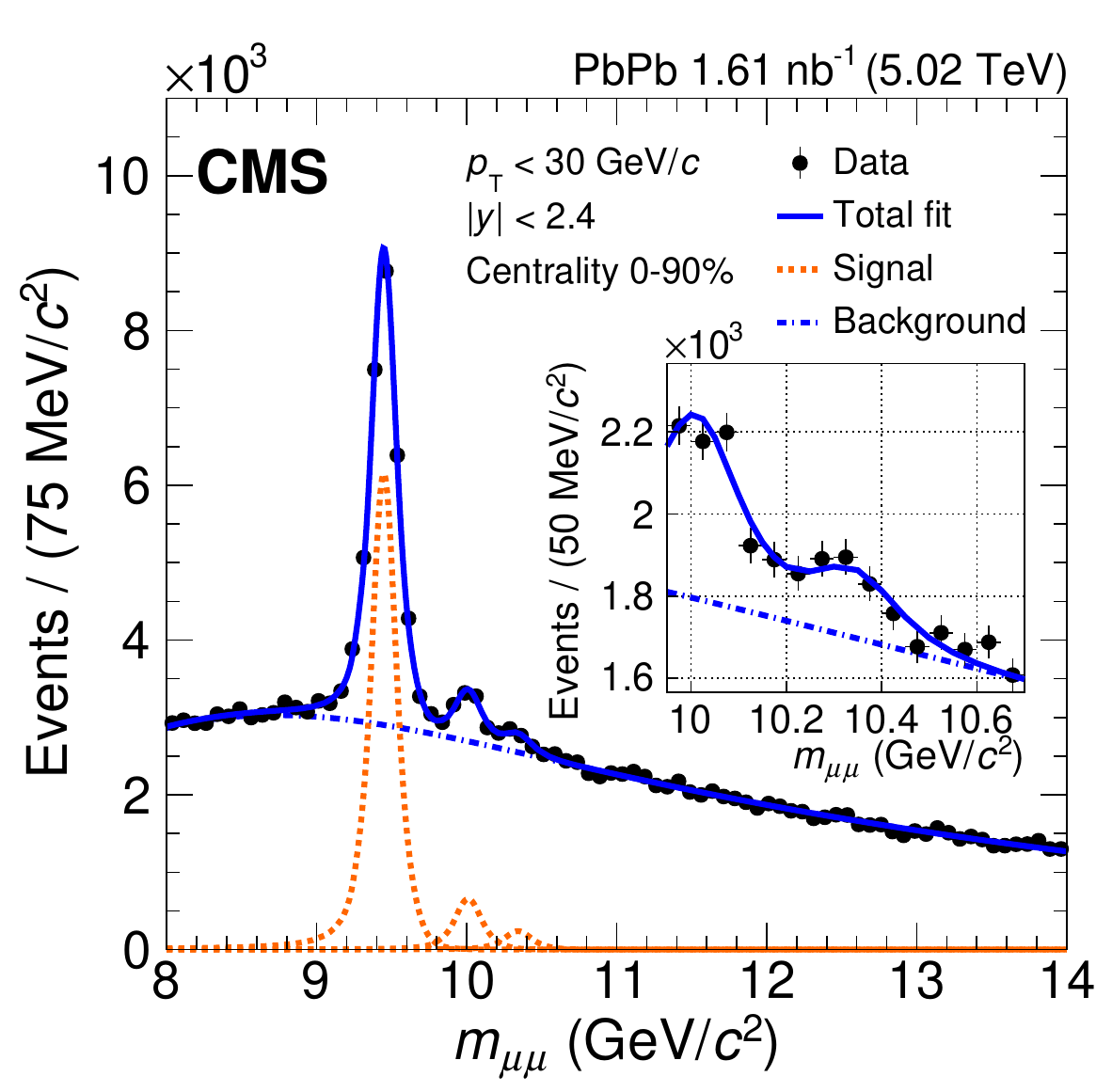}
\caption{The dimuon invariant mass distribution measured in \PbPb collisions when integrating over the full kinematic range of $\pt<30\GeV$ and $\abs{y}<2.4$. The solid curves show the fit results, and the orange dashed and blue dash-dotted curves display the three \PGU states and the background, respectively. The inset shows the region around the \PGUP{3S} meson mass. \FigureCompiledSingular{HIN-21-007}}
\label{fig:Upsilon3S}
\end{figure}

It is worth emphasizing here that the measurement of \PGUP{3S} production represents a remarkable \emph{tour de force}, given its small production rate.
In a previous CMS publication~\cite{HIN-16-023}, the \RAA of the \PGUP{1S} and \PGUP{2S} states could be properly studied, 
while only upper limits were reported for the \PGUP{3S} state.
Benefiting from the larger luminosity recorded in 2018 (Table~\ref{tab:tabHIN}), the new analysis~\cite{HIN-21-007} succeeds in observing the \PGUP{3S} peak in the dimuon mass distribution with a significance well above five standard deviations, as shown in Fig.~\ref{fig:Upsilon3S},
thanks to a state-of-the-art analysis method that uses boosted decision trees
to reduce the large yield of background muon pairs and, hence, 
obtain a signal-enriched dimuon sample.
Another important aspect that makes this measurement feasible is the 
rather good dimuon mass resolution (0.6\% at midrapidity),
enabling the observation of well-resolved invariant mass peaks for the \PGUP{2S} and \PGUP{3S} excited states.
Finally, we should not forget that the bottomonium production cross sections 
are much smaller than those of the charmonium states, 
so that their studies also require large integrated luminosities, efficient triggers, 
and the allocation of suitable DAQ (permanent storage) bandwidths.
All of the above points explain why CMS is particularly well suited, 
among the LHC experiments, 
to probe the physics of bottomonium production in HI collisions.
An additional challenge in the observation of the \PGUP{2S} and \PGUP{3S} states 
in \PbPb collisions is that their production rates are considerably suppressed, 
much more than the \PGUP{1S} state, 
with respect to the linear scaling with \ncoll from \pp collisions.
This is especially true for the \PGUP{3S} state and is most pronounced in the most central \PbPb collisions, as shown in the lower right panel of Fig.~\ref{fig:quarkonia-RAA}.

Looking at all the patterns shown in Fig.~\ref{fig:quarkonia-RAA} 
allows us to clearly see that the \Pgy meson is particularly fragile, 
not surprisingly if we consider that it is the most weakly bound state,
with a binding energy of only 44\MeV, barely 1\% of the meson's mass~\cite{ParticleDataGroup:2022pth}.
On the other hand, the \JPsi and \PGUP{1S} states are the least suppressed ones,
presumably also related to their much larger binding energies of 633 and 1099\MeV, respectively.
As previously mentioned, a thorough examination of the \RAA patterns of the \emph{five} quarkonia
can only be made by also accounting for the presence of poorly known feed-down decays stemming from S- and P-wave quarkonium states, and their respective binding energies.
Nevertheless, the present measurements provide a strong indication that we indeed see signs of nuclear suppression effects that have a stronger 
(sequential) effect on the more weakly bound states.

\subsubsection{Other \texorpdfstring{\JPsi}{JPsi} production measurements}

\begin{figure}[t]
\centering
\includegraphics[width=0.48\linewidth]{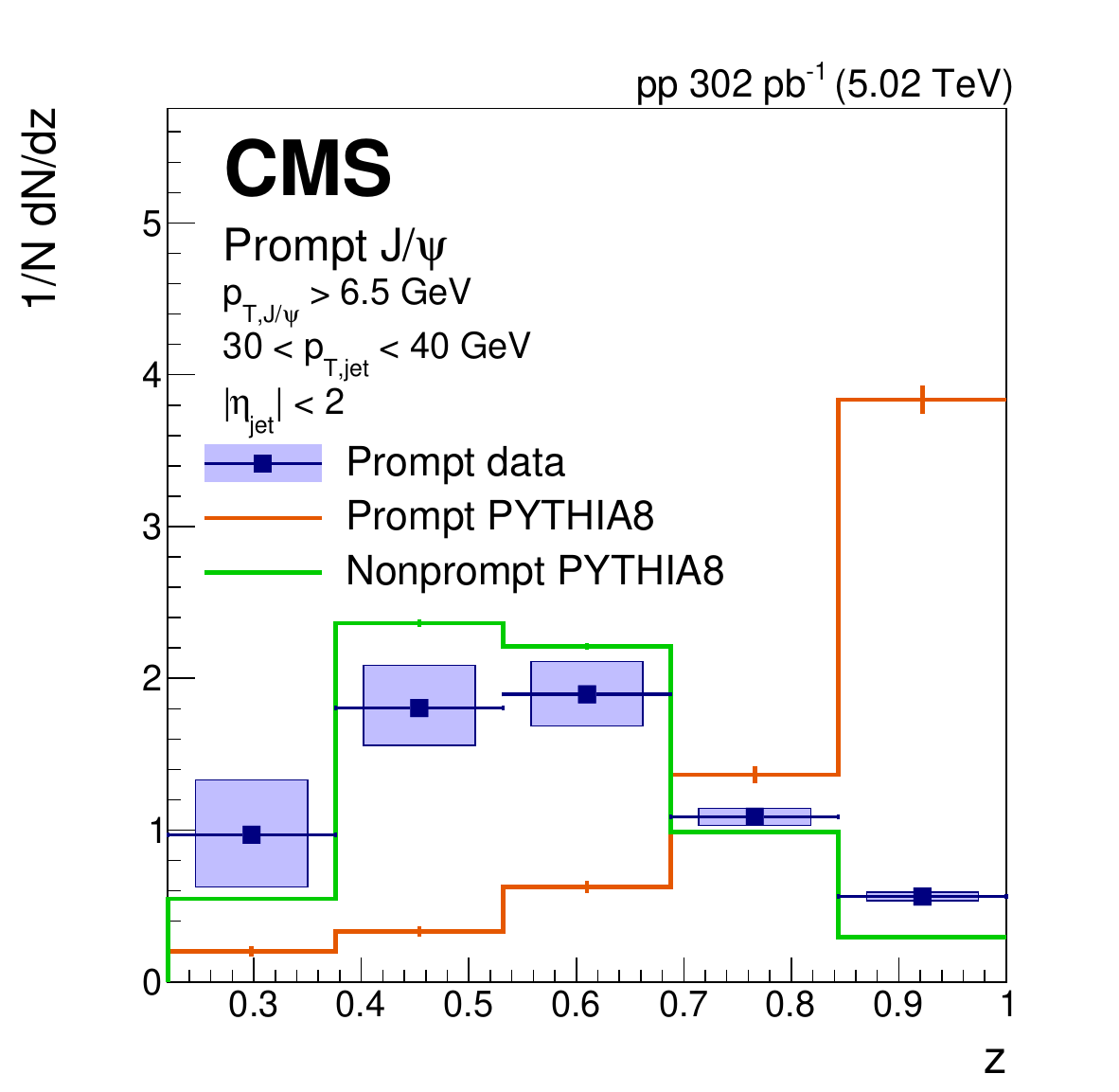}
\caption{Normalized $z$ distribution of \JPsi mesons in jets measured in \pp collisions at 5.02\TeV, compared to prompt and nonprompt \JPsi in \PYTHIA{}8. \FigureFrom{CMS:2021puf}}
\label{fig:JPsi-frag-pp}
\end{figure}

To understand quarkonium production, it is also important to know if there is a parton shower contribution, in addition to the standard (mostly gluon fusion) production term. 
For such studies, CMS measured distributions of the jet
fragmentation variable $z$, the ratio of the \JPsi \pt to the jet \pt, in both \pp and \PbPb collisions. The normalized $z$ distribution of prompt \JPsi in \pp collisions is shown in Fig.~\ref{fig:JPsi-frag-pp}. Unlike what is seen from a sample of prompt \JPsi particles generated with \PYTHIA{}8 (red line), where the mesons are produced in the initial-state partonic scattering, the measured distribution (black squares) shows a larger yield at low $z$ values, where the surrounding jet activity is more important. The $z$ distribution in data more closely resembles that of the nonprompt \JPsi\ \PYTHIA{}8 simulation (green line), which contains a larger jet-like component from fragmentation, as well as other products of the \PQb-hadron decay. This suggests a significant parton shower contribution to \JPsi production, indicating that parton energy loss in the QGP should also contribute to the suppression of the observed quarkonium yield.

\begin{figure}[b]
\centering
\includegraphics[width=0.48\linewidth]{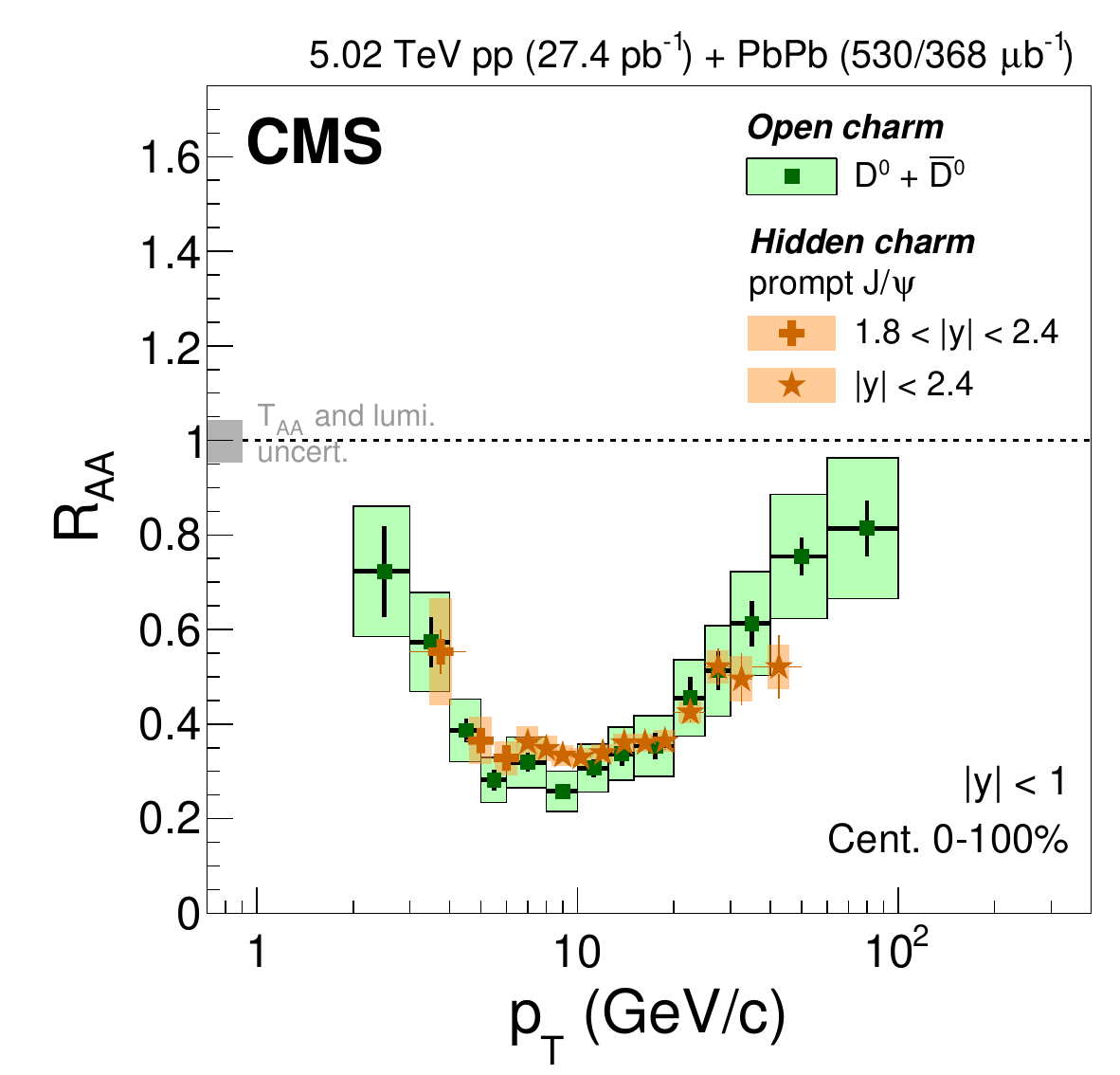}
\includegraphics[width=0.48\linewidth]{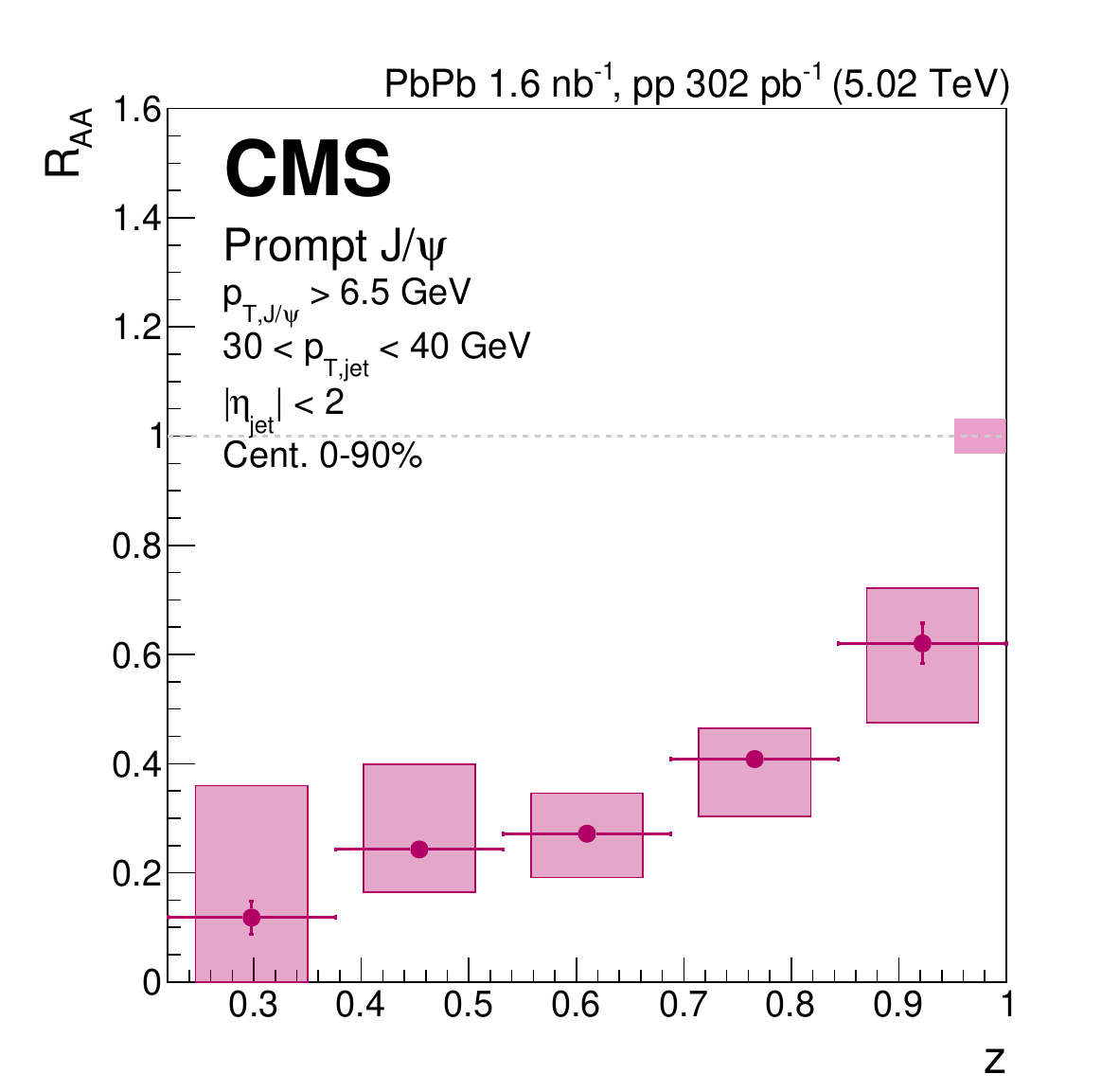}
\caption{Nuclear modification factors \RAA for the promptly-produced \JPsi, as a function of \pt, compared with \PDz mesons (left) and as a function of $z$ (right),
as measured from \pp and \PbPb data at 5.02\TeV. \FigureCompiled{CMS:2017qjw,CMS:2017uuv,CMS:2021puf}}
\label{fig:JPsi-raa-z}
\end{figure}

Moreover, insights into the comparison between open and hidden charm particles are sought by examining the prompt \JPsi and \PDz \RAA values. As shown in the left panel of Fig.~\ref{fig:JPsi-raa-z}, the prompt \JPsi and \PDz mesons have similar \RAA patterns as a function of \pt, suggesting a similar jet quenching mechanism. The \RAA values of the prompt \JPsi mesons as a function of $z$, shown in the right panel of Fig.~\ref{fig:JPsi-raa-z}, present a consistent picture: the suppression is stronger at small $z$, where the large parton multiplicity is expected to result in a large degree of interaction with the QGP.

\begin{figure}[ht]
\centering
\includegraphics[width=0.95\linewidth]{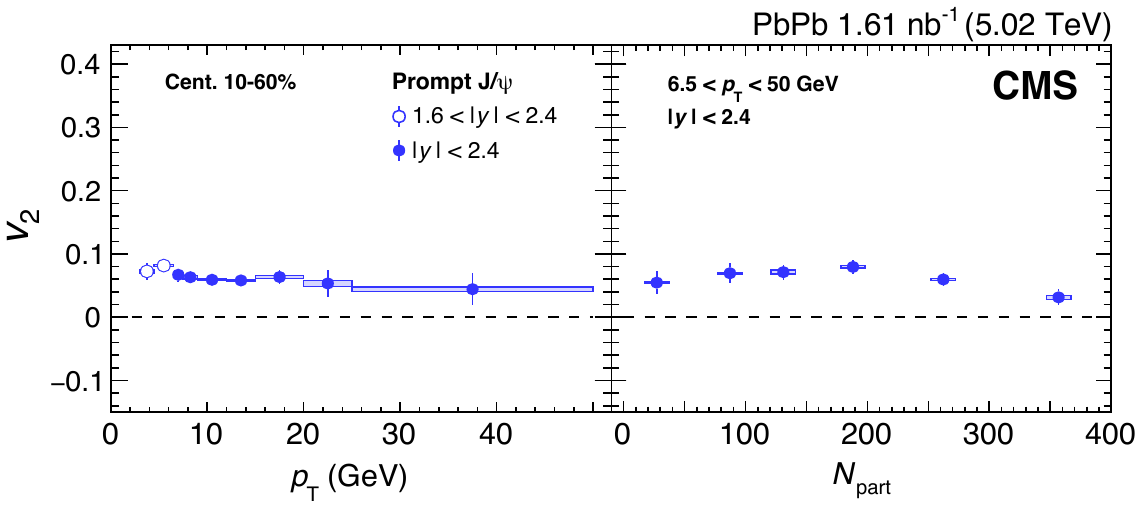}
\caption{The prompt \JPsi \vTwo as a function of \pt (left) and \npart (right), in \PbPb collisions at 5.02\TeV. 
\FigureAdaptedFrom{CMS:2023mtk}}
\label{fig:JPsi-flow}
\end{figure}

Figure~\ref{fig:JPsi-flow} shows that prompt \JPsi mesons have significant and positive \vTwo values, over a wide kinematic range. At low \pt, the \JPsi \vTwo values can be attributed to collective hydrodynamic flow, as found for charged hadrons, while at high \pt, where the hydrodynamical effects are expected to vanish, the non-zero \vTwo values suggest a source of path-length-dependent parton energy loss.

\subsection{Summary of hard probes in the QGP}
\label{sec:HardProbesConclusions}

The CMS Collaboration has used hard scatterings in HI collisions as a powerful toolset for probing the QGP at short length scales. High-\pt objects such as jets, hadrons, electroweak bosons, and heavy-flavor quarks, along with more complex event types such as dijets, heavy-flavor jets, and photon-jet pairs, have been instrumental in revealing the intricate dynamics of the QGP. The extensive data collected during LHC Runs 1 and 2, combined with the broad tracking and calorimeter coverage of the CMS detector, have facilitated a comprehensive exploration of these hard probes.

Early results using data from LHC Run~1 for \PbPb collisions  identified significant dijet \pt asymmetries and suppressed jet and hadron nuclear modification factors, confirming the presence of and expanding the available data related to jet quenching in the QGP. Subsequent studies have deepened our understanding of the path-length dependence of parton energy loss, although a clear observation of the color-charge dependence of jet quenching remains elusive. Events with back-to-back electroweak boson-jet pairs have allowed the determination of absolute jet energy loss, while also highlighting potential selection biases when comparing jets in \pp and \PbPb collisions.

The internal structure of jets is heavily modified by the QGP, as revealed by measurements of jet fragmentation functions and shapes. These results suggest that jet quenching not only reduces jet energy, but also redistributes it from high-\pt constituents to softer particles and from small to large angles relative to the jet axis. Newer jet grooming techniques have offered insights into the earliest stages of parton shower evolution within the QGP, indicating that the groomed jet mass remains relatively unmodified, potentially linking jet evolution to initial parton splittings.

The CMS Collaboration has systematically studied the mass dependence of quark energy loss, comparing \RAA and \vTwo for light, charm, and beauty hadrons across a wide \pt range. At high \pt, these hadrons exhibit similar suppression, consistent with radiative energy loss processes. However, at lower \pt, a flavor hierarchy emerges, indicating increased quark diffusion and elastic collisions. Heavy-flavor hadronization has been investigated through ratios of baryon to meson yields, revealing that coalescence effects are minimal for $\pt>6\GeV$, although beauty mesons with strange or charm quarks show slightly enhanced \RAA values, hinting at possible coalescence effects.

Quarkonium suppression studies have provided crucial insights into the sequential suppression conjecture, which links quarkonium suppression in the QGP to their binding energies. Significant suppression of \Pgy and \PGUP{3S} mesons, even in peripheral \PbPb collisions, highlights their fragility, while \JPsi and \PGUP{1S} states exhibit milder suppression because of their stronger binding. These findings advance our understanding of how quarkonium binding energies influence suppression in high-energy nuclear collisions.

The ongoing increase in LHC luminosity will enable more detailed studies of hard probes, particularly in quarkonium production, for which the CMS detector is exceptionally well-suited. The hard scales associated with these probes provide vital connections to pQCD theory, aiding in the theoretical interpretation of the observables and enhancing our understanding of high-density QCD.

\clearpage

\section{Studies of high-density QCD in small collision systems}
\label{sec:SmallSystems}

Before the first LHC data became available, the primary objective of studying small collision systems, such as \pp and \pA collisions, was to provide essential reference measurements that represent interactions in the absence of QGP formation. The data obtained from \pp collisions offered valuable insights into particle production and hadronization without the complexities introduced by a nuclear initial state. Similarly, \pA reference data were used to extract information about CNM effects by comparing them to results from \pp collisions.
However, in 2010, the CMS experiment made an unexpected breakthrough. By analyzing two-particle correlations in \pp collisions with high multiplicities using a specially designed trigger, a long-range near-side ridge signal was found. This discovery challenged the prevailing understanding of \pp collisions and suggested the potential existence of collective behavior within these systems (discussed in Refs.~\cite{dEnterria:2010xip,Cunqueiro:2009zem}). Moreover, the observation of collectivity in \pPb collisions, using multiparticle correlations, has significantly broadened the scope of flow-like correlation studies. The comprehensive exploration of collectivity signals in various small systems revealed a remarkable similarity between high-multiplicity \pp, \pPb, and \PbPb collisions. 

These groundbreaking discoveries demonstrated the emergence of collectivity in small collision systems and offered possible indications of QGP formation. This unexpected connection between high-multiplicity \pp, \pPb, and \PbPb collisions led to a substantial paradigm shift in our understanding of the prerequisites for QGP formation. 
The investigations of flow-like phenomena with CMS expanded to smaller systems like \PhotonP in UPC \pPb collisions. These studies were complemented by examining archived ALEPH data for similar phenomena in \EE collisions, where recent claims suggest possible indications of flow-like effects, and in $\Pe\mathrm{A}$ collisions at HERA. Nonetheless, no definitive collectivity signal has been detected in these latter searches.

This section summarizes CMS results in small collision systems, detailing first the particle production and hadronization studies, followed by investigations into collectivity with soft probes, analysis of quarkonia production in \pp and \pPb collisions, and finally a discussion of the search for jet quenching signals.

\subsection{Particle production and hadronization}
\label{sec:SmallSystems_ParticleProduction}

The CMS Collaboration has conducted several measurements and comparisons in \pp and \pPb collisions at the LHC. For the studies of inclusive particle production in small collision systems, pseudorapidity distributions of primary charged hadrons have been measured in \pp, \pPb, \XeXe, and \PbPb collisions at various collision energies. These measurements are intriguing not only for the insights they provide about particle production but also for their vital role in calibrating other physics objects, such as jets and isolated photons. 

\begin{figure}[ht]
    \centering
    \includegraphics[width=0.49\linewidth]{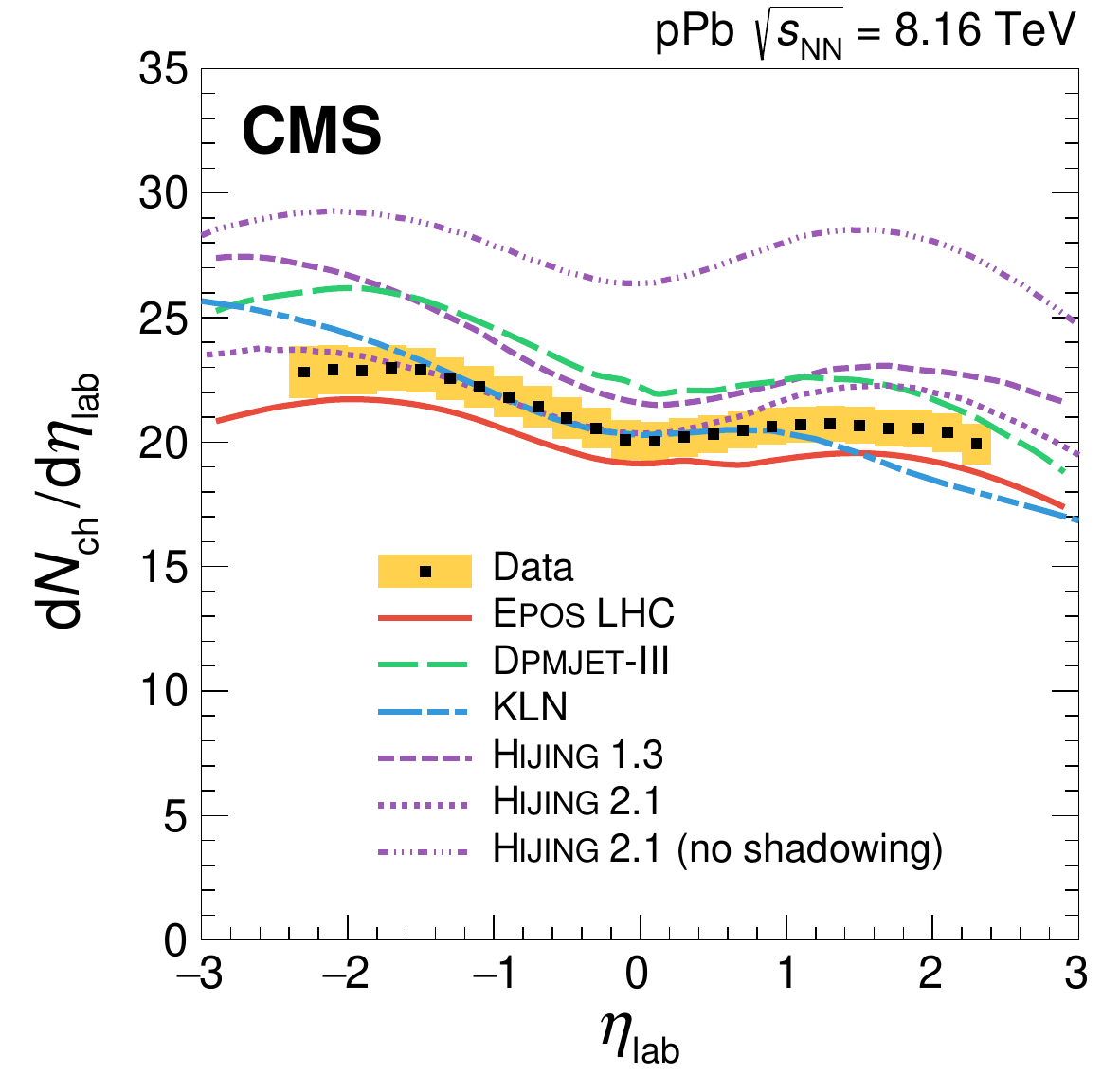}
    \includegraphics[width=0.49\linewidth]{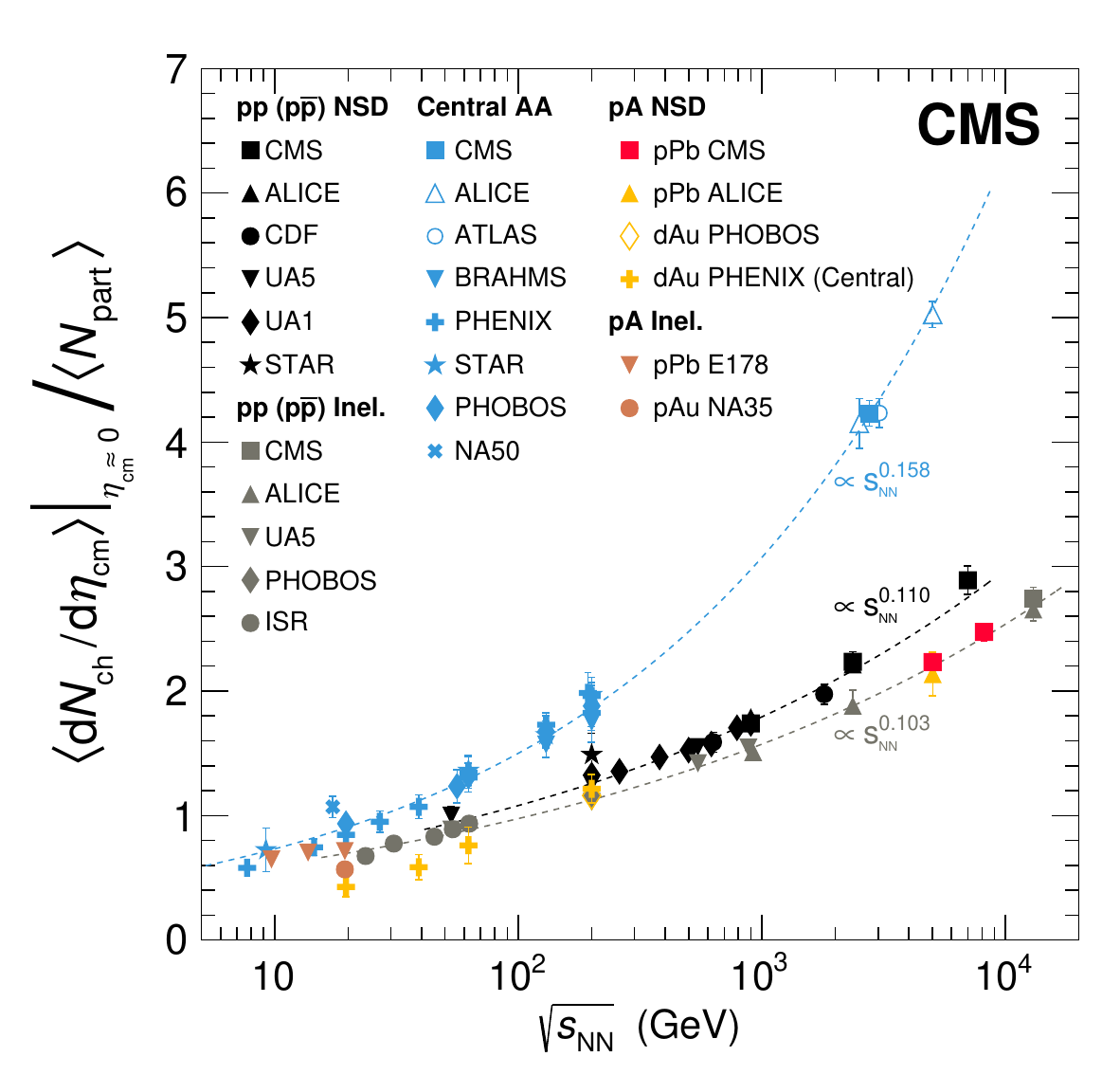}
    \caption{Left: Pseudorapidity density of charged hadrons in the range $\abs{\eta_{\mathrm{lab}}}<2.4$ in \pPb collisions at 8.16\TeV. The results are compared to predictions from the MC event generators \textsc{epos lhc}~\cite{Werner:2005jf,Pierog:2013ria} (v3400), \HIJING~\cite{Wang:1991hta} (versions 1.3~\cite{Gyulassy:1994ew} and 2.1~\cite{Xu:2012au}), and \textsc{dpmjet-iii}~\cite{Roesler:2000he}, as well as from the KLN model~\cite{Dumitru:2011wq}. The shaded boxes around the data points indicate their systematic uncertainties. The proton beam goes in the positive $\eta_{\mathrm{lab}}$ direction. Right: Comparison of the measured density at midrapidity, scaled by \npart in \pPb~\cite{ALICE:2012xs,Elias:1979cp}, $\Pp\mathrm{Au}$~\cite{Alber:1997sn},
\dAu~\cite{Back:2003hx,Adare:2015bua,PHENIX:2017nae} and
central heavy ion collisions~\cite{Alver:2010ck,Abreu:2002fw,Adler:2001yq,
Bearden:2001xw,Bearden:2001qq,Adcox:2000sp,Back:2000gw,Back:2001bq,
Back:2002wb,Aamodt:2010pb,ATLAS:2011ag,Chatrchyan:2011pb,Adare:2015bua,
Abelev:2009bw,Abelev:2008ab}, as well as
NSD~\cite{Albajar:1989an,Alpgard:1982kx,Alner:1986xu,Abelev:2008ab,
Abe:1989td,Khachatryan:2010xs,Khachatryan:2010us} and
inelastic~\cite{Thome:1977ky,Alner:1986xu,Aamodt:2010ft,Alver:2010ck,
Khachatryan:2015jna} \pp collisions. The dashed curves, included to
guide the eye, are from Refs.~\cite{Aamodt:2010ft,Aamodt:2010pb}.}
    \label{fig:dNdetaInPPPPb}
\end{figure}

As shown in Fig.~\ref{fig:dNdetaInPPPPb}, the pseudorapidity spectra in \pPb collisions feature an asymmetrical shape with a higher density in the lead-going direction. Although all theoretical models largely capture the asymmetric shape, the magnitude is better described by KLN, \textsc{epos lhc}, and \HIJING~2.1 with shadowing. The comparison to the \HIJING generator underscores the importance of including the shadowing effect. 
The pseudorapidity density normalized by the number of participating nucleons in \pPb collisions can be compared to \pp and \AonA data. The non-single-diffractive \pPb results at 5.02 and 8.16\TeV align with the results from inclusive \pp collisions, which are significantly lower than the NSD \pp and \AonA results. The data illustrate that \AonA collisions have a higher efficiency in converting energy into charged particles than \pp and \pPb collisions.

\begin{figure}[ht!]
    \centering
    \includegraphics[width=0.49\linewidth]{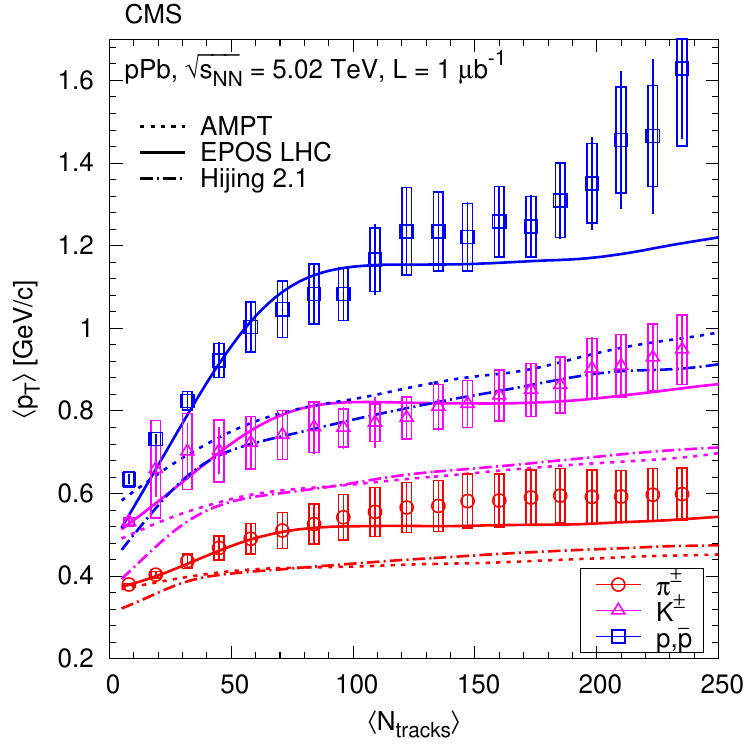}
    \includegraphics[width=0.49\linewidth]{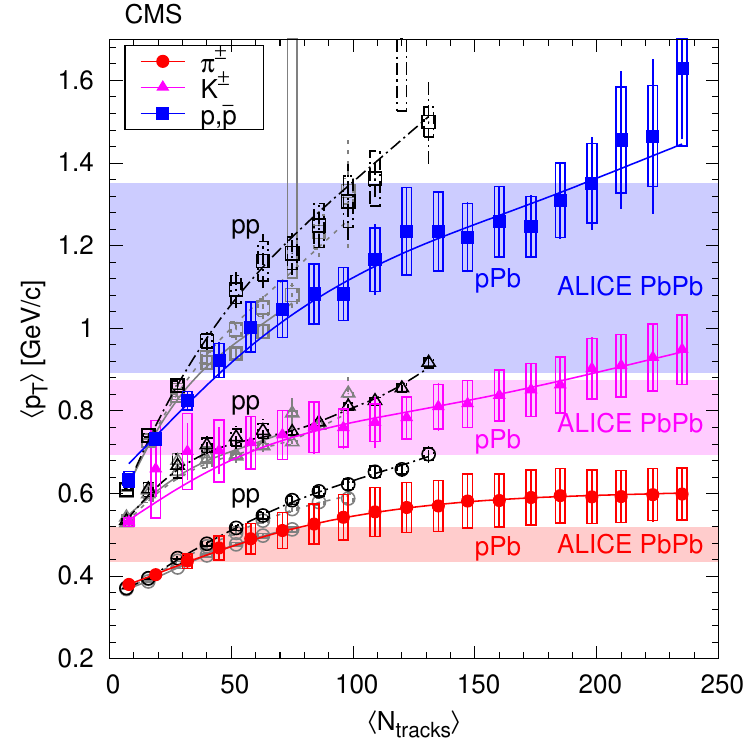}
    \caption{Average transverse momentum of identified charged hadrons in the range $\abs{y}<1$ as a function of the corrected track multiplicity for $\abs{\eta}<2.4$, for \pp collisions (open symbols) at several energies~\cite{CMS:2012xvn} and for \pPb collisions (filled symbols) at $\sqrtsNN = 5.02\TeV$. Left: Results compared to predictions from event generators. Right: Comparison of \pp, \pPb, and \PbPb data. The ranges of $\langle\pt\rangle$ values measured by ALICE in various centrality \PbPb collisions at $\sqrtsNN=2.76\TeV$~\cite{Abelev:2013vea} are indicated with horizontal bands.}
    \label{fig:meanPtInPPPPb}
\vglue4mm
    \centering
    \includegraphics[width=\linewidth]{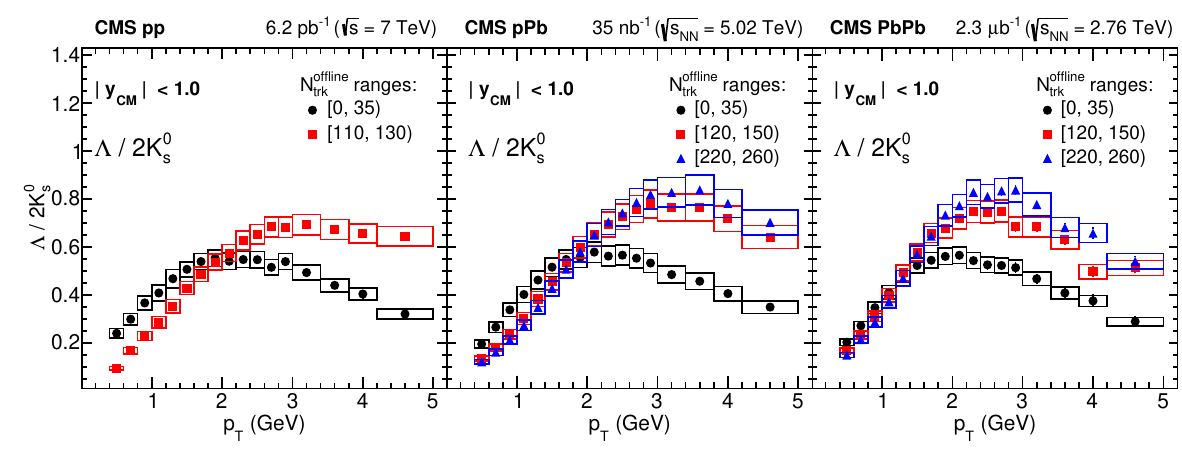}
    \caption{Ratios of \pt spectra for $\PGL/2\PKzS$ in the center-of-mass rapidity range $\abs{y_\mathrm{cm}} < 1.0$ for \pp collisions at $\roots = 7\TeV$ (left),
  \pPb collisions at $\roots = 5.02\TeV$ (middle), and \PbPb collisions at $\rootsNN = 2.76\TeV$ (right). Two (for \pp) or three (for \pPb and \PbPb) representative multiplicity intervals are presented.
  \FigureFrom{CMS:2016zzh}}
    \label{fig:ratioVsSystem}
\end{figure}

To explore the system size dependence of particle production, characterized by the final state particle multiplicity, spectra of identified charged hadrons were measured in \pPb collisions at 5.02\TeV. As shown in Fig.~\ref{fig:meanPtInPPPPb}, the average \pt was observed to increase with particle mass and charged particle multiplicity, with heavier hadrons exhibiting a more pronounced increase. Comparisons with MC event generators revealed that \textsc{epos lhc}, which incorporates additional hydrodynamic evolution of the created system, could reproduce most of the data features, unlike \HIJING and \textsc{ampt}. The study also conducted comparisons of the \pPb \pt spectra and integrated yields to those in \pp and \PbPb collisions, revealing an intriguing similarity between different collision systems at the same particle multiplicity.

The \ET distribution was also measured in \pPb collisions at 5.02\TeV, using the hermetic coverage of the CMS detector~\cite{CMS:2018xfv}. The study covered a wide pseudorapidity range and leveraged the presence of the CASTOR calorimeter. The results demonstrated a strong centrality dependence, characterized by a significantly greater increase of \ddinline{\ET}{\eta} in more central events on the lead-going side compared to the proton-going side. Predictions from \textsc{epos lhc}, \textsc{qgsjet~ii}~\cite{Ostapchenko:2004ss}, and \HIJING were compared to the data, but none could fully encompass all aspects of the $\eta$ and centrality dependence.

Measurements of transverse momentum spectra of strange hadrons (\PKzS, $\PGL+\PAGL$, and $\PgXm+\PagXp$) in \pp, \pPb, and \PbPb collisions at different collision energies have been performed with CMS~\cite{CMS:2016zzh}, extending the measurements beyond studies with light-flavor hadrons. These measurements are sensitive to medium-induced modifications of the final-state particle composition. Figure~\ref{fig:ratioVsSystem} summarizes the ratio of \pt spectra for $\PGL/2\PKzS$ in \pp, \pPb, and \PbPb collisions as a function of \pt. 
In a radial flow picture, we expect the $\PGL$ baryon, which contains three constituent quarks, to receive a larger boost, in transverse momentum, compared to the $\PKzS$ meson, resulting in the kind of modification pattern that we see in the high multiplicity data.
An enhancement of the ratio is observed in all collision systems at intermediate to high \pt and high multiplicity, indicating a similarity in the multiplicity and \pt dependence of the ratio.

The average transverse kinetic energy~(\KETavg) of strange hadrons is observed to rise with multiplicity, with a more pronounced increase noted for heavier particles across all collision systems. Furthermore, when comparing results at similar multiplicities, the difference in \KETavg among various strange-particle species is more substantial in \pp and \pPb events than in \PbPb events. In \pPb collisions, the average transverse kinetic energy is slightly larger in the Pb-going direction than in the p-going direction for events featuring high particle multiplicities. 

In conclusion, the observed patterns, especially the mass-dependent rise in \KETavg and the enhanced $\PGL/2\PKzS$ ratio, align with expectations from radial flow, suggesting additional evidence of collectivity in these systems. 

\subsection{Studies of collectivity in small systems}
\label{sec:SmallSystems_Collectivity}

\subsubsection{Exploring small-system collectivity using light-flavor particles}

Well before the LHC began operating, the presence of collectivity in \AonA collisions at RHIC and elsewhere had been well established. Flow coefficients were extracted using a variety of methods, including measurements of the azimuthal anisotropy of particle yields with respect to the event plane and measurements of two-particle and multiparticle correlations~\cite{Voloshin:2008dg}. 
The dominant source of the second order coefficient \vTwo was understood as hydrodynamic flow driven by the asymmetric shape of the overlap region of the two nuclei. Higher order harmonics were also understood by that time as resulting from asymmetries created by fluctuations in the collision geometry~\cite{Alver:2010gr}. 

The LHC started its operation with \pp collisions, in which collectivity was not expected to exist. However, it was speculated that for sufficiently high charged particle multiplicities, collectivity might be observed if fluctuations in the collision geometry could also create initial-state anisotropies in these collisions. As a measure of the necessary multiplicities, the values of \dnchdeta\ for AuAu collisions at 200\GeV in the 60--70\% and 70--80\% centrality ranges are $45\pm3$ and $22\pm2$, respectively~\cite{Abelev:2008ab}. Clear \vTwo signals were observed in both of these \AonA collision cases~\cite{STAR:2008ftz}. 

\begin{figure}[ht]
    \centering
    \includegraphics[width=\linewidth]{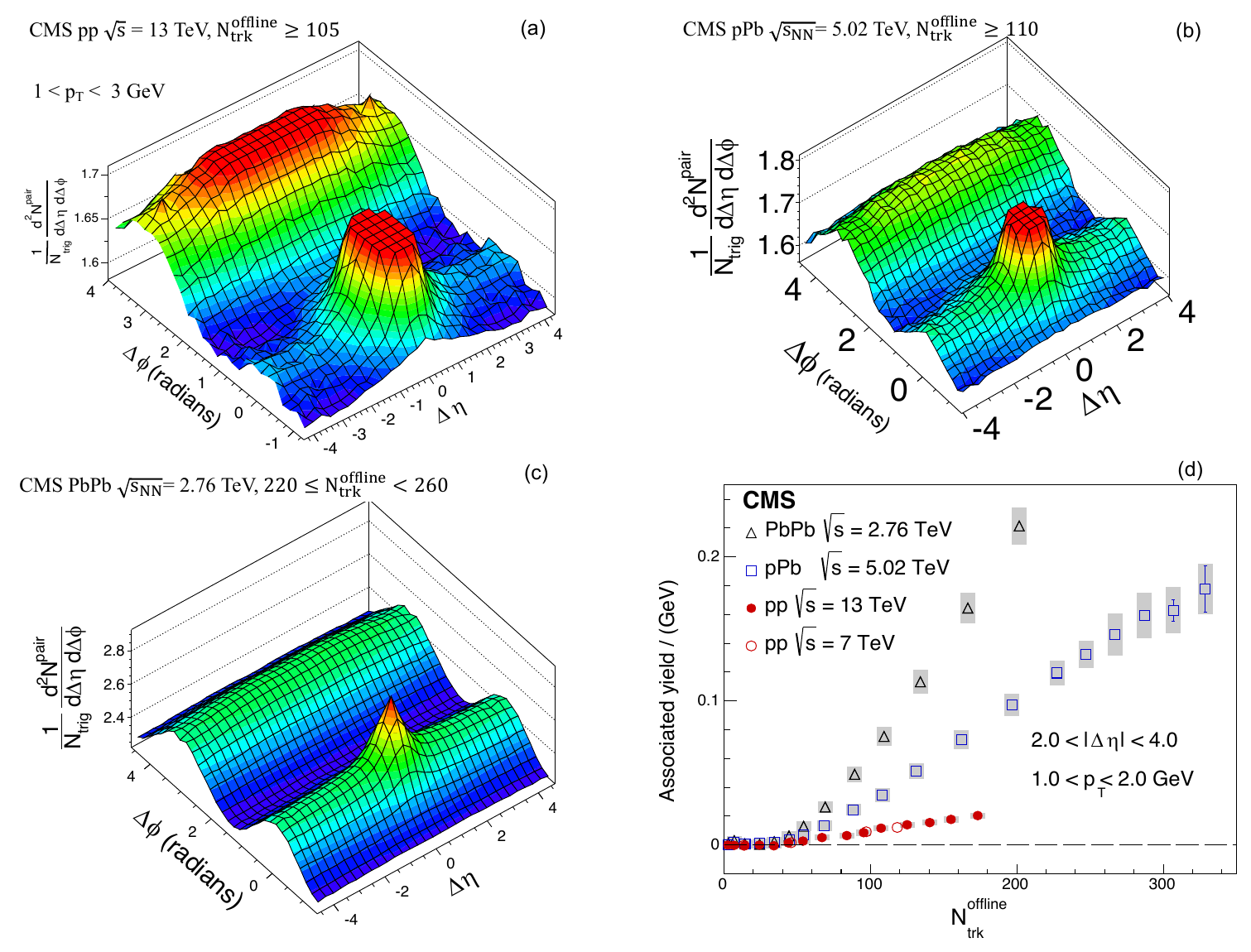}
    \caption{Panels (a), (b), and (c) show the 2D two-particle correlation functions for pairs of charged particles with $1<\pt<3\GeV$ for high multiplicity events in \pp at 7\TeV and \pPb at 5.02\TeV, as well as peripheral \PbPb collisions at 2.76\TeV. Panel (d) displays the ridge yield as a function of multiplicity in \pp, \pPb, and \PbPb collisions. The vertical bars and shaded boxes denote the statistical and systematic uncertainties, respectively. \FigureCompiled{CMS:2010ifv, CMS:2012qk, Chatrchyan:2013nka, CMS:2015fgy}}
    \label{fig:small:ridge_LF_4panels}
\end{figure}

Even at LHC energies, the average \dnchdeta\ for MB \pp collisions is much smaller than these values. However, during the first \pp run, the CMS Collaboration developed a way to trigger on high-multiplicity \pp events. Using this trigger in \pp collisions at 7\TeV, the average number of charged particles with $\pt>0.4\GeV$ and $\abs{\eta}<2.4$ corrected for tracking efficiency, \ntrcorr, was about 136 for events selected with the number of observed tracks reconstructed offline $\noff \geq 110$~\cite{CMS:2010ifv}. This corresponds to $\dnchdeta\approx 28$ with $\pt>0.4 \GeV$, which is comparable to the values mentioned above for peripheral AuAu collisions. In two-particle correlations (as described in Section~\ref{sec:TransportProperties}), a striking long-range ($\abs{\deta}>2$) ridge-like structure was observed on the near-side ($\dphi \approx 0$) for $1 < \pt < 3\GeV$ with $\noff > 110$, as shown in the upper left panel of Fig.~\ref{fig:small:ridge_LF_4panels}. This discovery motivated further studies of small collision systems at RHIC, as well as \pp, \pPb, and peripheral \PbPb collisions at the LHC. This section describes some of the CMS analyses of small systems using various flow analysis techniques, including multiparticle correlations, and new observables such as cumulant ratios and symmetric cumulants. 

To further understand the origins of the ridge, the \pt and multiplicity dependence of its yield, as well as flow coefficients, \vN, were studied using two-particle correlations.
Panels (a), (b), and (c) of Fig.~\ref{fig:small:ridge_LF_4panels} compare 2D two-particle correlation functions for pairs of charged particles with $1<\pt<3\GeV$ for \pp at 7\TeV with $\noff \geq 110$, \pPb at 5.02\TeV with $\noff \geq 110$, and peripheral \PbPb collisions at 2.76\TeV with $220 \leq \noff < 260$, respectively. To investigate the long-range, near-side correlations
in detail, 1D distributions in \dphi are found
by averaging the two-dimensional distributions over
$2 < \abs{\deta} < 4$. The ridge yield is then calculated by integrating over the region $\abs{\dphi} < 1.2$, with the results for the three systems shown in panel (d) of Fig.~\ref{fig:small:ridge_LF_4panels} as a function of multiplicity. The ridge yields show an approximately linear increase for $\noff \gtrsim 40$, which corresponds to $\ntrcorr \gtrsim 53$.
Although the shape of the multiplicity dependence is qualitatively similar for \pp, \pPb, and \PbPb collisions, a significantly higher yield per trigger particle is seen in \PbPb than for \pPb collisions, which is itself larger than for \pp collisions at a given multiplicity.

\begin{figure}[ht]
    \centering
    \includegraphics[width=\linewidth]{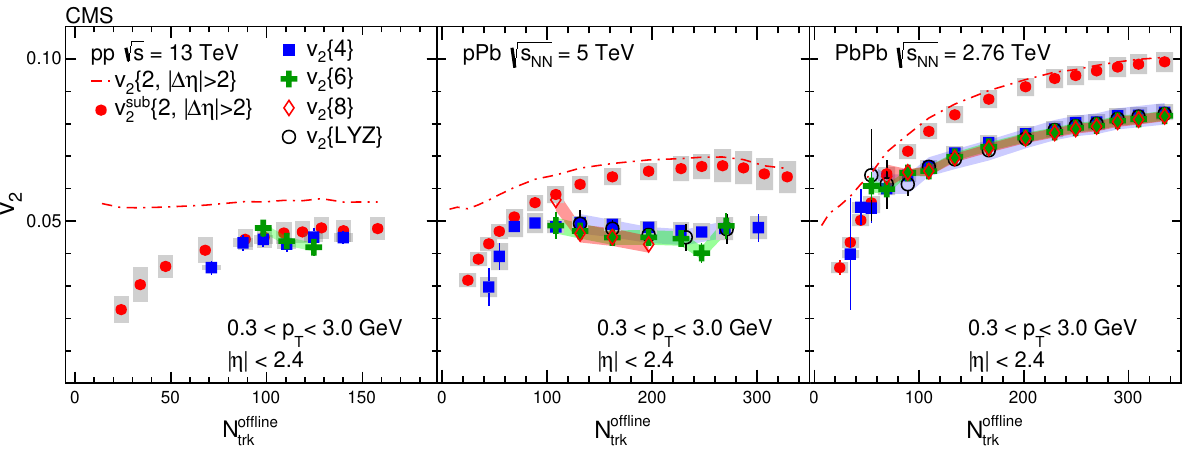}
    \caption{Left: The $\vtwo{2, \abs{\deta}>2}$, $\vTwo^\text{sub}\{2, \abs{\deta}>2\}$, $\vtwo{4}$, and $\vtwo{6}$ values as functions of \noff for charged particles, averaged over $0.3 < \pt < 3.0\GeV$
     and $\abs{\eta}<2.4$, in \pp collisions at 13\TeV. 
     Middle: The $\vtwo{2, \abs{\deta}>2}$, $\vTwo^\text{sub}\{2, \abs{\deta}>2\}$, $\vtwo{4}$, $\vtwo{6}$, $\vtwo{8}$,
     and $\vtwo{\mathrm{LYZ}}$ values in \pPb collisions at 5\TeV.
     Right: The $\vtwo{2, \abs{\deta}>2}$, $\vTwo^\text{sub}\{2, \abs{\deta}>2\}$, $\vtwo{4}$, $\vtwo{6}$, $\vtwo{8}$, and $\vtwo{\mathrm{LYZ}}$ values in \PbPb collisions at 2.76\TeV. The vertical bars and shaded boxes for $\vTwo^\text{sub}\{2, \abs{\deta}>2\}$ and $\vtwo{4}$ denote the statistical and systematic uncertainties, respectively, with the former generally being smaller than the symbols. For $\vtwo{6}$, $\vtwo{8}$, and $\vtwo{\mathrm{LYZ}}$, vertical bars show statistical uncertainties and systematic uncertainties are shown by green, red, and gray shaded bands, respectively.~\FigureFrom{CMS:2016fnw}}
    \label{fig:small:CMS-HIN-16-010_Figure_aux001}
\end{figure}

Flow coefficients can be extracted via a Fourier decomposition of the long-range two-particle \dphi correlation function described in Section~\ref{sec:TransportProperties}. However, back-to-back nonflow correlations, which are more significant for \pp and \pA than in \AonA collisions, are still present in the away-side region of these long-range distributions. This contribution can be suppressed by performing low-multiplicity subtractions~\cite{Chatrchyan:2013nka}. The \vTwo values before and after this subtraction, $\vtwo{2, \abs{\deta}>2}$ and $\vTwo^\text{sub}\{2, \abs{\deta}>2\}$, are shown as a dot-dash line and red circles, respectively, in Fig.~\ref{fig:small:CMS-HIN-16-010_Figure_aux001}. At lower multiplicity, the nonflow contributions increase and a reliable extraction of the flow signal becomes model-dependent. 

Another way to suppress nonflow effects is to use multiparticle correlation methods~\cite{Chatrchyan:2013nka, CMS:2015yux, CMS:2016fnw}. The flow coefficient values from 4-, 6-, and 8-particle cumulants, \vtwo{4}, \vtwo{6}, and \vtwo{8}, as well as all-particle correlations, $\vtwo{\mathrm{LYZ}}$, are also shown in Fig.~\ref{fig:small:CMS-HIN-16-010_Figure_aux001}. Within experimental uncertainties, the \vTwo values from all of the multiparticle correlation methods are
consistent with each other. This provides strong
evidence for the collective nature of the long-range correlations observed in these small systems.

\begin{figure}[ht]
    \centering
    \includegraphics[width=0.45\linewidth]{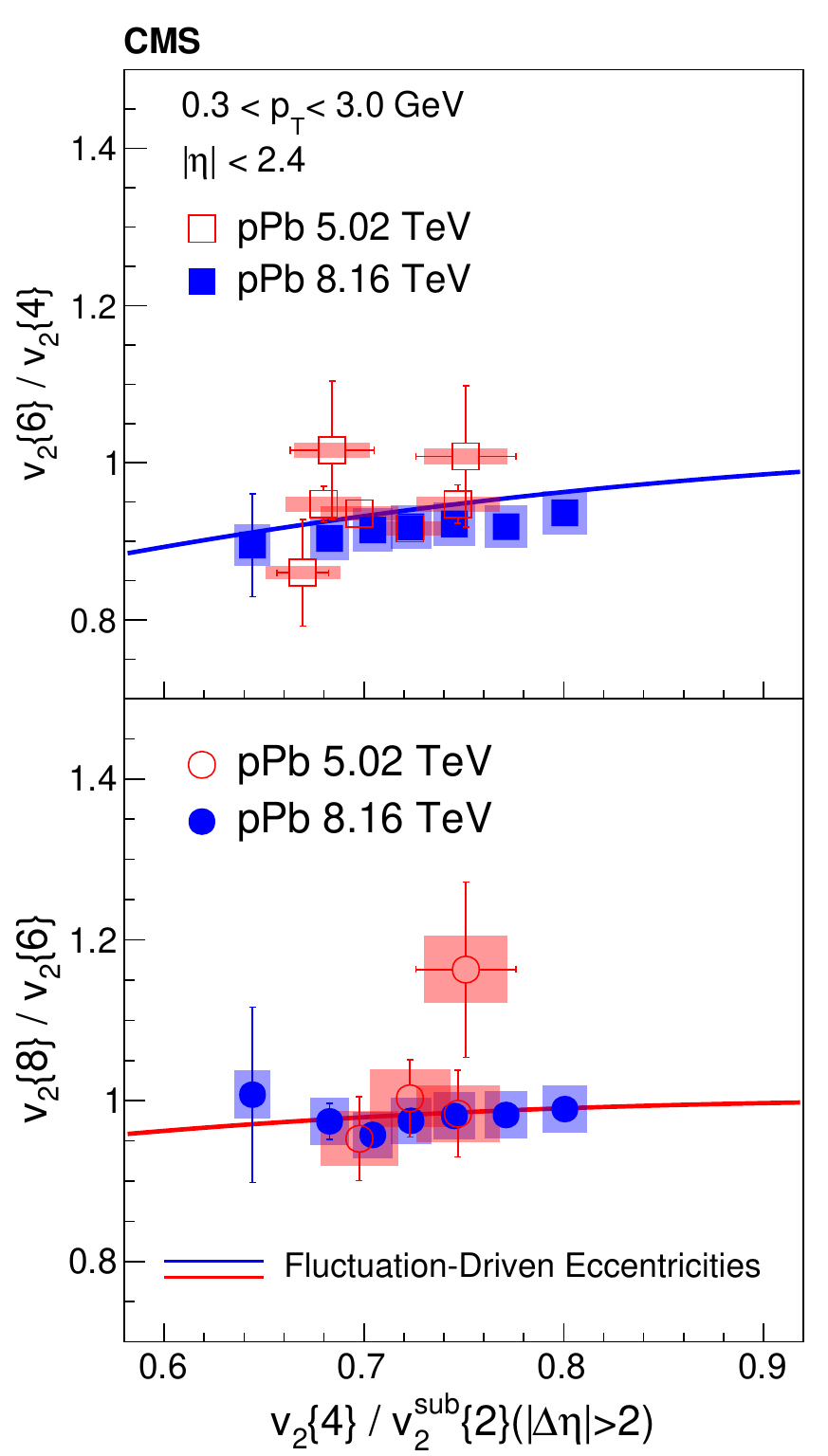}
    \caption{Cumulant ratios \vtworatio{6}{4} (upper) and \vtworatio{8}{6} (lower) as functions of $\vtwo{4}/v^{\text{sub}}_2\{2\}$ in \pPb collisions at 5.02 and 8.16\TeV.
        The solid curves show the expected behavior based on a hydrodynamics-motivated study of the role of initial-state fluctuations~\cite{Yan:2013laa}. \FigureFrom{CMS:2019wiy}}
    \label{fig:small:CMS-HIN-17-004_Figure_003}
\end{figure}

\begin{figure}[ht]
    \centering
    \includegraphics[width=\linewidth]{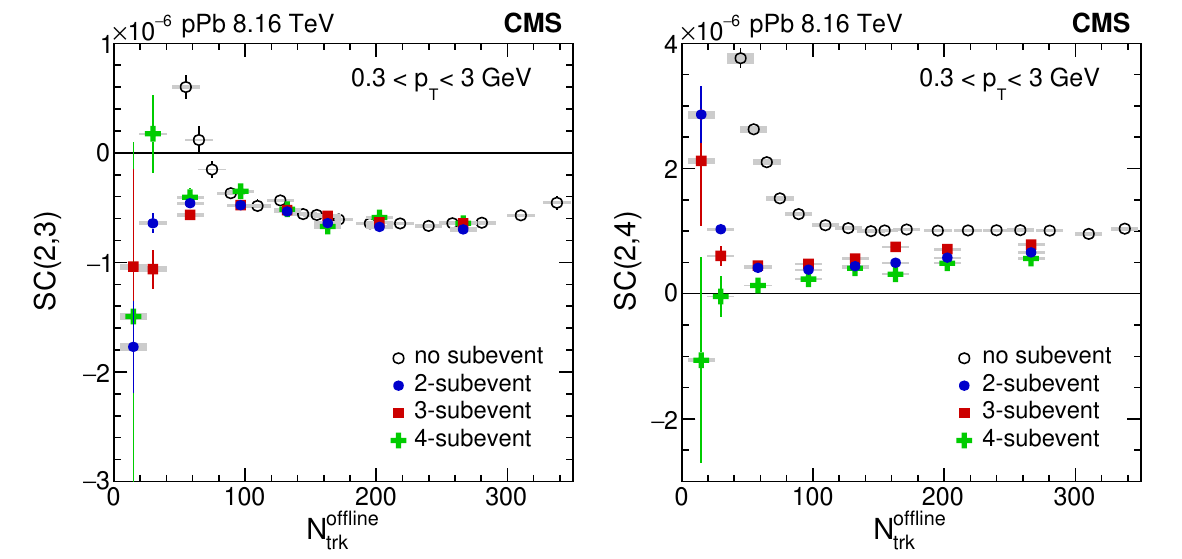}
    \caption{The \SCnm{2}{3} (left panel) and \SCnm{2}{4} (right panel) distributions as functions of \noff from methods using no (open black circles), 2 (full blue circles), 3 (red squares), and 4 (green crosses) subevents  for \pPb at 8.16\TeV. Statistical and systematic uncertainties are shown by vertical bars and shaded boxes, respectively.~\FigureFrom{CMS:2019lin}}
    \label{fig:small:CMS-HIN-18-015_Figure_001}
\end{figure}

The differences between cumulants of different orders originate from fluctuations in the eccentricity distribution in the initial state~\cite{Yan:2013laa}.
To further investigate whether the flow coefficients in small systems are directly related to the geometry of the initial stage, as is the case for larger \AonA systems, the ratios \vtworatio{6}{4} and \vtworatio{8}{6} as functions of the ratio \vtworatio{4}{2} are measured. These values can then be compared with the same ratios found using initial-state eccentricities resulting from geometry fluctuations. Figure~\ref{fig:small:CMS-HIN-17-004_Figure_003} shows these comparisons for \pPb collisions at 5.02 and 8.16\TeV. The agreement of the calculations with the data shows that the differences found among the multiparticle cumulant results for the \vTwo values can be described by initial-state fluctuations~\cite{Yan:2013laa}. These results confirm the hypothesis that multiparticle correlations originate from the multiplication of single-particle correlations with respect to symmetry planes. These single-particle correlations stem from source fluctuations related to the overall collision geometry, similar to what is observed in larger collision systems~\cite{CMS:2019wiy}. 

The values of the initial-state eccentricities $\epsilon_{2}$ and $\epsilon_{3}$ quantify the degree to which the initial state of an event has an elliptic or triangular geometry, respectively. Assuming that the flow coefficients \vTwo and \vThree are proportional to these eccentricities, the event-by-event correlation between \vTwo and \vThree should be negative, so long as the hydrodynamic evolution of the system maintains this proportionality. One technique to extract this correlation is by measuring the symmetric cumulant \SC, which correlates the Fourier coefficients of order $m$ and $n$,
\begin{linenomath}
    \begin{equation}
\label{eq:small:sc}
\SC = \left<\vN^2v^{2}_{m}\right> - \left<\vN^2\right>\left<v^{2}_{m}\right>,
\end{equation}
\end{linenomath}
where $\left<\ldots\right>$ denotes the average over all events.
To remove nonflow effects, \SC can be measured using different subevent methods~\cite{Jia:2017hbm, CMS:2017kcs, CMS:2019lin}. In the subevent approach, every event is subdivided into multiple subevents, each of which spans a distinct rapidity range. 
A negative correlation between \vTwo and \vThree has been observed in large collision systems~\cite{ATLAS:2015qwl, 1604.07663}. Figure~\ref{fig:small:CMS-HIN-18-015_Figure_001} shows \SCnm{2}{3} (left panel) and \SCnm{2}{4} (right panel) as functions of \noff using no, 2, 3, and 4 subevents for \pPb at 8.16\TeV. Nonflow contributions are suppressed by using multiple subevents. A clear anticorrelation is observed
between the single-particle anisotropy harmonics \vTwo and \vThree, while \vTwo and \vFour are positively correlated. These results provide further evidence for the onset of long-range collective behavior in high
multiplicity events in small systems. 

Significant progress has been made, both theoretically and experimentally, towards understanding collectivity in small systems~\cite{Dusling:2015gta, Nagle:2018nvi}. In addition to explanations using hydrodynamic models, there are alternative interpretations such as parton scattering~\cite{Xu:2007ns, Bzdak:2014dia} and initial-state momentum correlation~\cite{Dumitru:2010iy, Dusling:2012iga}. 
Several additional observables have been proposed to distinguish between the various interpretations, including correlating \vTwo and mean \pt values~\cite{Giacalone:2020byk} and studying QCD collectivity in a single-parton system propagating in vacuum~\cite{2104.11735}. Future experiments, and possibly new observables, are expected to further enhance our understanding of the origins of azimuthal correlations and collectivity in small collision systems.

\subsubsection{Exploring small system collectivity using heavy-flavor particles}

As a consequence of their large masses, heavy quarks (charm and bottom) are primarily produced in the early stages of collisions. If a QGP is formed, heavy-quark interactions with the medium will probe its entire evolution~\cite{Braun-Munzinger:2007fth}. 
Flow measurements for heavy-flavor mesons in HI collisions at RHIC~\cite{Adamczyk:2017xur} 
and the LHC~\cite{Abelev:2014ipa,Acharya:2017qps,Sirunyan:2017plt} suggest that charm quarks develop a strong collective behavior, 
similar to that for light-flavor particles, which are primarily produced from the bulk of the QGP. 
In small systems, collective flow of heavy-flavor mesons, and especially 
the comparison to results for light hadrons, can impose further constraints 
on interpretations of the origin of the observed collectivity. 

\begin{figure}[t]
    \centering
    \includegraphics[width=0.6\linewidth]{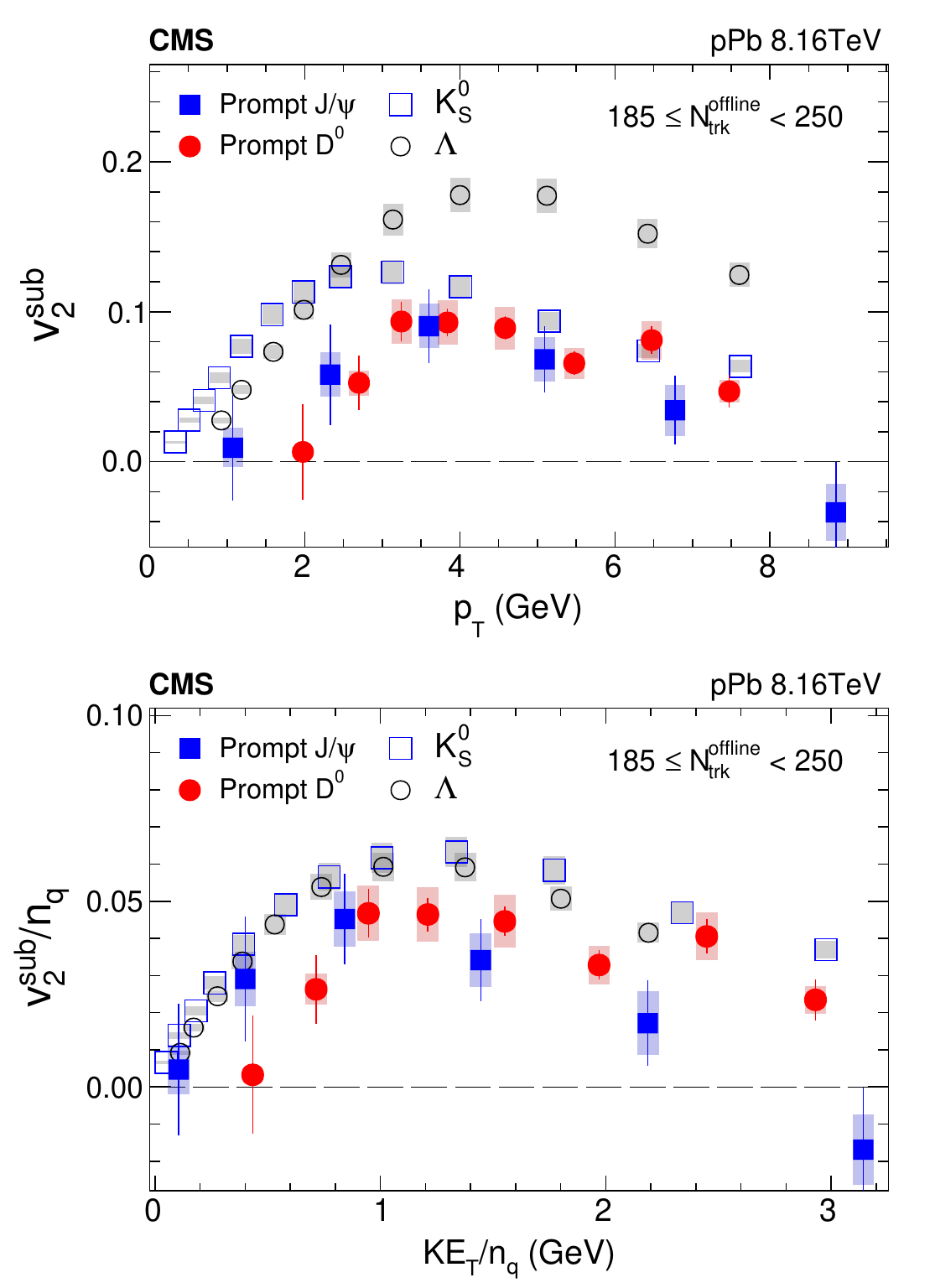}
    \caption{Upper: The \vTwoSub values for prompt \cPJgy\ mesons at forward rapidities ($-2.86<\ycm<-1.86$ or $0.94<\ycm<1.94$), as well as for \PKzS\ and \PGL hadrons, and prompt \PDz\ mesons at midrapidity ($-1.46<\ycm<0.54$), as a function of \pt\ for \pPb\ collisions at $\sqrtsNN = 8.16\TeV$ with $185 \leq \noff < 250$. Lower: The \nq-normalized \vTwoSub results.
    The vertical bars correspond to statistical uncertainties, while the shaded boxes denote the systematic uncertainties. \FigureFrom{CMS:2018duw}}
    \label{fig:CMS-HIN-18-010_Figure_003}
\end{figure}

Collective flow measurements have been performed 
for \PDz and \JPsi mesons using CMS data of \pPb collisions at 
$\sqrtsNN = 8.16\TeV$ in 2016~\cite{CMS:2018loe,CMS:2018duw} and of \pp collisions at $\sqrts = 13\TeV$ 
in 2017 and 2018~\cite{CMS:2020qul}. 
Because of the asymmetric beam energies in \pPb collisions at 
$\sqrtsNN = 8.16\TeV$ (6.5\TeV for the protons and
2.56\TeV per nucleon for the lead nuclei),
particles selected with the laboratory rapidity \ylab 
have a corresponding nucleon-nucleon center-of-mass
frame rapidity $\ycm = \ylab - 0.46$.
The upper panel of Fig.~\ref{fig:CMS-HIN-18-010_Figure_003} shows elliptic flow results after subtracting jet correlations ({\vTwoSub}) for 
prompt \cPJgy\ mesons at forward rapidities 
($-2.86<\ycm<-1.86$ and $0.94<\ycm<1.94$ combined), 
as well as for \PKzS\ and \PGL hadrons and prompt \PDz\ mesons at midrapidity 
($-1.46<\ycm<0.54$), 
for high-multiplicity ($185 \leq \noff < 250$) \pPb\ collisions, as a function of \pt from 0.2 to 10\GeV.
Positive \vTwoSub values are observed for prompt \PDz\ and \cPJgy\ mesons, with an initial increase 
up to $\pt \approx 4\GeV$, and then a slow decrease toward higher \pt. 
Over the full \pt range, the \vTwoSub values for these two mesons
are consistent with each other within uncertainties, and are smaller than those for \PKzS\ and \PGL hadrons. This observation indicates 
that charm quarks develop a collective response to the bulk medium 
in this small system, albeit weaker than that for light quarks.

A recent model calculation of \cPJgy\ meson \vTwo in \pPb collisions, based on 
final-state interactions between produced charm quarks and a QGP medium, 
predicts far smaller values than seen in the data~\cite{Du:2018wsj}. 
This calculation suggests that additional contributions, \eg, those
from initial-state interactions, may be needed to account for the observed 
\vTwo signal for prompt \cPJgy\ mesons in high-multiplicity \pPb\ events.

Motivated by the quark coalescence model~\cite{Molnar:2003ff,Greco:2003xt,Fries:2003vb}, 
collective flow at the partonic level is investigated by studying the scaling 
properties of \vTwoSub divided by the number of constituent quarks ({\nq}), 
as a function of the transverse kinetic energy per constituent quark ($\ket/\nq$, where $\ket = \sqrt{\smash[b]{m^2 + \pt^2}} - m$).
The lower panel of Fig.~\ref{fig:CMS-HIN-18-010_Figure_003} shows the same data as the upper panel, but now as a function of \ket\ with both \vTwoSub and \ket\ normalized by \nq.
The observed similarity of \nq-normalized \vTwoSub values 
for the \PKzS\ meson and \PGL baryon is known as 
number-of-constituent-quark~(NCQ) scaling~\cite{Khachatryan:2014jra,Adams:2003am,STAR:2007afq,PHENIX:2006dpn}, 
indicating that collective behavior is first developed among the partons,
which later recombine into final-state hadrons.
The values of ($\vTwoSub/\nq$) for 
prompt \PDz\ mesons are consistently smaller than those for the \PKzS\ meson and \PGL baryon. 
For \cPJgy\ mesons, $\vTwoSub/\nq$ values are consistent with 
those of \PKzS and \PGL hadrons within statistical uncertainties at lower $\ket/\nq$, while for $\ket/\nq \gtrsim 1\GeV$, the results are consistently smaller than those for the other two particles. 

\begin{figure}[t]
    \centering
    \includegraphics[width=0.6\linewidth]{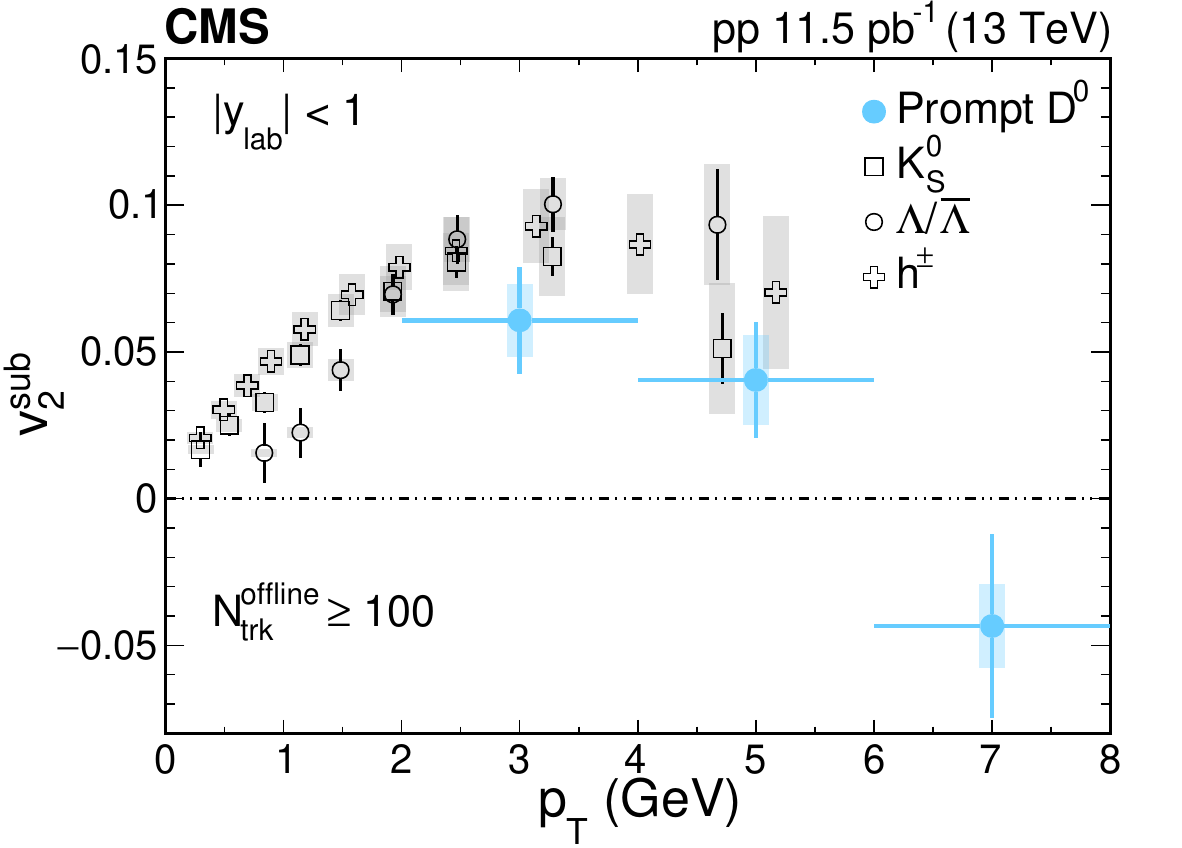}
    \caption{Results of \vTwoSub for prompt \PDz mesons, as a function of \pt for $\abs{\ylab}<1$, with $\noff \geq 100$ in \pp collisions at $\roots = 13\TeV$. The results for charged particles, \PKzS mesons, and \PGL baryons are shown for comparison. Vertical bars correspond to the statistical uncertainties, while the shaded boxes denote the systematic uncertainties. The horizontal bars represent the width of the \pt bins for prompt \PDz mesons. \FigureFrom{CMS:2020qul}}
    \label{fig:CMS-HIN-19-009_Figure_004}
\end{figure}

To investigate whether collective behavior of heavy-flavor quarks exists 
in even smaller systems, 
similar measurements have been performed for prompt \PDz mesons from \pp\ collisions at 
$\roots = 13\TeV$, with the \vTwoSub distribution presented in Fig.~\ref{fig:CMS-HIN-19-009_Figure_004} 
as a function of \pt for \PDz rapidity $\abs{\ylab}<1$ and event multiplicity $\noff \geq 100$.
The positive \vTwo signal ($0.061\pm0.018 \stat \pm0.013 \syst$) 
over a \pt range of ${\sim}2\text{--}4\GeV$ (with a declining trend toward higher \pt) provides an indication 
of collectivity for charm quarks in \pp collisions. 
The \vTwo magnitude for prompt \PDz mesons is found to be compatible with that for light-flavor hadron species, 
which suggests that collectivity is comparable (or slightly weaker) for charm hadrons than that for light-flavor hadrons in 
high-multiplicity \pp collisions.

\begin{figure}[ht]
    \centering
    \includegraphics[width=0.5\linewidth]{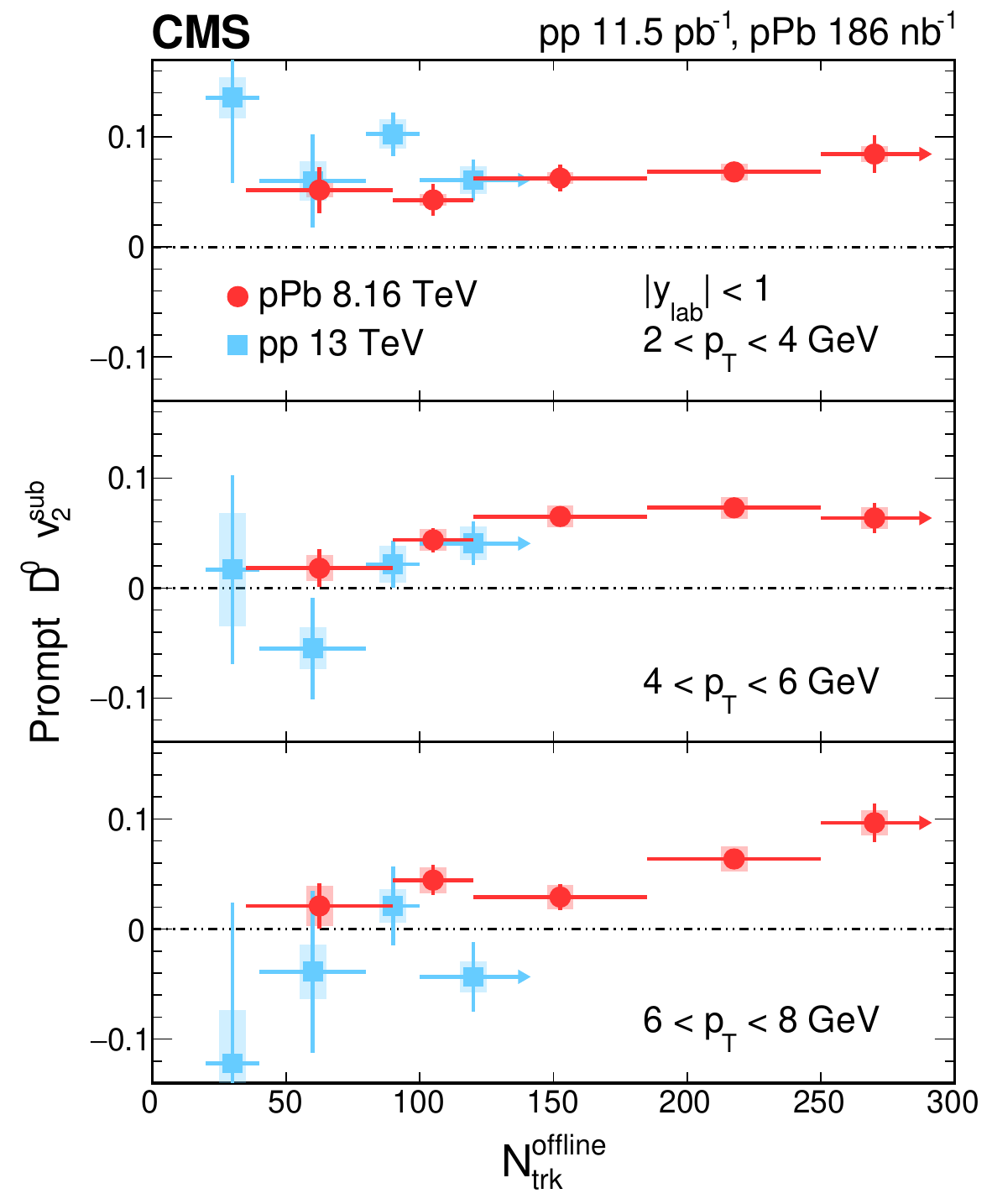}
    \caption{Results of \vTwoSub for prompt \PDz mesons, as a function of event multiplicity for three different \pt ranges, with $\abs{\ylab}<1$ in \pp collisions at $\roots = 13\TeV$, and \pPb collisions at $\sqrtsNN  = 8.16\TeV$. The vertical bars correspond to statistical uncertainties, while the shaded boxes denote the systematic uncertainties. Vertical bars extending beyond the y-axis are symmetric with respect to the central values. The horizontal bars represent the width of the \noff bins. The right-most points with right-hand arrows correspond to $\noff\geq 100$ for \pp collisions and $\noff\geq250$ for \pPb collisions. \FigureFrom{CMS:2020qul}}
    \label{fig:CMS-HIN-19-009_Figure_005}
\end{figure}

To further investigate the possible system size dependence of collectivity for 
charm hadrons in small colliding systems, 
\vTwo for prompt \PDz mesons in \pPb and \pp collisions are measured in 
different multiplicity classes. 
The prompt \PDz meson \vTwo as a function of event multiplicity for three different 
\pt ranges: $2<\pt<4\GeV$, $4<\pt<6\GeV$, and $6<\pt<8\GeV$ are presented 
in Fig.~\ref{fig:CMS-HIN-19-009_Figure_005}. 
At similar multiplicities of $\noff \sim 100$, 
the prompt \PDz \vTwo values are found to be comparable
within uncertainties in \pp and \pPb systems. 
For $2<\pt<4\GeV$, the measurement for prompt \PDz mesons
provides indications of positive \vTwo down to $\noff \sim 50$ with a significance of
more than $2.4$ standard deviations, while for $6<\pt<8\GeV$
clearly positive signals are only present in the higher-multiplicity 
region. Because of the large 
uncertainties, especially at low multiplicities, no clear multiplicity dependence can be determined for \pp results.

\begin{figure}[ht!]
    \centering
    \includegraphics[width=0.55\linewidth]{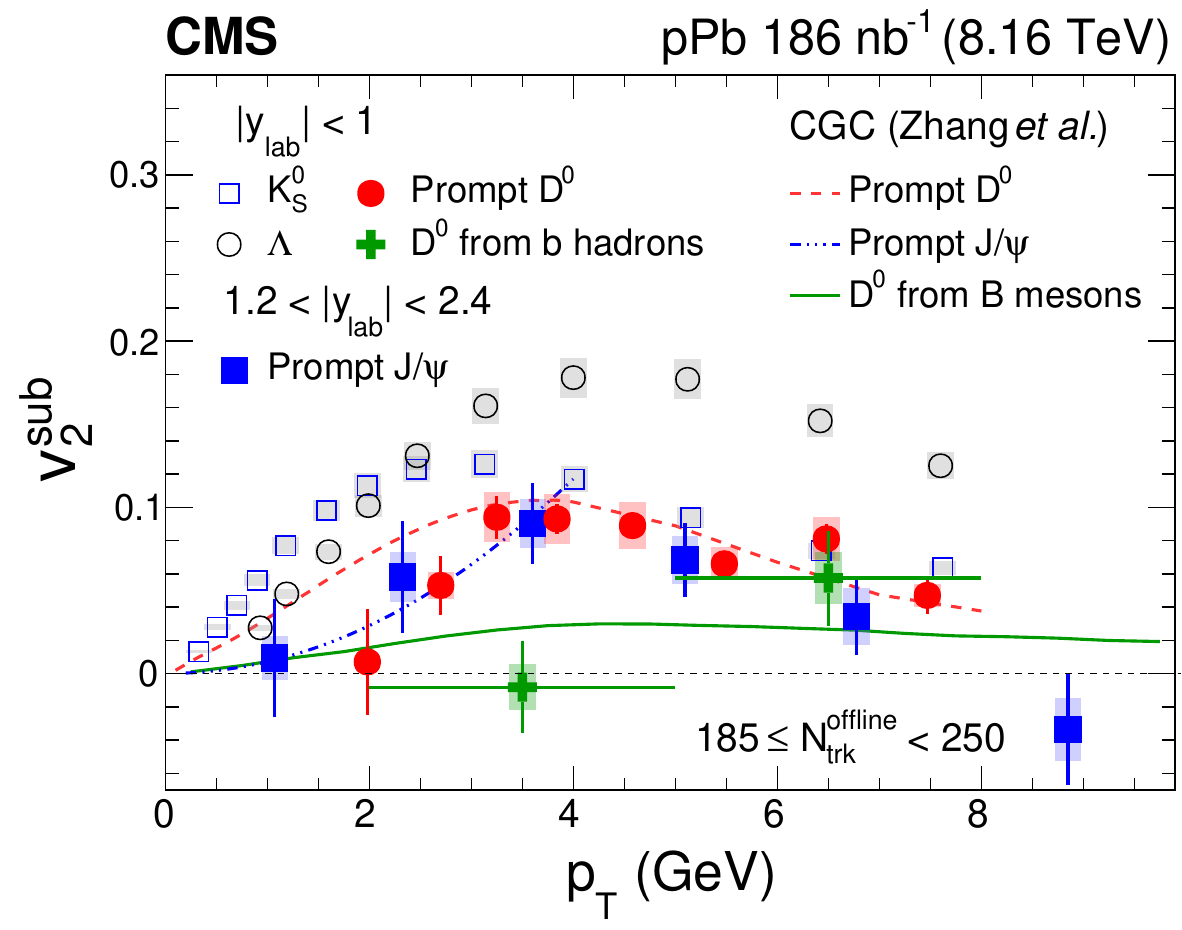}
    \caption{
    Results of \vTwoSub for prompt and nonprompt \PDz mesons, as well as \PKzS mesons, \PGL baryons for $\abs{\ylab}<1$, and prompt \PJGy\ mesons for $1.2<\abs{\ylab}<2.4$, as a function of \pt with $185 \leq \noff < 250$ in \pPb collisions at $\sqrtsNN  = 8.16\TeV$. 
    The vertical bars correspond to statistical uncertainties, while the shaded boxes denote the systematic uncertainties. The horizontal bars represent the width of the nonprompt \PDz \pt bins.
    The 
    red dashed, blue dash-dotted, and green solid lines show the theoretical calculations for prompt \PDz, \PJGy, and nonprompt \PDz mesons, respectively, within the CGC framework~\cite{Zhang:2019dth,Zhang:2020ayy}. \FigureFrom{CMS:2020qul}}
    \label{fig:CMS-HIN-19-009_Figure_006}
\vglue4mm
\centering
\includegraphics[width=0.65\textwidth]{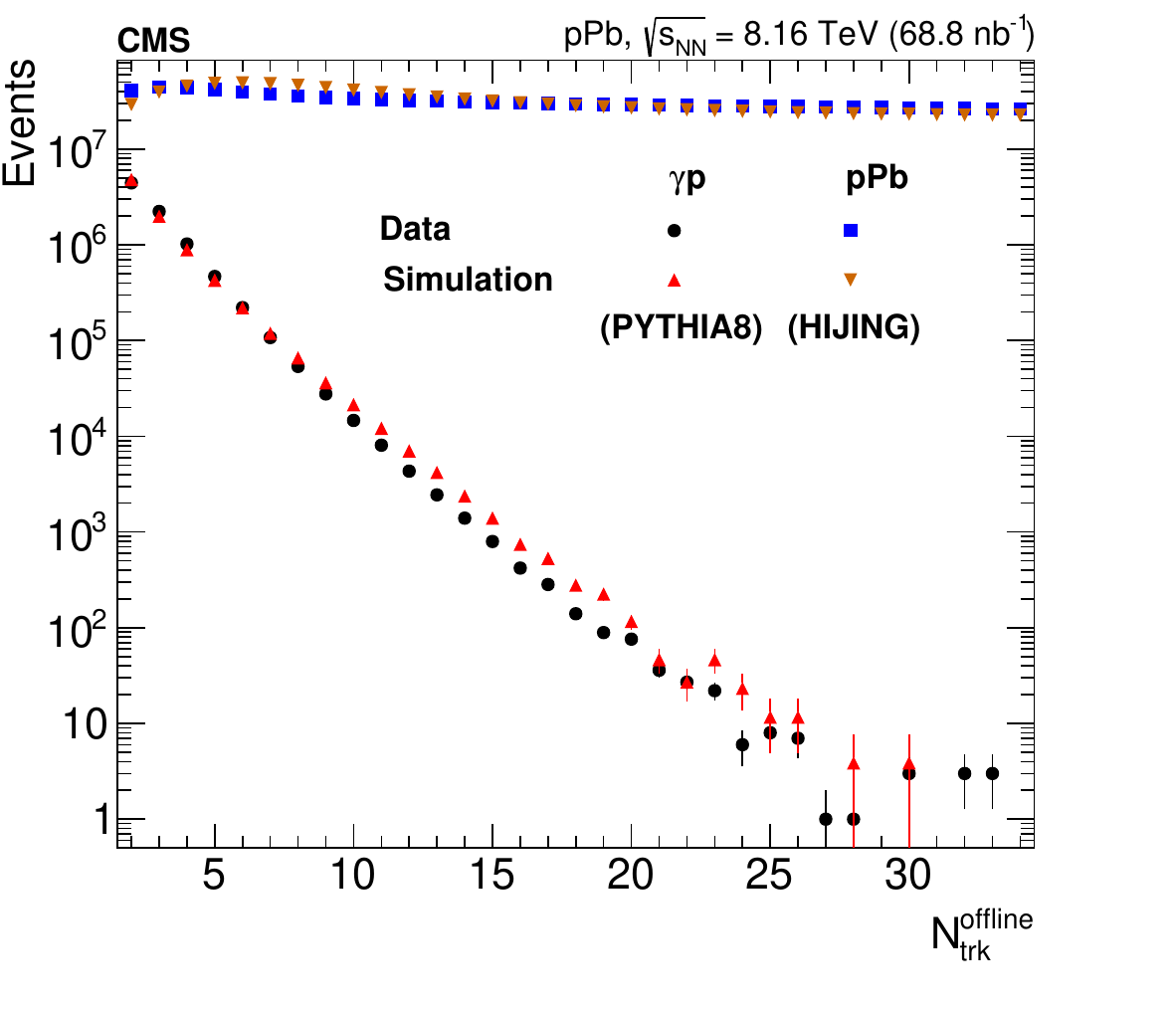}
\caption{The \noff spectra for \PhotonP and minimum bias \pPb samples. The simulated \noff distribution for \PhotonP events has been normalized to the same event yield as the \PhotonP-enhanced data sample.}
\label{fig:spectra}
\end{figure} 

Figure~\ref{fig:CMS-HIN-19-009_Figure_006} adds \vTwoSub values for nonprompt \PDz mesons from bottom hadron decays to the results shown in the upper panel of Fig.~\ref{fig:CMS-HIN-18-010_Figure_003}
for \pPb collisions at 8.16\TeV\ with $185 \leq \noff < 250$. 
At low \pt, the nonprompt \PDz \vTwo is consistent with zero, while at high \pt, 
a hint of a positive \vTwo value for \PDz mesons from \cPqb hadron decays is suggested.
At $\pt \sim 2\text{--}5\GeV$, the nonprompt \PDz meson \vTwo from bottom hadron decays 
is observed to be smaller than that for prompt \PDz mesons with a significance of 2.7 standard deviations, suggesting a flavor hierarchy of the collectivity signal 
that tends to diminish for the heavier bottom hadrons.
This is qualitatively consistent with the scenario of \vTwo being generated via final-state rescatterings, where heavier quarks tend to develop a weaker collective \vTwo signal~\cite{Dong:2019byy}.

Correlations at the initial stage of the collision between partons originating 
from projectile protons and dense gluons in the lead nucleus 
are able to generate sizable elliptic flow in the CGC framework~\cite{Dusling:2015gta,Zhang:2019dth,Zhang:2020ayy}.
These CGC calculations of \vTwo signals for prompt \PJGy mesons, as well as 
prompt and nonprompt (from \cPqb hadron decay) \PDz mesons, are also shown in Fig.~\ref{fig:CMS-HIN-19-009_Figure_006}.
Note that the parameterizations used in the CGC model in Ref.~\cite{Zhang:2020ayy} are unable to describe \vTwo for the full \PJGy meson \pt spectrum.
The qualitative agreement between data and theory suggests that 
initial-state effects may play an important role in the generation of collectivity 
for these particles in \pPb collisions. 

\subsubsection{Search for collective behavior in the smallest system limit}
\label{sec:SmallSystems_HIN18008}

Fourier coefficients ({\VnDelta}) of the azimuthal distributions of charged hadrons emitted in photon-proton ({\PhotonP}) interactions were measured using \pPb UPCs at $\sqrtsNN = 8.16\TeV$. Lead ions produce a flux of photons that interact with oncoming protons. This \PhotonP system provides a unique set of initial conditions with a multiplicity lower than that in photon-lead collisions but comparable to recent electron-positron and electron-proton data ~\cite{alepCorr:2019,alepCorr:2023, Belle:2022fvl,Belle:2022ars, zeus:Dec2019_ep, ZEUS:2021qzg, alepCorr:2023}.

Figure~\ref{fig:spectra} shows the \noff spectra for the \PhotonP-enhanced and MB data samples along with simulations from the {\PYTHIA{}8} and \HIJING event generators. The \noff average is 2.9 for the \PhotonP sample and 16.6 for the \pPb sample.
For a given multiplicity range, the mean \pt of charged particles is smaller in \PhotonP than in \pPb collisions.
For both the \PhotonP and \pPb samples, \VoneDelta is negative, \VtwoDelta is positive, and \VthreeDelta is consistent with 0. 

\begin{figure}
\centering
\includegraphics[width=.8\textwidth]{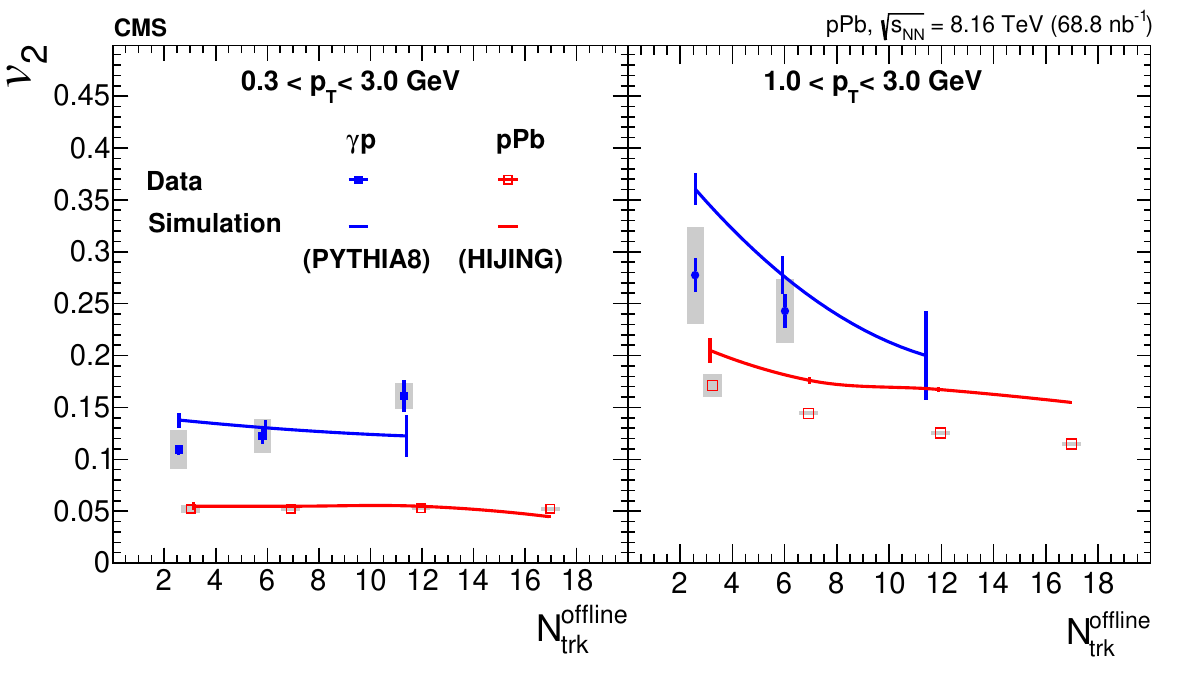}
\caption{ Single-particle azimuthal anisotropy \vTwo versus \noff for \PhotonP-enhanced and \pPb samples in two \pt regions. The systematic uncertainties are shown by the shaded bars in the two panels. 
Predictions from the {\PYTHIA{}8} and \HIJING generators are shown for the \PhotonP and MB \pPb samples respectively. For the \PhotonP events, the same \noff bin arrangement as in Fig.~\ref{fig:spectra} is kept, while for \pPb the bins $[2, 5)$, $[5, 10)$, $[10, 15)$, and $[15, 20)$ are used.}
\label{fig:v2Ntrack}
\end{figure}

Figure~\ref{fig:v2Ntrack} shows the single-particle $\vTwo=\sqrt{\VtwoDelta}$ as a function of \noff and two \pt regions for both \PhotonP and MB data sets. For $0.3 < \pt < 3.0\GeV$, the MB results are consistent with the previously published CMS results \cite{CMS:2017kcs}. Predictions from the \PYTHIA{}8 and \HIJING generators are also shown for \PhotonP and MB \pPb interactions, respectively. None of the models incorporates collective effects. An increase of \vTwo with \pt is evident in both the data measurements and the simulations, as shown in Fig.~\ref{fig:v2Ntrack}. However, both generators slightly exceed the data at higher \pt. It is noticeable that, for a given \pt and \noff, \vTwo is larger for \PhotonP than for \pPb interactions.

The \PhotonP data are consistent with model predictions that do not have collective effects. This suggests that the data are dominated by noncollective effects, \eg, back-to-back dijet production. Within the scope of the current experimental sensitivity, no substantial signal of collectivity is observed.
 
\subsection{Modification of quarkonium production in small systems}

\label{sec:SmallSystems_Quarkonia}

\begin{figure}[t]
    \centering
    \includegraphics[width=0.45\linewidth]{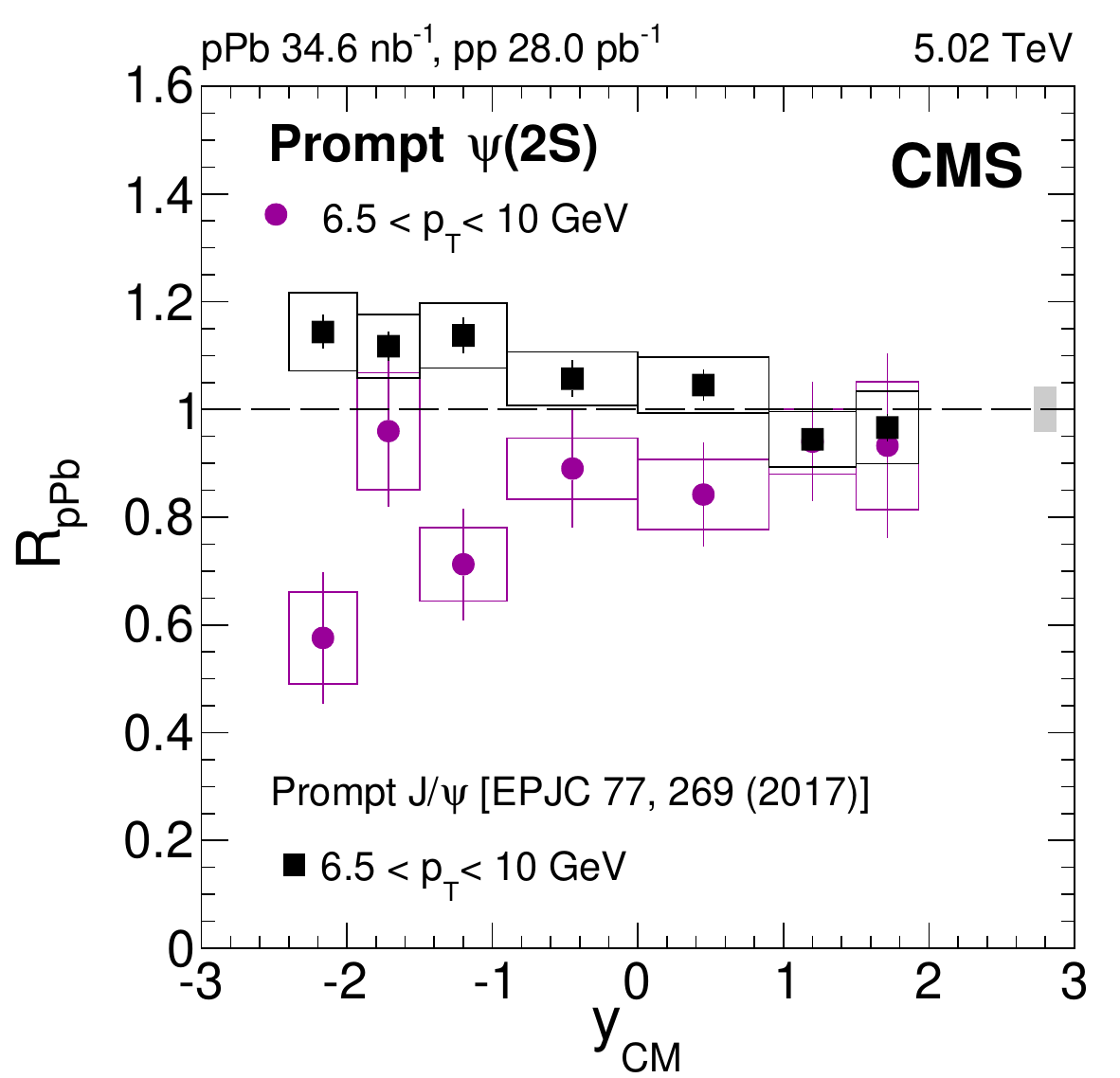}
    \includegraphics[width=0.45\linewidth]{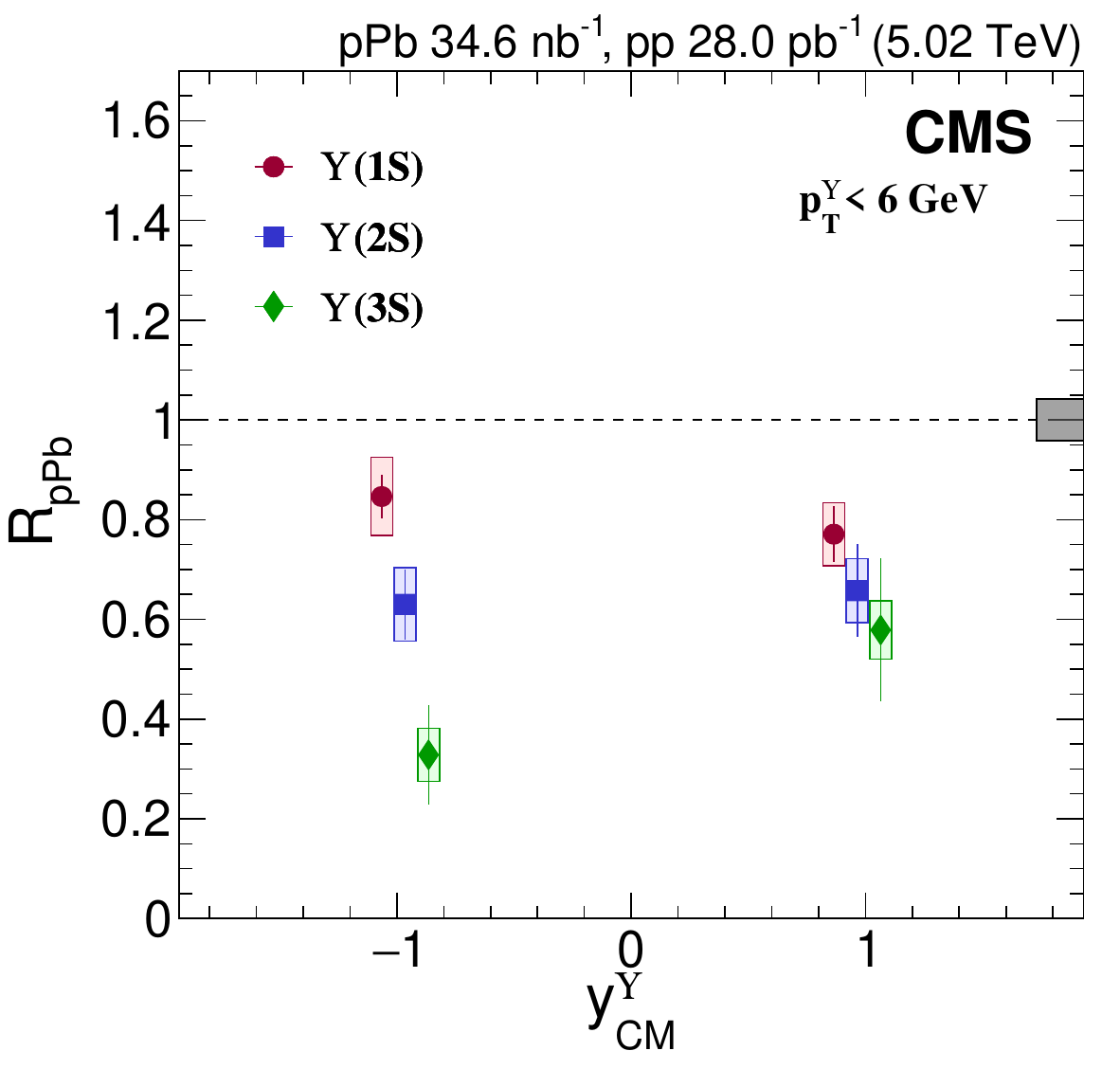}
    \caption{Left: Rapidity dependence of \RpPb for prompt \Pgy meson in the \pt range $6.5 <\pt< 10 \GeV$. For comparison, the prompt \JPsi meson nuclear modification factor is also shown.~\FigureCompiled{CMS:2017exb, CMS:2022wfi} Right: Nuclear modification factor of \PGUP{1S} (red dots), \PGUP{2S} (blue squares), and \PGUP{3S} (green diamonds) at forward and backward rapidity~\cite{CMS:2022wfi}. For both panels, statistical and systematic uncertainties are represented with vertical bars and boxes, respectively. The fully correlated global uncertainty of 4.2\%, affecting both charmonia equally, is displayed as the gray box around $\RpPb= 1$.}
    \label{fig:RpPbQuarkonia}
\end{figure}

As discussed in Section~\ref{sec:Quarkonia}, quarkonium states are powerful probes of the QGP that span its dynamic evolution from the early stage onward. The interaction of heavy quarks with the QGP generated in \PbPb collisions alters the yields of quarkonium states, depending on their binding energies. Understanding the initial- and final-state CNM effects is essential for interpreting the \PbPb data. To address this issue, the CMS Collaboration has conducted studies on quarkonium production in \pPb collisions. For charmonia, as depicted in the left panel of Fig.~\ref{fig:RpPbQuarkonia}, CMS identified differences in the suppression levels between the excited and ground states in the backward (lead-going) rapidity region~\cite{CMS:2017exb}. One interpretation posits that, as the charmonia are surrounded by more comoving particles (as is the case with the higher \dndeta in the backward direction) and the interaction probability rises, there is enhanced dissociation of the excited states. In contrast, in the forward (proton-going) region, these CNM effects diminish, and the nuclear modifications to both the ground and excited states are more similar. An analogous observation was made for bottomonia nuclear modification factors~\cite{CMS:2022wfi}, as shown in the right panel of Fig.~\ref{fig:RpPbQuarkonia}. These results suggest the presence of final-state effects in \pPb collisions, in line with predictions of models that include disintegration of bound quarkonium states via interactions with comoving particles from the underlying event.

In addition to these studies using inclusive \pPb collisions, the production cross section ratios of the excited \PGUP{2S} and \PGUP{3S} mesons relative to the \PGUP{1S} ground state have been examined as a function of the number of charged particles in \pp collisions at 7\TeV~\cite{CMS:2020fae}. These ratios were observed to decrease as the particle multiplicity increases, especially at low meson \pt values. Events including a \PGUP{1S} meson exhibited a multiplicity higher than that for the excited states, a discrepancy that cannot be solely attributed to feed-down contributions. Events were also categorized by sphericity, with high sphericity indicating a uniform, sphere-like emission and low sphericity indicating a narrow, jet-like emission. For \PGUP{nS} mesons with $\pt>7\GeV$, ratios of their production relative to that for \PGUP{1S} were seen to be independent of multiplicity in jet-like events (which have small sphericity). Furthermore, in jet-like events, the average number of charged particles per event remained consistent across all three \PGU states, suggesting that the variation in associated particle counts is not directly tied to mass differences between these states.

These measurements illustrate that interpreting the sequential disappearance of quarkonia in HI collisions requires a deep understanding of their elementary production processes and of the effect of the surrounding multiplicity in small systems. 

\subsection{Searches for jet quenching in small systems}
\label{sec:SmallSystems_JetQuenching}

\begin{figure}[ht]
    \centering
    \includegraphics[width=0.6\linewidth]{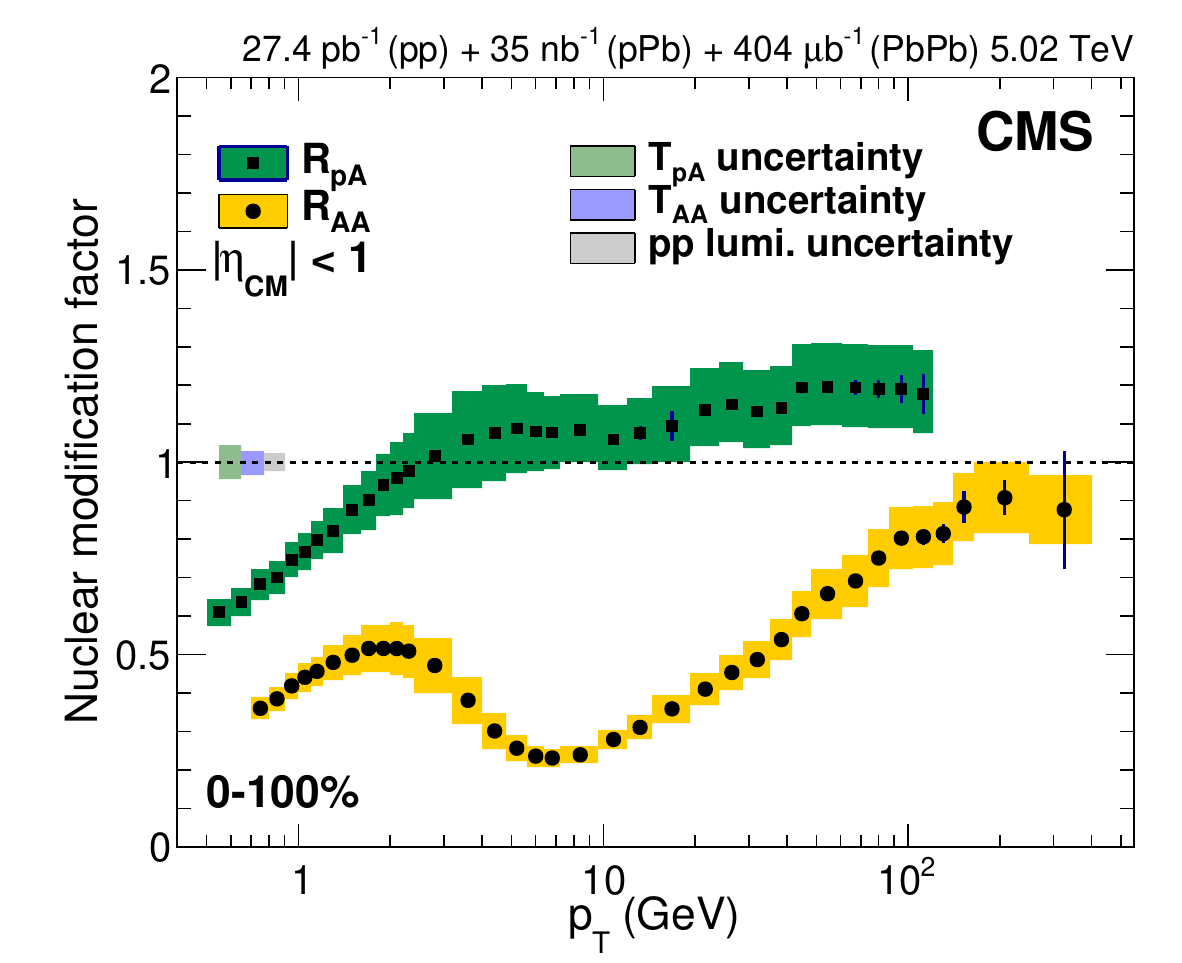}
    \caption{Nuclear modification factors versus \pt for an inclusive centrality selection for both \PbPb and \pPb collisions. The green and orange boxes show the systematic uncertainties for \RpA and \RAA, respectively, while the $T_{\Pp\mathrm{A}}$, \TAA, and \pp integrated luminosity uncertainties are shown as grey boxes around unity at low \pt. Statistical uncertainties are shown as vertical bars. \FigureFrom{CMS:2016xef}}
    \label{fig:RpPb}
\end{figure}

In Section~\ref{sec:hardQGP}, studies of jet quenching, a phenomenon
sensitive to the formation of a QGP, were discussed for HI collisions. As detailed in Section~\ref{sec:SmallSystems_Collectivity}, analyzing particle correlations in small collision systems revealed a significant flow-like signal, similar to observations that have been linked to QGP formation in larger systems. Moreover, measurements of quarkonium \RpA, presented in Section~\ref{sec:SmallSystems_Quarkonia}, suggest final-state effects in \pPb collisions. 
Consequently, these observations have motivated the investigation of the presence of jet quenching in small systems~\cite{Huss:2020whe}.

The simplest observable used to study jet quenching is the inclusive charged-particle nuclear modification factor, denoted as \RpA. Figure~\ref{fig:RpPb} shows both \RAA and \RpA for events integrated in all centralities as functions of \pt~\cite{CMS:2016xef}. The observation of an \RpA smaller than unity at $\pt<2\GeV$ can be attributed to initial-state effects, such as the nuclear shadowing and saturation effects discussed in Section~\ref{sec:InitialState}. On the other hand, at high $\pt$ the charged-particle \RpA is above unity, consistent with mild antishadowing effects in the intermediate-$x$ region~\cite{Huss:2020whe}. Since the data are consistent with models that include only initial-state effects, these results show no indication of jet quenching within uncertainties in inclusive hadronic \pPb collisions at 5.02\TeV. Moreover, this finding reinforces the conclusion that the smaller than unity charged-particle \RAA observed in \PbPb collisions is primarily due to jet quenching in the QGP.

\begin{figure}[ht]
    \centering
    \includegraphics[width=0.7\linewidth]{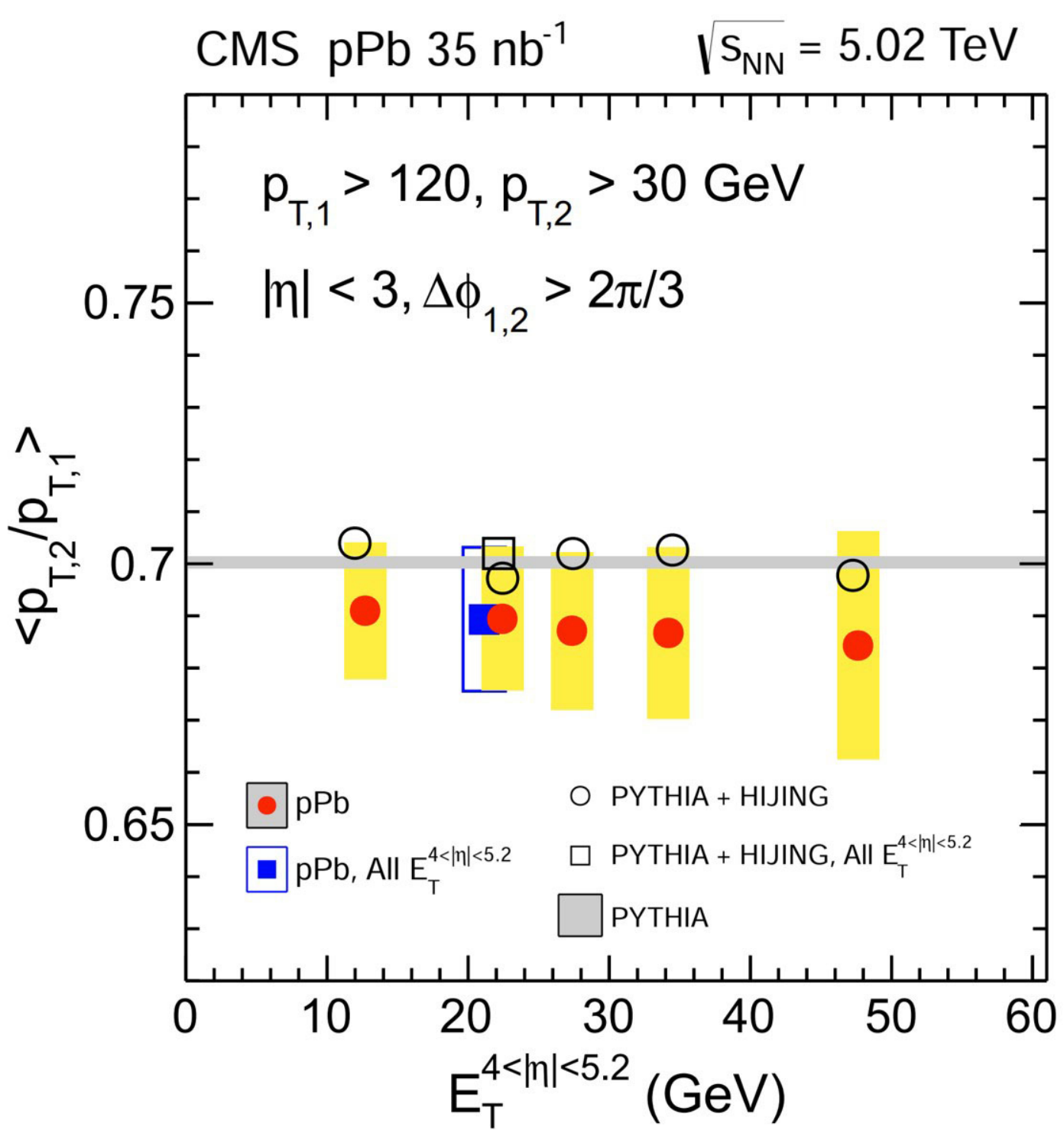}
    \caption{Average ratios of jet transverse momenta as a function of \ETfour. The inclusive HF activity results for \pPb and \textsc{pythia}+\textsc{hijing} are shown as blue solid and black empty squares, respectively. The systematic (statistical) uncertainties are indicated by the yellow, grey, and blue boxes (vertical bars). Various theoretical calculations are shown by the open square and circles and the grey band at about 0.7. \FigureFrom{CMS:2014qvs}}
    \label{fig:DijetInPPb}
\end{figure}

Additionally, dijet \pt asymmetry studies in \pPb collisions at 5.02\TeV~\cite{CMS:2014qvs} have been performed to explore the possibility of larger dijet \pt asymmetry than in the \pp reference, as observed in \PbPb collisions (discussed in Section~\ref{sec:InclusiveJetObservables}). The dijet $\pt$ balance \XJ, which is the ratio of the subleading ($\ptOne$) to leading ($\ptTwo$) jet \pt, is sensitive to differences between the jet quenching effects on the leading and subleading jets in the event. 
Selected MB and dijet events were divided into HF activity classes according to the raw transverse energy measured in the HF detectors within the pseudorapidity interval $4.0<\abs{\eta}<5.2$, denoted as \ETfour. This pseudorapidity interval is chosen to separate the event class selection and dijet measurements by a pseudorapidity gap of at least one unit ($3.0<\abs{\eta}<4.0$).
For all \ETfour classes, no significant modification of \XJ has been detected. In Fig.~\ref{fig:DijetInPPb}, the mean of \XJ in different event activity classes is compared to a \textsc{pythia+hijing} reference with no jet quenching effect. Even for events with the largest \ETfour, no significant deviation from the \textsc{pythia}+\textsc{hijing} reference is observed. This result provides valuable insights into the allowed size of any quenching effects. Furthermore, studies of jet fragmentation functions, as well as inclusive and charm jet nuclear modification factors in \pPb collisions, indicate no significant nuclear modifications when compared to \pp references~\cite{CMS:2016svx,CMS:2016wma}. This consistency indicates that, within the current experimental uncertainties, no significant modifications are observed in the jet fragmentation patterns in \pPb collisions.

An alternative approach to investigate jet quenching involves studying the azimuthal angle anisotropy of high-\pt hadrons through two-particle correlation functions. Nevertheless, it is crucial to acknowledge that measuring the high-\pt \vTwo in \pPb collisions presents challenges as a result of significant nonflow contributions. To address this issue, various techniques, such as selecting events with large rapidity gaps and subtracting low multiplicity events, have been employed to mitigate the impact of nonflow effects. Using these techniques, the CMS analysis reveals positive high-\pt \vTwo values up to $\pt = 8\GeV$, as depicted in Fig.~\ref{fig:CMS-HIN-19-009_Figure_006}. These findings suggest possible room for jet quenching effect in \pPb collisions at high \pt. However, it is important to note that the magnitude of \vTwo heavily depends on the nonflow subtraction method, which currently hinders reaching a conclusion when using \vTwo measurements in the search for jet quenching in small systems. In summary, the reported high-\pt jet and hadron results for \pPb collisions do not exhibit clear indications of jet quenching, setting important constraints on the size of any such effects. 

\subsection{Summary of results for small systems}

Studies of inclusive particle production across \pp, \pPb, and \PbPb collisions have demonstrated that the \PbPb system, at the same \rootsNN, converts energy into charged particles more efficiently than either the \pp or \pPb systems. In high-multiplicity \pPb collisions, transverse energy distributions and the mean \pt of charged particles are better described by the \textsc{epos lhc} generator, which incorporates hydrodynamical evolution, 
than by models without hydrodynamics. This underscores the importance of hydrodynamic effects in these systems.

The CMS Collaboration has observed long-range correlations indicative of charged-hadron collectivity in high-multiplicity \pp and \pPb collisions (and with some indication in low-multiplicity \pp collisions), similar to those seen in HI collisions. Multiparticle correlation analyses across \pp, \pPb, and \PbPb collisions provide strong evidence of collectivity. These studies have been extended to the \PhotonPb system using UPCs, where the data align with models excluding collective effects, highlighting the varying nature of collectivity in different collision systems.

Heavy-flavor meson collectivity has also been explored, with measurements of prompt \PDz and \cPJgy mesons in \pPb and \pp collisions suggesting a weaker collectivity signal for charm quarks compared to light quarks. Charm quarks exhibit positive elliptic flow even in low-multiplicity \pPb collisions, while bottom hadrons show weaker collective signals (albeit better measurements are needed to confirm this indication), pointing to a mass hierarchy in quark collectivity.

In the realm of quarkonia, CMS has identified distinct final-state interactions in \pPb collisions, particularly in the suppression patterns of excited versus ground states of \ccbar and \bbbar systems. These suppression patterns are more pronounced in the lead-going rapidity region, likely a consequence of increased interactions with comoving particles leading to greater dissociation of excited states. In contrast, the forward (proton-going) region shows similar suppression for both states, aligning with studies of \PGU mesons, where particle multiplicity influences the production ratios of excited-to-ground states. Additionally, in high-multiplicity \pp collisions, events with ground state \PGUP{nS} mesons tend to have more charged-particle tracks than those with excited states, suggesting that factors beyond feed-down contributions affect particle production.

Lastly, CMS has conducted searches for jet quenching in small collision systems. Studies using various observables, including inclusive charged-particle \RpA and jet fragmentation functions, have shown no detectable jet quenching in \pPb collisions at high \pt.
However, more precise measurements are needed to clarify, with improved significance, if jet quenching is completely absent or simply below the levels that can be probed with the current samples of \pPb collisions.

\cleardoublepage

\section{Tests of the electroweak sector and searches for new physics}
\label{sec:QEDBSM}

In addition to the nuclear hadronic interactions discussed in previous sections, EM interactions can also be studied using the CMS detector. In UPCs 
that occur at large separations of the colliding nuclei in the transverse plane (\ie, such that the strong interaction does not dominate the collision 
dynamics), very large EM fields are possible~\cite{Baur:2001tj,Bertulani:2005ru,Baltz:2007kq}. In HI collisions, the field of each ion can interact 
with that of the other ion, leading to particle production via \gaga and \PhotonA interactions. 

Two-photon interactions are fundamental processes that have previously been studied, in particular, at HERA and LEP~\cite{Baur:2002vr}. In general, \gaga 
measurements using UPC at the LHC have focused on QED processes and probing new physics phenomena. In HI collisions, studies of QED processes 
with strong EM fields benefit from a background-free environment due to the dominance of the \gaga process over central diffraction, both 
of which are characterized by substantial rapidity gaps~\cite{Engel:1996aa}. Many final states have been measured in UPC \gaga interactions 
of proton and/or lead beams at the LHC, as described in Refs.~\cite{CMS:2018erd,CMS:2020skx,CMS:2022arf} and references therein, including lepton 
pair production ($\gaga \to \Pep\Pem,\,\mumu,\,\tautau$). In combination with the identification of interactions in which at least 
one of the lead nuclei is excited, based on neutrons detected by the ZDC detectors (discussed in Section~\ref{sec:InitialState_VM}), these processes 
are studied over a wide range of nuclear impact parameters. More specifically, these effects may be enhanced in events with a higher number of 
neutrons emitted or depleted in events with a lower number of emitted neutrons~\cite{Baltz:2002pp}. At higher photon energies, where the photon 
flux is large, QCD two-photon processes are also of great interest, \eg, double vector meson production~\cite{Azevedo:2023vsz}, which complements 
multiparton scattering studies in \pp, \pPb, and \PbPb collisions~\cite{dEnterria:2017yhd}, and investigations of charmonium states to constrain 
their decay widths~\cite{Esposito:2021ptx,Fariello:2023uvh}. Since UPC calculations can be extended to include collisions with partial nuclear 
overlap, where dynamics related to the strong interaction is present, \gaga interactions have also been studied in peripheral nuclear collisions, 
as discussed in Ref.~\cite{Aaboud:2018eph} and references therein.

In the case of ultraperipheral \pA collisions, the proton can also interact with the EM field generated by the heavy ion. The \pA measurements 
extend the energy range accessible in photoproduction studies at HERA for several important processes. As discussed in Section~\ref{sec:InitialState}, 
the nonlinear QCD dynamics in heavy nuclei at small-$x$ gluon densities can be studied through heavy quark production by photon-gluon fusion when 
the gluon originates from the nucleus, or via diffractive dynamics when the gluon comes from a Pomeron~\cite{Roldao:2000ze}. Although these studies 
are also important for understanding the \AonA collision dynamics in the framework of collinear factorization at NLO in pQCD~\cite{Thomas:2016oms,Eskola:2022vpi} 
and in the dipole picture~\cite{Mantysaari:2022kdm}, this section focuses on interesting \gaga interaction processes in \PbPb collisions.

Stronger experimental limits on increasingly larger masses of BSM particles have made the potential discovery of these particles in \gaga 
processes at the LHC more challenging~\cite{CidVidal:2018eel}. Still, there are interesting regions of parameter space for BSM production that 
can be explored. At the LHC, higher \gaga invariant masses are accessible with \pp collisions and lower masses can be explored 
with \PbPb collisions~\cite{Bruce:2018yzs,dEnterria:2022sut}.
In \pp collisions, protons that have lost a few percent of their energy can be tagged~\cite{Fichet:2013gsa}, which makes it possible to 
study processes involving, for example, EW bosons ($\gaga \to \PWp\PWm,\,\PZ\PZ,\,\gaga$)~\cite{Begel:2022kwp}. 
Although the \PGg spectrum falls less rapidly for smaller charges---favoring proton over nuclear beams in the production of large invariant 
mass diphoton systems---each photon flux scales with the squared charge of the hadron, $Z^2$, such that the effective \gaga luminosities 
are significantly enhanced for ion beams (\eg, in the case of \PbPb collisions, the enhancement factor is $Z^4 = 5\ten{7}$). The Lorentz factor 
of the Pb beam at the LHC dictates the maximum quasireal
photon energy of approximately 80\GeV, leading to \gaga collisions of energies up to $\sqrt{s}\approx 160\GeV$. This is comparable to the 
center-of-mass energy achieved at LEP~\cite{Shao:2022cly} but with $Z^4$ enhanced production cross sections.

\begin{figure}[ht]
    \centering
    \includegraphics[width=0.7\textwidth]{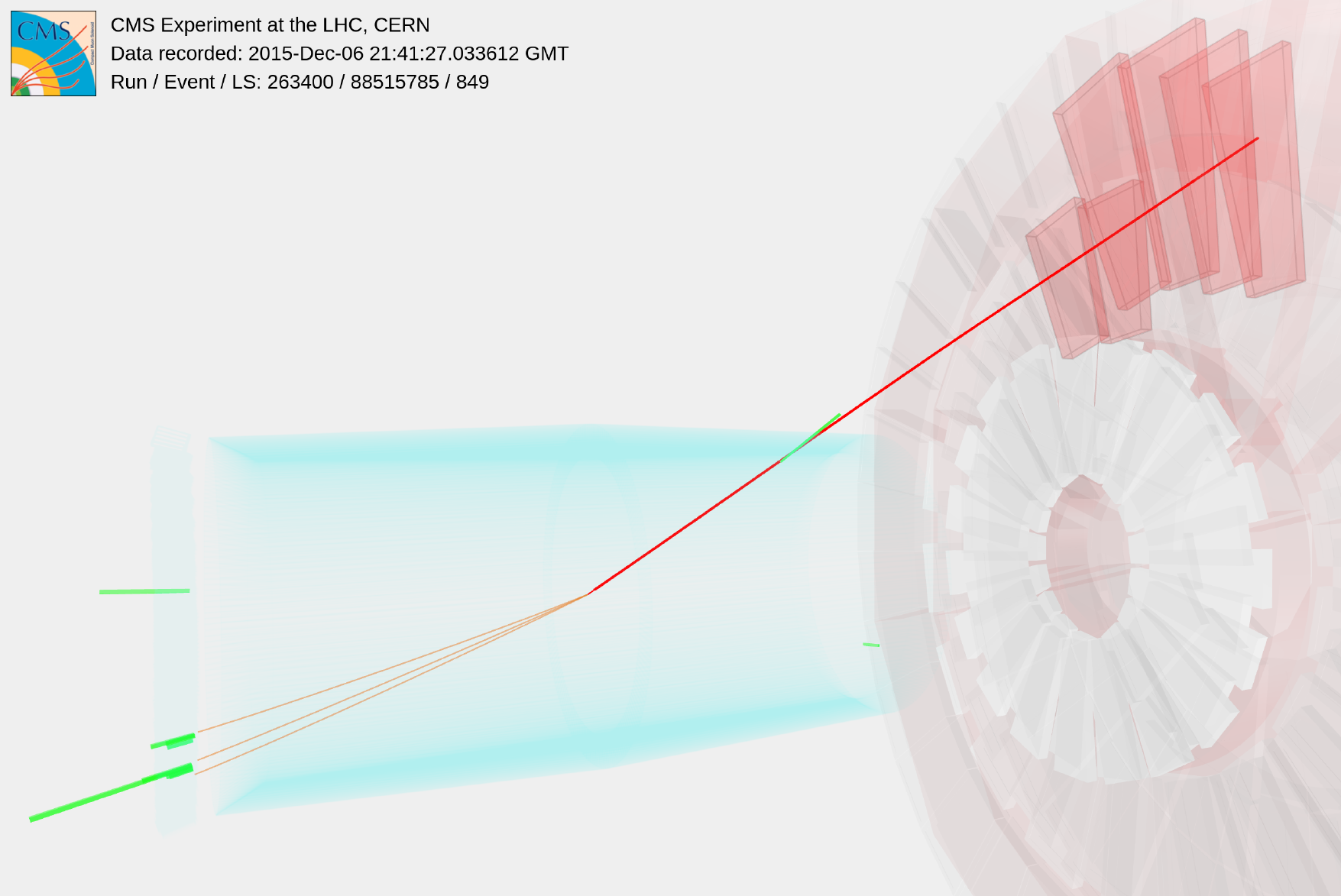}
    \caption{Event display of a candidate \gammagammatautau event measured in a \PbPb UPC at CMS. The event is reconstructed as corresponding 
    to a leptonic \PGt decay (red), \taumunu, and a hadronic \PGt decay (yellow), \tautripi. \FigureFrom{CMS:2022arf}}
    \label{fig:display}
\end{figure}

Therefore, a wide range of processes can be studied using \gaga interactions in UPCs. In the following, some examples of photon-induced 
processes using \PbPb UPC are described, including exclusive high-mass dilepton ($\mleplep\gtrsim5\GeV$) production (Sections~\ref{sec:SMAndBSM_SM_1} 
and~\ref{sec:SMAndBSM_SM_2}), the rare processes of light-by-light~(\LbL) scattering and \PGt lepton production (Section~\ref{sec:SMAndBSM_BSM1}), 
and BSM searches, \eg, for axion-like particles~(ALPs, Section~\ref{sec:SMAndBSM_BSM2}). The final-state signature of these studies is exceptionally 
clean. Figure~\ref{fig:display}, as an example, shows the interaction $\PbPb\to\gaga\to \mathrm{Pb}^{(\ast)}\tautau\mathrm{Pb}^{(\ast)}$, 
with a leptonic \PGt decay (red), \taumunu, and a hadronic \PGt decay (yellow), \tautripi. Otherwise, 
the central part of the detector is empty. Typically, outgoing Pb ions survive the interaction, whereas neutrons originating from a potential 
electromagnetic excitation (denoted by the superscript ``($\ast$)'') are detected at very high $\abs{\eta}$. Interestingly, the $\gaga \to \Pep\Pem$ 
production process, in which the electron is captured in a bound state with one of the ions (``bound-free pair production''), is the dominant 
beam-physics effect restricting the maximum \PbPb luminosity at the LHC~\cite{Bruce:2007zza}.

\subsection{The QED production of an exclusive muon pair}
\label{sec:SMAndBSM_SM_1} 

An experimental handle is essential to determine the impact parameter dependence of lepton pair production in UPCs~\cite{Baltz:2009jk}. As discussed 
in Section~\ref{sec:InitialState_VM}, the impact parameter of the UPC can be controlled by using forward-emitted neutron multiplicities from the 
electromagnetic dissociation~(EMD) of one or both of the Pb ions. In this way, we can disentangle possible contributions from initial-state (associated 
with the QED field strength) and final-state (\eg, multiple scattering in the QGP) effects that are both strongly dependent on the impact parameter. More 
specifically, a broadening of lepton pair azimuthal angle correlations (or, equivalently, an increase in the lepton pair \pt) is observed in hadronic 
collisions compared to those from UPCs. Alternative origins of this modification have been proposed, including final-state EM modifications of 
lepton pairs within a QGP medium~\cite{Adam:2018tdm,Aaboud:2018eph} or an impact parameter dependence of the initial photon \pt value~\cite{Zha:2018tlq,Li:2019sin,Wang:2021kxm}, 
or both processes combined. 

\begin{figure*}[t]
    \centering
    \includegraphics[width=\textwidth]{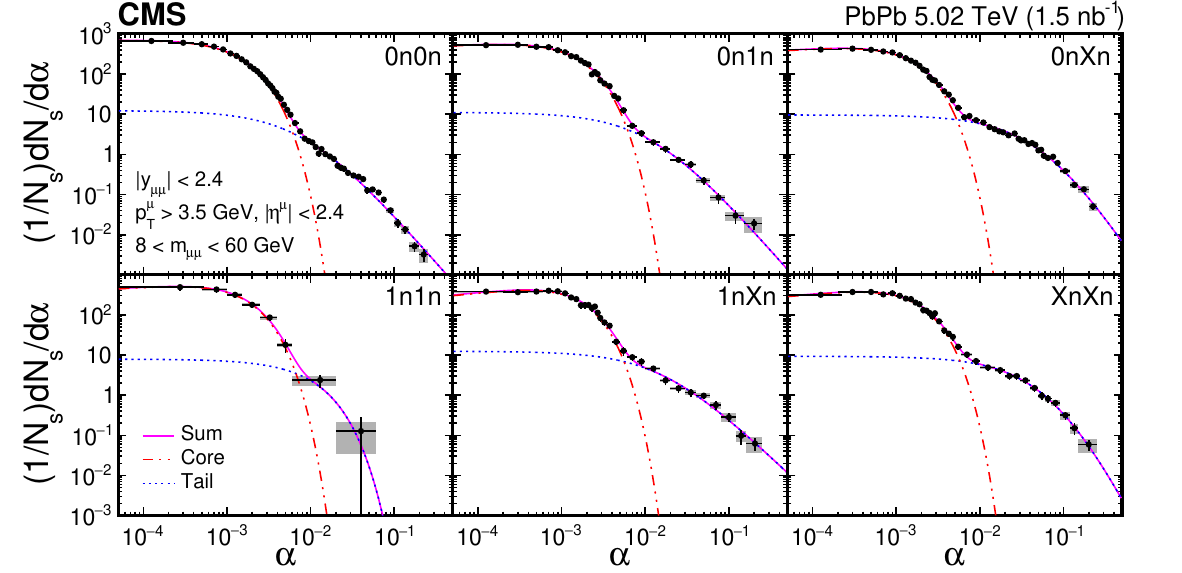}
    \caption{Neutron multiplicity dependence of acoplanarity distributions from \gammagammamumu in ultraperipheral \PbPb collisions 
    at $\sqrtsNN = 5.02\TeV$. The $\alpha$ distributions are normalized to unit integral over their measured range. The dot-dot-dashed and dotted 
    lines indicate the core and tail contributions, respectively. The vertical lines on data points depict the statistical uncertainties, while the 
    systematic uncertainties and horizontal bin widths are shown as gray boxes. \FigureFrom{CMS:2020skx}}
    \label{fig:alphaSpec}
\end{figure*}

Figure~\ref{fig:alphaSpec} shows the distributions of the acoplanarity, $\alpha = 1 - \abs{\phi^+ - \phi^-}/\pi$, of \mumu pairs for six 
neutron multiplicity classes in \PbPb UPCs at $\sqrtsNN = 5.02\TeV$. Here, $\phi^{\pm}$ represents the azimuthal angle of the positive and negative 
muons in the laboratory frame, so $\alpha$ characterizes the deviation from back-to-back azimuthal separation of the muon pair. The $0\Pn 0\Pn$ 
class corresponds to no Coulomb breakup of either nucleus, and the $1\Pn X\Pn$ ($X \geq 2$) class corresponds to one neutron emitted from one 
nucleus and at least two neutrons emitted from the other nucleus. Each $\alpha$ distribution is characterized by a narrow core close to zero (note 
the logarithmic horizontal scale) and a long tail. The core component originates mainly from LO \gammagammamumu scattering. In the tail 
component, however, higher order processes dominate, \eg, extra photon radiation from the produced lepton(s), multiple-photon interactions, and 
scattering of (one or both) photons emitted from one of the protons inside the nucleus~\cite{Baur:2007fv,ATLAS:2020epq}.

\begin{figure}[t]
    \centering
    \includegraphics[width=0.5\textwidth]{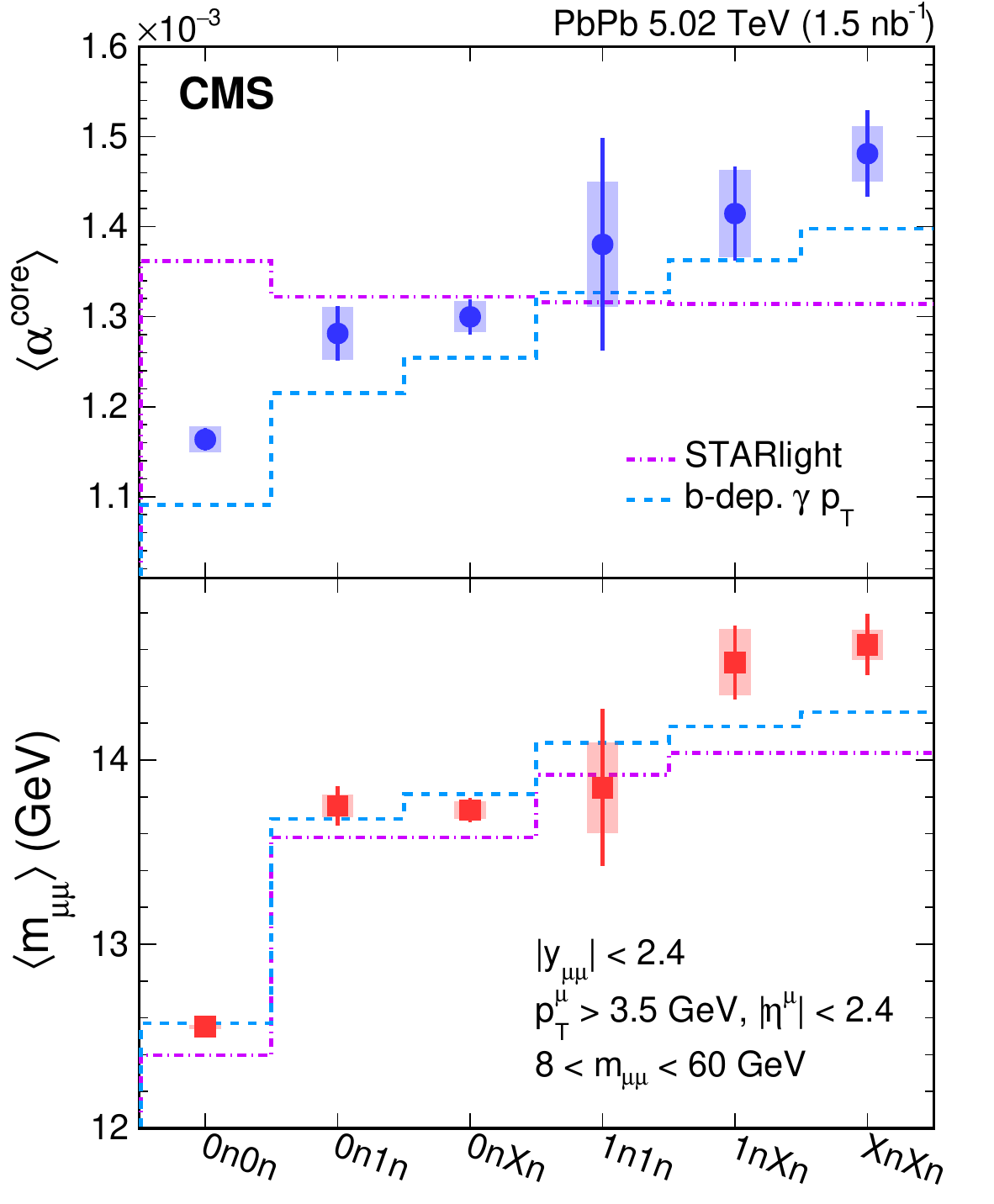}
    \caption{Neutron multiplicity dependence of the (upper) average acoplanarity $\langle \alpha^\text{core} \rangle$ and (lower) average invariant 
    mass $\langle m_{\PGm\PGm} \rangle$ of \mumu pairs in ultraperipheral \PbPb collisions at $\sqrtsNN = 5.02\TeV$. The vertical lines on 
    data points depict the statistical uncertainties while the systematic uncertainties of the data are shown as shaded areas. The dot-dashed line 
    shows the \Starlight MC prediction and the dashed line corresponds to the LO QED calculation of Ref.~\cite{Brandenburg:2020ozx}. The calculation 
    incorporating Sudakov radiative corrections is also compared to data in Ref.~\cite{CMS:2020skx}, leading to an overall better agreement. \FigureFrom{CMS:2020skx}}
    \label{fig:meanAlphaAndMassvsNeuNum}
\end{figure}

To investigate a possible dependence of the initial photon \pt value on the impact parameter, the core contribution to the $\alpha$ distribution 
is decoupled from the tail contribution using a two-component empirical fit function~\cite{CMS:2020skx}. The average acoplanarity of \mumu 
pairs from the core component~($\langle \alpha^\text{core} \rangle$) is then determined using the fit function. The neutron multiplicity dependence 
of $\langle \alpha^\text{core} \rangle$ for \mumu pairs is shown in Fig.~\ref{fig:meanAlphaAndMassvsNeuNum} (upper). A strong neutron multiplicity 
dependence of $\langle \alpha^\text{core} \rangle$ is clearly observed, while $\langle \alpha^\text{core} \rangle$ predicted by the \Starlight MC 
generator~\cite{Klein:2016yzr}, shown as the dot-dashed line in Fig.~\ref{fig:meanAlphaAndMassvsNeuNum} (upper), is almost constant. In contrast, 
the $\langle \alpha^\text{core} \rangle$ value in the data increases as the multiplicity of neutrons emitted increases. A constant value of $\langle \alpha^\text{core} \rangle$ 
as a function of the neutron multiplicity is rejected with a $p$ value corresponding to 5.7 standard deviations. A LO QED calculation~\cite{Brandenburg:2020ozx}, 
which incorporates an impact parameter dependence of the initial photon \pt, can qualitatively describe the increasing trend of $\langle \alpha^\text{core} \rangle$, 
as shown by the dashed line in Fig.~\ref{fig:meanAlphaAndMassvsNeuNum} (upper). This observation suggests that the \pt values of the initial photons 
producing \mumu pairs have a significant dependence on the impact parameter, which affects both the \pt and the acoplanarity of the muon 
pairs in the final state. This initial-state contribution must be properly taken into account when exploring possible final-state EM effects arising 
from a hot QGP medium formed in hadronic heavy ion collisions~\cite{Adam:2018tdm, Aaboud:2018eph}. 

In Fig.~\ref{fig:meanAlphaAndMassvsNeuNum} (lower), the average invariant mass $\langle m_{\PGm\PGm} \rangle$ of muon pairs is shown as a function 
of the neutron multiplicity. A clear neutron multiplicity dependence of $\langle m_{\PGm\PGm} \rangle$ is observed, with the $\langle m_{\PGm\PGm} \rangle$ 
value measured in $X\Pn X\Pn$ events being greater than that in $0\Pn 0\Pn$ events with a significance exceeding 5 standard deviations. This trend 
of $\langle m_{\PGm\PGm} \rangle$ can be qualitatively described by both model calculations. As the muon pair invariant mass is largely determined 
by the initial photon energy, this observation suggests that the energy of the photons is, on average, larger in collisions with a smaller impact 
parameter, a conclusion similar to that previously drawn for the initial photon \pt value.

\subsection{The QED production of an exclusive electron-positron pair}
\label{sec:SMAndBSM_SM_2}

\begin{figure*}[t]
\centering
 \includegraphics[width=\textwidth]{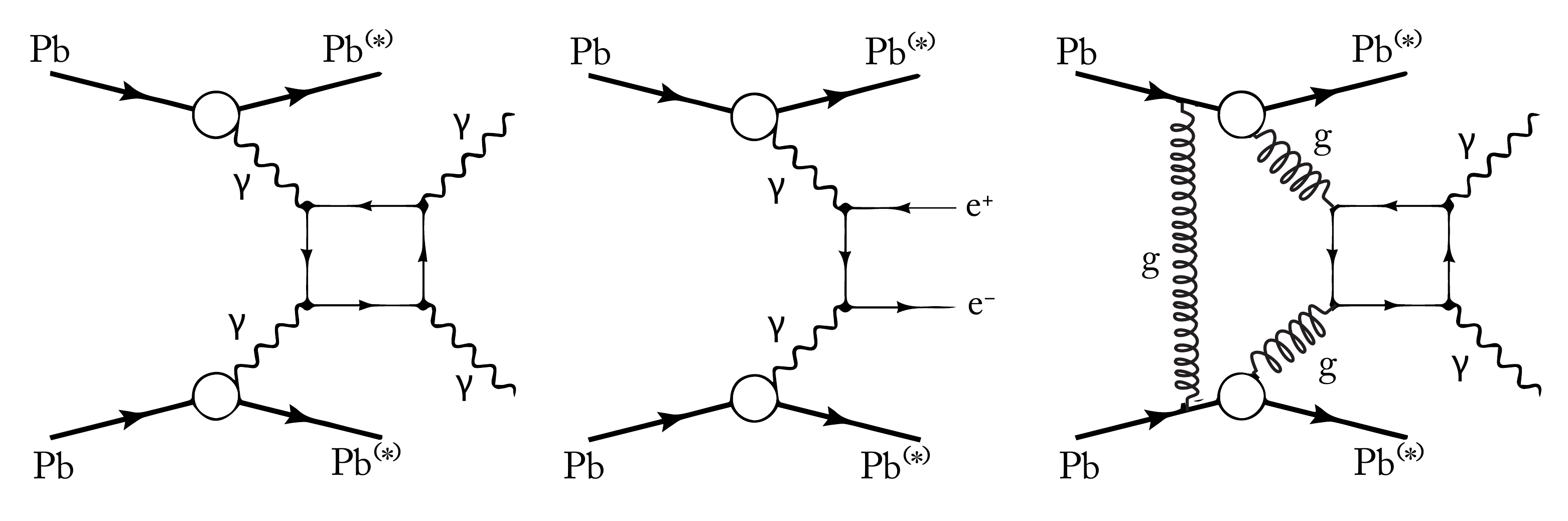}
 \caption{Schematic diagrams of light-by-light scattering ($\gaga \to \gaga$, left), QED dielectron
($\gaga \to \Pep\Pem$, center), and central exclusive diphoton ($\Pg\Pg \to \gaga$, right) production
in ultraperipheral \PbPb collisions. The ``($\ast$)'' superscript indicates a potential electromagnetic excitation
of the outgoing ions. \FigureFrom{CMS:2018erd}\label{fig:feynman}}
\end{figure*}

One of the possible backgrounds in the $\gaga \to \gaga$ final state (shown schematically in 
Fig.~\ref{fig:feynman}, left) is the QED production of an exclusive electron-positron pair (Fig.~\ref{fig:feynman}, center) and the gluon-induced 
central exclusive production~(CEP) (Fig.~\ref{fig:feynman}, right). Exclusive $\gaga\to\Pep\Pem$ events can be misidentified as $\gaga \to \gaga$ 
scattering in the case that neither electron track is reconstructed or when both electrons undergo hard bremsstrahlung. Given that the cross 
section for the $\gaga \to \Pep\Pem$ process is four to five orders of magnitude larger than that for $\gaga \to \gaga$ scattering, 
and its identification relies on physics objects (electrons) that closely resemble those of the signal (\PGg), a thorough analysis of the exclusive 
dielectron background is undertaken. This aims not only to estimate event-level efficiencies that are common for the dielectron and diphoton 
final states, but also to determine a $\gaga \to \gaga\,/\,\gaga \to \Pep\Pem$ production cross section ratio with reduced common uncertainties. 

In addition to the low-\pt (low-\et) online event selection and to the selection of physics objects (discussed in Section~\ref{sec:ExperimentalMethods}), 
the offline analysis includes a series of additional requirements to increase the signal component coming from exclusive \gaga production. More 
specifically, the so-called ``neutral and charged exclusivity'' selection criteria are applied to reject events that have additional
activity in the $\abs{\eta} < 5.2$ range.
For the neutral exclusivity criteria, events must have no activity in the calorimeters above energy noise thresholds, with noise thresholds determined 
from no- or single-bunch crossing events and separately in the barrel and endcap regions (discussed in Section~\ref{subsec:CMSapparatus}). For 
the charged exclusivity criteria, events with additional reconstructed charged-particle tracks above a certain low-\pt threshold (\eg, 0.1\GeV) 
are removed from further consideration. To further eliminate nonexclusive backgrounds, characterized by a final state with a larger vector sum of the
\pt values and larger diphoton acoplanarities $\aphi = (1-\Delta \phi^{\Pep\Pem}/\pi)$
than the back-to-back exclusive events, the transverse momentum and acoplanarity of the reconstructed systems are required to satisfy $\pt^{\Pep\Pem}< 1\GeV$ 
and $\aphi < 0.01$. These values are motivated by initial phenomenological studies~\cite{dEnterria:2013zqi} and further optimized based 
on similar CMS studies of exclusive dilepton production~\cite{Chatrchyan:2011ci,Chatrchyan:2012tv}.

\begin{figure}[t]
 \centering
  \includegraphics[width=0.55\textwidth]{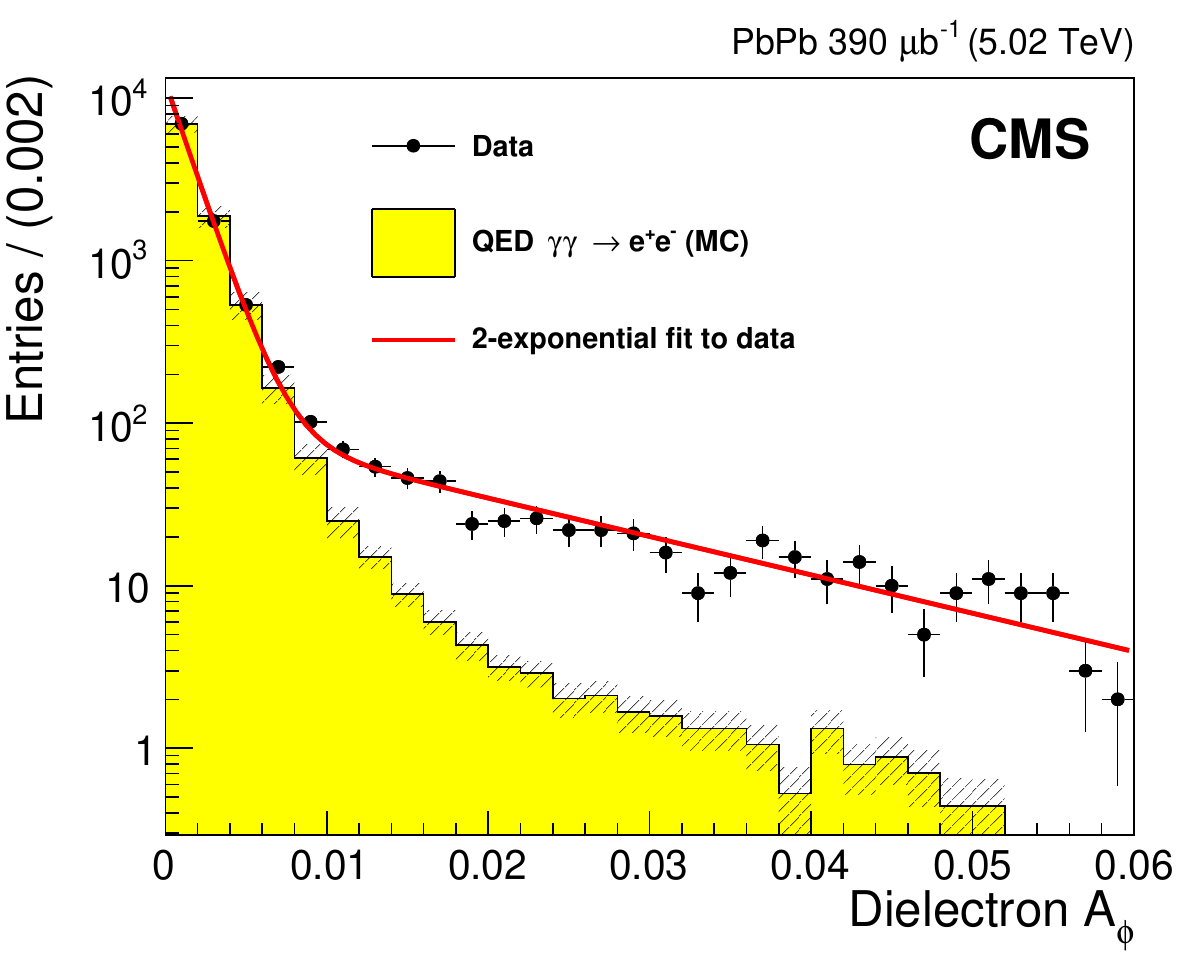}
\caption{\label{fig:qed_ee_acop} Acoplanarity distribution of exclusive $\Pep\Pem$ events measured in data (circles),
compared to the expected QED $\Pep\Pem$ spectrum in a LO MC simulation (histogram). The curve shows a $\chi^2$ fit to the sum of two exponential 
distributions, corresponding to exclusive $\Pep\Pem$ plus any residual (nonacoplanar) background pairs.
The error bars represent statistical uncertainties while the hashed bands around the histogram
represent the systematic and MC statistical uncertainties added in quadrature. The horizontal bars indicate the bin size. \FigureFrom{CMS:2018erd}}
\end{figure}

Figure~\ref{fig:qed_ee_acop} shows the $\Pep\Pem$ acoplanarity distribution measured in the data, compared to MC expectations. The curve is a 
binned $\chi^2$ fit of the data to the sum of two exponential functions, representing
the exclusive QED $\Pep\Pem$ production plus any residual background in the high-\aphi tail. It should be noted that using the 2015 \PbPb 
data set (Table~\ref{tab:tabHIN} in Section~\ref{sec:ExperimentalMethods_trigger}) approximately ten thousand dielectron events are reconstructed 
in the signal-dominated region of $\aphi<0.01$ with a purity of almost unity (as obtained
from the ratio of amplitudes of the two exponential functions fitted to the data). The yellow histogram shows the same distribution, obtained 
directly from a LO QED MC simulation. A small difference is found between the average $\text{A}_{\phi}$ value obtained from the data and the MC 
prediction, resulting from the higher experimental yields for events with $\aphi>0.01$. This is probably the result of $\gaga\to\Pep\Pem$ 
events where one (or both) electrons radiate an extra soft photon, which are not explicitly simulated with LO MC event generators, and/or any 
residual background surviving the event selection. When integrated over the whole range of the distributions, these discrepancies modify the measurements 
below the current level of precision, and hence do not significantly alter the interpretation of the data. However, their influence on the accuracy 
of the extracted cross sections will gradually increase with the accumulation of a larger amount of luminosity. Some recent progress has been 
reported towards understanding higher-order QED corrections in more detail, particularly those resulting from
final-state photon radiation from the leptons~\cite{Harland-Lang:2023ohq}. 

\begin{figure*}[t]
 \centering
  \includegraphics[width=0.4\textwidth]{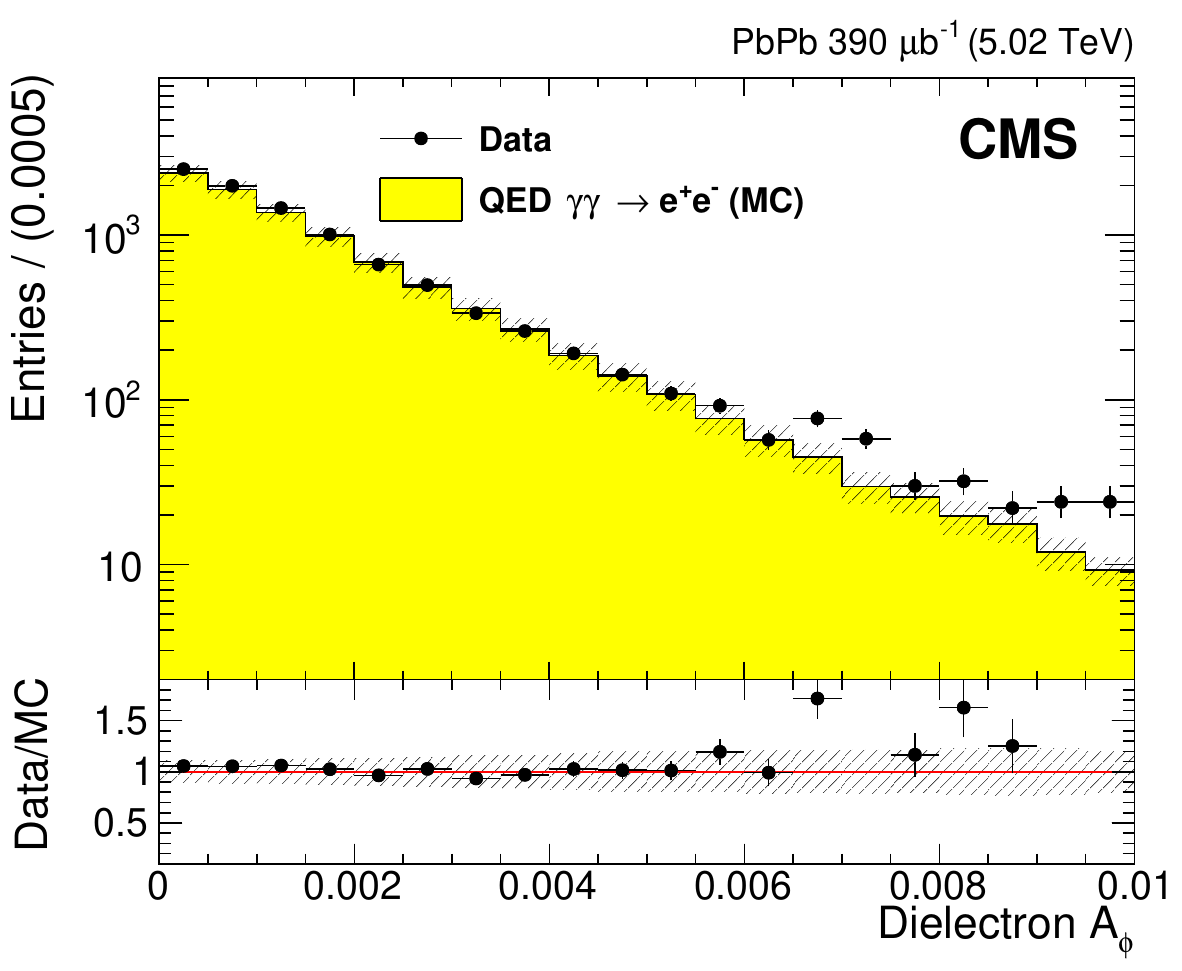}
  \includegraphics[width=0.4\textwidth]{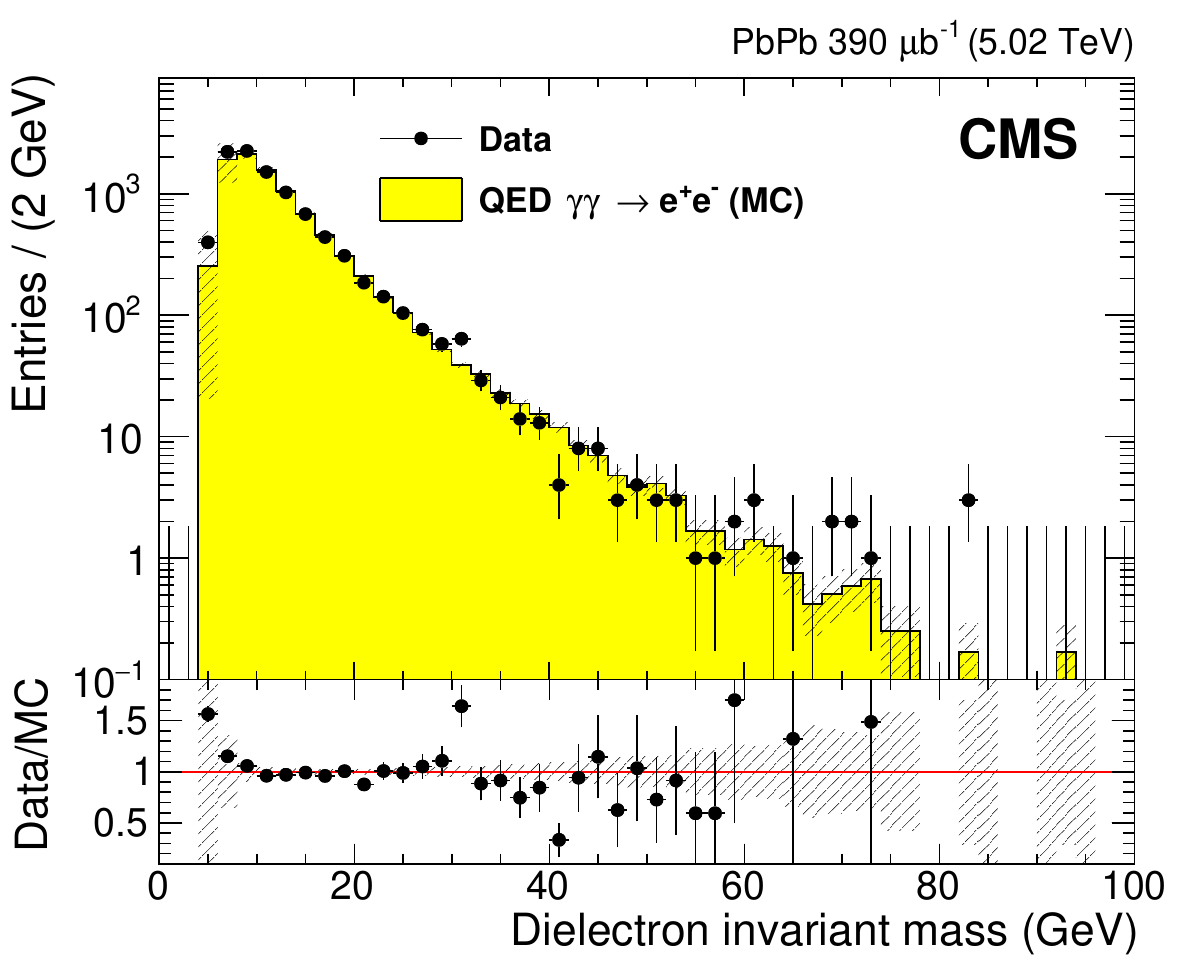}
  \includegraphics[width=0.4\textwidth]{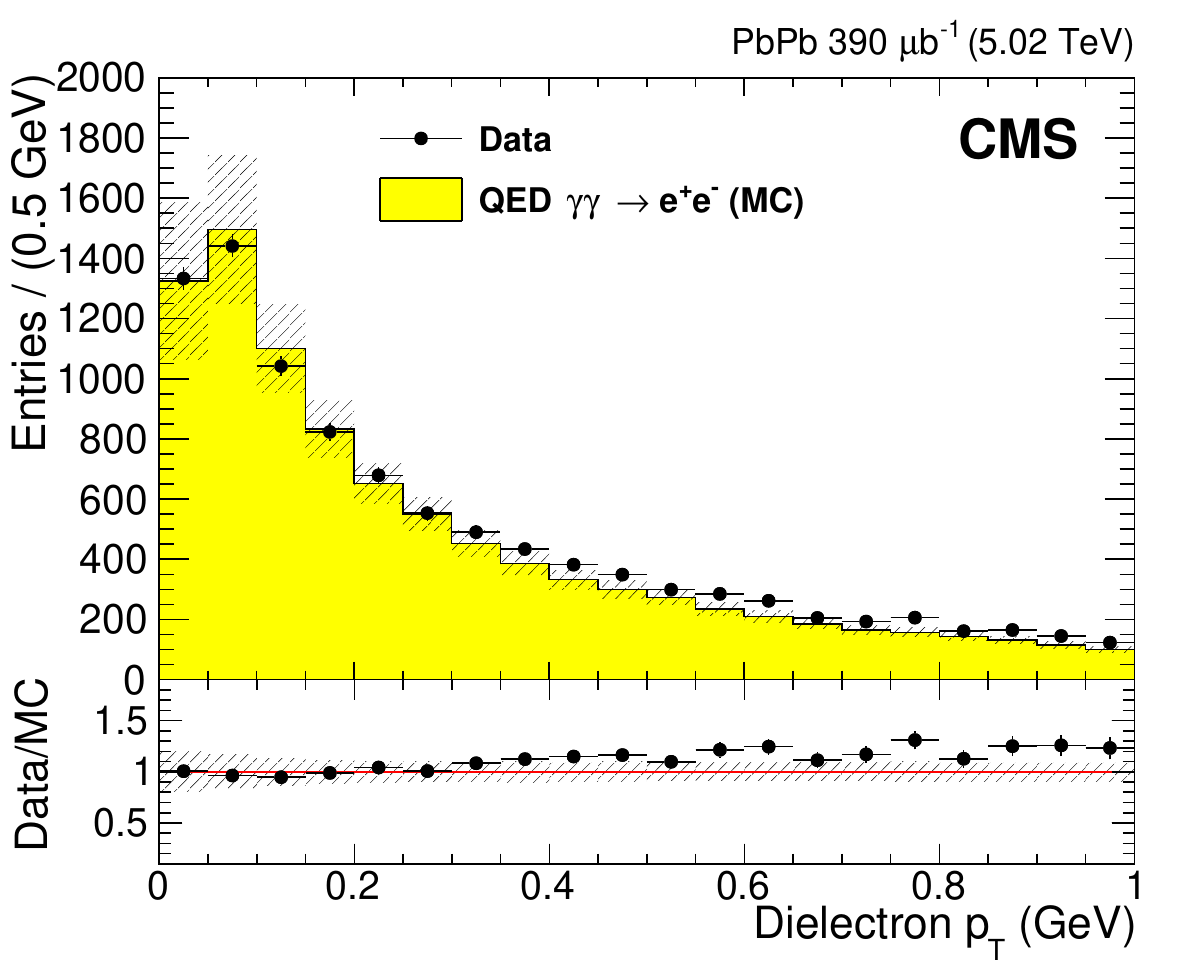}
  \includegraphics[width=0.4\textwidth]{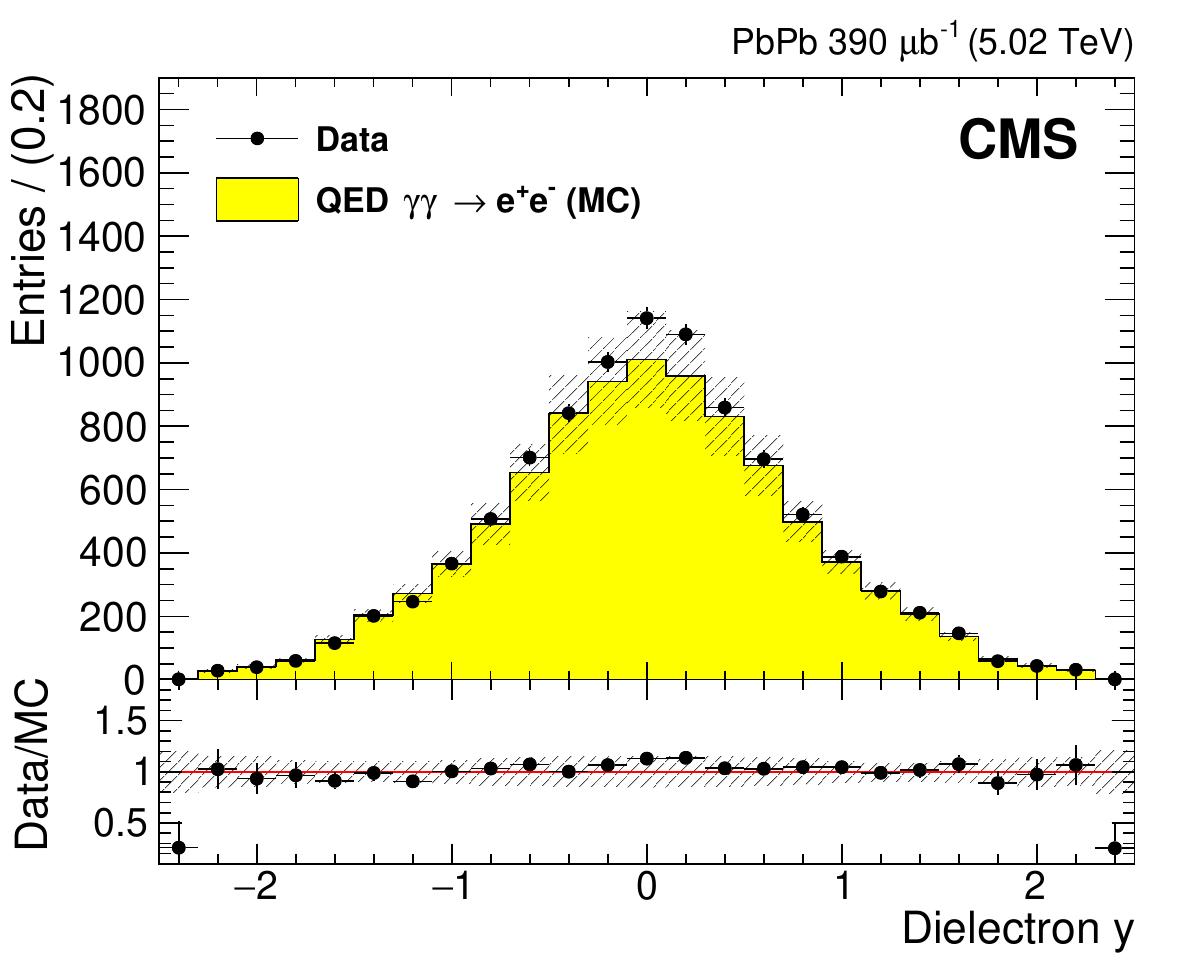}
\caption{\label{fig:qed_ee_kine} Comparison of data (circles) and MC expectation (histogram)
for the exclusive $\Pep\Pem$ events passing the selection criteria, as a function of dielectron acoplanarity (upper left), mass (upper right),
\pt (lower left), and rapidity $y$ (lower right). The error bars around the data points represent
statistical uncertainties, while the hashed bands around the histograms represent the systematic and MC
statistical uncertainties added in quadrature. The horizontal bars 
indicate the bin size. The ratio of the data to the MC expectation is shown in the lower panels. \FiguresFrom{CMS:2018erd}}
\end{figure*}

Some corresponding kinematic distributions of the selected $\gaga \to \Pep\Pem$ events in the $\aphi < 0.01$ region
are shown in Fig.~\ref{fig:qed_ee_kine}, together with the corresponding MC predictions. Good agreement between data and simulations is found, 
thereby confirming the MC predictions for exclusive particle production in \PbPb UPCs at the LHC, as well as illustrating the quality of the EM 
particle reconstruction and the exclusive event selection criteria in CMS.

\subsection{Light-by-light scattering and \texorpdfstring{$\PGt$}{PGt} lepton pair production }
\label{sec:SMAndBSM_BSM1}

As indicated in Fig.~\ref{fig:feynman} (left), the elastic \LbL scattering that occurs in HI collisions is a purely quantum-mechanical process
that, to LO in the QED coupling constant $\alpha$, proceeds via virtual box diagrams~\cite{dEnterria:2013zqi,Klusek-Gawenda:2016euz}. The QED 
box diagram involves contributions from either charged fermions or the \PWpm bosons. The direct observation of \LbL scattering in the laboratory has
remained elusive until recently due to a very suppressed production cross section, proportional to $\alpha^{4}\approx 3\times 10^{-9}$. However, 
based on \PbPb collision data recorded in 2015, both the ATLAS~\cite{ATLAS:2017fur} and CMS~\cite{CMS:2018erd} Collaborations have found direct 
evidence of \LbL scattering. The ATLAS Collaboration subsequently analyzed a larger \PbPb data sample, obtained in 2018~\cite{ATLAS:2020hii}. More 
recently, an aggregate analysis was performed~\cite{Krintiras:2023axs}, further improving the experimental uncertainty by approximately 10\% compared 
to the individual analyses.

The \LbL signal is extracted by applying the same selection criteria, including full exclusivity, as described above for the QED \Pep\Pem events, 
with the main requirement corresponding to having two photons (rather than electrons), each with $\et > 2\GeV$ and $\abs{\eta} < 2.4$, and having 
a diphoton invariant mass larger than 5\GeV. In the analysis, photons falling in the 
range $1.444<\abs{\eta}<1.566$, corresponding to a 
gap region between the barrel and endcaps of the ECAL detector (discussed in Section~\ref{subsec:CMSapparatus}), are missed. We observe 14 \LbL scattering candidates,
to be compared with the $9.0 \pm 0.9\thy$ \Starlight MC generator prediction for the \LbL scattering signal. The (conservative) 10\% uncertainty 
in the \LbL theoretical prediction covers different implementations of the nonhadronic overlap condition for varying Pb radius and
\NN cross section values, as well as neglected NLO corrections. The overall data-to-simulation agreement is very good, given the small residual 
diphoton background: $3.0\pm 1.1\stat$ from CEP and $1.0\pm 0.3\stat$ from misidentified QED $\Pep\Pem$ events. 

\begin{figure*}[ht]
\centering
 \includegraphics[width=0.45\textwidth]{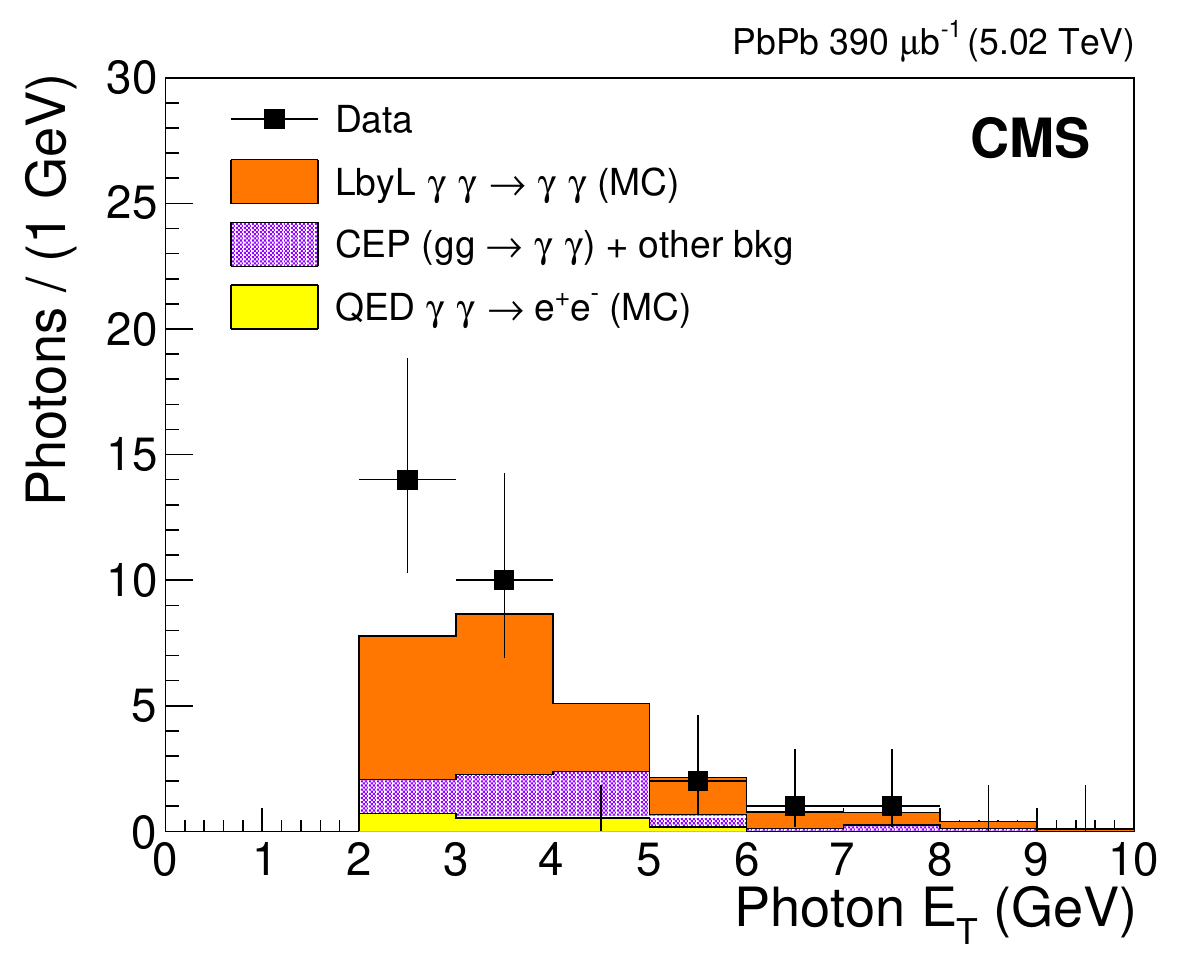}
 \includegraphics[width=0.45\textwidth]{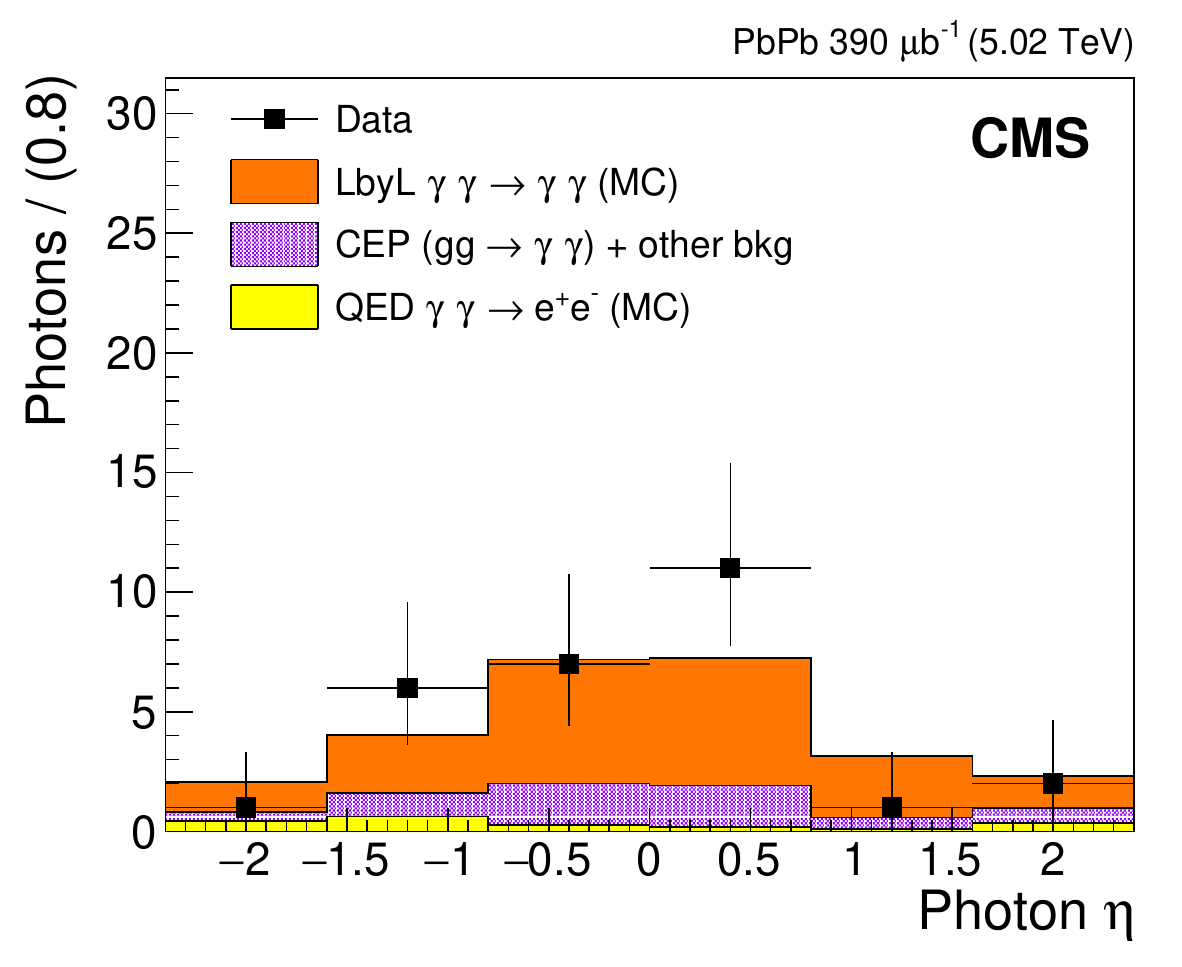}
 \includegraphics[width=0.45\textwidth]{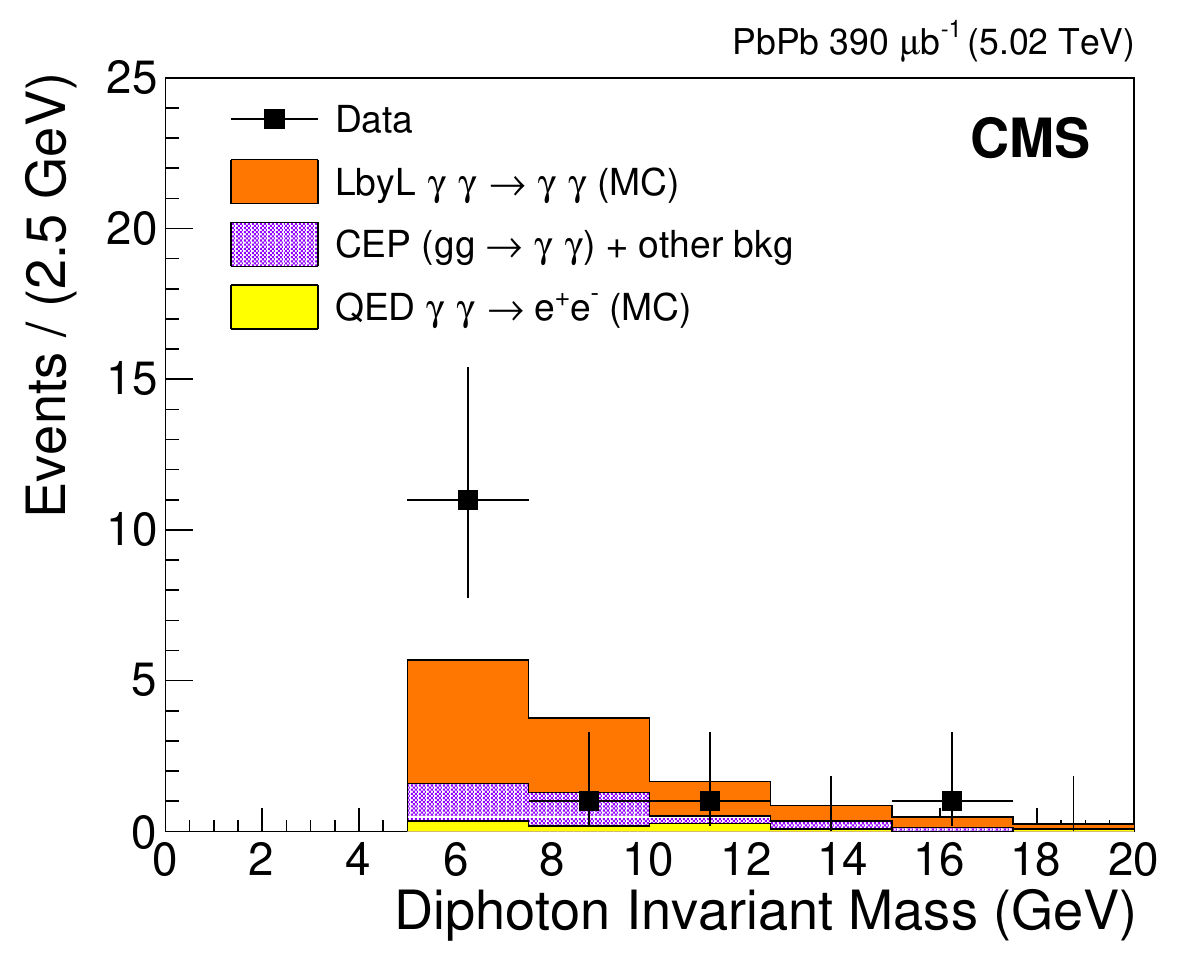}
 \includegraphics[width=0.45\textwidth]{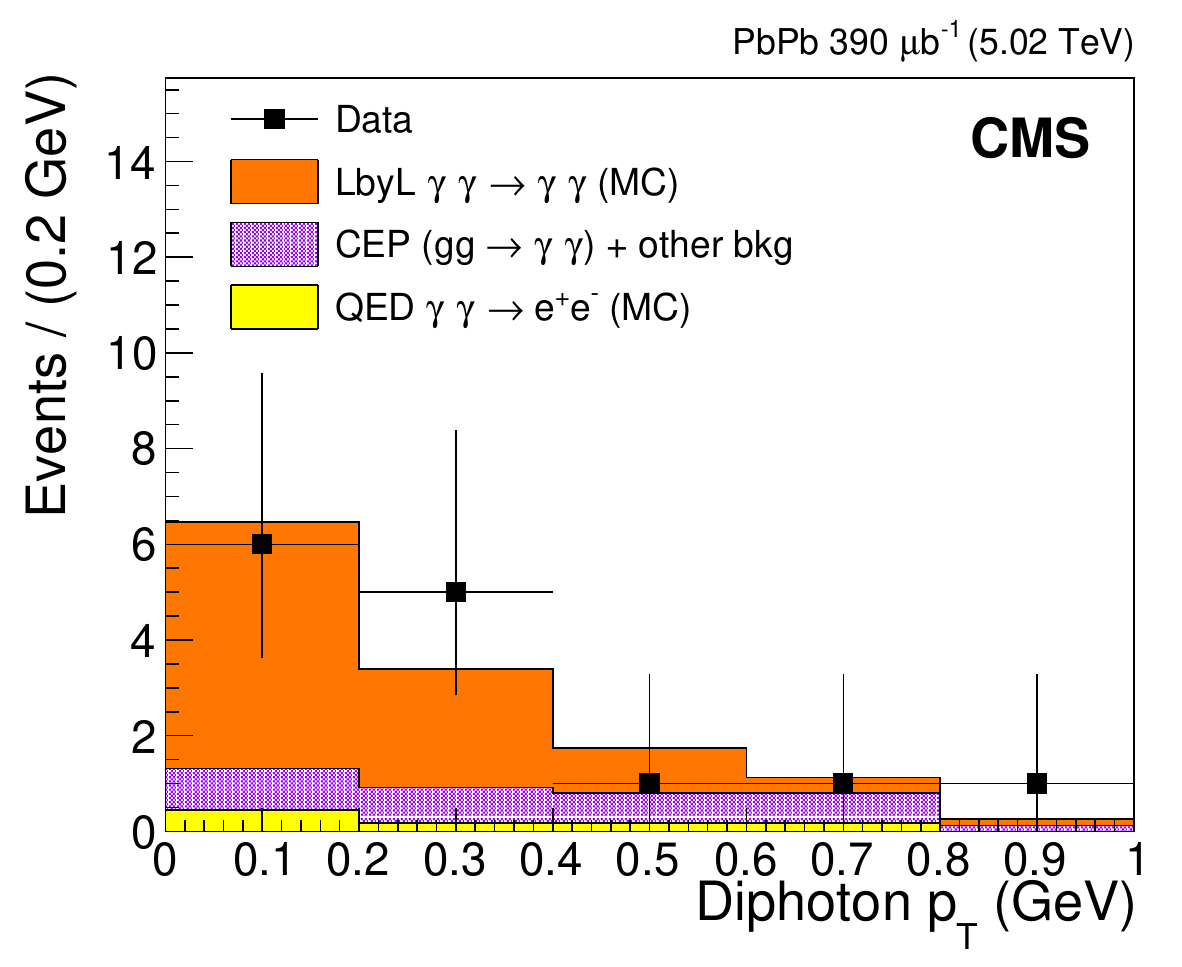}
 \caption{Distributions of the single photon \et (upper left) and $\eta$ (upper right), as well as diphoton invariant mass (lower left) and \pt (lower 
 right), measured for the exclusive events passing the selection criteria (squares), compared to the expectations of \LbL scattering signal (orange), QED \Pep\Pem MC 
 generator predictions (yellow), and the CEP background (light blue). The error bars indicate statistical uncertainties. \FiguresFrom{CMS:2018erd}}
\label{fig:data_mc_ged_af_scale}
\end{figure*}

Figure~\ref{fig:data_mc_ged_af_scale} compares the measured and simulated distributions for the
single photon \pt and $\eta$, and for the diphoton invariant mass and \pt. Similarly to the measured yields, the kinematic distributions are also 
in agreement with the combination of the \LbL scattering signal plus the background expectation. It should be noted that the overall diphoton 
cross section efficiency is approximately 20\%, compared with about 10\% for dielectrons. The lower efficiency results from each individual electron
having a relatively large probability of losing energy by bremsstrahlung before reaching the ECAL, thereby causing some losses by not satisfying 
the trigger selection threshold. 

As noted in Ref.~\cite{CMS:2022arf}, the possibility of observing photon-induced \PGt lepton production in UPC events at a HI collider was considered 
well before the LHC era~\cite{delAguila:1991rm}. Recently, theoretical studies have proposed that the kinematic properties of \PGt lepton pairs 
produced in UPCs at the LHC can be used to constrain the electromagnetic couplings of the \PGt lepton~\cite{Beresford:2019gww,Dyndal:2020yen,Burmasov:2023cwv}. These 
constraints allow for fundamental tests of QED and searches for BSM physics. This has motivated the use of novel experimental approaches to observe 
this process, as undertaken in recent measurements by the ATLAS~\cite{ATLAS:2022ryk} and CMS~\cite{CMS:2022arf} Collaborations.

\begin{figure*}[t!]
\centering
\includegraphics[width=0.45\textwidth]{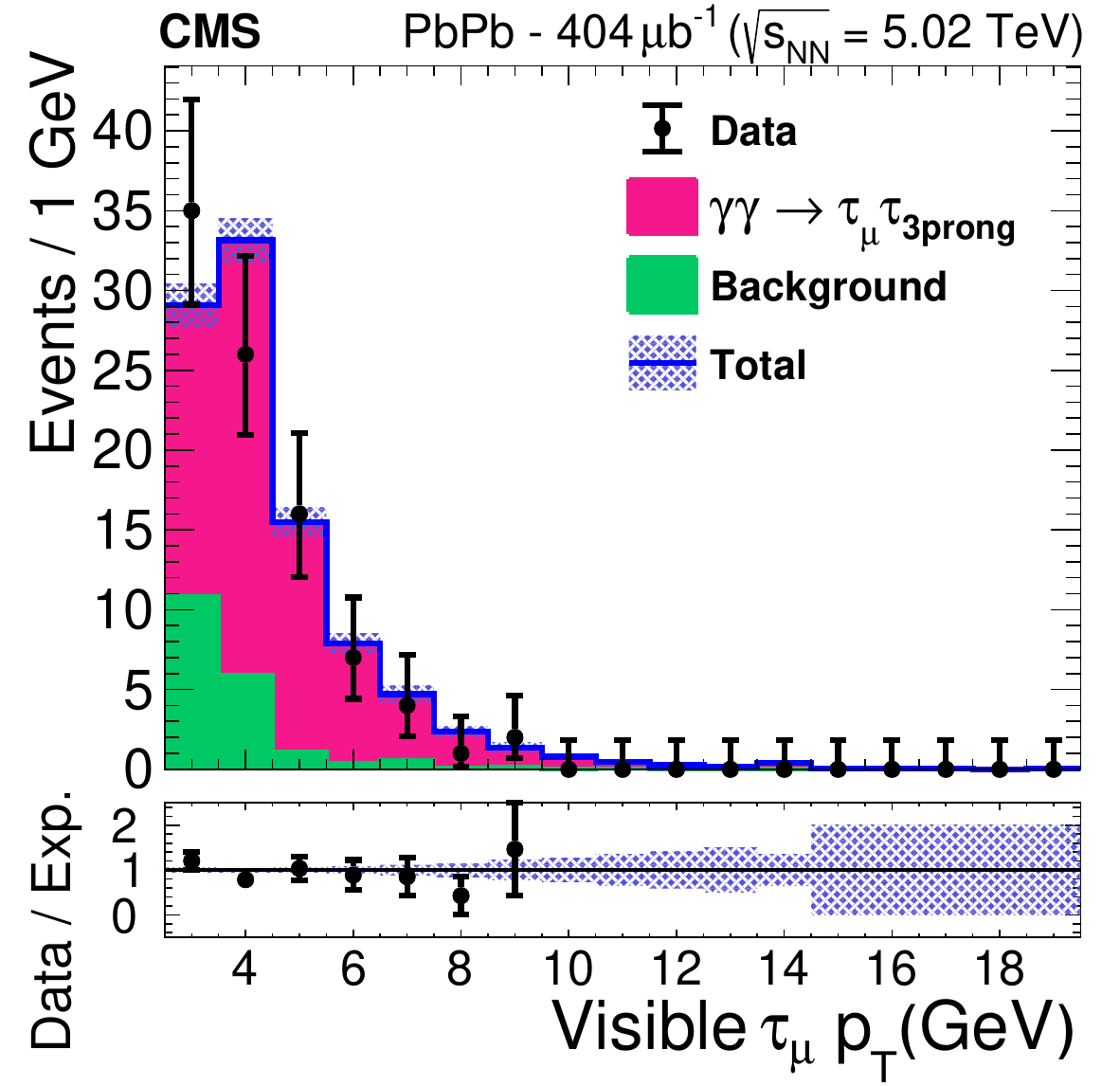}
\includegraphics[width=0.45\textwidth]{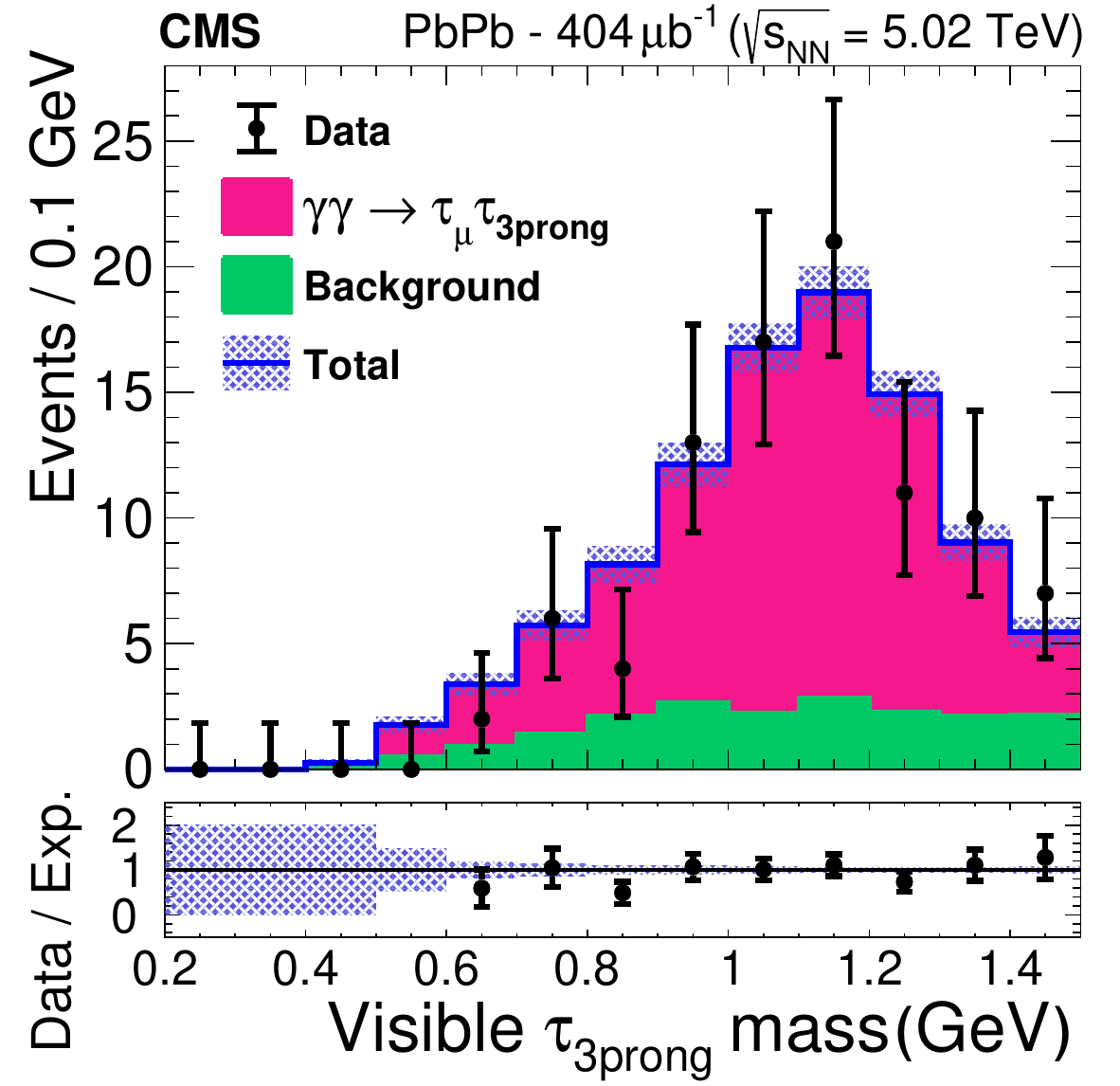}\\
\includegraphics[width=0.45\textwidth]{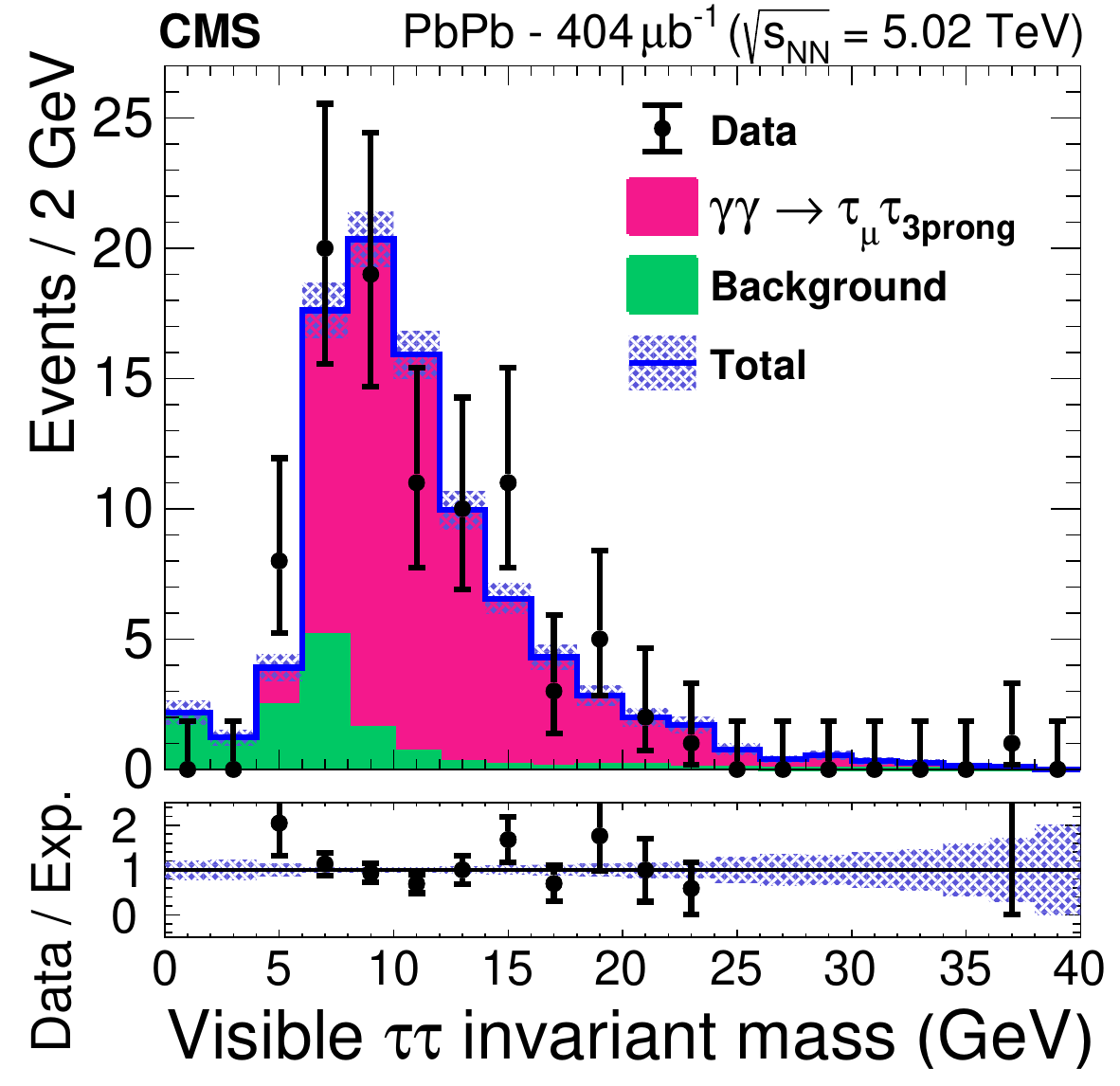}
\includegraphics[width=0.45\textwidth]{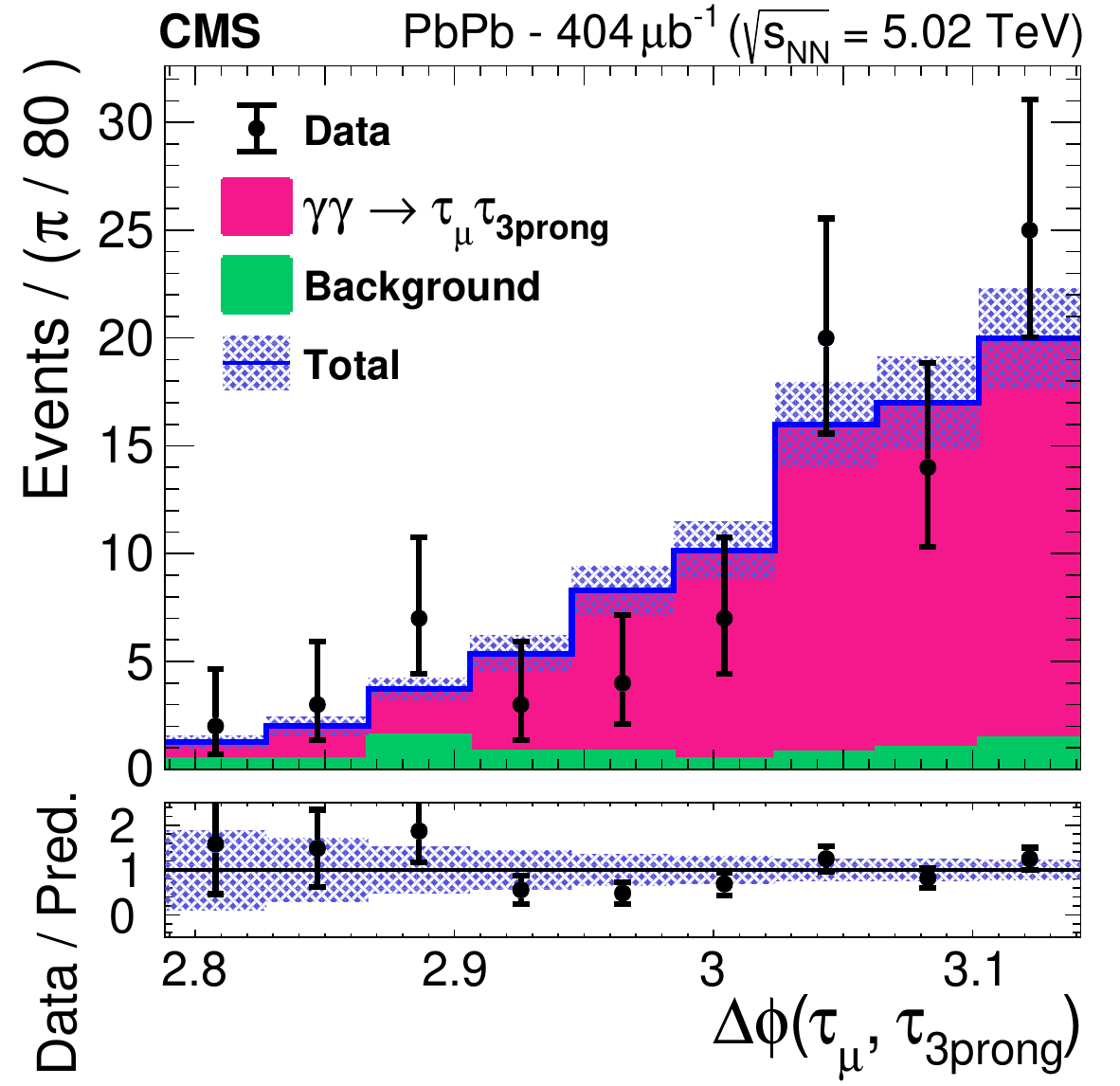}
\caption{Transverse momentum of the muon originating from the \muonic candidate (upper left). Invariant mass of the three pions forming 
the \threeprong candidate (upper right). Invariant mass of the \tautau system (lower left). The $\dphi(\muonic,\threeprong)$ 
azimuthal difference (lower right). In all plots, the signal component (magenta histogram) is stacked on top of the background component (green 
histogram). The sum of signal and background is displayed by a blue line and the shaded area shows the statistical uncertainty. The data are represented 
with black points and the uncertainty is statistical only. The lower panels show the ratios of data to the signal-plus-background prediction and 
the shaded bands represent the statistical uncertainty in the prefit expectation. \FiguresFrom{CMS:2022arf}}
\label{fig:muonKinematicsDataMC}
\end{figure*}

Based on the 2015 \PbPb data sample, CMS has observed \PGt lepton pairs in UPC \PbPb collisions, \gammagammatautau, in events that 
may contain excitations of the outgoing Pb ions. One \PGt lepton (\muonic) is reconstructed through its decay to one muon and two neutrinos, 
while the other (\threeprong) is reconstructed through its ``3 pronged'' decay into hadrons plus a neutrino~\cite{ParticleDataGroup:2022pth}. A 
typical event display is shown in Fig.~\ref{fig:display}. This choice of final state offers a clean experimental signature, with the muon used 
for online selection and the hadronically decaying \PGt candidate providing discrimination against dimuon photoproduction and thus providing an 
unambiguous reconstruction of the \PGt lepton decay. Kinematic distributions showing the \gammagammatautau signal process, as well 
as the background model based on control samples in the data, are shown in Fig.~\ref{fig:muonKinematicsDataMC}. Good agreement is observed between the measured
distributions and the sum of the signal simulation and background estimation.

A maximum likelihood (binned) fit of the signal and background components is used for the signal extraction. The fit is performed on the distribution 
of the difference in azimuthal opening angle between the \muonic and \threeprong candidates, $\dphi(\muonic,\threeprong)$, 
exploiting the fact that the two signal \PGt leptons are produced azimuthally back-to-back in UPCs. We measure $77\pm 12$ \gammagammatautau 
signal events as the integral of the postfit signal component. The signal and background postfit $\dphi(\muonic,\threeprong)$ 
templates, along with the data, are also shown in Fig.~\ref{fig:muonKinematicsDataMC} (lower right).

\begin{figure}[t]
  \centering
    \includegraphics[width=0.5\textwidth]{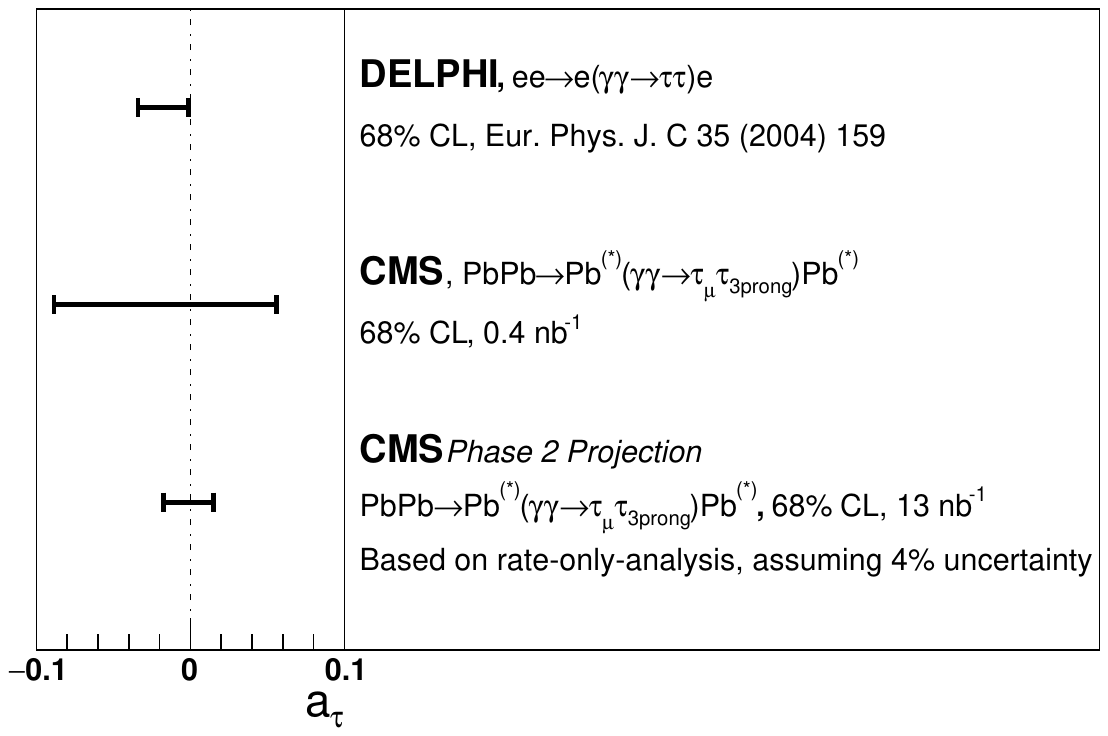}
    \caption{The $\sigma(\gammagammatautau)$ cross section, measured in a fiducial phase space region at $\sqrtsNN = 5.02\TeV$. The theoretical 
    predictions~\cite{Beresford:2019gww,Dyndal:2020yen} are computed with leading order accuracy in QED and are represented by the vertical solid 
    lines, which can be compared with the vertical dotted line representing this measurement. The outer blue (inner red) error bars represent the 
    total (statistical) uncertainties, whereas the green hatched bands correspond to the uncertainty in the theoretical predictions, as described 
    in the text. The potential electromagnetic excitation of the outgoing Pb ions is denoted by $(^{\ast})$. \FigureFrom{CMS:2022arf}}
    \label{fig:sum}
\end{figure}

The measured fiducial cross section is shown in Fig.~\ref{fig:sum}, in good agreement with LO QED predictions~\cite{Beresford:2019gww,Dyndal:2020yen}. The 
analytical calculation from Ref.~\cite{Dyndal:2020yen} results in a cross section that is 20\% higher than that found in Ref.~\cite{Beresford:2019gww}. This 
is explained in Ref.~\cite{Dyndal:2020yen} as mainly stemming from the different requirements applied in the modeling of single-photon fluxes. In 
both cases, although further advancements in theory are needed for a proper uncertainty evaluation, a conservative uncertainty of 10\% is reported, 
following the approach of Ref.~\cite{CMS:2018erd}.

\subsection{Exclusion limits on axion-like particle production and anomalous \texorpdfstring{$\PGt$}{PGt} lepton magnetic moment}
\label{sec:SMAndBSM_BSM2}

A contributing factor in the coupling of the lepton~(\Pell) to the photon~(\PGg) is the anomalous magnetic moment $a_{\Pell} = (g-2)_{\Pell} / 2$, 
with the $g$-factor being the proportionality constant that relates the magnetic moment to the spin of the lepton. Although the predicted value 
of \atau is $0.00117721\,(5)$, with the number in parentheses denoting the uncertainty in the last digit, its best measured value is $-0.018\pm0.017$, 
from the DELPHI Collaboration~\cite{DELPHI:2003nah} (other existing limits on \atau can be found in Ref.~\cite{ParticleDataGroup:2022pth}). The 
larger uncertainty in \atau compared to the measurements of $a_{\PGm}$ and $a_{\Pe}$ mainly results from the short \PGt lepton lifetime, which 
is of the order of $10^{-13}\unit{s}$, such that \PGt leptons cannot be stored to measure their \atau-dependent precession in a magnetic field. A 
more precise \atau determination would facilitate tighter constraints on BSM physics models, in which additional particles with mass $M$ contribute 
with terms typically proportional to $(m_\Pl/M)^2$. 

\begin{figure*}[ht]
\centering
\includegraphics[width=0.65\textwidth]{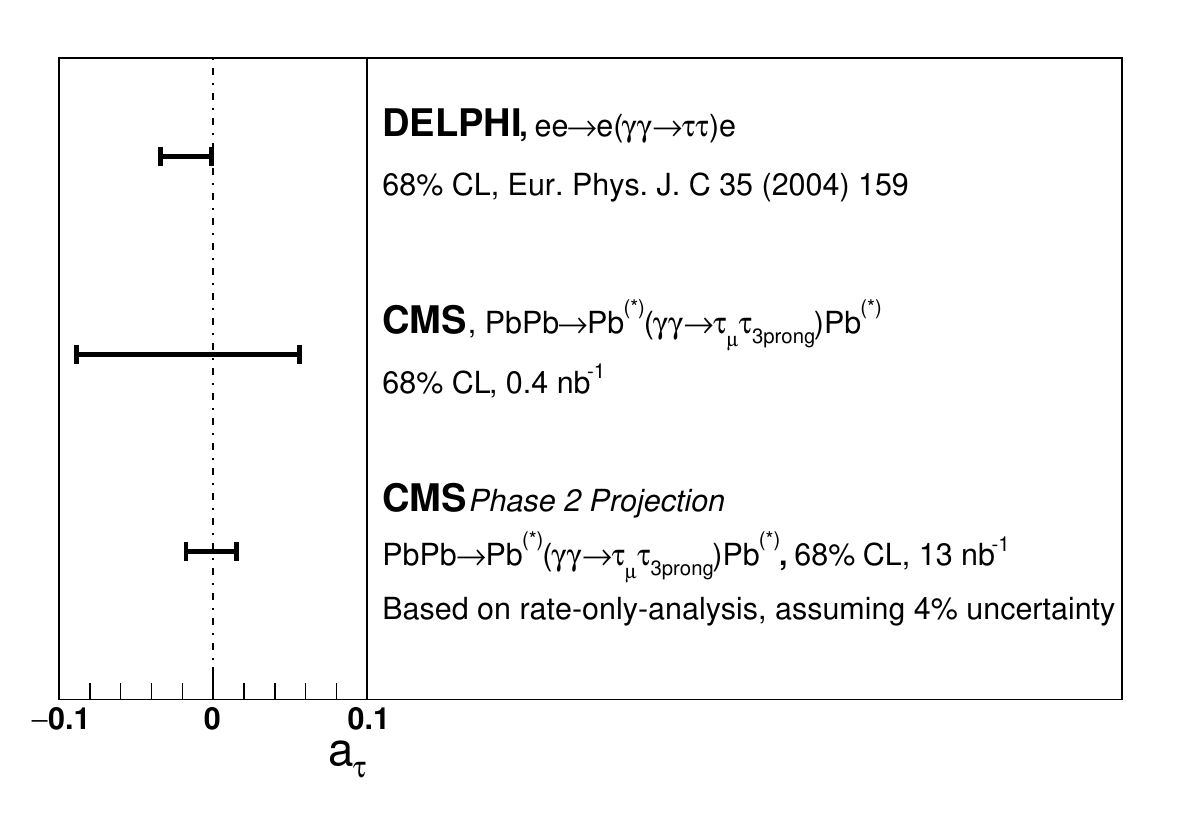}
\caption{Comparison of the constraints on \atau at 68\% \CL from the analysis in Ref.~\cite{CMS:2022arf} and the DELPHI experiment at LEP~\cite{DELPHI:2003nah}. 
The projection to the integrated \PbPb luminosity expected from the high-luminosity LHC program is included. \FigureFrom{CMS:2022arf}}
\label{fig:a_tau}
\end{figure*}

Thus, more recent calculations have evaluated the impact of BSM processes on the \gammagammatautau cross section. The BSM coupling variations 
in \atau can change the expected cross section and alter, \eg, the \PGt lepton \pt spectrum~\cite{Beresford:2019gww,Dyndal:2020yen}. In Ref.~\cite{CMS:2022arf}, 
the dependence of the total \xsec on \atau~\cite{Beresford:2019gww} was used to extract a model-dependent value of \atau at the LHC, as shown 
in Fig.~\ref{fig:a_tau} at 68\% \CL. The projection to the integrated \PbPb luminosity expected from the high-luminosity LHC program is also shown~\cite{CMS-PAS-FTR-22-001}.

As noted in Ref.~\cite{CMS:2018erd}, the \LbL process has been proposed as a particularly sensitive channel for studying BSM physics. Modifications 
of the \LbL scattering rates can occur if, \eg, new heavy particles, such as magnetic monopoles,
vector-like fermions or dark-sector particles, contribute to the virtual corrections of
the box depicted in Fig.~\ref{fig:feynman} (left). Other new spin-even particles, such as ALPs~\cite{Knapen:2016moh} or gravitons~\cite{Ahern:2000jn,dEnterria:2023npy}, 
can also contribute to the \LbL scattering continuum or to new diphoton resonances. In addition, \LbL cross sections are sensitive to Born--Infeld 
extensions of QED~\cite{Ellis:2017edi}, and anomalous quartic gauge couplings~\cite{Chapon:2009hh}.

\begin{figure*}[ht]
\centering
\includegraphics[width=0.75\textwidth]{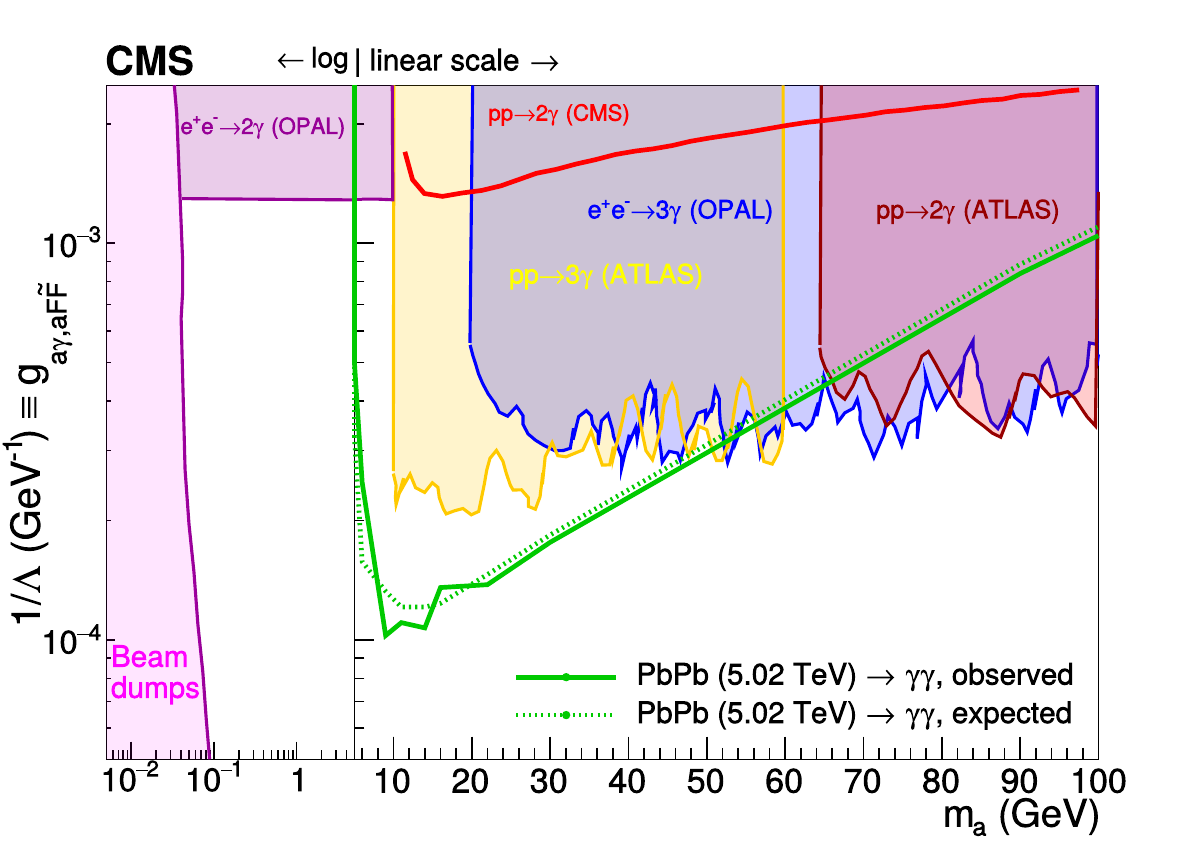}\hfill
\caption{Exclusion limits at 95\% \CL in the ALP-photon coupling $g_{\Pa\PGg}$ vs. ALP
mass $m_\Pa$ plane, for the operators $\Pa F\widetilde{F}/4\Lambda$ assuming
ALP coupling to photons only, derived in Refs.~\cite{Knapen:2016moh,Jaeckel:2015jla}
from measurements at beam dumps~\cite{Dobrich:2015jyk}, in $\Pep\Pem$ collisions at LEP 1~\cite{Jaeckel:2015jla} and LEP 2~\cite{OPAL:2002vhf},
and in \pp collisions at the LHC~\cite{Chatrchyan:2012tv,ATLAS:2014jdv,ATLAS:2015rsn}, and compared to the 
limits obtained from Ref.~\cite{CMS:2018erd}. \FigureFrom{CMS:2018erd}}
\label{fig:axion_aFFBB}
\end{figure*}

The measured invariant mass distribution (Fig.~\ref{fig:data_mc_ged_af_scale}, lower left) has been used to search for possible narrow diphoton 
resonances, such as pseudoscalar ALPs produced in the process $\gaga\to\Pa\to\gaga$.
All other processes, \ie, \LbL, QED, and CEP, are considered as background in this search. Fully simulated \Starlight MC samples for
ALP masses, $m_\Pa$, ranging from 5 to 90\GeV are reconstructed. A binned maximum likelihood fit of the ALP signal and background contributions 
is performed on the data. A profile likelihood ratio is used as a test statistic based on the \CLs criterion~\cite{CLS1,CLS2} to extract exclusion 
limits at 68 and 95\% confidence levels~(\CL): first, in the $\sigma(\gaga\to\Pa\to\gaga)$ cross section; and then, in the $g_{\Pa\PGg}$ 
\vs $m_\Pa$ plane, where $g_{\Pa\PGg}\equiv 1/\Lambda$ is the ALP coupling to photons (with $\Lambda$ being the energy scale associated with 
the underlying U(1) symmetry whose spontaneous breaking generates the ALP mass). Two scenarios are considered where
the ALP couples to photons $F^{\mu\nu}$ alone (shown in Fig.~\ref{fig:axion_aFFBB}) or also to hypercharge. The derived constraints on the ALP 
mass and its coupling to photons in Fig.~\ref{fig:axion_aFFBB} are also compared to those obtained from various experiments (available up to the 
time of publication of Ref.~\cite{CMS:2018erd}),
assuming a 100\% ALP decay branching fraction to diphotons. Despite the updated \LbL\ measurement in Ref.~\cite{ATLAS:2020hii} our exclusion limits 
still remain competitive over the $m_\Pa \approx 5\text{--}10\GeV$ mass range~\cite{ATLAS:2020hii}, regardless of the sensitivity to the EM current alone 
or of extra ALP couplings to EW currents.

\subsection{Summary of QED results and BSM searches with UPC}
\label{sec:SMAndBSM_Summary}

A broad range of photon-induced processes have been investigated using UPCs of lead ions at a center-of-mass energy per nucleon pair of 5.02 TeV. These studies span from standard QED tests to searches for BSM physics.

High-rate, exclusive, high-mass dilepton production ($\mleplep\gtrsim5\GeV$) from the 2018 data run is compared with QED expectations, providing a rigorous test of QED predictions. Rare processes, such as \LbL scattering and \PGt lepton production, have been explored to extend QED tests and to assess the potential of these channels to uncover BSM physics. The clean final-state signatures in these rare processes make such searches
possible.

The CMS Collaboration conducted the first measurements of \gammagammamumu production as a function of forward neutron multiplicity in \PbPb UPCs. A notable broadening of back-to-back azimuthal correlations was observed, increasing with the multiplicity of forward neutrons. This trend is qualitatively reproduced by a LO QED calculation that accounts for the influence of the impact parameter on the average \pt value of the photon. In the region of near back-to-back emission, the $\gaga \to \Pep\Pem$ ratio shows good agreement between theory and data, confirming the quality of electromagnetic particle reconstruction and event selection criteria for exclusive QED production in \PbPb UPCs.

Evidence for LbL scattering in \PbPb UPC data was reported by CMS using data from 2015. The observed total yields and kinematic distributions align with expectations for the LbL scattering signal, with a small residual background primarily from misidentified exclusive dielectron and gluon-induced central exclusive processes. The exclusive diphoton invariant mass distribution was employed to set new exclusion limits on the production of pseudoscalar ALPs through the process $\gaga \to \Pa \to \gaga$, covering the mass range of 5--90 GeV.

Additionally, CMS observed the production of \PGt lepton pairs in \PbPb UPCs using the 2015 data. Events featuring a final state with one muon and three charged hadrons were reconstructed, reaching a statistical significance exceeding five standard deviations with respect to the background-only expectation. The measured kinematical distributions and the extracted cross section both agree with LO QED predictions. From these measurements, a model-depen\-dent value of the anomalous magnetic moment of the \PGt lepton, \atau, was estimated, offering a novel experimental probe of \atau through heavy ion collisions at the LHC.

\clearpage
\section{Summary}
\label{sec:Summary}

\subsection{Discoveries and insights from the CMS heavy-ion physics program}
\label{sec:Summary_1}

This review presents the first comprehensive summary of results from the CMS heavy ion physics program using data collected during the first two running periods of the LHC: 2010--2013 and 2015--2018. 
After having successfully addressed many experimental challenges (Section~\ref{sec:ExperimentalMethods}), in particular thanks to major advances in the areas of online event selection and offline physics object reconstruction, CMS performed a series of measurements that covered and extended those initially anticipated (Section~\ref{sec:IntroductionQCD}).
Those experimental results, reviewed in the previous sections of this paper, covered several topics, including high-density quantum chromodynamics, precision quantum electrodynamics, and even novel searches for phenomena beyond the standard model.

These studies provide detailed macroscopic and microscopic probes of the quark-gluon plasma created at LHC energies, achieving the highest temperature and smallest baryon-chemical potential ever reached in a laboratory. The results have yielded groundbreaking insights across a wide range of quantum chromodynamics phenomena, representing some of the most important and novel findings in the history of the field. For example, CMS discovered that small collision systems, such as \pp and \pPb, can exhibit signs of collectivity, a phenomenon previously only associated with larger collision systems, such as \PbPb. This discovery opened new avenues for understanding how fluidity and plasma-like properties emerge in QCD matter. Additionally, jet quenching measurements with fully reconstructed jets have set new standards, allowing us to experimentally assess medium modifications of entire parton showers beyond leading-hadron observables and to extract information about the medium response to hard probes. Studies of the nuclear modifications of the production yields of (fully reconstructed) beauty and charm hadrons, as well as of all five S-wave quarkonium states, including the rarely produced \PGUP{3S}, have addressed long-standing questions in the field. Furthermore, evidence for gluon antishadowing and saturation, along with novel results from rare QED processes and beyond standard model searches, vastly expanded the scope of these studies. The three-dimensional evolution of the QGP has been explored, and signals of chiral magnetic effects have been excluded to a large extent. In the following paragraphs, we offer more details on these achievements, which highlight the significant contributions of CMS to the progress of our understanding of high-density QCD.

The study of the collectivity of charged hadrons in high-multiplicity \pp and \pPb collisions (Section~\ref{sec:SmallSystems}) has provided the 
first observations of long-range correlations similar to those seen in HI collisions. The CMS Collaboration has offered further evidence of 
collectivity through multiparticle correlation and heavy-flavor meson analyses. The study of multiparticle correlations has been extended to smaller 
collision systems using ultraperipheral collisions, where the separation of the ions in the transverse plane strongly reduces the role of 
interactions mediated by quarks and gluons. One of the motivations for the small collision system studies was to search for evidence of jet 
quenching in these systems, to compare with the results obtained in collisions involving two heavy ions. Jet quenching effects have not been 
observed in \pPb collisions.

The initial state of the nucleons and nuclei before
a HI collision strongly influences the subsequent evolution of the created medium. The density of quarks and gluons within a nucleon, 
as a function of the fraction of the nucleon momentum ($x$) carried by each parton and the squared transverse momentum transfer~(\QTwo), 
is parameterized in terms of parton distribution functions. When the nucleon is embedded in a nucleus, this density is expressed as nuclear PDFs. 
Proton-lead collision data have been used to constrain the quark and gluon nuclear densities through measurements of the cross section of 
electroweak gauge bosons, dijets, and top quark pairs (Section~\ref{sec:InitialState}). Some of these results have been used as input to the latest nPDF fits, 
leading to a significant improvement in the precision across an extended phase space region. For studying the small-$x$ region, which is primarily driven 
by the evolution of the gluon density, the measurements of forward inclusive jet cross sections in \pPb collisions and the cross sections for 
exclusive vector meson production in \pPb and \PbPb collisions have been used. 
As part of these studies, a technique has been developed to use forward neutron multiplicities
in order to unfold the cross sections for exclusive vector meson production in the photon-nucleus frame, giving unprecedented access to the small-$x$ regime.

As expected, the LHC collaborations find a significant increase in the charged particle density and average transverse energy per charged particle 
compared to those found at RHIC energies, indicating a denser and hotter medium formed at the LHC. The CMS Collaboration has an extensive program 
for studying such bulk properties of the quark-gluon plasma in ultrarelativistic nuclear collisions and searching for novel phenomena 
(Section~\ref{sec:softQGP}). Taking advantage of the wide pseudorapidity coverage of the CMS apparatus, long-range collective particle 
correlations (``flow'') are observed with unprecedented high precision. At the same time, factorization breaking in flow harmonics 
($v_n$) has been observed and studied for the first time by the CMS Collaboration and has been shown to have a strong sensitivity to the granularity 
of initial-state fluctuations. The observation of an $\eta$-dependent factorization breaking has provided sensitivity to the longitudinal dynamics of 
the QGP. In addition, the shape and size of the systems produced in different colliding systems and at various LHC energies were also investigated 
via femtoscopic correlation measurements.
In relativistic HI collisions leading to QGP formation, the resulting medium may experience intense magnetic fields produced by the colliding ions. 
If net chiral (left- or right-handed) quarks are present, a localized current can be generated, leading to a charge separation known as the chiral 
magnetic effect and, as a separate process, a long-wavelength collective excitation known as a chiral magnetic wave. 
The CMS Collaboration has unambiguously shown that the CME and CMW signals are too small 
to be observed with the currently available data sample.

The experimental use of hard probes as a way to study the short-wavelength structure of the QGP has greatly advanced during the LHC Runs~1 and~2 
(Section~\ref{sec:hardQGP}). With the initial studies, the depletion of particles with high transverse momentum observed in two-particle 
correlations, at BNL RHIC was confirmed to be the result of jet quenching with LHC measurements of dijet asymmetries using fully reconstructed jets. 
Further evidence comes from the suppression of jet and hadron yields in HI collisions compared to those expected by scaling up the results from \pp collisions. 
The yield suppression is generally expressed in terms of the nuclear modification factor and can be associated with parton energy loss. 
Subsequent detailed studies of hadrons and jets have provided information regarding the path-length dependence of parton energy loss. 
The associated production of jets with electroweak bosons has made possible the determination of the absolute magnitude of the jet energy 
loss and these studies are now applied to test the survivor bias in inclusive jet samples. A multitude of measurements, including those of 
jet fragmentation functions and jet shapes, have established a qualitative picture in which quenching redistributes jet energy from the 
high-\pt jet constituents to softer particles, and from small to large angles relative to the jet axis. Novel background subtraction algorithms 
and jet grooming techniques (which remove wide-angle soft radiation from a jet) allow the investigation of the early stages (early vacuum) of a 
parton shower in the QGP, well before its later medium-modified stage. These studies suggest that jet modifications can be sensitive to the 
earliest splittings in the evolution of the parton shower. However, further investigations are needed to properly account for a bias when 
selecting broader early-vacuum structures, and hence more heavily quenched jet momenta.

The CMS Collaboration has also performed systematic studies of the mass dependence of quark energy loss by comparing 
the \RAA and $v_2$ results for fully reconstructed light- and heavy-flavor (charm and beauty) hadrons over an unprecedentedly large \pt range:
the production yields of both light and heavy (high-\pt) quarks are seemingly suppressed in the QGP, 
the dependence on the quark mass decreasing as \pt increases, 
as expected in the context of radiative energy loss. 
These studies led to unique measurements of \PB~mesons in heavy ion collisions.
The hadronization of heavy-flavor particles has also been examined in detail using various ratios of their yields, including, for the first time, 
details of the internal structure of exotic hadrons in the presence of the QGP.

The suppression patterns of the five S-wave quarkonia (\PJGy, \PGyP{2S}, and $\PGUP{n\mathrm{S}}$, \mbox{$n=1$--$3$}), never previously measured in a single experiment, strongly indicate that the nuclear suppression effects follow a sequential hierarchy reflecting the binding energy of the quarkonium state, as expected if the bound state is broken apart by the QGP medium.

In addition to nuclear hadronic interactions, electromagnetic interactions can also be studied in ultraperipheral collisions (Section~\ref{sec:QEDBSM}) since heavy 
ions with energies of several~\TeV per nucleon can interact through very intense electromagnetic fields. The Lorentz factor of the Pb beam at 
the LHC determines the maximum quasireal photon energy of approximately 80\GeV, leading to photon-photon collisions of center-of-mass energies 
up to 160\GeV, \ie, similar to those reached at LEP 2 but with $Z^4$ enhanced production cross sections. A broad range of precision SM and BSM 
processes has been studied in these photon-induced interactions, including exclusive high-mass dilepton ($\mleplep\gtrsim5\GeV$) production 
as well as the rare processes of light-by-light scattering and \PGt lepton production.

\subsection{Future physics opportunities at CMS for high-density QCD measurements}
\label{sec:Summary_2}

The QCD theory, a cornerstone of the standard model, remains a crucial aspect in our understanding of the strong interaction, albeit with 
lingering questions. The large values of strong coupling~(\alpS) at low \QTwo render the traditional small-\alpS perturbation theory inapplicable, 
such that collective phenomena in nuclei are nonperturbative. However, a coordinated application of the QCD parton model for conventional hadrons, 
an effort to grasp the exotic hadron spectroscopy, and advances from lattice QCD calculations hold promise of a fundamentally improved understanding 
of the characteristics of nuclei and their interactions and how deconfinement arises.

Many unresolved questions remain regarding the precise nature of the initial state from which thermal QCD matter potentially emerges. How the parton 
density varies across the broad nuclear $(x, \QTwo)$ phase space is still only partially known and, in particular, no unambiguous 
evidence has yet been found to mark the onset of parton saturation. 
Additionally, it is not yet quantitatively understood how the collective properties of the quark-gluon plasma emerge at a microscopic level from the 
interactions among the individual quarks and gluons that make up this medium.
Therefore, a crucial aspect of nuclear studies is the exploitation of future opportunities for high-density QCD studies with ion and proton beams. 
This will allow for the study of cold nuclear matter effects, the onset of nuclear saturation, and the emergence of long-range correlations. 
Examination of high-\pt hadrons, fully reconstructed jets, heavy quarkonia, open heavy-flavor particles, as well as novel tools~\cite{Andrews:2018jcm} 
to investigate more detailed aspects~\cite{Andres:2022ovj} of jet quenching, will provide additional information about the strongly coupled QGP, 
complementing the bulk and collective observables of the soft sector. Long-term initiatives, such as the use of top quarks to unravel the intricacies 
of jet quenching at different time scales of the QGP evolution, are in their early stages and are projected to rapidly progress with the increased 
luminosity anticipated in the LHC Run~3 (2022--2026) and beyond. A pilot run of oxygen-oxygen and proton-oxygen collisions will help answer the key 
prerequisite conditions for the onset of hot-medium effects~\cite{Brewer:2021kiv}. It is also important to understand the level at which these effects 
could be phenomenologically limited by knowledge of nPDFs. At present, there is a lack of experimental oxygen data for comprehensive global nPDF fitting, 
underscoring the importance of proton-oxygen collisions in ensuring the accuracy of nPDFs for lighter ions. This also has far-reaching implications 
for modeling ultrahigh-energy (cosmic ray) phenomena, and is crucial for addressing significant unresolved questions in this field~\cite{Albrecht:2021cxw}.

In addition to the larger luminosity, the detector upgrades planned for the CMS experiment in the LHC Run~4 (starting in year 2030) will 
significantly benefit the HI program. In particular, the increased $\eta$ acceptance for charged particles resulting from tracker 
upgrades~\cite{CMS:2017lum} will be very beneficial for bulk particle measurements. The upgraded Zero Degree Calorimeters~\cite{Longo:2023ubo} 
will further improve the existing triggering and identification of UPCs. The addition of time-of-flight particle identification capability, 
enabled by the Minimum Ionizing Particle Timing Detector~\cite{CMS:2667167}, will allow identification between low-momentum charged hadrons, 
such as pions, kaons, and protons, which will improve the measurements of heavy-flavored particles and neutral strange hadrons, 
while improving the prospects for identified jet substructure
measurements~\cite{CMS-DP-2021-037}.

Proton-nucleus collisions have been an integral part of the LHC program since the 2011 and 2012 pilot runs. Within collinear
factorization, constraints on our knowledge of the nuclear wave functions were extended at high \QTwo using dijet, heavy gauge boson, and top 
quark production processes available for the first time in nuclear collisions. Further insights have been gained at lower \QTwo with heavy-flavor 
production based on the assumption that the nuclear modification of their yields can be accurately incorporated in global analyses of nPDFs. 
In Run~2, the increased luminosity and detector improvements allowed for increased statistical precision, expanding the kinematic reach to 
encompass a broader range of accessible processes. Following the discoveries of
collective-like effects in small collision systems, an order of magnitude higher integrated luminosity target for \pPb collisions is set for 
Runs~3 and~4, including a large sample of \pp collisions at the highest LHC energy, but with moderate pileup to reach the largest possible 
multiplicities over a full range of hadronic colliding systems.

The large \PbPb integrated luminosity in Runs~3 and~4, coupled with high-accuracy theoretical QED calculations and several detector upgrades, 
will maximize the potential of UPC measurements. Collectively, these factors will broaden the phase space region and overall scope of physics 
exploration in the studies of low-mass resonances, the continuum, and heavy-flavor mesons in UPC events. The primary goal will be to cover a much 
wider range of masses: the expected spectrum obtainable by CMS for a 13\nbinv integrated luminosity run can extend to masses up to about 200\GeV, 
bridging the gap for BSM searches between \PbPb and \pp collisions (in the latter case, by employing the forward proton tagging technique) and overall 
extending the physics reach not only for (pseudo)scalar but also for tensor resonances~\cite{dEnterria:2023npy}. Interestingly, these high-mass pairs 
correspond to two-photon interactions in, or in close proximity to the two nuclei, enhancing the effects owing to interactions with the medium and magnetic 
fields associated with the QGP. Lower masses should be accessible with looser requirements for track and electron \pt and their overall identification 
quality~\cite{DP-2024-011}. Exclusive dimuon production can offer a precision measurement of photon fluxes associated with ion beams, and as such can 
be used to constrain predictions for all other UPC processes. Additional \LbL scattering data will also be crucial in determining the nature of newly 
discovered resonant structures, such as the \HepParticleResonance{X}{6900}{}{} state~\cite{Biloshytskyi:2022dmo}. 

Continuing the LHC HI physics program into the HL-LHC era~\cite{Citron:2018lsq,Achenbach:2023pba} offers the opportunity to collide intermediate-mass 
nuclei (\eg, oxygen and argon), facilitating the study of the initial stage of ion collisions, small-$x$ physics, and the determination of nPDFs. 
Furthermore, higher luminosities will allow vastly improved access to rare probes of the QGP. At the same time, it complements other key research 
efforts in the nuclear physics QCD community (\eg, ongoing efforts at RHIC~\cite{PAC_2023} and the upcoming Electron-Ion Collider~\cite{Accardi:2012qut}), 
as well as technical developments in the high-energy and cosmic-ray~\cite{dEnterria:2022sut} physics communities. Collectively, 
these initiatives will be pivotal in deepening our understanding of both QCD and QED, illuminating the intricate nature of matter in the early microseconds of the universe.

\vfill\newpage

\begin{acknowledgments}
\hyphenation{Bundes-ministerium Forschungs-gemeinschaft Forschungs-zentern Rachada-pisek} We congratulate our colleagues in the CERN accelerator departments for the excellent performance of the LHC and thank the technical and administrative staffs at CERN and at other CMS institutes for their contributions to the success of the CMS effort. In addition, we gratefully acknowledge the computing centers and personnel of the Worldwide LHC Computing Grid and other centers for delivering so effectively the computing infrastructure essential to our analyses. Finally, we acknowledge the enduring support for the construction and operation of the LHC, the CMS detector, and the supporting computing infrastructure provided by the following funding agencies: the Armenian Science Committee, project no. 22rl-037; the Austrian Federal Ministry of Education, Science and Research and the Austrian Science Fund; the Belgian Fonds de la Recherche Scientifique, and Fonds voor Wetenschappelijk Onderzoek; the Brazilian Funding Agencies (CNPq, CAPES, FAPERJ, FAPERGS, and FAPESP); the Bulgarian Ministry of Education and Science, and the Bulgarian National Science Fund; CERN; the Chinese Academy of Sciences, Ministry of Science and Technology, the National Natural Science Foundation of China, and Fundamental Research Funds for the Central Universities; the Ministerio de Ciencia Tecnolog\'ia e Innovaci\'on (MINCIENCIAS), Colombia; the Croatian Ministry of Science, Education and Sport, and the Croatian Science Foundation; the Research and Innovation Foundation, Cyprus; the Secretariat for Higher Education, Science, Technology and Innovation, Ecuador; the Estonian Research Council via PRG780, PRG803, RVTT3 and the Ministry of Education and Research TK202; the Academy of Finland, Finnish Ministry of Education and Culture, and Helsinki Institute of Physics; the Institut National de Physique Nucl\'eaire et de Physique des Particules~/~CNRS, and Commissariat \`a l'\'Energie Atomique et aux \'Energies Alternatives~/~CEA, France; the Shota Rustaveli National Science Foundation, Georgia; the Bundesministerium f\"ur Bildung und Forschung, the Deutsche Forschungsgemeinschaft (DFG), under Germany's Excellence Strategy -- EXC 2121 ``Quantum Universe" -- 390833306, and under project number 400140256 - GRK2497, and Helmholtz-Gemeinschaft Deutscher Forschungszentren, Germany; the General Secretariat for Research and Innovation and the Hellenic Foundation for Research and Innovation (HFRI), Project Number 2288, Greece; the National Research, Development and Innovation Office (NKFIH), Hungary; the Department of Atomic Energy and the Department of Science and Technology, India; the Institute for Studies in Theoretical Physics and Mathematics, Iran; the Science Foundation, Ireland; the Istituto Nazionale di Fisica Nucleare, Italy; the Ministry of Science, ICT and Future Planning, and National Research Foundation (NRF), Republic of Korea; the Ministry of Education and Science of the Republic of Latvia; the Research Council of Lithuania, agreement No.\ VS-19 (LMTLT); the Ministry of Education, and University of Malaya (Malaysia); the Ministry of Science of Montenegro; the Mexican Funding Agencies (BUAP, CINVESTAV, CONACYT, LNS, SEP, and UASLP-FAI); the Ministry of Business, Innovation and Employment, New Zealand; the Pakistan Atomic Energy Commission; the Ministry of Education and Science and the National Science Center, Poland; the Funda\c{c}\~ao para a Ci\^encia e a Tecnologia, grants CERN/FIS-PAR/0025/2019 and CERN/FIS-INS/0032/2019, Portugal; the Ministry of Education, Science and Technological Development of Serbia; MCIN/AEI/10.13039/501100011033, ERDF ``a way of making Europe", Programa Estatal de Fomento de la Investigaci{\'o}n Cient{\'i}fica y T{\'e}cnica de Excelencia Mar\'{\i}a de Maeztu, grant MDM-2017-0765, projects PID2020-113705RB, PID2020-113304RB, PID2020-116262RB and PID2020-113341RB-I00, and Plan de Ciencia, Tecnolog{\'i}a e Innovaci{\'o}n de Asturias, Spain; the Ministry of Science, Technology and Research, Sri Lanka; the Swiss Funding Agencies (ETH Board, ETH Zurich, PSI, SNF, UniZH, Canton Zurich, and SER); the Ministry of Science and Technology, Taipei; the Ministry of Higher Education, Science, Research and Innovation, and the National Science and Technology Development Agency of Thailand; the Scientific and Technical Research Council of Turkey, and Turkish Energy, Nuclear and Mineral Research Agency; the National Academy of Sciences of Ukraine; the Science and Technology Facilities Council, UK; the US Department of Energy, and the US National Science Foundation.

Individuals have received support from the Marie-Curie program and the European Research Council and Horizon 2020 Grant, contract Nos.\ 675440, 724704, 752730, 758316, 765710, 824093, 101115353,101002207, and COST Action CA16108 (European Union) the Leventis Foundation; the Alfred P.\ Sloan Foundation; the Alexander von Humboldt Foundation; the Belgian Federal Science Policy Office; the Fonds pour la Formation \`a la Recherche dans l'Industrie et dans l'Agriculture (FRIA-Belgium); the Agentschap voor Innovatie door Wetenschap en Technologie (IWT-Belgium); the F.R.S.-FNRS and FWO (Belgium) under the ``Excellence of Science -- EOS" -- be.h project n.\ 30820817; the Beijing Municipal Science \& Technology Commission, No. Z191100007219010; the Ministry of Education, Youth and Sports (MEYS) of the Czech Republic; the Shota Rustaveli National Science Foundation, grant FR-22-985 (Georgia); the Hungarian Academy of Sciences, the New National Excellence Program - \'UNKP, the NKFIH research grants K 131991, K 133046, K 138136, K 143460, K 143477, K 146913, K 146914, K 147048, 2020-2.2.1-ED-2021-00181, and TKP2021-NKTA-64 (Hungary); the Council of Scientific and Industrial Research, India; ICSC -- National Research Center for High Performance Computing, Big Data and Quantum Computing and FAIR -- Future Artificial Intelligence Research, funded by the EU NexGeneration program (Italy); the Latvian Council of Science; the Ministry of Education and Science, project no. 2022/WK/14, and the National Science Center, contracts Opus 2021/41/B/ST2/01369 and 2021/43/B/ST2/01552 (Poland); the Funda\c{c}\~ao para a Ci\^encia e a Tecnologia, grant FCT CEECIND/01334/2018; the National Priorities Research Program by Qatar National Research Fund; the Programa Estatal de Fomento de la Investigaci{\'o}n Cient{\'i}fica y T{\'e}cnica de Excelencia Mar\'{\i}a de Maeztu, grant MDM-2017-0765 and projects PID2020-113705RB, PID2020-113304RB, PID2020-116262RB and PID2020-113341RB-I00, and Programa Severo Ochoa del Principado de Asturias (Spain); the Chulalongkorn Academic into Its 2nd Century Project Advancement Project, and the National Science, Research and Innovation Fund via the Program Management Unit for Human Resources \& Institutional Development, Research and Innovation, grant B37G660013 (Thailand); the Kavli Foundation; the Nvidia Corporation; the SuperMicro Corporation; the Welch Foundation, contract C-1845; and the Weston Havens Foundation (USA).    
\end{acknowledgments}

\vfill\newpage

\appendix
\section{Glossary}
\label{app:Glossary}

\begin{longtable}{ll}
AA & Ion-ion collision system \\
AGS & Alternating Gradient Synchrotron \\
AJ & Dijet asymmetry \\
ALEPH & Apparatus for LEP PHysics \\
ALICE & A Large Ion Collider Experiment \\
ALP & Axion Like Particles \\
ATLAS & A Toroidal LHC Apparatus \\
BDT & Boosted Decision Tree \\
BEC & Bose--Einstein Correlations \\
BFKL & Balitsky--Fadin--Kuraev--Lipatov \\
BNL & Brookhaven National Laboratory \\
BRAHMS & Broad RAnge Hadron Magnetic Spectrometers \\
BSC & Beam Scintillation Counter \\
BSM & Beyond the Standard Model \\
CASTOR & Centauro And STrange Object Research \\
CDF & Collider Detector at Fermilab \\
CEP & Central Exclusive Production \\
CERN & European Organization for Nuclear Research \\
CGC & Color Glass Condensate \\
CKF & Combinatorial Kalman filter \\
CL & Confidence Level \\
CM & Center of mass \\
CME & Chiral Magnetic Effect \\
CMS & Compact Muon Solenoid \\
CMW & Chiral Magnetic Wave \\
CNM & Cold Nuclear Matter \\
CPU & Central Processing Units \\
CS & Constituent Subtraction \\
CSC & Cathode Strip Chambers \\
CSE & Chiral Separation Effect \\
DAQ & Data Acquisition \\
DELPHI & DEtector with Lepton, Photon and Hadron Identification \\
DESY & Deutsches Elektronen-Synchrotron \\
DGLAP & Dokshitzer--Gribov--Lipatov--Altarelli--Parisi \\
DT & Drift Tubes \\
DY & Drell--Yan \\
ECAL & Electromagnetic Calorimeter \\
EM & Electromagnetic \\
EMC & European Muon Collaboration \\
EMD & Electromagnetic Dissociation \\
ESE & Event Shape Engineering \\
ET & Transverse Energy \\
ETA & Pseudorapidity \\
EW & Electroweak \\
HCAL & Hadron Hadronic Calorimeter \\
HERA & Hadron-Electron Ring Accelerator \\
HF & Forward Hadron Calorimeter \\
HI & Heavy Ions \\
HLT & High-Level Trigger \\
HQ & Heavy Quark \\
ID & Identification \\
IP & Impact Parameter \\
ISR & Initial-State Radiation \\
KET & Transverse Kinetic Energy \\
LEP & Large Electron--Positron Collider \\
LHC & Large Hadron Collider \\
LO & Leading order \\
LPM & Landau--Pomeranchuck--Migdal \\
MB & Minimum Bias \\
MC & Monte Carlo \\
NAA & Corresponding yield of the particle species of interest in AA collisions \\
NCQ & Number Constituent Scaling \\
NLO & Next to Leading Order \\
NN & Nucleon Nucleon \\
NNLL & Next-to-Next-to-Leading Logarithmic \\
NNLO & Next-to-Next-to-Leading Order \\
NPDF & Nuclear PDFs \\
NSD & Non-Single-Diffractive \\
OPAL & Omni-Purpose Apparatus for LEP \\
OS & Opposite Sign \\
PDFs & Parton Distribution Functions \\
PF & Particle Flow \\
PHENIX & Pioneering High Energy Nuclear Interaction eXperiment \\
PHOBOS & One of the initial suite of four detectors installed at RHIC \\
POWHEG & Positive Weight Hardest Emission Generator \\
PP & Proton-proton collision system \\
PT & Transverse Momentum \\
PU & Pileup \\
PV & Primary Vertex \\
PYTHIA & Event generator \\
QCD & Quantum Chromodynamics \\
QED & Quantum Electrodynamics \\
QGP & Quark-Gluon Plasma \\
RAA & Nuclear Modification Factor \\
RHIC & Relativistic Heavy Ion Collider \\
RMS & Root-mean-squared \\
RPC & Resistive-Plate Chambers \\
SM & Standard Model \\
SPS & Super Proton Synchrotron \\
SR & Signal Region \\
SS & Same Sign \\
STAR & Solenoidal Tracker at RHIC \\
STAT & Statistical Uncertainty \\
SYST & Systematic Uncertainty \\
TAA & Nuclear Overlap Function \\
TMVA & Toolkit for Multivariate Data Analysis \\
UE & Underlying Event \\
UPC & Ultraperipheral Collisions \\
VM & Vector Meson \\
ZDC & Zero Degree Calorimeter \\
ZEUS & Particle Detector at HERA \\
\end{longtable}

\clearpage
\bibliography{auto_generated}
\cleardoublepage \section{The CMS Collaboration \label{app:collab}}\begin{sloppypar}\hyphenpenalty=5000\widowpenalty=500\clubpenalty=5000
\cmsinstitute{Yerevan Physics Institute, Yerevan, Armenia}
{\tolerance=6000
A.~Hayrapetyan, A.~Tumasyan\cmsAuthorMark{1}\cmsorcid{0009-0000-0684-6742}
\par}
\cmsinstitute{Institut f\"{u}r Hochenergiephysik, Vienna, Austria}
{\tolerance=6000
W.~Adam\cmsorcid{0000-0001-9099-4341}, J.W.~Andrejkovic, T.~Bergauer\cmsorcid{0000-0002-5786-0293}, S.~Chatterjee\cmsorcid{0000-0003-2660-0349}, K.~Damanakis\cmsorcid{0000-0001-5389-2872}, M.~Dragicevic\cmsorcid{0000-0003-1967-6783}, P.S.~Hussain\cmsorcid{0000-0002-4825-5278}, M.~Jeitler\cmsAuthorMark{2}\cmsorcid{0000-0002-5141-9560}, N.~Krammer\cmsorcid{0000-0002-0548-0985}, A.~Li\cmsorcid{0000-0002-4547-116X}, D.~Liko\cmsorcid{0000-0002-3380-473X}, I.~Mikulec\cmsorcid{0000-0003-0385-2746}, J.~Schieck\cmsAuthorMark{2}\cmsorcid{0000-0002-1058-8093}, R.~Sch\"{o}fbeck\cmsorcid{0000-0002-2332-8784}, D.~Schwarz\cmsorcid{0000-0002-3821-7331}, M.~Sonawane\cmsorcid{0000-0003-0510-7010}, S.~Templ\cmsorcid{0000-0003-3137-5692}, W.~Waltenberger\cmsorcid{0000-0002-6215-7228}, C.-E.~Wulz\cmsAuthorMark{2}\cmsorcid{0000-0001-9226-5812}
% ...
\par}
\vskip\cmsinstskip
\dag:~Deceased\\
$^{1}$Also at Yerevan State University, Yerevan, Armenia\\
$^{2}$Also at TU Wien, Vienna, Austria\\
% ...
\end{sloppypar}
\end{document}